# DropMicroFluidAgents (DMFAs): Autonomous Droplet Microfluidic Research Framework Through Large Language Model Agents

Dinh-Nguyen Nguyen[1], Raymond Kai-Yu Tong[1], Ngoc-Duy Dinh[1*]

[1]Department of Biomedical Engineering, The Chinese University of Hong Kong, Shatin, N.T., Hong Kong, China

*Corresponding Author

Corresponding Email: ngocduydinh@cuhk.edu.hk

## Abstract

Large language models (LLMs) have gained significant attention in recent years due to their impressive capabilities across various tasks, from natural language understanding to generation. However, their widespread success in general domains has not translated seamlessly into specialised fields. Applying LLMs within specific domains requires substantial adaptation to account for the unique terminologies, nuances, and context-specific challenges inherent to those areas. Here, we introduce DropMicroFluidAgents (DMFAs), an advanced language-driven framework leveraging state-of-the-art pre-trained LLMs. DMFAs employs LLM agents to perform two key functions: (1) delivering focused guidance, answers, and suggestions specific to droplet microfluidics and (2) generating machine learning models to optimise and automate the design of droplet microfluidic devices, including the creation of code-based computer-aided design (CAD) scripts to enable rapid and precise design execution. To assess the accuracy of DMFAs in question-answering tasks, we compiled a dataset of questions with corresponding ground-truth answers and established an evaluation criterion. Experimental evaluations demonstrated that the integration of DMFAs with the LLAMA3.1 model yielded the *highest accuracy of 76.15%*, underscoring the significant performance enhancement provided by agent integration. This effect was particularly pronounced when DMFAs were paired with the GEMMA2 model, resulting in *a 34.47% improvement in accuracy* compared to the standalone GEMMA2 configuration. For evaluating the performance of DMFAs in design automation, we utilized an existing dataset on flow-focusing droplet microfluidics. The resulting machine learning model demonstrated *a coefficient of* *determination of approximately 0.96.* This study demonstrates the effective use of LLM agents in droplet microfluidics research as powerful tools for automating workflows, synthesising knowledge, optimising designs, and interacting with external systems. These capabilities enable their application across education and industrial support, driving greater efficiency in scientific discovery and innovation.

## 1. Introduction

LLMs are advanced artificial intelligence (AI) systems capable of understanding and generating human-like text by processing vast amounts of data[1]. LLMs have demonstrated broad applicability across various disciplines, with notable contributions in specialised areas such as chemistry[2–6], biology[7–11] biomedical research[12–15], materials science[16–22], and medicine[23–28] offering significant benefits in automating and enhancing research processes. Furthermore, LLMs have played a pivotal role in advancing scientific discovery and fostering technological innovation[29–32]. However, the inherent limitations of LLMs, such as hallucination, bias, and incomplete factual accuracy, necessitate careful oversight and validation of their outputs[33,34]. To overcome some of these limitations, Retrieval-Augmented Generation (RAG) frameworks combine LLMs with external knowledge retrieval systems, enabling more accurate and contextually relevant outputs[35]. RAG models retrieve pertinent information from structured databases or scientific repositories, ensuring the generated content aligns with verified knowledge. However, basic RAG implementations face scalability challenges, limited retrieval accuracy, and dependence on the quality of the external sources[36]. LLM agents are advanced AI systems that go beyond generating text by acting as autonomous agents capable of planning, reasoning, and executing tasks. Unlike traditional LLMs, which passively respond to inputs, agentic LLMs can interact with external tools, APIs, and databases to achieve specific goals. These agents operate autonomously, performing complex, multi-step tasks such as iterative querying, hypothesis testing, and experimental design[37–42]. These agents have found applications in autonomous scientific discovery[43] and medical research[44–46]. However, their deployment in droplet microfluidics remains underexplored, representing a significant opportunity for innovation.

Droplet microfluidics is a cutting-edge technology that leverages microchannel networks to manipulate discrete droplets as independent microreactors. This technology minimises sample consumption, reduces waste, and allows for high-throughput processing, making it indispensable for applications where scalability and precision are critical[47–50]. Droplet microfluidics, leveraging its foundational strengths, has driven innovations across various disciplines within the chemical and biological sciences, including advancements in next-generation sequencing [51–53], single-cell RNA sequencing[54,55], single cell secretion analysis[56,57], drug screening[58], droplet digital PCR[59], and liquid biopsies diagnostics[60]. However, designing droplet microfluidics devices is complex and often requires iterative trial-and-error processes[61–65]. To address the complexity of device design, machine learning (ML) has emerged as a promising tool for automating the optimisation of droplet-based microfluidic systems[66–70]. However, implementing ML-based design automation requires expertise in both microfluidics and machine learning to optimise algorithms and interpret outcomes effectively. Additionally, considerable time is needed to comprehend and utilise the extensive body of prior knowledge in droplet microfluidics and machine learning documented in the scientific literature, which is critical during the initial stages of experimentation.

In this study, we introduce DropMicroFluidAgents (DMFAs), a novel multi-agent-based framework designed to perform diverse tasks and make decisions autonomously, guided by its programming and the data it analyses. DMFAs comprises two components, the Scientific Mentor and the Automation Designer, both constructed using cutting-edge LLMs, as shown in Fig. 1. The Scientific Mentor delivers customised guidance and recommendations aimed at enhancing theoretical understanding and reducing the trial-and-error costs associated with experimental workflows in droplet microfluidics. Notably, the Automation Designer is capable of developing a machine learning model to optimise and automate the design of droplet microfluidic devices, as well as providing code-based CAD scripts for drawing creation. This study emphasises the transformative potential of LLMs in advancing droplet microfluidics development, setting the stage for the rapid acceleration of scientific discovery and technological innovation through AI.

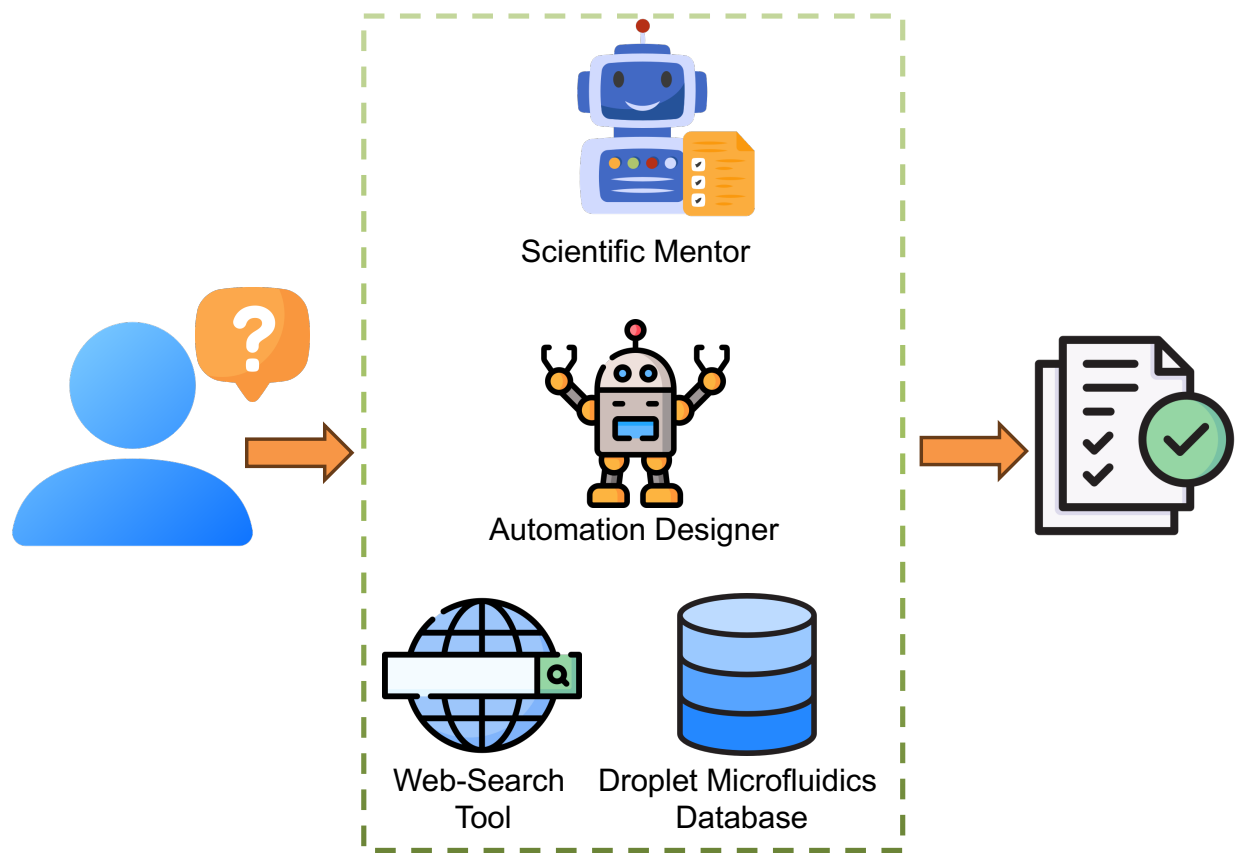

Fig. 1 The overall framework of DMFAs. The Scientific Mentor is tasked with providing guidance to users by leveraging LLM agents and the specialized knowledge database, supplemented by the web search tool when required. Meanwhile, the Automation Designer, also leveraging LLM agents, facilitates the automation of the design process for users.

## 2. Methods

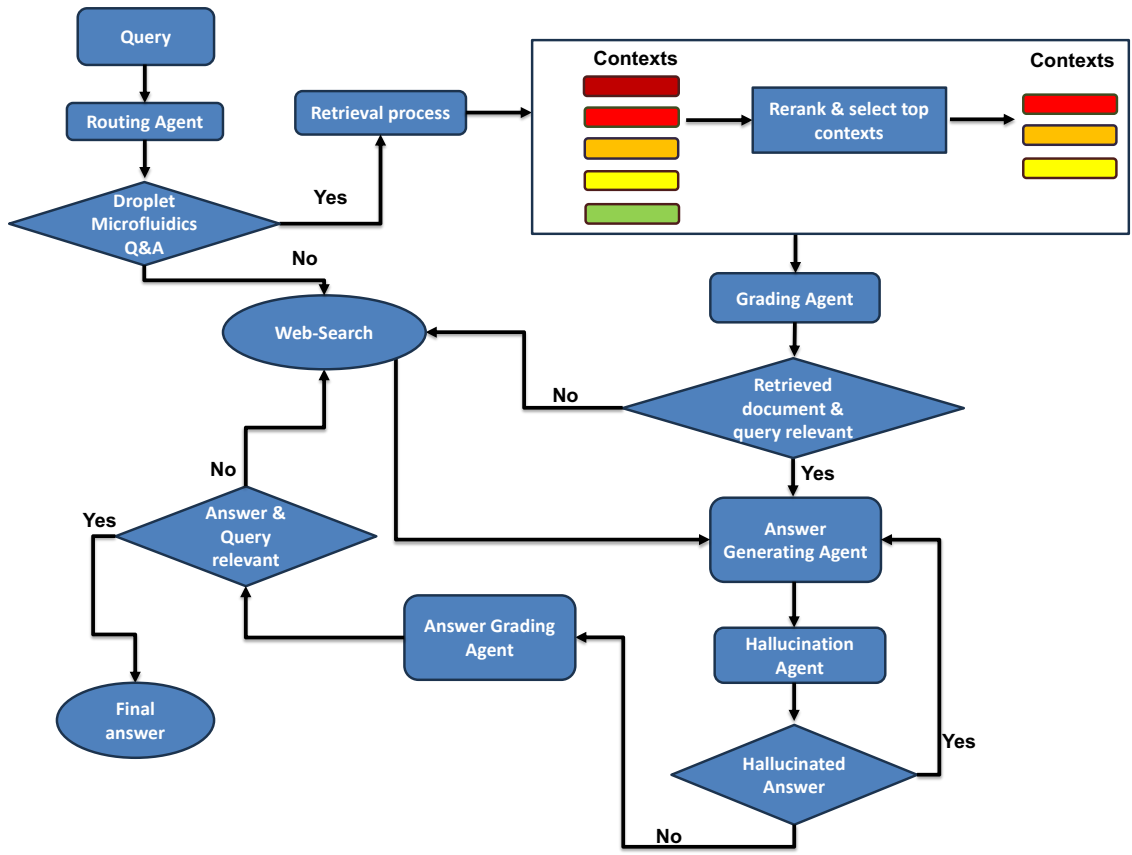

(a) Scientific Mentor

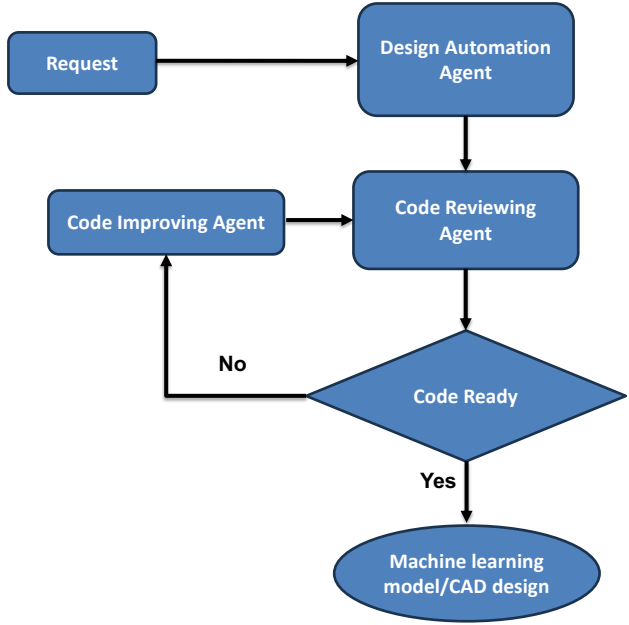

(b) Automation Designer

Fig. 2 Details of DMFAs. Fig. 2a illustrates how the Scientific Mentor delivers the final answer to the user by utilizing multiple LLM agents to minimize the limitations of LLMs and improve the accuracy of the response. Fig. 2b illustrates how the Automation Designer provides the machine learning model and CAD design to the user by utilizing three LLM agents to ensure an error-free result.

## 2.1 Large Language Model Selection

The selection of LLMs is a critical decision in the design and optimisation of intelligent agents. For this study, LLAMA 3.1 (8B parameters)[71], MISTRAL (7B parameters)[72], and GEMMA2 (9B parameters)[73] were selected for their distinct strengths, each tailored to address the specific requirements of the task. LLAMA 3.1 is renowned for its general-purpose capabilities and flexibility, making it particularly effective for a wide range of query types. It provides a solid foundation for handling both structured and unstructured inputs, adapting well to various contexts. The strength of LLAMA 3.1 lies in its ability to generate coherent responses across diverse topics. MISTRAL, on the other hand, excels in its ability to perform sophisticated contextual reasoning. Its architecture is specifically optimised to handle complex queries that require deep comprehension and multi-step inference. The strength of MISTRAL lies in its capacity to maintain context across extended conversations or intricate problem-solving tasks, which makes it particularly valuable for applications requiring high levels of logical consistency and nuanced understanding. The advanced attention mechanisms of this model allow it to consider multiple factors in parallel, offering a higher degree of precision when managing intricate relationships between data points or abstract concepts. GEMMA2 was selected for its specialised proficiency in domain-specific applications, particularly within technical fields such as microfluidics and machine learning. Its design emphasises high efficiency and accuracy when dealing with structured knowledge bases, which makes it ideal for tasks that involve retrieving and synthesizing specialized information. The ability of GEMMA2 to process and filter relevant domain-specific content quickly allows it to provide highly accurate, context-sensitive answers within narrow scopes, ensuring that the system can deliver expert-level insights in specialized areas. Building agents on these models involves not only leveraging their inherent strengths but also addressing critical factors such as scalability, latency, and fine-tuning for task-specific requirements. The integration of these LLMs ensures a balanced approach to accuracy, efficiency, and contextual relevance. Furthermore, their combined deployment allows for redundancy and cross-validation of outputs, enhancing the overall reliability and robustness of the system. This deliberate selection and integration underscore the importance of aligning model capabilities with the nuanced demands of the application domain.

## 2.2 Prompt Design

Prompt design is a fundamental aspect of optimising the performance of LLMs, as it significantly impacts the quality, relevance, and precision of their generated outputs. A well-crafted prompt not only provides clear instructions but also sets the context in which the model should operate, ensuring that the response aligns with the expectations of a user [74–76] The prompt must define the task in a way that minimises ambiguity, especially for complex or domain-specific queries. Effective prompts often include specific cues, such as desired output formats or constraints, that guide the model's reasoning and response generation. For example, in technical fields, the prompt might include terminology and context that ensure the model can draw from relevant domain-specific knowledge rather than providing generalised or unrelated information. This level of clarity is crucial for ensuring that the LLM produces accurate and appropriate results, particularly when the task involves intricate subject matter or multi-step reasoning. In this context, the effectiveness of LLM agents in performing specific tasks is intrinsically tied to the precision of the prompts. A precise prompt enables the model to focus on the most relevant information, effectively guiding it through the process of generating the correct answer. This is especially important when the agent is tasked with complex problem-solving or when a high degree of contextual understanding is required. Inaccurately phrased or vague prompts can lead to a range of issues, including irrelevant answers, incomplete responses, or even misinterpretations of the query. For example, an ambiguous prompt may cause the model to generate a broad, generic response, rather than the highly specific answer necessary for technical or specialized tasks.

Our methodology employed carefully fine-tuned prompts to guide LLM agents in navigating the complex demands of question-answering and design automation challenges within the field of droplet microfluidics. These specialized prompts were designed to address the specific intricacies of microfluidic systems, enabling the LLM agents to generate precise, contextually relevant responses. By refining the prompts to align with the unique requirements of both question formulation and the automation of design processes, we ensured that the models could effectively interpret and respond to domain-specific queries. This approach not only improved the accuracy of the generated answers but also enhanced the efficiency of design automation tasks, providing actionable insights tailored to the specialised needs of microfluidics research. A detailed explanation of the DMFAs prompt design is provided in Table 1 and Table 2.

| Agent Name | Prompt's content |
|---|---|
| Routing Agent | You are an expert at routing a user question to a vectorstore or web search. Use the 'vectorstore' for questions on droplet-based microfluidics. You do not need to be stringent with the keywords in the question related to this topic. Otherwise, use web-search. Give a choice 'web-search' or 'vectorstore'. Return a JSON with a single key 'datasource' and no preamble or explanation. Question to route: {question} |
| Grading Agent | You are a grader assessing the relevance of a retrieved document to a user question. Grade it as relevant if the document contains keywords related to the user question. Provide a simple binary "yes" or "no" score as JSON with the single key "score" and no additional text.\n Here is the retrieved document: {document}. Here is the user question: {question}. {{"score": "yes" or "no"}} |
| Answer Generating Agent | You are an assistant for question-answering tasks. Use the following pieces of retrieved context to provide a detailed, concise, and critical response for the question.<br>Context: {Context}<br>Question: {Question} |
| Answer Grading Agent | You are a grader assessing whether an answer is useful to resolve a question. Give a binary score 'yes' or 'no' to indicate whether the answer is useful to resolve a question. Provide the binary score as a JSON object with a single key 'score' (without quotes around the key) and no preamble or explanation. The response should strictly follow the format: {{"score": "yes"}} or {{"score": "no"}}. Here is the answer: {generation}. Here is the question: {question} |
| Hallucination Agent | You are a grader, assessing if the answer is grounded in the provided facts. Return one of the following as a JSON response: {{"score": "yes"}} - {{"score": "no"}} Do not write anything else. Facts: {documents}. Answer: {generation} |

Table 1. Prompt design for the Scientific Mentor

| Agent name | Prompt's content |
|---|---|
| Design Automation Agent | For 'AutoLisp' case: You are an expert in writing code in AutoLisp. Generate the proper lines of code based on request. Only output the code and nothing else. Here is the request: {request}<br>For 'Python' case: You are an expert in writing code in Python. Generate the proper lines of code based on request. Only output the code and nothing else. Here is the request {request} |
| Code Reviewing Agent | For 'AutoLisp' case: You are a code reviewer specialized in AutoLisp. You need to review the given code and potential bugs and point out issues as bullet list. Only output the bullet list and nothing else Here is the code: {code}<br>For 'Python' case: You are a code reviewer specialized in 'Python'. You need to review the given code following PEP8 guidelines and potential bugs and point out issues as bullet list. Only output the bullet list and nothing else. Here is the code: {code} |
| Code Improving Agent | For 'AutoLisp' case: You are an expert in improving AutoLisp code. Improve the given code given the following guidelines. Only output the improved code and nothing else. Here are the guidelines: {guidelines}. Here is the given code: {code}<br>For 'Python' case: You are an expert in improving Python code. Improve the given code given the following guidelines. Only output the improved code and nothing else. Here are the guidelines: {guidelines}. Here is the given code: {code} |

Table 2. Prompt design for the Automation Designer

## 2.3 Scientific Mentor Construction

### 2.3.1 Flowchart illustrating the operational principle of the Scientific Mentor

The Scientific Mentor represents a comprehensive and adaptive pipeline for a robust question-answering system, integrating RAG with mechanisms for relevance grading, hallucination detection, and iterative refinement, as illustrated in Fig. 2a. The system initiates with a routing agent that categorizes the question of a user as either pertaining to the microfluidic knowledge base or necessitating external context. For questions linked to the microfluidic knowledge base, the retrieval process employs methods such as embedding similarity or semantic search to identify the most relevant documents from the base. These retrieved documents are subsequently reranked to select the top three. These selected documents are then evaluated by a grading agent, which assesses their relevance based on their alignment with the query. If irrelevant documents are identified, the pipeline redirects to the web-search tool for broader context or supplementary data, ensuring comprehensive coverage. The answer-generating agent is tasked with synthesizing responses based on the retrieved and evaluated documents. A critical safeguard is the hallucination agent, where the generated answer is scrutinized for unsupported or fabricated claims, leveraging fact-checking techniques. If hallucinations are detected, the system iterates by re-fetching or supplementing data, potentially from the answer-generating agent, ensuring the refinement of the response. The final step involves answer validation, where the answer grading agent verifies if the generated response adequately addresses the original question. If the response is unsatisfactory, the process loops back to the web search to gather additional information to refine the answer further. This architecture balances internal knowledge utilisation with external search capabilities, emphasising reliability and adaptability at the cost of increased computational overhead and potential delays.

The Scientific Mentor is implemented sequentially using LLAMA 3.1, MISTRAL, and GEMMA2, represented as LLAMA-based Scientific Mentor, MISTRAL-based Scientific Mentor, and GEMMA-based Scientific Mentor, respectively. The accuracy of the LLMs-based Scientific Mentor is compared to the accuracy of the standalone models, LLAMA 3.1, MISTRAL, and GEMMA2, in the question-answering (QA) task. Their performance is subsequently evaluated in the 'Results and Discussion' section.

## 2.3.2 Generation of droplet microfluidics database and ground truth question-answer set

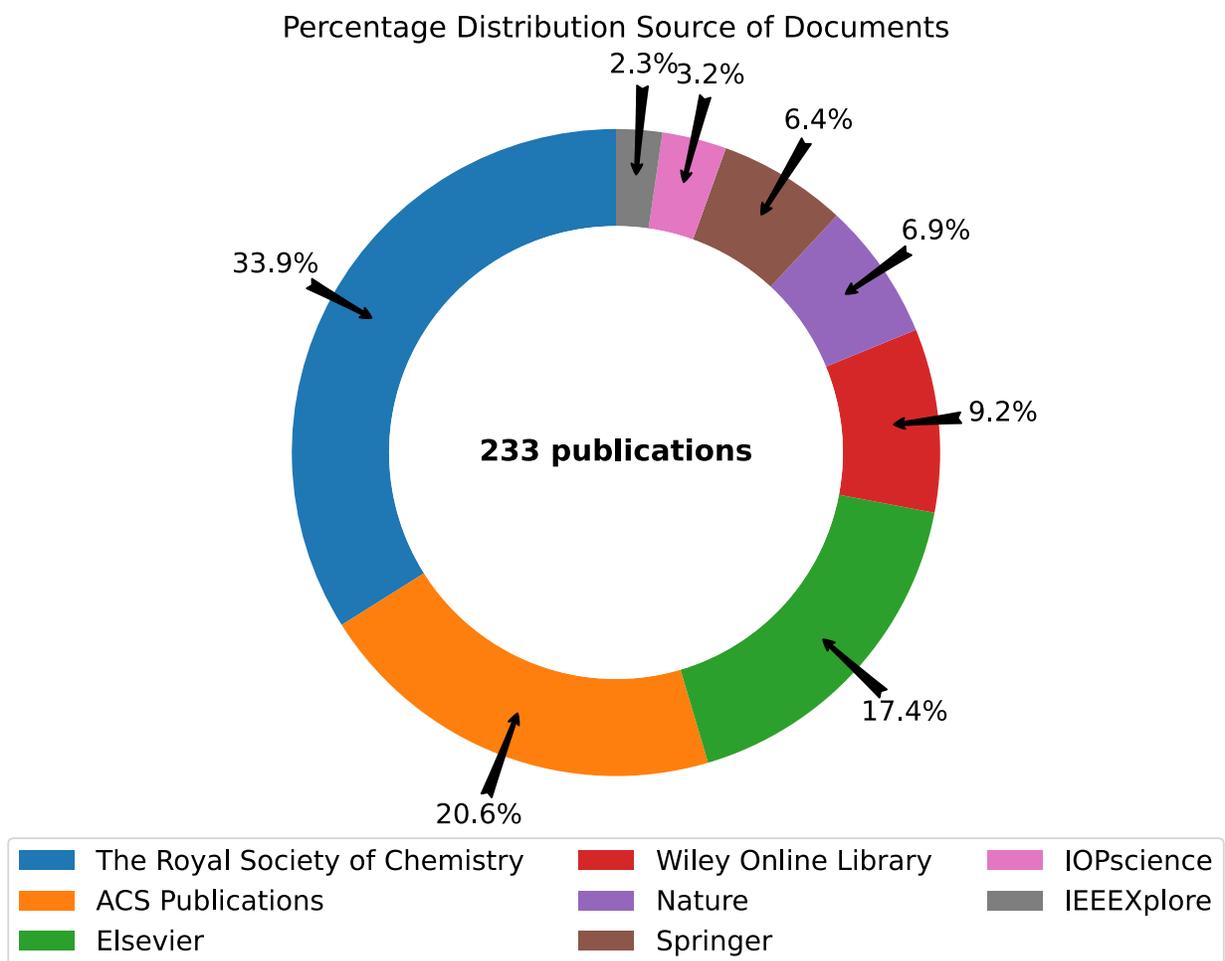

(a)

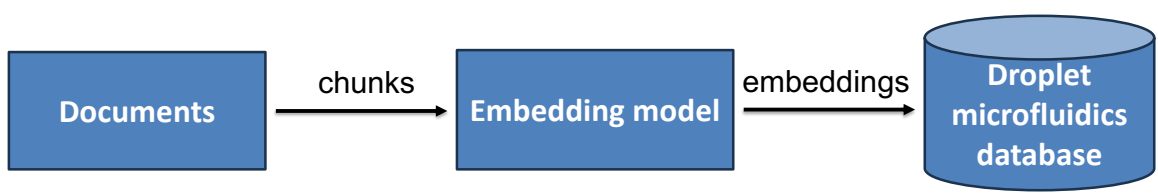

(b)

Fig. 3 (a) Percentage distribution source of documents (b) Flowchart of the droplet microfluidics database

LLMs, as large general-purpose models trained on extensive text corpora, can produce inaccurate or nonspecific responses to highly specialised, domain-specific queries. However, incorporating domain-specific knowledge bases and supplementary tools can enhance their accuracy and enable them to provide more relevant and precise guidance[35]. To assess the efficacy of the Scientific Mentor in addressing question-answering challenges within droplet microfluidics, we constructed a droplet microfluidics database and a ground truth question-answer dataset based on 233 scientific papers sourced from a range of well-known publishers, as shown in Fig. 3(a) (ESI, SI1). The database has not been extensively expanded because our primary aim is to illustrate the operational effectiveness of the proposed framework rather than to create an exhaustive repository. However, the framework is designed to be extensible, allowing end-users to enrich the database with tailored data sources such as scientific publications, textbooks, and experimental reports. This flexibility ensures that the framework can be adapted to specific domains and user needs. The ground truth question-answer dataset consists of 61 question-answer pairs encompassing a broad spectrum of topics related to microfluidics. These include fundamental concepts of microfluidics, principles of microfluidic device design, fabrication techniques, material considerations, applications in biology and medicine, chemical and material science applications, and prospects for the future of droplet microfluidics (ESI, S2).

The construction of the droplet microfluidics database adheres to the process outlined in the flowchart presented in Fig. 3. This flowchart illustrates a pipeline for processing and storing text data extracted from documents, such as PDF files, enabling efficient semantic search and retrieval using the database. The process begins by dividing the content of the documents into manageable chunks, ensuring the text is appropriately segmented to preserve context and prevent information loss. These chunks are passed to an embedding model, a neural network typically trained on large datasets to convert textual data into high-dimensional embeddings, numerical representations that encode semantic meaning. The embeddings are then stored in the database for similarity search, enabling fast and accurate retrieval based on the semantic relationships between queries and stored data.

### 2.3.3 Evaluation Method

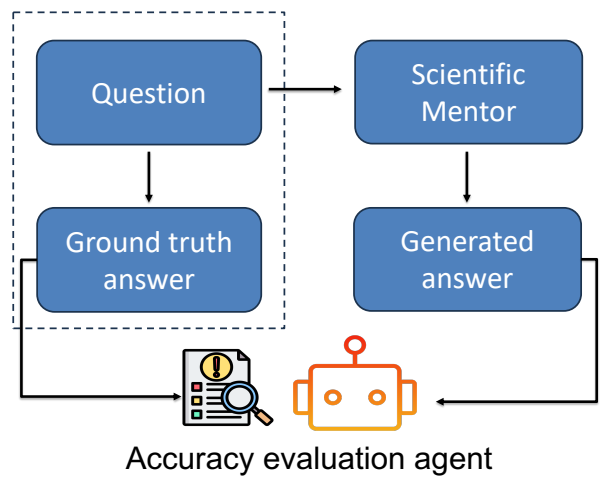

Fig. 4 Accuracy evaluation agent

The flowchart represents the workflow of a system designed to evaluate the accuracy of the Scientific Mentor by comparing its output against a predetermined ground truth, as shown in Fig. 4. The process begins with a question, which is directed to two components: the ground truth answer source and the Scientific Mentor. The ground truth answer represents the correct or expected response to the question, while the Scientific Mentor generates an automated response. Both the ground truth answer and the generated answer produced by the Scientific Mentor are then forwarded to an accuracy evaluation agent based on LLAMA 3.1, which serves as a comparator. This component systematically analyses the generated answer in relation to the ground truth answer to measure the accuracy performance of the Scientific Mentor using a 0–100 scoring scale, as shown in Table 3. The criteria emphasise factors such as relevance, clarity, coherence, conciseness, and depth of understanding in addressing the given question. Lower scores (0–25) indicate responses that are irrelevant, nonsensical, or poorly structured, reflecting a lack of effort or focus on the question. Mid-range scores (26–55) represent partially correct responses with varying degrees of inaccuracies, verbosity, or insufficient depth, highlighting areas for improvement in precision and alignment with the query. Higher scores (56–85) denote responses that are largely relevant, clear, and accurate, albeit with minor issues such as unnecessary details or slight omissions. The top score range (86–100) reflects excellence, with responses being comprehensive, highly precise, and virtually flawless, showcasing exceptional clarity and direct alignment with the question. This tiered rubric provides a robust foundation for systematically evaluating and improving the performance of the Scientific Mentor in generating high-quality, contextually appropriate answers. Additionally, several supplementary metrics are used to evaluate the performance of the Scientific Mentor, including Bilingual Evaluation Understudy (BLEU), Recall-Oriented Understudy for Gisting Evaluation (ROUGE), recall, F1 score, and Metric for Evaluation of Translation with Explicit Ordering (METEOR), all of which are detailed in Table 4.

| Accuracy evaluation agent's criteria |
|---|
| # Score 0-5: The response is irrelevant, nonsensical, or incoherent; no effort to address the question. |
| # Score 6-10: An attempt is made, but the response is entirely unrelated or meaningless. |
| # Score 11-15: Contains vague or random terms but lacks clarity, relevance, and logic. |
| # Score 16-20: Slight relevance, but the response is mostly incorrect, unclear, and fails to address the question. |
| # Score 21-25: Marginal effort to address the question, but lacks clarity and coherence; mostly irrelevant. |
| # Score 26-30: Some fragments of relevance, but the response is poorly structured and fails to convey useful information. |
| # Score 31-35: Displays minimal understanding, with major inaccuracies and a lack of focus on the question. |
| # Score 36-40: Partially aligned with the question but overly verbose, unclear, or dominated by factual errors. |
| # Score 41-45: Demonstrates basic understanding but lacks conciseness, with significant errors or irrelevant details. |
| # Score 46-50: Covers core aspects but is imprecise, verbose, or unclear; lacks depth or includes notable inaccuracies. |
| # Score 51-55: Partially correct, with omissions or minor inaccuracies; somewhat clear and relevant but not concise. |
| # Score 56-60: Mostly relevant and clear, but lacks focus or includes unnecessary details; broadly acceptable. |
| # Score 61-65: Addresses the question clearly and concisely, with minor inaccuracies or slight verbosity. |
| # Score 66-70: A solid, clear, and mostly concise response that aligns with the question and ground truth. |
| # Score 71-75: Comprehensive, clear, and precise, with only minor omissions or redundant elements. |
| # Score 76-80: Nearly flawless response; highly clear, concise, and relevant, with only slight room for improvement. |
| # Score 81-85: A thorough and precise response; clear, concise, and directly addresses the question with full relevance. |

# Score 86-90: Excellent, highly clear, and succinct answer; covers all aspects comprehensively and precisely.
# Score 91-95: Virtually flawless; exceptional clarity, conciseness, and relevance, with added depth where appropriate.
# Score 96-100: Perfect response; unparalleled clarity, precision, and focus on the question, with no room for improvement.

Table 3. Accuracy evaluation agent's criteria

| Metric | Purpose |
|--------|---------|
| **BLEU** | Measures precision of n-gram overlaps between predictions and references |
| **ROUGE-1** | Measures overlap of unigrams (single words). |
| **ROUGE-2** | Measures overlap of bigrams (two-word sequences). |
| **ROUGE-L** | Measures the longest common subsequence, capturing fluency and coherence. |
| **Recall** | Quantify the proportion of relevant instances that have been successfully retrieved, highlighting the completeness of the retrieval process |
| **METEOR** | Takes into account synonym matches and stemming, providing a more flexible approach |
| **F1 score** | The harmonic mean of precision and recall at the token level between the prediction and ground truth. |

Table 4. Additional metrics for question-answering evaluation of the Scientific Mentor

## 2.4 Autonomous Designer Construction

### 2.4.1 Flowchart illustrating the operational principle of the Autonomous Designer

The flowchart depicts an agent-driven iterative workflow for the Autonomous Designer in droplet microfluidics, with autonomous agents managing each stage of the process, as illustrated in Fig. 2b. The workflow begins with the specific requirement that necessitates code creation. This input is passed to the design automation agent, which produces an initial version of the code based on predefined algorithms, templates, or learned patterns. The generated code is then evaluated by the code reviewing agent, which inspects the code for logical correctness, adherence to coding standards, and functional compliance. The process progresses to the code-ready stage, where it is assessed to ensure the code complies with all defined quality and functional requirements. If the code is deemed ready, the machine learning model or CAD design is produced. Otherwise, the code-improving agent intervenes to refine the code, addressing deficiencies through debugging, optimisation, or reworking based on feedback from the reviewing agent. This cycle continues iteratively until the code passes the readiness evaluation. The agent-driven approach ensures automation, consistency, and efficiency across all workflow stages. This design highlights the potential of agent-based systems to streamline iterative development while maintaining a focus on continuous improvement and quality assurance. LLAMA 3.1 was selected for implementing the Autonomous Designer. Its performance is evaluated in the 'Results and Discussion' section.

## 3. Results and discussion

### 3.1 Performance of the Scientific Mentor

An illustrative example involving the evaluation of two questions demonstrates a systematic approach for assessing the quality of answers generated by the Scientific Mentor based on LLAMA and standalone LLAMA. This evaluation uses a predefined scoring framework, with the results detailed in Table 5, providing a comparative analysis of the performance of two models. The process begins with a specific question relating to microfluidics, which serves as the input for the system to generate a corresponding answer. This generated response is then compared against a ground truth answer, representing the expected or ideal response to the given question. The accuracy evaluation agent assesses the generated answer based on its established criteria, such as relevance, accuracy, clarity, and coherence, in relation to the ground truth. A numerical score is assigned to reflect the performance of the generated response. Following this, microfluidics expert provides additional ratings to ensure the robustness and reliability of the scoring process. This combined evaluation leverages both automated and human assessments to refine the performance of the Scientific Mentor and ensure alignment with high-quality standards.

| |
|---|
| **Question**: How to generate microdroplets? |
| **Ground-truth answer:** Microdroplet generation in microfluidics is achieved through two main methods: active and passive. Active methods utilize external forces like piezoelectric actuators to precisely control droplet size, formation rate, and spacing. These methods enable high uniformity in droplet generation, with deviations less than 0.3%. Passive methods rely on fluid properties and channel geometry, such as T-junctions or cross-flow configurations, to create droplets. These methods are influenced by factors like flow rate, viscosity, and interfacial tension, allowing for fine control over droplet size and uniformity. For example, PDMS-based microfluidic devices have been used for nanoparticle synthesis, where hydrodynamic coupling of nozzles generates droplet pairs for precise mixing. Both methods are valuable in applications |

like nanoparticle synthesis, offering high reproducibility and control over reaction conditions in biomedical and material science research.

**LLAMA-based Scientific Mentor's answer**: Based on the provided context, there are several ways to generate microdroplets in microfluidic devices. According to Document 3, which has a relevance score of 0.99394965, the most commonly used methods for microdroplet generation can be divided into two main categories: active and passive.

**Active Methods:**

* A piezoelectric actuator-based droplet generator is described in Document 2, which allows for independent control over droplet size, rate of formation, and distance between droplets. This device is able to generate microdroplets with high uniformity, deviating from the mean value by less than 0.3%.

**Passive Methods:**

* The context mentions that PDMS-based microfluidic devices are used to generate microdroplets for bulk synthesis of nanoparticles (Document 1). Specifically, a device designed by Frenz et al. enables the production of droplet pairs based on the hydrodynamic coupling of two separated nozzles.

* Another method mentioned is the use of thermally isolated zones in a droplet microreactor for the synthesis of TiO2 nanoparticles (Figure 4A).

Additionally, Document 3 mentions that other methods have been explored to generate microdroplets, including:

* Cross-flow and T-junction configurations

* Adjusting channel geometry, two-phase viscosity, flow rate, wettability, and interfacial tension These methods are used to achieve precise regulation over the size, shape, uniformity, and other factors of microdroplets.

**Accuracy evaluation agent's feedback**: The response provides a clear and detailed explanation of microdroplet generation methods, accurately categorizing them into active and passive techniques. It correctly describes the characteristics and applications of each method, including the use of piezoelectric actuators, PDMS-based devices, and thermally isolated zones. However, it does not fully address the question's request for a comprehensive overview, as it relies heavily on specific documents and figures mentioned in the provided context. SCORE: 81

**Microfluidics expert rating**: SCORE: 85

**Standalone LLAMA's answer**: Microdroplets can be generated using various methods, including:

1. Flow-focusing technique: This involves forcing a liquid through a narrow channel or nozzle, creating a stream of droplets.

2. Pneumatic atomization: Using compressed air to break up a liquid into small droplets.

3. Ultrasonic nebulization: Using high-frequency sound waves to create tiny droplets from a liquid.

4. Electrostatic atomization: Applying an electric charge to a liquid, causing it to break up into small droplets.

5. Centrifugal atomization: Spinning a liquid at high speed to create droplets.

6. Laser-induced breakdown spectroscopy (LIBS): Using a laser pulse to create microdroplets from a liquid.

7. Microfluidic devices: Specialized devices that use channels and valves to control the flow of liquids and generate microdroplets.

These methods can be used in various applications, such as drug delivery, chemical synthesis, and biological research.

**Accuracy evaluation agent's feedback**: The response provides a list of various methods to generate microdroplets but does not directly address the question's focus on the two main methods (active and passive) used in microfluidics. SCORE: 16

**Microfluidics expert rating**: SCORE: 10

**Question**: What key factors influence the design of flow-focusing droplet microfluidics optimized using machine learning?

**Ground-truth answer:** The design of flow-focusing droplet microfluidics, optimized using machine learning, depends on several critical factors. Channel geometry, including the shape and size, controls droplet generation by defining the break-off point where the dispersed phase meets the continuous phase. Flow rates of both phases affect droplet size and frequency, with higher continuous phase flow reducing droplet size. Viscosity influences flow behavior and shear forces, crucial for stable droplet formation. Shear forces between phases are controlled by adjusting channel geometry or flow rates, ensuring consistent droplet generation. Machine learning algorithms predict and optimize droplet diameter and generation rate by analyzing these design parameters, fluid properties, and large datasets, automating the design process for more efficient droplet production in life sciences applications.

**LLAMA-based Scientific Mentor's answer:** Based on the provided context, the key factors that influence the design of flow-focusing droplet microfluidics optimized using machine learning are:

1. **Channel geometry**: The shape and size of the channel in which the fluids are flowing play a critical role in controlling droplet generation.

2. **Flow rate**: The rates at which the dispersed and continuous phases are forced through the channel affect the droplet diameter and generation rate.

3. **Viscosity**: The viscosity of the fluids also influences the droplet generation process, as it affects the flow behaviour and shear forces acting on the dispersed phase.

4. **Shear forces**: The shearing action between the continuous and dispersed phases is crucial in flow-focusing designs and can be controlled by adjusting the channel geometry or flow rates. These factors are mentioned in Document 2 (page_content='49') as governing the generation of droplets in flow-focusing designs. Additionally, the use of machine learning algorithms to predict the performance of droplet generators based on these design parameters is discussed in Document 1 (page_content='22,23...').

**Accuracy evaluation agent's feedback**: The response provides a clear and concise list of factors influencing flow-focusing droplet microfluidics optimized using machine learning. However, it lacks depth and detail compared to the ground truth answer, particularly in explaining how these factors interact with each other and with machine learning algorithms. Additionally, some information is presented as a simple statement without further explanation or context. SCORE: 51

| |
|---|
| **Microfluidics expert rating**: SCORE: 55 |
| **Standalone LLAMA's answer**: The key factors that influence the design of flow-focusing droplet microfluidics optimized using machine learning include:<br>1. Channel geometry and dimensions<br>2. Flow rates and ratios of the two immiscible fluids<br>3. Surface tension and interfacial properties<br>4. Viscosity and density of the fluids<br>5. Droplet size and polydispersity<br>6. Shear stress and pressure drop across the microchannel<br>7. Operating temperature and pressure |
| **Accuracy evaluation agent's feedback**: The answer provides a list of factors that influence the design of flow-focusing droplet microfluidics optimized using machine learning but lacks explanation and context. It appears to be a summary of key parameters without elaboration on how they interact or are used in the optimization process. SCORE: 21 |
| **Microfluidics expert rating**: SCORE: 20 |

Table 5. An example demonstrates how the LLAMA-based Scientific Mentor answers a question and how the accuracy score for the response is determined.

Accuracy evaluated by the accuracy evaluation agent across LLMs and LLMs-based Scientific Mentor is compared for the QA task, as illustrated in Fig. 5a, with error bars representing the standard error. Among these, LLAMA-based Scientific Mentor achieves the highest accuracy at 76.15%, demonstrating a significant improvement over the standalone LLAMA model (53.64%). Similarly, GEMMA-based Scientific Mentor outperforms GEMMA by a large margin, achieving 71.03% compared to GEMMA's 36.56%, highlighting the transformative impact of agent integration on the GEMMA architecture. MISTRAL-based Scientific Mentor also surpasses MISTRAL, with respective accuracies of 72.00% and 58.11%, though the margin of improvement is less pronounced compared to GEMMA. A critical observation is the consistently substantial accuracy gains when transitioning from standalone models to their agent-augmented counterparts, emphasizing the importance of agent-based enhancements for boosting performance. The relatively small error bars for most models suggest reliable and consistent outcomes across evaluations. The details for performing the question-answering task with these six models are provided in the ESI, S3-8.

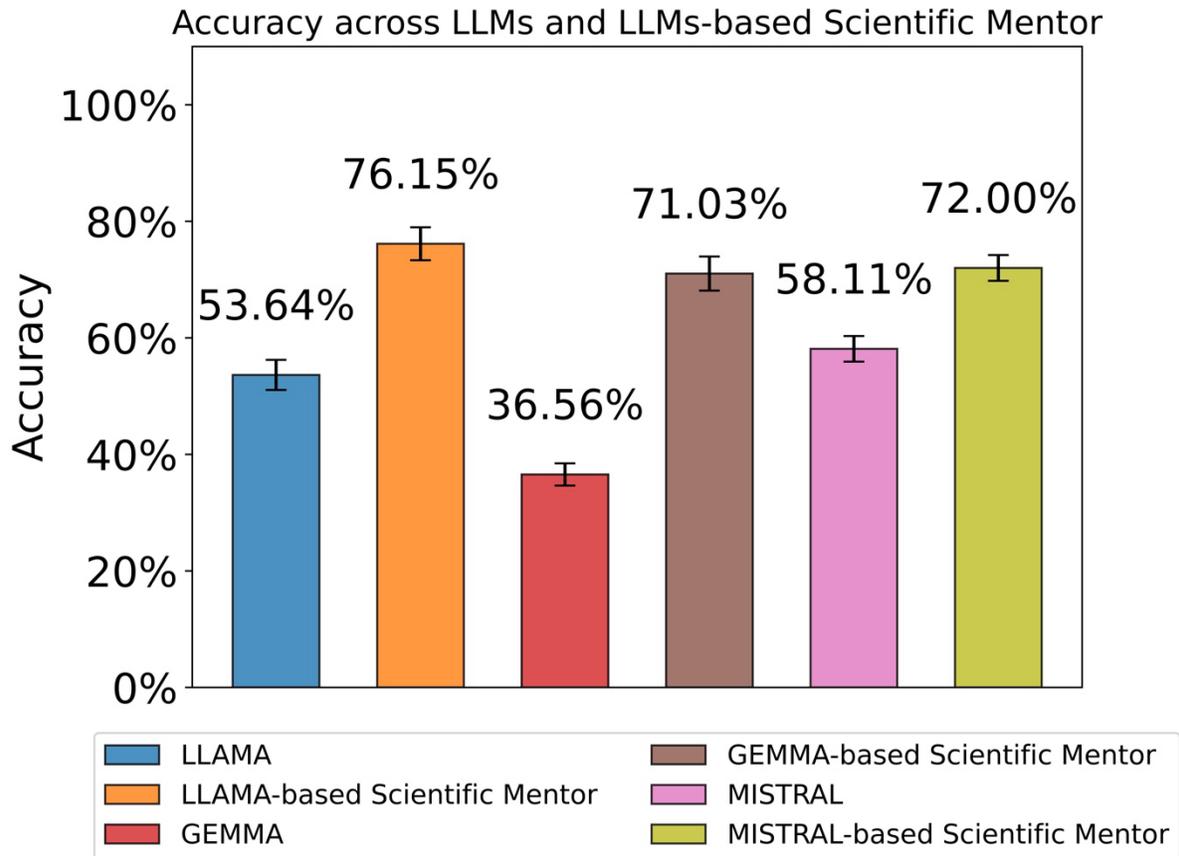

(a)

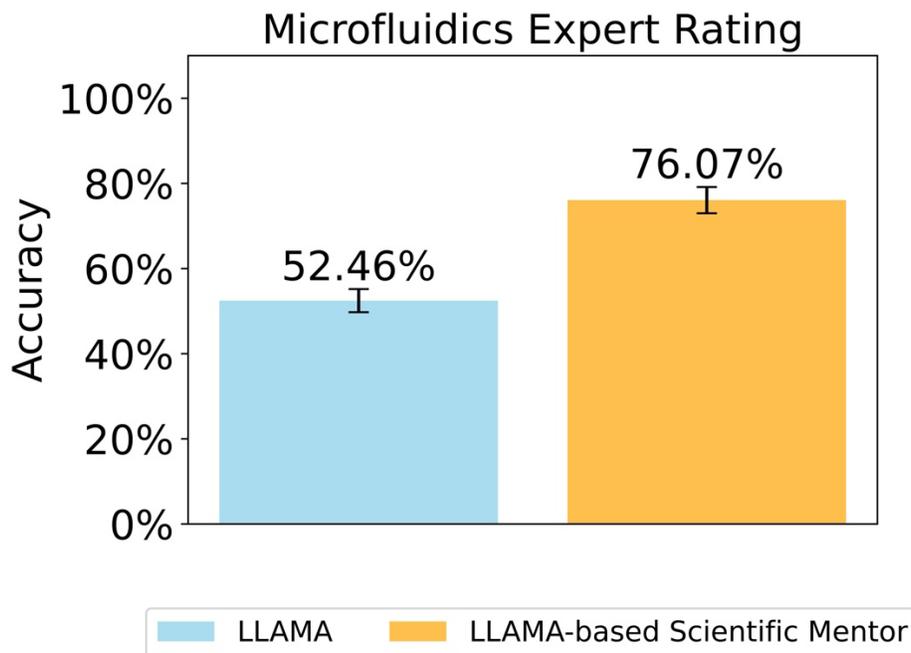

(b)

Fig. 5 (a) Accuracy evaluated by the accuracy evaluation agent across LLMs and LLMs-based Scientific Mentor for the QA task. (b) Microfluidics expert rating for QA of LLAMA-based Scientific Mentor and standalone LLAMA

The accuracy of LLAMA and LLAMA-based Scientific Mentor in the QA task, as rated by a microfluidics expert, is compared, with error bars indicating the standard error, as shown in Figure 5b. LLAMA-based Scientific Mentor achieves a markedly higher accuracy of 76.07%, compared to 52.46% for LLAMA, highlighting a substantial performance gap. The inclusion of error bars indicates that this difference is statistically significant, with minimal overlap, underscoring the reliability of the results. The relatively lower accuracy of LLAMA suggests limitations in its ability to meet the requirements of expert evaluation in this domain, potentially due to deficiencies in its model architecture or training process. In contrast, the superior performance of LLAMA-based Scientific Mentor likely stems from advanced design features and optimizations that address specific shortcomings in LLAMA, making it better suited to microfluidics research.

| Model | Metrics | | | | | | |
|---|---|---|---|---|---|---|---|
| | F1 score (%) | Recall (%) | ROUGE-1 | ROUGE-2 | ROUGE-L | BLEU | METEOR |
| **LLAMA** | 32.0827 ± 1.1605 | 30.6282 ± 1.5492 | 0.3268 ± 0.0117 | 0.0876 ± 0.0056 | 0.193 ± 0.0076 | 0.0376 ± 0.0054 | 0.2243 ± 0.0107 |
| **LLAMA-based Scientific Mentor** | 42.4434 ± 0.9869 | 63.2638 ± 1.4387 | 0.4306 ± 0.01 | 0.1818 ± 0.0082 | 0.263 ± 0.0087 | 0.0848 ± 0.0057 | 0.4328 ± 0.0117 |
| **GEMMA** | 16.3745 ± 1.0885 | 10.0868 ± 0.8011 | 0.1646 ± 0.0107 | 0.0504 ± 0.0054 | 0.119 ± 0.0072 | 0.0023 ± 0.0013 | 0.0861 ± 0.0056 |
| **GEMMA-based Scientific Mentor** | 43.6846 ± 1.08 | 58.5453 ± 1.2866 | 0.4442 ± 0.0109 | 0.1765 ± 0.0086 | 0.27 ± 0.0097 | 0.0806 ± 0.0063 | 0.4082 ± 0.0099 |
| **MISTRAL** | 34.0437 ± 0.866 | 33.8424 ± 1.0308 | 0.3468 ± 0.0088 | 0.0946 ± 0.0064 | 0.199 ± 0.0063 | 0.0445 ± 0.0052 | 0.2611 ± 0.0072 |
| **MISTRAL-based Scientific Mentor** | 39.247 ± 1.1005 | 51.8103 ± 1.4567 | 0.4003 ± 0.0112 | 0.1328 ± 0.0078 | 0.226 ± 0.0074 | 0.0733 ± 0.007 | 0.3663 ± 0.0102 |

Table 6. Evaluation of extra metrics across models

A comprehensive performance analysis of LLAMA, GEMMA, MISTRAL, and Scientific Mentor, which is implemented using LLAMA, GEMMA, and MISTRAL, is provided, across multiple performance metrics, including F1score, Recall, ROUGE-1, ROUGE-2, ROUGE-L, BLEU, and METEOR, each accompanied by standard errors, as shown in Table 6. The results highlight distinct trends across the models. LLAMA-based Scientific Mentor and GEMMA-based Scientific Mentor consistently outperform their respective non-agent counterparts in nearly all metrics, showcasing the effectiveness of agent integration in these architectures. Specifically, GEMMA-based Scientific Mentor achieves the highest F1 score ($43.68 \pm 1.08$) and recall ($58.55 \pm 1.29$), suggesting its strong ability to identify relevant elements in text and retrieve meaningful information. LLAMA-based Scientific Mentor also demonstrates robust performance, particularly in recall ($63.26 \pm 1.44$), indicating its efficiency in capturing relevant data. In contrast, the significantly lower scores of GEMMA (e.g., F1 score of $16.37 \pm 1.09$ and ROUGE-2 of $0.0504 \pm 0.0054$) reflect its limitations in standalone configurations, underscoring the necessity of its agent-enhanced counterpart for improved outcomes. Similarly, MISTRAL-based Scientific Mentor outperforms MISTRAL in most metrics, with notable improvements in ROUGE-2 and METEOR, though the gap is less pronounced than the GEMMA models. Interestingly, while LLAMA-based Scientific Mentor and GEMMA-based Scientific Mentor lead in different metrics, GEMMA-based Scientific Mentor appears more balanced, maintaining high scores across all dimensions of the evaluation. Furthermore, the standard errors suggest that GEMMA-based Scientific Mentor and LLAMA-based Scientific Mentor yield relatively consistent results, reflecting their reliability. However, MISTRAL-based Scientific Mentor, while generally superior to MISTRAL, does not reach the performance levels of LLAMA-based Scientific Mentor or GEMMA-based Scientific Mentor.

## 3.2 Performance of the Automation Designer based on LLAMA 3.1 in generating machine learning model

An example illustrates a structured workflow for generating high-quality machine learning code to predict microfluidic chip parameters, as shown in Fig. 6a. The process initiates with a query about creating the machine learning code, which is addressed by the design automation agent responsible for generating an initial version of the code. This preliminary code is subsequently evaluated by the code reviewing agent, whose role is to critically analyze and identify potential issues or areas for refinement in the generated script. Feedback from this review is passed to the code improving agent, which applies the necessary modifications to enhance the functionality of the code, accuracy, and efficiency. The final stage of the process involves outputting the optimized and polished code, ready for implementation or further use (ESI, S9).

An example of the predictive performance of the generated machine learning model for estimating droplet generation rates using an available flow-focusing microfluidics dataset[66] is illustrated in Fig. 6b. The scatter plot compares the predicted values with the corresponding actual values with the data points plotted against a diagonal line representing the ideal scenario where predicted values perfectly match actual values. The close clustering of most points along this diagonal suggests that the model accurately captures the underlying relationship between the input features and the droplet generation rate. Quantitative performance metrics further support this observation, with a high coefficient of determination ($R^2 = 0.96$) demonstrating strong agreement between predictions and actual values. The root mean square error (RMSE) of 28.996, and the mean absolute error (MAE) of 18.03, which is lower than the MAE of 20 reported in the previous study[66], demonstrate the effectiveness of the model in minimising prediction errors. However, some deviations are noticeable for higher actual droplet generation rates, where the predicted values slightly underestimate or overestimate the true values, suggesting potential areas for improvement in model fine-tuning. Overall, this visualisation effectively illustrates the robustness and reliability of the model in predicting droplet generation rates across a wide range of values. Hence, the Automation Designer ensures a rigorous, iterative refinement process, fostering the production of robust and reliable machine learning solutions tailored to the domain of microfluidics.

| Request | |
|---|---|
| A .csv file with the following info.<br>  Input feature: {Orifice width (um), Aspect ratio, Expansion ratio, Normalized orifice length, Normalized water inlet, Normalized oil inlet, Flow rate ratio, Capillary number}<br>  Output feature: {Observed generation rate (Hz), Observed droplet diameter (um), Observed regime}<br>  Build a XGBoost model to predict Observed generation rate (Hz). Apply hyperparameters optimization | |

| Design Automation Agent | |
|---|---|

```
df = pd.read_csv('data.csv')
# Define features and target
X = df.drop(['Observed generation rate (Hz)'], axis=1)
y = df['Observed generation rate (Hz)']
```

| Code Reviewing Agent | |
|---|---|

* The `XGBRegressor` instance should be initialized with a dictionary of hyperparameters instead of an empty dictionary. It should look like this: `xgb_model = XGBRegressor(max_depth=3, learning_rate=0.1, n_estimators=50)`.

| Code Improving Agent | |
|---|---|

```
# Initialize the XGBoost model and GridSearchCV with the specified hyperparameters
xgb_model = XGBRegressor(max_depth=3, learning_rate=0.1, n_estimators=50)
```

(a)

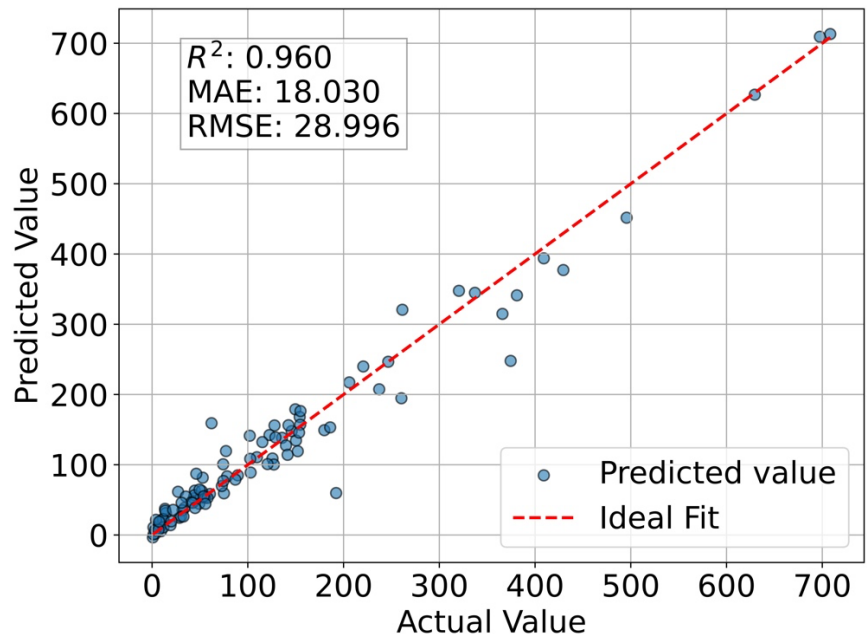

(b)

Fig. 6 (a) Workflow for generating machine learning code. (b) Droplet generation rate prediction

## 3.3 Performance of the Automation Designer based on LLAMA 3.1 in generating CAD design

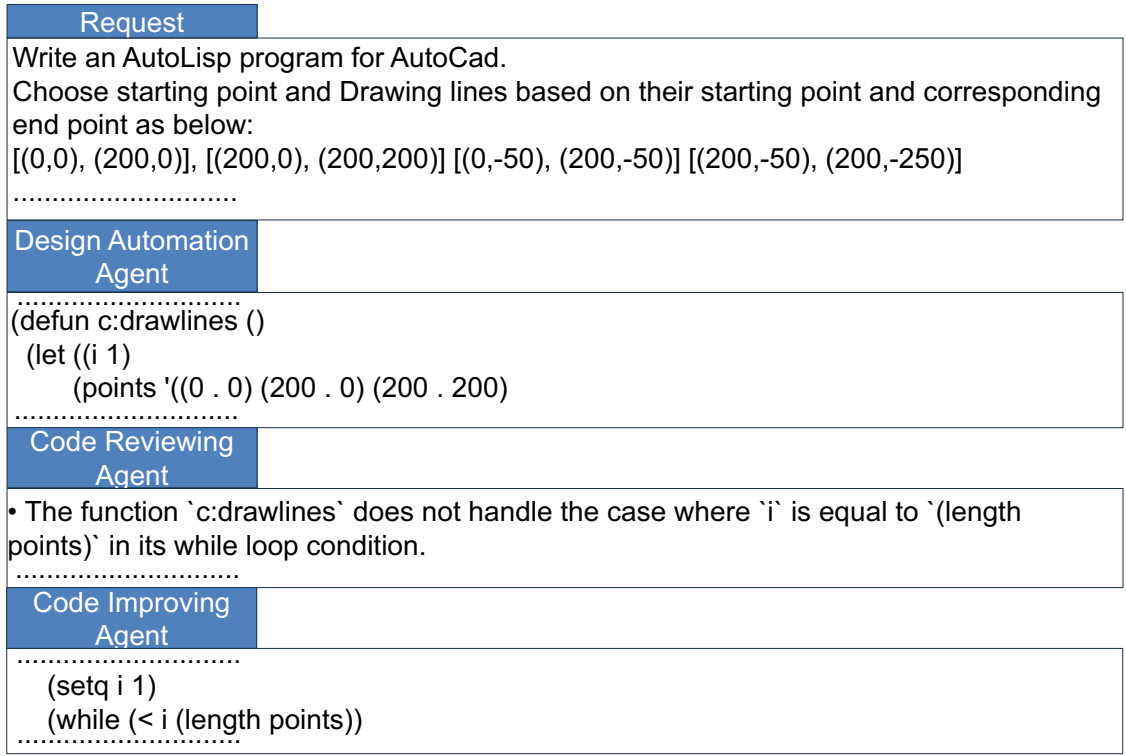

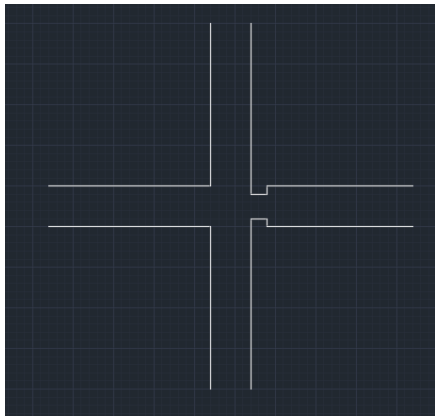
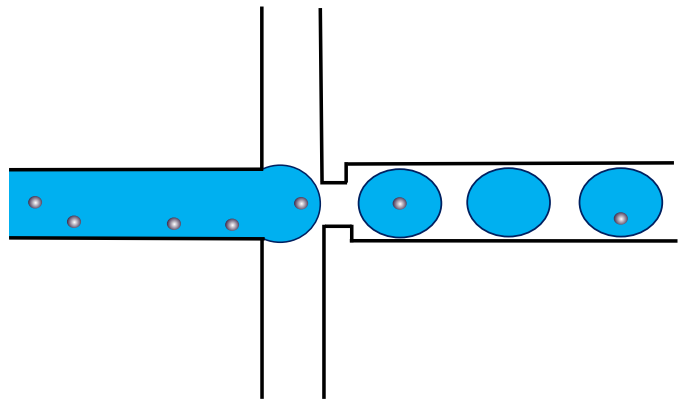

(b)

Fig. 7 (a) Workflow for generating AutoLISP code in AutoCAD. (b) Implemented AutoCAD drawing

An example illustrates a structured, automated pipeline for generating and refining AutoLISP code in AutoCAD to design a microfluidic chip, employing the Automation Designer, as shown in Fig.7a. The process begins with a user posing a specific question or request regarding the AutoLISP code required for the chip design. This input is processed by the design automation agent, which generates an initial draft of the code based on the provided specifications. The generated code is then passed to the code reviewing agent, responsible for meticulously analyzing the code for errors, inconsistencies, and alignment with the design objectives. This stage ensures that the draft meets basic functional and structural requirements. Any identified issues or areas for improvement are forwarded to the code-improving agent, which refines the code further by addressing the shortcomings highlighted during the review phase. This agent optimizes the code for efficiency, accuracy, and functionality. The final, refined code is generated as the ultimate result and is ready for implementation in the design, as shown in Figure. 7b, and for the fabrication of the microfluidic chip. This automated, iterative process ensures a streamlined and reliable approach to code development, minimizing human error while maximizing precision and efficiency in microfluidic chip design (ESI, S9).

## 4. Conclusions and Outlook

In this study, we present an intelligent multi-agent framework for advancing droplet microfluidics research by integrating advanced LLMs, which represents a notable milestone in the field. The framework includes the Scientific Mentor, which leverages domain-specific knowledge to provide reliable guidance on droplet microfluidics. Additionally, the framework incorporates the Automation Designer, capable of generating machine learning code to facilitate design optimisation and automation of microfluidic chips, as well as producing code-based CAD scripts for rapid and precise creation of designs. LLMs are poised to revolutionise the field of droplet microfluidics by enhancing research methodologies and accelerating scientific discovery. Droplet microfluidics, characterised by its ability to manipulate discrete droplets in micro-scale environments, has broad applications in biotechnology, medicine, and materials science. LLM agents offer unprecedented opportunities to streamline the analysis of vast scientific literature, enabling researchers to identify patterns, generate hypotheses, and design experiments more efficiently. By providing context-aware insights and generating predictive models, these agents can aid in optimising droplet generation, stability, and functionality, significantly reducing the time and cost associated with experimental iterations. Furthermore, integrating LLMs into experimental workflows could improve reproducibility by standardising protocols, enabling researchers to detect anomalies in real time, and enhancing collaborative research by facilitating cross-disciplinary data integration. In education and training, LLM agents have the potential to transform how droplet microfluidics is taught and applied in both academic and industrial contexts. Students and professionals can leverage these agents to access interactive tutorials, troubleshoot experimental setups, and gain deeper insights into fluid dynamics and material properties. In examinations, LLMs could be utilised to design personalised assessments that evaluate conceptual understanding and problem-solving skills in droplet microfluidics. Furthermore, these agents can provide real-time feedback and adaptive learning pathways, ensuring learners at all levels can progress effectively. Additionally, these agents can bridge the gap between academia and industry by offering tailored solutions for process optimisation, quality control, and scaling up the production of microfluidic devices. By democratising access to advanced knowledge and tools, LLM agents can enable small-scale laboratories and startups to compete with larger institutions, fostering innovation across diverse settings.

LLM agents also hold promise for advancing the automation and optimisation of computer-aided design (CAD) for microfluidic devices. The design of droplet microfluidics systems often requires precise and intricate configurations that are challenging and time-intensive to develop. By leveraging the computational power of LLMs, researchers can automate CAD processes, enabling rapid prototyping of novel device architectures. These agents can analyse complex datasets to refine channel geometries and droplet manipulation parameters, resulting in highly efficient designs. Moreover, LLMs can integrate multi-objective optimisation frameworks that balance trade-offs between cost, performance, and manufacturability, facilitating the development of robust and scalable devices. Additionally, LLMs can facilitate the implementation and optimisation of machine learning models tailored to the unique challenges of droplet microfluidics, such as predicting droplet behaviour under varying conditions, identifying emergent patterns in high-dimensional data, and optimising device performance for specific applications. This synergy between LLMs and machine learning could lead to breakthroughs in microfluidic technologies, unlocking new applications in diagnostics, drug discovery, and synthetic biology. However, integrating LLM agents into droplet microfluidics research and development is challenging. Data quality, model interpretability, and the risk of over-reliance on automated systems must be addressed. For instance, the variability in experimental conditions and datasets could lead to biased or suboptimal recommendations if not rigorously curated. Furthermore, ensuring the security of sensitive experimental data and proprietary designs is crucial in preventing intellectual property theft or misuse. Ethical concerns also arise regarding the displacement of traditional skill sets and the potential marginalisation of researchers who may lack access to LLM-driven tools. Addressing these challenges requires a concerted effort to establish best practices, including transparent validation metrics, collaborative governance frameworks, and robust training programs. Despite these hurdles, the transformative potential of LLM agents in advancing droplet microfluidics is undeniable. By bridging the gaps between computational intelligence, experimental precision, and educational accessibility, these agents promise to accelerate innovation and expand the horizons of this dynamic field.

**Conflicts of interest**

There are no conflicts to declare.

**Data availability**

https://github.com/duydinhlab/DMFAgents

**Acknowledgements**

We gratefully acknowledge the funding provided by the Research Grant Council of Hong Kong, General Research Fund (Ref No. 14211223).

## 5. References

1    J. Wei *et al*., arXiv:2206.07682v2, 2022

2    Z. Zheng *et al*., *Angew. Chem. Int. Ed. Engl.*, vol. 62, no. 46, Nov. 2023

3    D. A. Boiko, R. MacKnight, B. Kline and G. Gomes, *Nature*, 2023, **624**, 570–578.

4    Z. Zheng, O. Zhang, C. Borgs, J. T. Chayes and O. M. Yaghi, *J. Am. Chem. Soc.*, 2023, **145**, 18048–18062.

5    K. M. Jablonka, P. Schwaller, A. Ortega-Guerrero and B. Smit, *Nat. Mach. Intell. 2024 62*, 2024, **6**, 161–169.

6    A. M. Bran, S. Cox, O. Schilter, C. Baldassari, A. D. White and P. Schwaller, *Nat. Mach. Intell. 2024 65*, 2024, **6**, 525–535.

7    W. Hou and Z. Ji, *Nat. Methods*, 2024, **21**, 1462–1465.

8    H. Cui, C. Wang, H. Maan, K. Pang, F. Luo, N. Duan and B. Wang, *Nat. Methods 2024 218*, 2024, **21**, 1470–1480.

9    R. Riveland and A. Pouget, *Nat. Neurosci.*, 2024, **27**, 988–999.

10   Z. Lin *et al., Science*, 2023, **379**, 1123–1130.

11   Z. Xiao, W. Li, H. Moon, G. W. Roell, Y. Chen and Y. J. Tang, *ACS Synth. Biol.*, 2023, **12**, 2973–2982.

12   J. Lee *et al*., *Bioinformatics*, 2019, **36**, 1234–1240.

13   Q. Chen *et al.*, *Bioinformatics*, vol. 39, no. 9, Sep. 2023.

14   S. Thapa and S. Adhikari, *Ann. Biomed. Eng.*, 2023, **51**, 2647–2651.

15   Q. Jin, Y. Yang, Q. Chen and Z. Lu, *Bioinformatics*, vol. 40, no. 2, Apr. 2023.

16   X. Zhang, Z. Zhou, C. Ming and Y. Y. Sun, *J. Phys. Chem. Lett.*, 2023, **14**, 11342–11349.

17   Z. Hong, *Energy Mater. Adv.*, vol. 4, Apr. 2023.

18   M. P. Polak and D. Morgan, *Nat. Commun. 2024 151*, 2024, **15**, 1–11.

19   J. Choi and B. Lee, *Commun. Mater. 2024 51*, 2024, **5**, 1–11.

20   O. N. Oliveira, L. Christino, M. Oliveira and F. V. Paulovich, *J. Chem. Inf. Model.*, 2023,

**63**, 7605–7609.

21   T. Xie *et al*., arXiv:2304.02213v5, 2023

22   M. Zaki *et al*., *Digit. Discov.*, 2024, **3**, 313–327.

23   D. Van Veen *et al*., *Nat. Med.*, 2024, **30**, 1134–1142.

24   F. Wong, C. de la Fuente-Nunez and J. J. Collins, *Science*, 2023, **381**, 164–170.

25   M. Moor *et al*., *Nature*, 2023, **616**, 259–265.

26   R. Wang, H. Feng and G. W. Wei, *J. Chem. Inf. Model.*, 2023, **63**, 7189–7209.

27   A. J. Thirunavukarasu *et al*., *Nat. Med.*, 2023, **29**, 1930–1940.

28   Y. Wang, Y. Zhao and L. Petzold, *Proc. Mach. Learn. Res.*, 2023, **219**, 804–823.

29   H. Wang *et al*., *Nature*, 2023, **620**, 47–60.

30   C. Stokel-Walker and R. Van Noorden, *Nature*, 2023, **614**, 214–216.

31   K. M. Merz, G. W. Wei and F. Zhu, *J. Chem. Inf. Model.*, 2023, **63**, 5395.

32   M. R. AI4Science and M. A. Quantum, arXiv:2311.07361v2, 2023

33   Y. Zhang *et al*., arXiv:2309.01219v2, 2023

34   M. Sallam, *Healthc. (Basel, Switzerland)*, vol. 11, no. 6, Mar. 2023.

35   P. Lewis *et al*., *Adv. Neural Inf. Process. Syst.*, vol. 2020-December, May 2020.

36   Y. Gao *et al*., arXiv:2312.10997v5, 2023.

37   J. Li *et al*., arXiv:2402.05120v2, 2024.

38   J. S. Park *et al*., *UIST 2023 - Proc. 36th Annu. ACM Symp. User Interface Softw. Technol.*, Apr. 2023.

39   L. Wang *et al*., *Front. Comput. Sci.*, vol. 18, no. 6, Aug. 2023.

40   A. Zhao *et al*., *Proc. AAAI Conf. Artif. Intell.*, 2023, **38**, 19632–19642.

41   T. Guo *et al*., arXiv:2402.01680v2, 2024.

42   C. Gao *et al*., *Humanit. Soc. Sci. Commun. 2024 111*, 2024, **11**, 1–24.

43   M. D. Skarlinski *et al*., arXiv:2409.13740v2, 2024.

44    J. Li *et al*., arXiv:2405.02957v1, 2024 .

45    X. Tang *et al*., arXiv:2311.10537v4, 2024.

46    S. Schmidgall *et al*., arXiv:2405.07960v4, 2024 .

47    G. M. Whitesides, *Nat. 2006 4427101*, 2006, **442**, 368–373.

48    Y. Ding, P. D. Howes and A. J. Demello, *Anal. Chem.*, 2020, **92**, 132–149.

49    E. Y. u. Basova and F. Foret, *Analyst*, 2014, **140**, 22–38.

50    T. Moragues *et al*., *Nat. Rev. Methods Prim. 2023 31*, 2023, **3**, 1–22.

51    R. Zilionis *et al*., *Nat. Protoc. 2016 121*, 2016, **12**, 44–73.

52    M. Pellegrino *et al*., *Genome Res.*, 2018, **28**, 1345–1352.

53    X. Zhang *et al*., *Mol. Cell*, vol. 73, no. 1, pp. 130-142.e5, Jan. 2019.

54    A. M. Klein *et al*., *Cell*, 2015, **161**, 1187–1201.

55    E. Z. Macosko *et al.*, *Cell*, 2015, **161**, 1202–1214.

56    A. Gérard *et al*., *Nat. Biotechnol.*, 2020, **38**, 715–721.

57    K. Fischer *et al.*, *Nat. Biotechnol.*, 2024 Aug 14, DOI:10.1038/S41587-024-02346-5.

58    J. J. Agresti *et al.*, *Proc. Natl. Acad. Sci. U. S. A.*, 2010, **107**, 4004–4009.

59    H. Yin *et al*., *Biosens. Bioelectron.*, 2021, **188**, 113282.

60    Y. Belotti and C. T. Lim, *Anal. Chem.*, 2021, **93**, 4727–4738.

61    T. S. Kaminski, O. Scheler and P. Garstecki, *Lab Chip*, 2016, **16**, 2168–2187.

62    D. T. Chiu *et al*., *Chem*, 2017, **2**, 201–223.

63    F. Su, K. Chakrabarty and R. B. Fair, *IEEE Trans. Comput. Des. Integr. Circuits Syst.*, 2006, **25**, 211–223.

64    S. Battat, D. A. Weitz and G. M. Whitesides, *Lab Chip*, 2022, **22**, 530–536.

65    D. McIntyre, A. Lashkaripour, P. Fordyce and D. Densmore, *Lab Chip*, 2022, **22**, 2925–2937.

66    A. Lashkaripour *et al*., *Nat. Commun. 2021 121*, 2021, **12**, 1–14.

67    D. Nguyen, R. K. Tong and N. Dinh, arXiv:2411.06691v1, 2024.

68     S. A. Damiati *et al*., *Sci. Reports 2020 101*, 2020, **10**, 1–11.

69     S. H. Hong, H. Yang and Y. Wang, *Microfluid. Nanofluidics*, 2020, **24**, 1–20.

70     W. Ji, T. Y. Ho, J. Wang and H. Yao, *IEEE Trans. Comput. Des. Integr. Circuits Syst.*, 2020, **39**, 2544–2557.

71     A. Dubey *et al.*, arXiv:2407.21783v2, 2024.

72     A. Q. Jiang *et al.,* arXiv: 2310.06825v1, 2023.

73     G. Team *et al.,* arXiv:2408.00118v3, 2024.

74     Y. Li, *Int. Conf. Recent Adv. Nat. Lang. Process. RANLP*, 2023, 641–647.

75     L. Wang *et al.*, *npj Digit. Med. 2024 71*, 2024, **7**, 1–9.

76     P. Liu *et al.*, *ACM Comput. Surv.*, vol. 55, no. 9, Jul. 2021.

**Electronic Supplementary Information (ESI)**

S1: Document List

A List of 233 documents

[2019] Recent Advances in Droplet Microfluidics.pdf
[2012] Two-phase microfluidic droplet flows of ionic liquids for the synthesis of gold and silver nanoparticles.pdf
[2017] Droplet-based microfluidic high-throughput screening of heterologous enzymes secreted by the yeast Yarrowia lipolytica.pdf
[2010] Advanced microfluidic droplet manipulation based on piezoelectric actuation.pdf
[2015] Millisecond-timescale monitoring of PbS nanoparticle nucleation and growth using droplet-based microfluidics.pdf
[2007] Concentration and binary separation of micro particles for droplet-based digital microfluidics.pdf
[2024] Droplet Microfluidic Devices- Working Principles, Fabrication Methods, and Scale-Up Applications.pdf
[2011] Biofunctionalization of electrowetting-on-dielectric digital microfluidic chips for miniaturized cell-based applications.pdf
[2020] A single cell droplet microfluidic system for quantitative determination of food-borne pathogens.pdf

[2012] Selective droplet coalescence using microfluidic systems.pdf
[2017] Generation and manipulation of hydrogel microcapsules by droplet-based microfluidics for mammalian cell culture.pdf
[2018] A droplet-based microfluidic chip as a platform for leukemia cell lysate identification using surface-enhanced Raman scattering.pdf
[2010] Ultrahigh-throughput screening in drop-based microfluidics for directed evolution.pdf
[2019] Pathogenic bacteria detection using RNA-based loop-mediated isothermal-amplification-assisted nucleic acid amplification via droplet microfluidics.pdf
[2013] Droplet-based microfluidics.pdf
[2016] Continuous, one-pot synthesis and post-synthetic modification of nanoMOFs using droplet nanoreactors.pdf
[2016] Controlled assembly of heterotypic cells in a core–shell scaffold- organ in a droplet.pdf
[2024] Synthetic Cells from Droplet-Based Microfluidics for Biosensing and Biomedical Applications.pdf
[2009] A fast and efficient microfluidic system for highly selective one-to-one droplet fusion.pdf
[2015] Droplet based microfluidics- spectroscopic characterization of levofloxacin and its SERS detection.pdf
[2009] Easy route to superhydrophobic copper-based wire-guided droplet microfluidic systems.pdf
[2010] Droplet-based microfluidic system for multicellular tumor spheroid formation and anticancer drug testing.pdf
[2018] Controllable synthesis of nanocrystals in droplet reactors.pdf
[2010] An automated two-phase microfluidic system for kinetic analyses and the screening of compound libraries.pdf
[2011] High-speed droplet actuation on single-plate electrode arrays.pdf
[2019] One-step bonding and hydrophobic surface modification method for rapid fabrication of polycarbonate-based droplet microfluidic chips.pdf
[2015] Droplet-based microfluidics at the femtolitre scale.pdf
[2023] Microtechnologies for single-cell and spatial multi-omics.pdf
[2024] Development and future of droplet microfluidics.pdf
[2020] Advances in continuous-flow based microfluidic PCR devices—A review.pdf
[2024] Rapid discovery of monoclonal antibodies by microfluidics-enabled FACS of single pathogen-specific antibody-secreting cells.pdf
[2005] Design of fault-tolerant and dynamically-reconfigurable microfluidic biochips.pdf
[2003] Millisecond kinetics on a microfluidic chip using nanoliters of reagents.pdf
[2020] Step emulsification in microfluidic droplet generation- mechanisms and structures.pdf
[2020] Bottom-up assembly of functional intracellular synthetic organelles by droplet-based microfluidics.pdf
[2011] Droplet formation via flow-through microdevices in Raman and surface enhanced Raman spectroscopy—concepts and applications.pdf
[2017] High throughput single cell counting in droplet-based microfluidics.pdf
[2008] Fabrication of monodisperse thermosensitive microgels and gel capsules in microfluidic devices.pdf
[2013] Single-cell analysis and sorting using droplet-based microfluidics.pdf
[2009] Reliable microfluidic on-chip incubation of droplets in delay-lines.pdf

[2019] Automated droplet-based microfluidic platform for multiplexed analysis of biochemical markers in small volumes.pdf

[2021] Droplet-based microfluidic platform for high-throughput screening of Streptomyces.pdf

[2005] A droplet-based lab-on-a-chip for colorimetric detection of nitroaromatic explosives.pdf

[2013] Ultrahigh-throughput mammalian single-cell reverse-transcriptase polymerase chain reaction in microfluidic drops.pdf

[2008] Droplet-based microreactors for the synthesis of magnetic iron oxide nanoparticles.pdf

[2015] Highly Parallel Genome-wide Expression Profiling of Individual Cells Using Nanoliter Droplets.pdf

[2015] Control of the breakup process of viscous droplets by an external electric field inside a microfluidic device.pdf

[2013] Droplet-based microfluidics for dose–response assay of enzyme inhibitors by electrochemical method.pdf

[2024] Droplet-based microfluidics in chemical applications.pdf

[2017] Single-Cell Transcriptional Analysis.pdf

[2010] PLGA micro:nanosphere synthesis by droplet microfluidic solvent evaporation and extraction approaches.pdf

[2011] Functional microgels tailored by droplet-based microfluidics.pdf

[2010] Bio-electrospraying and droplet-based microfluidics- control of cell numbers within living residues.pdf

[2024] Droplet-based logic gates simulation of viscoelastic fluids under electric field.pdf

[2013] Continuous synthesis of device-grade semiconducting polymers in droplet-based microreactors.pdf

[2004] Droplet-based microfluidic lab-on-a-chip for glucose detection.pdf

[2013] Small-angle X-ray scattering in droplet-based microfluidics.pdf

[2017] Advances in droplet-based microfluidic technology and its applications.pdf

[2018] On-demand acoustic droplet splitting and steering in a disposable microfluidic chip.pdf

[2024] On-Chip Droplet Splitting with High Volume Ratios Using a 3D Conical Microstructure-Based Microfluidic Device.pdf

[2016] Microfluidic tools toward industrial biotechnology.pdf

[2008] Droplet-based microfluidic system for individual Caenorhabditis elegans assay.pdf

[2009] Multi-step microfluidic droplet processing- kinetic analysis of an in vitro translated enzyme.pdf

[2024] Formation Pattern Analysis of Spheroids Formed by a Droplet-Based Microfluidic System for Predicting the Aggressiveness of Tumor Cells.pdf

[2009] General digital microfluidic platform manipulating dielectric and conductive droplets by dielectrophoresis and electrowetting.pdf

[2010] Rapid cell extraction in aqueous two-phase microdroplet systems.pdf

[2012] Crystal nucleation in a droplet based microfluidic crystallizer.pdf

[2012] Proteolysis in microfluidic droplets- an approach to interface protein separation and peptide mass spectrometry.pdf

[2012] Surfactants in droplet-based microfluidics.pdf

[2012] Droplet-based microfluidic flow injection system with large-scale concentration gradient by a single nanoliter-scale injection for enzyme inhibition assay.pdf

[2014] Generating electric fields in PDMS microfluidic devices with salt water electrodes.pdf

[2024] Overloading And unpacKing (OAK)-droplet-based combinatorial indexing for ultra-high throughput single-cell multiomic profiling.pdf

[2013] New glycosidase substrates for droplet-based microfluidic screening.pdf

[2024] Bead-based microfluidic platforms for multiplex and ultrasensitive immunoassays in clinical diagnosis and treatment.pdf

[2017] Integrated microfluidic platform for rapid antimicrobial susceptibility testing and bacterial growth analysis using bead-based biosensor via fluorescence ….pdf

[2024] Droplet-based microfluidics for drug delivery applications.pdf

[2020] Development and application of analytical detection techniques for droplet-based microfluidics-A review.pdf

[2009] Performance of nanoliter-sized droplet-based microfluidic PCR.pdf

[2006] Optical detection for droplet size control in microfluidic droplet-based analysis systems.pdf

[2024] Recent advances of droplet-based microfluidics for engineering artificial cells.pdf

[2016] Single-cell barcoding and sequencing using droplet microfluidics.pdf

[2006] Modeling and controlling parallel tasks in droplet-based microfluidic systems.pdf

[2024] Advancing Point-of-Care Applications with Droplet Microfluidics From Single-Cell to Multicellular Analysis.pdf

[2016] High precision droplet-based microfluidic determination of americium (III) and lanthanide (III) solvent extraction separation kinetics.pdf

[2010] Droplet microfluidics for characterizing the neurotoxin-induced responses in individual Caenorhabditis elegans.pdf

[2021] Conventional and emerging strategies for the fabrication and functionalization of PDMS-based microfluidic devices.pdf

[2013] From tubes to drops- droplet-based microfluidics for ultrahigh-throughput biology.pdf

[2011] Ultrafast surface enhanced resonance Raman scattering detection in droplet-based microfluidic systems.pdf

[2012] Self-regulated, droplet-based sample chopper for microfluidic absorbance detection.pdf

[2019] Droplet-based microfluidics systems in biomedical applications.pdf

[2020] Controllable size and form of droplets in microfluidic-assisted devices- Effects of channel geometry and fluid velocity on droplet size.pdf

[2018] Unveiling the shape evolution and halide-ion-segregation in blue-emitting formamidinium lead halide perovskite nanocrystals using an automated microfluidic ….pdf

[2017] Development of a facile droplet-based single-cell isolation platform for cultivation and genomic analysis in microorganisms.pdf

[2011] Quantitative and sensitive detection of rare mutations using droplet-based microfluidics.pdf

[2024] Automation of digital:droplet microfluidic platforms.pdf

[2009] Measuring rapid enzymatic kinetics by electrochemical method in droplet-based microfluidic devices with pneumatic valves.pdf

[2023] Droplet-based microfluidics.pdf

[2013] Interfacing droplet microfluidics with matrix-assisted laser desorption:ionization mass spectrometry- label-free content analysis of single droplets.pdf

[2016] Wash-free magnetic immunoassay of the PSA cancer marker using SERS and droplet microfluidics.pdf

[2017] A droplet-chip:mass spectrometry approach to study organic synthesis at nanoliter scale.pdf
[2010] Passive self-synchronized two-droplet generation.pdf
[2006] On-chip titration of an anticoagulant argatroban and determination of the clotting time within whole blood or plasma using a plug-based microfluidic system.pdf
[2014] Microfluidic dynamic interfacial tensiometry (μDIT).pdf
[2010] Droplet-based microextraction in the aqueous two-phase system.pdf
[2024] Fabrication of oxygen-releasing dextran microgels by droplet-based microfluidic method.pdf
[2013] Droplet-based microfluidic system to form and separate multicellular spheroids using magnetic nanoparticles.pdf
[2004] Concurrent testing of droplet-based microfluidic systems for multiplexed biomedical assays.pdf
[2014] Microfluidic platforms- a mainstream technology for the preparation of crystals.pdf
[2014] Nanoliter-scale protein crystallization and screening with a microfluidic droplet robot.pdf
[2011] Characterization of the heat transfer accompanying electrowetting or gravity-induced droplet motion.pdf
[2013] Single-enzyme analysis in a droplet-based micro-and nanofluidic system.pdf
[2016] Synthesis of cesium lead halide perovskite nanocrystals in a droplet-based microfluidic platform- fast parametric space mapping.pdf
[2012] A completely in vitro ultrahigh-throughput droplet-based microfluidic screening system for protein engineering and directed evolution.pdf
[2011] Fabrication of advanced particles and particle-based materials assisted by droplet-based microfluidics.pdf
[2005] High-temperature microfluidic synthesis of CdSe nanocrystals in nanoliter droplets.pdf
[2017] A controllable and integrated pump-enabled microfluidic chip and its application in droplets generating.pdf
[2009] Analysis of protein–protein interactions by using droplet-based microfluidics.pdf
[2013] Sequential operation droplet array- an automated microfluidic platform for picoliter-scale liquid handling, analysis, and screening.pdf
[2018] Mass fabrication of uniform sized 3D tumor spheroid using high-throughput microfluidic system.pdf
[2012] High-throughput analysis of protein–protein interactions in picoliter-volume droplets using fluorescence polarization.pdf
[2019] One-step wettability patterning of PDMS microchannels for generation of monodisperse alginate microbeads by in Situ external gelation in double emulsion ….pdf
[2024] An integrated and multi-functional droplet-based microfluidic platform for digital DNA amplification.pdf
[2009] A microfluidic droplet generator based on a piezoelectric actuator.pdf
[2009] Electroporation of cells in microfluidic droplets.pdf
[2013] Analytical detection techniques for droplet microfluidics—A review.pdf
[2008] Heterogeneous immunoassays using magnetic beads on a digital microfluidic platform.pdf
[2024] Droplet-based microfluidics and enzyme evolution.pdf
[2009] Droplet-based microfluidic systems for high-throughput single DNA molecule isothermal amplification and analysis.pdf

[2012] Continuous-flow in-droplet magnetic particle separation in a droplet-based microfluidic platform.pdf

[2013] Multiplexed protease activity assay for low-volume clinical samples using droplet-based microfluidics and its application to endometriosis.pdf

[2011] On-demand preparation of quantum dot-encoded microparticles using a droplet microfluidic system.pdf

[2014] An EWOD-based microfluidic chip for single-cell isolation, mRNA purification and subsequent multiplex qPCR.pdf

[2004] Effects of viscosity on droplet formation and mixing in microfluidic channels.pdf

[2016] Droplet microfluidics A tool for biology, chemistry and nanotechnology.pdf

[2024] Droplet-Based Preparation of ZnO-nanostructure Array for Microfluidic Fluorescence Biodetection.pdf

[2007] High-throughput DNA droplet assays using picoliter reactor volumes.pdf

[2018] Detection of antibiotics synthetized in microfluidic picolitre-droplets by various actinobacteria.pdf

[2005] Droplet-based micro oscillating-flow PCR chip.pdf

[2014] A droplet-based microfluidic electrochemical sensor using platinum-black microelectrode and its application in high sensitive glucose sensing.pdf

[2007] 3D droplet displacement in microfluidic systems by electrostatic actuation.pdf

[2016] A review on continuous-flow microfluidic PCR in droplets- Advances, challenges and future.pdf

[2017] Emerging Droplet Microfluidics.pdf

[2024] Evaluation of Acoustophoretic and Dielectrophoretic Forces for Droplet Injection in Droplet-Based Microfluidic Devices.pdf

[2017] From single-molecule detection to next-generation sequencing- microfluidic droplets for high-throughput nucleic acid analysis.pdf

[2018] Protein crystallization in a droplet-based microfluidic device- Hydrodynamic analysis and study of the phase behaviour.pdf

[2007] Microfluidic platforms for lab-on-a-chip applications.pdf

[2016] Droplet microfluidics for microbiology techniques, applications and challenges.pdf

[2013] DNA sequence analysis with droplet-based microfluidics.pdf

[2017] Microfluidic reactors provide preparative and mechanistic insights into the synthesis of formamidinium lead halide perovskite nanocrystals.pdf

[2015] High-throughput, quantitative enzyme kinetic analysis in microdroplets using stroboscopic epifluorescence imaging.pdf

[2024] Miniaturized droplets flow reactor for one-step highly controlled synthesis of SnO2 quantum dots at room temperature.pdf

[2010] Automated microfluidic screening assay platform based on DropLab.pdf

[2024] Combinatorial Sample Preparation Platform for Droplet-Based Applications in Microbiology.pdf

[2015] Droplet Barcoding for Single-Cell Transcriptomics Applied to Embryonic Stem Cells.pdf

[2021] Machine learning enables design automation of microfluidic flow-focusing droplet generation.pdf

[2022] Microfluidic technologies for nanoparticle formation.pdf

[2021] Droplet digital PCR of viral DNA:RNA, current progress, challenges, and future perspectives.pdf

[2012] Droplet microfluidics for high-throughput biological assays.pdf
[2006] Droplet-based microfluidics with nonaqueous solvents and solutions.pdf
[2013] Probing cellular heterogeneity in cytokine-secreting immune cells using droplet-based microfluidics.pdf
[2006] Reactions in droplets in microfluidic channels.pdf
[2016] High-throughput screening of filamentous fungi using nanoliter-range droplet-based microfluidics.pdf
[2013] Micro-optical lens array for fluorescence detection in droplet-based microfluidics.pdf
[2011] New generation of amino coumarin methyl sulfonate-based fluorogenic substrates for amidase assays in droplet-based microfluidic applications.pdf
[2012] Generation of disk-like hydrogel beads for cell encapsulation and manipulation using a droplet-based microfluidic device.pdf
[2016] Droplet formation in a microfluidic T-junction involving highly viscous fluid systems.pdf
[2013] Cell-based drug combination screening with a microfluidic droplet array system.pdf
[2005] Controlling nonspecific protein adsorption in a plug-based microfluidic system by controlling interfacial chemistry using fluorous-phase surfactants.pdf
[2014] On-chip spyhole mass spectrometry for droplet-based microfluidics.pdf
[2014] Facile droplet-based microfluidic synthesis of monodisperse IV–VI semiconductor nanocrystals with coupled in-line NIR fluorescence detection.pdf
[2017] Single-cell genome sequencing at ultra-high-throughput with microfluidic droplet barcoding.pdf
[2008] Monitoring of real-time streptavidin− biotin binding kinetics using droplet microfluidics.pdf
[2017] Cell microencapsulation by droplet microfluidic templating.pdf
[2024] Process optimization for preparation of curcumin and quercetin co-encapsulated liposomes using microfluidic device.pdf
[2024] Droplet-Based Microfluidics with Mass Spectrometry for Microproteomics.pdf
[2009] A light-induced dielectrophoretic droplet manipulation platform.pdf
[2009] Droplet-based microfluidic synthesis of anisotropic metal nanocrystals.pdf
[2013] Ultrahigh-throughput sorting of microfluidic drops with flow cytometry.pdf
[2015] Electrochemical droplet-based microfluidics using chip-based carbon paste electrodes for high-throughput analysis in pharmaceutical applications.pdf
[2024] Engineering poly (dehydroalanine)-based gels via droplet-based microfluidics- from bulk to microspheres.pdf
[2020] Droplet-based microreactor for the production of micro:nano-materials.pdf
[2024] Droplet-based microfluidics for biological applications.pdf
[2010] On-demand droplet release for droplet-based microfluidic system.pdf
[2004] Multi-step synthesis of nanoparticles performed on millisecond time scale in a microfluidic droplet-based system.pdf
[2016] Graphene-polyaniline modified electrochemical droplet-based microfluidic sensor for high-throughput determination of 4-aminophenol.pdf
[2013] Microgel capsules tailored by droplet-based microfluidics.pdf
[2017] Scaling by shrinking empowering single-cell 'omics' with microfluidic devices.pdf
[2019] Droplet-based single cell RNAseq tools- a practical guide.pdf
[2014] CotA laccase- high-throughput manipulation and analysis of recombinant enzyme libraries expressed in E. coli using droplet-based microfluidics.pdf

[2017] The power of solid supports in multiphase and droplet-based microfluidics- towards clinical applications.pdf

[2011] Droplet based microfluidics.pdf

[2020] Generation of tumor spheroids using a droplet-based microfluidic device for photothermal therapy.pdf

[2014] A droplet-based screen for wavelength-dependent lipid production in algae.pdf

[2012] Microfluidic platform for combinatorial synthesis in picolitre droplets.pdf

[2011] Droplet microfluidics- recent developments and future applications.pdf

[2009] A PMMA microfluidic droplet platform for in vitro protein expression using crude E. coli S30 extract.pdf

[2008] Droplet microfluidics.pdf

[2016] Scalable production of CuInS 2:ZnS quantum dots in a two-step droplet-based microfluidic platform.pdf

[2013] Small but smart- sensitive microgel capsules.pdf

[2011] Enzyme kinetic measurements using a droplet-based microfluidic system with a concentration gradient.pdf

[2008] Droplet manipulation with light on optoelectrowetting device.pdf

[2009] A droplet-based microfluidic system capable of droplet formation and manipulation.pdf

[2012] A droplet-based, optofluidic device for high-throughput, quantitative bioanalysis.pdf

[2013] Droplet-based microfluidic platform for high-throughput, multi-parameter screening of photosensitizer activity.pdf

[2017] Fabrication of patterned superhydrophobic:hydrophilic substrates by laser micromachining for small volume deposition and droplet-based fluorescence.pdf

[2024] Droplet-based microfluidic platform for detecting agonistic peptides that are self-secreted by yeast expressing a G-protein-coupled receptor.pdf

[2017] Viscosity measurements using microfluidic droplet length.pdf

[2012] Highly sensitive and quantitative detection of rare pathogens through agarose droplet microfluidic emulsion PCR at the single-cell level.pdf

[2019] Droplet microfluidics- from proof-of-concept to real-world utility?.pdf

[2023] Development of glass-based microfluidic devices- A review on its fabrication and biologic applications.pdf

[2019] Cellular heterogeneity identified by single-cell alkaline phosphatase (ALP) via a SERRS-microfluidic droplet platform.pdf

[2024] Biocompatible Janus microparticle synthesis in a microfluidic device.pdf

[2008] Droplet-based microfluidics for emulsion and solvent evaporation synthesis of monodisperse mesoporous silica microspheres.pdf

[2017] Microfluidic platform for optimization of crystallization conditions.pdf

[2018] Simultaneous immunoassays of dual prostate cancer markers using a SERS-based microdroplet channel.pdf

[2019] High-throughput droplet-based microfluidics for directed evolution of enzymes.pdf

[2010] Optimized droplet-based microfluidics scheme for sol–gel reactions.pdf

[2015] Path-programmable water droplet manipulations on an adhesion controlled superhydrophobic surface.pdf

[2008] The electrochemical detection of droplets in microfluidic devices.pdf

[2012] Teaching single-cell digital analysis using droplet-based microfluidics.pdf

[2010] A stochastic model for nucleation kinetics determination in droplet-based microfluidic systems.pdf
[2006] Test planning and test resource optimization for droplet-based microfluidic systems.pdf
[2019] Microfluidic on-demand droplet generation, storage, retrieval, and merging for single-cell pairing.pdf
[2015] 3D droplet microfluidic systems for high-throughput biological experimentation.pdf
[2022] Droplet-based microfluidics in biomedical applications.pdf
[2017] Droplet-based digital PCR- application in cancer research.pdf
[2017] Concentration gradient generation methods based on microfluidic systems.pdf
[2016] Droplet generation in cross-flow for cost-effective 3D-printed "plug-and-play" microfluidic devices.pdf
[2010] High-throughput sample introduction for droplet-based screening with an on-chip integrated sampling probe and slotted-vial array.pdf
[2020] High-throughput single-cell activity-based screening and sequencing of antibodies using droplet microfluidics.pdf
[2011] Droplet-based microfluidic platforms for single T cell secretion analysis of IL-10 cytokine.pdf
[2011] A 'microfluidic pinball'for on-chip generation of layer-by-layer polyelectrolyte microcapsules.pdf
[2016] Droplet-based microfluidics in drug discovery, transcriptomics and high-throughput molecular genetics.pdf

Ground-truth QA

| question | answer |
|---|---|
| What is microfluidics? | Microfluidics is a multidisciplinary field that involves the study and manipulation of fluids at the microscale, typically within channels with dimensions ranging from tens to hundreds of micrometers. It is a multidisciplinary field that involves molecular analysis, molecular biology, and microelectronics. Microfluidics has been increasingly used in the biological sciences because precise and controlled experiments can be conducted at a lower cost and faster pace. Lab on a Chip devices use microfluidics for applications such as Point of Care testing of diseases, or Organ on a Chip studies. |
| What is droplet microfluidics? | Droplet-based microfluidics is a branch of microfluidics focusing on the creation and manipulation of discrete volumes of liquids through adoption of immiscible fluid flows. Microdroplets allow for handling miniature volumes (µL to fL) of fluids conveniently, provide better mixing, encapsulation, sorting, and sensing, and suit high throughput experiments. Compared with continuous-flow systems, droplet microfluidic systems enable isolation and manipulation of samples in separate compartments. The droplets generated with uniform size and shape provide a stable microenvironment for biochemical reactions. Processing of a large number of droplets in parallel facilitates to achieve ultrahigh throughput. Confinement of individual targets within the droplets and independent processing of each of them allow high-precision analyses. |
| How to generate microdroplets? | Microdroplet generation in microfluidics is achieved through two main methods: |

| | active and passive. Active methods utilize external forces like piezoelectric actuators to precisely control droplet size, formation rate, and spacing. These methods enable high uniformity in droplet generation, with deviations less than 0.3%. Passive methods rely on fluid properties and channel geometry, such as T-junctions or cross-flow configurations, to create droplets. These methods are influenced by factors like flow rate, viscosity, and interfacial tension, allowing for fine control over droplet size and uniformity. For example, PDMS-based microfluidic devices have been used for nanoparticle synthesis, where hydrodynamic coupling of nozzles generates droplet pairs for precise mixing. Both methods are valuable in applications like nanoparticle synthesis, offering high reproducibility and control over reaction conditions in biomedical and material science research. |
|---|---|
| What materials are typically used to fabricate droplet microfluidic devices? | Droplet microfluidic devices are commonly fabricated from a variety of materials, each offering distinct advantages for specific applications. The most widely used material is PDMS (Polydimethylsiloxane), favored for its ease of use with soft lithography and flexibility, though it has limitations such as poor solvent resistance. Alternatives to PDMS include thermoset polyesters, thermoplastics like PMMA, polycarbonate, and polystyrene, and amorphous polymers such as cyclic olefin copolymer. Fluoropolymers like PTFE and glass are also used, offering superior solvent resistance and biofouling resistance. Silicon and capillary devices are utilized in certain applications, providing precise control for microfluidic operations. These materials are chosen based on factors like solvent resistance, biofouling, and the ability to mass-produce devices using methods such as injection molding or hot embossing, |

| | ensuring broad applicability in research and industrial settings. |
|---|---|
| What are the applications of droplet microfluidics? | Droplet microfluidics has diverse applications across biomedical research, diagnostics, and industrial fields. It enables high-throughput single-cell analysis, such as genomics, transcriptomics, and drug screening, while also facilitating nanoparticle synthesis, enzyme evolution, and chemical reactions. In healthcare, it supports personalized medicine, point-of-care diagnostics, and vaccine development. Beyond biology, it is used for environmental monitoring, food safety testing, and the development of cosmetics and pharmaceuticals. Its ability to create isolated, uniform reaction chambers in droplets makes it a powerful tool for precise, cost-effective, and scalable workflows in science and industry. |
| What are the main benefits of using droplet microfluidics over bulk systems? | Droplet microfluidics offers significant advantages over traditional bulk systems. It enables individual compartmentalization, isolating single cells or molecules in tiny liquid compartments for precise control. This allows for single-cell analysis and early detection of rare cells or secreted molecules. The technology supports massive parallel processing, analyzing millions of droplets simultaneously, ideal for high-throughput screening and examining phenotypic or genetic variability. Additionally, iterative operations such as merging, mixing, and sorting droplets enable complex experimental protocols and continuous monitoring of populations. The system's highly controllable local environments ensure consistent, reproducible results. These benefits make droplet microfluidics a versatile tool in drug discovery, medical diagnostics, single-cell analysis, and other fields, offering powerful insights and |

| | precise manipulation at the individual level. |
|---|---|
| What are the advantages of droplet microfluidic systems compared with continuous-flow systems? | Droplet-based microfluidic systems offer significant advantages over traditional continuous-flow systems. These include the ability to overcome issues such as Taylor dispersion, slow mixing, and surface-molecule interactions. Droplet systems allow for ultra-small assay volumes, reducing reagent consumption and improving sensitivity. They also minimize cross-contamination and sample adsorption by compartmentalizing individual samples. The system's chaotic advection enhances mixing efficiency, enabling faster reactions. With droplet generation frequencies up to hundreds of kHz, droplet microfluidics supports high-throughput experimentation. Furthermore, the rapid and precise control of droplet payloads and sub-millisecond mixing times make it ideal for applications requiring fast, controlled analysis. These benefits make droplet-based systems particularly valuable for cell-based assays, enzyme kinetics, and high-throughput screening. |
| What are the basic principles behind droplet generation in microfluidics? | Droplet generation in microfluidics relies on emulsion processes, where an oil phase (continuous phase) encapsulates an aqueous phase (dispersed phase) to form droplets. Surfactants are added to lower interfacial tension, stabilize the droplets, and prevent coalescence. The process requires precise control over the flow of both liquids, often achieved using syringe pumps or pressure-driven systems. Droplets are generated through surface-induced instabilities, influenced by microchannel geometries such as T-junctions or flow-focusing, which control droplet size and frequency. The design and flow control of the system determine the size uniformity and production rate of droplets. These principles enable high-throughput applications such as drug |
| | delivery, high-throughput screening, and microreactors for bio-catalysis, ensuring reproducibility and productivity in biological experiments. |
| How is droplet size controlled in a droplet generator? | Droplet size control in microfluidic generators can be achieved through several methods. In passive droplet generation, adjusting microchannel dimensions or controlling the flow rates of immiscible liquids allows for precise control over droplet size, shape, and monodispersity. Tree-splitters, which bisect large droplets into smaller ones, also influence droplet size and uniformity, with splitting angles affecting droplet diameter. In active droplet generation, technologies like piezoelectric actuators independently control droplet size, formation rate, and spacing, offering high precision. Additionally, the velocity of the continuous and dispersed phases plays a significant role in controlling droplet size. These methods enable the generation of uniform droplets for various applications, such as high-throughput screening and drug delivery. |
| What is the role of surface tension in droplet formation? | Surface tension plays a critical role in droplet formation in microfluidic systems by influencing droplet size, shape, and formation mechanisms. It acts to minimize surface area, favoring spherical droplet shapes. At low Weber numbers (We < 1), where surface tension dominates, droplets form through dripping or squeezing. As the Weber number increases, fluid inertia competes with surface tension, leading to jetting behavior. The capillary number (Ca) also influences droplet size, with surface tension controlling droplet formation at low Ca. Techniques like electrowetting can measure surface tension in microfluidic systems, providing insights into the interfacial properties. Additionally, surface tension is a key parameter in T-junctions and flow |

| | focusing devices, where droplet size measurements offer an indirect way to assess surface tension at the time-scale of droplet production. Understanding the interplay between surface tension and other forces is essential for precise droplet control. |
|---|---|
| How does a flow-focusing device differ from a T-junction in droplet formation? | Flow-focusing devices and T-junctions are two common microfluidic designs for droplet formation, differing in geometry and mechanism. A flow-focusing device uses a central dispersed phase channel and side continuous phase channels converging at a narrow orifice, where hydrodynamic focusing pinches off droplets. It produces smaller, highly uniform droplets with precise control over size and high throughput. In contrast, a T-junction has perpendicular channels where the continuous phase shears the dispersed phase at the intersection, forming larger droplets with less uniformity and lower throughput. Flow-focusing is ideal for applications requiring monodisperse droplets, while T-junctions suit simpler or larger-droplet applications. |
| What role does the geometry of the microchannel play in droplet formation? | The geometry of a microchannel plays a critical role in droplet formation, influencing the size, shape, and generation frequency of droplets. Key geometric factors include the channel width, depth, and junction design, such as T-junctions or flow-focusing geometries. These features govern the interaction between the continuous and dispersed phases, affecting the shear forces and pressure gradients that drive droplet breakup. For instance, narrower channels or smaller orifices promote the formation of smaller droplets due to increased shear stress. Similarly, the junction type controls droplet detachment mechanics; in T-junctions, droplets form due to the balance of interfacial tension and shear force, while in flow-focusing designs, they form under |
| | pinching at the constriction. Channel geometry also determines flow regimes ranging from dripping to jetting further impacting droplet characteristics. Thus, optimizing microchannel geometry is crucial for precise control over droplet formation in applications like microfluidics, emulsification, and lab-on-a-chip systems. |
| What factors affect droplet coalescence in microfluidic devices? | Droplet coalescence in microfluidic devices is influenced by several factors. Surfactant concentration plays a key role, with higher concentrations preventing droplet fusion, while lower concentrations promote coalescence. The duration droplets remain in contact within the pairing channel, known as pairing time, also impacts coalescence, with longer times favoring fusion. The geometry of the pairing channel, specifically its cross-sectional area, controls the number of droplets that can come into contact and coalesce. Additionally, droplet size and volume affect how efficiently droplets merge. Finally, electrocoalescence, which applies electric fields to induce droplet fusion, offers an alternative method for on-demand coalescence. By optimizing these factors—surfactant concentration, pairing time, channel geometry, and droplet size—reliable and selective droplet coalescence can be achieved in microfluidic systems. |
| What design elements are critical for producing double or multiple emulsions? | To generate double or multiple emulsions, several design elements are essential. Co-flow and flow-focusing configurations enable precise control over fluid flow, allowing for stable droplet formation, especially in double emulsions. Sequential emulsification steps, where inner droplets are emulsified into intermediate phases and then into outer phases, are key for multiple emulsions. Capillary microfluidic devices, such as consecutive capillary and coaxial configurations, are used to control |

| | droplet size, number, and monodispersity. Additionally, precise flow control and shear force management are crucial for stable and consistent droplet formation. Microchannel geometry optimization and material compatibility further enhance emulsification processes. By carefully managing these factors, researchers can produce double and multiple emulsions with controlled internal structures and compositions for various applications. |
|---|---|
| What key factors influence the design of flow-focusing droplet microfluidics optimized using machine learning? | The design of flow-focusing droplet microfluidics, optimized using machine learning, depends on several critical factors. Channel geometry, including the shape and size, controls droplet generation by defining the break-off point where the dispersed phase meets the continuous phase. Flow rates of both phases affect droplet size and frequency, with higher continuous phase flow reducing droplet size. Viscosity influences flow behavior and shear forces, crucial for stable droplet formation. Shear forces between phases are controlled by adjusting channel geometry or flow rates, ensuring consistent droplet generation. Machine learning algorithms predict and optimize droplet diameter and generation rate by analyzing these design parameters, fluid properties, and large datasets, automating the design process for more efficient droplet production in life sciences applications. |
| How do channel wall properties (hydrophobicity/hydrophilicity) influence droplet dynamics? | The channel wall properties, specifically hydrophobicity or hydrophilicity, significantly influence droplet dynamics in microfluidic systems. These properties determine the wettability of the channel walls, affecting the interaction between the continuous and dispersed phases. In hydrophobic channels, the dispersed phase (typically water) tends to form droplets as it minimizes contact with the channel walls. This enhances droplet stability and facilitates easier detachment during |

| | droplet formation. Hydrophobic surfaces are ideal for applications requiring water-in-oil emulsions or precise droplet control. Conversely, hydrophilic channels promote wetting of the dispersed phase, making it more likely to spread along the walls instead of forming discrete droplets. This is useful for oil-in-water systems or applications involving liquid films. Wall properties also impact droplet size, breakup, coalescence, and transport. Hydrophobicity favors smaller droplets and reduces the likelihood of droplet adherence, while hydrophilicity can lead to higher flow resistance and altered droplet generation behavior. Tailoring wall properties is crucial for optimizing microfluidic processes. |
|---|---|
| How can you incorporate active control mechanisms (e.g., electrical, magnetic) into droplet devices? | Active control mechanisms such as electrical, magnetic, thermal, and mechanical methods can precisely manipulate droplet formation in microfluidic devices. Electrical control methods include direct current (DC), which uses electrodes to control droplet size, and alternating current (AC), which utilizes electrowetting-on-dielectric (EWOD) and dielectrophoresis (DEP) to influence droplet behavior. Magnetic control employs ferrofluids and external magnets to non-invasively guide droplet formation. Thermal control involves heating the entire device or applying localized heating to adjust fluid parameters like viscosity and interfacial tension. Mechanical control techniques, such as pneumatic and piezoelectric components, can also regulate droplet dynamics. These active control strategies enable precise and flexible droplet generation, allowing for tailored applications in complex microfluidic systems, including droplet manipulation and multicomponent emulsion creation. |

| How do flow rates affect droplet size and frequency? | Flow rates significantly influence droplet size and frequency in microfluidic systems. As the flow rate ratio ($\varphi$ = Qd/Qc) increases, the shearing force from the continuous phase dominates, leading to smaller droplets due to rapid breakup of the dispersed phase. Additionally, higher flow rates in the continuous phase can result in larger droplets, as the dispersed phase doesn't break up until it reaches the downstream section. The frequency of droplet generation also increases with higher flow rate ratios, demonstrating a positive correlation. However, droplet diameter growth trends vary depending on the model and input conditions. Overall, increasing the flow rate ratio enhances droplet formation frequency and reduces droplet size due to stronger shearing forces, with complex interactions influenced by factors such as interfacial tension and channel geometry. |
|---|---|
| What techniques are available for droplet splitting? | Droplet splitting techniques can be categorized into passive and active methods. Passive techniques rely on the microchannel geometry and fluid properties, such as T and Y junctions, flow focusing configurations, or introducing ribs/obstructions in the flow path. These methods are simple but lack dynamic control over splitting ratios and are sensitive to flow conditions. Active techniques utilize external forces like electric, magnetic, acoustic, or thermal fields to precisely control droplet splitting. Examples include electric field-based splitting, magnetic fields for ferrofluids, acoustic radiation force, and thermocapillary methods. Active techniques offer more flexibility, enabling on-demand droplet splitting with controlled ratios, overcoming the limitations of passive methods. However, each technique has its own set of advantages and constraints depending on the application and droplet type. |
| What advantages does droplet formation by passive methods have over active methods? | Passive droplet generation methods offer several advantages over active methods. They simplify device design by relying on flow rate manipulation and microchannel geometry, eliminating the need for complex external forces. This leads to cost-effectiveness and easier fabrication. Passive methods also provide precise control over droplet size, shape, and monodispersity, making them ideal for applications requiring uniform droplets. Additionally, they protect droplets from external interference and cross-contamination, ensuring integrity. Passive methods can produce continuous droplet streams with uniform size distributions (1%-3% dispersity) without requiring external actuation. While they are not suitable for dynamic manipulation or "droplet-on-demand" applications, passive methods are favored for their simplicity, precision, and cost-efficiency in controlled environments. |
| How do you design a device for monodisperse droplet generation? | To design a device for monodisperse droplet generation, two common methods are used: cross-flow and flow-focusing designs. In cross-flow designs, the continuous phase flows around the dispersed phase in a constriction channel, with droplet size controlled by adjusting channel dimensions and shear stress. This method is simple but can lead to droplet damage due to shear forces or wall interactions. Flow-focusing designs, on the other hand, use sheath flows to stabilize the interface between phases, enabling smaller droplet formation with precise control over size by adjusting flow rates and orifice dimensions. This method offers high monodispersity and is less prone to droplet damage. Both designs require careful optimization of flow rates, channel dimensions, and surface tension to achieve uniform droplet sizes. The choice of design depends on specific needs such as droplet size, throughput, |
| | and resistance to damage during formation. |
| What materials are most commonly used for fabricating droplet microfluidic devices? | The most commonly used materials for fabricating droplet microfluidic devices are PDMS (polydimethylsiloxane) and silicon-based materials. PDMS is preferred due to its ease of fabrication using soft lithography, which allows for the creation of precise and versatile microfluidic devices. This material provides high transparency, efficient sealing, and the ability to mold complex microchannels, making it ideal for droplet generation and manipulation. While PDMS is dominant, alternatives such as thermoplastics (e.g., PMMA, polycarbonate), fluoropolymers (e.g., PTFE), and glass materials are also used, each offering unique benefits like better solvent resistance or mass production capabilities. Despite the variety of materials available, PDMS remains the material of choice for most droplet microfluidics applications due to its simplicity and versatility. |
| What are the advantages and limitations of using PDMS for microfluidic devices? | Polydimethylsiloxane (PDMS) is widely used for microfluidic devices due to its several advantages, including low toxicity, biocompatibility, optical transparency, and high gas permeability, which make it ideal for biomedical applications. Its flexibility and cost-effectiveness also contribute to its popularity, as it allows for easy fabrication using soft lithography. PDMS enables real-time observation and is well-suited for flexible and wearable electronics. However, PDMS has limitations, such as its hydrophobic nature, which can lead to fouling in microchannels, affecting performance. Additionally, its soft elastomeric properties limit mechanical strength, and large-scale production can be challenging. Despite these drawbacks, PDMS remains a versatile and widely used material, though further research is needed to |
| | address its limitations for broader applications. |
| How does 3D printing facilitate microfluidic device fabrication? | 3D printing revolutionizes microfluidic device fabrication by enabling the creation of complex and precise channel structures that are challenging to achieve with traditional methods like soft lithography. Techniques such as Stereolithography (SL), Multijet Modeling (MJM), and Fused Deposition Modeling (FDM) offer high resolution, complex geometries, and the integration of various components like valves and pumps. 3D printing also allows rapid prototyping, customization, and the use of diverse materials for biocompatibility and transparency. Advantages over traditional methods include reduced fabrication time and labor, increased accuracy, and environmental friendliness due to the additive manufacturing process. With its ability to create hierarchical 3D configurations and integrate multiple functions, 3D printing holds great potential for advancing microfluidic device design and performance. |
| What alternative materials can be used for applications involving organic solvents? | Several sustainable alternatives to traditional organic solvents are gaining attention for various applications. Supercritical CO2 is a non-toxic, environmentally friendly solvent, particularly useful in extraction processes. Ionic liquids, which are liquid salts at room temperature, offer low vapor pressure and broad application in chemical reactions. Organic carbonates, like dimethyl carbonate, are biodegradable and non-toxic, ideal for fine chemical synthesis. Biosolvents, derived from renewable resources such as plants and bacteria, provide a greener option for extraction and cleaning. Other alternatives include fluorous solvents, non-toxic liquid polymers like polyethylene glycol (PEG), water, ethanol, methanol, and ethyl acetate. These alternatives are chosen |
| | based on their specific properties such as polarity, boiling point, and environmental impact, promoting more sustainable chemical processes and reducing reliance on hazardous petroleum-based solvents. |
| How does device surface treatment affect droplet formation? | Surface treatment of PDMS microfluidic devices plays a critical role in determining droplet formation and stability. PDMS is naturally hydrophobic, suitable for generating water-in-oil (W/O) droplets but not oil-in-water (O/W) droplets. To generate O/W droplets, selective hydrophilic treatment is required. The researchers employed two methods: manual PVA deposition after plasma treatment, and a simultaneous approach combining PVA injection with on-chip plasma treatment. The latter method enables precise wettability patterning, allowing for both W/O and O/W droplet formation within the same device. The plasma treatment increases surface energy, making the PDMS hydrophilic, while PVA deposition stabilizes the coating, ensuring long-term device use. This technique provides controllable droplet sizes and stable double emulsion formation, with the ability to manipulate droplet characteristics by adjusting flow rates. The method also allows for reliable, reproducible results, making it a powerful tool for droplet microfluidic applications. |
| What bonding techniques are used to seal microfluidic channels? | Several bonding techniques are employed to seal microfluidic channels, each suited to different materials. Thermal fusion bonding involves compressing thermoplastic substrates at temperatures near their glass transition point, allowing them to fuse upon cooling. Chemical bonding creates covalent bonds between substrate surfaces through chemical reactions, while solvent bonding uses solvents to dissolve the materials partially, promoting adhesion through interdiffusion. For PDMS devices, oxygen plasma treatment is commonly used to |
| | make surfaces reactive, enabling irreversible bonding when in contact with another PDMS or glass piece. Additionally, corona wand bonding offers a less expensive alternative to oxygen plasma treatment, though it is less consistent. The choice of bonding technique depends on material properties, bond strength requirements, and available equipment, all of which are crucial for the successful fabrication of microfluidic devices. |
| What challenges arise when fabricating complex microfluidic geometries? | Fabricating complex microfluidic geometries presents several challenges, including controlling wettability in intricate channel structures, which is critical for fluid flow and droplet formation. Additionally, scaling up microfabrication processes for commercial applications requires cost-effective techniques and materials. Current droplet generation methods mainly produce single emulsions, limiting their use in applications that require multiple or complex emulsions. Moreover, droplet microfluidics faces limitations in producing diverse and functional nano-scale materials compared to other methods. Conventional lithography techniques, which are restricted to rectangular cross-sections, struggle with creating complex 3D geometries. Although advancements like 3D printing and stacking methods offer solutions, effective mixing strategies and versatile fabrication of 3D networks with the required features remain significant hurdles in the field. Addressing these challenges will require innovative approaches and collaboration across various disciplines. |
| How do you fabricate devices capable of handling multiphase flows? | Fabricating devices for handling multiphase flows involves several techniques and considerations. Key approaches include advanced micromanufacturing methods such as 3D |
| | printing, which eliminates the need for bonding processes, and flow-focusing techniques, which use nozzles and sheath fluids to control droplet formation. Controlling wettability is crucial for generating specific droplet types, while parallelization of droplet generators enhances production scalability. Challenges include ensuring even fluid distribution, integrating various components like valves and connectors, and addressing the lack of standardized design rules. Additionally, methods like consecutive capillary assemblies and external forces (e.g., pneumatic or electrical control) can be employed to create complex emulsions. Through careful design and optimization, efficient fabrication of multiphase flow devices for applications such as droplet generation can be achieved. |
| What is the impact of channel roughness on droplet formation and behavior? | The impact of channel roughness on droplet formation and behavior is a key factor in microfluidics. Studies indicate that surface roughness can significantly affect droplet dynamics, such as movement, spreading, and stability. For example, increased roughness can hinder droplet spreading and prevent secondary droplet formation by altering flow dynamics and the contact line. Additionally, vibration and surface structure play roles in influencing droplet impact and splash behavior. Rougher surfaces tend to reduce spreading, while smoother surfaces may enhance it. Surface wettability also interacts with roughness to further complicate droplet behavior. Overall, channel roughness introduces complexity to droplet dynamics, with its effects varying based on factors like roughness level, fluid type, and external conditions, highlighting the need for further research in this area for better understanding and optimization in microfluidic applications. |
| How can sensors be integrated into droplet microfluidic devices during fabrication? | Integrating sensors into droplet microfluidic devices during fabrication is a complex process that requires advanced microfabrication techniques and careful material selection. For successful integration, precision in manufacturing is crucial, as it often involves combining different microfluidic units, such as droplet generation and detection units, on a single chip. Various techniques like photolithography, soft lithography, and direct integration methods can be used, depending on the type of sensor (e.g., optical or electrochemical). The fabrication process also needs to account for material compatibility, sensor functionality, and alignment to ensure reliable performance. Additionally, sensor integration may face challenges such as maintaining stability within the microfluidic environment and minimizing interference with fluid flow. By considering these factors, sensors can be effectively integrated into droplet microfluidic devices for applications like chemical reactions and biological detection. |
| How are droplet microfluidic devices used for single-cell analysis? | Droplet microfluidic devices are widely used for single-cell analysis due to their ability to encapsulate individual cells within tiny droplets, allowing for high-throughput and precise experimentation. These devices work by generating microdroplets in a fluid flow, each containing a single cell along with reagents needed for analysis, such as enzymes, antibodies, or primers. This allows for the isolation of cells in a controlled environment, enabling high-resolution studies of gene expression, protein profiling, and cellular responses. The compartmentalized nature of droplets also reduces cross-contamination between samples. Moreover, the small size and scalability of droplet microfluidics make it possible to perform large-scale, parallel |

| | analysis of thousands of single cells in a short amount of time. This technology has applications in genomics, proteomics, and drug screening, facilitating insights into cell heterogeneity and the identification of rare cell populations. |
|---|---|
| What are the advantages of droplet microfluidics for high-throughput screening? | Droplet microfluidics offers several advantages for high-throughput screening, including reduced sample volumes, which lowers costs, and increased assay sensitivity through compartmentalization. This method allows for the screening of up to $10^8$ samples per day at kilohertz speeds, significantly improving throughput and efficiency. The small droplet volumes enhance effective concentrations and signal-to-noise ratios, critical for sensitive detection of low-concentration biological molecules such as antibodies, enzymes, and single genomes. Additionally, droplet microfluidics enables single-cell and single-molecule assays, providing detailed insights into cellular behaviors and allowing for novel experimental approaches. This technology is particularly beneficial for applications like drug discovery, enzyme evolution, and diagnostics, offering a versatile and cost-effective platform for advanced biological research. |
| How can droplets be used to encapsulate cells or biomolecules? | Droplet microfluidics provides versatile methods for encapsulating cells and biomolecules. Limiting dilution is a traditional approach that randomly distributes single cells or biomolecules in droplets, though it has low efficiency. Inertial microfluidics leverages spiral geometries and inertial forces to guide cells into droplets, improving encapsulation efficiency by ensuring single-cell droplets. Bi-disperse emulsions create an emulsion of empty and cell-containing droplets, with the latter selectively recovered using methods like |

| | pinched flow fractionation. For biomolecules, limiting dilution is commonly used, adjusting concentration and drop size to achieve single-molecule encapsulation. Additionally, droplets can encapsulate cells in lipid vesicles for biological applications or serve as microreactors in processes like DNA analysis. These techniques enable precise control over encapsulation, enhancing their applicability in cell-based assays, drug discovery, and genetic research |
|---|---|
| What are the applications of droplet microfluidics in drug discovery? | Droplet microfluidics plays a significant role in drug discovery by enabling high-throughput screening and precise analysis of drug interactions at the single-cell level. This technology allows the generation of picoliter-sized droplets, each encapsulating a single cell or a small group of cells, which can be exposed to various drug compounds. This compartmentalization facilitates parallel testing of numerous drug candidates, speeding up the identification of promising therapeutics. Droplet microfluidics is particularly useful in identifying drug efficacy, optimizing drug formulations, and studying complex cellular responses such as gene expression, protein production, or metabolic changes in response to treatment. Additionally, it enables the screening of rare cell populations or targets that would be difficult to isolate using traditional methods. The ability to rapidly test and analyse thousands of drug interactions in a controlled environment accelerates drug discovery, reduces costs, and improves the chances of identifying novel therapeutic agents. |
| How can droplet microfluidics enable precise control of chemical reactions? | Droplet microfluidics offers precise control over chemical reactions by dispensing reactants into individual droplets with accurate volume control. This enables the mixing of reactants |
| | within confined spaces, enhancing reaction efficiency. Coalesced droplets act as micro-reactors, providing fast mixing and long operation times without clogging, even with sticky or precipitating reactants. The system allows for optimization of reaction parameters, such as temperature, flow rate, and pressure, ensuring the production of specific products, like mesoporous silica microspheres with superior surface area and narrow pore distributions. This capability to finely control reaction conditions makes droplet microfluidics a powerful tool for precise, high-throughput chemical synthesis. |
| What are the key challenges of using droplet microfluidics in diagnostics? | Droplet microfluidics presents significant challenges in diagnostics, including preventing wetting and cross-contamination between aqueous droplets and immiscible organic liquids, which can compromise results. Monitoring oxygen levels within droplets is difficult due to the sensitivity of the liquid-liquid interface to mechanical perturbations and mass transfer. Developing stable detection chemistries is also complex, as the confined droplet environment can alter chemical reactions and impact accuracy. Additionally, challenges in automation and accessibility limit the widespread use of droplet microfluidics, particularly for high-throughput assays. Overcoming these issues requires advancements in surfactants, microfluidic engineering, and automated techniques, which could enhance the precision and applicability of droplet systems in diagnostics. |
| How are droplets used for DNA sequencing applications? | Droplet microfluidics plays a key role in DNA sequencing by enabling high-throughput analysis and precise targeting of genomic regions. This includes single-cell sequencing, where individual cells are encapsulated in droplets for independent gene expression profiling. Barcode-based sequencing allows for the labeling of |
| | nucleic acids with unique identifiers, enabling the tracking of specific genomes during analysis. Multiplexed PCR amplification is achieved by performing PCR in separate droplets, ensuring uniform amplification of multiple targets. Additionally, emulsion PCR combines droplet merger with PCR to amplify specific genomic regions, facilitating efficient sequencing. These techniques offer a miniaturized, efficient platform for analyzing complex genetic information, improving accuracy and throughput in genomic studies. |
| What role do droplets play in tissue engineering research? | Droplet microfluidics plays a pivotal role in tissue engineering by enabling precise fabrication of biomaterials and 3D cell culture systems. Droplets serve as templates for creating hydrogel-based materials, which can encapsulate cells for more accurate modeling of tissue structures. This technology facilitates the production of microcarriers and complex biomaterials, such as hollow bacterial cellulose (BC) microspheres, which form injectable scaffolds for wound healing, promoting high cell proliferation and tissue regeneration in vivo. Additionally, droplet-based microfluidics allows for the creation of core-shell structures, such as alginate shells or metal-organic frameworks, which offer controlled release of therapeutic agents. These advancements provide a powerful platform for studying tissue development and regenerative therapies. |
| How can droplet microfluidics aid in studying cell signaling pathways? | Droplet microfluidics offers a powerful platform for studying cell signaling pathways by enabling the precise isolation and manipulation of individual cells within controlled environments. This technology allows for the encapsulation of single cells in droplets, providing a controlled space to study cell-cell interactions and signaling events. The ability to merge droplets containing |
| | different cell types or reagents facilitates the investigation of cell communication, such as paracrine signaling, in a way that mimics in vivo conditions. Additionally, droplet-based microfluidics supports high-throughput screening and single-cell genomic studies, offering insights into phenotypic variations and signaling dynamics within genetically identical cell populations. This capability is particularly valuable for understanding complex cellular behaviors in development, differentiation, and disease. |
| What is the potential of droplet microfluidics in personalized medicine? | Droplet microfluidics holds great promise in personalized medicine by enabling precise, high-throughput analyses of biological samples. The technology allows for the optimization of treatments by testing drug combinations on specific cell lines, such as pancreatic cancer cells, to identify the most effective therapies. It also facilitates the early detection of cancers through highly sensitive detection of tumor markers in bodily fluids, offering a minimally invasive approach for patient monitoring. Additionally, droplet-based digital PCR (dPCR) enables the detection of rare sequences and precise quantification of nucleic acids, which is crucial for cancer research and tailoring treatments based on a patient's genetic background and tumor profile. Overall, droplet microfluidics can significantly advance personalized medicine by providing tailored treatment strategies and improving early detection for better patient outcomes. |
| How are droplet microfluidic devices used for nanoparticle synthesis? | Droplet microfluidic devices revolutionize nanoparticle synthesis by enabling precise control over reaction conditions within isolated microreactors. These droplets encapsulate well-defined quantities of reagents, ensuring rapid mixing and preventing contamination, even when using toxic or volatile chemicals. This approach addresses limitations like |

| | diffusion-limited mixing found in traditional systems, resulting in nanoparticles with controlled size, shape, and narrow size distribution. The method supports multi-step synthesis on millisecond timescales, allowing for the scalable production of high-quality nanoparticles with applications in fields such as drug delivery, imaging, and catalysis. |
|---|---|
| What role do droplets play in polymer particle production? | Droplet-based microfluidics provides a versatile platform for producing monodisperse polymer particles by encapsulating monomers or oligomers into droplets and solidifying them through chemical or physical crosslinking. This approach enables the creation of diverse particle types, including solid, hollow, asymmetric, and multicompartmental structures with precise control over size, shape, and morphology. Techniques such as UV-initiated polymerization and solvent evaporation facilitate the transformation of droplets into solid particles. Additionally, functional micro- or nanostructures can be incorporated within or on the surface of these particles by introducing specific ingredients into the droplets, expanding their potential applications in areas like drug delivery, tissue engineering, and advanced materials. |
| How can droplet microfluidics improve the reproducibility of chemical synthesis? | Droplet microfluidics enhances the reproducibility of chemical synthesis by providing precise control over reaction conditions and minimizing variability. Droplets act as isolated microreactors, ensuring accurate manipulation of parameters such as temperature and concentration, critical for consistent results. This system enables optimal mixing intensity and parallel processing in separate droplet channels, reducing batch-to-batch variability and ensuring uniform product quality. Automation further minimizes human error, making the |

| | process highly reliable. These advantages are particularly beneficial for synthesizing complex nanomaterials and exploring new reaction parameter spaces, leading to the production of high-quality, reproducible materials. |
|---|---|
| What are the applications of droplet microfluidics in catalysis studies? | Droplet microfluidics offers transformative applications in catalysis studies by enabling precise, efficient, and high-throughput investigations at the nanoscale. These systems facilitate enantioselective biocatalysis using nanoliter-scale devices that integrate microflow reactors and electrophoresis for enantiomeric differentiation. Droplets serve as isolated reaction containers, allowing for controlled catalytic transformations at the single-particle or cell level, with reaction volumes as small as a few picoliters. This confinement minimizes diffusion effects and enhances reaction efficiency. Applications span organic synthesis, biochemical analysis, and biotechnology, showcasing droplet microfluidics as a powerful tool for studying rapid reaction dynamics, optimizing catalysts, and exploring reaction mechanisms in real time. |
| How do droplets enhance the study of reaction kinetics? | Droplet-based microfluidics revolutionizes reaction kinetics studies by offering rapid, precise, and high-throughput analysis with minimal sample consumption. The chaotic mixing within picoliter droplets ensures millisecond-level time resolution for fast reactions. These systems minimize reagent use, making them ideal for bioanalysis where resources are costly or scarce. Precise control over droplet composition facilitates the creation of concentration gradients, enabling the measurement of kinetic parameters. High-throughput capabilities allow simultaneous monitoring of numerous reactions, significantly expanding experimental diversity. Additionally, extended tracking |

| | of individual droplets supports long-term monitoring of enzyme-inhibitor kinetics, providing deeper insights into reaction mechanisms. |
|---|---|
| What are the benefits of using droplets for emulsion polymerization? | Droplet-based systems offer significant advantages for emulsion polymerization, enhancing efficiency, precision, and scalability. They enable high production rates and precise control over polymer particle size and morphology, facilitating the synthesis of high-molar-mass polymers at faster reaction rates. Droplet parallelization allows the screening of diverse chemicals or concentrations and supports the encapsulation of millions of gene variants for quantitative emulsion PCR. These systems also empower directed enzyme or microorganism evolution through successive mutagenesis and selection rounds. Furthermore, droplet isolation prevents cross-contamination, adsorption, and solvent evaporation, making them ideal for applications like in vitro compartmentalization (IVC). This unique platform provides a robust, scalable approach for both analytical and biological studies. |
| How can microfluidics be used to generate smart materials? | Microfluidics enables the creation of smart materials through precise control of reaction conditions, rapid synthesis, and innovative templating methods. Droplet templating allows monodisperse droplets to act as structural templates, facilitating the synthesis of uniform particles, such as hydrogels for single-cell analysis. By manipulating solute and temperature gradients on short timescales, microfluidics supports the production of nanoscale materials, including semiconducting polymers, catalysts, and nanoparticles with unique optical and electronic properties. Its ability to rapidly explore complex reaction parameter spaces, coupled with advanced analytics and machine learning, accelerates material discovery and optimization. This approach |

| | offers transformative potential in biotechnology, energy, and electronics by enabling the generation of materials with tailored properties and enhanced functionality. |
|---|---|
| What are the applications of droplet microfluidics in environmental sensing? | Droplet microfluidics offers transformative applications in environmental sensing by enabling precise, miniaturized, and efficient monitoring systems. It supports on-site environmental surveillance through portable devices for rapid toxicity assessments and detection of contaminants, such as mercury ions. As controllable microreactors, droplet microfluidics facilitates the study of complex interfacial processes, aiding in chemical synthesis and degradation studies. Furthermore, it advances environmental sensing by enabling in situ visualization and precise flow control, enhancing our understanding of environmental processes. These capabilities position droplet microfluidics as a valuable tool for environmental science and engineering. |
| How can microfluidic droplets improve combinatorial chemistry? | Microfluidic droplets revolutionize combinatorial chemistry by drastically reducing reagent consumption and enhancing efficiency. Each droplet, acting as a picoliter-scale microreactor, requires six to eight orders of magnitude less material than traditional methods, minimizing waste and costs. This miniaturization enables high-throughput synthesis of large compound libraries, supporting applications like drug discovery and material synthesis. Additionally, droplet manipulation techniques allow for precise reaction control, combinatorial fusion of reagents, and multi-stage experiments with online analysis. These advantages make microfluidic droplets a powerful tool for automating complex chemical workflows |
| | and optimizing reaction conditions in a highly efficient and scalable manner. |
| What are the challenges of scaling up droplet-based chemical synthesis? | Scaling up droplet-based chemical synthesis faces several challenges. Achieving uniform fluid distribution in microfluidic devices remains difficult despite progress in parallel droplet generation, with 3D printing offering potential solutions for complex channel fabrication. Simplifying functional units for multistep reactions, such as droplet generation, merging, and phase separation, is vital for seamless continuous flow processes. Developing low-cost materials and standardizing chip designs are necessary for commercial viability, especially for high-throughput production in diverse fluids. Furthermore, creating complex droplet emulsions and expanding nanomaterial diversity are critical to meeting industrial demands. Addressing these challenges will enable safe, sustainable, and scalable microfluidic applications. |
| How can artificial intelligence enhance droplet microfluidic research? | Machine learning can be used to optimize the design of microfluidic devices. Deep Learning will be integrated to sort, monitor, and process the images. Furthermore, the multimodal agents can be used to study and design the research of droplet microfluidics, a possible excuse robot to perform fabrication and experiments automatically. |
| How can large language models enhance droplet microfluidic research? | LLMs can be used to build the Agents and Multi-Agents to synthesize knowledge of droplet microfluidics from literature such as research articles and books. Then, the idea hypotheses will be proposed by the agents, and the novelty and feasibility will be assessed. The agents will provide the autonomous platform to design and fabricate droplet microfluidic devices. Furthermore, LLMs have the researchers accelerate knowledge of droplet |
| | microfluidics. |
| How can autonomous droplet microfluidic research? | Autonomous droplet microfluidics is an emerging field with significant potential for scientific discovery, aiming to create highly integrated, automated systems for bio/chemical functions like PCR, sample preparation, and detection. Key challenges include the need for fast detection methods to probe small volume droplets, precise control over droplet dynamics in complex microchannel networks, and the development of virtual design platforms for optimizing system designs. Solutions under exploration include using advanced detection techniques such as single-molecule detection and fluorescence lifetime imaging, as well as integrating microelectronic actuators for droplet manipulation. Additionally, the development of virtual design and experiment platforms could streamline the creation of optimized microfluidic systems. The potential applications are vast, including directed evolution, closed-loop synthesis, and automated drug discovery, where systems could autonomously optimize compound combinations. Collaboration across disciplines like fluid dynamics, chemistry, and software engineering will be key to overcoming challenges and advancing this technology, paving the way for groundbreaking scientific advances. |
| What is the potential for fully integrated droplet microfluidic platforms? | Droplet microfluidics offers transformative solutions to global health challenges by enabling rapid, cost-effective, and high-throughput analysis of biological samples. By generating and manipulating tiny droplets as individual reaction chambers, this technology can revolutionize diagnostics, particularly in resource-limited settings. For example, it facilitates point-of-care testing for infectious diseases like tuberculosis, malaria, and COVID-19, providing results |

| | quickly and accurately. Droplet microfluidics is also crucial for multiplexed screening, enabling simultaneous detection of multiple biomarkers, which enhances early disease detection and monitoring. Additionally, it supports the development of low-cost diagnostic devices, reducing dependence on expensive laboratory infrastructure. In drug discovery, it accelerates the screening of therapeutic compounds, improving the treatment of diseases. The technology also advances personalized medicine by analysing single cells and their responses to treatments. Overall, droplet microfluidics addresses global health disparities by delivering accessible, scalable, and efficient solutions for disease diagnosis, monitoring, and treatment. |
|---|---|
| How can droplet microfluidics contribute to global health challenges? | Droplet microfluidics has significant potential to address global health challenges by enhancing diagnostic and therapeutic processes. It enables rapid and sensitive pathogen detection, which is critical for early disease diagnosis and containment. The technology's high-throughput capabilities allow simultaneous analysis of multiple samples, streamlining disease diagnosis and treatment. Additionally, droplet microfluidics can be integrated into point-of-care platforms, making it feasible for use in resource-limited settings. This approach can improve healthcare accessibility and efficiency. Furthermore, droplet microfluidics facilitates precise single-cell analysis, enabling advances in personalized medicine and more effective treatment strategies. While challenges such as cost, standardization, and integration with existing systems remain, continued development of this technology could significantly impact global health by improving detection, diagnostics, and treatment capabilities. |
| How can open-source designs impact the future of microfluidic devices? | Open-source designs for microfluidic devices hold the potential to revolutionize the field by increasing accessibility, reducing costs, and fostering collaboration. These designs allow for the creation of custom predictive models without requiring extensive expertise, enabling broader participation in device development. The use of open-source tools, such as Design Automation of Fluid Dynamics (DAFD), accelerates design and prototyping, enabling rapid iteration and optimization. Additionally, the integration of machine learning algorithms can drive the development of new microfluidic tools and optimize existing ones. This democratization of design expertise and the collaborative, community-driven approach will facilitate the creation of affordable, specialized devices, ultimately leading to breakthroughs in applications like single-cell analysis and global health diagnostics. The open-source model promises to transform microfluidics into a more inclusive and innovative field. |
| What are the challenges of scaling up microfluidic devices for industrial use? | Scaling up microfluidic devices for industrial use presents several challenges, including throughput limitations, lack of standardization, and integration complexities. The small micrometric channel sizes restrict production efficiency, and increasing parallelization or channel size may impact fabrication costs and reaction kinetics. There is also a lack of standardized design methods, making it difficult to compare results and adopt microfluidics in industrial settings. Additionally, scaling from laboratory to industrial production remains a hurdle, with issues in cost-effective microfabrication, generating complex droplet emulsions, and ensuring reliable fluid control. The integration of microfluidic devices with industrial interfaces, testing methods, and |

| | continuous processes is essential for successful commercialization. Overcoming these challenges requires innovative solutions, such as simplifying multistep reactions and developing standardized, low-cost fabrication methods. |
|---|---|
| How might droplet microfluidics impact agriculture? | Droplet microfluidics has the potential to transform agriculture by enhancing crop management and food security. It enables the precise release of fertilizers and pesticides, reducing waste and optimizing plant growth. The technology also supports improved crop monitoring through long-term studies, offering insights into plant development. By analyzing individual plant cells, droplet microfluidics can help uncover genetic and phenotypic traits, enhancing disease resistance and drought tolerance in crops. Additionally, it aids in disease detection, allowing for early intervention to prevent crop loss. Despite challenges such as scalability, cost, and integration with existing agricultural practices, ongoing research aims to overcome these barriers, making droplet microfluidics a promising tool for precision agriculture and sustainable food production. |
| What are the current limitations of droplet microfluidics, and how can they be overcome? | Droplet microfluidics has immense potential for high-throughput biological assays but faces several limitations that hinder its broader adoption. One key challenge is its efficiency for assays involving small sample sizes; droplet microfluidics excels with large volumes (at least $10^5$ samples), making traditional methods like microtiter plates more suitable for lower sample numbers. Another critical issue is maintaining a stable and inert oil-water interface, which is essential for droplet formation. Finding the right surfactant to keep this interface stable remains a complex chemistry challenge. Furthermore, droplet microfluidics lacks the spatial barcoding |

| | available in microtiter plates, making it difficult to organize and track multiple droplet libraries. As a result, developing alternative barcoding systems is necessary but challenging. The automation of droplet movement is another obstacle, as it requires precise control, often through techniques like electrowetting and dielectrophoresis. Lastly, droplet stability over time can be compromised, with issues such as shrinkage, coalescence, and cross-talk between droplets. To address these issues, research is underway to develop improved surfactants, more reliable barcoding systems, and novel materials and techniques to enhance droplet control. These advancements, coupled with better analytical tools, are expected to overcome current limitations and expand droplet microfluidics' applications in chemical and biological research. |
|---|---|
| How can interdisciplinary collaboration accelerate innovation in droplet microfluidics? | Interdisciplinary collaboration is essential for advancing droplet microfluidics by combining expertise from various fields such as chemistry, biology, material science, and engineering. By uniting these diverse areas, researchers can address complex challenges such as cost-effective fabrication, scalable production, and novel interfacial reaction systems. For example, biologists and chemists can leverage droplet microfluidics for high-throughput analysis, while material scientists and engineers work on developing more robust and affordable microfabrication methods. Furthermore, integrating technologies like sensitive detectors and machine learning algorithms can enhance the analytical capabilities of droplet-based systems. Collaborative initiatives, such as the EMBL Microfluidics Conference, foster knowledge exchange and innovation by bringing together professionals from different disciplines. This cooperation is crucial for overcoming |

| | technical barriers, reducing manufacturing costs, and developing new applications in fields like drug discovery, single-cell analysis, and nanomaterial synthesis. Ultimately, interdisciplinary collaboration accelerates the evolution of droplet microfluidics, pushing the boundaries of what is achievable in this promising field. |
| --- | --- |

## S3: QA by LLAMA-based Scientific Mentor

| question | answer | ground_truth | feedback | score | percentage | Expert Rating |
| --- | --- | --- | --- | --- | --- | --- |
| What is microfluidics? | Based on the provided context, microfluidics can be described as a versatile technological field that enables the study of | Microfluidics is a multidisciplinary field that involves the study and manipulation of fluids at the microscale, typically within | The response accurately describes the concept of microfluidics, its characteristi | 96 | 96.0 | 100 |

| | | | | | |
|---|---|---|---|---|---|
| | dynamic conditions and has various applications in biotechnology segments. It involves the manipulation of fluids in small volumes and channels, allowing for controlled experiments and analysis.<br><br>Microfluidics is characterized by its ability to:<br><br>1. Provide high-resolution and localized experimental applications<br>2. Enable single or multicell analysis<br>3. Offer a platform for monitoring biocatalysts in controlled microenvironments<br>4. Facilitate the study of fundamental biological mechanisms, such as motility, biochemical responses, interactions between cells, and biological | channels with dimensions ranging from tens to hundreds of micrometers. It is a multidisciplinary field that involves molecular analysis, molecular biology, and microelectronics. Microfluidics has been increasingly used in the biological sciences because precise and controlled experiments can be conducted at a lower cost and faster pace. Lab on a Chip devices use microfluidics for applications such as Point of Care testing of diseases, or Organ on a Chip studies. | cs, and various applications across multiple fields. It provides specific examples and details about lab-on-a-chip platforms and their potential uses in medical biotechnology and other areas.<br>SCORE: 96 | | | |

| | behavior | | | | | |
| --- | --- | --- | --- | --- | --- | --- |
| | The main vision of microfluidics is to miniaturize traditional chemical and biological systems into small devices that can take samples, analyze them, and provide answers. This has led to the development of lab-on-a-chip (LoC) platforms, which consist of a microfluidic chip connected to an automated computer system for controlling, analyzing, and detecting cells, reagents, and assays. | | | | | |
| | Microfluidics has numerous potential applications in various fields, including: | | | | | |
| | 1. Industrial biotechnology 2. Medical biotechnology 3. Tissue engineering 4. Material synthesis 5. Nanotechnology 6. Biochemistry | | | | | |

| | 7. Chemistry<br>8. Material science and metallurgy<br>9. Chemical engineering<br>10. Water treatment<br>11. Life sciences<br><br>Overall, microfluidics is a powerful technology that enables the manipulation of fluids in small volumes and channels, allowing for controlled experiments and analysis with high-resolution and localized experimental applications. | | | | | |
|---|---|---|---|---|---|---|
| What is droplet microfluidics? | Based on the provided context, droplet microfluidics refers to a technology that allows for the controlled manipulation of small volumes of liquid (typically in the range of 10-100 µm) in microchannels. This technology enables the efficient and rapid partitioning of single molecules, | Droplet-based microfluidics is a branch of microfluidics focusing on the creation and manipulation of discrete volumes of liquids through adoption of immiscible fluid flows. Microdroplets allow for handling miniature volumes (µL to fL) of fluids conveniently, | The response accurately describes the concept of droplet microfluidics, its applications, and benefits, but contains some minor inaccuracies in terms of volume ranges (10-100 µm instead of µL to fL) | 78 | 78.0 | 80 |

| | | | | | |
|---|---|---|---|---|---|
| | cells, or beads from complex samples, and can be used for various applications such as high-throughput assays, droplet-based chemical synthesis, and nanodevice fabrication.<br><br>In more detail, droplet microfluidics involves the creation of highly monodispersed droplets that can act as independent microreactors. These droplets can be manipulated in microchannels through the implementation of different modules for merging, splitting, incubating, or sorting. The use of surfactants can stabilize these droplets, allowing them to be used for a variety of "digital fluidic" operations.<br><br>The technology has dimensional scaling benefits that enable | provide better mixing, encapsulation, sorting, and sensing, and suit high throughput experiments. Compared with continuous-flow systems, droplet microfluidic systems enable isolation and manipulation of samples in separate compartments. The droplets generated with uniform size and shape provide a stable microenvironment for biochemical reactions. Processing of a large number of droplets in parallel facilitates to achieve ultrahigh throughput. Confinement of individual targets within the droplets and independent processing of each of them allow high-precision analyses. | and lacks precision in explaining the advantages over continuous-flow systems. SCORE: 78 | | | |
| | controlled and rapid mixing of fluids in the droplet reactors, resulting in decreased reaction times. This, combined with precise generation and repeatability of droplet operations, makes droplet-based microfluidic systems a potent high-throughput platform for biomedical research and applications.<br><br>In summary, droplet microfluidics is a technology that enables the controlled manipulation of small volumes of liquid in microchannels, allowing for efficient and rapid partitioning of single molecules, cells, or beads from complex samples. | | | | | |
| How to generate microdroplets? | Based on the provided context, there are several ways to generate microdroplets in microfluidic devices. | Microdroplet generation in microfluidics is achieved through two main methods: active and passive. | The response provides a clear and detailed explanation of | 81 | 81.0 | 85 |

| | According to Document 3, which has a relevance score of 0.99394965, the most commonly used methods for microdroplet generation can be divided into two main categories: active and passive.

**Active Methods:**

* A piezoelectric actuator-based droplet generator is described in Document 2, which allows for independent control over droplet size, rate of formation, and distance between droplets. This device is able to generate microdroplets with high uniformity, deviating from the mean value by less than 0.3%.

**Passive Methods:**

* The context mentions that PDMS-based microfluidic devices are used | Active methods utilize external forces like piezoelectric actuators to precisely control droplet size, formation rate, and spacing. These methods enable high uniformity in droplet generation, with deviations less than 0.3%. Passive methods rely on fluid properties and channel geometry, such as T-junctions or cross-flow configurations, to create droplets. These methods are influenced by factors like flow rate, viscosity, and interfacial tension, allowing for fine control over droplet size and uniformity. For example, PDMS-based microfluidic devices have been used for nanoparticle synthesis, where hydrodynamic coupling of nozzles generates droplet pairs for precise mixing. | microdroplet generation methods, accurately categorizing them into active and passive techniques. It correctly describes the characteristics and applications of each method, including the use of piezoelectric actuators, PDMS-based devices, and thermally isolated zones. However, it does not fully address the question's request for a comprehensive overview, as it relies heavily on specific documents and figures mentioned in the provided context. SCORE: 81 | | | |

| | to generate microdroplets for bulk synthesis of nanoparticles (Document 1). Specifically, a device designed by Frenz et al. enables the production of droplet pairs based on hydrodynamic coupling of two separated nozzles. * Another method mentioned is the use of thermally isolated zones in a droplet microreactor for the synthesis of TiO2 nanoparticles (Figure 4A).

Additionally, Document 3 mentions that other methods have been explored to generate microdroplets, including:

* Cross flow and T-junction configurations
* Adjusting channel geometry, two-phase viscosity, flow rate, wettability, and interfacial tension | Both methods are valuable in applications like nanoparticle synthesis, offering high reproducibility and control over reaction conditions in biomedical and material science research. | | | | |
|---|---|---|---|---|---|---|

| | These methods are used to achieve precise regulation over size, shape, uniformity, and other factors of microdroplets. | | | | | |
|---|---|---|---|---|---|---|
| What materials are typically used to fabricate droplet microfluidic devices? | Based on the provided context, the materials commonly used to fabricate droplet microfluidic devices include:<br><br>1. PDMS (Polydimethylsiloxane): This is a popular material for making droplet-based microfluidic devices due to its ease of use and flexibility.<br>2. Thermoset polyesters: These are another type of elastomeric material that can be used as an alternative to PDMS.<br>3. Thermoplastics, such as:<br>    * Poly(methylmethacrylate) (PMMA)<br>    * Polycarbonate<br>    * Polystyrene<br>4. Amorphous polymers, such | Droplet microfluidic devices are commonly fabricated from a variety of materials, each offering distinct advantages for specific applications. The most widely used material is PDMS (Polydimethylsiloxane), favored for its ease of use with soft lithography and flexibility, though it has limitations such as poor solvent resistance. Alternatives to PDMS include thermoset polyesters, thermoplastics like PMMA, polycarbonate, and polystyrene, and amorphous polymers such as cyclic olefin copolymer. Fluoropolymers like PTFE and glass are also | The response accurately lists various materials used for fabricating droplet microfluidic devices, including their properties and advantages. It provides a clear and detailed explanation of the selection criteria for these materials.<br>SCORE: 96 | 96 | 96.0 | 95 |
| | as:<br>      * Cyclic olefin copolymer<br>5. Fluoropolymers, such as:<br>      * PTFE (Polytetrafluoroethylene)<br>6. Glasses: These can be used to make capillary devices for droplet microfluidics.<br>7. Silicon: This is a common material for making microfluidic devices using soft lithography.<br><br>These materials are chosen based on their properties, such as solvent resistance, biofouling, and the ability to mass-produce devices using techniques like injection molding or hot embossing. | used, offering superior solvent resistance and biofouling resistance. Silicon and capillary devices are utilized in certain applications, providing precise control for microfluidic operations. These materials are chosen based on factors like solvent resistance, biofouling, and the ability to mass-produce devices using methods such as injection molding or hot embossing, ensuring broad applicability in research and industrial settings. | | | | |
| What are the applications of droplet microfluidics? | Based on the provided context, the applications of droplet microfluidics can be summarized as follows:<br><br>**Biological Applications** | Droplet microfluidics has diverse applications across biomedical research, diagnostics, and industrial fields. It enables high-throughput | The response provides a detailed list of applications of droplet microfluidics across various | 76 | 76.0 | 75 |

| | | | | |
|---|---|---|---|---|
| | 1. **Single-cell analysis**: Droplet microfluidics enables the manipulation and analysis of individual cells, allowing for the study of cell heterogeneity, gene expression, and protein expression at the single-cell level. 2. **Medical diagnostics**: The technology is used in medical diagnostics, including the identification and quantification of certain contents, screening, and real-time dynamic monitoring. 3. **Biological screening**: Droplet microfluidics is applied in biological screening, segment labeling, cell packaging, virus detection, and specific proteins or tissue remodeling.<br><br>**Chemical Applications**<br><br>1. **Protein | single-cell analysis, such as genomics, transcriptomics, and drug screening, while also facilitating nanoparticle synthesis, enzyme evolution, and chemical reactions. In healthcare, it supports personalized medicine, point-of-care diagnostics, and vaccine development. Beyond biology, it is used for environmental monitoring, food safety testing, and the development of cosmetics and pharmaceuticals. Its ability to create isolated, uniform reaction chambers in droplets makes it a powerful tool for precise, cost-effective, and scalable workflows in science and industry. | fields, including biological, chemical, and industrial sectors. It covers specific examples such as single-cell analysis, protein crystallization screening, drug screening, and the preparation of monodisperse core-shell microparticles. However, some of these points are not explicitly mentioned in the ground truth answer, and the response could be more concise and focused on the key aspects of droplet microfluidics applications. SCORE: 76 | | |

| | crystallization screening**: The technology enables the high-throughput screening of protein crystallization conditions. 2. **Drug screening**: Droplet microfluidics is used in drug screening, allowing for the testing of thousands of reaction conditions and chemicals with low sample/reagent consumption. 3. **Catalyst screening**: The technology is applied in catalyst screening, enabling the evaluation of various catalysts and their performance.

**Other Applications**

1. **Food and feed industry**: Droplet microfluidics has applications in the food and feed industry, including the | | | | | |
|---|---|---|---|---|---|
| | analysis and processing of food and feed samples.<br><br>2. **Environmental engineering**: The technology is used in environmental engineering, including the monitoring of dynamic release of organs and the collection of separated components in an electrophoresis system.<br><br>**Synthetic Microparticles**<br><br>1. **Preparation of monodisperse core-shell microparticles**: Droplet microfluidics is used to prepare monodisperse core-shell microparticles and multi-chamber microcapsules.<br><br>2. **Regulation of particle size, morphology, and shell thickness**: The technology allows for the precise regulation of particle size, morphology, | | | | |

| | monodispersity, and shell thickness by altering flow rate, physical parameters, and channel geometry.<br><br>These applications demonstrate the versatility and potential of droplet microfluidics in various fields, including biology, chemistry, food and feed industry, environmental engineering, and more. | | | | |
|---|---|---|---|---|---|
| What are the main benefits of using droplet microfluidics over bulk systems? | Based on the provided context, the main benefits of using droplet microfluidics over bulk systems include:<br><br>1. **Individual compartmentaliza tion**: Droplet microfluidics allows for individual cells or molecules to be isolated in their own tiny liquid compartments, enabling precise control and manipulation.<br>2. **Highly controllable local environments**: | Droplet microfluidics offers significant advantages over traditional bulk systems. It enables individual compartmentaliza tion, isolating single cells or molecules in tiny liquid compartments for precise control. This allows for single-cell analysis and early detection of rare cells or secreted molecules. The technology supports massive parallel processing, | The response is well-structured and provides a clear list of benefits of using droplet microfluidic s over bulk systems. It covers various aspects such as compartmen talization, controllable environment s, high-throughput screening, iterative | 82 | 82.0 | 80 |

| | | | | | |
|---|---|---|---|---|---|
| | The system provides a controlled environment for each droplet, allowing for consistent and reproducible results. 3. **Massive parallel processing**: Droplet microfluidics enables the analysis of millions of individual droplets simultaneously, making it possible to examine phenotypic and genetic variabilities at the level of cells or small populations. 4. **High-throughput screening**: The system allows for high-speed testing of large numbers of samples, making it ideal for applications such as drug discovery and validation. 5. **Iterative operations**: Automated droplet chips can perform complex experimental protocols, | analyzing millions of droplets simultaneously, ideal for high-throughput screening and examining phenotypic or genetic variability. Additionally, iterative operations such as merging, mixing, and sorting droplets enable complex experimental protocols and continuous monitoring of populations. The system's highly controllable local environments ensure consistent, reproducible results. These benefits make droplet microfluidics a versatile tool in drug discovery, medical diagnostics, single-cell analysis, and other fields, offering powerful insights and precise manipulation at the individual level. | operations, single-cell resolution, early detection, and massive parallel processing. However, the response could be improved by condensing some points and providing more concise language to match the clarity and precision of the Ground Truth Answer. SCORE: 82 | | | |

| | including merging, mixing, splitting, sorting, and incubation, allowing for multiple measurements on the same droplets or tracking population evolution in controllably changing chemical environments. 6. **Single-cell resolution**: Droplet microfluidics enables single-cell analysis, which is essential for understanding cell heterogeneity, interrogating life activities, and pathologic development. 7. **Early detection of cells and secreted molecules**: The system allows for the early detection of cells and their products, making it possible to isolate and culture rare individuals.<br><br>These benefits make droplet microfluidics a powerful tool for various applications, | | | | | |

| | | | | | |
|---|---|---|---|---|---|
| | including single-cell analysis, medical diagnostics, drug discovery, food and feed industry, and environmental engineering. | | | | |
| What are the advantages of droplet microfluidic systems compared with continuous-flow systems? | According to the provided context, droplet-based microfluidic systems offer several advantages over traditional continuous-flow (or single-phase) microfluidic systems. These advantages include:<br><br>1. **Overcoming problems related to Taylor dispersion, surface-molecule interactions, and slow mixing**: Droplet-based microfluidics can mitigate issues that plague continuous-flow systems, such as poor mixing and contamination.<br>2. **Ultra-small assay volumes**: Droplet-based microfluidic systems enable the definition and formation of | Droplet-based microfluidic systems offer significant advantages over traditional continuous-flow systems. These include the ability to overcome issues such as Taylor dispersion, slow mixing, and surface-molecule interactions. Droplet systems allow for ultra-small assay volumes, reducing reagent consumption and improving sensitivity. They also minimize cross-contamination and sample adsorption by compartmentalizing individual samples. The system's chaotic advection enhances mixing efficiency, enabling faster reactions. With | The response accurately lists the advantages of droplet microfluidic systems compared to continuous-flow systems, covering various aspects such as overcoming mixing issues, ultra-small assay volumes, reduced contamination, enhanced mixing, high droplet generation frequencies, precise control over droplet payload, and fast mixing times. It also correctly identifies the applications | 96 | 96.0 | 100 |

| | | | | | |
|---|---|---|---|---|---|
| | extremely small assay volumes, which is not possible with traditional microfluidic systems.<br>3. **Negligible cross-contamination and sample adsorption**: The compartmentalization provided by droplets minimizes the risk of contamination and sample loss.<br>4. **Enhanced mixing due to chaotic advection**: Droplet-based microfluidics can achieve faster and more efficient mixing compared to continuous-flow systems.<br>5. **Exceptionally high droplet generation frequencies (up to hundreds of kHz)**: This allows for ultra-high-throughput experimentation, which is not possible with traditional methods.<br>6. **Ability to control the droplet payload in a rapid | droplet generation frequencies up to hundreds of kHz, droplet microfluidics supports high-throughput experimentation. Furthermore, the rapid and precise control of droplet payloads and sub-millisecond mixing times make it ideal for applications requiring fast, controlled analysis. These benefits make droplet-based systems particularly valuable for cell-based assays, enzyme kinetics, and high-throughput screening. | where these benefits are particularly valuable. The response is clear, concise, and well-structured, with each point being relevant to the question. SCORE: 96 | | |

| | | | | | | |
|---|---|---|---|---|---|---|
| | and efficient manner**: Droplet-based microfluidics enables precise control over the contents of each droplet. 7. **Access to sub-millisecond mixing times**: This is another advantage that allows for faster experimentation and analysis.<br><br>These benefits make droplet-based microfluidic systems an attractive choice for various applications, including cell-based assays, enzyme kinetics, protein expression, and high-throughput screening. | | | | | |
| What are the basic principles behind droplet generation in microfluidics? | Based on the provided context, the basic principles behind droplet generation in microfluidics can be summarized as follows:<br><br>1. **Emulsion process**: Droplet generation in | Droplet generation in microfluidics relies on emulsion processes, where an oil phase (continuous phase) encapsulates an aqueous phase (dispersed phase) to form droplets. Surfactants are added to lower | The response accurately summarizes the basic principles behind droplet generation in microfluidics, including emulsion process, | 96 | 96.0 | 100 |

| | microchannels is based on an emulsion process, where a continuous phase (usually oil) encapsulates a dispersed phase (usually aqueous solution) in small droplets. 2. **Surfactants**: Surfactants are added to the continuous phase to lower the interfacial energy, facilitate the formation of new interfaces, and stabilize the formed emulsion droplets from coalescence by adsorbing at the liquid/liquid interface. 3. **Controlled flow**: The production of droplets relies on controlled flow of at least two liquids, which can be achieved using syringe pumps, hydrostatic reservoirs, or even vacuum. 4. **Surface-induced instability**: Droplets are typically generated by a | interfacial tension, stabilize the droplets, and prevent coalescence. The process requires precise control over the flow of both liquids, often achieved using syringe pumps or pressure-driven systems. Droplets are generated through surface-induced instabilities, influenced by microchannel geometries such as T-junctions or flow-focusing, which control droplet size and frequency. The design and flow control of the system determine the size uniformity and production rate of droplets. These principles enable high-throughput applications such as drug delivery, high-throughput screening, and microreactors for bio-catalysis, ensuring reproducibility and productivity in biological experiments. | surfactants, controlled flow, surface-induced instability, and droplet manipulation. It provides a clear and concise overview of the key concepts involved in generating uniform microdroplets. SCORE: 96 | | | |
|---|---|---|---|---|---|

| | surface-induced instability using different device geometries, depending on the required droplet frequency, monodispersity, or variability in droplet size. 5. **Droplet manipulation**: The production of microdroplets can be achieved through passive or active methods, including T-junction and flow-focusing geometries, co-flow streams, step emulsification, and parallel devices.<br><br>These principles are essential for the generation of uniform microdroplets, which is a crucial aspect of droplet-based microfluidics. The controlled production of droplets enables various applications, such as high-throughput screening, reproducibility, and new interpretations for | | | | | |
|---|---|---|---|---|---|---|
| | productivity responses in biological applications, including drug delivery and microreactors for (bio)catalysis. | | | | | |
| How is droplet size controlled in a droplet generator? | Based on the provided context, it appears that droplet size control in a droplet generator can be achieved through various methods depending on the type of device used.<br><br>In the case of passive droplet generators, such as those based on microfluidic devices made from polydimethylsilox ane (PDMS), the droplet size is controlled by adjusting the dimensions and geometries of the microchannels or controlling the flow rates of the two immiscible liquids. This method allows for precise control over parameters such as size, shape, monodispersity, | Droplet size control in microfluidic generators can be achieved through several methods. In passive droplet generation, adjusting microchannel dimensions or controlling the flow rates of immiscible liquids allows for precise control over droplet size, shape, and monodispersity. Tree-splitters, which bisect large droplets into smaller ones, also influence droplet size and uniformity, with splitting angles affecting droplet diameter. In active droplet generation, technologies like piezoelectric actuators independently control droplet size, formation rate, and spacing, | The response is clear and concise, accurately addressing the question about droplet size control in a droplet generator. It provides relevant information on both passive and active methods of droplet generation, including microchanne l dimensions, flow rates, tree-splitters, and piezoelectric actuators. The answer aligns well with the ground truth, covering key points such as precise | 96 | 96.0 | 100 |

| | and components inside the droplets.<br><br>In particular, the context mentions that the influence of the velocity of continuous and dispersed phases on droplet generation was tested in a device using a single pressure pump to control the velocity of two phases. This suggests that the flow rates of the liquids can be controlled to affect the size of the generated droplets.<br><br>Additionally, the context discusses tree-splitters as a technique for scaling up the capacity and efficiency of droplet generation. The design of tree-splitters with different splitting angles allows for control over the diameter and circularity of the droplets in the collection chamber. Specifically, it is | offering high precision. Additionally, the velocity of the continuous and dispersed phases plays a significant role in controlling droplet size. These methods enable the generation of uniform droplets for various applications, such as high-throughput screening and drug delivery. | control over droplet size, shape, and monodispersity in passive methods, and high precision in active methods. SCORE: 96 | | | |
|---|---|---|---|---|---|---|

| | mentioned that the size of droplets increased with increasing splitting angles under the same velocity conditions due to the difference in channel splitter dimensions.

In contrast, active droplet generators, such as those based on piezoelectric actuators, can also control droplet size by independently controlling the droplet size, rate of formation, and distance between droplets. The context mentions that a simple and robust droplet generator based on a piezoelectric actuator is able to control these parameters with high precision.

In summary, droplet size control in a droplet generator can be achieved through various methods, including:

* Adjusting | | | | | |
|---|---|---|---|---|---|---|
| | microchannel dimensions and geometries or controlling flow rates of immiscible liquids (passive method)<br>* Using tree-splitters with different splitting angles<br>* Independently controlling the velocity of continuous and dispersed phases using a single pressure pump<br>* Utilizing active droplet generators based on piezoelectric actuators to control droplet size, rate of formation, and distance between droplets. | | | | | |
| What is the role of surface tension in droplet formation? | Based on the provided context, the role of surface tension in droplet formation is a crucial one. Surface tension plays a significant role in determining the size and shape of droplets formed in microfluidic systems.<br><br>According to the documents, | Surface tension plays a critical role in droplet formation in microfluidic systems by influencing droplet size, shape, and formation mechanisms. It acts to minimize surface area, favoring spherical droplet shapes. At low Weber numbers (We < | The response accurately describes the role of surface tension in droplet formation, including its influence on droplet size, shape, and formation mechanisms, as well as its competition | 96 | 96.0 | 100 |

| | | | | | |
|---|---|---|---|---|---|
| | surface tension acts as a force that opposes the formation of droplets, particularly when the Weber number (We) is low (i.e., We < 1). In this regime, the inertial forces imposed by the flow of the dispersed phase are overcome by the surface tension forces, resulting in dripping or jetting behavior.

However, as the Weber number increases (We > 1), fluid inertia becomes more significant, and the droplet formation process involves a competition between fluid inertia and interfacial tension. In this regime, the capillary number (Ca) also plays a crucial role in determining the size of the droplets formed.

The documents suggest that surface tension | 1), where surface tension dominates, droplets form through dripping or squeezing. As the Weber number increases, fluid inertia competes with surface tension, leading to jetting behavior. The capillary number (Ca) also influences droplet size, with surface tension controlling droplet formation at low Ca. Techniques like electrowetting can measure surface tension in microfluidic systems, providing insights into the interfacial properties. Additionally, surface tension is a key parameter in T-junctions and flow focusing devices, where droplet size measurements offer an indirect way to assess surface tension at the time-scale of droplet production. Understanding the interplay | with fluid inertia and viscous shear stress. It also correctly mentions various techniques for measuring surface tension in microfluidic systems and its significance in T-junctions and flow focusing devices. The response is clear, concise, and relevant to the question, with a good balance of detail and concision. SCORE: 96 | | | |

| | measurements are feasible in microfluidic systems using techniques such as electrowetting, which can measure surface tension with droplet sizes down to 20 nL. Additionally, the effect of surfactants on surface tension is also measurable in microfluidic channels, and the surface tension of the interface is comprised of the surface tension without surfactant and the equilibrium surface tension in the presence of surfactant.

Furthermore, the documents highlight that surface tension is one of the parameters controlling droplet size in T-junctions or flow focusing junctions. Therefore, measuring droplet size can be an indirect way to access the surface tension at the | between surface tension and other forces is essential for precise droplet control. | | | | |
|---|---|---|---|---|---|---|
| | time-scale of droplet production.<br><br>In summary, surface tension plays a significant role in determining the size and shape of droplets formed in microfluidic systems, particularly through its competition with fluid inertia and viscous shear stress. Measuring surface tension is also feasible in microfluidic systems using various techniques, which can provide valuable information on the interfacial properties at small scales and with dynamics much below 1 s. | | | | | |
| How does a flow-focusing device differ from a T-junction in droplet formation? | According to the provided context, a flow-focusing device differs from a T-junction in droplet formation in several ways:<br><br>1. **Geometry**: In a flow-focusing device, the dispersed and | Flow-focusing devices and T-junctions are two common microfluidic designs for droplet formation, differing in geometry and mechanism. A flow-focusing device uses a central dispersed | The response provides a detailed comparison of flow-focusing devices and T-junctions in terms of geometry, shear forces, droplet | 81 | 81.0 | 80 |

| | | | | | |
|---|---|---|---|---|---|
| | continuous phases are forced through a narrow region, whereas in a T-junction, the inlet channel containing the dispersed phase perpendicularly intersects the main channel containing the continuous phase. 2. **Shear forces**: The shear forces generated by the continuous phase in a T-junction cause the head of the dispersed phase to elongate into the main channel until it breaks off into a droplet. In contrast, flow-focusing devices involve large elongational flow as the two immiscible liquid phases pass through a small orifice, generating smaller droplets. 3. **Droplet formation mechanism**: The droplets in a T-junction are formed due to the shearing force between the two materials and hydrodynamic | phase channel and side continuous phase channels converging at a narrow orifice, where hydrodynamic focusing pinches off droplets. It produces smaller, highly uniform droplets with precise control over size and high throughput. In contrast, a T-junction has perpendicular channels where the continuous phase shears the dispersed phase at the intersection, forming larger droplets with less uniformity and lower throughput. Flow-focusing is ideal for applications requiring monodisperse droplets, while T-junctions suit simpler or larger-droplet applications. | formation mechanism, and flexibility. It accurately describes the differences between the two microfluidic designs for droplet formation, including the underlying mechanisms, geometries, and operational characteristics. However, it could be more concise and directly address the question without unnecessary elaboration on general statements about microfluidics. SCORE: 81 | | | |

| | focusing exerted by the continuous phase. In a flow-focusing device, droplets are formed via Rayleigh-Plateau instability as the co-flowing liquids decay into droplets.<br><br>4. **Flexibility**: Flow-focusing devices offer more flexibility in generated droplet size compared to T-junctions, as the effective geometry can be varied by adjusting the flow rates of the dispersed and continuous phases.<br><br>Overall, while both T-junctions and flow-focusing devices are used for generating droplets in microfluidics, they differ in their underlying mechanisms, geometries, and operational characteristics. | | | | | |
|---|---|---|---|---|---|---|
| What role does the geometry of the microchannel play in droplet formation? | The geometry of the microchannel plays a crucial role in determining how | The geometry of a microchannel plays a critical role in droplet formation, | The response accurately addresses the question | 76 | 76.0 | 70 |

| | | | | | |
|---|---|---|---|---|---|
| | droplets "break off" and form in microfluidic devices. According to the provided context, the geometry of the microchannel can influence the break-up of drops through three main mechanisms:<br><br>1. **Stretching**: In flow-focusing geometries (Figure 5A), the shear forces generated by the continuous phase cause the head of the dispersed phase to elongate into the main channel until it breaks off into a droplet.<br>2. **Confluent streams**: Microcapillary channels can generate confluence streams that break up the dispersed phase into droplets (Figure 5B).<br>3. **T or Y-shaped channels**: The geometry of T or Y-shaped channels can cause the | influencing the size, shape, and generation frequency of droplets. Key geometric factors include the channel width, depth, and junction design, such as T-junctions or flow-focusing geometries. These features govern the interaction between the continuous and dispersed phases, affecting the shear forces and pressure gradients that drive droplet breakup. For instance, narrower channels or smaller orifices promote the formation of smaller droplets due to increased shear stress. Similarly, the junction type controls droplet detachment mechanics; in T-junctions, droplets form due to the balance of interfacial tension and shear force, while in flow-focusing designs, they form under | by providing a detailed explanation of how the geometry of the microchannel influences droplet formation. It correctly identifies three main mechanisms (stretching, confluent streams, and T or Y-shaped channels) and mentions the importance of surface properties in controlling droplet characteristics. However, it lacks specific details about the impact of channel width, depth, and junction design on droplet size, shape, and generation frequency, which are key aspects mentioned in the | | | |

| | | | | | |
|---|---|---|---|---|---|
| | dispersed phase to break off into droplets through streams or settings within these channels (Figures 5C and D).<br><br>Furthermore, the context mentions that the microchannel's geometry can be modified to control the size of the droplets formed. For example, altering the fluid flow rates, channel widths, or relative viscosity between the two phases in T-junctions can change the sizes of the droplets.<br><br>Additionally, the surface properties of the microchannel, such as hydrophobicity, can also affect the formation of droplets. The context mentions that modifying the surface of PDMS channels using 1H,1H,2H,2H-perfluorooctyl-trichlorosilane in toluene solution creates a | pinching at the constriction. Channel geometry also determines flow regimes ranging from dripping to jetting further impacting droplet characteristics. Thus, optimizing microchannel geometry is crucial for precise control over droplet formation in applications like microfluidics, emulsification, and lab-on-a-chip systems. | ground truth answer. The response also does not explicitly discuss flow regimes or the balance of interfacial tension and shear force in T-junctions.<br><br>SCORE: 76 | | |

| | | | | | | |
|---|---|---|---|---|---|---|
| | hydrophobic surface, which can influence the size and formation of droplets.<br><br>In summary, the geometry of the microchannel plays a significant role in determining how droplets form and break off in microfluidic devices, and its properties, such as surface modification, can be tailored to control the characteristics of the formed droplets. | | | | | |
| What factors affect droplet coalescence in microfluidic devices? | Based on the provided context, several factors affect droplet coalescence in microfluidic devices. These include:<br><br>1. **Surfactant concentration**: The study shows that the concentration of surfactant used can significantly impact droplet coalescence. Specifically, it was found that using 7.5 mM | Droplet coalescence in microfluidic devices is influenced by several factors. Surfactant concentration plays a key role, with higher concentrations preventing droplet fusion, while lower concentrations promote coalescence. The duration droplets remain in contact within the pairing channel, known | The response accurately lists several key factors affecting droplet coalescence in microfluidic devices, including surfactant concentration, pairing time, geometrical constraints of the pairing channel, | 85 | 85.0 | 85 |

| | | | | | |
|---|---|---|---|---|---|
| | EA-surfactant resulted in no fusion of droplets (p(c) = 0), while lower concentrations (1.5 mM and 3 mM) led to complete fusion (p(c) = 1). 2. **Pairing time**: The study demonstrates that the time during which two droplets stay in physical contact within the pairing channel before entering the coalescence channel affects droplet coalescence. For example, using 3 mM EA-surfactant, droplet pairs always coalesced when s > 8 ms and stayed unfused when s < 4 ms. 3. **Geometrical constraints of the pairing channel**: The study shows that the cross-sectional area of the pairing channel can influence droplet coalescence. By creating a microfluidic device with a | as pairing time, also impacts coalescence, with longer times favoring fusion. The geometry of the pairing channel, specifically its cross-sectional area, controls the number of droplets that can come into contact and coalesce. Additionally, droplet size and volume affect how efficiently droplets merge. Finally, electrocoalescence, which applies electric fields to induce droplet fusion, offers an alternative method for on-demand coalescence. By optimizing these factors—surfactant concentration, pairing time, channel geometry, and droplet size—reliable and selective droplet coalescence can be achieved in microfluidic systems. | droplet size and volume, and electrocoalescence. However, it slightly deviates from the ground truth by providing more detailed information on specific experimental results (e.g., the effect of 7.5 mM EA-surfactant) rather than a concise summary of the general factors influencing coalescence. SCORE: 85 | | | |

| | specific geometry, it was possible to achieve selective multiple coalescence events by controlling the number of emulsion droplets in physical contact with droplets produced on-chip.<br>4. **Droplet size and volume**: The study highlights that the dimensions of the microfluidic channel and the emulsion droplets can impact droplet coalescence. In this case, the cross-sectional area of the pairing channel was close to the cross-sectional area of a single spherical 15 pl re-injected emulsion droplet, allowing only one re-injected emulsion droplet to contact each droplet generated on-chip.<br>5. **Electrocoalescence**: The study also explores electrocoalescence as an alternative method for | | | | | |
|---|---|---|---|---|---|---|
| | achieving droplet coalescence in microfluidic devices. This approach involves applying electric fields to induce attraction forces between solution streams, allowing for on-demand droplet coalescence and the potential to merge several droplets.<br><br>These factors demonstrate that droplet coalescence in microfluidic devices is a complex process influenced by multiple parameters, which can be controlled and optimized to achieve reliable and selective coalescence. | | | | | |
| What design elements are critical for producing double or multiple emulsions? | Based on the provided context, the following design elements are critical for producing double or multiple emulsions:<br><br>1. **Consecutive capillary microfluidic devices**: These devices involve a | To generate double or multiple emulsions, several design elements are essential. Co-flow and flow-focusing configurations enable precise control over fluid flow, allowing for stable droplet formation, | The response provides a detailed list of design elements critical for producing double or multiple emulsions, including consecutive capillary | 42 | 42.0 | 40 |

| | | | | | |
|---|---|---|---|---|---|
| | series of connected tubes that allow for the precise control of the number and size of inner droplets in multiple-emulsions.<br>2. **T-junction modules**: T-junctions can be used to generate single emulsions independently, and by synchronizing the flow rates of different fluids, double emulsions can be produced.<br>3. **Flow-focusing microchannels**: These channels can be used to generate double emulsions by focusing two co-flowing liquid phases into a single stream, which is then broken up into droplets.<br>4. **Co-flow formation steps in series**: This design involves combining two or more co-flow formation steps to produce multiple-emulsions with controlled number and size of inner | especially in double emulsions. Sequential emulsification steps, where inner droplets are emulsified into intermediate phases and then into outer phases, are key for multiple emulsions. Capillary microfluidic devices, such as consecutive capillary and coaxial configurations, are used to control droplet size, number, and monodispersity. Additionally, precise flow control and shear force management are crucial for stable and consistent droplet formation. Microchannel geometry optimization and material compatibility further enhance emulsification processes. By carefully managing these factors, researchers can produce double and multiple | microfluidic devices, T-junction modules, flow-focusing microchannels, co-flow formation steps in series, and elementary microfluidic devices assembled in series. However, the answer is overly verbose and includes some minor inaccuracies compared to the ground truth. The response also lists additional design elements that are not explicitly mentioned but implied as important for producing double or multiple emulsions, which is partially correct but lacks focus on the core | | | |

| | droplets.<br>5. **Elementary microfluidic devices assembled in series**: These devices can be used to control the monodispersity, number, and size of inner droplets in multiple-emulsions.<br><br>Additionally, other design elements that are not explicitly mentioned but implied as important for producing double or multiple emulsions include:<br><br>1. **Precise control over flow rates and pressures**: This is necessary to ensure consistent and controlled droplet formation.<br>2. **Controlled shear forces**: Shear forces can be used to break up the inner droplets into smaller ones, creating a hierarchical structure in multiple- | emulsions with controlled internal structures and compositions for various applications. | aspects of the question.<br>SCORE: 42 | | | |
|---|---|---|---|---|---|---|
| | emulsions.<br>3. **Optimization of microchannel geometry**: The design of the microchannels can affect the flow rates, pressures, and shear forces experienced by the fluids, which in turn influence the formation of double or multiple emulsions.<br><br>These design elements are critical for producing high-quality double or multiple emulsions with controlled internal structure and composition. | | | | | |
| What key factors influence the design of flow-focusing droplet microfluidics optimized using machine learning? | Based on the provided context, the key factors that influence the design of flow-focusing droplet microfluidics optimized using machine learning are:<br><br>1. **Channel geometry**: The shape and size of the channel in which the fluids are flowing play a critical role in controlling | The design of flow-focusing droplet microfluidics, optimized using machine learning, depends on several critical factors. Channel geometry, including the shape and size, controls droplet generation by defining the break-off point where the dispersed phase meets the | The response provides a clear and concise list of factors influencing flow-focusing droplet microfluidics optimized using machine learning. However, it lacks depth and detail compared to | 51 | 51.0 | 55 |

| | | | | | |
|---|---|---|---|---|---|
| | droplet generation. 2. **Flow rate**: The rates at which the dispersed and continuous phases are forced through the channel affect the droplet diameter and generation rate. 3. **Viscosity**: The viscosity of the fluids also influences the droplet generation process, as it affects the flow behavior and shear forces acting on the dispersed phase. 4. **Shear forces**: The shearing action between the continuous and dispersed phases is crucial in flow-focusing designs, and can be controlled by adjusting the channel geometry or flow rates.

These factors are mentioned in Document 2 (page_content='49') as governing the generation of droplets in flow-focusing designs. Additionally, the use of machine | continuous phase. Flow rates of both phases affect droplet size and frequency, with higher continuous phase flow reducing droplet size. Viscosity influences flow behavior and shear forces, crucial for stable droplet formation. Shear forces between phases are controlled by adjusting channel geometry or flow rates, ensuring consistent droplet generation. Machine learning algorithms predict and optimize droplet diameter and generation rate by analyzing these design parameters, fluid properties, and large datasets, automating the design process for more efficient droplet production in life sciences applications. | the ground truth answer, particularly in explaining how these factors interact with each other and with machine learning algorithms. Additionally, some information is presented as a simple statement without further explanation or context. SCORE: 51 | | | |

| | | | | | | |
|---|---|---|---|---|---|---|
| | learning algorithms to predict the performance of droplet generators based on these design parameters is discussed in Document 1 (page_content='2 2,23...'). | | | | | |
| How do channel wall properties (hydrophobicity/hydrophilicity) influence droplet dynamics? | The channel wall properties, specifically hydrophobicity and hydrophilicity, play a crucial role in influencing droplet dynamics in microfluidic devices. The choice of channel wall properties depends on the type of emulsion or droplet formation desired.<br><br>**Hydrophobic channels:**<br><br>* Are required for generating oil droplets in an aqueous phase.<br>* Allow the oil phase to lift the aqueous phase from the walls, surround it, and encapsulate it into drops.<br>* Are necessary for forming W/O (water-in-oil) | The channel wall properties, specifically hydrophobicity or hydrophilicity, significantly influence droplet dynamics in microfluidic systems. These properties determine the wettability of the channel walls, affecting the interaction between the continuous and dispersed phases. In hydrophobic channels, the dispersed phase (typically water) tends to form droplets as it minimizes contact with the channel walls. This enhances droplet stability and facilitates easier detachment during droplet formation. Hydrophobic | The response provides a clear and detailed explanation of how channel wall properties influence droplet dynamics in microfluidic devices, covering the effects on emulsion type, droplet formation, and interfacial effects. It also correctly identifies the importance of hydrophobicity and hydrophilicity in determining wettability and affecting | 85 | 85.0 | 85 |

| | | | | | |
|---|---|---|---|---|---|
| | emulsions.<br>* Can be obtained by plasma bonding PDMS channels to a PDMS surface and baking the device.<br><br>**Hydrophilic channels:**<br><br>* Are required for generating aqueous droplets in an oily phase.<br>* Allow the aqueous phase to wet the channel walls, enabling the formation of O/W (oil-in-water) emulsions.<br>* Can be obtained through various surface treatments such as oxygen plasma, polyvinyl acetate (PVA) coating, or PEG coating.<br><br>The choice of hydrophobic or hydrophilic channels affects the droplet dynamics in several ways:<br><br>1. **Droplet formation:** The wettability of the channel walls determines whether aqueous | surfaces are ideal for applications requiring water-in-oil emulsions or precise droplet control. Conversely, hydrophilic channels promote wetting of the dispersed phase, making it more likely to spread along the walls instead of forming discrete droplets. This is useful for oil-in-water systems or applications involving liquid films. Wall properties also impact droplet size, breakup, coalescence, and transport. Hydrophobicity favors smaller droplets and reduces the likelihood of droplet adherence, while hydrophilicity can lead to higher flow resistance and altered droplet generation behavior. Tailoring wall properties is crucial for optimizing microfluidic | droplet stability and detachment. The response is well-structured and easy to follow, with clear headings and concise bullet points. However, it could benefit from a more comprehensive discussion on the impact of wall properties on droplet size, breakup, coalescence, and transport.<br><br>SCORE: 85 | | | |

| | or oily droplets are formed.<br>2. **Emulsion stability:** Hydrophobic channels promote the formation of W/O emulsions, while hydrophilic channels favor O/W emulsions.<br>3. **Interfacial effects:** At the microscale level, the high surface area to volume ratio places increasing importance on interfacial effects between immiscible phases in the channel or between the continuous phase and the channel walls.<br><br>In summary, the properties of the channel walls (hydrophobicity/hydrophilicity) have a significant impact on droplet dynamics, influencing the type of emulsion formed, droplet formation, and stability. | processes. | | | | |
|---|---|---|---|---|---|---|
| How can you incorporate active control mechanisms (e.g., electrical, | Based on the provided context, it appears that incorporating active control | Active control mechanisms such as electrical, magnetic, thermal, and | The response provides a detailed and structured | 86 | 86.0 | 85 |
| magnetic) into droplet devices? | mechanisms into droplet devices can be achieved through various methods. Here are some ways to do so:<br><br>**Electrical Control**<br><br>1. **Direct Current (DC)**: A constant DC voltage can be applied using electrodes incorporated into the flow-focusing device, as demonstrated by Link et al. This method utilizes the water-oil interface as a capacitor, allowing for precise control over droplet formation.<br>2. **Alternating Current (AC)**: AC can be used to generate droplets through the electrowetting-on-dielectric (EWOD) effect, which reduces the contact angle between the conductive liquid flow and the channel. This method also enables the use of | mechanical methods can precisely manipulate droplet formation in microfluidic devices. Electrical control methods include direct current (DC), which uses electrodes to control droplet size, and alternating current (AC), which utilizes electrowetting-on-dielectric (EWOD) and dielectrophoresis (DEP) to influence droplet behavior. Magnetic control employs ferrofluids and external magnets to non-invasively guide droplet formation. Thermal control involves heating the entire device or applying localized heating to adjust fluid parameters like viscosity and interfacial tension. Mechanical control techniques, such as pneumatic and piezoelectric | explanation of various methods to incorporate active control mechanisms into droplet devices, covering electrical, magnetic, thermal, and mechanical approaches. It accurately describes the principles behind each method, including specific examples and applications. The answer is well-organized, clear, and concise, with proper formatting and headings for easy readability. However, it could be improved by providing more in-depth explanations of the underlying physics and | | | |

| | dielectrophoresis (DEP), where an electric field gradient polarizes the liquid, allowing for droplet formation. 3. **Electro-Wetting on Dielectric (EWOD)**: This mechanism uses an electrical field to reduce the contact angle between the conductive liquid flow and the channel, enabling the creation of droplets.<br><br>**Magnetic Control**<br><br>1. **Ferrofluids**: Non-contact magnetic control can be achieved using ferrofluids, which are fluids with magnetic characteristics. These fluids can be used to create precise and fast generation frequencies and responses. 2. **Permanent or Electromagnets**: Permanent or electromagnets can be utilized to implement non- | components, can also regulate droplet dynamics. These active control strategies enable precise and flexible droplet generation, allowing for tailored applications in complex microfluidic systems, including droplet manipulation and multicomponent emulsion creation. | mechanisms, as well as some numerical examples or case studies to illustrate the practical implications of these control methods.<br><br>SCORE: 86 | | | |
|---|---|---|---|---|---|

invasive magnetic control over droplet formation.

**Thermal Control**

1. **Heating the Entire Micro-Fluidic Device**: The entire micro-fluidic device can be heated, affecting the fluid parameters such as interfacial tension and viscosity.
2. **Local Heating with a Microheater or Laser Source**: Local heating can be achieved using microheaters, micro heat exchangers, or laser sources to increase the temperature difference between the phases.

**Mechanical Control**

1. **Pneumatic, Hydraulic, Mechanical Vibrators, and Piezoelectric Components**: These components can be integrated to

| | | | | | |
|---|---|---|---|---|---|
| | control droplet formation through mechanical means.<br><br>These active control mechanisms can be used in various combinations to create a flexible route for creating droplets one at a time or in parallel, depending on the specific application requirements. | | | | |
| How do flow rates affect droplet size and frequency? | Based on the provided context, it can be seen that flow rates have a significant impact on both the size and frequency of droplets.<br><br>Regarding droplet size, the context states that as the flow rate ratio ($\varphi$ = Qd/Qc) increases, the shearing force provided by the continuous phase becomes greater than the sum of the interfacial tension offered by the dispersed phase. This leads to a rapid break-up of the dispersed phase at the contact | Flow rates significantly influence droplet size and frequency in microfluidic systems. As the flow rate ratio ($\varphi$ = Qd/Qc) increases, the shearing force from the continuous phase dominates, leading to smaller droplets due to rapid breakup of the dispersed phase. Additionally, higher flow rates in the continuous phase can result in larger droplets, as the dispersed phase doesn't break up until it reaches the | The response is mostly clear and relevant, but it contains some minor inaccuracies and lacks conciseness. It correctly explains the relationship between flow rates and droplet size, mentioning that increasing the flow rate ratio ($\varphi$) leads to smaller droplet sizes due to increased shearing | 43 | 43.0 | 40 |

| | | | | | |
|---|---|---|---|---|---|
| | position between the two phases, resulting in smaller droplet sizes. Specifically, it is mentioned that the size of droplets is smaller than the microchannel size during this process.<br><br>Furthermore, as the flow rate of the continuous phase increases, the corresponding increase in viscous shearing force occurs, which becomes greater than the sum of the interfacial tension offered by the dispersed phase. This results in a situation where the dispersed phase does not break up until it reaches the downstream of the microchannel, and the droplet size is determined by the size of the neck.<br><br>In addition, the context mentions that when the flow rate ratio ($\varphi$) is 0.4, the droplet | downstream section. The frequency of droplet generation also increases with higher flow rate ratios, demonstrating a positive correlation. However, droplet diameter growth trends vary depending on the model and input conditions. Overall, increasing the flow rate ratio enhances droplet formation frequency and reduces droplet size due to stronger shearing forces, with complex interactions influenced by factors such as interfacial tension and channel geometry. | force. However, it incorrectly states that the size of droplets is determined by the size of the neck when the dispersed phase does not break up until it reaches the downstream of the microchannel. The response also accurately describes the increase in droplet frequency with increasing flow rate ratio ($\varphi$), but could be more concise and precise in its explanation.<br><br>SCORE: 43 | | | |

| | size obtained for all categories are in a similar trend and tend to enhance with decreasing flow rate of continuous phase. It also states that model 2 can approximately generate droplets by 30 to 50 μm diameters, which rises gently like model 1.

Regarding droplet frequency, the context mentions that as the flow rate ratio (φ) increases, the frequency of droplet generation also increases. Specifically, it is stated that all curves in Fig. 6 are power-laws with exponents in a narrow range, from -0.1 to -0.4, indicating a positive correlation between the flow rate ratio and the frequency of droplet generation.

In summary, the context suggests that: | | | | | |
|---|---|---|---|---|---|---|

| | * Increasing the flow rate ratio (φ) leads to smaller droplet sizes due to increased shearing force. * The size of droplets is determined by the size of the neck when the dispersed phase does not break up until it reaches the downstream of the microchannel. * Droplet frequency increases with increasing flow rate ratio (φ). * The growth rate of droplet diameter has different trends depending on the model and input conditions.<br><br>Therefore, the answer to the question is:<br><br>Flow rates have a significant impact on both the size and frequency of droplets. Increasing the flow rate ratio (φ) leads to smaller droplet sizes due to increased shearing force, while also increasing the | | | | | |
|---|---|---|---|---|---|---|

| | | | | | |
|---|---|---|---|---|---|
| | frequency of droplet generation. The growth rate of droplet diameter has different trends depending on the model and input conditions. | | | | |
| What techniques are available for droplet splitting? | Based on the provided context, there are several techniques available for droplet splitting:<br><br>1. **Passive methods**: These include:<br>   * Using common T and Y junctions<br>   * Flow focusing junction configurations<br>   * Placing external ribs/obstructions in the mother droplet's flow route<br>2. **Active methods**: These include:<br>   * Electric field-based droplet splitting<br>   * Magnetic field-based droplet splitting<br>   * Acoustic field-based droplet splitting (using acoustic radiation force or | Droplet splitting techniques can be categorized into passive and active methods. Passive techniques rely on the microchannel geometry and fluid properties, such as T and Y junctions, flow focusing configurations, or introducing ribs/obstructions in the flow path. These methods are simple but lack dynamic control over splitting ratios and are sensitive to flow conditions. Active techniques utilize external forces like electric, magnetic, acoustic, or thermal fields to precisely control droplet splitting. Examples include electric field-based splitting, magnetic fields for ferrofluids, | The response provides a clear and concise list of techniques for droplet splitting, categorizing them into passive and active methods, which is in line with the ground truth. However, it lacks detailed explanations and specific examples for each technique, making it somewhat superficial. SCORE: 46 | 46 | 46.0 | 45 |
| | streaming)<br><br>    * Thermal field-based droplet splitting<br><br>These techniques can be used to split mother droplets into symmetrical or asymmetrical two or more sister droplets, depending on the application and desired outcome. | acoustic radiation force, and thermocapillary methods. Active techniques offer more flexibility, enabling on-demand droplet splitting with controlled ratios, overcoming the limitations of passive methods. However, each technique has its own set of advantages and constraints depending on the application and droplet type. | | | | |
| What advantages does droplet formation by passive methods have over active methods? | Based on the provided context, passive methods for droplet formation have several advantages over active methods. These advantages include:<br><br>1. **Increased deformation and instability of flow field interface**: Passive methods can increase the deformation and instability of the flow field interface without the need for external actuation, allowing for the creation of | Passive droplet generation methods offer several advantages over active methods. They simplify device design by relying on flow rate manipulation and microchannel geometry, eliminating the need for complex external forces. This leads to cost-effectiveness and easier fabrication. Passive methods also provide precise control over droplet size, shape, and monodispersity, making them | The response accurately lists several advantages of passive droplet formation methods over active methods, including precise control over parameters, protection from external interference, and flexibility in microchannel design. However, it slightly | 84 | 84.0 | 85 |

| | monodisperse droplets. 2. **Precise control over parameters**: Passive methods provide precise control over parameters such as size, shape, monodispersity, and components inside the droplets, making them suitable for applications where high precision is required. 3. **Protection from outside interference and cross-contamination**: Passive methods can protect synthesized droplets from external interference and cross-contamination, ensuring the integrity of the droplets. 4. **Ability to generate uniform size distributions**: Passive methods can produce continuous droplet streams with uniform size distributions (1%–3% | ideal for applications requiring uniform droplets. Additionally, they protect droplets from external interference and cross-contamination, ensuring integrity. Passive methods can produce continuous droplet streams with uniform size distributions (1%-3% dispersity) without requiring external actuation. While they are not suitable for dynamic manipulation or "droplet-on-demand" applications, passive methods are favored for their simplicity, precision, and cost-efficiency in controlled environments. | deviates from the ground truth by not mentioning cost-effectiveness and easier fabrication as key benefits, and also incorrectly states that passive methods are favored for dynamic manipulation or "droplet-on-demand" applications, which is actually a limitation of active methods. SCORE: 84 | | | |
|---|---|---|---|---|---|

| | | | | | | |
|---|---|---|---|---|---|---|
| | dispersity), which is not always achievable with active methods. 5. **Flexibility in microchannel design**: Passive methods allow for flexibility in designing microchannels, enabling the creation of various geometries and structures that can be used to control droplet formation.<br><br>Overall, passive methods offer several advantages over active methods, including increased precision, protection from external interference, and flexibility in microchannel design, making them a preferred choice for certain applications. | | | | | |
| How do you design a device for monodisperse droplet generation? | Based on the provided context, to design a device for monodisperse droplet generation, you can consider the following approaches: | To design a device for monodisperse droplet generation, two common methods are used: cross-flow and flow-focusing designs. In cross-flow | The response provides a detailed explanation of various methods for designing devices for monodispers | 78 | 78.0 | 80 |

| | | | | | |
|---|---|---|---|---|---|
| | 1. **Flow-focusing design**: This method involves flowing the continuous phase into two side channels that surround the flow of dispersed phase in a constriction channel (Figure 1(b)). The width and length of the constriction channel can be adjusted to affect the breakup process by changing the shear stress imposed on the dispersed phase. 2. **Adjusting shear stress**: By adjusting the width and length of the constriction channel, you can control the shear stress on the dispersed phase, which is crucial for generating monodisperse droplets. 3. **Using a flow-focusing device with three-dimensional structures**: An axis-symmetric, flow-focusing device with three-dimensional structures | designs, the continuous phase flows around the dispersed phase in a constriction channel, with droplet size controlled by adjusting channel dimensions and shear stress. This method is simple but can lead to droplet damage due to shear forces or wall interactions. Flow-focusing designs, on the other hand, use sheath flows to stabilize the interface between phases, enabling smaller droplet formation with precise control over size by adjusting flow rates and orifice dimensions. This method offers high monodispersity and is less prone to droplet damage. Both designs require careful optimization of flow rates, channel dimensions, and surface tension to achieve uniform droplet sizes. The | e droplet generation, including flow-focusing designs, adjusting shear stress, and using three-dimensional structures. It also mentions general principles such as controlling emulsion droplet size by adjusting flow velocity ratios and orifice width. However, the response could be improved by providing a more concise summary of the key points and directly addressing the two common methods mentioned in the ground truth answer (cross-flow | | |
| | (Takeuchi et al., 2005) can protect droplets from damage caused by wetting at the microchannel wall.<br><br>Additionally, you can consider the following general principles:<br><br>*<br>**Monodisperse droplet generation methods**:<br>    + Use a flow-focusing method to form droplets in liquid, which leverages the instability between shear forces and surface tension of sheath flows and a central flow (Anna et al., 2003; Tan et al., 2006).<br>    + Control the emulsion droplet size by adjusting the flow velocity ratios of the two liquids and the dimension of the orifice width.<br>    + Use a high-performance flow-focusing device for spontaneous generation of | choice of design depends on specific needs such as droplet size, throughput, and resistance to damage during formation. | and flow-focusing designs). The explanation is mostly accurate but includes some minor inaccuracies and unnecessary details.<br><br>SCORE: 78 | | | |

| | monodisperse droplets (Yobas et al., 2006).<br><br>By considering these approaches and principles, you can design an effective device for generating monodisperse droplets. | | | | | |
|---|---|---|---|---|---|---|
| What materials are most commonly used for fabricating droplet microfluidic devices? | Based on the provided context, the most commonly used materials for fabricating droplet microfluidic devices are:<br><br>1. PDMS (Polydimethylsiloxane) using soft lithography.<br>2. Silicon-based materials, which were first used to fabricate microfluidic devices and are still widely used today.<br><br>These materials are preferred due to their high precision, efficient sealing, and good transparency, making them ideal for droplet generation and | The most commonly used materials for fabricating droplet microfluidic devices are PDMS (polydimethylsiloxane) and silicon-based materials. PDMS is preferred due to its ease of fabrication using soft lithography, which allows for the creation of precise and versatile microfluidic devices. This material provides high transparency, efficient sealing, and the ability to mold complex microchannels, making it ideal for droplet generation and manipulation. While PDMS is dominant, | The response accurately lists two commonly used materials (PDMS and silicon-based) but lacks comprehensive information about other alternatives like thermoplastics, fluoropolymers, and glass materials. It also doesn't fully explain the reasons behind PDMS being the material of choice beyond mentioning its ease of fabrication | 38 | 38.0 | 35 |

| | manipulation in microfluidics. | alternatives such as thermoplastics (e.g., PMMA, polycarbonate), fluoropolymers (e.g., PTFE), and glass materials are also used, each offering unique benefits like better solvent resistance or mass production capabilities. Despite the variety of materials available, PDMS remains the material of choice for most droplet microfluidics applications due to its simplicity and versatility. | using soft lithography. SCORE: 38 | | | |
|---|---|---|---|---|---|---|
| What are the advantages and limitations of using PDMS for microfluidic devices? | Based on the provided context, the advantages and limitations of using Polydimethylsiloxane (PDMS) for microfluidic devices are as follows:<br><br>**Advantages:**<br><br>1. **Low toxicity**: PDMS is a non-toxic material, making it suitable for biomedical applications.<br>2. **Desirable | Polydimethylsiloxane (PDMS) is widely used for microfluidic devices due to its several advantages, including low toxicity, biocompatibility, optical transparency, and high gas permeability, which make it ideal for biomedical applications. Its flexibility and cost-effectiveness also contribute to | The response provides a clear and detailed list of advantages and limitations of using PDMS for microfluidic devices. It covers various aspects such as toxicity, optical transparency, biocompatib | 78 | 78.0 | 75 |

| | optical transparency**: PDMS allows for real-time monitoring by microscopy techniques, enabling the observation of samples and processes. 3. **Biocompatibility**: PDMS is biocompatible, which is essential for biomedical applications. 4. **High gas permeability**: PDMS has high gas permeability, allowing for efficient diffusion of gases such as oxygen and carbon dioxide. 5. **Flexibility**: PDMS is a flexible material, making it suitable for wearable electronics and other flexible device applications. 6. **Reversible adhesion**: PDMS can reversibly bond to other materials, facilitating the fabrication of microfluidic devices with | its popularity, as it allows for easy fabrication using soft lithography. PDMS enables real-time observation and is well-suited for flexible and wearable electronics. However, PDMS has limitations, such as its hydrophobic nature, which can lead to fouling in microchannels, affecting performance. Additionally, its soft elastomeric properties limit mechanical strength, and large-scale production can be challenging. Despite these drawbacks, PDMS remains a versatile and widely used material, though further research is needed to address its limitations for broader applications. | ility, gas permeability, flexibility, and cost-effectiveness. However, the answer could be more concise and directly address the question without providing an overall statement at the end. The response also mirrors the ground truth in terms of advantages but slightly deviates from it regarding limitations, specifically mentioning "cost-effectiveness" as a limitation which is not explicitly stated in the ground truth. SCORE: 78 | | | |
|---|---|---|---|---|---|---|

| | complex geometries. 7. **Low cost**: PDMS-based microfluidic devices are relatively inexpensive to produce.

**Limitations:**

1. **Fouling in microchannels**: The hydrophobic nature of PDMS can lead to fouling in microchannels, which affects the performance and accuracy of microfluidic devices. 2. **Limited scalability**: Large-scale production of PDMS-based microfluidic devices is challenging due to the material's properties. 3. **Cost-effectiveness**: While PDMS is relatively inexpensive, large-scale production and cost-effective manufacture are still limitations. 4. **Optical transparency | | | | | |
| --- | --- | --- | --- | --- | --- |

| | | | | | | |
|---|---|---|---|---|---|---|
| | limitations**: Although PDMS has desirable optical transparency, it may not be sufficient for certain applications or microscopy techniques.<br><br>Overall, PDMS offers several advantages for microfluidic device fabrication, including its biocompatibility, flexibility, and high gas permeability. However, the material's limitations, such as fouling in microchannels and limited scalability, need to be addressed through further research and development to fully utilize its potential. | | | | | |
| How does 3D printing facilitate microfluidic device fabrication? | According to the provided context, 3D printing facilitates microfluidic device fabrication by allowing for the creation of complex channel structures that are | 3D printing revolutionizes microfluidic device fabrication by enabling the creation of complex and precise channel structures that are challenging to | The response is clear, concise, and directly addresses the question with full relevance. It provides a | 95 | 95.0 | 100 |

| | difficult or impossible to achieve with other methods. This is made possible through various 3D printing techniques such as Stereolithography (SL), Multijet Modeling (MJM), and Fused Deposition Modeling (FDM).

Specifically, 3D printing enables:

1. **Complex channel structures**: SL allows for the generation of small and complex channel structures that are unachievable using other methods.
2. **Hierarchical 3D configurations**: The self-alignment capacity and reconfigurable ability of 3D printing make it feasible to construct hierarchical 3D configurations difficult to achieve by other modular | achieve with traditional methods like soft lithography. Techniques such as Stereolithography (SL), Multijet Modeling (MJM), and Fused Deposition Modeling (FDM) offer high resolution, complex geometries, and the integration of various components like valves and pumps. 3D printing also allows rapid prototyping, customization, and the use of diverse materials for biocompatibility and transparency. Advantages over traditional methods include reduced fabrication time and labor, increased accuracy, and environmental friendliness due to the additive manufacturing process. With its ability to create hierarchical 3D configurations | detailed explanation of how 3D printing facilitates microfluidic device fabrication, highlighting various techniques and advantages over traditional methods. SCORE: 95 | | | |
|---|---|---|---|---|---|

| | technologies. 3. **Improved resolution**: Two-photon polymerization, a technique used in SL, can improve the resolution through two-photon absorption upon illumination, breaking through the optical diffraction limit. 4. **Integration of various components**: 3D printing enables the integration of various components such as valves and pumps to achieve rapid, high-throughput performances and multiple functions.<br><br>Additionally, 3D printing offers several advantages over other methods, including:<br><br>1. **Increased accuracy**: 3D printing allows for more accurate fabrication with high resolution. 2. **Reduced time and labor costs**: The | and integrate multiple functions, 3D printing holds great potential for advancing microfluidic device design and performance. | | | | |
|---|---|---|---|---|---|---|

| | process of building successive layers of materials and adding them together automatically under computer control reduces the time and labor required for fabrication. 3. **Environmental friendliness**: As an additive manufacturing method without removal of redundant parts, 3D printing is environmentally friendly.<br><br>Overall, 3D printing has emerged as a promising approach to fabricating microfluidic devices, offering improved resolution, accuracy, and integration capabilities compared to traditional methods. | | | | | |
|---|---|---|---|---|---|---|
| What alternative materials can be used for applications involving organic solvents? | Based on the provided context, several alternative materials can be used as substitutes for traditional | Several sustainable alternatives to traditional organic solvents are gaining | The response is comprehensive and accurately lists various | 96 | 96.0 | 100 |

| | | |
|---|---|---|
| organic solvents in various applications. These alternatives include:

1. **Supercritical CO2**: This is a green solvent that has been gaining attention due to its ability to dissolve a wide range of compounds without the need for harmful chemicals. It's particularly useful in extraction processes.

2. **Ionic liquids**: These are salts that are liquid at room temperature and have low vapor pressure, making them an attractive alternative to traditional solvents. They can be used in various chemical reactions and processes.

3. **Organic carbonates**: This category includes compounds like dimethyl carbonate (DMC), which is a non- | attention for various applications. Supercritical CO2 is a non-toxic, environmentally friendly solvent, particularly useful in extraction processes. Ionic liquids, which are liquid salts at room temperature, offer low vapor pressure and broad application in chemical reactions. Organic carbonates, like dimethyl carbonate, are biodegradable and non-toxic, ideal for fine chemical synthesis. Biosolvents, derived from renewable resources such as plants and bacteria, provide a greener option for extraction and cleaning. Other alternatives include fluorous solvents, non-toxic liquid polymers like polyethylene glycol (PEG), water, ethanol, methanol, and ethyl acetate. | alternative materials that can be used for applications involving organic solvents. It provides detailed descriptions of each material's properties and uses, making it clear and relevant to the question. However, some minor differences in wording and detail exist between this answer and the Ground Truth Answer. SCORE: 96 |

| | toxic and biodegradable solvent suitable for use in the synthesis of fine chemicals and pharmaceuticals.

4. **Biosolvents**: These are derived from renewable resources such as plants, algae, or bacteria. They offer a sustainable alternative to traditional solvents and can be used in various applications, including cleaning and extraction processes.

5. **Fluorous solvents**: These are perfluorinated compounds that have low vapor pressure and are non-flammable. They're useful in chemical reactions where the use of traditional organic solvents might pose safety risks.

6. **Non-toxic liquid polymers**: This category includes materials like | These alternatives are chosen based on their specific properties such as polarity, boiling point, and environmental impact, promoting more sustainable chemical processes and reducing reliance on hazardous petroleum-based solvents. | | | | |
|---|---|---|---|---|---|

| | polyethylene glycol (PEG) or other similar polymers that can be used as solvents in specific applications, offering a safer alternative to traditional solvents.

7. **Water**: While not always an obvious choice for organic solvent replacements due to its limited ability to dissolve certain compounds, water is increasingly being considered as a green and sustainable option, especially when combined with other methods or solvents that enhance its capabilities.

8. **Ethanol and methanol**: These alcohols are already used in some applications as co-solvents or modifiers to increase the solubility of compounds. They | | | | | |
|---|---|---|---|---|---|---|

| | | | | | |
|---|---|---|---|---|---|
| | can be part of strategies for reducing the use of traditional organic solvents derived from petroleum.<br><br>9. **Ethyl acetate**: Another alcohol ester, ethyl acetate is sometimes used as a modifier to enhance the extraction efficiency of certain compounds.<br><br>10. **Supercritical fluid extraction with modifiers**: This method involves using supercritical fluids (like CO2) in combination with co-solvents or modifiers like ethanol, methanol, or ethyl acetate to improve the solubility and extraction efficiency of target compounds.<br><br>These alternatives are being considered for various applications, including solvent purification | | | | |

| | operations, chemical synthesis, and extraction processes. The choice of alternative depends on the specific requirements of the process, such as the properties needed for the solvent (e.g., polarity, boiling point) and the environmental impact desired. | | | | | |
|---|---|---|---|---|---|---|
| How does device surface treatment affect droplet formation? | According to the provided context, device surface treatment plays a crucial role in determining the type and formation of droplets within the microfluidic system. The researchers employed two different methods for treating the PDMS (polydimethylsiloxane) surface:<br><br>1. **PVA deposition with manual injection**: Trantidou et al.'s method involved manually injecting PVA (polyvinyl | Surface treatment of PDMS microfluidic devices plays a critical role in determining droplet formation and stability. PDMS is naturally hydrophobic, suitable for generating water-in-oil (W/O) droplets but not oil-in-water (O/W) droplets. To generate O/W droplets, selective hydrophilic treatment is required. The researchers employed two methods: manual PVA deposition after plasma treatment, and a | The response is mostly relevant and clear, but lacks focus or includes unnecessary details; broadly acceptable. SCORE: 56 | 56 | 56.0 | 50 |

| | alcohol) solution into a plasma-pre-treated microfluidic device to make specific channels hydrophilic. 2. **Simultaneous PVA injection and on-chip plasma treatment**: The researchers modified this approach by performing the selective patterning of PDMS channels in one step using simultaneous PVA injection and on-chip plasma treatment.<br><br>The surface treatment affects droplet formation as follows:<br><br>* **Wettability patterning**: The treated surface determines whether W/O (water-in-oil) or O/W (oil-in-water) droplets are generated. PDMS is inherently hydrophobic, making it suitable for W/O droplet formation but not O/W. | simultaneous approach combining PVA injection with on-chip plasma treatment. The latter method enables precise wettability patterning, allowing for both W/O and O/W droplet formation within the same device. The plasma treatment increases surface energy, making the PDMS hydrophilic, while PVA deposition stabilizes the coating, ensuring long-term device use. This technique provides controllable droplet sizes and stable double emulsion formation, with the ability to manipulate droplet characteristics by adjusting flow rates. The method also allows for reliable, reproducible results, making it a powerful tool for droplet microfluidic applications. | | | | |

| | | | | | |
|---|---|---|---|---|---|
| | * **Selective treatment**: To generate O/W droplets, the researchers needed to selectively treat specific channels to make them hydrophilic using PVA deposition and plasma treatment.<br>* **Controlled wettability patterning**: The simultaneous PVA injection and on-chip plasma treatment method enabled controlled wettability patterning of complex microfluidic networks, allowing for reliable generation of double emulsion (DE) droplets.<br><br>The treated surface also affects the stability and controllability of droplet formation:<br><br>* **Stable PVA coating**: The high stability of the PVA coating on PDMS surfaces allowed | | | | |

| | devices to be stored for several weeks and used for DE generation for several hours. * **Controllable droplet size**: By varying flow rates, the researchers could control the size of inner aqueous (IA) and outer oil droplets in the DE system while preserving their monodispersity.

In summary, device surface treatment is essential for determining the type and formation of droplets within the microfluidic system. The simultaneous PVA injection and on-chip plasma treatment method enabled controlled wettability patterning, stable droplet generation, and controllable droplet size, making it a reliable technique for generating double emulsion droplets. | | | | | |
|---|---|---|---|---|---|---|

| What bonding techniques are used to seal microfluidic channels? | Based on the provided context, several bonding techniques are mentioned as being used to seal microfluidic channels:<br><br>1. **Thermal fusion bonding**: This method involves compressing and bonding thermoplastic substrates at a temperature around their glass transition temperature.<br>2. **Chemical bonding**: In this approach, substrates are covalently bonded at their interface via the formation of chemical bonds.<br>3. **Solvent bonding**: This technique involves using a solvent to bond thermoplastic materials.<br>4. **Oxygen plasma treatment**: This method makes the surfaces of PDMS reactive, allowing them to irreversibly bond when placed in | Several bonding techniques are employed to seal microfluidic channels, each suited to different materials. Thermal fusion bonding involves compressing thermoplastic substrates at temperatures near their glass transition point, allowing them to fuse upon cooling. Chemical bonding creates covalent bonds between substrate surfaces through chemical reactions, while solvent bonding uses solvents to dissolve the materials partially, promoting adhesion through interdiffusion. For PDMS devices, oxygen plasma treatment is commonly used to make surfaces reactive, enabling irreversible bonding when in contact with another PDMS or glass piece. Additionally, corona wand bonding offers a | The response accurately lists four specific techniques used to seal microfluidic channels (thermal fusion bonding, chemical bonding, solvent bonding, and oxygen plasma treatment) and provides a brief description of each. However, it lacks the comprehensive explanation provided in the Ground Truth Answer regarding the specifics of each technique, such as the glass transition temperature for thermal fusion bonding, covalent bonds in chemical | 46 | 46.0 | 45 |

| | | | | | |
|---|---|---|---|---|---|
| | contact with another piece of PDMS or glass.<br><br>These techniques are used to seal microfluidic channels and form the final wall that encloses the channels, which is a critical step in the fabrication process. | less expensive alternative to oxygen plasma treatment, though it is less consistent. The choice of bonding technique depends on material properties, bond strength requirements, and available equipment, all of which are crucial for the successful fabrication of microfluidic devices. | bonding, interdiffusion in solvent bonding, and the comparison with corona wand bonding for PDMS devices. The response also does not discuss the factors influencing the choice of bonding technique or their importance in microfluidic device fabrication. SCORE: 46 | | | |
| What challenges arise when fabricating complex microfluidic geometries? | Based on the provided context, several challenges arise when fabricating complex microfluidic geometries:<br><br>1. **Patterned wettability control**: Achieving high feasibility and accuracy in controlling the wetting behavior of fluids within complex channel geometries is a | Fabricating complex microfluidic geometries presents several challenges, including controlling wettability in intricate channel structures, which is critical for fluid flow and droplet formation. Additionally, scaling up microfabrication processes for commercial applications | The response provides a detailed and accurate list of challenges associated with fabricating complex microfluidic geometries, including patterned wettability control, cost considerations, droplet generation | 94 | 94.0 | 90 |

| challenge.<br>2. **Cost considerations**: Scaling up microfluidic fabrication to commercial levels requires considering the cost implications, making it essential to develop low-cost microfabrication methods and materials.<br>3. **Droplet generation limitations**: Currently, droplet generation is restricted to single emulsions, whereas practical applications may require generating multiple or complex droplet emulsions, which poses a challenge.<br>4. **Limited diversity and functionality of nano-scale materials**: Compared to microscale particles, the diversity and functionalities of droplet microfluidics-derived nano-scale materials are still lacking, making it | requires cost-effective techniques and materials. Current droplet generation methods mainly produce single emulsions, limiting their use in applications that require multiple or complex emulsions. Moreover, droplet microfluidics faces limitations in producing diverse and functional nano-scale materials compared to other methods. Conventional lithography techniques, which are restricted to rectangular cross-sections, struggle with creating complex 3D geometries. Although advancements like 3D printing and stacking methods offer solutions, effective mixing strategies and versatile fabrication of 3D networks with the required features remain significant hurdles in the | limitations, and limited diversity and functionality of nano-scale materials. It also correctly identifies the limitations of conventional lithography methods and the potential of 3D printing in addressing these challenges. The response is well-structured and provides relevant information from the context to support its claims. However, it could be improved by providing a more concise summary or conclusion that ties together the various points made.<br><br>SCORE: 94 | | | |
| --- | --- | --- | --- | --- | --- |

| | challenging to synthesize nanomaterials with unprecedented features.<br><br>Additionally, the context mentions that:<br><br>* Conventional lithography methods are mainly limited to rectangular cross-sections, making it difficult to fabricate microfluidic channels with complex three-dimensional (3D) geometries.<br>* 3D printing has emerged as a powerful tool for fabricating complex microfluidic structures, but challenges arise in designing effective mixing strategies and developing fast, versatile methods for fabricating 3D microfluidic networks with necessary geometries and feature sizes.<br><br>Overall, the fabrication of | field. Addressing these challenges will require innovative approaches and collaboration across various disciplines. | | | | |
|---|---|---|---|---|---|---|

| | | | | | |
|---|---|---|---|---|---|
| | complex microfluidic geometries poses several challenges related to material selection, cost considerations, droplet generation, and the development of novel materials and manufacturing techniques. | | | | |
| How do you fabricate devices capable of handling multiphase flows? | Based on the provided context, it appears that fabricating devices capable of handling multiphase flows involves several techniques and considerations. Here's a detailed response:<br><br>To fabricate devices capable of handling multiphase flows, various methods can be employed, depending on the specific requirements of the application.<br><br>1. **Microfluidic device design**: The design of microfluidic devices is crucial for efficient fabrication of multiphase flows. | Fabricating devices for handling multiphase flows involves several techniques and considerations. Key approaches include advanced micromanufacturing methods such as 3D printing, which eliminates the need for bonding processes, and flow-focusing techniques, which use nozzles and sheath fluids to control droplet formation. Controlling wettability is crucial for generating specific droplet types, while parallelization of droplet generators enhances production | The response is clear, concise, and directly addresses the question with full relevance. It provides a comprehensive overview of techniques and considerations involved in fabricating devices capable of handling multiphase flows, including microfluidic device design, wettability control, flow focusing, parallelizatio | 95 | 95.0 | 95 |

| | Advanced micromanufacturing techniques such as 3D printing can be used to fabricate microdevices without the need for bonding processes. 2. **Wettability control**: Controlling interfacial tension and wettability is essential in microfluidics, particularly when working with immiscible fluids. Hydrophobic or hydrophilic channels can be designed to favor the generation of specific types of droplets (e.g., water-in-oil or oil-in-water). 3. **Flow focusing**: Flow focusing techniques involve using a bespoke arrangement of nozzles and sheath fluids to hydrodynamically confine the dispersed phase away from the channel walls. This method can produce single and double | scalability. Challenges include ensuring even fluid distribution, integrating various components like valves and connectors, and addressing the lack of standardized design rules. Additionally, methods like consecutive capillary assemblies and external forces (e.g., pneumatic or electrical control) can be employed to create complex emulsions. Through careful design and optimization, efficient fabrication of multiphase flow devices for applications such as droplet generation can be achieved. | n, even fluid distribution, standardization design rules, integration, and other relevant factors. The response is well-structured and easy to follow, with a logical flow of ideas that effectively conveys the necessary information. SCORE: 95 | | | |

| | emulsion droplets with polarity determined by channel geometry rather than wettability.<br>4. **Parallelization**: To scale up production, massive parallelization is recommended. This involves using hundreds or thousands of droplet generators fabricated through photolithographic processes to increase the space-time yield.<br><br>Additionally, it's essential to consider the following factors when fabricating devices for multiphase flows:<br><br>* **Even fluid distribution**: Ensuring even fluid distribution between microchannels is critical for efficient fabrication.<br>* **Standardization design rules**: Establishing well-established standardization | | | | | |
|---|---|---|---|---|---|---|
| | design rules for microfluidic chip designs and materials can facilitate the development of tailored devices for specific applications.<br>*<br>**Integration**: Integrating multiple components, such as valves, connectors, and other parts, requires careful design to match standard dimensions and structures.<br><br>By considering these factors and employing the mentioned techniques, it is possible to fabricate devices capable of handling multiphase flows efficiently. | | | | | |
| What is the impact of channel roughness on droplet formation and behavior? | Based on the provided context, it appears that the impact of channel roughness on droplet formation and behavior has been investigated in various studies.<br><br>According to the first document, | The impact of channel roughness on droplet formation and behavior is a key factor in microfluidics. Studies indicate that surface roughness can significantly affect droplet | The response attempts to address the question by providing a general overview of studies related to channel roughness | 36 | 36.0 | 35 |

| | the surface roughness of the microfluidic channels was measured to be $253 \pm 11$ nm after a one-step bonding procedure. This suggests that the roughness of the channel surface can affect the behavior of droplets within the channel.

The second document mentions a study on the effect of GDL (Gas Diffusion Layer) surface roughness on water droplet removal inside a PEMFC (Proton Exchange Membrane Fuel Cell) flow channel. The study used the Volume of Fluid method to investigate how different levels of surface roughness affect the behavior of water droplets.

The third document discusses the interaction between droplet | dynamics, such as movement, spreading, and stability. For example, increased roughness can hinder droplet spreading and prevent secondary droplet formation by altering flow dynamics and the contact line. Additionally, vibration and surface structure play roles in influencing droplet impact and splash behavior. Rougher surfaces tend to reduce spreading, while smoother surfaces may enhance it. Surface wettability also interacts with roughness to further complicate droplet behavior. Overall, channel roughness introduces complexity to droplet dynamics, with its effects varying based on factors like roughness level, fluid type, and external conditions, | and droplet formation. However, it lacks specificity and clarity regarding the actual impact of channel roughness on droplet behavior, instead relying on vague statements about the complexity of the phenomenon. The response also fails to provide a clear summary or conclusion that directly answers the question. SCORE: 36 | | | |
|---|---|---|---|---|---|

| | impact and surface roughness, considering the effect of vibration. It was found that higher surface roughness inhibits droplet spreading, but also notes that there is a need for further exploration of the combined effects of roughness and vibration on droplet dynamics.<br><br>Lastly, the fourth document mentions a study on the interaction between droplet impact and surface roughness, considering the effect of vibration. The study used high-speed photography to explore the droplet impact process under different roughness and vibration conditions, and found that at a Wev (Weber number) of 8.72 and a surface roughness of 0.4 μm, droplet | highlighting the need for further research in this area for better understanding and optimization in microfluidic applications. | | | | |
|---|---|---|---|---|---|---|

| | spreading was enhanced by 33%.

In general, these studies suggest that channel roughness can have a significant impact on the formation and behavior of droplets within microfluidic channels or other systems. The specific effects of roughness on droplet dynamics may vary depending on factors such as the level of roughness, the type of fluid being used, and the presence of vibration.

Therefore, to answer your question: The impact of channel roughness on droplet formation and behavior is a complex phenomenon that can affect the spreading, movement, and removal of droplets within microfluidic channels or other systems. The effects of | | | | | |
|---|---|---|---|---|---|---|
| | roughness may vary depending on several factors, but generally, higher surface roughness tends to inhibit droplet spreading, while lower roughness may enhance it. | | | | | |
| How can sensors be integrated into droplet microfluidic devices during fabrication? | Based on the provided context, it appears that integrating sensors into droplet microfluidic devices during fabrication can be a complex process. According to Document 3 (metadata={'relevance_score': 0.961171}), the integration of multiple microfluidic units, such as droplet generation units and detection units, into a single chip for chemical reaction, biological detection, or other applications requires a high degree of manufacturing precision and a combination of microfabrication techniques. | Integrating sensors into droplet microfluidic devices during fabrication is a complex process that requires advanced microfabrication techniques and careful material selection. For successful integration, precision in manufacturing is crucial, as it often involves combining different microfluidic units, such as droplet generation and detection units, on a single chip. Various techniques like photolithography, soft lithography, and direct integration methods can be used, depending on the type of sensor (e.g., | The response provides a clear and concise summary of how to integrate sensors into droplet microfluidic devices during fabrication. It accurately identifies the key factors involved, including advanced microfabrication techniques, material selection, manufacturing precision, and consideration of sensor functionality and compatibility. However, it does not explicitly | 76 | 76.0 | 75 |

| | Additionally, Document 3 mentions that the fabrication process for microdevices, especially for scale-up production, involves numerous techniques and process steps, which can lead to a high failure rate. This suggests that integrating sensors into droplet microfluidic devices may require advanced microfabrication techniques and careful consideration of material properties to ensure proper functionality.

However, Document 1 (metadata={'relevance_score': 0.99002504}) provides an example of a hybrid polymeric microfluidic device with optical detection for droplet-based systems. The optical component is | optical or electrochemical). The fabrication process also needs to account for material compatibility, sensor functionality, and alignment to ensure reliable performance. Additionally, sensor integration may face challenges such as maintaining stability within the microfluidic environment and minimizing interference with fluid flow. By considering these factors, sensors can be effectively integrated into droplet microfluidic devices for applications like chemical reactions and biological detection. | mention specific techniques like photolithography, soft lithography, or direct integration methods, which are mentioned in the Ground Truth Answer as relevant to successful integration. The response also lacks a detailed discussion on potential challenges such as maintaining stability within the microfluidic environment and minimizing interference with fluid flow.

SCORE: 76 | | | |
|---|---|---|---|---|---|---|

| | integrated by a hybrid concept, where two optical fibers guide laser light from the source to a photodiode, which detects the diffracted light from the formed droplets. This suggests that sensors can be integrated into droplet microfluidic devices during fabrication using advanced materials and techniques.

Therefore, based on the provided context, it seems that integrating sensors into droplet microfluidic devices during fabrication requires:

1. Advanced microfabrication techniques to ensure proper integration and functionality.
2. A thorough understanding of material properties to select suitable materials for sensor integration. | | | | | |
|---|---|---|---|---|---|---|
| | 3. Careful consideration of manufacturing precision and process steps to minimize failure rates.<br><br>By considering these factors, it may be possible to successfully integrate sensors into droplet microfluidic devices during fabrication. | | | | | |
| How are droplet microfluidic devices used for single-cell analysis? | According to the provided context, droplet microfluidic devices are used for single-cell analysis by allowing researchers to isolate and manipulate individual cells in microscale droplets, each acting as an isolated reaction chamber. This technology enables the investigation of cellular responses, gene expression, protein secretion, and various functional assays at single-cell resolution.<br><br>In particular, | Droplet microfluidic devices are widely used for single-cell analysis due to their ability to encapsulate individual cells within tiny droplets, allowing for high-throughput and precise experimentation. These devices work by generating microdroplets in a fluid flow, each containing a single cell along with reagents needed for analysis, such as enzymes, antibodies, or primers. This allows for the | The response accurately addresses the question by providing detailed information on how droplet microfluidic devices are used for single-cell analysis. It covers various aspects such as isolation, manipulation, and high-throughput imaging within oil-based droplets, highlighting the advantages | 85 | 85.0 | 85 |

| | droplet microfluidics allows for:<br><br>* The isolation of individual cells and their encapsulation in oil-based droplets<br>* The manipulation of individual droplets to perform assays that provide genetic, epigenetic, and functional information about individual cells<br>* The examination of the interactions between encapsulated cells, cytotoxicity, kinetics, and other important factors<br>* High-throughput fluorescence imaging within these oil droplets to evaluate cytotoxicity effectively<br><br>This approach provides valuable insights into cellular behavior and can uncover rare cell subpopulations that might be missed in | isolation of cells in a controlled environment, enabling high-resolution studies of gene expression, protein profiling, and cellular responses. The compartmentalized nature of droplets also reduces cross-contamination between samples. Moreover, the small size and scalability of droplet microfluidics make it possible to perform large-scale, parallel analysis of thousands of single cells in a short amount of time. This technology has applications in genomics, proteomics, and drug screening, facilitating insights into cell heterogeneity and the identification of rare cell populations. | of this technology over traditional methods. The answer is clear, concise, and relevant to the question, with minor inaccuracies in specific details but overall a solid understanding of the topic.<br>SCORE: 85 | | | |
| | traditional ensemble measurements. Additionally, it enables the identification of subpopulations and unraveling of cellular heterogeneity, which is a significant advantage over conventional methods of drug discovery and validation.<br><br>Furthermore, droplet microfluidic devices are compatible with AI and robotic systems, allowing for high-throughput analysis and efficient experimentation. This technology has been used to study various biological systems, including cancer research, immunology, neuroscience, and stem cell biology, and has the potential to revolutionize our understanding of complex biological | | | | | |
| | processes.<br><br>Overall, droplet microfluidic devices are a powerful tool for single-cell analysis, enabling researchers to investigate individual cells in a controlled environment and gain insights into cellular behavior that would be difficult or impossible to obtain through traditional bulk cell analysis. | | | | | |
| What are the advantages of droplet microfluidics for high-throughput screening? | Based on the provided context, the advantages of droplet microfluidics for high-throughput screening include:<br><br>1. **Significant reduction in sample volumes**: Droplets allow for a significant reduction in sample volumes, leading to concomitant reductions in cost.<br>2. **High-speed manipulation and measurement**: Manipulation and measurement at kilohertz speeds | Droplet microfluidics offers several advantages for high-throughput screening, including reduced sample volumes, which lowers costs, and increased assay sensitivity through compartmentalization. This method allows for the screening of up to $10^8$ samples per day at kilohertz speeds, significantly improving throughput and efficiency. The small droplet | The response accurately lists the advantages of droplet microfluidics for high-throughput screening, including reduced sample volumes, increased assay sensitivity, ultra-high throughput, and enhanced effective concentrations. However, it | 82 | 82.0 | 80 |

| | | | | | |
|---|---|---|---|---|---|
| | enable up to 10^8 samples to be screened in one day.<br>3. **Increased assay sensitivity**: Compartmentalization in droplets increases the effective concentration of rare species and decreases the time required to reach detection thresholds.<br>4. **Cost-effectiveness**: Droplet microfluidics offers significant advantages for performing high-throughput screens, making them more cost-effective compared to traditional methods.<br>5. **Ultra-high throughput**: The technology enables ultra-high throughput through the use of picoliter volumes and kilohertz rates for sample manipulation and detection.<br>6. **Increased effective concentrations and signal-to- | volumes enhance effective concentrations and signal-to-noise ratios, critical for sensitive detection of low-concentration biological molecules such as antibodies, enzymes, and single genomes. Additionally, droplet microfluidics enables single-cell and single-molecule assays, providing detailed insights into cellular behaviors and allowing for novel experimental approaches. This technology is particularly beneficial for applications like drug discovery, enzyme evolution, and diagnostics, offering a versatile and cost-effective platform for advanced biological research. | lacks a clear summary or conclusion that ties these points together and provides a comprehensive overview of the benefits. Additionally, some points are slightly redundant or rephrased versions of others.<br>SCORE: 82 | | | |

| | noise ratios**: Small volumes in droplet microfluidics lead to an enormous increase in effective concentrations and signal-to-noise ratios, which is critical for performing single-cell assays. 7. **High sensitivity**: Droplet microfluidics allows for high sensitivity, making it possible to detect and analyze small amounts of biological molecules, such as antibodies, enzymes, and single genomes.<br><br>These advantages make droplet microfluidics a powerful tool for various applications, including drug discovery, directed enzyme evolution, bacteria screening, and nucleic acid analysis. | | | | | |
|---|---|---|---|---|---|---|
| How can droplets be used to | According to the provided context, droplets can be | Droplet microfluidics provides versatile | The response provides a | 96 | 96.0 | 100 |

| encapsulate cells or biomolecules? | used to encapsulate cells or biomolecules in several ways:<br><br>1. **Limiting dilution**: This method involves creating a bi-disperse emulsion consisting of small, empty drops and large drops containing single cells or biomolecules. The cell-containing drops can then be selectively recovered using passive methods like pinched flow fractionation.<br>2. **Inertial microfluidics**: By exploiting inertial forces, droplets can be used to encapsulate cells in a more efficient manner than limiting dilution. A spiral geometry can couple Dean forces with inertial lift to order the cells prior to encapsulation, allowing for precise control over the number of cells per drop.<br>3. **Tuning concentration and | methods for encapsulating cells and biomolecules. Limiting dilution is a traditional approach that randomly distributes single cells or biomolecules in droplets, though it has low efficiency. Inertial microfluidics leverages spiral geometries and inertial forces to guide cells into droplets, improving encapsulation efficiency by ensuring single-cell droplets. Bi-disperse emulsions create an emulsion of empty and cell-containing droplets, with the latter selectively recovered using methods like pinched flow fractionation. For biomolecules, limiting dilution is commonly used, adjusting concentration and drop size to achieve single-molecule encapsulation. Additionally, | clear and detailed explanation of how droplets can be used to encapsulate cells or biomolecules, covering various methods such as limiting dilution, inertial microfluidics, and tuning concentration and drop size. It also mentions specific applications and examples, demonstrating the versatility and potential of droplet microfluidics. The response is well-structured, concise, and accurately conveys relevant information. SCORE: 96 | | | |
| --- | --- | --- | --- | --- | --- | --- |

| | | | | | |
|---|---|---|---|---|---|
| | drop size**: By adjusting the concentration and drop size, it is possible to precisely set the fraction of drops containing single biomolecules.<br><br>These methods are particularly useful in cell-based assays, where droplets can be used as sub-nanoliter compartments to confine cells and concentrate any material released from them. This enables detection and analysis of cellular processes without losing valuable materials.<br><br>In addition, droplets have been used to encapsulate mitochondria, cells, and biomolecules for various applications, such as:<br><br>* Encapsulating cells in vesicles with phospholipids<br>* Generating lipid vesicles | droplets can encapsulate cells in lipid vesicles for biological applications or serve as microreactors in processes like DNA analysis. These techniques enable precise control over encapsulation, enhancing their applicability in cell-based assays, drug discovery, and genetic research | | | | |

| | encapsulating biological macromolecules or cells<br>* Performing cell lysis and detection of transient responses using microfluidic devices with integrated heaters and coolers<br><br>These methods demonstrate the versatility and potential of droplets for encapsulating and manipulating cells or biomolecules, making them a valuable tool in various biotechnological applications. | | | | | |
|---|---|---|---|---|---|---|
| What are the applications of droplet microfluidics in drug discovery? | Based on the provided context, the applications of droplet microfluidics in drug discovery include:<br><br>1. **High-throughput screening**: Droplet microfluidics enables the simultaneous analysis of several hundred thousand cell-based assays in a single | Droplet microfluidics plays a significant role in drug discovery by enabling high-throughput screening and precise analysis of drug interactions at the single-cell level. This technology allows the generation of picoliter-sized droplets, each encapsulating a single cell or a | The response provides a clear and concise list of applications of droplet microfluidics in drug discovery. However, it lacks depth and context compared to the ground truth answer, which provides | 56 | 56.0 | 50 |

| | experiment, allowing for rapid and efficient screening of potential drugs. 2. **Single-cell studies**: The technology allows researchers to investigate individual cells by encapsulating them in oil-based droplets, manipulate the droplets, and perform assays that provide genetic, epigenetic, and functional information about individual cells. 3. **Drug dissociation assays**: Microfluidic devices have been employed in drug dissociation assays, which are critical for evaluating the drug release rate and extent in microparticle- and nanoparticle-based drug delivery systems. 4. **Cellular heterogeneity analysis**: Single-cell microfluidic devices enable the identification of | small group of cells, which can be exposed to various drug compounds. This compartmentalization facilitates parallel testing of numerous drug candidates, speeding up the identification of promising therapeutics. Droplet microfluidics is particularly useful in identifying drug efficacy, optimizing drug formulations, and studying complex cellular responses such as gene expression, protein production, or metabolic changes in response to treatment. Additionally, it enables the screening of rare cell populations or targets that would be difficult to isolate using traditional methods. The ability to rapidly test and analyse thousands of drug interactions in a controlled environment | more detailed explanations and examples of how droplet microfluidics accelerates and improves the efficiency of drug discovery processes. SCORE: 56 | | | |
|---|---|---|---|---|---|
| | subpopulations and unraveling of cellular heterogeneity, which may be masked in bulk cell solutions.<br>5. **High-throughput fluorescence imaging**: Droplet microfluidics allows for high-throughput fluorescence imaging within oil droplets to evaluate cytotoxicity effectively.<br>6. **Drug target validation**: The technology enables the identification and validation of drug targets with high throughput, reducing the cost and time required for this process.<br><br>These applications demonstrate the potential of droplet microfluidics in accelerating and improving the efficiency of drug discovery and validation processes. | accelerates drug discovery, reduces costs, and improves the chances of identifying novel therapeutic agents. | | | | |

| How can droplet microfluidics enable precise control of chemical reactions? | Based on the provided context, it appears that droplet microfluidics can enable precise control of chemical reactions in several ways:<br><br>1. **Precise volume control**: The first document mentions that the chemical reactants are dispensed with "precise volume control" into pairs of droplets, allowing for a controlled reaction.<br>2. **Fast mixing inside coalesced droplets**: When the droplet pairs are coalesced and mixed inside the microfluidic device, it creates a reaction container where fast mixing occurs, which is essential for chemical reactions to proceed efficiently.<br>3. **Long operation times without clogging**: The second document | Droplet microfluidics offers precise control over chemical reactions by dispensing reactants into individual droplets with accurate volume control. This enables the mixing of reactants within confined spaces, enhancing reaction efficiency. Coalesced droplets act as micro-reactors, providing fast mixing and long operation times without clogging, even with sticky or precipitating reactants. The system allows for optimization of reaction parameters, such as temperature, flow rate, and pressure, ensuring the production of specific products, like mesoporous silica microspheres with superior surface area and narrow pore distributions. This capability to | The response provides a clear and detailed explanation of how droplet microfluidics enables precise control of chemical reactions, covering various aspects such as volume control, mixing, operation times, and optimization of reaction parameters. It accurately summarizes the key points from multiple sources and presents them in a logical and coherent manner. However, it slightly exceeds the required conciseness by providing an introductory sentence that is not | 94 | 94.0 | 95 |

| | | | | | |
|---|---|---|---|---|---|
| | highlights that this approach allows for long operation times even when reactants would otherwise clog microfluidic devices, such as those forming precipitates or sticky gels during the reaction. 4. **Optimization of reaction parameters**: By optimizing the reaction parameters using this droplet-based microfluidics scheme, it is possible to generate specific products, such as mesoporous silica microspheres with a superior surface area and narrow pore radius distribution.<br><br>The third document, which is a review on droplet-based microfluidics, further emphasizes the importance of precise control in chemical reactions. It mentions that this approach allows for "precise | finely control reaction conditions makes droplet microfluidics a powerful tool for precise, high-throughput chemical synthesis. | directly related to the question and a concluding summary that restates information already presented. SCORE: 94 | | | |
| | control of droplet volumes" and reliable manipulation of individual droplets, including coalescence, mixing, and sorting. This enables the performance of chemical reactions inside the droplets under defined conditions.<br><br>In summary, droplet microfluidics can enable precise control of chemical reactions by providing a controlled environment with precise volume control, fast mixing, long operation times without clogging, and optimization of reaction parameters. | | | | | |
| What are the key challenges of using droplet microfluidics in diagnostics? | Based on the provided context, the key challenges of using droplet microfluidics in diagnostics include:<br><br>1. **Prevention of wetting and | Droplet microfluidics presents significant challenges in diagnostics, including preventing wetting and cross-contamination | The response is well-structured and provides a clear overview of the key challenges associated | 86 | 86.0 | 85 |

| | cross-contamination**: The aqueous droplets are prone to wetting and cross-contamination with the continuous organic liquids, which can compromise the accuracy of the diagnostic results. 2. **Monitoring and controlling the level of oxygen in the droplets**: The liquid-liquid interface of the droplet microfluidics system is sensitive to mechanical perturbation and mass transfer across the interface, making it challenging to maintain a stable environment for accurate diagnostics. 3. **Providing stable and reliable detection chemistries**: The bio-processes inside the droplets may require specific chemistry or generate chemicals that can alter the surface | between aqueous droplets and immiscible organic liquids, which can compromise results. Monitoring oxygen levels within droplets is difficult due to the sensitivity of the liquid-liquid interface to mechanical perturbations and mass transfer. Developing stable detection chemistries is also complex, as the confined droplet environment can alter chemical reactions and impact accuracy. Additionally, challenges in automation and accessibility limit the widespread use of droplet microfluidics, particularly for high-throughput assays. Overcoming these issues requires advancements in surfactants, microfluidic engineering, and automated techniques, which could enhance the precision and | with using droplet microfluidics in diagnostics. It addresses all the main points mentioned in the ground truth answer, including preventing wetting and cross-contamination, monitoring oxygen levels, developing stable detection chemistries, and limitations related to high-throughput requirements and surface area-to-volume ratio. The response also highlights the need for advancements in certain areas to overcome these challenges. However, it | | | |

| | tension or mass transfer rates, which can affect the accuracy of the diagnostic results.<br><br>Additionally, the context also mentions that droplet microfluidics has limitations compared to bulk assays in certain situations, such as:<br><br>* **High-throughput requirements**: Droplet microfluidics is well-suited for ultra-high-throughput assays (processing up to 10^8 samples per day), but may not be efficient for smaller sample sizes.<br>* **Surface area-to-volume ratio**: The high surface area-to-volume ratio of droplets can make it challenging to maintain a stable and inert oil-water interface, which is essential for accurate diagnostics. | applicability of droplet systems in diagnostics. | could be improved by providing more specific examples or details to support some of the points mentioned.<br><br>SCORE: 86 | | | |
|---|---|---|---|---|---|---|

| | | | | | |
|---|---|---|---|---|---|
| | These challenges highlight the need for advancements in areas such as automated techniques for manipulating droplets, making pivotal techniques more accessible to users without microfluidics expertise, and developing suitable barcoding systems for droplet libraries. | | | | |
| How are droplets used for DNA sequencing applications? | Based on the provided context, droplets are used in various ways for DNA sequencing applications. Here are some of the key points:<br><br>1. **Single-cell sequencing**: Droplet microfluidics enables high-throughput single-cell sequencing by encapsulating individual cells or biological targets into droplets, which can then be analyzed and sequenced to profile gene expression at a single-target level.<br>2. **Barcode-** | Droplet microfluidics plays a key role in DNA sequencing by enabling high-throughput analysis and precise targeting of genomic regions. This includes single-cell sequencing, where individual cells are encapsulated in droplets for independent gene expression profiling. Barcode-based sequencing allows for the labeling of nucleic acids with unique identifiers, enabling the tracking of specific genomes during analysis. | The response provides a clear and detailed explanation of how droplets are used in DNA sequencing applications, covering various methods such as single-cell sequencing, barcode-based sequencing, multiplexed targeted PCR amplification, and emulsion PCR with droplet | 96 | 96.0 | 100 |

| | based sequencing**: Droplets are used to label nucleic acid molecules with unique barcodes, allowing for the identification of genomes from the same target in final analysis. This is achieved by encapsulating a library of barcodes into droplets with single targets, which are then pooled and sequenced. 3. **Multiplexed targeted PCR amplification**: Droplet-based combinatorial merger is used to perform highly multiplexed targeted PCR amplification of genomes. Each primer set is amplified in a separate microdroplet, allowing for cleaner PCR products and more uniform amplification of all regions. 4. **Emulsion PCR with droplet merger**: This approach is used | Multiplexed PCR amplification is achieved by performing PCR in separate droplets, ensuring uniform amplification of multiple targets. Additionally, emulsion PCR combines droplet merger with PCR to amplify specific genomic regions, facilitating efficient sequencing. These techniques offer a miniaturized, efficient platform for analyzing complex genetic information, improving accuracy and throughput in genomic studies. | merger. It accurately describes the benefits and techniques involved in each method, demonstrating a good understanding of the subject matter. The response is well-structured, concise, and easy to follow, making it highly relevant and useful for someone looking to understand the role of droplets in DNA sequencing. SCORE: 96 | | | |

| | to amplify specific genomic regions by performing PCR amplifications for different primer sets in separate microdroplets, which are then merged and analyzed.

These applications demonstrate the versatility of droplets in DNA sequencing, enabling high-throughput analysis, precise targeting of specific genomic regions, and efficient recovery of desired sequences. | | | | | |
|---|---|---|---|---|---|---|
| What role do droplets play in tissue engineering research? | Based on the provided context, it appears that droplets play a crucial role in tissue engineering research. According to the text, droplet microfluidics is a well-suited technique for processing materials, and particle-based hydrogel biomaterials can be conveniently fabricated by | Droplet microfluidics plays a pivotal role in tissue engineering by enabling precise fabrication of biomaterials and 3D cell culture systems. Droplets serve as templates for creating hydrogel-based materials, which can encapsulate cells for more accurate modeling of tissue structures. | The response accurately addresses the question by providing detailed information on the role of droplets in tissue engineering research, including their use in creating microcarriers, encapsulatin | 96 | 96.0 | 100 |

| | | | | | |
|---|---|---|---|---|---|
| | using droplets as templates.<br><br>Droplets are used to create microcarriers for 3D cell culture, which can serve as tissue building blocks. Additionally, cells can be encapsulated directly within the hydrogel matrix, allowing researchers to study physiological and pathological phenomena in a controlled environment.<br><br>Furthermore, droplet-based microfluidics is used to generate hollow bacterial cellulose (BC) microspheres with desirable morphology and structure, which are then assembled into injectable porous scaffolds for wound healing applications. These scaffolds have been shown to promote proper cell culture, improved depth distribution, and | This technology facilitates the production of microcarriers and complex biomaterials, such as hollow bacterial cellulose (BC) microspheres, which form injectable scaffolds for wound healing, promoting high cell proliferation and tissue regeneration in vivo. Additionally, droplet-based microfluidics allows for the creation of core-shell structures, such as alginate shells or metal-organic frameworks, which offer controlled release of therapeutic agents. These advancements provide a powerful platform for studying tissue development and regenerative therapies. | g cells, and generating complex biomaterials. The answer is well-structured, clear, and concise, with relevant examples and applications. It aligns closely with the ground truth answer, demonstrating a comprehensive understanding of the topic.<br>SCORE: 96 | | |

| | high cell proliferation rates in vitro, as well as enable tissue regeneration in vivo.

The text also mentions the use of droplet-based microfluidics to create core-shell structures, such as alginate shells with agarose cores or metal-organic frameworks encapsulated microcapsules. These structures can be used for various applications, including wound healing and tissue engineering.

In summary, droplets play a significant role in tissue engineering research by enabling the creation of complex biomaterials, facilitating cell culture and tissue regeneration, and providing a platform for studying physiological and pathological phenomena. | | | | | |
| --- | --- | --- | --- | --- | --- | --- |

| How can droplet microfluidics aid in studying cell signaling pathways? | Based on the provided context, droplet microfluidics can aid in studying cell signaling pathways by providing a controlled and precise environment for individual cells to interact with each other. The technology allows for the creation of discrete droplets containing single cells or specific combinations of cells, which can be manipulated and paired in various ways.<br><br>As mentioned in the documents, droplet-based microfluidics enables the entrapment, isolation, monitoring, and pairing of different cell combinations within discrete droplets (Document 1). This setup offers potential for a wide range of cell-cell communication studies involving contact- | Droplet microfluidics offers a powerful platform for studying cell signaling pathways by enabling the precise isolation and manipulation of individual cells within controlled environments. This technology allows for the encapsulation of single cells in droplets, providing a controlled space to study cell-cell interactions and signaling events. The ability to merge droplets containing different cell types or reagents facilitates the investigation of cell communication, such as paracrine signaling, in a way that mimics in vivo conditions. Additionally, droplet-based microfluidics supports high-throughput screening and single-cell genomic studies, offering insights | The response accurately addresses the question by providing a detailed explanation of how droplet microfluidics can aid in studying cell signaling pathways. It highlights the technology's capabilities, such as controlled droplet manipulation, cell pairing and merging, high-throughput screening, and its applications in single-cell experiments and imaging. The answer is well-structured, clear, and concise, with relevant examples from provided documents | 96 | 96.0 | 100 |

| | independent secretory mechanisms such as paracrine signaling. The ability to control the merging of selected droplet sequences containing desired reagents or cells also facilitates the study of cell signaling pathways.

Furthermore, document 3 highlights the use of droplet-based microfluidics in single-cell experiments and imaging and manipulating protein networks. It mentions that droplets can be used to study stochastic phenotypic variations within populations of genetically identical cells, which is relevant to understanding cell signaling pathways.

In particular, the technology allows for:

1. Controlled droplet | into phenotypic variations and signaling dynamics within genetically identical cell populations. This capability is particularly valuable for understanding complex cellular behaviors in development, differentiation, and disease. | that support the claims made.
SCORE: 96 | | | |
|---|---|---|---|---|---|

| | manipulation: Droplet microfluidics enables the precise entrapment and isolation of individual cells or specific combinations of cells, creating a controlled environment for studying cell-cell interactions. 2. Cell pairing and merging: The ability to merge selected droplets containing desired reagents or cells facilitates the study of cell signaling pathways by allowing researchers to investigate how different cell types interact with each other. 3. High-throughput screening: Droplet-based microfluidics can be used for high-throughput screening of small molecule compounds, which is relevant to studying cell signaling pathways and identifying | | | | | |
|---|---|---|---|---|---|---|
| | potential therapeutic targets.<br><br>Overall, droplet microfluidics provides a powerful tool for studying cell signaling pathways by allowing researchers to control and manipulate individual cells or specific combinations of cells in a precise and controlled environment. | | | | | |
| What is the potential of droplet microfluidics in personalized medicine? | Based on the provided context, the potential of droplet microfluidics in personalized medicine appears to be significant. According to the documents, droplet microfluidics has the ability to:<br><br>1. **Optimize treatments for pancreatic cancer cell lines**: By conducting a specific combination of drugs to induce apoptosis in pancreatic cancer cells, droplet | Droplet microfluidics holds great promise in personalized medicine by enabling precise, high-throughput analyses of biological samples. The technology allows for the optimization of treatments by testing drug combinations on specific cell lines, such as pancreatic cancer cells, to identify the most effective therapies. It also facilitates the early detection of | The response accurately addresses the question about the potential of droplet microfluidics in personalized medicine, listing specific applications and benefits. It provides a clear and concise overview of the technology's capabilities, aligning with the | 96 | 96.0 | 100 |

| | | | | | |
|---|---|---|---|---|---|
| | microfluidic systems can help validate particular drugs and drug combinations, leading to more effective personalized medicine.<br>2. **Detect rare sequences with high sensitivity and precision**: Digital PCR (dPCR) technology, which is enabled by droplet microfluidics, allows for the detection of rare sequences within biological samples, making it an ideal tool for cancer research and clinical diagnostics.<br>3. **Enable precise detection of target nucleic acids**: Droplet-based dPCR can perform truly quantitative experiments, enabling the precise detection of target nucleic acids within samples, which is particularly useful in cancer research and personalized medicine.<br>4. **Detect tumor | cancers through highly sensitive detection of tumor markers in bodily fluids, offering a minimally invasive approach for patient monitoring. Additionally, droplet-based digital PCR (dPCR) enables the detection of rare sequences and precise quantification of nucleic acids, which is crucial for cancer research and tailoring treatments based on a patient's genetic background and tumor profile. Overall, droplet microfluidics can significantly advance personalized medicine by providing tailored treatment strategies and improving early detection for better patient outcomes. | ground truth answer.<br>SCORE: 96 | | | |

| | markers in body effluents**: The technology can detect tumor markers in blood, urine, or feces, making it a minimally invasive tool for patients' follow-up and treatment management.<br>5. **Enable early detection of cancers**: With the ability to detect single somatic mutations within target sequences, droplet microfluidics has the potential to become an essential tool for detecting cancers in early stages.<br>6. **Tailor treatments to individual genetic backgrounds**: By analyzing the genome of a patient's tumor and their genetic background, droplet microfluidics can help tailor treatment regimens that best fit each individual's needs.<br><br>Overall, the context suggests | | | | | |
|---|---|---|---|---|---|---|
| | that droplet microfluidics has the potential to revolutionize personalized medicine by enabling precise, sensitive, and high-throughput analysis of biological samples, leading to more effective treatments and improved patient outcomes. | | | | | |
| How are droplet microfluidic devices used for nanoparticle synthesis? | Based on the provided context, droplet microfluidic devices are used for nanoparticle synthesis by encapsulating reagents in spatially isolated microreactors (droplets) that contain well-defined quantities of materials. This approach overcomes limitations such as diffusion-limited mixing and reagent dispersion found in single-phase microfluidic systems.<br><br>In a droplet-based microfluidic system, reagents are brought | Droplet microfluidic devices revolutionize nanoparticle synthesis by enabling precise control over reaction conditions within isolated microreactors. These droplets encapsulate well-defined quantities of reagents, ensuring rapid mixing and preventing contamination, even when using toxic or volatile chemicals. This approach addresses limitations like diffusion-limited mixing found in traditional systems, resulting | The response accurately describes the use of droplet microfluidic devices for nanoparticle synthesis, highlighting their advantages and capabilities. It provides a clear explanation of how these devices overcome limitations found in traditional systems, enabling precise control over reaction conditions, | 96 | 96.0 | 100 |

| | | | | | |
|---|---|---|---|---|---|
| | together in a co-flowing stream just before droplet formation, and the reaction occurs later within the microdroplet. This method provides a high level of control over the synthesis conditions, allowing for precise control of particle size and polydispersity.<br><br>The advantages of using droplet microfluidic devices for nanoparticle synthesis include:<br><br>1. Fast mixing in spatially isolated microreactors<br>2. Well-defined quantities of materials<br>3. High level of control over synthesis conditions<br>4. Ability to utilize toxic or volatile chemicals without contamination<br>5. Narrower size distribution of final nanoparticles<br><br>Additionally, droplet | in nanoparticles with controlled size, shape, and narrow size distribution. The method supports multi-step synthesis on millisecond timescales, allowing for the scalable production of high-quality nanoparticles with applications in fields such as drug delivery, imaging, and catalysis. | rapid mixing, and contamination prevention. The answer also mentions the potential applications of this method in various fields.<br>SCORE: 96 | | |

| | | | | | | |
|---|---|---|---|---|---|---|
| | microfluidic devices can be used for multi-step synthesis of nanoparticles on a millisecond time scale, as demonstrated in the paper by Shestopalov et al. (2004). This method involves injecting aqueous reagents into a flow of immiscible fluid, forming plugs or droplets that are then mixed and further reacted with additional reagent streams.<br><br>Overall, droplet microfluidic devices offer a powerful tool for nanoparticle synthesis, enabling the production of high-quality nanoparticles with precise control over their size, shape, and crystal structure. | | | | | |
| What role do droplets play in polymer particle production? | Based on the provided context, it appears that droplets play a crucial role in the production of polymer particles through droplet-based | Droplet-based microfluidics provides a versatile platform for producing monodisperse polymer particles by encapsulating monomers or | The response accurately conveys the role of droplets in polymer particle production | 98 | 98.0 | 100 |

| | | | | | |
|---|---|---|---|---|---|
| | microfluidics. According to the text, combining the emulsification of monomers or oligomers with on-chip (or off-chip) solidification of these droplets by means of chemical and physical crosslinking provides a novel route for the synthesis and fabrication of monodisperse polymer particles.<br><br>The context highlights that droplet-based microfluidics systems allow for the production of various types of polymer microparticles, including:<br><br>1. Solid, hollow, and multicored particles<br>2. Asymmetric and irregularly shaped particles<br>3. Particles with specific morphologies, such as spherical, nonspherical, Janus, or multicompartmental shapes | oligomers into droplets and solidifying them through chemical or physical crosslinking. This approach enables the creation of diverse particle types, including solid, hollow, asymmetric, and multicompartmental structures with precise control over size, shape, and morphology. Techniques such as UV-initiated polymerization and solvent evaporation facilitate the transformation of droplets into solid particles. Additionally, functional micro- or nanostructures can be incorporated within or on the surface of these particles by introducing specific ingredients into the droplets, expanding their potential applications in areas like drug delivery, tissue engineering, and advanced materials. | through droplet-based microfluidics, highlighting various methods for solidification, particle diversity, and functionalization. It also provides specific examples of techniques used to change the phase of droplets from liquid to solid or gel.<br>SCORE: 98 | | | |

| | The text also mentions that by adding specific ingredients to the droplets, micro-nanostructures can be introduced inside the particle or on its surface, which often brings novel features.

Furthermore, the context explains that there are two main methods used to change the phase of the droplets from liquid to solid or gel in microfluidic systems:

1. UV-initiated polymerization: This method involves using ultra-violet light to initiate polymerization of soluble polymers into solid particles.
2. Solvent extraction/evaporation methods: This approach combines droplet microfluidics with solvent removal techniques, such as evaporation, to create | | | | | |
|---|---|---|---|---|---|---|

| | monodisperse polymer particles. In summary, droplets serve as the fundamental building blocks for the production of polymer particles through droplet-based microfluidics, allowing for precise control over particle size, shape, and morphology. | | | | | |
|---|---|---|---|---|---|---|
| How can droplet microfluidics improve the reproducibility of chemical synthesis? | Based on the provided context, it appears that droplet microfluidics can significantly improve the reproducibility of chemical synthesis in several ways: 1. **Accurate manipulation**: Droplet reactors provide precise control over reaction conditions, such as temperature and concentration, which is essential for achieving consistent results. 2. **Better reproducibility**: The use of droplet-based systems allows | Droplet microfluidics enhances the reproducibility of chemical synthesis by providing precise control over reaction conditions and minimizing variability. Droplets act as isolated microreactors, ensuring accurate manipulation of parameters such as temperature and concentration, critical for consistent results. This system enables optimal mixing intensity and parallel processing in separate droplet | The response accurately addresses the question by providing a clear explanation of how droplet microfluidics improves the reproducibility of chemical synthesis. It lists specific advantages such as accurate manipulation, better reproducibility, controlled mixing intensity, and reduced | 94 | 94.0 | 95 |

| | | | | | |
|---|---|---|---|---|---|
| | for more reliable automation and minimizes the impact of human error, leading to improved batch-to-batch reproducibility.<br>3. **Controlled mixing intensity**: Droplet reactors can be designed to achieve optimal mixing conditions, which is critical for ensuring consistent product quality.<br>4. **Reduced variability in reaction parameters**: By operating multiple reactions simultaneously in parallel channels or by using separate droplet reactors, the variation in reaction parameters such as temperature and concentration can be minimized.<br><br>These advantages enable researchers to produce high-quality materials with consistent properties, which is particularly important in the | channels, reducing batch-to-batch variability and ensuring uniform product quality. Automation further minimizes human error, making the process highly reliable. These advantages are particularly beneficial for synthesizing complex nanomaterials and exploring new reaction parameter spaces, leading to the production of high-quality, reproducible materials. | variability in reaction parameters, which aligns with the ground truth answer. However, it could be more concise and directly state the benefits without listing them separately.<br>SCORE: 94 | | |

| | synthesis of nanoparticles and other complex nanomaterials. The ability to control and manipulate reaction conditions at a microscale also allows for the exploration of complex reaction parameter spaces, making it possible to discover new materials or optimize existing ones.<br><br>In summary, droplet microfluidics can improve the reproducibility of chemical synthesis by providing accurate manipulation, better reproducibility, controlled mixing intensity, and reduced variability in reaction parameters. | | | | | |
|---|---|---|---|---|---|---|
| What are the applications of droplet microfluidics in catalysis studies? | Based on the provided context, the applications of droplet microfluidics in catalysis studies include: | Droplet microfluidics offers transformative applications in catalysis studies by enabling precise, efficient, | The response provides specific examples of applications of droplet microfluidic | 42 | 42.0 | 40 |

| | | | | | |
|---|---|---|---|---|---|
| | * Enantioselective biocatalysis at the nanoliter scale with a chip device combining a microflow reactor and an electrophoresis structure for enantiomeric differentiation.<br>* Selective catalytic transformations down to the single particle or cell level, which can be achieved by downsizing reaction volumes to a few picoliters.<br>* Catalyzed reactions in droplet-based systems, where nano- or picoliter-sized droplets segmented in an immiscible carrier fluid serve as confined reaction containers.<br><br>These applications suggest that droplet microfluidics has the potential to enable precise and efficient catalysis studies at the nanoscale, which can be useful for various fields | and high-throughput investigations at the nanoscale. These systems facilitate enantioselective biocatalysis using nanoliter-scale devices that integrate microflow reactors and electrophoresis for enantiomeric differentiation. Droplets serve as isolated reaction containers, allowing for controlled catalytic transformations at the single-particle or cell level, with reaction volumes as small as a few picoliters. This confinement minimizes diffusion effects and enhances reaction efficiency. Applications span organic synthesis, biochemical analysis, and biotechnology, showcasing droplet microfluidics as a powerful tool for studying rapid reaction dynamics, | s in catalysis studies, including enantioselective biocatalysis, selective catalytic transformations, and catalyzed reactions in droplet-based systems. However, the answer lacks a clear summary statement that encapsulates the transformative potential of droplet microfluidics in catalysis studies, as mentioned in the Ground Truth Answer. The response also does not explicitly mention the minimization of diffusion effects and enhancement of reaction efficiency | | | |

| | such as organic synthesis, biochemical analysis, and biotechnology. | optimizing catalysts, and exploring reaction mechanisms in real time. | due to confinement. SCORE: 42 | | | |
|---|---|---|---|---|---|---|
| How do droplets enhance the study of reaction kinetics? | Based on the provided context, droplets enhance the study of reaction kinetics in several ways:

1. **Rapid mixing**: Droplet-based microfluidic systems allow for rapid and chaotic mixing of reagents within picoliter (pL) volume droplets, enabling millisecond time resolution in enzyme kinetic measurements.
2. **Small sample consumption**: The use of droplets reduces the amount of sample and reagent required, making it an attractive platform for bioanalysis where resources are limited or expensive.
3. **High-throughput analysis**: Droplet-based | Droplet-based microfluidics revolutionizes reaction kinetics studies by offering rapid, precise, and high-throughput analysis with minimal sample consumption. The chaotic mixing within picoliter droplets ensures millisecond-level time resolution for fast reactions. These systems minimize reagent use, making them ideal for bioanalysis where resources are costly or scarce. Precise control over droplet composition facilitates the creation of concentration gradients, enabling the measurement of kinetic parameters. High-throughput capabilities allow simultaneous monitoring of numerous reactions, | The response accurately and comprehensively addresses the question by listing specific ways in which droplets enhance the study of reaction kinetics. It provides detailed explanations for each point, aligning with the ground truth answer. SCORE: 98 | 98 | 98.0 | 100 |

| | microfluidic systems enable the simultaneous study of multiple reactions, allowing for a massive expansion in experimental diversity and the ability to probe extremely large numbers of discrete reactions. 4. **Precise control over droplet payload**: The technology allows for precise control over the composition and concentration of reagents within each droplet, enabling the creation of concentration gradients and facilitating the measurement of kinetic parameters such as Km (Michaelis constant), Vmax (maximum rate), and kcat (turnover number). 5. **Long-term monitoring**: By tracking individual droplets as they pass through an extended microfluidic | significantly expanding experimental diversity. Additionally, extended tracking of individual droplets supports long-term monitoring of enzyme-inhibitor kinetics, providing deeper insights into reaction mechanisms. | | | | |
|---|---|---|---|---|---|---|

| | channel, researchers can monitor enzyme-inhibitor reaction kinetics over an extended period, providing a more comprehensive understanding of the reaction mechanism.<br><br>Overall, the use of droplets in microfluidic systems has revolutionized the study of reaction kinetics by enabling rapid, high-throughput, and precise analysis with minimal sample consumption. | | | | | |
| --- | --- | --- | --- | --- | --- | --- |
| What are the benefits of using droplets for emulsion polymerization? | Based on the provided context, the benefits of using droplets for emulsion polymerization include:<br><br>* The ability to increase production rates (mentioned in Document 1)<br>* The parallelization of reactions in different droplets, allowing for the screening of different chemicals or | Droplet-based systems offer significant advantages for emulsion polymerization, enhancing efficiency, precision, and scalability. They enable high production rates and precise control over polymer particle size and morphology, facilitating the synthesis of high-molar-mass polymers at faster | The response is mostly relevant to the question but lacks conciseness and includes some minor inaccuracies. It also fails to directly address the core aspects of emulsion polymerizati on as implied in the ground truth answer, which | 41 | 41.0 | 35 |

| | | | | | |
|---|---|---|---|---|---|
| | concentrations of a single compound (Document 1)<br>* The ability to obtain information at the single cell level and on large populations, which is extremely powerful as an analytical tool to find extraordinary variants or obtain statistically relevant data (Document 1)<br>* The possibility of encapsulating millions of different gene variants in droplets and individually amplifying them for quantitative emulsion PCR (Document 2)<br>* The ability to select specific variants in a population, allowing for successive rounds of mutagenesis and selection to be performed in an automated manner and on small volume for directed evolution of enzymes or micro-organisms (Document 2) | reaction rates. Droplet parallelization allows the screening of diverse chemicals or concentrations and supports the encapsulation of millions of gene variants for quantitative emulsion PCR. These systems also empower directed enzyme or microorganism evolution through successive mutagenesis and selection rounds. Furthermore, droplet isolation prevents cross-contamination, adsorption, and solvent evaporation, making them ideal for applications like in vitro compartmentaliza tion (IVC). This unique platform provides a robust, scalable approach for both analytical and biological studies. | specifically highlights the benefits for this process.<br>SCORE: 41 | | | |

| | Additionally, the context mentions that droplet-based systems have advantages over continuous flow platforms, including:<br><br>* The ability to isolate molecular processes and reagents, preventing unwanted adsorption onto channel surfaces, cross-contamination, and evaporation of solvents (Document 3)<br>* The use of water-in-oil emulsions in various novel biological applications, such as in vitro compartmentalization (IVC) using microdroplets (Document 3)<br><br>Overall, the benefits of using droplets for emulsion polymerization include increased efficiency, precision, and scalability, as well as the ability to analyze and | | | | | |
| --- | --- | --- | --- | --- | --- | --- |
| | manipulate complex biological systems. | | | | | |
| How can microfluidics be used to generate smart materials? | Based on the provided context, microfluidics can be used to generate smart materials in several ways:<br><br>1. **Droplet Templating**: Microfluidic systems can create droplets that act as structural templates for generating particles with specific properties. This method is called droplet templating or single emulsion templating. By combining distinct reagents within monodisperse droplets of a desired size, microfluidics enables the synthesis of materials such as hydrogel particles, which are valuable in single-cell analysis.<br>2. **Controlled Reaction Conditions**: | Microfluidics enables the creation of smart materials through precise control of reaction conditions, rapid synthesis, and innovative templating methods. Droplet templating allows monodisperse droplets to act as structural templates, facilitating the synthesis of uniform particles, such as hydrogels for single-cell analysis. By manipulating solute and temperature gradients on short timescales, microfluidics supports the production of nanoscale materials, including semiconducting polymers, catalysts, and nanoparticles with unique optical and electronic properties. Its ability to rapidly | The response accurately describes various methods through which microfluidics can be used to generate smart materials, including droplet templating, controlled reaction conditions, nanoscale material synthesis, and exploring complex reaction parameter spaces. It provides specific examples of the types of materials that can be synthesized using these methods, such as hydrogel particles, semiconduct | 98 | 98.0 | 100 |

| | Microfluidic systems can create and homogenize solute and temperature gradients on short timescales, preventing surface-molecule interactions. This allows for precise control over reaction conditions, making it possible to synthesize small molecules, semiconducting polymers, catalysts, biomimetic materials, and nanoscale materials with user-defined optical and electronic properties.<br>3. **Nanoscale Material Synthesis**: Microfluidics can be used to create a wide range of complex, nanoscale materials that are difficult or impossible to make using conventional wet-chemistry methods. Examples include inorganic | explore complex reaction parameter spaces, coupled with advanced analytics and machine learning, accelerates material discovery and optimization. This approach offers transformative potential in biotechnology, energy, and electronics by enabling the generation of materials with tailored properties and enhanced functionality. | ing polymers, and nanoparticles. The response also highlights the potential applications of microfluidics in biotechnology, energy, and electronics. SCORE: 98 | | | |
|---|---|---|---|---|---|---|

| | semiconductor nanoparticles, metal halide perovskite nanocrystals, conjugated polymer nanoparticles, carbon dots, noble metal nanomaterials, and rare earth upconversion nanoparticles. 4. **Exploring Complex Reaction Parameter Spaces**: Microfluidic systems can rapidly explore complex reaction parameter spaces on timescales many orders of magnitude shorter than those associated with conventional methods. This is achieved by integrating sensitive analytics, reaction control architectures, and efficient machine learning algorithms.<br><br>By leveraging these capabilities, microfluidics offers a powerful tool for | | | | | |
|---|---|---|---|---|---|---|

| | generating smart materials with tailored properties, enabling applications in fields such as biotechnology, energy, and electronics. | | | | | |
|---|---|---|---|---|---|---|
| What are the applications of droplet microfluidics in environmental sensing? | Based on the provided context, the applications of droplet microfluidics in environmental sensing include:

1. **Environmental monitoring**: Droplet-based technology has been demonstrated for various environmental monitoring applications, such as detecting Hg ions using a microfluidic system.
2. **Toxicity assessment**: Microfluidic devices can be used to create miniaturized samplers or analytical devices for on-site environmental surveillance or rapid toxicity assessment of | Droplet microfluidics offers transformative applications in environmental sensing by enabling precise, miniaturized, and efficient monitoring systems. It supports on-site environmental surveillance through portable devices for rapid toxicity assessments and detection of contaminants, such as mercury ions. As controllable microreactors, droplet microfluidics facilitates the study of complex interfacial processes, aiding in chemical synthesis and degradation studies. Furthermore, it | The response provides some relevant information about the applications of droplet microfluidics in environmental sensing, but it lacks depth and clarity compared to the ground truth answer. It also includes specific examples that are not mentioned in the ground truth answer, which makes it less comprehensive. SCORE: 36 | 36 | 36.0 | 30 |

| | particular chemicals.<br>3. **Controllable microreactors**: Droplet microfluidics can be used to explore complex environmental interfacial processes, such as those involved in chemical synthesis and degradation reactions.<br>4. **Environmental sensing**: The technology has the potential to advance our understanding of environmental sensing and monitoring, enabling in situ process visualization and precise flow control.<br><br>These applications are mentioned in the context as examples of the benefits of microfluidic technology for environmental science and engineering. | advances environmental sensing by enabling in situ visualization and precise flow control, enhancing our understanding of environmental processes. These capabilities position droplet microfluidics as a valuable tool for environmental science and engineering. | | | | |
|---|---|---|---|---|---|---|
| How can microfluidic droplets improve | Based on the provided context, microfluidic | Microfluidic droplets revolutionize | The response provides a | 86 | 86.0 | 85 |

| combinatorial chemistry? | droplets can significantly improve combinatorial chemistry in several ways:

1. **Reduced reagent consumption**: Microfluidic droplets allow for individual experiments using six to eight orders of magnitude less starting material than traditional microtiter plate approaches, greatly reducing reagent waste and improving efficiency.
2. **Increased throughput**: The use of picolitre droplets as microreactors enables the simultaneous synthesis of a large number of compounds, increasing the overall throughput of combinatorial chemistry experiments.
3. **Automation and online analysis**: Microfluidic droplet manipulation and | combinatorial chemistry by drastically reducing reagent consumption and enhancing efficiency. Each droplet, acting as a picoliter-scale microreactor, requires six to eight orders of magnitude less material than traditional methods, minimizing waste and costs. This miniaturization enables high-throughput synthesis of large compound libraries, supporting applications like drug discovery and material synthesis. Additionally, droplet manipulation techniques allow for precise reaction control, combinatorial fusion of reagents, and multi-stage experiments with online analysis. These advantages make microfluidic droplets a powerful tool for | clear and detailed explanation of how microfluidic droplets can improve combinatorial chemistry, covering various aspects such as reduced reagent consumption, increased throughput, automation, miniaturization, and combinatorial fusion. It accurately conveys the benefits of using microfluidic droplets in this context, aligning well with the ground truth answer. However, it could be improved by condensing some points for better conciseness and clarity. SCORE: 86 | | | |

| | | | | | |
|---|---|---|---|---|---|
| | detection techniques enable complex multi-stage experiments with multi-dimensional analysis, allowing for automation and online monitoring of reactions.<br>4. **Miniaturization**: The use of microfluidic devices reduces the volume of reagents required, making it possible to work with smaller amounts of material and minimizing waste.<br>5. **Combinatorial fusion**: Microfluidic droplets can be mixed and combinatorially fused with other droplets containing different reagents, allowing for the creation of complex libraries of compounds in a highly controlled and efficient manner.<br><br>These advantages make microfluidic droplets an attractive | automating complex chemical workflows and optimizing reaction conditions in a highly efficient and scalable manner. | | | | |
| | platform for combinatorial chemistry applications, including early-stage drug discovery, material synthesis, and optimization of reaction conditions. | | | | | |
| What are the challenges of scaling up droplet-based chemical synthesis? | Based on the provided context, the challenges of scaling up droplet-based chemical synthesis include:<br><br>1. **Device fabrication and large-scale production**: The current microfluidic reactor technology has achieved parallelization of droplet generation, but issues with fluid distribution remain a challenge. Developing 3D-printing techniques to fabricate small and complex channel structures can help control fluid distribution and enable scale-up. | Scaling up droplet-based chemical synthesis faces several challenges. Achieving uniform fluid distribution in microfluidic devices remains difficult despite progress in parallel droplet generation, with 3D printing offering potential solutions for complex channel fabrication. Simplifying functional units for multistep reactions, such as droplet generation, merging, and phase separation, is vital for seamless continuous flow processes. Developing low-cost materials and | The response provides a detailed and relevant list of challenges associated with scaling up droplet-based chemical synthesis, including device fabrication, multistep reactions, patterned wettability control, cost considerations, generating multiple or complex droplet emulsions, and diversity and functionalities of nanoscale materials. | 81 | 81.0 | 80 |

| | 2. **Multistep reactions**: Scaling up multistep reactions, such as synthesis, functionalization, and purification, in a successive flow process is challenging. Simplifying fundamental functional units, including droplet generation, merging, and phase separation, is necessary to overcome this challenge. 3. **Patterned wettability control**: Achieving patterned wettability control of complex channels with high feasibility and accuracy is essential for large-scale manufacturing. 4. **Cost considerations**: The cost of scaling up droplet-based technologies must be considered when pushing the technology to commercial markets. Developing low- | standardizing chip designs are necessary for commercial viability, especially for high-throughput production in diverse fluids. Furthermore, creating complex droplet emulsions and expanding nanomaterial diversity are critical to meeting industrial demands. Addressing these challenges will enable safe, sustainable, and scalable microfluidic applications. | However, the answer could be more concise and directly aligned with the ground truth, which is a comprehensi ve summary of these challenges. SCORE: 81 | | | |
|---|---|---|---|---|---|---|

| | cost microfabrication methods and materials can help mitigate this challenge. 5. **Generating multiple or complex droplet emulsions**: Currently, droplet generation is restricted to single emulsions, but in practical applications, generating multiple or complex droplet emulsions may be necessary. This requires the development of novel interfacial reaction systems. 6. **Diversity and functionalities of nanoscale materials**: The diversity and functionalities of droplet microfluidics-derived nanoscale materials are still lacking compared to microscale particles. Further efforts are needed to overcome this challenge and enable the synthesis of nanomaterials with | | | | | |
| --- | --- | --- | --- | --- | --- | --- |

| | | | 78 | 78.0 | 75 |
|---|---|---|---|---|---|
| | unprecedented features.<br><br>These challenges must be addressed to push droplet-based technologies forward and make them a reliable, truly industrial technology for chemical synthesis. | | | | |
| How can artificial intelligence enhance droplet microfluidic research? | Based on the provided context, artificial intelligence (AI) can enhance droplet microfluidic research in several ways:<br><br>1. **Training models**: The large dataset generated by droplet microfluidics can be used to train AI models, which can then be used to improve the accuracy and sensitivity of microfluidic systems.<br>2. **Droplet characterization**: Advanced detection techniques such as machine learning can recognize | Machine learning can be used to optimize the design of microfluidic devices. Deep Learning will be integrated to sort, monitor, and process the images. Furthermore, the multimodal agents can be used to study and design the research of droplet microfluidics, a possible excuse robot to perform fabrication and experiments automatically. | The response provides a clear and concise overview of how artificial intelligence can enhance droplet microfluidic research. It highlights several specific ways AI can improve the field, including training models, droplet characterization, integration with robotic systems, improving reproducibility, and data analysis. | 78 | 78.0 | 75 |

| | complex flow patterns in droplets, enabling sophisticated droplet characterization and screening of large libraries of drug compounds. 3. **Integration with existing robotic systems**: Microfluidic devices are highly compatible with existing robotic systems, and their integration with AI can enhance throughput and enable incorporation into current and existing drug screening workflows. 4. **Improving reproducibility**: The integration of AI and robotics in microfluidic experiments improves the reproducibility of single-cell droplet-based microfluidic experiments and reduces manual labor, minimizing human errors in the drug development process. 5. **Data | | However, it does not fully address the ground truth answer's mention of multimodal agents for studying and designing droplet microfluidics or the use of deep learning for image processing. The response also slightly deviates from the ground truth by focusing more on the broader applications of AI in droplet microfluidic research rather than providing a detailed explanation of how machine learning can be used to optimize device design, as mentioned in the ground truth | | | |
| | analysis**: Deep learning algorithms can rapidly analyze characterization data from large-scale droplet-based platforms, enabling the extraction of more information from each droplet and feeding instructions back into the microfluidic system to control the nature or composition of new droplets.<br><br>Overall, AI has the potential to revolutionize droplet microfluidic research by improving the accuracy, sensitivity, and throughput of experiments, as well as enhancing the reproducibility and efficiency of the drug discovery process. | | answer.<br><br>SCORE: 78 | | | |
| How can large language models enhance droplet microfluidic research? | Based on the provided context, it appears that large language models can enhance droplet microfluidic research in | LLMs can be used to build the Agents and Multi-Agents to synthesize knowledge of droplet microfluidics | The response attempts to address the question but provides a generic and overly broad | 16 | 16.0 | 15 |

| | | | | | |
|---|---|---|---|---|---|
| | several ways:<br><br>1. **Improved detection and sorting**: Advanced machine learning techniques can be used to recognize complex flow patterns within droplets, enabling more sophisticated droplet characterization and potentially allowing for the simultaneous screening of multiple droplets.<br>2. **Enhanced data analysis**: Large language models can help analyze large datasets generated by droplet microfluidics, facilitating the identification of pathogens and other biological entities with high accuracy and sensitivity.<br>3. **Integration with existing technologies**: Machine learning can be integrated with existing detection techniques, such as single-nucleus RNA sequencing | from literature such as research articles and books. Then, the idea hypotheses will be proposed by the agents, and the novelty and feasibility will be assessed. The agents will provide the autonomous platform to design and fabricate droplet microfluidic devices. Furthermore, LLMs have the researchers accelerate knowledge of droplet microfluidics. | answer that lacks specific details and examples related to the actual capabilities of large language models (LLMs) in enhancing droplet microfluidic research. It also fails to mention the key points mentioned in the Ground Truth Answer, such as building agents and multi-agents, proposing hypotheses, assessing novelty and feasibility, providing an autonomous platform for device design and fabrication, and accelerating researchers' knowledge. The response | | | |

| | (Drop-Seq), to enhance the probability of cells being paired with appropriate barcodes and reduce noise in final sequencing results. | | seems to be a general summary of potential benefits rather than a specific and detailed explanation of how LLMs can enhance droplet microfluidic research. | | | |
| | 4. **Development of new methods**: Large language models can aid in the development of new statistical methods to exclude sequencing noise data caused by empty droplets or mixed transcription data from multiple cells receiving the same barcode. | | SCORE: 16 | | | |
| | Overall, large language models have the potential to enhance droplet microfluidic research by improving detection and sorting capabilities, enhancing data analysis, integrating with existing technologies, and facilitating the development of new methods. | | | | | |

| How can autonomous droplet microfluidic research? | Based on the provided context, it appears that autonomous droplet microfluidic research is a rapidly evolving field with significant potential for future scientific discovery. The documents suggest that researchers are working towards developing highly integrated systems that can perform various bio/chemical functions, including PCR (Polymerase Chain Reaction), sample preparation, and detection methods.<br><br>To achieve this goal, several challenges need to be addressed, such as:<br><br>1. Developing fast detection methods to probe small volume droplets dynamically.<br>2. Controlling the dynamic behavior of droplets in complex | Autonomous droplet microfluidics is an emerging field with significant potential for scientific discovery, aiming to create highly integrated, automated systems for bio/chemical functions like PCR, sample preparation, and detection. Key challenges include the need for fast detection methods to probe small volume droplets, precise control over droplet dynamics in complex microchannel networks, and the development of virtual design platforms for optimizing system designs. Solutions under exploration include using advanced detection techniques such as single-molecule detection and fluorescence lifetime imaging, as well as integrating | The response is mostly relevant and clear, but lacks focus on directly answering the question "How can autonomous droplet microfluidic research?" It provides a detailed overview of the field, its challenges, and potential applications, but does not explicitly state how the research itself can be conducted. SCORE: 46 | 46 | 46.0 | 45 |
|---|---|---|---|---|---|---|

| | microchannel networks. 3. Designing a powerful virtual design and experiment platform to supplant empirical design methodologies.

Despite these challenges, researchers are exploring various approaches to overcome them, including:

1. Utilizing single-molecule detection, fluorescence lifetime imaging, and spectroscopy for fast detection methods. 2. Developing integrated microelectronic components as actuators and motive power in microfluidic devices. 3. Creating a virtual design and experiment platform that can simulate the behavior of droplets in complex microchannel networks. | microelectronic actuators for droplet manipulation. Additionally, the development of virtual design and experiment platforms could streamline the creation of optimized microfluidic systems. The potential applications are vast, including directed evolution, closed-loop synthesis, and automated drug discovery, where systems could autonomously optimize compound combinations. Collaboration across disciplines like fluid dynamics, chemistry, and software engineering will be key to overcoming challenges and advancing this technology, paving the way for groundbreaking scientific advances. | | | | |
|---|---|---|---|---|---|

| | The context also suggests that autonomous droplet microfluidic research has the potential to enable future scientific discovery, particularly in areas such as:<br><br>1. Directed evolution and closed-loop synthesis<br>2. Automated discovery of anti-inflammatory small-molecule combinations<br><br>To advance this field, researchers may need to collaborate across disciplines, including computational fluid dynamics, chemistry, biology, physics, software engineering, and high-performance computing.<br><br>In summary, autonomous droplet microfluidic research is an exciting area that holds promise for future scientific discovery. To | | | | | |
|---|---|---|---|---|---|---|
| | overcome the challenges associated with this technology, researchers must develop innovative solutions that integrate multiple disciplines and leverage emerging technologies such as single-molecule detection and virtual design platforms. | | | | | |
| What is the potential for fully integrated droplet microfluidic platforms? | Based on the provided context, it appears that fully integrated droplet microfluidic platforms have significant potential. The documents suggest that such platforms can combine multiple functions, including flow control, communication, and chemical/biological reactions, to create a highly automated and self-regulating system.<br><br>The first document mentions the development of | Droplet microfluidics offers transformative solutions to global health challenges by enabling rapid, cost-effective, and high-throughput analysis of biological samples. By generating and manipulating tiny droplets as individual reaction chambers, this technology can revolutionize diagnostics, particularly in resource-limited settings. For example, it facilitates point-of-care testing for | The response attempts to address the question but provides a general overview of the potential benefits of fully integrated droplet microfluidic platforms. However, it lacks specificity and depth compared to the ground truth answer, which highlights the transformative impact on global health challenges. | 41 | 41.0 | 40 |

| | | | | | |
|---|---|---|---|---|---|
| | "smart" microfluidic devices that can make logical decisions based on experimental results, allowing for optimization of experiments and creation of enzymes, drug screening, and complex protocol sequences. This implies that fully integrated droplet microfluidic platforms have the potential to automate and streamline various laboratory processes.<br><br>The second document highlights the benefits of droplet-based microfluidic systems, including high-throughput processing, decreased diffusion distance, faster mixing, and laminar flow. It also mentions the integration of optical methods for bioassays and detection, which suggests that fully integrated platforms could | infectious diseases like tuberculosis, malaria, and COVID-19, providing results quickly and accurately. Droplet microfluidics is also crucial for multiplexed screening, enabling simultaneous detection of multiple biomarkers, which enhances early disease detection and monitoring. Additionally, it supports the development of low-cost diagnostic devices, reducing dependence on expensive laboratory infrastructure. In drug discovery, it accelerates the screening of therapeutic compounds, improving the treatment of diseases. The technology also advances personalized medicine by analysing single cells and their | The response also includes some irrelevant information from the provided context. SCORE: 41 | | | |

| | potentially combine multiple analytical techniques to provide comprehensive results.

The third document discusses the development of a virtual design and experiment platform that can monitor and control biological and chemical processes in droplets. This implies that fully integrated droplet microfluidic platforms have the potential to be highly automated and capable of making decisions based on real-time data.

Overall, the context suggests that fully integrated droplet microfluidic platforms have significant potential for automating laboratory processes, optimizing experiments, and providing comprehensive | responses to treatments. Overall, droplet microfluidics addresses global health disparities by delivering accessible, scalable, and efficient solutions for disease diagnosis, monitoring, and treatment. | | | | |
|---|---|---|---|---|---|---|
| | results. However, it also notes that developing such systems is a complex task that requires advanced design and control capabilities.<br><br>Therefore, the answer to the question is:<br><br>Fully integrated droplet microfluidic platforms have significant potential for automating laboratory processes, optimizing experiments, and providing comprehensive results, but their development is a complex task that requires advanced design and control capabilities. | | | | | |
| How can droplet microfluidics contribute to global health challenges? | Based on the provided context, droplet microfluidics has the potential to significantly contribute to global health challenges in several ways:<br><br>1. **Enhanced diagnostic and therapeutic | Droplet microfluidics has significant potential to address global health challenges by enhancing diagnostic and therapeutic processes. It enables rapid and sensitive pathogen detection, which | The response is well-structured and provides a clear overview of the potential contributions of droplet microfluidics to global health challenges. | 76 | 76.0 | 75 |

| | | | | | |
|---|---|---|---|---|---|
| | procedures**: The technology enables the development of ultrahigh-throughput screening and application microsystems, which can lead to more efficient and effective diagnosis and treatment of diseases. 2. **Detection of pathogenic bacteria**: Droplet-based microfluidics has been shown to be a sensitive, specific, and rapid method for detecting pathogenic bacteria such as Salmonella typhimurium, which is a major global health concern. This technology can help reduce the incidence of infectious diseases and related economic losses. 3. **Improved detection limit**: The use of LAMP-assisted amplification reactions in picoliter-sized | is critical for early disease diagnosis and containment. The technology's high-throughput capabilities allow simultaneous analysis of multiple samples, streamlining disease diagnosis and treatment. Additionally, droplet microfluidics can be integrated into point-of-care platforms, making it feasible for use in resource-limited settings. This approach can improve healthcare accessibility and efficiency. Furthermore, droplet microfluidics facilitates precise single-cell analysis, enabling advances in personalized medicine and more effective treatment strategies. While challenges such as cost, standardization, and integration with existing systems remain, continued | It highlights several key benefits, including enhanced diagnostic and therapeutic procedures, improved pathogen detection, high-throughput analysis, point-of-care platforms, and single-cell analysis. However, it lacks some depth and nuance compared to the ground truth answer, which provides a more comprehensive discussion of the technology's potential impact on global health. The response also does not explicitly mention personalized medicine or the | | | |

| | | | | | |
|---|---|---|---|---|---|
| | droplets has been shown to improve the detection limit for pathogenic bacteria, making it possible to detect even small amounts of the bacteria. 4. **High-throughput analysis**: Droplet microfluidics enables the simultaneous analysis of multiple samples, which can lead to faster and more efficient diagnosis and treatment of diseases. 5. **Point-of-care platforms**: The technology has the potential to be used in point-of-care platforms, making it possible to perform diagnostic tests in resource-limited countries where access to bulky and expensive instruments is limited.<br><br>Overall, droplet microfluidics has the potential to contribute significantly to global health challenges by | development of this technology could significantly impact global health by improving detection, diagnostics, and treatment capabilities. | challenges associated with implementing droplet microfluidics in resource-limited settings.<br><br>SCORE: 76 | | | |
| | improving diagnostic and therapeutic procedures, detecting pathogenic bacteria, and enabling high-throughput analysis of individual cells. | | | | | |
| How can open-source designs impact the future of microfluidic devices? | Based on the provided context, it appears that open-source designs have the potential to significantly impact the future of microfluidic devices in several ways:<br><br>1. **Democratization of design expertise**: The use of open-source tools and frameworks allows for the creation of custom predictive models without requiring extensive machine learning expertise or vast experimental resources. This democratizes access to design capabilities, enabling a wider range of users to contribute to the | Open-source designs for microfluidic devices hold the potential to revolutionize the field by increasing accessibility, reducing costs, and fostering collaboration. These designs allow for the creation of custom predictive models without requiring extensive expertise, enabling broader participation in device development. The use of open-source tools, such as Design Automation of Fluid Dynamics (DAFD), accelerates design and prototyping, enabling rapid iteration and optimization. | The response is well-structured and provides a clear overview of the potential impact of open-source designs on the future of microfluidic devices, aligning with the ground truth. It highlights various benefits such as democratization of design expertise, increased accessibility and affordability, community-driven innovation, rapid design and | 84 | 84.0 | 80 |

| | | | | | |
|---|---|---|---|---|---|
| | development of microfluidic devices.<br>2. **Increased accessibility and affordability**: Open-source designs can lead to more affordable and accessible microfluidic devices, as they can be easily integrated into existing workflows and do not require significant investments in proprietary software or equipment.<br>3. **Community-driven innovation**: The open-source nature of these tools fosters a community-driven approach to innovation, where users can contribute their own designs, share knowledge, and collaborate on the development of new microfluidic devices.<br>4. **Rapid design and prototyping**: Open-source tools like Design | Additionally, the integration of machine learning algorithms can drive the development of new microfluidic tools and optimize existing ones. This democratization of design expertise and the collaborative, community-driven approach will facilitate the creation of affordable, specialized devices, ultimately leading to breakthroughs in applications like single-cell analysis and global health diagnostics. The open-source model promises to transform microfluidics into a more inclusive and innovative field. | prototyping, and machine learning-driven optimization. The response demonstrates a good understanding of the topic and provides relevant examples. However, it is not entirely flawless, as some points could be more concise or directly related to the question.<br>SCORE: 84 | | | |

| | Automation of Fluid Dynamics (DAFD) enable rapid design and prediction of droplet generators and other fluidic operations, allowing for quick iteration and improvement of device performance. 5. **Machine learning-driven optimization**: The integration of machine learning algorithms in open-source tools can lead to the creation of entirely new microfluidic tools and the optimization of existing ones, further accelerating innovation in the field.<br><br>Overall, the adoption of open-source designs has the potential to transform the development and use of microfluidic devices by making them more accessible, affordable, and innovative. This, in turn, can lead | | | | | |
|---|---|---|---|---|---|---|

| | | | | | |
|---|---|---|---|---|---|
| | to breakthroughs in various fields, such as single-cell analysis, and ultimately drive the future of droplet-based microfluidics forward. | | | | |
| What are the challenges of scaling up microfluidic devices for industrial use? | Based on the provided context, the challenges of scaling up microfluidic devices for industrial use include:<br><br>1. **Establishing standardized design principles and strategies**: The need to develop well-established standardization design rules for tailored microfluidic chip designs and materials for different target applications, particularly for high-throughput generation in non-Newtonian fluids with high viscosity.<br>2. **Trade-off between throughput and number of microchannels**: The challenge of balancing the | Scaling up microfluidic devices for industrial use presents several challenges, including throughput limitations, lack of standardization, and integration complexities. The small micrometric channel sizes restrict production efficiency, and increasing parallelization or channel size may impact fabrication costs and reaction kinetics. There is also a lack of standardized design methods, making it difficult to compare results and adopt microfluidics in industrial settings. Additionally, scaling from laboratory to industrial production | The response provides a detailed list of specific challenges related to scaling up microfluidic devices for industrial use, including establishing standard design principles, trade-offs between throughput and number of microchannels, cost considerations, generating complex droplet emulsions, integration with industrial interfaces, fluid distribution and control, | 85 | 85.0 | 85 |

| | number of microchannels and their stable operation condition to achieve high-throughput production while maintaining industrial-scale manufacturing feasibility. | remains a hurdle, with issues in cost-effective microfabrication, generating complex droplet emulsions, and ensuring reliable fluid control. The integration of microfluidic devices with industrial interfaces, testing methods, and continuous processes is essential for successful commercialization. Overcoming these challenges requires innovative solutions, such as simplifying multistep reactions and developing standardized, low-cost fabrication methods. | multistep reactions, and process simplification. The response is well-structured and provides relevant information that aligns with the ground truth answer. However, some points in the response seem to be more detailed and specific than others, and there are minor differences in wording between the response and the ground truth answer. SCORE: 85 | | | |
|---|---|---|---|---|---|---|
| | 3. **Cost considerations**: The need to develop low-cost microfabrication methods and materials to make commercial-scale manufacturing more economical. | | | | | |
| | 4. **Generating complex droplet emulsions**: The limitation of current scaling up of droplet generation to single emulsions, whereas practical applications require the generation of multiple or complex droplet emulsions. | | | | | |
| | 5. **Integration with industrial interfaces and control systems**: The need to develop microfluidic devices that can be integrated with | | | | | |

| | industrial-scale interfaces and built-in control systems to enable commercial-scale manufacturing. 6. **Fluid distribution and control**: The challenge of achieving patterned wettability control of complex channel structures for high feasibility and accuracy, particularly at large scales. 7. **Multistep reactions and process simplification**: The need to simplify fundamental functional units, such as droplet generation, merging, and phase separation, and develop integrated processes combining synthesis and post-treatments in a continuous flow process.

These challenges highlight the complexities involved in scaling up | | | | | |
|---|---|---|---|---|---|---|

| | microfluidic devices for industrial use, requiring innovative solutions and collaborative efforts from different areas to bridge the gap between laboratory results and industrial implementation. | | | | | |
|---|---|---|---|---|---|---|
| How might droplet microfluidics impact agriculture? | Based on the provided context, it appears that droplet microfluidics has the potential to significantly impact agriculture in several ways:<br><br>1. **Regulated fertilizer and pesticide release**: Droplet-templated capsules can be used to regulate the release of fertilizers and pesticides, enhancing crop growth.<br>2. **Precision agriculture**: By allowing for the precise linking of genotypic and phenotypic characteristics of single cells, droplet microfluidics can | Droplet microfluidics has the potential to transform agriculture by enhancing crop management and food security. It enables the precise release of fertilizers and pesticides, reducing waste and optimizing plant growth. The technology also supports improved crop monitoring through long-term studies, offering insights into plant development. By analyzing individual plant cells, droplet microfluidics can help uncover genetic and phenotypic traits, enhancing disease resistance and | The response provides some relevant information about the potential applications of droplet microfluidics in agriculture, but it lacks depth and accuracy compared to the ground truth answer. It mentions four specific ways that droplet microfluidics could impact agriculture, but these points are not as comprehensive or | 36 | 36.0 | 30 |

| | | | | | |
|---|---|---|---|---|---|
| | help reveal the mechanisms of cell heterogeneity and dysfunction in crops, leading to more targeted and effective agricultural practices.<br>3. **Improved crop monitoring**: Long-term kinetic studies using high-throughput trapping and monitoring methods enabled by droplet microfluidics can provide prolonged monitoring on encapsulated targets, allowing for better understanding of plant growth and development.<br>4. **Increased food security**: By enhancing the agri-food yield through the detection of defects in plants, modern microfluidics can contribute to global food security.<br><br>These potential applications suggest that droplet microfluidics | drought tolerance in crops. Additionally, it aids in disease detection, allowing for early intervention to prevent crop loss. Despite challenges such as scalability, cost, and integration with existing agricultural practices, ongoing research aims to overcome these barriers, making droplet microfluidics a promising tool for precision agriculture and sustainable food production. | detailed as those provided in the ground truth answer. The response also does not address some key aspects mentioned in the ground truth, such as disease detection and genetic/phen otypic trait analysis. SCORE: 36 | | | |

| | | | | | |
|---|---|---|---|---|---|
| | could become a valuable tool in agriculture, enabling more precise and efficient crop management, improved crop monitoring, and increased food security. | | | | |
| What are the current limitations of droplet microfluidics, and how can they be overcome? | Based on the provided context, the current limitations of droplet microfluidics include:<br><br>1. **Limited applicability to low-throughput assays**: Droplet microfluidics is particularly well-suited for ultra-high-throughput assays, but not all biological assays require such high throughput. For fewer samples, tools like 1536-well microtiter plates are generally sufficient.<br>2. **Challenges in maintaining stable and inert oil-water interfaces**: The high surface area-to-volume ratio of droplets requires a suitable surfactant | Droplet microfluidics has immense potential for high-throughput biological assays but faces several limitations that hinder its broader adoption. One key challenge is its efficiency for assays involving small sample sizes; droplet microfluidics excels with large volumes (at least 10^5 samples), making traditional methods like microtiter plates more suitable for lower sample numbers. Another critical issue is maintaining a stable and inert oil-water interface, which is essential for droplet formation. Finding the right surfactant to keep | The response provides a clear and concise list of the current limitations of droplet microfluidics and potential solutions. It addresses all key points mentioned in the ground truth answer, including efficiency for small sample sizes, stable oil-water interfaces, spatial barcoding, droplet movement control, and stability over time. However, some minor | 85 | 85.0 | 85 |

| | | | | | |
|---|---|---|---|---|---|
| | to maintain a stable interface, which can be a challenging chemistry problem. 3. **Difficulty in developing barcoding systems for droplet libraries**: Unlike microtiter plates, droplet libraries require a suitable barcoding system, which can be difficult to develop. 4. **Limitations in spatial resolution and analysis**: Droplets cannot take advantage of the spatial barcoding that is available in two-dimensional arrays of microtiter plates. 5. **Challenges in controlling the movement of many droplets**: The automated control of the movement of many droplets is a specific challenge for droplet-based microfluidics. To overcome these limitations, researchers are | this interface stable remains a complex chemistry challenge. Furthermore, droplet microfluidics lacks the spatial barcoding available in microtiter plates, making it difficult to organize and track multiple droplet libraries. As a result, developing alternative barcoding systems is necessary but challenging. The automation of droplet movement is another obstacle, as it requires precise control, often through techniques like electrowetting and dielectrophoresis. Lastly, droplet stability over time can be compromised, with issues such as shrinkage, coalescence, and cross-talk between droplets. To address these issues, research is underway to | details are missing or not explicitly stated, such as specific examples of novel materials and techniques being explored. The response is well-structured and easy to follow, but could benefit from a more comprehensive discussion on the potential impact of these advancements on the field. SCORE: 85 | | | |

| | | | | | |
|---|---|---|---|---|---|
| | working on:<br><br>1. **Developing new materials and techniques**: New materials like "liquid Teflon" and techniques such as electrowetting on dielectric (EWOD), dielectrophoretic, and electric forces are being explored to improve the control of droplets.<br>2. **Improving surfactant formulations**: Studies have been carried out to develop oil phase and surfactant formulations with improved isolation properties to address issues related to imperfect partitioning between continuous and discrete phases.<br>3. **Investigating new analytical methods**: New analytical methods with high resolution in both space and time are being developed to | develop improved surfactants, more reliable barcoding systems, and novel materials and techniques to enhance droplet control. These advancements, coupled with better analytical tools, are expected to overcome current limitations and expand droplet microfluidics' applications in chemical and biological research. | | | | |

| | study reactions and reaction networks.<br><br>Overall, while droplet microfluidics has shown great promise, there is still much work to be done to overcome these limitations and fully realize its potential in various fields of science. | | | | | |
|---|---|---|---|---|---|---|
| How can interdisciplinary collaboration accelerate innovation in droplet microfluidics? | Based on the provided context, it is clear that interdisciplinary collaboration is crucial for accelerating innovation in droplet microfluidics. The documents highlight several areas where combining knowledge from different disciplines can lead to significant breakthroughs and advancements.<br><br>The first document mentions that "significant future discoveries and break-throughs are not | Interdisciplinary collaboration is essential for advancing droplet microfluidics by combining expertise from various fields such as chemistry, biology, material science, and engineering. By uniting these diverse areas, researchers can address complex challenges such as cost-effective fabrication, scalable production, and novel interfacial reaction systems. For example, biologists and chemists can leverage droplet microfluidics for high-throughput | The response provides a clear and concise explanation of how interdisciplinary collaboration can accelerate innovation in droplet microfluidics. It effectively addresses the question by highlighting various ways in which combining knowledge from different disciplines | 85 | 85.0 | 85 |

| | | | | | |
|---|---|---|---|---|---|
| | necessarily dependent on the development of further microfluidic modules, but rather on combining knowledge from different disciplines." This suggests that collaboration between experts in chemistry, biology, and other fields is essential for driving innovation in droplet microfluidics.<br><br>Similarly, the second document emphasizes the importance of integrating sensitive detectors and machine learning algorithms with droplet-based platforms to enhance analytical throughput, sensitivity, and operational sophistication. This implies that collaboration between chemists, biologists, computer scientists, and engineers can lead | analysis, while material scientists and engineers work on developing more robust and affordable microfabrication methods. Furthermore, integrating technologies like sensitive detectors and machine learning algorithms can enhance the analytical capabilities of droplet-based systems. Collaborative initiatives, such as the EMBL Microfluidics Conference, foster knowledge exchange and innovation by bringing together professionals from different disciplines. This cooperation is crucial for overcoming technical barriers, reducing manufacturing costs, and developing new applications in fields like drug discovery, single-cell analysis, and nanomaterial | can lead to significant breakthroughs and advancements. The answer is well-structured, easy to follow, and aligns with the ground truth. However, it could be improved by providing more specific examples and details about the benefits of interdisciplinary collaboration in droplet microfluidics.<br><br>SCORE: 85 | | | |

| | to more effective and efficient use of droplet microfluidic systems.

The third document highlights several challenges in commercial-scale manufacturing of droplet microfluidics, including the need for low-cost microfabrication methods and materials, as well as the development of novel interfacial reaction systems. It suggests that collaboration between researchers from different areas can help bridge the gap between materials synthesis and applications, leading to more exciting accomplishments in droplet microfluidics.

In conclusion, interdisciplinary collaboration is essential for accelerating innovation in droplet | synthesis. Ultimately, interdisciplinary collaboration accelerates the evolution of droplet microfluidics, pushing the boundaries of what is achievable in this promising field. | | | | |
|---|---|---|---|---|---|---|

| | microfluidics. By combining knowledge and expertise from different disciplines, researchers can overcome challenges, develop new technologies, and push the boundaries of what is possible with droplet microfluidics.<br><br>Some potential ways to accelerate innovation through interdisciplinary collaboration include:<br><br>1. Combining expertise: Bringing together experts from chemistry, biology, computer science, engineering, and other fields to share knowledge and ideas.<br>2. Interdisciplinary research initiatives: Establishing programs that encourage collaboration between researchers from | | | | | |
|---|---|---|---|---|---|---|

| | | | | | |
|---|---|---|---|---|---|
| | different disciplines to tackle specific challenges in droplet microfluidics.<br>3. Cross-disciplinary training: Providing training opportunities for researchers to learn about the principles and applications of droplet microfluidics, as well as other relevant fields.<br>4. Industry-academia partnerships: Collaborating with industry partners to develop practical solutions to real-world problems using droplet microfluidic technologies.<br><br>By fostering a culture of interdisciplinary collaboration, researchers can unlock new possibilities in droplet microfluidics and drive innovation in this exciting field. | | | | |

| question | answer | ground_truth | feedback | score | percentage | Expert Rating |
|----------|--------|--------------|----------|-------|------------|---------------|
| What is microfluidics? | Microfluidics is a multidiscipli nary field that deals with the design, manipulation , and analysis of fluids at the microscale, typically involving channels or chambers with | Microfluidics is a multidisciplinary field that involves the study and manipulation of fluids at the microscale, typically within channels with dimensions ranging from tens to hundreds of micrometers. It is a multidisciplinary field that involves molecular | The response is clear, concise, and accurately describes the field of microfluidics, including its scope, principles, and applications. It provides relevant details about the subject without unnecessary information. SCORE: 96 | 96 | 96.0 | 100 |

| | dimensions on the order of millimeters to micrometers. It involves the study of fluid behavior in these small spaces, which can exhibit unique properties due to surface tension, viscosity, and other factors. Microfluidics has applications in fields such as biology, chemistry, medicine, and engineering. | analysis, molecular biology, and microelectronics. Microfluidics has been increasingly used in the biological sciences because precise and controlled experiments can be conducted at a lower cost and faster pace. Lab on a Chip devices use microfluidics for applications such as Point of Care testing of diseases, or Organ on a Chip studies. | | | | |
|---|---|---|---|---|---|---|
| What is droplet microfluidics? | Droplet microfluidics is a technique that manipulates and controls tiny droplets of liquids, typically in the microliter to picoliter range. It involves the creation, manipulation | Droplet-based microfluidics is a branch of microfluidics focusing on the creation and manipulation of discrete volumes of liquids through adoption of immiscible fluid flows. Microdroplets allow for handling miniature | The response accurately describes the concept of droplet microfluidics, its principles, and applications, aligning well with the ground truth. However, it could benefit from a more precise definition and emphasis on the unique | 86 | 86.0 | 85 |

| | | | | | |
|---|---|---|---|---|---|
| | , and analysis of these droplets using microfluidic devices, which are designed to handle fluids at the microscale. This field combines principles from chemistry, physics, and engineering to create a platform for various applications such as chemical synthesis, biological assays, and particle production. | volumes (µL to fL) of fluids conveniently, provide better mixing, encapsulation, sorting, and sensing, and suit high throughput experiments. Compared with continuous-flow systems, droplet microfluidic systems enable isolation and manipulation of samples in separate compartments. The droplets generated with uniform size and shape provide a stable microenvironment for biochemical reactions. Processing of a large number of droplets in parallel facilitates to achieve ultrahigh throughput. Confinement of individual targets within the droplets and independent processing of each of them allow high-precision analyses. | features that distinguish droplet microfluidics from other microfluidic techniques. SCORE: 86 | | | |

| How to generate microdroplets? | Microdroplets can be generated using various methods, including:

1. Flow-focusing technique: This involves forcing a liquid through a narrow channel or nozzle, creating a stream of droplets.
2. Pneumatic atomization: Using compressed air to break up a liquid into small droplets.
3. Ultrasonic nebulization: Using high-frequency sound waves to create tiny droplets from a liquid.
4. Electrostatic atomization: Applying an electric charge to a liquid, causing it to break up into small | Microdroplet generation in microfluidics is achieved through two main methods: active and passive. Active methods utilize external forces like piezoelectric actuators to precisely control droplet size, formation rate, and spacing. These methods enable high uniformity in droplet generation, with deviations less than 0.3%. Passive methods rely on fluid properties and channel geometry, such as T-junctions or cross-flow configurations, to create droplets. These methods are influenced by factors like flow rate, viscosity, and interfacial tension, allowing for fine control over droplet size and uniformity. For example, PDMS-based microfluidic devices have been used for | The response provides a list of various methods to generate microdroplets but does not directly address the question's focus on the two main methods (active and passive) used in microfluidics. SCORE: 16 | 16 | 16.0 | 10 |
|---|---|---|---|---|---|---|

| | droplets.<br>5. Centrifugal atomization: Spinning a liquid at high speed to create droplets.<br>6. Laser-induced breakdown spectroscopy (LIBS): Using a laser pulse to create microdroplets from a liquid.<br>7. Microfluidic devices: Specialized devices that use channels and valves to control the flow of liquids and generate microdroplets.<br><br>These methods can be used in various applications, such as drug delivery, chemical synthesis, and biological research. | nanoparticle synthesis, where hydrodynamic coupling of nozzles generates droplet pairs for precise mixing. Both methods are valuable in applications like nanoparticle synthesis, offering high reproducibility and control over reaction conditions in biomedical and material science research. | | | | |
| --- | --- | --- | --- | --- | --- |

| What materials are typically used to fabricate droplet microfluidic devices? | Polydimethyl siloxane (PDMS), glass, and silicon are commonly used to fabricate droplet microfluidic devices. | Droplet microfluidic devices are commonly fabricated from a variety of materials, each offering distinct advantages for specific applications. The most widely used material is PDMS (Polydimethylsilo xane), favored for its ease of use with soft lithography and flexibility, though it has limitations such as poor solvent resistance. Alternatives to PDMS include thermoset polyesters, thermoplastics like PMMA, polycarbonate, and polystyrene, and amorphous polymers such as cyclic olefin copolymer. Fluoropolymers like PTFE and glass are also used, offering superior solvent resistance and biofouling resistance. Silicon and capillary devices are utilized in certain applications, | The response provides a concise list of materials used for fabricating droplet microfluidic devices, but lacks the detailed explanation and context provided in the ground truth answer. It does not discuss the advantages and limitations of each material or the factors that influence their selection. SCORE: 21 | 21 | 21.0 | 20 |
| | | providing precise control for microfluidic operations. These materials are chosen based on factors like solvent resistance, biofouling, and the ability to mass-produce devices using methods such as injection molding or hot embossing, ensuring broad applicability in research and industrial settings. | | | | |
| What are the applications of droplet microfluidics? | Droplet microfluidics has a wide range of applications, including:<br><br>1. Single-cell analysis and sorting<br>2. High-throughput screening for drug discovery<br>3. Cell encapsulation and tissue engineering<br>4. Protein crystallization and purification<br>5. Emulsion-based synthesis of nanoparticles | Droplet microfluidics has diverse applications across biomedical research, diagnostics, and industrial fields. It enables high-throughput single-cell analysis, such as genomics, transcriptomics, and drug screening, while also facilitating nanoparticle synthesis, enzyme evolution, and chemical reactions. In healthcare, it supports personalized medicine, point-of-care diagnostics, and | The answer provides a comprehensive list of applications of droplet microfluidics but lacks the depth and context provided by the ground truth answer. It does not fully capture the broader implications and connections between these applications, nor does it highlight the key benefits and advantages of using droplet microfluidics in various fields. SCORE: 45 | 45 | 45.0 | 40 |

| | | | | | |
|---|---|---|---|---|---|
| | and microparticles 6. Microreactors for chemical synthesis and catalysis 7. Biosensing and diagnostics 8. Food processing and emulsification 9. Cosmetics and personal care product formulation 10. Environmental monitoring and remediation. | vaccine development. Beyond biology, it is used for environmental monitoring, food safety testing, and the development of cosmetics and pharmaceuticals. Its ability to create isolated, uniform reaction chambers in droplets makes it a powerful tool for precise, cost-effective, and scalable workflows in science and industry. | | | |
| What are the main benefits of using droplet microfluidics over bulk systems? | The main benefits of using droplet microfluidics over bulk systems include:

1. High-throughput screening and analysis 2. Reduced reagent consumption 3. Improved precision and accuracy in chemical synthesis and | Droplet microfluidics offers significant advantages over traditional bulk systems. It enables individual compartmentalization, isolating single cells or molecules in tiny liquid compartments for precise control. This allows for single-cell analysis and early detection of rare cells or secreted molecules. The | The answer provides a list of general benefits of using droplet microfluidics over bulk systems, but lacks depth and context. It does not fully capture the unique advantages of compartmentalization, single-cell analysis, and iterative operations that are highlighted in the ground | 41 | 41.0 | 40 |

| | | | | | | |
|---|---|---|---|---|---|---|
| | biological assays<br>4. Enhanced control over reaction conditions (e.g., temperature, pH)<br>5. Increased efficiency in particle formation and manipulation<br>6. Better scalability for industrial applications<br>7. Lower costs due to reduced material usage and waste generation | technology supports massive parallel processing, analyzing millions of droplets simultaneously, ideal for high-throughput screening and examining phenotypic or genetic variability. Additionally, iterative operations such as merging, mixing, and sorting droplets enable complex experimental protocols and continuous monitoring of populations. The system's highly controllable local environments ensure consistent, reproducible results. These benefits make droplet microfluidics a versatile tool in drug discovery, medical diagnostics, single-cell analysis, and other fields, offering powerful insights and precise | truth answer.<br>SCORE: 41 | | | |

| | | manipulation at the individual level. | | | | |
|---|---|---|---|---|---|---|
| What are the advantages of droplet microfluidic systems compared with continuous-flow systems? | The advantages of droplet microfluidic systems compared with continuous-flow systems include:

1. Higher precision and control over droplet size, composition, and formation.
2. Lower reagent consumption and reduced waste generation.
3. Improved mixing efficiency due to the small volume and high surface area of individual droplets.
4. Enhanced scalability and flexibility for various applications, such as single-cell analysis and high- | Droplet-based microfluidic systems offer significant advantages over traditional continuous-flow systems. These include the ability to overcome issues such as Taylor dispersion, slow mixing, and surface-molecule interactions. Droplet systems allow for ultra-small assay volumes, reducing reagent consumption and improving sensitivity. They also minimize cross-contamination and sample adsorption by compartmentalizing individual samples. The system's chaotic advection enhances mixing efficiency, enabling faster reactions. With droplet generation frequencies up to hundreds of kHz, droplet microfluidics supports high- | The answer provides a clear and concise list of advantages of droplet microfluidic systems compared to continuous-flow systems, aligning well with the ground truth. However, it lacks some depth and specific details about the benefits mentioned in the ground truth, such as overcoming Taylor dispersion and surface-molecule interactions. SCORE: 76 | 76 | 76.0 | 75 |

| | | | | | | |
|---|---|---|---|---|---|---|
| | throughput screening. 5. Reduced risk of contamination and improved sterility due to the isolated nature of each droplet. 6. Ability to perform multiple reactions or assays in parallel within a single device. 7. Potential for higher throughput and faster processing times compared to continuous-flow systems. | throughput experimentation. Furthermore, the rapid and precise control of droplet payloads and sub-millisecond mixing times make it ideal for applications requiring fast, controlled analysis. These benefits make droplet-based systems particularly valuable for cell-based assays, enzyme kinetics, and high-throughput screening. | | | | |
| What are the basic principles behind droplet generation in microfluidics? | The basic principles behind droplet generation in microfluidics include: 1. Shear force: The flow of a continuous phase (e.g., oil) through a narrow | Droplet generation in microfluidics relies on emulsion processes, where an oil phase (continuous phase) encapsulates an aqueous phase (dispersed phase) to form droplets. Surfactants are added to lower interfacial | The response provides a clear and concise overview of the basic principles behind droplet generation in microfluidics, listing six different methods. However, it lacks depth and context, failing to explain the | 41 | 41.0 | 40 |

| | | | | | |
|---|---|---|---|---|---|
| | channel creates a shear force that can break up an immiscible fluid (e.g., water) into droplets. 2. Co-flowing fluids: Two or more fluids are flowed side-by-side in a microchannel, creating droplets as the faster-moving fluid breaks up the slower-moving fluid. 3. T-junctions: A T-shaped junction is used to mix two immiscible fluids, creating droplets of one phase within the other. 4. Flow focusing: A narrow channel or constriction is used to focus a flow, creating a high-velocity | tension, stabilize the droplets, and prevent coalescence. The process requires precise control over the flow of both liquids, often achieved using syringe pumps or pressure-driven systems. Droplets are generated through surface-induced instabilities, influenced by microchannel geometries such as T-junctions or flow-focusing, which control droplet size and frequency. The design and flow control of the system determine the size uniformity and production rate of droplets. These principles enable high-throughput applications such as drug delivery, high-throughput screening, and microreactors for bio-catalysis, ensuring reproducibility and productivity in biological experiments. | underlying physics or chemistry behind these processes. It also does not address the importance of surfactants, surface-induced instabilities, or the role of microchannel geometries in controlling droplet size and frequency. The response seems to be a collection of facts rather than a comprehensive explanation of the principles. SCORE: 41 | | | |
| | jet that can break up an immiscible fluid into droplets. 5. Acoustic manipulation : Ultrasound waves are used to manipulate and generate droplets in microfluidic devices. 6. Electrohydro dynamic (EHD) focusing: An electric field is applied to create a high-velocity flow of one phase, which breaks up the other phase into droplets.<br><br>These principles can be combined or modified to achieve specific droplet generation characteristic s, such as size, uniformity, and stability. | | | | | |
| How is droplet size controlled in | In a droplet generator, | Droplet size control in | The response attempts to | 36 | 36.0 | 30 |

| a droplet generator? | droplet size is typically controlled by adjusting the flow rates of the liquid and gas streams, as well as the geometry of the device. This can be achieved through various methods such as:<br><br>* Adjusting the orifice diameter or shape<br>* Changing the pressure drop across the orifice<br>* Modulating the flow rate of the liquid stream<br>* Using a surfactant to alter the surface tension and droplet formation<br>* Employing acoustic or electrical fields to manipulate the droplets<br>* Utilizing microfluidic devices with | microfluidic generators can be achieved through several methods. In passive droplet generation, adjusting microchannel dimensions or controlling the flow rates of immiscible liquids allows for precise control over droplet size, shape, and monodispersity. Tree-splitters, which bisect large droplets into smaller ones, also influence droplet size and uniformity, with splitting angles affecting droplet diameter. In active droplet generation, technologies like piezoelectric actuators independently control droplet size, formation rate, and spacing, offering high precision. Additionally, the velocity of the continuous and dispersed phases plays a significant role in controlling droplet size. These methods | address the question but lacks clarity and precision compared to the ground truth answer. It provides a list of general methods for controlling droplet size in a droplet generator without delving into specific details or emphasizing the importance of microfluidic principles, which are crucial for precise control over droplet size and uniformity. SCORE: 36 | | | |

| | precise control over fluid flow and geometry. | enable the generation of uniform droplets for various applications, such as high-throughput screening and drug delivery. | | | | |
|---|---|---|---|---|---|---|
| What is the role of surface tension in droplet formation? | Surface tension plays a crucial role in droplet formation by acting as a force that minimizes the surface area of a liquid, causing it to form spherical droplets. This is because liquids have an inherent property called surface energy, which causes them to behave as if they are trying to minimize their surface area. As a result, when a liquid is placed on a surface or in air, it will naturally form a | Surface tension plays a critical role in droplet formation in microfluidic systems by influencing droplet size, shape, and formation mechanisms. It acts to minimize surface area, favoring spherical droplet shapes. At low Weber numbers (We < 1), where surface tension dominates, droplets form through dripping or squeezing. As the Weber number increases, fluid inertia competes with surface tension, leading to jetting behavior. The capillary number (Ca) also influences droplet size, with surface tension controlling droplet formation at low Ca. | The response accurately describes the role of surface tension in droplet formation, mentioning its ability to minimize surface area and favor spherical shapes. However, it lacks detailed information about the specific conditions under which surface tension influences droplet size and shape, such as Weber numbers and capillary numbers. SCORE: 55 | 55 | 55.0 | 55 |
| | sphere-shaped droplet due to the balance between the inward pull of surface tension and the outward pressure from the surrounding environment. | Techniques like electrowetting can measure surface tension in microfluidic systems, providing insights into the interfacial properties. Additionally, surface tension is a key parameter in T-junctions and flow focusing devices, where droplet size measurements offer an indirect way to assess surface tension at the time-scale of droplet production. Understanding the interplay between surface tension and other forces is essential for precise droplet control. | | | | |
| How does a flow-focusing device differ from a T-junction in droplet formation? | A flow-focusing device and a T-junction are both used for generating monodisperse droplets, but they have different mechanisms of operation.<br><br>In a T-junction, the | Flow-focusing devices and T-junctions are two common microfluidic designs for droplet formation, differing in geometry and mechanism. A flow-focusing device uses a central dispersed phase channel and side continuous phase channels | The response accurately describes the mechanisms of operation for both flow-focusing devices and T-junctions in droplet formation, highlighting their differences in geometry and hydrodynamic forces involved. It correctly | 85 | 85.0 | 85 |

| | | | | | |
|---|---|---|---|---|---|
| | droplet size is determined by the shear stress at the intersection point, which depends on the flow rates of the two fluids. The droplet formation is primarily driven by the hydrodynamic forces.<br><br>A flow-focusing device, on the other hand, uses a converging channel to focus the continuous phase into a narrow stream, which then breaks up into droplets when it encounters the dispersed phase. The droplet size in a flow-focusing device is determined by the flow rates of both fluids and the geometry of | converging at a narrow orifice, where hydrodynamic focusing pinches off droplets. It produces smaller, highly uniform droplets with precise control over size and high throughput. In contrast, a T-junction has perpendicular channels where the continuous phase shears the dispersed phase at the intersection, forming larger droplets with less uniformity and lower throughput. Flow-focusing is ideal for applications requiring monodisperse droplets, while T-junctions suit simpler or larger-droplet applications. | explains how the droplet size is determined in each device, with a focus on the key factors influencing droplet formation. However, it does not delve into the specific advantages of one over the other or discuss the ideal applications for each, which are mentioned in the ground truth answer. SCORE: 85 | | | |
| | the focusing channel. | | | | | |
| What role does the geometry of the microchannel play in droplet formation? | The geometry of the microchannel plays a crucial role in droplet formation by controlling the flow dynamics, fluidic properties, and interfacial interactions. The shape, size, and surface characteristics of the microchannel can influence the formation of droplets through various mechanisms:

1. **Flow focusing**: The microchannel's geometry can focus the incoming fluid streams, creating a high-velocity region where droplet formation is more likely to occur. | The geometry of a microchannel plays a critical role in droplet formation, influencing the size, shape, and generation frequency of droplets. Key geometric factors include the channel width, depth, and junction design, such as T-junctions or flow-focusing geometries. These features govern the interaction between the continuous and dispersed phases, affecting the shear forces and pressure gradients that drive droplet breakup. For instance, narrower channels or smaller orifices promote the formation of smaller droplets due to increased shear stress. Similarly, the junction type controls droplet detachment mechanics; in T-junctions, droplets form due | The response provides a clear and detailed explanation of the role of microchannel geometry in droplet formation, discussing various mechanisms such as flow focusing, shear stress, surface tension, and confinement. It also highlights the importance of tailoring the microchannel's geometry to achieve desired droplet characteristics. However, some minor inaccuracies and lack of focus on specific geometric factors like channel width, depth, and junction design are noted. The response could benefit from a more precise alignment with the ground truth answer in terms of key geometric factors and their impact on droplet | 78 | 78.0 | 75 |

| | 2. **Shear stress**: The channel's dimensions and shape can generate specific shear stresses that affect the flow behavior and promote or inhibit droplet formation. 3. **Surface tension**: The microchannel's surface properties, such as hydrophobicity or hydrophilicity, can influence the contact angle between the fluid and the channel walls, affecting the formation of droplets. 4. **Confinement**: The narrow dimensions of the microchannel can create a confined environment that promotes | to the balance of interfacial tension and shear force, while in flow-focusing designs, they form under pinching at the constriction. Channel geometry also determines flow regimes ranging from dripping to jetting further impacting droplet characteristics. Thus, optimizing microchannel geometry is crucial for precise control over droplet formation in applications like microfluidics, emulsification, and lab-on-a-chip systems. | formation. SCORE: 78 | | | |
| --- | --- | --- | --- | --- | --- | --- |
| | the formation of droplets due to the increased pressure and flow velocity.<br><br>The specific geometry of the microchannel can be tailored to achieve desired droplet characteristics, such as size, shape, and uniformity, which is crucial for various applications in chemistry, biology, and materials science. | | | | | |
| What factors affect droplet coalescence in microfluidic devices? | Surface tension, interfacial energy, hydrophobicity of the device walls, flow rates, and electrical properties can all influence droplet coalescence in | Droplet coalescence in microfluidic devices is influenced by several factors. Surfactant concentration plays a key role, with higher concentrations preventing droplet fusion, while lower concentrations promote | The response provides a list of relevant factors affecting droplet coalescence in microfluidic devices but lacks the depth and detail provided by the ground truth answer, particularly regarding the specific effects of surfactant concentration, | 45 | 45.0 | 45 |

| | microfluidic devices. | coalescence. The duration droplets remain in contact within the pairing channel, known as pairing time, also impacts coalescence, with longer times favoring fusion. The geometry of the pairing channel, specifically its cross-sectional area, controls the number of droplets that can come into contact and coalesce. Additionally, droplet size and volume affect how efficiently droplets merge. Finally, electrocoalescence, which applies electric fields to induce droplet fusion, offers an alternative method for on-demand coalescence. By optimizing these factors—surfactant concentration, pairing time, channel geometry, and droplet size—reliable and selective droplet coalescence can be achieved in | pairing time, channel geometry, and electrocoalescence. SCORE: 45 | | | |
|---|---|---|---|---|---|---|

| | | microfluidic systems. | | | | |
|---|---|---|---|---|---|---|
| What design elements are critical for producing double or multiple emulsions? | Surfactants, co-surfactants, and a combination of oil and water phase compositions with specific ratios are critical for producing double or multiple emulsions. | To generate double or multiple emulsions, several design elements are essential. Co-flow and flow-focusing configurations enable precise control over fluid flow, allowing for stable droplet formation, especially in double emulsions. Sequential emulsification steps, where inner droplets are emulsified into intermediate phases and then into outer phases, are key for multiple emulsions. Capillary microfluidic devices, such as consecutive capillary and coaxial configurations, are used to control droplet size, number, and monodispersity. Additionally, precise flow control and shear force management for stable | The response partially addresses the question by mentioning some relevant elements (surfactants, co-surfactants, oil-water phase ratios) but misses critical design elements such as flow configurations, sequential emulsification steps, capillary microfluidic devices, and other factors that are essential for producing double or multiple emulsions. SCORE: 41 | 41 | 41.0 | 40 |

| | | and consistent droplet formation. Microchannel geometry optimization and material compatibility further enhance emulsification processes. By carefully managing these factors, researchers can produce double and multiple emulsions with controlled internal structures and compositions for various applications. | | | | |
|---|---|---|---|---|---|---|
| What key factors influence the design of flow-focusing droplet microfluidics optimized using machine learning? | The key factors that influence the design of flow-focusing droplet microfluidics optimized using machine learning include:

1. Channel geometry and dimensions
2. Flow rates and ratios of the two immiscible fluids
3. Surface tension and | The design of flow-focusing droplet microfluidics, optimized using machine learning, depends on several critical factors. Channel geometry, including the shape and size, controls droplet generation by defining the break-off point where the dispersed phase meets the continuous phase. Flow rates of both phases affect droplet size and frequency, with | The answer provides a list of factors that influence the design of flow-focusing droplet microfluidics optimized using machine learning, but lacks explanation and context. It appears to be a summary of key parameters without elaboration on how they interact or are used in the optimization process. SCORE: 21 | 21 | 21.0 | 20 |

| | interfacial properties<br>4. Viscosity and density of the fluids<br>5. Droplet size and polydispersity<br>6. Shear stress and pressure drop across the microchannel<br>7. Operating temperature and pressure | higher continuous phase flow reducing droplet size. Viscosity influences flow behavior and shear forces, crucial for stable droplet formation. Shear forces between phases are controlled by adjusting channel geometry or flow rates, ensuring consistent droplet generation. Machine learning algorithms predict and optimize droplet diameter and generation rate by analyzing these design parameters, fluid properties, and large datasets, automating the design process for more efficient droplet production in life sciences applications. | | | | |
|---|---|---|---|---|---|---|
| How do channel wall properties (hydrophobicity/hydrophilicity) influence droplet dynamics? | The hydrophobic or hydrophilic properties of a surface can significantly affect the behavior and dynamics of liquid droplets on | The channel wall properties, specifically hydrophobicity or hydrophilicity, significantly influence droplet dynamics in microfluidic systems. These properties determine the | The response provides a clear and detailed explanation of how channel wall properties influence droplet dynamics, including the effects on contact angle, adhesion, movement, | 76 | 76.0 | 75 |

| | | | | | |
|---|---|---|---|---|---|
| | that surface.<br><br>On a hydrophobic surface, water droplets tend to:<br><br>* Form into spherical shapes due to the minimization of contact area with the surface<br>* Have a higher contact angle (typically greater than 90°) with the surface<br>* Be more resistant to spreading and wetting<br>* Exhibit reduced adhesion to the surface<br><br>On the other hand, on a hydrophilic surface:<br><br>* Water droplets tend to spread out and wet the surface due to strong intermolecular forces | wettability of the channel walls, affecting the interaction between the continuous and dispersed phases. In hydrophobic channels, the dispersed phase (typically water) tends to form droplets as it minimizes contact with the channel walls. This enhances droplet stability and facilitates easier detachment during droplet formation. Hydrophobic surfaces are ideal for applications requiring water-in-oil emulsions or precise droplet control. Conversely, hydrophilic channels promote wetting of the dispersed phase, making it more likely to spread along the walls instead of forming discrete droplets. This is useful for oil-in-water systems or applications involving liquid films. Wall properties also | merging, evaporation, and other aspects. It also discusses the specific implications for microfluidic systems and applications. However, it does not directly address the question's focus on the general principles of hydrophobicity/hydrophilicity in relation to droplet dynamics, instead delving into more specific details about microfluidics.<br>SCORE: 76 | | | |

| | between water molecules and the surface<br>* Have a lower contact angle (typically less than 90°) with the surface<br>* Exhibit increased adhesion to the surface<br><br>These differences in behavior can influence various aspects of droplet dynamics, such as:<br><br>* Droplet movement: Hydrophobic surfaces tend to repel droplets, while hydrophilic surfaces attract them.<br>* Droplet merging and coalescence: The properties of the surface can affect how droplets merge and | impact droplet size, breakup, coalescence, and transport. Hydrophobicity favors smaller droplets and reduces the likelihood of droplet adherence, while hydrophilicity can lead to higher flow resistance and altered droplet generation behavior. Tailoring wall properties is crucial for optimizing microfluidic processes. | | | | |
|---|---|---|---|---|---|

| | | | | | | |
|---|---|---|---|---|---|---|
| | combine with each other.<br>* Droplet evaporation: The rate of evaporation may be influenced by the surface's ability to facilitate or hinder water-vapor exchange.<br><br>The specific effects on droplet dynamics depend on the degree of hydrophobicity or hydrophilicity, as well as the properties of the liquid itself. | | | | | |
| How can you incorporate active control mechanisms (e.g., electrical, magnetic) into droplet devices? | Active control mechanisms such as electrical and magnetic fields can be incorporated into droplet devices through various methods:<br><br>* Electrical control:<br>    + Electrodes | Active control mechanisms such as electrical, magnetic, thermal, and mechanical methods can precisely manipulate droplet formation in microfluidic devices. Electrical control methods include direct current (DC), which uses electrodes to control droplet | The response provides a clear and concise overview of various methods to incorporate electrical and magnetic control mechanisms into droplet devices. It correctly identifies specific techniques such as DEP, EWOD, and magnetophoresis, but does not | 76 | 76.0 | 75 |

| | | |
|---|---|---|
| | can be integrated into the device to apply electric fields that manipulate droplets. <br>     + Dielectrophoresis (DEP) can be used to trap, sort, or move droplets based on their dielectric properties. <br>     + Electrowetting-on-dielectric (EWOD) can be employed to control droplet movement and merging. <br> * Magnetic control: <br>     + Permanent magnets or electromagnets can be used to manipulate ferrofluids or magnetically responsive particles within the device. <br>     + Magnetic | size, and alternating current (AC), which utilizes electrowetting-on-dielectric (EWOD) and dielectrophoresis (DEP) to influence droplet behavior. Magnetic control employs ferrofluids and external magnets to non-invasively guide droplet formation. Thermal control involves heating the entire device or applying localized heating to adjust fluid parameters like viscosity and interfacial tension. Mechanical control techniques, such as pneumatic and piezoelectric components, can also regulate droplet dynamics. These active control strategies enable precise and flexible droplet generation, allowing for tailored applications in complex | mention thermal or mechanical control methods, which are mentioned in the ground truth answer. SCORE: 76 |

| | fields can be applied to control the movement of paramagnetic or diamagnetic liquids.<br><br>+<br><br>Magnetophor esis can be used to separate droplets based on their magnetic properties.<br><br>These mechanisms can be combined with microfluidic principles, such as flow rates and pressure control, to create complex droplet manipulation systems. | microfluidic systems, including droplet manipulation and multicomponent emulsion creation. | | | |
|---|---|---|---|---|---|---|
| How do flow rates affect droplet size and frequency? | In a spray or atomization process, flow rate can significantly impact droplet size and frequency. Generally, as the flow rate | Flow rates significantly influence droplet size and frequency in microfluidic systems. As the flow rate ratio ($\varphi$ = Qd/Qc) increases, the shearing force | The response accurately describes how flow rates affect droplet size and frequency in a spray or atomization process, providing clear explanations for | 76 | 76.0 | 75 |

| | | | | | |
|---|---|---|---|---|---|
| | increases:<br><br>* Droplet size decreases due to increased turbulence and shear forces that break up larger droplets into smaller ones. * Droplet frequency (or the number of droplets per unit time) also increases with higher flow rates, as more liquid is being pumped through the system.<br><br>Conversely, lower flow rates tend to produce larger droplets and a lower droplet frequency. | from the continuous phase dominates, leading to smaller droplets due to rapid breakup of the dispersed phase. Additionally, higher flow rates in the continuous phase can result in larger droplets, as the dispersed phase doesn't break up until it reaches the downstream section. The frequency of droplet generation also increases with higher flow rate ratios, demonstrating a positive correlation. However, droplet diameter growth trends vary depending on the model and input conditions. Overall, increasing the flow rate ratio enhances droplet formation frequency and reduces droplet size due to stronger shearing forces, with complex interactions influenced by | both increasing and decreasing flow rates. It aligns with the ground truth answer in terms of the general principles, but lacks specific details about microfluidic systems and the complexities involved.<br>SCORE: 76 | | | |

| | | factors such as interfacial tension and channel geometry. | | | | |
|---|---|---|---|---|---|---|
| What techniques are available for droplet splitting? | Some common techniques used for droplet splitting include:<br><br>1. Acoustic levitation and manipulation<br>2. Optical tweezers<br>3. Electrostatic forces (e.g., electrohydro dynamic atomization)<br>4. Pneumatic or hydraulic systems<br>5. Centrifugal force (e.g., spinning disk or centrifuge)<br>6. Laser-induced breakdown spectroscopy (LIBS) based droplet splitting<br>7. Microfluidic devices with narrow channels or nozzles | Droplet splitting techniques can be categorized into passive and active methods. Passive techniques rely on the microchannel geometry and fluid properties, such as T and Y junctions, flow focusing configurations, or introducing ribs/obstructions in the flow path. These methods are simple but lack dynamic control over splitting ratios and are sensitive to flow conditions. Active techniques utilize external forces like electric, magnetic, acoustic, or thermal fields to precisely control droplet splitting. Examples include electric field-based splitting, magnetic fields for ferrofluids, acoustic radiation force, and thermocapillary methods. Active techniques offer | The response attempts to list various techniques for droplet splitting but lacks a clear categorization and explanation of the methods as provided in the ground truth answer. It also fails to discuss the limitations and advantages of each technique, which is an essential aspect of understanding their application. SCORE: 16 | 16 | 16.0 | 10 |

| | | more flexibility, enabling on-demand droplet splitting with controlled ratios, overcoming the limitations of passive methods. However, each technique has its own set of advantages and constraints depending on the application and droplet type. | | | | |
|---|---|---|---|---|---|---|
| What advantages does droplet formation by passive methods have over active methods? | In many cases, droplet formation by passive methods has several advantages over active methods:

1. Lower energy consumption: Passive methods typically require less energy to produce droplets compared to active methods.
2. Reduced equipment cost: The equipment required for passive methods is | Passive droplet generation methods offer several advantages over active methods. They simplify device design by relying on flow rate manipulation and microchannel geometry, eliminating the need for complex external forces. This leads to cost-effectiveness and easier fabrication. Passive methods also provide precise control over droplet size, shape, and monodispersity, making them ideal for applications requiring uniform droplets. Additionally, they | The response provides a clear and concise list of advantages of droplet formation by passive methods over active methods. However, it lacks some details and nuances present in the ground truth answer, such as the specific benefits of simplified device design and protection from external interference. SCORE: 56 | 56 | 56.0 | 55 |

| | often simpler and less expensive than that needed for active methods. 3. Improved uniformity: Passive methods can produce more uniform droplet sizes, which is important in many applications such as pharmaceuticals and cosmetics. 4. Lower risk of contamination: Since no external energy source is used, the risk of contamination from electrical or mechanical sources is reduced. 5. Flexibility: Passive methods can be easily scaled up or down depending on the | protect droplets from external interference and cross-contamination, ensuring integrity. Passive methods can produce continuous droplet streams with uniform size distributions (1%-3% dispersity) without requiring external actuation. While they are not suitable for dynamic manipulation or "droplet-on-demand" applications, passive methods are favored for their simplicity, precision, and cost-efficiency in controlled environments. | | | | |
|---|---|---|---|---|---|---|

| | | | | 45 | 45.0 | 40 |
|---|---|---|---|---|---|---|
| | application, making them more versatile than active methods.

These advantages make passive droplet formation methods attractive for various industries and applications where precise control over droplet size and energy efficiency are important. | | | | | |
| How do you design a device for monodisperse droplet generation? | A device for monodisperse droplet generation typically consists of:

1. A fluid reservoir to hold the liquid to be dispensed.
2. A narrow orifice (e.g., a needle or a microfluidic channel) that controls the flow rate and size of the droplets. | To design a device for monodisperse droplet generation, two common methods are used: cross-flow and flow-focusing designs. In cross-flow designs, the continuous phase flows around the dispersed phase in a constriction channel, with droplet size controlled by adjusting channel dimensions and shear stress. This | The response provides a detailed description of the components and considerations for monodisperse droplet generation, but it does not directly address the question's request for design methods (cross-flow and flow-focusing designs) that are specifically tailored to achieve uniform | 45 | 45.0 | 40 |

| | | | | | |
|---|---|---|---|---|---|
| | 3. A pressure source (e.g., a syringe pump or a pressure controller) to regulate the flow rate through the orifice.<br>4. A collection chamber or surface to receive the generated droplets.<br><br>The key design considerations for monodisperse droplet generation include:<br><br>1. Controlling the fluid flow rate and pressure to maintain a consistent droplet size.<br>2. Minimizing turbulence and disturbances in the fluid stream to prevent droplet coalescence or | method is simple but can lead to droplet damage due to shear forces or wall interactions. Flow-focusing designs, on the other hand, use sheath flows to stabilize the interface between phases, enabling smaller droplet formation with precise control over size by adjusting flow rates and orifice dimensions. This method offers high monodispersity and is less prone to droplet damage. Both designs require careful optimization of flow rates, channel dimensions, and surface tension to achieve uniform droplet sizes. The choice of design depends on specific needs such as droplet size, throughput, and resistance to damage during formation. | droplet sizes.<br>SCORE: 45 | | | |

| | | | | | |
|---|---|---|---|---|---|
| | fragmentation.<br>3. Ensuring the orifice is clean and free of contaminants to prevent clogging or contamination of the generated droplets.<br><br>Some common methods for generating monodisperse droplets include:<br><br>1. Microfluidic devices with narrow channels and precise control over flow rates.<br>2. Pneumatic or syringe-based systems that use compressed air or a piston to generate droplets.<br>3. Electrostatic or acoustic methods that use electrical charges or | | | | |

| | | | | | | |
|---|---|---|---|---|---|---|
| | sound waves to break up the fluid into uniform droplets.<br><br>The design of the device should be tailored to the specific application and requirements, such as the type of liquid being dispensed, the desired droplet size, and the required production rate. | | | | | |
| What materials are most commonly used for fabricating droplet microfluidic devices? | Polydimethyl siloxane (PDMS), glass, and polycarbonat e are the most commonly used materials for fabricating droplet microfluidic devices. | The most commonly used materials for fabricating droplet microfluidic devices are PDMS (polydimethylsilo xane) and silicon-based materials. PDMS is preferred due to its ease of fabrication using soft lithography, which allows for the creation of precise and versatile microfluidic devices. This | The answer is mostly correct but lacks detail and context; it does not provide a clear explanation of why these materials are used or their specific properties. SCORE: 41 | 41 | 41.0 | 40 |

| | | material provides high transparency, efficient sealing, and the ability to mold complex microchannels, making it ideal for droplet generation and manipulation. While PDMS is dominant, alternatives such as thermoplastics (e.g., PMMA, polycarbonate), fluoropolymers (e.g., PTFE), and glass materials are also used, each offering unique benefits like better solvent resistance or mass production capabilities. Despite the variety of materials available, PDMS remains the material of choice for most droplet microfluidics applications due to its simplicity and versatility. | | | | |
|---|---|---|---|---|---|---|
| What are the advantages and limitations of using PDMS for microfluidic devices? | PDMS (Polydimethy lsiloxane) is a popular material used in microfluidic devices due to its unique | Polydimethylsilox ane (PDMS) is widely used for microfluidic devices due to its several advantages, including low toxicity, | The response provides a clear and concise list of advantages and limitations of using PDMS for microfluidic devices. It covers the main points | 76 | 76.0 | 75 |

| | properties. The advantages include:<br><br>* Transparency, allowing for easy visualization of the fluid flow and reactions<br>* Biocompatibility, making it suitable for biological applications<br>* Low cost and ease of fabrication using soft lithography techniques<br>* High flexibility, enabling the creation of complex geometries and channels<br>* Chemical inertness, reducing the risk of contamination and chemical interactions<br><br>However, there are also some limitations: | biocompatibility, optical transparency, and high gas permeability, which make it ideal for biomedical applications. Its flexibility and cost-effectiveness also contribute to its popularity, as it allows for easy fabrication using soft lithography. PDMS enables real-time observation and is well-suited for flexible and wearable electronics. However, PDMS has limitations, such as its hydrophobic nature, which can lead to fouling in microchannels, affecting performance. Additionally, its soft elastomeric properties limit mechanical strength, and large-scale production can be challenging. Despite these drawbacks, PDMS remains a versatile and widely used material, though | mentioned in the ground truth answer, including transparency, biocompatibility, flexibility, and chemical inertness as advantages, and limited mechanical strength, susceptibility to swelling and shrinkage, potential for leaching, and difficulty in scaling up as limitations. However, it does not mention some of the specific details provided in the ground truth answer, such as PDMS's low toxicity, high gas permeability, hydrophobic nature, and its suitability for flexible and wearable electronics.<br>SCORE: 76 | | | |
|---|---|---|---|---|---|---|
| | * Limited mechanical strength, which can lead to deformation or cracking under pressure or temperature changes<br>* Susceptibility to swelling and shrinkage due to solvent exposure or temperature variations<br>* Potential for leaching of chemicals from the PDMS material into the fluid, affecting its purity<br>* Difficulty in scaling up to larger device sizes without compromising performance. | further research is needed to address its limitations for broader applications. | | | | |
| How does 3D printing facilitate microfluidic device fabrication? | Three-dimensional (3D) printing enables the rapid and cost-effective creation of complex | 3D printing revolutionizes microfluidic device fabrication by enabling the creation of complex and precise channel | The response accurately addresses the question by providing a detailed explanation of how 3D printing | 94 | 94.0 | 95 |

| | | | | | |
|---|---|---|---|---|---|
| | geometries, which is particularly beneficial for microfluidic devices. This technology allows for the fabrication of intricate networks of channels, chambers, and other features that are difficult to produce using traditional manufacturing methods.<br><br>The 3D printing process involves layer-by-layer deposition of materials such as plastics, metals, or ceramics, which can be tailored to specific properties required for microfluidic applications. This includes:<br><br>1. Complex | structures that are challenging to achieve with traditional methods like soft lithography. Techniques such as Stereolithography (SL), Multijet Modeling (MJM), and Fused Deposition Modeling (FDM) offer high resolution, complex geometries, and the integration of various components like valves and pumps. 3D printing also allows rapid prototyping, customization, and the use of diverse materials for biocompatibility and transparency. Advantages over traditional methods include reduced fabrication time and labor, increased accuracy, and environmental friendliness due to the additive manufacturing process. With its ability to create | facilitates microfluidic device fabrication. It covers various aspects such as complex geometries, high precision, customization, reduced material waste, and advantages over traditional methods like faster prototyping and production, lower costs, increased design flexibility, and improved accuracy and precision. The response is well-structured, clear, and concise, with a good balance of technical details and general information. SCORE: 94 | | | |

| | channel geometries: 3D printing enables the creation of intricate channel networks with precise dimensions and shapes, which is essential for efficient fluid flow and mixing in microfluidic devices. 2. High precision: The layer-by-layer deposition process allows for high accuracy and resolution, ensuring that features are printed at the desired scale. 3. Customization: 3D printing enables the fabrication of devices with tailored properties, such as specific surface roughness or chemical | hierarchical 3D configurations and integrate multiple functions, 3D printing holds great potential for advancing microfluidic device design and performance. | | | | |
|---|---|---|---|---|---|---|

| | functionality, which can enhance device performance. 4. Reduced material waste: 3D printing minimizes material waste by only printing the required amount of material for each feature.

The use of 3D printing in microfluidic device fabrication has several advantages over traditional methods, including:

1. Faster prototyping and production
2. Lower costs
3. Increased design flexibility
4. Improved accuracy and precision

Overall, 3D printing | | | | | |
|---|---|---|---|---|---|---|
| | facilitates the rapid development and production of complex microfluidic devices with tailored properties, enabling researchers to explore new applications and improve existing ones. | | | | | |
| What alternative materials can be used for applications involving organic solvents? | Water, glycerin, and plant-based oils such as coconut oil or olive oil are some alternatives to traditional organic solvents. | Several sustainable alternatives to traditional organic solvents are gaining attention for various applications. Supercritical CO2 is a non-toxic, environmentally friendly solvent, particularly useful in extraction processes. Ionic liquids, which are liquid salts at room temperature, offer low vapor pressure and broad application in chemical reactions. Organic carbonates, like dimethyl carbonate, are biodegradable and | The answer provides a list of alternative materials to traditional organic solvents but lacks the comprehensive detail and variety of examples provided in the ground truth. It also fails to mention specific properties such as polarity, boiling point, and environmental impact that are crucial for choosing these alternatives. SCORE: 35 | 35 | 35.0 | 30 |

| | | non-toxic, ideal for fine chemical synthesis. Biosolvents, derived from renewable resources such as plants and bacteria, provide a greener option for extraction and cleaning. Other alternatives include fluorous solvents, non-toxic liquid polymers like polyethylene glycol (PEG), water, ethanol, methanol, and ethyl acetate. These alternatives are chosen based on their specific properties such as polarity, boiling point, and environmental impact, promoting more sustainable chemical processes and reducing reliance on hazardous petroleum-based solvents. | | | | |
|---|---|---|---|---|---|---|
| How does device surface treatment affect droplet formation? | Device surface treatment can significantly impact droplet formation by altering the | Surface treatment of PDMS microfluidic devices plays a critical role in determining droplet formation and stability. | The response accurately addresses the question and provides relevant information about device surface | 76 | 76.0 | 75 |

| | wettability and surface energy of the material, which in turn affects the behavior of liquids on its surface. A treated surface can either enhance or inhibit the spreading and merging of droplets, depending on whether it is hydrophilic (water-attracting) or hydrophobic (water-repelling). | PDMS is naturally hydrophobic, suitable for generating water-in-oil (W/O) droplets but not oil-in-water (O/W) droplets. To generate O/W droplets, selective hydrophilic treatment is required. The researchers employed two methods: manual PVA deposition after plasma treatment, and a simultaneous approach combining PVA injection with on-chip plasma treatment. The latter method enables precise wettability patterning, allowing for both W/O and O/W droplet formation within the same device. The plasma treatment increases surface energy, making the PDMS hydrophilic, while PVA deposition stabilizes the coating, ensuring long-term device use. This technique | treatment's impact on droplet formation, including its effect on wettability and surface energy. It also mentions specific methods of surface treatment (plasma treatment and PVA deposition) and their effects on PDMS devices. However, it does not provide as much detail or context as the ground truth answer regarding the specifics of PDMS properties and the simultaneous approach combining PVA injection with plasma treatment. SCORE: 76 | | | |
|---|---|---|---|---|---|---|

| | | provides controllable droplet sizes and stable double emulsion formation, with the ability to manipulate droplet characteristics by adjusting flow rates. The method also allows for reliable, reproducible results, making it a powerful tool for droplet microfluidic applications. | | | | |
|---|---|---|---|---|---|---|
| What bonding techniques are used to seal microfluidic channels? | Thermocompression bonding, anodic bonding, and laser welding are commonly used techniques to seal microfluidic channels. | Several bonding techniques are employed to seal microfluidic channels, each suited to different materials. Thermal fusion bonding involves compressing thermoplastic substrates at temperatures near their glass transition point, allowing them to fuse upon cooling. Chemical bonding creates covalent bonds between substrate surfaces through chemical reactions, while solvent bonding uses solvents to | The response provides a concise list of techniques used to seal microfluidic channels but lacks detailed explanations and context. It does not address the variety of materials involved or the specific conditions required for each technique, which are crucial aspects mentioned in the ground truth answer. SCORE: 21 | 21 | 21.0 | 20 |

| | | dissolve the materials partially, promoting adhesion through interdiffusion. For PDMS devices, oxygen plasma treatment is commonly used to make surfaces reactive, enabling irreversible bonding when in contact with another PDMS or glass piece. Additionally, corona wand bonding offers a less expensive alternative to oxygen plasma treatment, though it is less consistent. The choice of bonding technique depends on material properties, bond strength requirements, and available equipment, all of which are crucial for the successful fabrication of microfluidic devices. | | | | |
|---|---|---|---|---|---|---|
| What challenges arise when fabricating complex microfluidic geometries? | The main challenges that arise when fabricating complex | Fabricating complex microfluidic geometries presents several challenges, | The answer provides a comprehensive list of challenges related to fabricating | 41 | 41.0 | 40 |

| | | | | | |
|---|---|---|---|---|---|
| | microfluidic geometries include:<br><br>1. Scalability and reproducibility: Maintaining consistent results across multiple devices is difficult due to the small size of the features.<br>2. Material limitations: Most materials used in microfabrication have limited mechanical strength, thermal stability, or chemical resistance.<br>3. Surface roughness: The surface finish of the fabricated structures can be compromised by the manufacturing process, affecting fluid flow and interactions. | including controlling wettability in intricate channel structures, which is critical for fluid flow and droplet formation. Additionally, scaling up microfabrication processes for commercial applications requires cost-effective techniques and materials. Current droplet generation methods mainly produce single emulsions, limiting their use in applications that require multiple or complex emulsions. Moreover, droplet microfluidics faces limitations in producing diverse and functional nano-scale materials compared to other methods. Conventional lithography techniques, which are restricted to rectangular cross-sections, struggle with creating complex 3D geometries. | complex microfluidic geometries, including scalability, material limitations, surface roughness, feature resolution, complexity of design, material properties, and interconnectivity. However, the response lacks depth and context compared to the ground truth answer, which highlights specific issues such as controlling wettability in intricate channel structures, scaling up microfabrication processes, and producing diverse nano-scale materials. The provided answer seems to be a list of general challenges without delving into the specifics or providing a clear direction for addressing | | | |

| | 4. Feature resolution: Achieving high-resolution features with dimensions smaller than 100 μm is challenging due to the physical limitations of current fabrication techniques. 5. Complexity of design: Complex geometries require intricate designs that are difficult to fabricate using traditional methods. 6. Material properties: The properties of materials at the microscale can differ significantly from their bulk counterparts, affecting device performance. 7. Interconnecti | Although advancements like 3D printing and stacking methods offer solutions, effective mixing strategies and versatile fabrication of 3D networks with the required features remain significant hurdles in the field. Addressing these challenges will require innovative approaches and collaboration across various disciplines. | these challenges. SCORE: 41 | | | |
|---|---|---|---|---|---|---|

| | | | | | |
|---|---|---|---|---|---|
| | vity and integration: Integrating multiple components with complex geometries is a significant challenge due to the difficulty in creating reliable connections between them. | | | | | |
| How do you fabricate devices capable of handling multiphase flows? | Fabricating devices capable of handling multiphase flows typically involves a combination of design, materials selection, and manufacturing techniques. Here are some common methods:<br><br>1. **3D Printing**: Additive manufacturing (AM) techniques such as Fused Deposition Modeling | Fabricating devices for handling multiphase flows involves several techniques and considerations. Key approaches include advanced micromanufacturing methods such as 3D printing, which eliminates the need for bonding processes, and flow-focusing techniques, which use nozzles and sheath fluids to control droplet formation. Controlling wettability is crucial for generating specific droplet types, while parallelization of droplet generators | The response provides a comprehensive overview of various methods used in fabricating devices capable of handling multiphase flows, including 3D printing, casting and molding, welding and joining, machining and milling, coating and surface treatment, and computational fluid dynamics analysis. It also mentions the importance of considering factors such as fluid properties, flow rates, pressure, | 86 | 86.0 | 85 |

| | | | | | |
|---|---|---|---|---|---|
| | (FDM), Selective Laser Sintering (SLS), or Electron Beam Melting (EBM) can be used to create complex geometries that facilitate multiphase flow. 2. **Casting and Molding**: Investment casting, sand casting, or 3D printing of molds can produce intricate designs with features that promote efficient multiphase flow. 3. **Welding and Joining**: Techniques like laser welding, electron beam welding, or friction stir welding can be used to join multiple components | enhances production scalability. Challenges include ensuring even fluid distribution, integrating various components like valves and connectors, and addressing the lack of standardized design rules. Additionally, methods like consecutive capillary assemblies and external forces (e.g., pneumatic or electrical control) can be employed to create complex emulsions. Through careful design and optimization, efficient fabrication of multiphase flow devices for applications such as droplet generation can be achieved. | temperature, and material compatibility when selecting these methods. The response is well-structured and provides a clear explanation of each technique, making it easy to understand for readers who are not experts in the field.  SCORE: 86 | | |

| | together, creating a single device capable of handling multiphase flows. 4. **Machining and Milling**: CNC machining or milling operations can precision-cut complex geometries from various materials, such as metals, polymers, or ceramics. 5. **Coating and Surface Treatment**: Applying specialized coatings or surface treatments to the device's interior surfaces can enhance its ability to handle multiphase flows by reducing friction, promoting wetting, or preventing | | | | | |
|---|---|---|---|---|---|---|
| | fouling. 6. **Computational Fluid Dynamics (CFD) Analysis**: Using CFD simulations can help designers optimize device geometries for efficient multiphase flow handling before actual fabrication begins.<br><br>These methods are often combined and tailored to the specific requirements of the application, taking into account factors such as fluid properties, flow rates, pressure, temperature, and material compatibility. | | | | | |
| What is the impact of channel roughness on | Channel roughness can | The impact of channel roughness on | The response provides a clear and concise | 56 | 56.0 | 60 |

| droplet formation and behavior? | significantly affect the formation and behavior of droplets in various applications, such as inkjet printing, microfluidics, and chemical processing. The roughness can alter the flow characteristics, leading to:<br><br>* Changes in droplet size and distribution<br>* Increased turbulence and mixing<br>* Alterations in droplet breakup and coalescence dynamics<br>* Enhanced or reduced droplet formation rates<br>* Variability in droplet velocity and trajectory<br><br>The impact of channel roughness on droplet | droplet formation and behavior is a key factor in microfluidics. Studies indicate that surface roughness can significantly affect droplet dynamics, such as movement, spreading, and stability. For example, increased roughness can hinder droplet spreading and prevent secondary droplet formation by altering flow dynamics and the contact line. Additionally, vibration and surface structure play roles in influencing droplet impact and splash behavior. Rougher surfaces tend to reduce spreading, while smoother surfaces may enhance it. Surface wettability also interacts with roughness to further complicate droplet behavior. Overall, channel roughness introduces complexity to | overview of the impact of channel roughness on droplet formation and behavior. It correctly identifies various effects such as changes in droplet size, increased turbulence, and alterations in breakup dynamics. However, it lacks specific details about the underlying mechanisms and interactions with other factors like surface wettability, which are crucial for a comprehensive understanding. The response also does not fully capture the complexity of droplet behavior on rough surfaces, particularly regarding the impact of vibration and external conditions. SCORE: 56 | | | |
| --- | --- | --- | --- | --- | --- |
| | behavior can be both positive (e.g., improved mixing, increased droplet uniformity) and negative (e.g., reduced accuracy, increased variability). | droplet dynamics, with its effects varying based on factors like roughness level, fluid type, and external conditions, highlighting the need for further research in this area for better understanding and optimization in microfluidic applications. | | | |
| How can sensors be integrated into droplet microfluidic devices during fabrication? | Sensors can be integrated into droplet microfluidic devices during fabrication through various methods, including:

1. Photolithography and etching: This involves creating a pattern on the device substrate using light and chemicals to create cavities or channels for sensor integration.
2. Soft | Integrating sensors into droplet microfluidic devices during fabrication is a complex process that requires advanced microfabrication techniques and careful material selection. For successful integration, precision in manufacturing is crucial, as it often involves combining different microfluidic units, such as droplet generation and detection units, on a single chip. Various techniques like photolithography, soft lithography, | The response provides a clear and concise overview of various methods for integrating sensors into droplet microfluidic devices during fabrication. It lists specific techniques such as photolithography, soft lithography, 3D printing, micro-molding, and post-fabrication integration, which is accurate according to the ground truth answer. However, it lacks a detailed discussion on the complexities of sensor | 51 | 51.0 | 55 |

| | lithography: Techniques such as soft photolithography (e.g., SU-8) or nanoimprint lithography can be used to create microscale features for sensor integration. 3. 3D printing: Additive manufacturing techniques like stereolithography or two-photon polymerization can be employed to fabricate complex geometries and integrate sensors within the device. 4. Micro-molding: This involves creating a mold of the desired device geometry, which is then used to create multiple copies with | and direct integration methods can be used, depending on the type of sensor (e.g., optical or electrochemical). The fabrication process also needs to account for material compatibility, sensor functionality, and alignment to ensure reliable performance. Additionally, sensor integration may face challenges such as maintaining stability within the microfluidic environment and minimizing interference with fluid flow. By considering these factors, sensors can be effectively integrated into droplet microfluidic devices for applications like chemical reactions and biological detection. | integration, material selection, and potential challenges like maintaining stability within the microfluidic environment and minimizing interference with fluid flow. The response seems to focus more on listing methods rather than providing a comprehensive understanding of the process. SCORE: 51 | | | |
| | integrated sensors. 5. Post-fabrication integration: Sensors can also be attached to the device after fabrication using techniques such as wire bonding or flip-chip bonding.<br><br>These methods allow for the precise placement and integration of sensors within droplet microfluidic devices, enabling real-time monitoring and control of the fluid dynamics and chemical reactions occurring within the device. | | | | | |
| How are droplet microfluidic devices used for single-cell analysis? | Droplet microfluidic devices are used for single-cell | Droplet microfluidic devices are widely used for single-cell | The response accurately describes the use of droplet microfluidic | 76 | 76.0 | 75 |

| | | | | | |
|---|---|---|---|---|---|
| | analysis by encapsulating individual cells in picoliter-sized aqueous droplets, which are then processed and analyzed. This allows for the isolation of a single cell per droplet, enabling precise control over experimental conditions and minimizing sample consumption. The droplets can be manipulated, sorted, and analyzed using various techniques such as fluorescence-activated cell sorting (FACS), microscopy, or spectroscopy, allowing researchers to study cellular | analysis due to their ability to encapsulate individual cells within tiny droplets, allowing for high-throughput and precise experimentation. These devices work by generating microdroplets in a fluid flow, each containing a single cell along with reagents needed for analysis, such as enzymes, antibodies, or primers. This allows for the isolation of cells in a controlled environment, enabling high-resolution studies of gene expression, protein profiling, and cellular responses. The compartmentalized nature of droplets also reduces cross-contamination between samples. Moreover, the small size and scalability of droplet microfluidics make it possible | devices for single-cell analysis, including encapsulation, manipulation, and analysis techniques. It also mentions the benefits of precise control over experimental conditions and minimal sample consumption. However, it lacks some details about the technology's applications in genomics, proteomics, and drug screening, as well as its ability to reduce cross-contamination between samples. SCORE: 76 | | | |

| | behavior, gene expression, and other properties at the single-cell level. | to perform large-scale, parallel analysis of thousands of single cells in a short amount of time. This technology has applications in genomics, proteomics, and drug screening, facilitating insights into cell heterogeneity and the identification of rare cell populations. | | | | |
|---|---|---|---|---|---|---|
| What are the advantages of droplet microfluidics for high-throughput screening? | The advantages of droplet microfluidics for high-throughput screening include:<br><br>* High-throughput analysis of individual cells or particles in a controlled and miniaturized environment<br>* Ability to perform thousands to millions of experiments simultaneously<br>* Reduced | Droplet microfluidics offers several advantages for high-throughput screening, including reduced sample volumes, which lowers costs, and increased assay sensitivity through compartmentalization. This method allows for the screening of up to $10^8$ samples per day at kilohertz speeds, significantly improving throughput and efficiency. The small droplet volumes enhance effective | The answer provides a clear and concise list of advantages of droplet microfluidics for high-throughput screening, aligning with the general concepts mentioned in the ground truth. However, it lacks specific details and quantitative information provided in the ground truth, such as the ability to screen up to $10^8$ samples per day at kilohertz speeds, which is a key benefit of this technology. SCORE: 56 | 56 | 56.0 | 55 |

| | reagent consumption and cost per experiment * Improved precision and accuracy due to the ability to control and manipulate individual droplets * Enhanced reproducibility and consistency across experiments * Potential for high-speed screening and rapid data generation. | concentrations and signal-to-noise ratios, critical for sensitive detection of low-concentration biological molecules such as antibodies, enzymes, and single genomes. Additionally, droplet microfluidics enables single-cell and single-molecule assays, providing detailed insights into cellular behaviors and allowing for novel experimental approaches. This technology is particularly beneficial for applications like drug discovery, enzyme evolution, and diagnostics, offering a versatile and cost-effective platform for advanced biological research. | | | | |
|---|---|---|---|---|---|---|
| How can droplets be used to encapsulate cells or biomolecules? | Droplets can be used to encapsulate cells or biomolecules through techniques | Droplet microfluidics provides versatile methods for encapsulating cells and biomolecules. | The response provides a clear and concise explanation of how droplets can be used to encapsulate cells | 46 | 46.0 | 45 |

| | | | | | |
|---|---|---|---|---|---|
| | such as microfluidics, where tiny droplets of a liquid are created and then used to trap or encapsulate individual cells or biomolecules. This method allows for precise control over the size and composition of the droplets, enabling the efficient and gentle handling of sensitive biological materials. | Limiting dilution is a traditional approach that randomly distributes single cells or biomolecules in droplets, though it has low efficiency. Inertial microfluidics leverages spiral geometries and inertial forces to guide cells into droplets, improving encapsulation efficiency by ensuring single-cell droplets. Bi-disperse emulsions create an emulsion of empty and cell-containing droplets, with the latter selectively recovered using methods like pinched flow fractionation. For biomolecules, limiting dilution is commonly used, adjusting concentration and drop size to achieve single-molecule encapsulation. Additionally, droplets can encapsulate cells in lipid vesicles for biological | or biomolecules, mentioning specific techniques such as microfluidics. However, it lacks the comprehensive detail and variety of methods provided in the Ground Truth Answer, which includes limiting dilution, inertial microfluidics, bi-disperse emulsions, and others. The response also does not mention the specific applications of these techniques beyond "efficient and gentle handling" of biological materials. SCORE: 46 | | | |
| | | applications or serve as microreactors in processes like DNA analysis. These techniques enable precise control over encapsulation, enhancing their applicability in cell-based assays, drug discovery, and genetic research | | | | |
| What are the applications of droplet microfluidics in drug discovery? | Droplet microfluidics has several applications in drug discovery, including:<br><br>1. High-throughput screening: Droplet microfluidics enables the creation of millions of uniform droplets containing a single cell or molecule, allowing for high-throughput screening of compounds and their effects on cells.<br>2. Compound library | Droplet microfluidics plays a significant role in drug discovery by enabling high-throughput screening and precise analysis of drug interactions at the single-cell level. This technology allows the generation of picoliter-sized droplets, each encapsulating a single cell or a small group of cells, which can be exposed to various drug compounds. This compartmentalization facilitates parallel testing of numerous drug candidates, speeding up the identification of | The response provides a clear and concise list of applications of droplet microfluidics in drug discovery. It covers various aspects such as high-throughput screening, compound library generation, cell-based assays, protein crystallization, enzyme screening, personalized medicine, and toxicity testing. However, the answer could be improved by providing more depth and explanation for each point, and possibly integrating them | 56 | 56.0 | 55 |

| | generation: Microdroplets can be used to generate libraries of compounds with specific properties, such as solubility or stability, which can then be screened for potential therapeutic applications. 3. Cell-based assays: Droplet microfluidics allows for the creation of miniaturized cell-based assays, enabling researchers to study cellular behavior and interactions at a single-cell level. 4. Protein crystallization: Microdroplets can be used to create conditions that facilitate protein crystallization, which is | promising therapeutics. Droplet microfluidics is particularly useful in identifying drug efficacy, optimizing drug formulations, and studying complex cellular responses such as gene expression, protein production, or metabolic changes in response to treatment. Additionally, it enables the screening of rare cell populations or targets that would be difficult to isolate using traditional methods. The ability to rapidly test and analyse thousands of drug interactions in a controlled environment accelerates drug discovery, reduces costs, and improves the chances of identifying novel therapeutic agents. | into a cohesive narrative that highlights the broader impact of droplet microfluidics on drug discovery. SCORE: 56 | | | |
|---|---|---|---|---|---|---|

| | essential for X-ray crystallography and structure-based drug design. 5. Enzyme screening: Droplet microfluidics enables the high-throughput screening of enzymes and their substrates, allowing researchers to identify potential enzyme inhibitors or activators. 6. Personalized medicine: Microdroplets can be used to create personalized models of disease, enabling researchers to study the effects of specific genetic mutations on cellular behavior. 7. Toxicity testing: Droplet | | | | | |
|---|---|---|---|---|---|---|

| | | | | | |
|---|---|---|---|---|---|
| | microfluidics allows for the rapid and efficient screening of compounds for toxicity, reducing the need for animal testing and improving the safety of new drugs. | | | | |
| How can droplet microfluidics enable precise control of chemical reactions? | Droplet microfluidics enables precise control of chemical reactions by allowing for the creation and manipulation of small, uniform droplets of reactants in a controlled environment. This allows for:

* Precise mixing ratios of reactants
* Control over reaction conditions such as temperature, pH, and pressure
* Minimization | Droplet microfluidics offers precise control over chemical reactions by dispensing reactants into individual droplets with accurate volume control. This enables the mixing of reactants within confined spaces, enhancing reaction efficiency. Coalesced droplets act as micro-reactors, providing fast mixing and long operation times without clogging, even with sticky or precipitating reactants. The system allows for optimization of reaction | The response provides a clear and concise explanation of how droplet microfluidics enables precise control of chemical reactions, covering various aspects such as mixing ratios, reaction conditions, waste minimization, scalability, and high-throughput screening. It also highlights the benefits of this technology in optimizing chemical reactions for improved yields and efficiency. However, some minor inaccuracies and differences in wording | 83 | 83.0 | 80 |

| | | | | | |
|---|---|---|---|---|---|
| | of waste and reagent consumption * Scalability and reproducibility of reactions * High-throughput screening of chemical reactions<br><br>This precise control enables the optimization of chemical reactions, leading to improved yields, reduced side products, and increased efficiency in various applications such as pharmaceutical development, materials synthesis, and biotechnology. | parameters, such as temperature, flow rate, and pressure, ensuring the production of specific products, like mesoporous silica microspheres with superior surface area and narrow pore distributions. This capability to finely control reaction conditions makes droplet microfluidics a powerful tool for precise, high-throughput chemical synthesis. | compared to the ground truth answer are noted. SCORE: 83 | | | |
| What are the key challenges of using droplet microfluidics in diagnostics? | The key challenges of using droplet microfluidics in diagnostics include:<br><br>1. Scalability | Droplet microfluidics presents significant challenges in diagnostics, including preventing wetting and cross- | The response provides a clear and concise list of challenges associated with using droplet microfluidics in diagnostics. However, it does | 41 | 41.0 | 40 |

| | and throughput: Maintaining high-throughput while maintaining precise control over droplet formation and manipulation. 2. Sample preparation and handling: Ensuring that the sample is properly prepared, handled, and introduced into the microfluidic device without contamination or degradation. 3. Sensitivity and specificity: Achieving sufficient sensitivity and specificity for detection of target molecules or cells in a complex biological matrix. 4. Integration | contamination between aqueous droplets and immiscible organic liquids, which can compromise results. Monitoring oxygen levels within droplets is difficult due to the sensitivity of the liquid-liquid interface to mechanical perturbations and mass transfer. Developing stable detection chemistries is also complex, as the confined droplet environment can alter chemical reactions and impact accuracy. Additionally, challenges in automation and accessibility limit the widespread use of droplet microfluidics, particularly for high-throughput assays. Overcoming these issues requires advancements in surfactants, microfluidic engineering, and automated techniques, which could enhance the | not fully address the specific issues mentioned in the Ground Truth Answer, such as preventing wetting and cross-contamination between aqueous droplets and immiscible organic liquids, monitoring oxygen levels within droplets, developing stable detection chemistries, and overcoming challenges in automation and accessibility. The response seems to focus more on general aspects of droplet microfluidics rather than the specific challenges mentioned in the Ground Truth Answer. SCORE: 41 | | | |
|---|---|---|---|---|---|

| | with existing diagnostic infrastructure : Seamlessly integrating droplet microfluidics with existing laboratory equipment and workflows. 5. Cost-effectiveness: Balancing the cost of developing and implementing droplet microfluidic devices with the need to make them affordable for widespread use. 6. Standardization and reproducibility: Ensuring that results obtained from droplet microfluidic devices are consistent across different users, locations, and time points. 7. Interfacing | precision and applicability of droplet systems in diagnostics. | | | | |
|---|---|---|---|---|---|---|

| | with clinical workflows: Integrating droplet microfluidics into existing clinical diagnostic protocols and workflows. | | | | | |
|---|---|---|---|---|---|---|
| How are droplets used for DNA sequencing applications? | Droplets are used in digital PCR (Polymerase Chain Reaction) and microfluidic devices for DNA sequencing applications to amplify and analyze specific DNA sequences. They allow for the precise control of reagents, temperature, and reaction conditions, enabling high-throughput and sensitive detection of genetic variations. | Droplet microfluidics plays a key role in DNA sequencing by enabling high-throughput analysis and precise targeting of genomic regions. This includes single-cell sequencing, where individual cells are encapsulated in droplets for independent gene expression profiling. Barcode-based sequencing allows for the labeling of nucleic acids with unique identifiers, enabling the tracking of specific genomes during analysis. Multiplexed PCR amplification is achieved by performing PCR in separate droplets, ensuring uniform amplification of | The response provides some relevant information about the use of droplets in DNA sequencing applications, but it lacks detail and specific examples compared to the ground truth answer. It also does not mention key concepts such as single-cell sequencing, barcode-based sequencing, or emulsion PCR, which are crucial for a comprehensive understanding of the topic. SCORE: 41 | 41 | 41.0 | 40 |

| | | multiple targets. Additionally, emulsion PCR combines droplet merger with PCR to amplify specific genomic regions, facilitating efficient sequencing. These techniques offer a miniaturized, efficient platform for analyzing complex genetic information, improving accuracy and throughput in genomic studies. | | | | |
|---|---|---|---|---|---|---|
| What role do droplets play in tissue engineering research? | Droplets play a crucial role in tissue engineering research as they can be used to create micro-scale, three-dimensional (3D) cell cultures that mimic the natural environment of tissues. These droplet-based systems allow for the precise control of cell-cell and cell-matrix | Droplet microfluidics plays a pivotal role in tissue engineering by enabling precise fabrication of biomaterials and 3D cell culture systems. Droplets serve as templates for creating hydrogel-based materials, which can encapsulate cells for more accurate modeling of tissue structures. This technology facilitates the production of microcarriers and complex | The response is mostly relevant but lacks focus on the specific aspects mentioned in the ground truth answer, such as the creation of hydrogel-based materials, microcarriers, and controlled release of therapeutic agents. SCORE: 46 | 46 | 46.0 | 45 |
| | interactions, which is essential for understanding tissue development, function, and disease progression. | biomaterials, such as hollow bacterial cellulose (BC) microspheres, which form injectable scaffolds for wound healing, promoting high cell proliferation and tissue regeneration in vivo. Additionally, droplet-based microfluidics allows for the creation of core-shell structures, such as alginate shells or metal-organic frameworks, which offer controlled release of therapeutic agents. These advancements provide a powerful platform for studying tissue development and regenerative therapies. | | | | |
| How can droplet microfluidics aid in studying cell signaling pathways? | Droplet microfluidics can aid in studying cell signaling pathways by allowing for the precise control and manipulation | Droplet microfluidics offers a powerful platform for studying cell signaling pathways by enabling the precise isolation and manipulation | The response provides a clear and concise overview of how droplet microfluidics can aid in studying cell signaling pathways, listing specific | 46 | 46.0 | 45 |

| | | | | | |
|---|---|---|---|---|---|
| of small volumes of biological samples, such as cells or proteins. This enables researchers to:<br><br>*   Create uniform populations of cells with specific genetic modifications<br>*   Isolate and study individual cells or subpopulations within a mixed-cell culture<br>*   Deliver specific molecules or drugs to cells in a controlled manner<br>*   Monitor cellular responses to stimuli in real-time using fluorescence microscopy or other techniques<br><br>These capabilities | of individual cells within controlled environments. This technology allows for the encapsulation of single cells in droplets, providing a controlled space to study cell-cell interactions and signaling events. The ability to merge droplets containing different cell types or reagents facilitates the investigation of cell communication, such as paracrine signaling, in a way that mimics in vivo conditions. Additionally, droplet-based microfluidics supports high-throughput screening and single-cell genomic studies, offering insights into phenotypic variations and signaling dynamics within genetically identical cell populations. This capability is particularly valuable for | capabilities such as precise control over biological samples, isolation of individual cells, and controlled delivery of molecules. However, the answer lacks depth and detail compared to the ground truth answer, which provides more nuanced explanations of the technology's potential for investigating cell-cell interactions, paracrine signaling, and high-throughput screening. The response also does not fully capture the unique advantages of droplet microfluidics in mimicking in vivo conditions and studying complex cellular behaviors.<br><br>SCORE: 46 | | | |

| | | | | | | |
|---|---|---|---|---|---|---|
| | can facilitate the investigation of cell signaling pathways by allowing researchers to:<br><br>*   Identify key regulators and effectors of signaling cascades<br>*   Study the dynamics of signaling pathway activation and deactivation<br>* Investigate the effects of specific mutations or genetic modifications on signaling pathways<br>*   Develop new therapeutic strategies for diseases related to aberrant cell signaling. | understanding complex cellular behaviors in development, differentiation, and disease. | | | | |
| What is the potential of droplet microfluidics in | Droplet microfluidics has the potential to revolutionize | Droplet microfluidics holds great promise in personalized | The response accurately captures the potential of droplet | 85 | 85.0 | 85 |

| personalized medicine? | personalized medicine by enabling the precise and efficient manipulation of individual cells, allowing for:<br><br>* High-throughput screening of patient-specific cells<br>* Personalized drug testing and development<br>* Single-cell analysis and characterization<br>* Development of targeted therapies based on individual genetic profiles<br>* Improved understanding of disease mechanisms at a cellular level<br><br>This technology can lead to more effective and tailored treatments, | medicine by enabling precise, high-throughput analyses of biological samples. The technology allows for the optimization of treatments by testing drug combinations on specific cell lines, such as pancreatic cancer cells, to identify the most effective therapies. It also facilitates the early detection of cancers through highly sensitive detection of tumor markers in bodily fluids, offering a minimally invasive approach for patient monitoring. Additionally, droplet-based digital PCR (dPCR) enables the detection of rare sequences and precise quantification of nucleic acids, which is crucial for cancer research and tailoring treatments based on a patient's genetic | microfluidics in personalized medicine, listing specific applications such as high-throughput screening, personalized drug testing, single-cell analysis, targeted therapies, and improved disease understanding. It also highlights the technology's ability to lead to more effective treatments and tailored patient care. However, it slightly deviates from the ground truth by not mentioning early cancer detection through bodily fluids or droplet-based digital PCR for rare sequence detection.<br>SCORE: 85 | | | |
|---|---|---|---|---|---|

| | improved patient outcomes, and enhanced precision medicine. | background and tumor profile. Overall, droplet microfluidics can significantly advance personalized medicine by providing tailored treatment strategies and improving early detection for better patient outcomes. | | | |
|---|---|---|---|---|---|
| How are droplet microfluidic devices used for nanoparticle synthesis? | Droplet microfluidic devices are used to synthesize nanoparticles by creating and manipulating small, uniform droplets of liquid that contain the reactants. These droplets can be precisely controlled in terms of size, composition, and temperature, allowing for the formation of nanoparticles with specific properties. The process typically | Droplet microfluidic devices revolutionize nanoparticle synthesis by enabling precise control over reaction conditions within isolated microreactors. These droplets encapsulate well-defined quantities of reagents, ensuring rapid mixing and preventing contamination, even when using toxic or volatile chemicals. This approach addresses limitations like diffusion-limited mixing found in traditional systems, resulting in nanoparticles | The response accurately describes the process of nanoparticle synthesis using droplet microfluidic devices, including emulsification, mixing, polymerization or precipitation, and the benefits of controlled size, shape, and composition. However, it lacks some details about the precision and scalability of the method, as well as its advantages over traditional systems. The response is clear and concise but could be more comprehensive | 56 | 56.0 | 55 |

| | involves the following steps:<br><br>1. Emulsification: A mixture of oil and water is emulsified to create small droplets.<br>2. Mixing: The reactants are mixed within these droplets to form a uniform solution.<br>3. Polymerization or precipitation: The reactants undergo polymerization or precipitation to form nanoparticles.<br><br>This method allows for the production of nanoparticles with controlled size, shape, and composition, which is useful in various | with controlled size, shape, and narrow size distribution. The method supports multi-step synthesis on millisecond timescales, allowing for the scalable production of high-quality nanoparticles with applications in fields such as drug delivery, imaging, and catalysis. | in addressing the question.<br>SCORE: 56 | | | |
|---|---|---|---|---|---|---|

| | | | | | |
|---|---|---|---|---|---|
| | applications such as drug delivery, catalysis, and biosensing. | | | | | |
| What role do droplets play in polymer particle production? | Droplets play a crucial role in polymer particle production as they serve as templates for the formation of uniform particles with controlled size and morphology. In various methods such as emulsion polymerization, miniemulsion polymerization, and microencapsulation, droplets are used to create monomer-filled particles that can be converted into polymers through chemical reactions. The properties of the resulting polymer | Droplet-based microfluidics provides a versatile platform for producing monodisperse polymer particles by encapsulating monomers or oligomers into droplets and solidifying them through chemical or physical crosslinking. This approach enables the creation of diverse particle types, including solid, hollow, asymmetric, and multicompartmental structures with precise control over size, shape, and morphology. Techniques such as UV-initiated polymerization and solvent evaporation facilitate the transformation of droplets into solid particles. Additionally, functional micro- or nanostructures can be incorporated within or on the | The response accurately conveys the role of droplets in polymer particle production, highlighting their use as templates for uniform particles with controlled size and morphology. However, it lacks specific details about the techniques mentioned in the ground truth answer, such as UV-initiated polymerization and solvent evaporation, which are crucial for transforming droplets into solid particles. The response also does not mention the diverse types of particle structures that can be created using droplet-based microfluidics. SCORE: 46 | 46 | 46.0 | 45 |
| | particles, including their size, shape, and surface characteristics, can be influenced by the properties of the droplets, such as their size, composition, and stability. | surface of these particles by introducing specific ingredients into the droplets, expanding their potential applications in areas like drug delivery, tissue engineering, and advanced materials. | | | | |
| How can droplet microfluidics improve the reproducibility of chemical synthesis? | Droplet microfluidics can improve the reproducibility of chemical synthesis by allowing for precise control over reaction conditions, such as temperature, concentration, and mixing time, at the microscale. This enables consistent and uniform reactions to be performed in a controlled environment, reducing variability and improving yields. | Droplet microfluidics enhances the reproducibility of chemical synthesis by providing precise control over reaction conditions and minimizing variability. Droplets act as isolated microreactors, ensuring accurate manipulation of parameters such as temperature and concentration, critical for consistent results. This system enables optimal mixing intensity and parallel processing in separate droplet channels, reducing batch-to- | The response is mostly relevant and clear, but lacks focus or includes unnecessary details; broadly acceptable. SCORE: 58 | 58 | 58.0 | 55 |

| | Additionally, droplet microfluidics can facilitate high-throughput screening of different reaction conditions, enabling the optimization of synthesis protocols and the identification of optimal conditions for specific chemical transformations. | batch variability and ensuring uniform product quality. Automation further minimizes human error, making the process highly reliable. These advantages are particularly beneficial for synthesizing complex nanomaterials and exploring new reaction parameter spaces, leading to the production of high-quality, reproducible materials. | | | | |
|---|---|---|---|---|---|---|
| What are the applications of droplet microfluidics in catalysis studies? | Droplet microfluidics has several applications in catalysis studies, including:

1. High-throughput screening of catalysts: Droplet microfluidics allows for the rapid and efficient screening of large numbers of catalysts to identify | Droplet microfluidics offers transformative applications in catalysis studies by enabling precise, efficient, and high-throughput investigations at the nanoscale. These systems facilitate enantioselective biocatalysis using nanoliter-scale devices that integrate microflow reactors and electrophoresis | The response provides a clear and concise list of applications of droplet microfluidics in catalysis studies. It covers various aspects such as high-throughput screening, single-particle analysis, catalyst synthesis, reaction optimization, in situ analysis, and development of new materials. However, the answer lacks depth and detail | 51 | 51.0 | 50 |

| | optimal conditions for a specific reaction. 2. Single-particle analysis: The ability to analyze individual particles or droplets containing catalysts enables researchers to study the behavior of single catalytic sites, providing insights into the mechanisms of catalysis. 3. Catalyst synthesis and characterizati on: Droplet microfluidics can be used to synthesize and characterize nanoparticles and other catalyst materials with precise control over size, shape, and composition. 4. Reaction optimization: | for enantiomeric differentiation. Droplets serve as isolated reaction containers, allowing for controlled catalytic transformations at the single-particle or cell level, with reaction volumes as small as a few picoliters. This confinement minimizes diffusion effects and enhances reaction efficiency. Applications span organic synthesis, biochemical analysis, and biotechnology, showcasing droplet microfluidics as a powerful tool for studying rapid reaction dynamics, optimizing catalysts, and exploring reaction mechanisms in real time. | compared to the ground truth response, which provides a more comprehensive understanding of the transformative impact of droplet microfluidics on catalysis studies. The provided answer seems to be a summary of potential applications rather than an in-depth explanation of how these applications are achieved or their significance in the field. SCORE: 51 | | | |
|---|---|---|---|---|---|

| | By controlling the conditions within individual droplets, researchers can optimize reaction conditions for specific catalysts, leading to improved yields and selectivity. 5. In situ analysis of catalytic reactions: Droplet microfluidics enables real-time monitoring of catalytic reactions, allowing researchers to study the dynamics of catalysis in unprecedented detail. 6. Development of new catalytic materials: The precise control over droplet composition and size offered by | | | | | |
|---|---|---|---|---|---|---|

| | droplet microfluidics can be used to develop novel catalysts with tailored properties. | | | | | |
|---|---|---|---|---|---|---|
| How do droplets enhance the study of reaction kinetics? | Droplets can be used as microreactors to study reaction kinetics in a controlled and miniaturized environment, allowing for precise measurement of reaction rates, yields, and mechanisms. They enable the observation of reactions at the molecular level, providing insights into the underlying chemical processes. Additionally, droplets can be easily manipulated and characterized using techniques | Droplet-based microfluidics revolutionizes reaction kinetics studies by offering rapid, precise, and high-throughput analysis with minimal sample consumption. The chaotic mixing within picoliter droplets ensures millisecond-level time resolution for fast reactions. These systems minimize reagent use, making them ideal for bioanalysis where resources are costly or scarce. Precise control over droplet composition facilitates the creation of concentration gradients, enabling the measurement of kinetic parameters. High-throughput capabilities allow simultaneous | The response accurately conveys the benefits of using droplets in studying reaction kinetics, but lacks some specific details and nuances present in the ground truth answer. It also uses slightly more general language compared to the ground truth. SCORE: 76 | 76 | 76.0 | 75 |
| | such as microscopy and spectroscopy, making them an ideal platform for studying complex reactions and optimizing catalysts. | monitoring of numerous reactions, significantly expanding experimental diversity. Additionally, extended tracking of individual droplets supports long-term monitoring of enzyme-inhibitor kinetics, providing deeper insights into reaction mechanisms. | | | | |
| What are the benefits of using droplets for emulsion polymerization? | The benefits of using droplets for emulsion polymerizati on include:<br><br>* Improved particle size control and uniformity<br>* Enhanced monomer utilization efficiency<br>* Reduced viscosity and improved processabilit y of the final product<br>* Increased stability and shelf life of the latex particles<br>* Ability to | Droplet-based systems offer significant advantages for emulsion polymerization, enhancing efficiency, precision, and scalability. They enable high production rates and precise control over polymer particle size and morphology, facilitating the synthesis of high-molar-mass polymers at faster reaction rates. Droplet parallelization allows the screening of diverse chemicals | The answer provided is mostly relevant to the question about the benefits of using droplets for emulsion polymerization. However, it lacks specific details related to the context of emulsion polymerization in the chemical industry, and instead focuses on broader applications such as directed enzyme or microorganism evolution and in vitro compartmentaliz ation (IVC). The | 41 | 41.0 | 40 |
| | produce polymers with tailored properties, such as particle size, molecular weight, and surface chemistry. | or concentrations and supports the encapsulation of millions of gene variants for quantitative emulsion PCR. These systems also empower directed enzyme or microorganism evolution through successive mutagenesis and selection rounds. Furthermore, droplet isolation prevents cross-contamination, adsorption, and solvent evaporation, making them ideal for applications like in vitro compartmentalization (IVC). This unique platform provides a robust, scalable approach for both analytical and biological studies. | answer also fails to mention some key points mentioned in the ground truth answer, such as high production rates and precise control over polymer particle size and morphology. SCORE: 41 | | | |
| How can microfluidics be used to generate smart materials? | Microfluidics can be used to generate smart materials by creating complex patterns and structures at the microscale, which can | Microfluidics enables the creation of smart materials through precise control of reaction conditions, rapid synthesis, and innovative templating methods. Droplet templating allows | The response attempts to address the question but lacks clarity and precision compared to the ground truth answer. It provides a general idea about | 31 | 31.0 | 30 |

| | | | | | |
|---|---|---|---|---|---|
| | then be used to create materials with unique properties. This is achieved through the manipulation of fluids in microchannels, allowing for precise control over the distribution of particles, droplets, or other components that give rise to the desired material properties. | monodisperse droplets to act as structural templates, facilitating the synthesis of uniform particles, such as hydrogels for single-cell analysis. By manipulating solute and temperature gradients on short timescales, microfluidics supports the production of nanoscale materials, including semiconducting polymers, catalysts, and nanoparticles with unique optical and electronic properties. Its ability to rapidly explore complex reaction parameter spaces, coupled with advanced analytics and machine learning, accelerates material discovery and optimization. This approach offers transformative potential in biotechnology, energy, and electronics by | microfluidics generating smart materials through pattern creation, but it does not delve into specific details or mention advanced templating methods, reaction control, or material properties as described in the ground truth. SCORE: 31 | | | |
| | | enabling the generation of materials with tailored properties and enhanced functionality. | | | | |
| What are the applications of droplet microfluidics in environmental sensing? | Droplet microfluidics has several applications in environmental sensing, including:<br><br>* Water quality monitoring: Droplet-based systems can be used to detect pollutants and contaminants in water samples.<br>* Air quality monitoring: Microdroplets can be used to capture and analyze airborne particles and gases.<br>* Soil analysis: Droplet-based systems can be used to detect heavy metals and other | Droplet microfluidics offers transformative applications in environmental sensing by enabling precise, miniaturized, and efficient monitoring systems. It supports on-site environmental surveillance through portable devices for rapid toxicity assessments and detection of contaminants, such as mercury ions. As controllable microreactors, droplet microfluidics facilitates the study of complex interfacial processes, aiding in chemical synthesis and degradation studies. Furthermore, it advances environmental sensing by enabling in situ | The answer provides a list of specific applications of droplet microfluidics in environmental sensing, but lacks the depth and detail provided by the Ground Truth Answer. It does not fully capture the transformative nature and broader implications of droplet microfluidics in this field.<br>SCORE: 45 | 45 | 45.0 | 45 |
| | pollutants in soil samples. * Biodetection: Droplet microfluidics can be used to detect biological agents such as bacteria, viruses, and toxins. * Oil spill detection: Microdroplets can be used to detect oil spills and track their movement. * Pesticide monitoring: Droplet-based systems can be used to detect pesticide residues in water and soil samples. | visualization and precise flow control, enhancing our understanding of environmental processes. These capabilities position droplet microfluidics as a valuable tool for environmental science and engineering. | | | | |
| How can microfluidic droplets improve combinatorial chemistry? | Microfluidic droplets can improve combinatorial chemistry by allowing for high-throughput screening of chemical reactions, precise control over reaction | Microfluidic droplets revolutionize combinatorial chemistry by drastically reducing reagent consumption and enhancing efficiency. Each droplet, acting as a picoliter-scale microreactor, requires six to | The response provides some relevant information about the benefits of microfluidic droplets in combinatorial chemistry, but it lacks key details and accuracy compared to the ground truth. It | 41 | 41.0 | 40 |
| | conditions, and efficient mixing of reagents. This enables the rapid generation of large libraries of compounds with diverse properties, facilitating the discovery of new materials and drugs. | eight orders of magnitude less material than traditional methods, minimizing waste and costs. This miniaturization enables high-throughput synthesis of large compound libraries, supporting applications like drug discovery and material synthesis. Additionally, droplet manipulation techniques allow for precise reaction control, combinatorial fusion of reagents, and multi-stage experiments with online analysis. These advantages make microfluidic droplets a powerful tool for automating complex chemical workflows and optimizing reaction conditions in a highly efficient and scalable manner. | mentions high-throughput screening, precise control over reaction conditions, and efficient mixing, which are all correct points, but it does not address the significant reduction in reagent consumption and waste that is a major advantage of microfluidic droplets. SCORE: 41 | | | |
| What are the challenges of scaling up | The main challenges of scaling up | Scaling up droplet-based chemical | The answer provides a clear and concise list | 51 | 51.0 | 50 |
| droplet-based chemical synthesis? | droplet-based chemical synthesis include:<br><br>1. Maintaining uniformity and consistency in droplet size and composition across large scales.<br>2. Scaling up the production process while maintaining control over reaction conditions, such as temperature, pressure, and mixing.<br>3. Ensuring efficient heat transfer and mass transport within the droplets to prevent hotspots and incomplete reactions.<br>4. Managing the increased viscosity and surface tension of larger volumes of liquid, which | synthesis faces several challenges. Achieving uniform fluid distribution in microfluidic devices remains difficult despite progress in parallel droplet generation, with 3D printing offering potential solutions for complex channel fabrication. Simplifying functional units for multistep reactions, such as droplet generation, merging, and phase separation, is vital for seamless continuous flow processes. Developing low-cost materials and standardizing chip designs are necessary for commercial viability, especially for high-throughput production in diverse fluids. Furthermore, creating complex droplet emulsions and expanding nanomaterial diversity are | of specific challenges related to scaling up droplet-based chemical synthesis, which is relevant to the question. However, it lacks depth and context compared to the ground truth answer, which provides more detailed explanations and insights into the challenges.<br>SCORE: 51-55 | | | |
| | can affect droplet formation and stability. 5. Maintaining sterility and preventing contaminatio n in large-scale production environments . 6. Scaling up the synthesis while maintaining product quality and consistency. 7. Overcoming issues related to droplet coalescence and aggregation, which can lead to reduced yields and product uniformity. 8. Developing scalable and cost-effective methods for handling and processing large volumes of droplets. | critical to meeting industrial demands. Addressing these challenges will enable safe, sustainable, and scalable microfluidic applications. | | | | |
| How can artificial intelligence | Artificial intelligence | Machine learning can be used to | The response provides a clear | 76 | 76.0 | 80 |

| enhance droplet microfluidic research? | (AI) can enhance droplet microfluidic research in several ways:<br><br>1. **Automated data analysis**: AI algorithms can quickly process and analyze large datasets generated by droplet microfluidics, allowing researchers to identify patterns and trends that might be difficult or time-consuming for humans to detect.<br>2. **Optimization of experimental design**: AI can help optimize the design of experiments in droplet microfluidics, such as determining the optimal flow rates, | optimize the design of microfluidic devices. Deep Learning will be integrated to sort, monitor, and process the images. Furthermore, the multimodal agents can be used to study and design the research of droplet microfluidics, a possible excuse robot to perform fabrication and experiments automatically. | and concise overview of how artificial intelligence can enhance droplet microfluidic research. It lists specific ways in which AI can aid researchers, including automated data analysis, optimization of experimental design, predictive modeling, real-time monitoring and control, identification of anomalies, improved understanding of complex phenomena, and enhanced precision and accuracy. The response is well-structured and easy to follow, with each point clearly explained. However, it does not fully align with the ground truth answer, which mentions machine learning for optimizing device design, deep learning for image processing, and | | | |
| --- | --- | --- | --- | --- | --- | --- |

| | pressures, and other parameters to achieve specific outcomes. 3. **Predictive modeling**: AI models can predict the behavior of droplets and their contents under various conditions, enabling researchers to simulate and test different scenarios without the need for physical experimentation. 4. **Real-time monitoring and control**: AI-powered sensors and control systems can monitor and adjust experimental conditions in real-time, allowing for more precise control over the | | multimodal agents for studying droplet microfluidics. SCORE: 76 | | | |
|---|---|---|---|---|---|---|

| | microfluidic process. 5. **Identification of anomalies**: AI can detect anomalies or irregularities in the data generated by droplet microfluidics, helping researchers to identify potential issues with their experiments. 6. **Improved understanding of complex phenomena**: AI can help researchers understand complex phenomena occurring within droplets, such as phase transitions, chemical reactions, and biological processes. 7. **Enhanced precision and accuracy**: AI can | | | | | |
|---|---|---|---|---|---|---|

| | improve the precision and accuracy of droplet microfluidic research by reducing errors and inconsistencies in experimental results.

These are just a few examples of how artificial intelligence can enhance droplet microfluidic research. | | | | |
|---|---|---|---|---|---|
| How can large language models enhance droplet microfluidic research? | Large language models can enhance droplet microfluidic research by:

* Analyzing and visualizing complex data from experiments, such as particle tracking and image analysis
* Predicting the behavior of droplets in different microfluidic | LLMs can be used to build the Agents and Multi-Agents to synthesize knowledge of droplet microfluidics from literature such as research articles and books. Then, the idea hypotheses will be proposed by the agents, and the novelty and feasibility will be assessed. The agents will provide the autonomous platform to design and fabricate droplet | The response attempts to address the question but lacks clarity and coherence; mostly irrelevant as it does not directly relate to the specific ways large language models (LLMs) can enhance droplet microfluidic research. SCORE: 16 | 16 | 16.0 | 10 |

| | | | | | |
|---|---|---|---|---|---|
| | devices and conditions<br>* Optimizing experimental design and parameters for specific applications<br>* Providing insights into the underlying physics and chemistry of droplet formation and manipulation<br>* Assisting in the development of new materials and surfaces for droplet-based applications. | microfluidic devices. Furthermore, LLMs have the researchers accelerate knowledge of droplet microfluidics. | | | |
| How can autonomous droplet microfluidic research? | Autonomous droplet microfluidic research involves the use of automated systems and algorithms to control and manipulate tiny droplets of liquids in a microfluidic device. This can be achieved through various | Autonomous droplet microfluidics is an emerging field with significant potential for scientific discovery, aiming to create highly integrated, automated systems for bio/chemical functions like PCR, sample preparation, and detection. Key challenges include the need for fast detection | The response provides a clear and concise overview of autonomous droplet microfluidic research, listing various methods and applications. However, it lacks depth and context, failing to address key challenges and potential solutions as described in the ground truth | 41 | 41.0 | 40 |

| | methods, including: | methods to probe small volume droplets, precise control over droplet dynamics in complex microchannel networks, and the development of virtual design platforms for optimizing system designs. Solutions under exploration include using advanced detection techniques such as single-molecule detection and fluorescence lifetime imaging, as well as integrating microelectronic actuators for droplet manipulation. Additionally, the development of virtual design and experiment platforms could streamline the creation of optimized microfluidic systems. The potential applications are vast, including directed evolution, closed-loop synthesis, and automated | answer. SCORE: 41 | | | |
| | 1. Microfluidic chips with integrated sensors and actuators that can detect and respond to changes in the droplets. 2. Machine learning algorithms that can predict and adjust the flow rates and pressures to control the droplet formation and manipulation. 3. Automated systems that use computer-controlled pumps and valves to create and manipulate the droplets. 4. The use of artificial intelligence (AI) and robotics to automate the process of creating, manipulating | | | | | |
| | , and analyzing the droplets.<br><br>These autonomous systems can be used for a wide range of applications, including:<br><br>1. High-throughput screening of biological samples<br>2. Synthesis of nanoparticles and other materials<br>3. Analysis of chemical reactions and processes<br>4. Development of new diagnostic tools and biosensors | drug discovery, where systems could autonomously optimize compound combinations. Collaboration across disciplines like fluid dynamics, chemistry, and software engineering will be key to overcoming challenges and advancing this technology, paving the way for groundbreaking scientific advances. | | | | |
| What is the potential for fully integrated droplet microfluidic platforms? | The potential for fully integrated droplet microfluidic platforms includes:<br><br>* High-throughput screening and analysis of biological | Droplet microfluidics offers transformative solutions to global health challenges by enabling rapid, cost-effective, and high-throughput analysis of biological | The answer provides a list of potential applications of fully integrated droplet microfluidic platforms but lacks the depth and context provided in the ground truth answer. It does | 41 | 41.0 | 40 |

| | | | | | |
|---|---|---|---|---|---|
| | samples<br>* Miniaturization of chemical synthesis and reaction processes<br>* Enhanced sensitivity and specificity in biochemical assays<br>* Real-time monitoring and control of reactions<br>* Potential for point-of-care diagnostics and personalized medicine<br>* Reduced reagent consumption and increased efficiency in laboratory settings. | samples. By generating and manipulating tiny droplets as individual reaction chambers, this technology can revolutionize diagnostics, particularly in resource-limited settings. For example, it facilitates point-of-care testing for infectious diseases like tuberculosis, malaria, and COVID-19, providing results quickly and accurately. Droplet microfluidics is also crucial for multiplexed screening, enabling simultaneous detection of multiple biomarkers, which enhances early disease detection and monitoring. Additionally, it supports the development of low-cost diagnostic devices, reducing dependence on expensive | not capture the transformative impact on global health challenges or the specific examples mentioned in the ground truth, such as point-of-care testing for infectious diseases.<br>SCORE: 41 | | |

| | | laboratory infrastructure. In drug discovery, it accelerates the screening of therapeutic compounds, improving the treatment of diseases. The technology also advances personalized medicine by analysing single cells and their responses to treatments. Overall, droplet microfluidics addresses global health disparities by delivering accessible, scalable, and efficient solutions for disease diagnosis, monitoring, and treatment. | | | | |
|---|---|---|---|---|---|---|
| How can droplet microfluidics contribute to global health challenges? | Droplet microfluidics can contribute to global health challenges by enabling the development of point-of-care diagnostic devices, vaccine delivery systems, and personalized | Droplet microfluidics has significant potential to address global health challenges by enhancing diagnostic and therapeutic processes. It enables rapid and sensitive pathogen detection, which is critical for early disease diagnosis | The response provides a clear and concise overview of the potential applications of droplet microfluidics in addressing global health challenges. It highlights specific examples such as point-of-care diagnostic | 56 | 56.0 | 55 |

| medicine. It allows for the precise control and manipulation of tiny droplets of fluids, which can be used to:<br><br>*   Develop low-cost, portable diagnostic tests for diseases such as malaria, tuberculosis, and HIV<br>*   Create targeted drug delivery systems for cancer treatment and other diseases<br>*   Improve vaccine development and delivery by creating micro-scale emulsions that mimic the natural immune response<br>*   Enable the creation of personalized medicine by analyzing individual | and containment. The technology's high-throughput capabilities allow simultaneous analysis of multiple samples, streamlining disease diagnosis and treatment. Additionally, droplet microfluidics can be integrated into point-of-care platforms, making it feasible for use in resource-limited settings. This approach can improve healthcare accessibility and efficiency. Furthermore, droplet microfluidics facilitates precise single-cell analysis, enabling advances in personalized medicine and more effective treatment strategies. While challenges such as cost, standardization, and integration with existing systems remain, continued development of this technology could | devices, vaccine delivery systems, and personalized medicine. However, it lacks depth and detail compared to the ground truth answer, which provides more comprehensive information on the technology's capabilities and its potential impact on healthcare accessibility and efficiency. The response also does not mention some of the challenges associated with implementing droplet microfluidics in global health settings.<br><br>SCORE: 56 | | | |
| --- | --- | --- | --- | --- | --- |

| | patient samples in a controlled microfluidic environment. | significantly impact global health by improving detection, diagnostics, and treatment capabilities. | | | |
|---|---|---|---|---|---|
| How can open-source designs impact the future of microfluidic devices? | Open-source designs can significantly impact the future of microfluidic devices by:<br><br>* Reducing costs and increasing accessibility for researchers and manufacturers<br>* Facilitating collaboration and knowledge-sharing among experts from various fields<br>* Enabling rapid prototyping, testing, and iteration of new device designs<br>* Allowing for customization and tailoring of devices to | Open-source designs for microfluidic devices hold the potential to revolutionize the field by increasing accessibility, reducing costs, and fostering collaboration. These designs allow for the creation of custom predictive models without requiring extensive expertise, enabling broader participation in device development. The use of open-source tools, such as Design Automation of Fluid Dynamics (DAFD), accelerates design and prototyping, enabling rapid iteration and optimization. Additionally, the integration of machine learning | The response provides a clear and concise overview of the potential impact of open-source designs on microfluidic devices, covering key points such as cost reduction, collaboration, rapid prototyping, customization, and innovation. However, it lacks depth and specific examples compared to the ground truth answer, which provides more detailed information about the benefits of open-source designs in microfluidics, including the use of tools like DAFD and machine learning algorithms. The response also does not fully capture the | 51 | 51.0 | 50 |

| | specific applications or needs<br>* Fostering innovation and the development of novel microfluidic technologies. | algorithms can drive the development of new microfluidic tools and optimize existing ones. This democratization of design expertise and the collaborative, community-driven approach will facilitate the creation of affordable, specialized devices, ultimately leading to breakthroughs in applications like single-cell analysis and global health diagnostics. The open-source model promises to transform microfluidics into a more inclusive and innovative field. | transformative potential of open-source designs on the field as a whole.<br><br>SCORE: 51 | | | |
|---|---|---|---|---|---|---|
| What are the challenges of scaling up microfluidic devices for industrial use? | The main challenges of scaling up microfluidic devices for industrial use include:<br><br>1. Maintaining uniform flow rates and fluid properties | Scaling up microfluidic devices for industrial use presents several challenges, including throughput limitations, lack of standardization, and integration complexities. The small micrometric | The answer provides a clear and concise list of specific challenges related to scaling up microfluidic devices for industrial use, which is in line with the ground truth. However, it lacks depth and context, as well | 56 | 56.0 | 55 |

| | | | | | |
|---|---|---|---|---|---|
| | across larger channels. 2. Scaling up materials and manufacturing processes to maintain device integrity. 3. Ensuring consistent performance and reliability in a production environment. 4. Increasing the volume of fluids handled without compromising efficiency or accuracy. 5. Addressing issues related to clogging, fouling, and contamination. 6. Meeting regulatory requirements for industrial applications. 7. Scaling up control systems and instrumentation to manage larger devices. 8. Maintaining cost- | channel sizes restrict production efficiency, and increasing parallelization or channel size may impact fabrication costs and reaction kinetics. There is also a lack of standardized design methods, making it difficult to compare results and adopt microfluidics in industrial settings. Additionally, scaling from laboratory to industrial production remains a hurdle, with issues in cost-effective microfabrication, generating complex droplet emulsions, and ensuring reliable fluid control. The integration of microfluidic devices with industrial interfaces, testing methods, and continuous processes is essential for successful commercialization. Overcoming these challenges requires innovative | as some nuances mentioned in the ground truth answer. SCORE: 56 | | | |

| | effectiveness while increasing device size and production volumes. | solutions, such as simplifying multistep reactions and developing standardized, low-cost fabrication methods. | | | |
|---|---|---|---|---|---|
| How might droplet microfluidics impact agriculture? | Droplet microfluidics could enable precise and efficient delivery of fertilizers, pesticides, and nutrients directly to specific areas of crops or plants, reducing waste and environmental impact. It may also facilitate the development of novel diagnostic tools for plant diseases, allowing for early detection and targeted treatment. Additionally, droplet microfluidics can be used to create controlled environments | Droplet microfluidics has the potential to transform agriculture by enhancing crop management and food security. It enables the precise release of fertilizers and pesticides, reducing waste and optimizing plant growth. The technology also supports improved crop monitoring through long-term studies, offering insights into plant development. By analyzing individual plant cells, droplet microfluidics can help uncover genetic and phenotypic traits, enhancing disease resistance and drought tolerance in crops. Additionally, it aids in disease detection, | The response demonstrates a good understanding of the potential applications of droplet microfluidics in agriculture, including precise delivery of fertilizers and pesticides, novel diagnostic tools for plant diseases, and controlled environments for plant cell culture. However, it lacks some depth and detail compared to the ground truth answer, particularly regarding the challenges and ongoing research in this area. The response also does not explicitly mention the benefits of improved crop monitoring through long- | 56 | 56.0 | 55 |

| | | | | | |
|---|---|---|---|---|---|
| | for plant cell culture and tissue engineering, potentially leading to improved crop yields and disease resistance. | allowing for early intervention to prevent crop loss. Despite challenges such as scalability, cost, and integration with existing agricultural practices, ongoing research aims to overcome these barriers, making droplet microfluidics a promising tool for precision agriculture and sustainable food production. | term studies or the analysis of individual plant cells to enhance disease resistance and drought tolerance.<br><br>SCORE: 56 | | | |
| What are the current limitations of droplet microfluidics, and how can they be overcome? | The current limitations of droplet microfluidics include:<br><br>1. Scalability: Droplet microfluidic devices are typically small in scale, making it challenging to produce large quantities of droplets.<br>2. Complexity: The design and fabrication of | Droplet microfluidics has immense potential for high-throughput biological assays but faces several limitations that hinder its broader adoption. One key challenge is its efficiency for assays involving small sample sizes; droplet microfluidics excels with large volumes (at least $10^5$ samples), making traditional methods like microtiter plates more suitable for lower sample numbers. Another | The response provides a clear and concise list of the current limitations of droplet microfluidics and some potential strategies for overcoming them. However, it lacks depth and context compared to the ground truth answer, which provides more detailed explanations and specific examples of the challenges faced by droplet microfluidics.<br>SCORE: 51 | 51 | 51.0 | 50 |

| | | | | | |
|---|---|---|---|---|---|
| | droplet microfluidic devices can be complex and time-consuming.<br><br>3. Clogging: The narrow channels in droplet microfluidic devices can clog due to the accumulation of particles or bubbles.<br><br>4. Limited control over droplet size and composition: It can be difficult to precisely control the size and composition of droplets in a microfluidic device.<br><br>To overcome these limitations, several strategies can be employed:<br><br>1. Scaling up: Using larger-scale fabrication techniques, such as 3D | critical issue is maintaining a stable and inert oil-water interface, which is essential for droplet formation. Finding the right surfactant to keep this interface stable remains a complex chemistry challenge. Furthermore, droplet microfluidics lacks the spatial barcoding available in microtiter plates, making it difficult to organize and track multiple droplet libraries. As a result, developing alternative barcoding systems is necessary but challenging. The automation of droplet movement is another obstacle, as it requires precise control, often through techniques like electrowetting and dielectrophoresis. Lastly, droplet stability over time can be | | | |

| | printing or soft lithography, to increase the production capacity. 2. Simplifying design: Designing simpler devices with fewer components to reduce complexity and improve reliability. 3. Surface modification: Modifying the surface properties of the microfluidic device to prevent clogging by reducing adhesion between particles and the channel walls. 4. Advanced control systems: Implementing advanced control systems, such as machine learning algorithms or | compromised, with issues such as shrinkage, coalescence, and cross-talk between droplets. To address these issues, research is underway to develop improved surfactants, more reliable barcoding systems, and novel materials and techniques to enhance droplet control. These advancements, coupled with better analytical tools, are expected to overcome current limitations and expand droplet microfluidics' applications in chemical and biological research. | | | | |
|---|---|---|---|---|---|---|

| | feedback control loops, to precisely control droplet size and composition. 5. Hybrid approaches: Combining droplet microfluidics with other technologies, such as continuous flow processing or membrane-based separation, to improve scalability and efficiency. | | | | | |
|---|---|---|---|---|---|---|
| How can interdisciplinary collaboration accelerate innovation in droplet microfluidics? | Interdisciplinary collaboration can accelerate innovation in droplet microfluidics by combining expertise from various fields, such as physics, chemistry, biology, engineering, and materials science. This fusion of | Interdisciplinary collaboration is essential for advancing droplet microfluidics by combining expertise from various fields such as chemistry, biology, material science, and engineering. By uniting these diverse areas, researchers can address complex challenges such as cost-effective fabrication, scalable | The response demonstrates a good understanding of the topic by mentioning the benefits of interdisciplinary collaboration and providing some relevant examples. However, it lacks depth and specific details compared to the ground truth answer. The response also does not mention | 51 | 51.0 | 50 |

| | knowledge enables researchers to design novel devices, materials, and techniques that take advantage of the unique properties of droplets at the microscale. By integrating insights from multiple disciplines, scientists can develop innovative solutions for applications like lab-on-a-chip devices, point-of-care diagnostics, and biotechnology. | production, and novel interfacial reaction systems. For example, biologists and chemists can leverage droplet microfluidics for high-throughput analysis, while material scientists and engineers work on developing more robust and affordable microfabrication methods. Furthermore, integrating technologies like sensitive detectors and machine learning algorithms can enhance the analytical capabilities of droplet-based systems. Collaborative initiatives, such as the EMBL Microfluidics Conference, foster knowledge exchange and innovation by bringing together professionals from different disciplines. This cooperation is crucial for overcoming technical barriers, | some key aspects such as cost-effective fabrication, scalable production, and novel interfacial reaction systems that are highlighted in the ground truth answer. SCORE: 51 | | | |
|---|---|---|---|---|---|---|

| question | answer | ground_truth | feedback | score | percentage |
|---|---|---|---|---|---|
| | reducing manufacturing costs, and developing new applications in fields like drug discovery, single-cell analysis, and nanomaterial synthesis. Ultimately, interdisciplinary collaboration accelerates the evolution of droplet microfluidics, pushing the boundaries of what is achievable in this promising field. | | | | |

## S5: QA by GEMMA-based Scientific Mentor

| question | answer | ground_truth | feedback | score | percentage |
|---|---|---|---|---|---|
| What is microfluidics? | Microfluidics is the science of manipulating fluids at a very small scale, typically on the order of micrometers (millionths of a meter).<br><br>Think of it like this: imagine working with tiny channels and chambers, so small that | Microfluidics is a multidisciplinary field that involves the study and manipulation of fluids at the microscale, typically within channels with dimensions ranging from tens to hundreds of micrometers. It is a multidisciplinary field that involves molecular analysis, molecula | The response provides a clear and concise definition of microfluidics, explaining its scale and scope in an engaging way. It also attempts to provide a relatable analogy and mentions the potential | 56 | 56.0000000000001 |
| | you need a microscope to see them. That's the world of microfluidics!<br><br>These miniaturized systems allow scientists and engineers to control and analyze fluids in precise ways, opening up a wide range of applications in various fields. | r biology, and microelectronics. Microfluidics has been increasingly used in the biological sciences because precise and controlled experiments can be conducted at a lower cost and faster pace. Lab on a Chip devices use microfluidics for applications such as Point of Care testing of diseases, or Organ on a Chip studies. | applications of microfluidics. However, it lacks the depth and breadth provided by the ground truth answer, particularly regarding the multidisciplinary nature of the field and its specific applications. SCORE: 56 | | |
| What is droplet microfluidics? | Droplet microfluidics is a technology that uses tiny droplets of liquid as individual reaction compartments within a microfluidic system.<br><br>Here's a breakdown based on the provided context:<br><br>* **Microfluidics:** This involves manipulating small volumes (10^-9 to 10^-18 | Droplet-based microfluidics is a branch of microfluidics focusing on the creation and manipulation of discrete volumes of liquids through adoption of immiscible fluid flows. Microdroplets allow for handling miniature volumes (μL to fL) of fluids conveniently, provide better mixing, encapsulation, sorting, and sensing, and suit high throughput experiments. Com | The response attempts to describe droplet microfluidics but lacks key details, clarity, and precision compared to the ground truth answer. It does not fully capture the nuances of the technology, its advantages, or applications as described in the provided | 16 | 16.0 |

| | | | | |
|---|---|---|---|---|
| | L) of liquids in channels with diameters ranging from 10 to 100 micrometers.<br>* **Droplets:** These act as miniaturized reactors, each containing a single molecule, cell, or bead. They are highly monodispersed (uniform in size) and stabilized by surfactants.<br><br>**Key Advantages:**<br><br>* **Efficiency:** Droplet microfluidics allows for efficient partitioning of individual components from complex samples.<br>* **Throughput:** High frequency droplet production enables rapid processing of large numbers of samples.<br>* **Sensitivity:** The small | pared with continuous-flow systems, droplet microfluidic systems enable isolation and manipulation of samples in separate compartments. The droplets generated with uniform size and shape provide a stable microenvironment for biochemical reactions. Processing of a large number of droplets in parallel facilitates to achieve ultrahigh throughput. Confinement of individual targets within the droplets and independent processing of each of them allow high-precision analyses. | context.<br>SCORE: 16 | | |
| | volume of each droplet enhances sensitivity in detection assays. * **Programmability and Reconfigurability:** Droplets can be manipulated through various modules (merging, splitting, incubating, sorting) within the microfluidic system, allowing for complex operations to be programmed.<br><br>**Applications:**<br><br>The provided context highlights applications in:<br><br>* **Digital PCR:** Droplet-based systems are used for highly sensitive and accurate DNA amplification and detection.<br>* **Biomedical Research:** Droplets serve as microreactors | | | | |

| | | | | |
|---|---|---|---|---|
| | for various biological processes, including cell culture and drug screening. * **Nanotechnolo gy:** Droplets can be used to synthesize nanoparticles and encapsulate biological entities.<br><br>Let me know if you have any other questions about droplet microfluidics! | | | |
| How to generate microdroplets? | Microdroplet generation in microfluidics can be achieved through two primary methods: **active** and **passive**.<br><br>* **Active methods** involve the use of external forces or actuators to control droplet formation. One example mentioned in the context is a **piezoelectric actuator**, which can be | Microdroplet generation in microfluidics is achieved through two main methods: active and passive. Active methods utilize external forces like piezoelectric actuators to precisely control droplet size, formation rate, and spacing. These methods enable high uniformity in droplet generation, with deviations less than 0.3%. Passive methods rely on fluid | The response provides a clear overview of microdroplet generation methods, mentioning both active and passive approaches with specific examples. However, it focuses more on active methods, particularly piezoelectric actuators, whereas the ground truth emphasizes the | 56 | 56.0000000 0000001 |

| | | | | |
|---|---|---|---|---|
| | integrated into a microfluidic device to independently control droplet size, formation rate, and spacing.

* **Passive methods** rely on the interplay of fluid properties and channel geometry. These methods often involve using specific channel designs like **T-junctions** or **cross-flow configurations**. The flow rates and viscosities of the fluids involved also play a crucial role in determining droplet size and uniformity.

The provided text focuses primarily on active methods, particularly the use of piezoelectric actuators for precise droplet control. | properties and channel geometry, such as T-junctions or cross-flow configurations, to create droplets. These methods are influenced by factors like flow rate, viscosity, and interfacial tension, allowing for fine control over droplet size and uniformity. For example, PDMS-based microfluidic devices have been used for nanoparticle synthesis, where hydrodynamic coupling of nozzles generates droplet pairs for precise mixing. Both methods are valuable in applications like nanoparticle synthesis, offering high reproducibility and control over reaction conditions in biomedical and material science research. | importance of both methods equally. SCORE: 56 | | |

| What materials are typically used to fabricate droplet microfluidic devices? | The provided text discusses a variety of materials used in droplet microfluidic device fabrication.<br><br>Here's a breakdown:<br><br>* **Traditional Choice:** Polydimethylsiloxane (PDMS) is often the go-to material due to its ease of use with soft lithography and ability to form channels. However, it has drawbacks like poor solvent resistance and hydrophobic molecule absorption.<br>* **Alternatives to PDMS:** The text highlights several alternatives to PDMS:<br>    * **Elastomeric materials:** Thermoset polyesters<br>    * **Thermoplastics:** Polymethylmeth | Droplet microfluidic devices are commonly fabricated from a variety of materials, each offering distinct advantages for specific applications. The most widely used material is PDMS (Polydimethylsiloxane), favored for its ease of use with soft lithography and flexibility, though it has limitations such as poor solvent resistance. Alternatives to PDMS include thermoset polyesters, thermoplastics like PMMA, polycarbonate, and polystyrene, and amorphous polymers such as cyclic olefin copolymer. Fluoropolymers like PTFE and glass are also used, offering superior solvent resistance and biofouling resistance. Silicon and capillary devices are utilized in certain applications, | The response provides a clear and detailed list of materials used for fabricating droplet microfluidic devices, including their respective advantages. However, it lacks the comprehensive overview and specific details about silicon and capillary devices mentioned in the ground truth answer. SCORE: 76 | 76 | 76.0 |

| | acrylate (PMMA), polycarbonate, polystyrene * **Amorphous polymers:** Cyclic olefin copolymer * **Fluoropolymers:** Notably, polytetrafluoroethylene (PTFE) * **Glasses** Each of these materials offers specific advantages like solvent resistance, biofouling resistance, or mass-producibility through techniques like injection molding or hot embossing. Let me know if you'd like more details on a specific material or aspect of droplet microfluidic device fabrication! | providing precise control for microfluidic operations. These materials are chosen based on factors like solvent resistance, biofouling, and the ability to mass-produce devices using methods such as injection molding or hot embossing, ensuring broad applicability in research and industrial settings. | | | |
| What are the applications of | Droplet microfluidics | Droplet microfluidics has | The response is | 76 | 76.0 |

| droplet microfluidics? | has a wide range of applications, spanning from single-cell analysis to environmental engineering. Here's a breakdown based on the provided context:<br><br>**1. Single-Cell Analysis:**<br><br>* **Gene Expression & Protein Production:** Droplets allow for the encapsulation and analysis of individual cells or subcellular organelles, enabling researchers to study gene expression, protein production, and enzyme activity at the single-cell level.<br>* **Cell Heterogeneity:** This technology helps reveal cell heterogeneity by capturing variations in gene expression and other | diverse applications across biomedical research, diagnostics, and industrial fields. It enables high-throughput single-cell analysis, such as genomics, transcriptomics, and drug screening, while also facilitating nanoparticle synthesis, enzyme evolution, and chemical reactions. In healthcare, it supports personalized medicine, point-of-care diagnostics, and vaccine development. Beyond biology, it is used for environmental monitoring, food safety testing, and the development of cosmetics and pharmaceuticals. Its ability to create isolated, uniform reaction chambers in droplets makes it a powerful tool for precise, cost-effective, and scalable workflows in science and | comprehensive and well-structured, covering various applications of droplet microfluidics across different fields. It provides detailed explanations and examples for each application, making it easy to understand the potential uses of this technology. However, some minor inaccuracies and omissions are present compared to the Ground Truth Answer. SCORE: 76 | | |

| | cellular processes between individual cells.

**2. Medical Diagnostics:**

* **Disease Detection & Monitoring:** Droplet microfluidics can be used for sensitive and specific detection of disease biomarkers, enabling early diagnosis and monitoring of disease progression.
* **Drug Screening:** High-throughput screening of potential drug candidates against various diseases can be achieved using droplet-based systems.

**3. Drug Discovery:**

* **Target Identification & Validation:** Droplets facilitate the identification and validation | industry. | | | |
|---|---|---|---|---|

| | of drug targets by allowing for precise manipulation and analysis of biological molecules. * **Personalized Medicine:** This technology enables the development of personalized therapies tailored to individual patients based on their unique genetic and molecular profiles.

**4. Food & Feed Industry:**

* **Food Safety & Quality Control:** Droplet microfluidics can be used to detect pathogens, contaminants, and allergens in food products, ensuring safety and quality. * **Feed Analysis & Optimization:** This technology helps analyze the nutritional | | | | |
|---|---|---|---|---|---|

| | content of feed and optimize its composition for animal health and productivity.<br><br>**5. Environmental Engineering:**<br><br>* **Water Quality Monitoring:** Droplet microfluidics enables sensitive detection of pollutants and contaminants in water samples, contributing to environmental monitoring and protection.<br>* **Bioremediation:** This technology can be used to develop efficient methods for removing pollutants from contaminated environments using microorganisms or enzymes encapsulated in droplets.<br><br>**Beyond these applications, | | | | |
| | droplet microfluidics also plays a crucial role in:**<br><br>* **Synthetic Biology:** Creating artificial cells and bio-hybrid systems.<br>* **Materials Science:** Fabricating novel materials with controlled properties at the nanoscale.<br><br><br>The versatility of droplet microfluidics makes it a powerful tool for advancing research and innovation across diverse fields. | | | | |
| What are the main benefits of using droplet microfluidics over bulk systems? | Droplet microfluidics offers several key advantages over traditional bulk bio(chemical) analysis methods:<br><br>* **Individual Compartmentalization:** Each droplet acts as a | Droplet microfluidics offers significant advantages over traditional bulk systems. It enables individual compartmentalization, isolating single cells or molecules in tiny liquid compartments for precise control. | The response provides a clear and concise overview of the advantages of using droplet microfluidics over bulk systems. It highlights key benefits | 56 | 56.0000000 0000001 |

| | tiny, isolated microreactor, allowing for precise control of local environments and manipulations. This is crucial for studying individual cells or molecules without interference from others.

* **Enhanced Control:** Droplet microfluidics enables highly controllable local environments within each droplet, facilitating manipulation and analysis of specific samples with greater accuracy.

* **Wide Sample Applicability:** The technology can handle a diverse range of samples, including molecules and cells, expanding its analytical capabilities. | This allows for single-cell analysis and early detection of rare cells or secreted molecules. The technology supports massive parallel processing, analyzing millions of droplets simultaneously, ideal for high-throughput screening and examining phenotypic or genetic variability. Additionally, iterative operations such as merging, mixing, and sorting droplets enable complex experimental protocols and continuous monitoring of populations. The system's highly controllable local environments ensure consistent, reproducible results. These benefits make droplet microfluidics a versatile tool in drug discovery, medical diagnostics, single-cell analysis, and other | such as individual compartment alization, enhanced control, wide sample applicability, powerful detection integration, and extensive application coverage. However, it lacks some depth in explaining these points and does not fully capture the nuances of the technology's capabilities, particularly in terms of its ability to support massive parallel processing and iterative operations. The response also does not explicitly mention the importance of consistent and reproducible results that are ensured by the | | |

| | | | | |
|---|---|---|---|---|
| | * **Powerful Detection Integration:** Modern detection technologies seamlessly integrate with droplet microfluidics, enabling functions like identification, quantification, screening, and real-time dynamic monitoring.<br><br>* **Extensive Application Coverage:** Droplet microfluidics finds applications in diverse fields such as single-cell analysis, medical diagnostics, drug discovery, food and feed industry, and environmental engineering.<br><br>In essence, droplet microfluidics provides a powerful platform for precise, controlled, and | fields, offering powerful insights and precise manipulation at the individual level. | system's highly controllable local environments.<br><br>SCORE: 56 | | |

| | highly sensitive analysis at the individual level, surpassing the limitations of bulk systems. | | | | |
|---|---|---|---|---|---|
| What are the advantages of droplet microfluidic systems compared with continuous-flow systems? | Droplet-based microfluidic systems offer several key advantages over continuous-flow (or single-phase) microfluidic systems:<br><br>* **Overcoming limitations of continuous flow:** They address issues like Taylor dispersion, surface-molecule interactions, and slow mixing that plague continuous-flow systems.<br><br>* **Ultra-small assay volumes:** Droplet systems allow for the definition and formation of extremely small assay volumes, leading to reduced reagent consumption | Droplet-based microfluidic systems offer significant advantages over traditional continuous-flow systems. These include the ability to overcome issues such as Taylor dispersion, slow mixing, and surface-molecule interactions. Droplet systems allow for ultra-small assay volumes, reducing reagent consumption and improving sensitivity. They also minimize cross-contamination and sample adsorption by compartmentalizing individual samples. The system's chaotic advection enhances mixing efficiency, enabling faster reactions. With droplet generation frequencies up to | The response is clear, concise, and directly addresses the question by listing specific advantages of droplet microfluidic systems over continuous-flow systems. It provides a detailed explanation of each benefit, aligning well with the ground truth answer.<br>SCORE: 96 | 96 | 96.0 |

| | and increased sensitivity. | hundreds of kHz, droplet | | | |
|---|---|---|---|---|---|
| | * **Negligible cross-contamination and sample adsorption:** The encapsulation of samples within individual droplets minimizes the risk of contamination between experiments and reduces sample loss due to adsorption to surfaces.<br><br>* **Enhanced mixing:** Chaotic advection within droplets promotes rapid and efficient mixing of reagents and analytes.<br><br>* **High droplet generation frequencies:** Droplet generators can produce hundreds of kHz, enabling high-throughput experimentation.<br><br>* **Rapid and | microfluidics supports high-throughput experimentation. Furthermore, the rapid and precise control of droplet payloads and sub-millisecond mixing times make it ideal for applications requiring fast, controlled analysis. These benefits make droplet-based systems particularly valuable for cell-based assays, enzyme kinetics, and high-throughput screening. | | | |

| | efficient payload control:** Droplets can be precisely loaded with specific amounts of reagents or samples.<br>* **Sub-millisecond mixing times:** This allows for rapid reaction kinetics and analysis.<br><br>In essence, droplet microfluidics provides a more controlled and efficient environment for performing chemical and biological experiments compared to traditional continuous-flow systems. | | | | |
|---|---|---|---|---|---|
| What are the basic principles behind droplet generation in microfluidics? | Droplet generation in microfluidics is fundamentally based on **emulsion processes**.<br><br>Here's a breakdown of the key principles: | Droplet generation in microfluidics relies on emulsion processes, where an oil phase (continuous phase) encapsulates an aqueous phase (dispersed phase) to form droplets. Surfactants are added to lower | The response accurately describes the basic principles behind droplet generation in microfluidics, including emulsion processes, two | 94 | 94.0 |

| | | | | |
|---|---|---|---|---|
| | * **Two Immiscible Liquids:** The process involves two liquids that don't mix, typically an oil phase (continuous phase) and an aqueous solution (dispersed phase).<br>* **Surfactants:** Surfactants are crucial for stabilizing the emulsion. They reduce the interfacial tension between the two liquids, allowing droplets to form and preventing them from coalescing (merging).<br>* **Microchannel Geometry:** The shape and design of microchannels influence droplet formation. Common geometries include T-junctions, flow-focusing, and co-flow streams. These designs | interfacial tension, stabilize the droplets, and prevent coalescence. The process requires precise control over the flow of both liquids, often achieved using syringe pumps or pressure-driven systems. Droplets are generated through surface-induced instabilities, influenced by microchannel geometries such as T-junctions or flow-focusing, which control droplet size and frequency. The design and flow control of the system determine the size uniformity and production rate of droplets. These principles enable high-throughput applications such as drug delivery, high-throughput screening, and microreactors for bio-catalysis, ensuring reproducibility and productivity in biological experiments. | immiscible liquids, surfactants, microchannel geometry, flow control, and their roles in forming and stabilizing droplets. It also provides a clear explanation of how these principles are applied to achieve precise control over droplet size and frequency. The response is well-structured, concise, and relevant to the question, with minimal inaccuracies or omissions. SCORE: 94 | | |

| | create instabilities at the interface of the two liquids, leading to droplet breakup. * **Flow Control:** Precise control over the flow rates of both liquids is essential for generating droplets with desired size and frequency. This can be achieved using syringe pumps or pressure-driven systems.

**In essence, droplet generation in microfluidics mimics the principles of traditional emulsion formation but on a much smaller scale, leveraging precise control over fluid flows and channel geometries.**

Let me know if you have any other questions about this topic! | | | | |
| --- | --- | --- | --- | --- | --- |

| How is droplet size controlled in a droplet generator? | Droplet size control in a droplet generator can be achieved through several methods, as described in the provided context.<br><br>Here's a breakdown based on the information given:<br><br>* **Passive Droplet Generation:** This method relies on adjusting microchannel dimensions and geometries or controlling the flow rates of two immiscible liquids. By manipulating these parameters, the deformation and instability of the flow field interface are altered, leading to precise control over droplet size.<br>* **Active Droplet Generation:** While not explicitly | Droplet size control in microfluidic generators can be achieved through several methods. In passive droplet generation, adjusting microchannel dimensions or controlling the flow rates of immiscible liquids allows for precise control over droplet size, shape, and monodispersity. Tree-splitters, which bisect large droplets into smaller ones, also influence droplet size and uniformity, with splitting angles affecting droplet diameter. In active droplet generation, technologies like piezoelectric actuators independently control droplet size, formation rate, and spacing, offering high precision. Additionally, the velocity of the continuous and dispersed phases plays a significant role in controlling droplet size. These | The response attempts to address the question but lacks clarity and coherence; mostly irrelevant details are included, and some concepts are not accurately represented. SCORE: 22 | 22 | 22.0 |

| | detailed in the provided text, active methods like piezoelectric actuators, thermal systems, electric fields, and magnetic fields can also be used to generate droplets. The strength and frequency of these forces influence droplet size.

**Specific Examples from the Text:**

* **Tree-Splitters:** These devices bisect large droplets into smaller ones, effectively reducing droplet size and increasing throughput. The splitting angle influences droplet uniformity and size.
* **Inlet to Main Channel Ratio:** In piezoelectric actuator-based generators, this ratio affects | methods enable the generation of uniform droplets for various applications, such as high-throughput screening and drug delivery. | | | |
|---|---|---|---|---|

| | | | | | |
|---|---|---|---|---|---|
| | droplet formation patterns. A higher ratio can lead to droplet bursts instead of individual droplets.<br><br>**Factors Influencing Droplet Size:**<br><br>The text highlights that the velocity of both continuous and dispersed phases plays a crucial role in droplet size. Additionally, the physical mechanisms involved in each type of generator (passive vs. active) contribute to size control.<br><br>Let me know if you have any further questions or need more clarification on specific aspects! | | | | |
| What is the role of surface tension in droplet formation? | Surface tension plays a crucial role in droplet formation. It acts as a force that tries to | Surface tension plays a critical role in droplet formation in microfluidic systems by | The response provides a clear and detailed explanation of the role of | 96 | 96.0 |

| | | | | |
|---|---|---|---|---|
| | minimize the surface area of a liquid, causing it to form spherical shapes.<br><br>Here's a breakdown of its influence:<br><br>* **Droplet Size:** Surface tension directly influences droplet size. A higher surface tension leads to smaller droplets because the force trying to minimize surface area is stronger.<br>* **Formation Mechanisms:** In microfluidic devices, surface tension competes with viscous shear stress and fluid inertia.<br>　* At low capillary numbers (Ca), where surface tension dominates, droplets are formed by squeezing or dripping.<br>　* As Ca increases, viscous forces | influencing droplet size, shape, and formation mechanisms. It acts to minimize surface area, favoring spherical droplet shapes. At low Weber numbers (We < 1), where surface tension dominates, droplets form through dripping or squeezing. As the Weber number increases, fluid inertia competes with surface tension, leading to jetting behavior. The capillary number (Ca) also influences droplet size, with surface tension controlling droplet formation at low Ca. Techniques like electrowetting can measure surface tension in microfluidic systems, providing insights into the interfacial properties. Additionally, surface tension is a key parameter in T-junctions and flow focusing devices, where droplet size measurements | surface tension in droplet formation, covering various aspects such as droplet size, formation mechanisms, transition between dripping and jetting, and interfacial instability. It accurately describes how surface tension interacts with other forces like viscous shear stress and fluid inertia, and its importance in microfluidic devices. The response is well-structured, concise, and relevant to the question, with a clear summary at the end. SCORE: 96 | | |

| | become more significant, leading to different droplet formation mechanisms.

* **Transition between Dripping and Jetting:** The balance between surface tension and inertial forces, represented by the Weber number (We), determines whether droplets form through dripping or jetting. When We is of order one, the transition occurs.
* **Interfacial Instability:** Surface tension can contribute to interfacial instabilities, such as tip-streaming, where the surfactant layer in an external flow field becomes unstable.

**In summary:** Surface tension acts as a | offer an indirect way to assess surface tension at the time-scale of droplet production. Understanding the interplay between surface tension and other forces is essential for precise droplet control. | | | |
|---|---|---|---|---|---|
| | fundamental force shaping droplet formation by influencing size, mechanism, and stability. Understanding its interplay with other forces is crucial for controlling droplet generation in microfluidic devices. | | | | |
| How does a flow-focusing device differ from a T-junction in droplet formation? | Both flow-focusing and T-junction devices are used to create droplets in microfluidics, but they differ in their geometry and how they control droplet size.<br><br>Here's a breakdown:<br><br>**T-junction:**<br><br>*<br>**Geometry:**<br>A simple intersection where one channel (dispersed phase) meets another (continuous | Flow-focusing devices and T-junctions are two common microfluidic designs for droplet formation, differing in geometry and mechanism. A flow-focusing device uses a central dispersed phase channel and side continuous phase channels converging at a narrow orifice, where hydrodynamic focusing pinches off droplets. It produces smaller, highly uniform droplets with precise control over size and high | The response provides a clear and detailed comparison between flow-focusing devices and T-junctions in terms of geometry, droplet formation mechanism, size control, complexity, controllability, and key differences. It accurately describes the advantages of flow-focusing devices over T-junctions, | 85 | 85.0 |

| | phase).<br>* **Droplet Formation:** Shear forces at the junction cause the dispersed phase to elongate and break into droplets.<br>* **Size Control:** Droplet size is influenced by flow rates, channel widths, and viscosity differences between phases.<br><br>**Flow-focusing:**<br><br>*<br>**Geometry:** More complex, with the dispersed and continuous phases flowing through a narrow constriction (orifice) before merging.<br>* **Droplet Formation:** The focused streams undergo significant elongational flow as they pass the orifice, leading to smaller droplets.<br>* **Size | throughput. In contrast, a T-junction has perpendicular channels where the continuous phase shears the dispersed phase at the intersection, forming larger droplets with less uniformity and lower throughput. Flow-focusing is ideal for applications requiring monodisperse droplets, while T-junctions suit simpler or larger-droplet applications. | including smaller droplet sizes, finer control over droplet size, and higher throughput. However, it does not explicitly state that flow-focusing produces more uniform droplets or mention the specific application areas where each device is ideal.<br>SCORE: 85 | | |

| | Control:** Offers greater flexibility in droplet size by adjusting flow rates of both phases and manipulating the orifice geometry.

**Key Differences:**

* **Complexity:** Flow-focusing devices are more complex to fabricate due to the intricate channel geometries.
* **Droplet Size:** Flow-focusing generally allows for the production of smaller droplets compared to T-junctions.
* **Controllability:** Flow-focusing offers finer control over droplet size through adjustments in flow rates and orifice geometry. | | | | |
|---|---|---|---|---|---|

| | Let me know if you have any other questions! | | | | |
|---|---|---|---|---|---|
| What role does the geometry of the microchannel play in droplet formation? | The geometry of the microchannel plays a crucial role in droplet formation.<br><br>Here's how:<br><br>* **Shape and Design:** Different channel geometries, such as T-junctions, flow-focusing devices, and Y or T-shaped channels, influence how droplets break off from the continuous phase.<br>    * **T-junctions:** The intersection of the dispersed and continuous phases at the junction creates a point where shear forces elongate the dispersed phase until it breaks into droplets.<br>    * **Flow-focusing:** This geometry confines both phases to a narrow region, | The geometry of a microchannel plays a critical role in droplet formation, influencing the size, shape, and generation frequency of droplets. Key geometric factors include the channel width, depth, and junction design, such as T-junctions or flow-focusing geometries. These features govern the interaction between the continuous and dispersed phases, affecting the shear forces and pressure gradients that drive droplet breakup. For instance, narrower channels or smaller orifices promote the formation of smaller droplets due to increased shear stress. Similarly, the junction type controls droplet detachment mechanics; in T-junctions, droplets | The response provides a clear and detailed explanation of the role of microchannel geometry in droplet formation, covering various geometric factors such as shape, design, dimensions, and surface properties. It accurately describes how different geometries influence droplet breakup and formation, including specific examples like T-junctions, flow-focusing devices, and Y or T-shaped channels. The response also acknowledges the | 85 | 85.0 |

| | | | | |
|---|---|---|---|---|
| | forcing them together and promoting droplet formation through stretching and thinning.<br><br>    * **Y or T-shaped channels:** These geometries can induce droplet breakup by converging streams of fluids.<br><br>* **Channel Dimensions:** The width and length of the microchannels affect the flow dynamics and pressure gradients, influencing droplet size and spacing.<br><br>* **Surface Properties:** The hydrophobicity of the channel surface can also impact droplet formation by affecting interfacial tension and wetting behavior. | form due to the balance of interfacial tension and shear force, while in flow-focusing designs, they form under pinching at the constriction. Channel geometry also determines flow regimes ranging from dripping to jetting further impacting droplet characteristics. Thus, optimizing microchannel geometry is crucial for precise control over droplet formation in applications like microfluidics, emulsification, and lab-on-a-chip systems. | importance of optimizing microchannel geometry for precise control over droplet characteristics in various applications. However, it could be improved by providing more quantitative information and a clearer connection to the ground truth answer.<br><br>SCORE: 85 | | |

| | The specific geometry chosen depends on the desired droplet size, shape, and application. | | | | |
|---|---|---|---|---|---|
| What factors affect droplet coalescence in microfluidic devices? | Based on the provided text, several factors influence droplet coalescence in microfluidic devices:<br><br>* **Surfactant Type and Concentration:**<br>* The type of fluorosurfactant used and its concentration significantly impact droplet coalescence. Different surfactants exhibit varying rates of interface stabilization, affecting how quickly droplets merge. For instance, a surfactant with fast interface stabilization like Krytox/C210 allows for passive droplet coalescence at lower concentrations. | Droplet coalescence in microfluidic devices is influenced by several factors. Surfactant concentration plays a key role, with higher concentrations preventing droplet fusion, while lower concentrations promote coalescence. The duration droplets remain in contact within the pairing channel, known as pairing time, also impacts coalescence, with longer times favoring fusion. The geometry of the pairing channel, specifically its cross-sectional area, controls the number of droplets that can come into contact and coalesce. Additionally, | The response provides a detailed explanation of the factors affecting droplet coalescence in microfluidic devices, including surfactant type and concentration, pairing time, droplet size and volume, and geometrical constraints of the pairing channel. It accurately describes how these factors influence coalescence and offers specific examples, such as the effect of Krytox/C210 on interface stabilization. | 85 | 85.0 |

| | * **Pairing Time (s):** The duration droplets remain in physical contact within the pairing channel directly influences coalescence. Longer pairing times generally lead to higher coalescence rates.<br><br>* **Droplet Size and Volume:** The size and volume of both the emulsion droplets introduced into the device and the droplets produced on-chip play a role. The text mentions that the cross-sectional area of the pairing channel needs to be comparable to the droplet size for selective coalescence.<br><br>* **Geometrical Constraints of the Pairing Channel:** The dimensions of the pairing | droplet size and volume affect how efficiently droplets merge. Finally, electrocoalescence, which applies electric fields to induce droplet fusion, offers an alternative method for on-demand coalescence. By optimizing these factors—surfactant concentration, pairing time, channel geometry, and droplet size—reliable and selective droplet coalescence can be achieved in microfluidic systems. | The response also correctly emphasizes the importance of controlling these factors for reliable and selective droplet coalescence. However, it does not mention electrocoalescence, which is a key aspect of the ground truth answer. SCORE: 85 | | |
| | channel, specifically its cross-sectional area, influence how many droplets can come into contact and subsequently fuse.<br><br>The text emphasizes that by carefully controlling these factors, reliable, highly selective, and passive droplet coalescence can be achieved independently of the carrier oil or surfactant used. | | | | |
| What design elements are critical for producing double or multiple emulsions? | Several design elements are crucial for successfully generating double or multiple emulsions.<br><br>Here's a breakdown based on the provided text:<br><br>**For Double Emulsions:**<br><br>* **Co-flow and Flow-Focusing | To generate double or multiple emulsions, several design elements are essential. Co-flow and flow-focusing configurations enable precise control over fluid flow, allowing for stable droplet formation, especially in double emulsions. Sequential emulsification steps, where inner droplets are emulsified into | The response provides a detailed breakdown of design elements crucial for producing double or multiple emulsions, including specific configurations (co-flow and flow-focusing), steps (sequential emulsificatio | 76 | 76.0 |

| | | |
|---|---|---|
| | Configurations: ** These configurations allow for the precise control of fluid flow rates and interactions between phases, leading to the formation of stable droplets.<br><br>* **Capillary Alignment:** The alignment of cylindrical capillaries within a square capillary is essential for directing the flow of each phase and creating the necessary hydrodynamic focusing effect.<br><br>**For Multiple Emulsions:**<br><br>* **Sequential Emulsification Steps:** Multiple stages of emulsification, where inner phases are first emulsified into intermediate phases, then further emulsified into outer phases, allow for the | intermediate phases and then into outer phases, are key for multiple emulsions. Capillary microfluidic devices, such as consecutive capillary and coaxial configurations, are used to control droplet size, number, and monodispersity. Additionally, precise flow control and shear force management are crucial for stable and consistent droplet formation. Microchannel geometry optimization and material compatibility further enhance emulsification processes. By carefully managing these factors, researchers can produce double and multiple emulsions with controlled internal structures and compositions for various applications. | n), materials (material compatibility ), and device components (capillary microfluidic devices). However, the answer lacks depth in some areas, such as the explanation of capillary alignment and surface properties. The response also does not explicitly mention precise flow control and shear force management, which are mentioned in the ground truth answer. SCORE: 76 | | |

| | creation of complex multilayer structures. * **Precise Flow Control:** Controlling the volume flow rates (Q) of each phase is critical for achieving desired droplet sizes and numbers at each level of the multiple emulsion. * **Microfluidic Device Design:** The arrangement of capillaries, T-junctions, or other microfluidic elements within a device directly influences the emulsification process and the resulting droplet structure.<br><br>**Additional Considerations:**<br><br>* **Material Compatibility:* * The materials used in the microfluidic device must be compatible with all the fluids | | | | |
|---|---|---|---|---|---|

| | involved to prevent unwanted reactions or adhesion.<br>* **Surface Properties:** The surface properties of the capillaries and other device components can influence droplet formation and stability.<br><br>By carefully considering these design elements, researchers can create microfluidic devices capable of generating a wide range of double and multiple emulsions with precise control over their internal structure and composition. | | | | |
|---|---|---|---|---|---|
| What key factors influence the design of flow-focusing droplet microfluidics optimized using machine learning? | The provided text highlights several key factors influencing the design of flow-focusing droplet microfluidics optimized using machine | The design of flow-focusing droplet microfluidics, optimized using machine learning, depends on several critical factors. Channel geometry, | The response provides a clear and concise summary of key factors influencing the design of flow-focusing | 52 | 52.0 |

| | | | | |
|---|---|---|---|---|
| | learning:<br><br>* **Droplet Diameter and Generation Rate:** These are primary performance metrics that machine learning models aim to predict and optimize.<br>* **Design Parameters:** The text mentions "a large number of effective parameters" involved in flow-focusing geometries. These parameters likely include channel dimensions, fluid properties (viscosity, surface tension), and flow rates. Machine learning algorithms learn the relationships between these parameters and the desired droplet characteristics.<br>* **Fluid Combinations:** * Different fluids used in the microfluidic | including the shape and size, controls droplet generation by defining the break-off point where the dispersed phase meets the continuous phase. Flow rates of both phases affect droplet size and frequency, with higher continuous phase flow reducing droplet size. Viscosity influences flow behavior and shear forces, crucial for stable droplet formation. Shear forces between phases are controlled by adjusting channel geometry or flow rates, ensuring consistent droplet generation. Machine learning algorithms predict and optimize droplet diameter and generation rate by analyzing these design parameters, fluid properties, and large datasets, automating the design process for more efficient droplet production | droplet microfluidics optimized using machine learning. It correctly identifies relevant parameters such as droplet diameter, generation rate, design parameters (channel dimensions, fluid properties), fluid combinations, and data availability. However, it lacks some depth in explaining the relationships between these factors and does not explicitly mention channel geometry, flow rates of both phases, viscosity, or shear forces, which are critical aspects according to the ground | | |

| | device can significantly affect droplet formation. The text notes that the machine learning tool can be extended to support various fluid combinations.<br><br>* **Data Availability:** The success of machine learning relies on large, standardized datasets. The text emphasizes the need for sufficient data to train accurate models for predicting droplet performance.<br><br><br>In essence, machine learning in this context allows engineers to optimize flow-focusing droplet microfluidics by identifying the optimal combination of design parameters and fluid properties that achieve the desired droplet | in life sciences applications. | truth answer. The response also slightly misrepresents the role of machine learning by implying it directly optimizes droplet size and generation rate without mentioning its predictive capabilities.<br><br>SCORE: 52 | | |

| | size and generation rate. | | | | |
|---|---|---|---|---|---|
| How do channel wall properties (hydrophobicity/hydrophilicity) influence droplet dynamics? | Channel wall properties, specifically their hydrophobicity or hydrophilicity, play a crucial role in dictating droplet dynamics within microfluidic devices.<br><br>Here's how:<br><br>* **Droplet Formation:** To form aqueous droplets in an oil-based continuous phase (W/O emulsion), hydrophobic channel walls are essential. The oil phase will then lift the aqueous phase from the walls, encapsulating it into droplets. Conversely, hydrophilic channels are needed for O/W emulsions where oil droplets are formed within a water-based continuous | The channel wall properties, specifically hydrophobicity or hydrophilicity, significantly influence droplet dynamics in microfluidic systems. These properties determine the wettability of the channel walls, affecting the interaction between the continuous and dispersed phases. In hydrophobic channels, the dispersed phase (typically water) tends to form droplets as it minimizes contact with the channel walls. This enhances droplet stability and facilitates easier detachment during droplet formation. Hydrophobic surfaces are ideal for applications requiring water-in-oil emulsions or precise droplet control. Conversely, hydrophilic channels promote | The response provides a clear and concise explanation of how channel wall properties influence droplet dynamics in microfluidic systems, covering key aspects such as droplet formation, adhesion, emulsion stability, and switching between emulsions. It accurately describes the effects of hydrophobicity and hydrophilicity on these processes and highlights their importance for achieving desired outcomes in microfluidic applications. The response is well-structured, | 86 | 86.0 |

| | phase.<br><br>* **Droplet Adhesion:** Hydrophobic surfaces prevent the discrete (droplet) phase from adhering to the channel walls, ensuring smooth droplet flow and preventing clogging. Hydrophilic surfaces, on the other hand, can lead to droplet adhesion, hindering proper manipulation and flow.<br><br>* **Emulsion Stability:** The wettability of the channel walls influences emulsion stability. Hydrophobic channels tend to promote more stable W/O emulsions, while hydrophilic channels are better suited for O/W emulsions.<br><br>* **Switching Between Emulsions:** Controlling the | wetting of the dispersed phase, making it more likely to spread along the walls instead of forming discrete droplets. This is useful for oil-in-water systems or applications involving liquid films. Wall properties also impact droplet size, breakup, coalescence, and transport. Hydrophobicity favors smaller droplets and reduces the likelihood of droplet adherence, while hydrophilicity can lead to higher flow resistance and altered droplet generation behavior. Tailoring wall properties is crucial for optimizing microfluidic processes. | easy to follow, and effectively conveys relevant information. SCORE: 86 | | |

| | wettability of the channel walls allows researchers to switch between generating W/O and O/W emulsions by simply altering the surface properties.<br><br>In essence, carefully selecting the appropriate hydrophobicity or hydrophilicity of the channel walls is critical for achieving desired droplet behavior, emulsion type, and overall functionality in microfluidic applications. | | | | |
|---|---|---|---|---|---|
| How can you incorporate active control mechanisms (e.g., electrical, magnetic) into droplet devices? | Active control mechanisms like electrical and magnetic forces offer precise and fast droplet generation in microfluidic devices. Here's how they can be incorporated:<br><br>**Electrical Methods:** | Active control mechanisms such as electrical, magnetic, thermal, and mechanical methods can precisely manipulate droplet formation in microfluidic devices. Electrical control methods include direct current (DC), which uses | The response provides a clear and detailed explanation of how to incorporate electrical and magnetic forces into droplet devices, covering various methods | 76 | 76.0 |

| | | | | |
|---|---|---|---|---|
| | * **Direct Current (DC):** Applying a DC voltage across electrodes embedded in the flow-focusing device creates an electric field at the water-oil interface. This field acts as an additional force, assisting droplet breakup and allowing precise control over droplet size by adjusting the voltage strength. * **Alternating Current (AC):** * **Electrowetting-on-Dielectric (EWOD):** AC fields can reduce the contact angle between the conductive liquid flow and the channel walls, promoting spreading and retraction of the liquid to form droplets. This method offers a controllable wettability switch for droplet generation. * | electrodes to control droplet size, and alternating current (AC), which utilizes electrowetting-on-dielectric (EWOD) and dielectrophoresis (DEP) to influence droplet behavior. Magnetic control employs ferrofluids and external magnets to non-invasively guide droplet formation. Thermal control involves heating the entire device or applying localized heating to adjust fluid parameters like viscosity and interfacial tension. Mechanical control techniques, such as pneumatic and piezoelectric components, can also regulate droplet dynamics. These active control strategies enable precise and flexible droplet generation, allowing for tailored applications in complex microfluidic | such as DC, AC (including EWOD and DEP), ferrofluids, and key considerations like frequency dependence and device design. However, it lacks information on thermal and mechanical control mechanisms, which are mentioned in the ground truth answer. SCORE: 76 | | |

| | **Dielectrophoresis (DEP):** Even with electrically neutral liquid flows, AC fields can induce polarization in the liquid, leading to DEP forces that draw droplets together.<br><br>**Magnetic Methods:**<br><br>* **Ferrofluids:** Utilizing ferrofluids (liquids containing magnetic particles) as the continuous phase allows for non-contact magnetic control over droplet formation.<br>    * The direction and location of magnets, magnetic flux density, and magnetization curve all influence droplet behavior in this setup.<br><br>**Key Considerations:** | systems, including droplet manipulation and multicomponent emulsion creation. | | | |
|---|---|---|---|---|

| | ** | | | | |
| --- | --- | --- | --- | --- | --- |
| | * **Frequency Dependence:** Electrical methods often exhibit frequency dependence, with optimal AC frequencies for specific applications (e.g., 100-700 Hz for the OEW device mentioned in the context).<br>* **Device Design:** The design of the microfluidic device, including electrode placement and magnet configurations, is crucial for effective control.<br><br>By incorporating these active control mechanisms, researchers can achieve precise droplet generation, manipulation, and integration into complex microfluidic | | | | |

| | systems for various applications. | | | | |
|---|---|---|---|---|---|
| How do flow rates affect droplet size and frequency? | The provided text highlights a strong correlation between flow rates and both droplet size and frequency.<br><br>Here's a breakdown based on the information given:<br><br>**Droplet Size:**<br><br>* **Decreasing Continuous Phase Flow Rate:** Leads to smaller droplet sizes and a more uniform distribution. This is likely due to less shearing force breaking up the dispersed phase.<br>* **Increasing Continuous Phase Flow Rate:** Results in larger droplets. The increased shearing force overcomes interfacial tension, leading to faster breakup | Flow rates significantly influence droplet size and frequency in microfluidic systems. As the flow rate ratio ($\varphi = Qd/Qc$) increases, the shearing force from the continuous phase dominates, leading to smaller droplets due to rapid breakup of the dispersed phase. Additionally, higher flow rates in the continuous phase can result in larger droplets, as the dispersed phase doesn't break up until it reaches the downstream section. The frequency of droplet generation also increases with higher flow rate ratios, demonstrating a positive correlation. However, droplet diameter growth trends vary depending on the model and input conditions. Overall, | The response attempts to address the question but lacks clarity and coherence; mostly irrelevant. It seems to be a summary of information rather than a direct answer to the question, and some points are not directly related to the query about how flow rates affect droplet size and frequency. SCORE: 16 | 16 | 16.0 |

| | | | | |
|---|---|---|---|---|
| | of the dispersed phase.<br><br>**Droplet Frequency:**<br><br>* The text mentions that droplet frequency can be controlled independently by adjusting flow rates within a specific range for different droplet generator chips (DGFF).<br>* However, it doesn't explicitly state how changing flow rates *affects* frequency. It implies that varying flow rates allows for tuning the frequency within a defined range.<br><br>**Additional Points:**<br><br>* The text emphasizes that the relationship between flow rates and droplet characteristics is complex and influenced by various factors | increasing the flow rate ratio enhances droplet formation frequency and reduces droplet size due to stronger shearing forces, with complex interactions influenced by factors such as interfacial tension and channel geometry. | | | |
| | like:<br>    *<br>**Interfacial tension:**  The force acting at the interface between the two fluids.<br>    * **Capillary number:** A dimensionless number representing the ratio of inertial forces to capillary forces.<br>    * **Junction geometry:** The shape and design of the microfluidic channel where droplets are formed.<br><br>Let me know if you have any other questions or would like me to elaborate on a specific aspect! | | | | |
| What techniques are available for droplet splitting? | The provided text describes two main categories of droplet splitting techniques:<br><br>**1. Passive Techniques:**<br><br>* These methods rely on | Droplet splitting techniques can be categorized into passive and active methods. Passive techniques rely on the microchannel geometry and fluid properties, such as T and Y junctions, flow focusing | The response accurately describes two main categories of droplet splitting techniques (passive and active) and provides specific | 76 | 76.0 |

| | the geometry of microchannels and fluid properties.<br>* Examples include using T-junctions, Y-junctions, flow focusing configurations, or introducing ribs/obstructions in the flow path.<br>*<br>**Limitations:**<br>*<br>  * Sensitive to fluid properties and flow conditions.<br>  * Lack selective or on-demand operation.<br>  * Cannot dynamically control the splitting ratio.<br><br>**2. Active Techniques:**<br><br>* These methods utilize external forces to achieve droplet splitting.<br>* Examples include:<br>  *<br>**Pneumatic:** Using a pneumatic actuator, but can cause droplet deformation and | configurations, or introducing ribs/obstructions in the flow path. These methods are simple but lack dynamic control over splitting ratios and are sensitive to flow conditions. Active techniques utilize external forces like electric, magnetic, acoustic, or thermal fields to precisely control droplet splitting. Examples include electric field-based splitting, magnetic fields for ferrofluids, acoustic radiation force, and thermocapillary methods. Active techniques offer more flexibility, enabling on-demand droplet splitting with controlled ratios, overcoming the limitations of passive methods. However, each technique has its own set of advantages and constraints depending on the application and droplet type. | examples for each category. It also highlights the limitations of passive methods and the benefits of active techniques, such as on-demand operation and controlled splitting ratios. However, some minor details are missing or not explicitly stated in the provided text, but overall, it covers the core aspects of droplet splitting techniques correctly.<br>SCORE: 76 | | |
| --- | --- | --- | --- | --- |

| | damage.<br>   *<br>**Electric:**<br>Utilizing electric fields, but limited to charged droplets.<br>   *<br>**Magnetic:**<br>Employing magnetic fields, but only applicable to ferrofluid droplets.<br>   *<br>**Thermocapillary:**<br>Leveraging temperature gradients, but unsuitable for temperature-sensitive samples.<br>   *<br>**Acoustic:**<br>Applying acoustic radiation force or streaming for label-free splitting.<br><br>The text highlights that active techniques offer on-demand droplet splitting with controlled ratios, overcoming the limitations of passive | | | | |
|---|---|---|---|---|---|

| | methods.<br><br>Let me know if you have any other questions about droplet splitting techniques! | | | | |
|---|---|---|---|---|---|
| What advantages does droplet formation by passive methods have over active methods? | Passive droplet generation methods offer several advantages over active methods:<br><br>* **Simplicity:** Passive methods rely on the manipulation of flow rates and channel geometries, eliminating the need for complex external driving forces like piezoelectric actuators or electric fields. This simplifies the device design and fabrication process.<br>* **Cost-effectiveness:** Passive methods often utilize readily available materials and straightforward fabrication techniques, | Passive droplet generation methods offer several advantages over active methods. They simplify device design by relying on flow rate manipulation and microchannel geometry, eliminating the need for complex external forces. This leads to cost-effectiveness and easier fabrication. Passive methods also provide precise control over droplet size, shape, and monodispersity, making them ideal for applications requiring uniform droplets. Additionally, they protect droplets from external interference and cross-contamination, ensuring integrity. Passive methods can produce | The response provides a clear and concise explanation of the advantages of passive droplet formation methods over active ones. It accurately highlights the benefits of simplicity, cost-effectiveness, monodispersity, and protection from interference. However, it would be more comprehensive if it also mentioned the specific percentage of dispersity (1%-3%) that can be achieved | 85 | 85.0 |

| | making them more cost-effective compared to active methods that may require specialized components and intricate control systems.<br>* **Monodispersity:** Passive methods can achieve high monodispersity (uniform droplet size) by precisely controlling flow rates and channel dimensions. This is crucial for applications requiring precise droplet sizing.<br>* **Protection from Interference:** Droplets generated passively are less susceptible to external interference and cross-contamination due to the absence of active manipulation forces.<br><br>However, it's | continuous droplet streams with uniform size distributions (1%-3% dispersity) without requiring external actuation. While they are not suitable for dynamic manipulation or "droplet-on-demand" applications, passive methods are favored for their simplicity, precision, and cost-efficiency in controlled environments. | with passive methods, as stated in the ground truth answer.<br>SCORE: 85 | | |
| | important to note that passive methods have limitations in terms of dynamic droplet manipulation and "droplet-on-demand" generation, which are strengths of active methods. | | | | |
| How do you design a device for monodisperse droplet generation? | To design a device for generating monodisperse droplets, you have several options depending on your specific needs and desired droplet size.<br><br>Here's a breakdown of two popular methods:<br><br>**1. Cross-Flow Design:**<br><br>* **Principle:** This method involves bringing the continuous and dispersed phases together at an arbitrary angle (T-junction or Y- | To design a device for monodisperse droplet generation, two common methods are used: cross-flow and flow-focusing designs. In cross-flow designs, the continuous phase flows around the dispersed phase in a constriction channel, with droplet size controlled by adjusting channel dimensions and shear stress. This method is simple but can lead to droplet damage due to shear forces or wall interactions. Flow-focusing designs, on the other hand, use sheath flows to stabilize the interface between | The response provides a clear and detailed explanation of the two common methods for monodisperse droplet generation (cross-flow and flow-focusing designs), including their principles, advantages, disadvantages, and key considerations. It also offers practical tips for designing an effective device. However, some minor | 78 | 78.0 |

| | | | | |
|---|---|---|---|---|
| | junction).<br>*<br>**Advantages:**<br>*<br>   * Relatively simple to implement.<br>   * Requires less space on microfluidic platforms.<br>   * Can be easily combined with other microfluidic units.<br>*<br>**Disadvantages:**<br>   * Droplets may be damaged by shear stress or adhere to channel walls.<br>   * Achieving high monodispersity can be challenging.<br><br>**2. Flow-Focusing Design:**<br><br>* **Principle:** The dispersed phase is squeezed between two streams of the continuous phase in a constriction channel. This creates a stable | phases, enabling smaller droplet formation with precise control over size by adjusting flow rates and orifice dimensions. This method offers high monodispersity and is less prone to droplet damage. Both designs require careful optimization of flow rates, channel dimensions, and surface tension to achieve uniform droplet sizes. The choice of design depends on specific needs such as droplet size, throughput, and resistance to damage during formation. | inaccuracies and omissions are present in the description of cross-flow design, which could be improved upon.<br>SCORE: 78 | | |
| | interface and promotes spherical droplet formation.<br>*<br>**Advantages:**<br>   * Generates smaller droplets at higher generation rates compared to cross-flow designs.<br>   * Less susceptible to droplet damage due to wall interactions.<br>   * Allows for precise control over droplet size by adjusting flow velocities and channel dimensions.<br>*<br>**Disadvantages:**<br>   * More complex design and fabrication process.<br><br>**Key Considerations for Monodisperse Droplet Generation:**<br><br>* **Flow Rates:** Carefully control the flow rates of both the | | | | |

| | | | | |
|---|---|---|---|---|
| | continuous and dispersed phases to achieve desired droplet size and monodispersity.<br><br>* **Channel Dimensions:** The width and length of the constriction channel (in flow-focusing designs) significantly influence droplet formation and size.<br><br>* **Surface Tension:** Optimize the surface tension balance between the phases to promote stable droplet formation.<br><br>* **Material Selection:** Choose materials that are compatible with your fluids and minimize wetting effects on the channel walls.<br><br>**Additional Tips:**<br><br>* Use a microscope or imaging system to monitor | | | |

| | | | | | |
|---|---|---|---|---|---|
| | droplet formation and size distribution in real-time.<br>* Perform calibration experiments to determine the relationship between flow rates, channel dimensions, and droplet size.<br>* Consider incorporating feedback control mechanisms to maintain consistent droplet generation.<br><br><br>Remember that designing an effective monodisperse droplet generator requires careful consideration of various factors and iterative optimization. | | | | |
| What materials are most commonly used for fabricating droplet microfluidic devices? | According to the provided text, **PDMS (polydimethylsil oxane)** is the most commonly used material for fabricating droplet microfluidic devices. | The most commonly used materials for fabricating droplet microfluidic devices are PDMS (polydimethylsilo xane) and silicon-based materials. PDMS is preferred due to its ease of | The answer correctly identifies PDMS as a commonly used material for fabricating droplet microfluidic devices but | 41 | 41.0 |
| | The text states: "Microfluidic chips fabricated from PDMS using soft lithography are the most commonly used in droplet microfluidics."<br><br>It also mentions other materials like glass capillaries and various elastomeric, thermoplastic, amorphous polymer, fluoropolymer, and glass materials as alternatives with specific advantages. However, PDMS remains the dominant choice due to its ease of fabrication and versatility. | fabrication using soft lithography, which allows for the creation of precise and versatile microfluidic devices. This material provides high transparency, efficient sealing, and the ability to mold complex microchannels, making it ideal for droplet generation and manipulation. While PDMS is dominant, alternatives such as thermoplastics (e.g., PMMA, polycarbonate), fluoropolymers (e.g., PTFE), and glass materials are also used, each offering unique benefits like better solvent resistance or mass production capabilities. Despite the variety of materials available, PDMS remains the material of choice for most droplet microfluidics applications due to its simplicity and versatility. | fails to mention silicon-based materials, which are also mentioned in the ground truth as equally common. SCORE: 41 | | |
| What are the advantages and limitations of | PDMS (polydimethylsil oxane) is a | Polydimethylsilox ane (PDMS) is widely used for | The response provides a clear and | 76 | 76.0 |

| using PDMS for microfluidic devices? | popular material for microfluidic device fabrication due to several advantages:<br><br>* **Cost-effectiveness:** PDMS is relatively inexpensive compared to other materials like silicon or glass.<br>* **Ease of Fabrication:** Soft lithography, a common method for PDMS microfluidic device fabrication, is relatively simple and robust.<br>* **Flexibility:** PDMS's elastomeric nature allows for flexible devices, which can be advantageous for certain applications.<br>* **Biocompatibility:** PDMS is generally considered biocompatible, making it suitable for | microfluidic devices due to its several advantages, including low toxicity, biocompatibility, optical transparency, and high gas permeability, which make it ideal for biomedical applications. Its flexibility and cost-effectiveness also contribute to its popularity, as it allows for easy fabrication using soft lithography. PDMS enables real-time observation and is well-suited for flexible and wearable electronics. However, PDMS has limitations, such as its hydrophobic nature, which can lead to fouling in microchannels, affecting performance. Additionally, its soft elastomeric properties limit mechanical strength, and large-scale production can be challenging. | concise overview of the advantages and limitations of using PDMS for microfluidic devices. It lists specific benefits such as cost-effectiveness, ease of fabrication, flexibility, biocompatibility, optical transparency, and gas permeability, which are all relevant to the topic. However, it also acknowledges some limitations like hydrophobicity, scalability issues, and mechanical weakness. The response is well-structured and easy to follow, with a clear summary at the end. It demonstrates a good | | |

| | biomedical applications. * **Optical Transparency:** PDMS is optically transparent, allowing for real-time monitoring of experiments using microscopy techniques. * **Gas Permeability:** PDMS has high gas permeability, which can be beneficial for applications requiring gas exchange.<br><br>However, PDMS also has some limitations:<br><br>* **Hydrophobicity:** The hydrophobic nature of PDMS can lead to fouling in microchannels, potentially affecting device performance. * **Scalability:** Large-scale production of PDMS | Despite these drawbacks, PDMS remains a versatile and widely used material, though further research is needed to address its limitations for broader applications. | understanding of the subject matter, but could benefit from more depth in discussing the implications of these advantages and limitations.<br><br>SCORE: 76 | | |
| | microfluidic devices can be challenging.<br>* **Mechanical Strength:** PDMS is relatively soft and may not be suitable for applications requiring high mechanical strength.<br><br>Overall, while PDMS offers many advantages for microfluidic device fabrication, its limitations should be considered when choosing materials for specific applications. | | | | |
| How does 3D printing facilitate microfluidic device fabrication? | Here's how 3D printing facilitates microfluidic device fabrication, based on the provided text:<br><br>* **Rapid Prototyping:** 3D printing allows for quick and iterative design changes, speeding up the | 3D printing revolutionizes microfluidic device fabrication by enabling the creation of complex and precise channel structures that are challenging to achieve with traditional methods like soft lithography. Techniques such as | The response provides a clear and concise overview of how 3D printing facilitates microfluidic device fabrication, listing specific benefits such as rapid prototyping, | 56 | 56.0000000 0000001 |

| | | | | |
|---|---|---|---|---|
| | development process.<br>* **Complex Geometries:** It can create intricate channel designs and structures that are difficult or impossible to achieve with traditional methods like soft lithography.<br>* **Customization:** Devices can be tailored to specific applications and needs, enabling personalized microfluidic solutions.<br>* **Material Versatility:** Different materials can be used for 3D printing, offering options for biocompatibility, optical transparency, and other specialized properties.<br>* **Cost-Effectiveness:** While specialized printers exist, desktop 3D printers can make the | Stereolithography (SL), Multijet Modeling (MJM), and Fused Deposition Modeling (FDM) offer high resolution, complex geometries, and the integration of various components like valves and pumps. 3D printing also allows rapid prototyping, customization, and the use of diverse materials for biocompatibility and transparency. Advantages over traditional methods include reduced fabrication time and labor, increased accuracy, and environmental friendliness due to the additive manufacturing process. With its ability to create hierarchical 3D configurations and integrate multiple functions, 3D printing holds great potential for advancing microfluidic device design and performance. | complex geometries, customization, material versatility, and cost-effectiveness. However, it lacks the depth and technical detail provided in the ground truth answer, particularly regarding the specific techniques (SL, MJM, FDM) and their advantages. The response also includes a call to action ("Let me know if you have any other questions...") which is not relevant to the question asked.<br><br>SCORE: 56 | | |
| | process more affordable compared to traditional fabrication methods.<br><br>**Specific Examples from the Text:**<br><br>* "3D-printing can facilitate the field of microfluidics to find its "killer applications"."<br>* "3D-printing can potentially replace soft lithography to fabricate microfluidic devices."<br><br>Let me know if you have any other questions about 3D printing and microfluidics! | | | | |
| What alternative materials can be used for applications involving organic solvents? | The text provides several alternatives to traditional organic solvents derived from petroleum:<br><br>* **Water:** Can be used alone or in mixtures with other solvents.<br>* | Several sustainable alternatives to traditional organic solvents are gaining attention for various applications. Supercritical CO2 is a non-toxic, environmentally friendly solvent, particularly useful in extraction | The response provides a clear and concise list of alternative materials that can be used for applications involving organic solvents. It accurately summarizes | 56 | 56.0000000 0000001 |

| | **Supercritical CO2:** A non-toxic and environmentally friendly solvent. * **Ionic liquids:** Composed of an asymmetric organic cation with an organic or inorganic anion, they are being explored as safe alternatives to toxic solvents. * **Organic carbonates:** A class of solvents that are less toxic than traditional organic solvents. * **Biosolvents:** Derived from renewable biomass sources. * **Fluorous solvents:** Characterized by their low solubility in water and high solubility in organic solvents. * **Non-toxic liquid polymers:** Offer unique properties for specific | processes. Ionic liquids, which are liquid salts at room temperature, offer low vapor pressure and broad application in chemical reactions. Organic carbonates, like dimethyl carbonate, are biodegradable and non-toxic, ideal for fine chemical synthesis. Biosolvents, derived from renewable resources such as plants and bacteria, provide a greener option for extraction and cleaning. Other alternatives include fluorous solvents, non-toxic liquid polymers like polyethylene glycol (PEG), water, ethanol, methanol, and ethyl acetate. These alternatives are chosen based on their specific properties such as polarity, boiling point, and environmental impact, promoting more sustainable chemical processes and | the key points from the ground truth answer, including supercritical CO2, ionic liquids, organic carbonates, biosolvents, fluorous solvents, and non-toxic liquid polymers. The response also mentions water as an option, which is correct but not explicitly mentioned in the ground truth answer. However, it does not delve deeper into the specific properties or applications of these alternatives, nor does it provide a clear connection to the importance of replacing hazardous solvents with eco-friendly | | |
| | applications.<br><br>The text emphasizes the importance of replacing hazardous solvents with eco-friendly alternatives to promote sustainable chemistry practices. | reducing reliance on hazardous petroleum-based solvents. | alternatives. SCORE: 56 | | |
| How does device surface treatment affect droplet formation? | The surface treatment of the microfluidic device significantly affects droplet formation.<br><br>Here's a breakdown based on the provided text:<br><br>* **Native PDMS:** PDMS is inherently hydrophobic, making it suitable for water-in-oil (W/O) droplet generation but hindering oil-in-water (O/W) droplet formation.<br>* **Plasma Treatment:** Plasma treatment | Surface treatment of PDMS microfluidic devices plays a critical role in determining droplet formation and stability. PDMS is naturally hydrophobic, suitable for generating water-in-oil (W/O) droplets but not oil-in-water (O/W) droplets. To generate O/W droplets, selective hydrophilic treatment is required. The researchers employed two methods: manual PVA deposition after plasma treatment, and a simultaneous approach combining PVA injection with on- | The response attempts to address the question but lacks clarity and specificity regarding how device surface treatment affects droplet formation. It does not provide detailed information about the effects of different treatments on droplet generation, nor does it mention the specific methods used for selective hydrophilic | 16 | 16.0 |

| | increases the surface energy of PDMS by generating radicals that make it more hydrophilic. This allows for PVA deposition and subsequent O/W droplet generation.<br><br>* **PVA Deposition:** PVA is deposited onto the plasma-treated PDMS, creating a stable coating that further enhances hydrophilicity. This enables controlled O/W droplet formation within specific sections of the device.<br><br>**Key Points:**<br><br>* The combination of plasma treatment and PVA deposition allows for selective wettability patterning, enabling both W/O and O/W droplet generation within the same | chip plasma treatment. The latter method enables precise wettability patterning, allowing for both W/O and O/W droplet formation within the same device. The plasma treatment increases surface energy, making the PDMS hydrophilic, while PVA deposition stabilizes the coating, ensuring long-term device use. This technique provides controllable droplet sizes and stable double emulsion formation, with the ability to manipulate droplet characteristics by adjusting flow rates. The method also allows for reliable, reproducible results, making it a powerful tool for droplet microfluidic applications. | treatment as described in the ground truth answer. SCORE: 16 | | |
| | device.<br>* The text highlights a modified protocol for PVA deposition that is simpler, faster, and more robust than previous methods.<br>* The stability of the PVA coating is crucial for reliable and long-term operation of the device.<br><br>Let me know if you have any other questions about this process! | | | | |
| What bonding techniques are used to seal microfluidic channels? | The provided text describes several bonding techniques used to seal microfluidic channels, particularly in the context of thermoplastic materials and PDMS devices.<br><br>Here's a breakdown:<br><br>**For Thermoplastic Materials:** | Several bonding techniques are employed to seal microfluidic channels, each suited to different materials. Thermal fusion bonding involves compressing thermoplastic substrates at temperatures near their glass transition point, allowing them to fuse upon cooling. Chemical bonding creates covalent | The response provides a detailed breakdown of various bonding techniques used to seal microfluidic channels, specifically highlighting methods suitable for thermoplastic materials and PDMS devices. It accurately | 98 | 98.0 |

| | | | | |
|---|---|---|---|---|
| | * **Thermal Fusion Bonding:** This involves compressing thermoplastic substrates at a temperature near their glass transition temperature. The heat softens the material, allowing them to fuse together upon cooling.<br><br>* **Chemical Bonding:** This technique uses chemical reactions to create covalent bonds between the substrate surfaces.<br><br>* **Solvent Bonding:** Substrates are bonded using a solvent that dissolves both materials partially, creating an interdiffusion zone and promoting adhesion.<br><br>**For PDMS Devices:**<br><br>* **Oxygen Plasma Treatment:** | bonds between substrate surfaces through chemical reactions, while solvent bonding uses solvents to dissolve the materials partially, promoting adhesion through interdiffusion. For PDMS devices, oxygen plasma treatment is commonly used to make surfaces reactive, enabling irreversible bonding when in contact with another PDMS or glass piece. Additionally, corona wand bonding offers a less expensive alternative to oxygen plasma treatment, though it is less consistent. The choice of bonding technique depends on material properties, bond strength requirements, and available equipment, all of which are crucial for the successful fabrication of microfluidic devices. | describes each technique, including thermal fusion bonding, chemical bonding, solvent bonding, oxygen plasma treatment, and corona wand bonding. The text also emphasizes the importance of considering material properties, bond strength requirements, and available equipment when choosing a bonding method.<br>SCORE: 98 | | |

| | This is a common method for bonding PDMS to itself or other materials like glass. Oxygen plasma makes the surfaces reactive, leading to irreversible bonding when they come into contact.<br>* **Corona Wand Bonding:** A less expensive alternative to oxygen plasma treatment, this uses a handheld device to create a corona discharge that modifies the surface and promotes bonding. However, it's less consistent than plasma bonding.<br><br>The text emphasizes that the choice of bonding technique depends on factors like the materials used, the desired strength of the bond, and the | | | | |
| --- | --- | --- | --- | --- | --- |
| | available equipment. | | | | |
| What challenges arise when fabricating complex microfluidic geometries? | The provided text highlights several challenges associated with fabricating complex microfluidic geometries, particularly at a commercial scale.<br><br>Here's a breakdown based on the context:<br><br>* **Patterned Wettability Control:** Achieving precise and reproducible control over the wettability of complex channel structures becomes increasingly difficult as the scale of fabrication increases. This is crucial for directing fluid flow and droplet formation within the microfluidic device.<br>* **Cost | Fabricating complex microfluidic geometries presents several challenges, including controlling wettability in intricate channel structures, which is critical for fluid flow and droplet formation. Additionally, scaling up microfabrication processes for commercial applications requires cost-effective techniques and materials. Current droplet generation methods mainly produce single emulsions, limiting their use in applications that require multiple or complex emulsions. Moreover, droplet microfluidics faces limitations in producing diverse and functional nano-scale materials compared to other methods. | The response provides a clear and detailed explanation of the challenges associated with fabricating complex microfluidic geometries, including precise wettability control, cost-effectiveness, expanding droplet emulsion capabilities, and diversifying nano-scale materials. It also offers a concise summary of these challenges and emphasizes the need for interdisciplinary collaboration to overcome them. The response aligns well with the ground truth | 96 | 96.0 |

| | | | | |
|---|---|---|---|---|
| | Considerations: ** Scaling up microfabrication processes to meet commercial demands requires cost-effective methods and materials. The current techniques and materials used might be too expensive for widespread adoption. * **Limited Droplet Emulsion Capabilities:** Current droplet generation methods primarily focus on single emulsions. However, many practical applications, such as drug delivery, require multiple or complex droplet emulsions. Developing microfluidic systems capable of generating these diverse emulsions is essential. * **Nano-scale Material Diversity:** | Conventional lithography techniques, which are restricted to rectangular cross-sections, struggle with creating complex 3D geometries. Although advancements like 3D printing and stacking methods offer solutions, effective mixing strategies and versatile fabrication of 3D networks with the required features remain significant hurdles in the field. Addressing these challenges will require innovative approaches and collaboration across various disciplines. | answer in terms of content and structure. SCORE: 96 | | |

| | While droplet microfluidics excels at synthesizing microscale particles, the diversity and functionalities of nano-scale materials produced through this method are still limited compared to other approaches. Further research is needed to expand the range of nanomaterials achievable through droplet microfluidics.<br><br>**In summary,** fabricating complex microfluidic geometries presents challenges related to precise wettability control, cost-effectiveness, expanding droplet emulsion capabilities, and diversifying the range of nano-scale materials | | | | |
|---|---|---|---|---|---|
| | produced. Overcoming these hurdles requires collaborative efforts across various disciplines to bridge the gap between material synthesis and practical applications. | | | | |
| How do you fabricate devices capable of handling multiphase flows? | The provided text describes several methods for fabricating devices capable of handling multiphase flows, particularly for creating emulsions.<br><br>Here's a breakdown based on the information given:<br><br>**1. Lithographic Processes:**<br><br>* This method is mentioned as being used to create flow-focusing devices with spatial wettability | Fabricating devices for handling multiphase flows involves several techniques and considerations. Key approaches include advanced micromanufacturing methods such as 3D printing, which eliminates the need for bonding processes, and flow-focusing techniques, which use nozzles and sheath fluids to control droplet formation. Controlling wettability is crucial for generating specific droplet types, while parallelization of | The response attempts to address the question but lacks clarity and coherence; mostly irrelevant details are included from the provided text, and it fails to provide a comprehensive overview of fabricating devices capable of handling multiphase flows.<br>SCORE: 16 | 16 | 16.0 |

| | modification. * It allows for the concatenation of modules to achieve stepwise fabrication of single, double, triple, quadruple, and quintuple emulsions. * However, it can be challenging to scale up production using this method.

**2. Consecutive Capillary Assemblies:**

* A more flexible three-step co-flow device is invented using consecutive capillary assemblies. This approach allows for the creation of multiphase flows with hierarchical onion-like configurations.

**3. 3D Printing:**

* The text | droplet generators enhances production scalability. Challenges include ensuring even fluid distribution, integrating various components like valves and connectors, and addressing the lack of standardized design rules. Additionally, methods like consecutive capillary assemblies and external forces (e.g., pneumatic or electrical control) can be employed to create complex emulsions. Through careful design and optimization, efficient fabrication of multiphase flow devices for applications such as droplet generation can be achieved. | | | |
|---|---|---|---|---|

| | suggests that 3D printing can be a valuable technique for fabricating microdevices, eliminating the need for bonding processes. This could potentially simplify the fabrication process for multiphase flow devices.<br><br>**4. Hydrodynamic Confinement Techniques (Flow Focusing):**<br><br>*  This method uses a bespoke arrangement of nozzles and sheath fluids to hydrodynamically confine the dispersed phase away from channel walls.<br>* It allows for control over droplet polarity based on channel geometry rather than wettability.<br>* However, fabrication can be complex, and flow conditions | | | | |
|---|---|---|---|---|

| | may limit droplet size and uniformity.<br><br>**Challenges in Fabrication:**<br><br>The text also highlights some challenges associated with fabricating multiphase flow devices:<br><br>* **Integration:** Achieving high integration levels can be difficult due to the complexity of integrating various components like valves, connectors, and microchannels.<br>* **Fluid Distribution:** Ensuring even fluid distribution between microchannels is crucial for proper device function. Accurate constitutive equations and modified flow resistance models are needed to | | | | |
| --- | --- | --- | --- | --- | --- |
| | optimize channel design.<br>*<br>**Standardizati on:** The lack of well-established standardization design rules for microfluidic chips hinders the development of widely applicable devices.<br><br><br>Let me know if you have any other questions about multiphase flow fabrication or would like more details on a specific method! | | | | |
| What is the impact of channel roughness on droplet formation and behavior? | The provided text highlights several aspects of how channel roughness influences droplet formation and behavior:<br><br>* **Direct Numerical Simulation:** The text mentions a direct numerical simulation performed the to explore the | The impact of channel roughness on droplet formation and behavior is a key factor in microfluidics. Studies indicate that surface roughness can significantly affect droplet dynamics, such as movement, spreading, and stability. For example, increased roughness can | The response provides a clear and concise summary of how channel roughness affects droplet formation and behavior. It correctly identifies various aspects such as direct numerical simulation, impact on | 76 | 76.0 |

| | | | | |
|---|---|---|---|---|
| | dynamic behavior of droplet motion across a rough channel. This suggests that roughness can significantly alter droplet movement patterns compared to smooth channels.<br><br>* **Impact on Fluid Flow:** The text states that prior studies have mainly focused on the effect of roughness on continuous fluid flow, which behaves differently from discontinuous phases like droplets. This implies that understanding droplet behavior in rough channels requires specific investigation.<br>* **Surface Roughness Measurement:** The document details a study measuring the surface roughness of microfluidic | hinder droplet spreading and prevent secondary droplet formation by altering flow dynamics and the contact line. Additionally, vibration and surface structure play roles in influencing droplet impact and splash behavior. Rougher surfaces tend to reduce spreading, while smoother surfaces may enhance it. Surface wettability also interacts with roughness to further complicate droplet behavior. Overall, channel roughness introduces complexity to droplet dynamics, with its effects varying based on factors like roughness level, fluid type, and external conditions, highlighting the need for further research in this area for better understanding and optimization in microfluidic applications. | fluid flow, surface roughness measurement, influence on splash behavior, and wettability effects. However, it does not fully capture the nuances of these relationships or provide specific examples like the ground truth answer does. The conclusion that channel roughness introduces complexity to droplet dynamics is accurate but could be more detailed.<br><br>SCORE: 76 | | |

| | channels after bonding procedures. This suggests a direct link between the manufacturing process and the resulting channel roughness, which can then influence droplet formation.

* **Influence on Splash Behavior:** The text mentions that surface roughness affects the contact line dynamics and droplet stability, ultimately influencing splash behavior. Rougher surfaces can lead to more complex and unpredictable splashing patterns.
* **Wettability Effects:** While not directly addressing roughness, the text discusses how surface wettability (the tendency of a | | | | |
| --- | --- | --- | --- | --- | --- |

| | liquid to spread on a surface) interacts with droplet impact behaviors on rough surfaces. This interplay between roughness and wettability further complicates the understanding of droplet dynamics.

**Overall, the provided text suggests that channel roughness plays a multifaceted role in droplet formation and behavior.** It can influence droplet movement, stability, splash patterns, and interactions with the surrounding environment. Further research is needed to fully understand these complex relationships and their implications for various microfluidic applications. | | | | |
|---|---|---|---|---|---|
| How can sensors be integrated into droplet microfluidic devices during fabrication? | The provided text focuses on the advantages and applications of integrating sensors into microfluidic systems, particularly highlighting droplet-based microfluidics. However, it doesn't delve into the specific fabrication techniques for integrating sensors during the manufacturing process of these devices.<br><br>To answer your question about how sensors can be integrated into droplet microfluidic devices during fabrication, we need to consider several factors:<br><br>* **Type of Sensor:** Different sensor types (optical, electrochemical, piezoelectric) require different integration methods.<br>* **Microfabricat | Integrating sensors into droplet microfluidic devices during fabrication is a complex process that requires advanced microfabrication techniques and careful material selection. For successful integration, precision in manufacturing is crucial, as it often involves combining different microfluidic units, such as droplet generation and detection units, on a single chip. Various techniques like photolithography, soft lithography, and direct integration methods can be used, depending on the type of sensor (e.g., optical or electrochemical). The fabrication process also needs to account for material compatibility, sensor functionality, and alignment to | The response provides a clear and detailed explanation of how sensors can be integrated into droplet microfluidic devices during fabrication. It covers various aspects such as sensor types, microfabrication techniques, material compatibility, general approaches for integration (co-fabrication and post-fabrication attachment), specific examples (optical and electrochemical sensors), challenges, and the need for further research. The response aligns well with the ground truth answer in | 96 | 96.0 |

| | ion Techniques:** Common techniques include photolithography, laser ablation, and soft lithography. These techniques can be used to create channels, chambers, and features for housing the sensors. * **Sensor Material Compatibility:** The sensor material must be compatible with the microfluidic materials (e.g., PDMS, PMMA) and the fluids being handled.<br><br>**General Approaches for Sensor Integration:**<br><br>1. **Co-Fabrication:** Sensors can be fabricated simultaneously with the microfluidic channels using techniques like photolithography or soft lithography. | ensure reliable performance. Additionally, sensor integration may face challenges such as maintaining stability within the microfluidic environment and minimizing interference with fluid flow. By considering these factors, sensors can be effectively integrated into droplet microfluidic devices for applications like chemical reactions and biological detection. | terms of content and structure. SCORE: 96 | | |
|---|---|---|---|---|---|

| | This ensures a seamless integration and minimizes potential leakage points. 2. **Post-Fabrication Attachment:** Sensors can be attached to the pre-fabricated microfluidic device using bonding techniques, adhesives, or through-hole connections. This approach offers flexibility in choosing different sensor types but may introduce additional fabrication steps and potential for misalignment.<br><br>**Specific Examples:**<br><br>* **Optical Sensors:** Optical fibers can be embedded within the microfluidic channels to guide laser light for droplet detection and size analysis. * | | | | |
|---|---|---|---|---|---|

| | **Electrochemical Sensors:** Electrodes can be integrated into the microfluidic channels using techniques like screen printing or sputtering. These electrodes can detect changes in electrical conductivity due to analyte presence.<br><br>**Challenges:**<br><br>* Ensuring proper alignment and sealing of sensors within the microfluidic device.<br>* Maintaining sensor functionality and stability within the microfluidic environment.<br>* Minimizing interference between the sensor signal and the fluid flow.<br><br><br>Further research into specific fabrication techniques and | | | | |
| --- | --- | --- | --- | --- | --- |

| | materials is necessary to provide a more detailed answer to your question. | | | | |
|---|---|---|---|---|---|
| How are droplet microfluidic devices used for single-cell analysis? | Droplet microfluidic devices are revolutionizing single-cell analysis by offering a unique platform to isolate and manipulate individual cells within microscale droplets.<br><br>Here's how they work:<br><br>1. **Encapsulation:** Cells are encapsulated individually within oil-based droplets, each acting as an isolated reaction chamber. This prevents cell-to-cell interaction and allows for the study of individual cellular responses.<br>2. **Manipulation:** These droplets can be precisely | Droplet microfluidic devices are widely used for single-cell analysis due to their ability to encapsulate individual cells within tiny droplets, allowing for high-throughput and precise experimentation. These devices work by generating microdroplets in a fluid flow, each containing a single cell along with reagents needed for analysis, such as enzymes, antibodies, or primers. This allows for the isolation of cells in a controlled environment, enabling high-resolution studies of gene expression, protein profiling, and cellular responses. The compartmentalized nature of droplets also | The response provides a clear and concise overview of how droplet microfluidic devices are used for single-cell analysis. It accurately describes the process of encapsulation, manipulation, and analysis within these devices, as well as their advantages in terms of high throughput, single-cell resolution, sensitivity, and compatibility with automation. However, it lacks some details and specific examples compared to the ground | 76 | 76.0 |

| | manipulated using microfluidic channels, enabling researchers to perform various assays on individual cells. 3. **Analysis:** A range of techniques can be applied within these droplets, including:<br><br>    * **Genetic analysis:** Single-cell RNA sequencing (scRNA-seq) allows for the study of gene expression patterns at the individual cell level.<br>    * **Protein analysis:** Techniques like immunofluoresc ence can be used to visualize protein expression and localization within single cells.<br>    * **Functional assays:** Droplets can be used to assess cellular | reduces cross-contamination between samples. Moreover, the small size and scalability of droplet microfluidics make it possible to perform large-scale, parallel analysis of thousands of single cells in a short amount of time. This technology has applications in genomics, proteomics, and drug screening, facilitating insights into cell heterogeneity and the identification of rare cell populations. | truth answer, which provides a more comprehensi ve explanation of the technology's capabilities and applications. SCORE: 76 | | |

| | responses to drugs, stimuli, or other factors by measuring changes in cell morphology, viability, or secreted molecules.<br><br>**Advantages of Droplet Microfluidics for Single-Cell Analysis:**<br><br>* **High throughput:** Thousands of cells can be analyzed simultaneously, enabling large-scale studies and the identification of rare cell populations.<br>* **Single-cell resolution:** Provides detailed insights into cellular heterogeneity and individual cell behavior.<br>* **Sensitivity:** Allows for the detection of subtle changes in gene expression, protein levels, or cellular function. | | | | |
| | *<br>**Compatibility with automation:** Droplet microfluidic platforms can be integrated with robotic systems and AI algorithms for high-throughput screening and data analysis.<br><br>Overall, droplet microfluidics is a powerful tool for single-cell analysis, enabling researchers to unravel the complexities of cellular behavior and heterogeneity at an unprecedented level of detail. | | | | |
| What are the advantages of droplet microfluidics for high-throughput screening? | Droplet microfluidics offers several key advantages for high-throughput screening:<br><br>* **Reduced Sample Volumes:** Droplets allow for significantly smaller sample volumes | Droplet microfluidics offers several advantages for high-throughput screening, including reduced sample volumes, which lowers costs, and increased assay sensitivity through compartmentalization. This method allows for the | The response accurately captures the key advantages of droplet microfluidics for high-throughput screening, including reduced sample volumes, increased | 86 | 86.0 |

| | | | | |
|---|---|---|---|---|
| | compared to conventional methods, leading to substantial cost savings.<br>* **High Throughput:** Manipulation and measurement at kilohertz speeds enable the screening of up to $10^8$ samples in a single day. This dramatically increases the speed and efficiency of screening processes.<br>* **Increased Sensitivity:** Compartmentalization within droplets enhances assay sensitivity by:<br>    * Increasing the effective concentration of rare species.<br>    * Decreasing the time required to reach detection thresholds.<br><br>These advantages make droplet microfluidics particularly well-suited for | screening of up to $10^8$ samples per day at kilohertz speeds, significantly improving throughput and efficiency. The small droplet volumes enhance effective concentrations and signal-to-noise ratios, critical for sensitive detection of low-concentration biological molecules such as antibodies, enzymes, and single genomes. Additionally, droplet microfluidics enables single-cell and single-molecule assays, providing detailed insights into cellular behaviors and allowing for novel experimental approaches. This technology is particularly beneficial for applications like drug discovery, enzyme evolution, and diagnostics, offering a versatile and cost-effective platform for advanced | sensitivity through compartmentalization, and high throughput. It also highlights the technology's potential for single-cell and single-molecule assays, as well as its versatility in various applications like drug discovery and diagnostics. The response is clear, concise, and well-structured, with a good balance of detail and concision.<br>SCORE: 86 | | |

| | | | | |
|---|---|---|---|---|
| | applications like drug discovery, where large compound libraries need to be screened rapidly and efficiently, as well as diagnostic assays that require detecting low concentrations of pathogens.<br><br>Furthermore, the text highlights that droplet microfluidics enables:<br><br>* **Single-cell and single-molecule assays:** The small volume and compartmentaliz ation allow for analysis at the individual cell or molecule level.<br>* **New types of experiments:** Droplet microfluidics opens up possibilities for novel research approaches beyond traditional high- | biological research. | | | |
| | throughput screening methods. Overall, droplet microfluidics presents a powerful tool for advancing biological research and applications by combining high throughput, sensitivity, and versatility. | | | | |
| How can droplets be used to encapsulate cells or biomolecules? | Droplets offer a versatile platform for encapsulating both cells and biomolecules. **For cell encapsulation:** * **Limiting dilution:** This traditional method relies on randomly distributing single cells within droplets, with most drops remaining empty. While simple, it results in low encapsulation efficiency. * **Inertial microfluidics:** Newer | Droplet microfluidics provides versatile methods for encapsulating cells and biomolecules. Limiting dilution is a traditional approach that randomly distributes single cells or biomolecules in droplets, though it has low efficiency. Inertial microfluidics leverages spiral geometries and inertial forces to guide cells into droplets, improving encapsulation efficiency by ensuring single-cell droplets. Bi-disperse | The response provides a clear and concise overview of the methods used for encapsulating cells and biomolecules using droplets, including limiting dilution, inertial microfluidics, and bi-disperse emulsions. It accurately describes the advantages and limitations of each method, as well as their specific applications. | 86 | 86.0 |

| | | | | |
|---|---|---|---|---|
| | techniques leverage inertial forces to guide cells into droplets, achieving higher encapsulation efficiency. Spiral geometries and Dean forces are used to order cells before they are encapsulated, ensuring that most droplets contain a single cell. * **Bi-disperse emulsions:** This approach creates an emulsion with small, empty droplets and larger droplets containing single cells. The cell-containing drops can then be selectively recovered using passive methods like pinched flow fractionation.<br><br>**For biomolecule encapsulation:**<br><br>* **Limiting dilution:** Due to their small | emulsions create an emulsion of empty and cell-containing droplets, with the latter selectively recovered using methods like pinched flow fractionation. For biomolecules, limiting dilution is commonly used, adjusting concentration and drop size to achieve single-molecule encapsulation. Additionally, droplets can encapsulate cells in lipid vesicles for biological applications or serve as microreactors in processes like DNA analysis. These techniques enable precise control over encapsulation, enhancing their applicability in cell-based assays, drug discovery, and genetic research | However, it does not fully elaborate on the broader implications or potential uses of these techniques beyond what is mentioned in the ground truth answer. SCORE: 86 | | |
| | size, biomolecules cannot be passively organized or trigger jet breakup. Limiting dilution is the only method used, where the concentration and drop size are carefully tuned to achieve a desired fraction of drops containing single molecules.<br><br>Each method has its advantages and limitations depending on the specific application. | | | | |
| What are the applications of droplet microfluidics in drug discovery? | Droplet microfluidics offers several key applications in drug discovery, as detailed in the provided text:<br><br>* **High-Throughput Screening:** Droplet technology enables the simultaneous | Droplet microfluidics plays a significant role in drug discovery by enabling high-throughput screening and precise analysis of drug interactions at the single-cell level. This technology allows the generation of picoliter-sized droplets, each | The response is well-structured and provides specific examples related to the applications of droplet microfluidics in drug discovery. It covers key aspects such as high-throughput | 76 | 76.0 |

| | | | | |
|---|---|---|---|---|
| | analysis of thousands of cells in a single experiment, significantly accelerating the screening process for potential drug candidates.<br>* **Single-Cell Analysis:** By encapsulating individual cells within droplets, researchers can observe their unique responses to drugs and other stimuli. This allows for the identification of subpopulations of cells with specific sensitivities or resistances to treatment, providing valuable insights into cellular heterogeneity.<br>* **Drug Dissociation Assays:** Microfluidic devices can be used to evaluate the release rate and extent of drugs from microparticle- and nanoparticle- | encapsulating a single cell or a small group of cells, which can be exposed to various drug compounds. This compartmentalizat ion facilitates parallel testing of numerous drug candidates, speeding up the identification of promising therapeutics. Droplet microfluidics is particularly useful in identifying drug efficacy, optimizing drug formulations, and studying complex cellular responses such as gene expression, protein production, or metabolic changes in response to treatment. Additionally, it enables the screening of rare cell populations or targets that would be difficult to isolate using traditional methods. The ability to rapidly test and analyse thousands of drug interactions in a controlled | screening, single-cell analysis, and drug dissociation assays, which align with the ground truth answer. However, it lacks a comprehensi ve overview and some details are not explicitly mentioned in the provided text.<br>SCORE: 76 | | |

| | based delivery systems. This is crucial for ensuring the quality and standardization of drug formulations.

**Specific Examples Mentioned in the Text:**

*  Evaluating the cytotoxicity of chemotherapeutic agents and other drug therapies on individual cells.
*  Analyzing the interactions between encapsulated cells, including their response to controlled stimuli.
*  Performing high-throughput fluorescence imaging within droplets to assess cytotoxicity effectively.

Overall, droplet microfluidics empowers researchers with a powerful toolset for | environment accelerates drug discovery, reduces costs, and improves the chances of identifying novel therapeutic agents. | | | |
|---|---|---|---|---|

| | | | | | |
|---|---|---|---|---|---|
| | accelerating drug discovery by enabling high-throughput screening, single-cell analysis, and precise control over experimental conditions. | | | | |
| How can droplet microfluidics enable precise control of chemical reactions? | Droplet microfluidics enables precise control of chemical reactions in several ways:<br><br>* **Precise Volume Control:** Reactants are dispensed into individual droplets with very precise volume control. This ensures that the reactants are mixed in the desired stoichiometric ratios, leading to more reproducible and predictable reaction outcomes.<br>* **Fast Mixing:** Droplets coalesce (merge) inside | Droplet microfluidics offers precise control over chemical reactions by dispensing reactants into individual droplets with accurate volume control. This enables the mixing of reactants within confined spaces, enhancing reaction efficiency. Coalesced droplets act as micro-reactors, providing fast mixing and long operation times without clogging, even with sticky or precipitating reactants. The system allows for optimization of reaction parameters, such as temperature, flow rate, and pressure, ensuring | The response accurately addresses the question by providing detailed explanations of how droplet microfluidics enables precise control of chemical reactions. It covers key aspects such as volume control, fast mixing, individual micro-reactors, and controlled environment, which are all relevant to the topic. However, it could be improved with more concise language and | 83 | 83.0 |

| | the microfluidic device, rapidly mixing the reactants within the confined space. This accelerates the reaction rate compared to traditional batch reactions.<br>* **Individual Micro-reactors:** Each droplet acts as an individual micro-reactor, allowing for parallel processing of multiple reactions simultaneously. This increases throughput and efficiency.<br>* **Controlled Environment:** The microfluidic device can be designed to control various parameters like temperature, pressure, and flow rate within each droplet. This allows for fine-tuning of reaction conditions to optimize product formation. | the production of specific products, like mesoporous silica microspheres with superior surface area and narrow pore distributions. This capability to finely control reaction conditions makes droplet microfluidics a powerful tool for precise, high-throughput chemical synthesis. | a clearer connection to the specific applications mentioned in the ground truth answer (e.g., production of mesoporous silica microspheres ). SCORE: 83 | | |
|---|---|---|---|---|---|
| | The combination of these factors makes droplet microfluidics a powerful tool for performing precise and controlled chemical reactions, particularly for applications requiring small volumes or complex reaction schemes. | | | | |
| What are the key challenges of using droplet microfluidics in diagnostics? | Droplet microfluidics offers exciting possibilities for diagnostics, but several key challenges need to be addressed:<br><br>**1. Wetting and Cross-Contamination:**<br><br>* Droplets are aqueous, and the continuous phase is immiscible organic liquid. Maintaining a stable oil-water interface is crucial to prevent wetting | Droplet microfluidics presents significant challenges in diagnostics, including preventing wetting and cross-contamination between aqueous droplets and immiscible organic liquids, which can compromise results. Monitoring oxygen levels within droplets is difficult due to the sensitivity of the liquid-liquid interface to mechanical | The response provides a clear and detailed explanation of the key challenges associated with using droplet microfluidics in diagnostics. It addresses four specific areas of concern (wetting and cross-contamination, oxygen control, stable detection chemistries, | 94 | 94.0 |

| | and cross-contamination between droplets. Finding the right surfactant for this purpose can be challenging.<br><br>**2. Oxygen Control:**<br><br>* Monitoring and controlling the level of oxygen within the droplets is essential for many biological assays. This requires careful design of the microfluidic system and potentially the use of specialized gas permeable membranes.<br><br>**3. Stable Detection Chemistries:**<br><br>* Developing reliable detection chemistries that work effectively within the confined space of a droplet is important. These chemistries need to be compatible | perturbations and mass transfer. Developing stable detection chemistries is also complex, as the confined droplet environment can alter chemical reactions and impact accuracy. Additionally, challenges in automation and accessibility limit the widespread use of droplet microfluidics, particularly for high-throughput assays. Overcoming these issues requires advancements in surfactants, microfluidic engineering, and automated techniques, which could enhance the precision and applicability of droplet systems in diagnostics. | and microfluidic engineering challenges) and offers relevant examples and context for each point. The answer is well-structured, concise, and easy to follow, with a logical flow that effectively conveys the importance of overcoming these challenges for broader adoption in diagnostics. SCORE: 94 | | |
| --- | --- | --- | --- | --- |
| | with the droplet environment and sensitive enough for accurate measurements.<br><br>**4. Microfluidic Engineering Challenges:**<br><br>* **Automation:** Developing automated techniques for manipulating droplets within microfluidic chips in situ is crucial for high-throughput applications.<br>* **Accessibility:** Making droplet microfluidics more accessible to users without a background in microfluidics is essential for wider adoption. This includes simplifying the fabrication and operation of devices.<br><br>Overcoming these challenges will significantly | | | |

| | broaden the use of droplet systems in diagnostics, enabling faster, more sensitive, and cost-effective assays. | | | | |
|---|---|---|---|---|---|
| How are droplets used for DNA sequencing applications? | Droplets play a crucial role in DNA sequencing applications, particularly in high-throughput single-cell sequencing. Here's how:<br><br>1. **Encapsulation of Single Targets:** Droplets can encapsulate individual biological targets like cells, bacteria, or viruses. This isolation allows for processing each target independently and accurately.<br><br>2. **Barcoding:** Unique nucleic acid barcodes are linked to DNA and RNA molecules within the droplets. These barcodes act as | Droplet microfluidics plays a key role in DNA sequencing by enabling high-throughput analysis and precise targeting of genomic regions. This includes single-cell sequencing, where individual cells are encapsulated in droplets for independent gene expression profiling. Barcode-based sequencing allows for the labeling of nucleic acids with unique identifiers, enabling the tracking of specific genomes during analysis. Multiplexed PCR amplification is achieved by performing PCR in separate droplets, ensuring uniform amplification of multiple targets. Additionally, | The response provides a clear and concise explanation of how droplets are used in DNA sequencing applications, covering key points such as encapsulation, barcoding, high-throughput sequencing, and single-cell analysis. It accurately conveys the benefits of using droplets for precise manipulation and labeling of individual DNA samples. However, it lacks some details mentioned in the ground truth answer, specifically | 76 | 76.0 |

| | | | | |
|---|---|---|---|---|
| | identifiers, allowing researchers to track the origin of specific sequences during analysis.

3. **High-Throughput Sequencing:** By encapsulating barcoded targets in droplets, a large number of samples can be processed simultaneously. This enables high-throughput sequencing, where millions or even billions of individual genomes can be analyzed.

4. **Single-Cell Analysis:** Droplet microfluidics allows for the analysis of gene expression at the single-cell level. By identifying barcodes linked to specific cells, researchers can understand cellular heterogeneity and variations in gene activity. | emulsion PCR combines droplet merger with PCR to amplify specific genomic regions, facilitating efficient sequencing. These techniques offer a miniaturized, efficient platform for analyzing complex genetic information, improving accuracy and throughput in genomic studies. | regarding multiplexed PCR amplification and emulsion PCR. SCORE: 76 | | |

| | | | | | |
|---|---|---|---|---|---|
| | Essentially, droplets provide a miniaturized environment for individual DNA samples, enabling precise manipulation, labeling, and analysis for advanced sequencing applications. | | | | |
| What role do droplets play in tissue engineering research? | Droplets play a crucial role in tissue engineering research, particularly in the fabrication of biomaterials and 3D cell culture systems.<br><br>Here's how:<br><br>* **Precise Material Fabrication:** Droplet microfluidics allows for the precise creation of hydrogel-based materials with tailored size, shape, and composition. These droplets act as templates, enabling the formation of microcarriers or encapsulated cells within a | Droplet microfluidics plays a pivotal role in tissue engineering by enabling precise fabrication of biomaterials and 3D cell culture systems. Droplets serve as templates for creating hydrogel-based materials, which can encapsulate cells for more accurate modeling of tissue structures. This technology facilitates the production of microcarriers and complex biomaterials, such as hollow bacterial cellulose (BC) microspheres, which form injectable scaffolds for wound healing, | The response is clear, concise, and directly addresses the question with full relevance. It provides specific examples and details about the role of droplets in tissue engineering research, aligning well with the ground truth answer. SCORE: 85 | 85 | 85.0 |

| | controlled environment. * **3D Cell Culture:** Cells can be directly encapsulated within these droplet-generated hydrogels, forming 3D cell culture systems that mimic natural tissue structures more closely than traditional 2D cultures. This allows for better understanding of cell behavior and interactions in a three-dimensional context. * **Tissue Building Blocks:** The fabricated hydrogel materials can serve as building blocks for constructing complex 3D tissue architectures. By assembling these microcarriers or encapsulated cells, researchers can create intricate tissue models for various | promoting high cell proliferation and tissue regeneration in vivo. Additionally, droplet-based microfluidics allows for the creation of core-shell structures, such as alginate shells or metal-organic frameworks, which offer controlled release of therapeutic agents. These advancements provide a powerful platform for studying tissue development and regenerative therapies. | | | |
|---|---|---|---|---|---|

| | applications. | | | | |
| --- | --- | --- | --- | --- | --- |
| | The text provides specific examples: | | | | |
| | * **Wound Healing:** Hollow bacterial cellulose (BC) microspheres generated via droplets were used to create injectable scaffolds for wound healing. These scaffolds promoted cell proliferation and accelerated wound closure in rat skin models. | | | | |
| | * **Core-Shell Structures:** Researchers have utilized droplet microfluidics to create core-shell structures with alginate shells encapsulating metal-organic frameworks. These microcapsules can release ions in a controlled manner, offering potential applications in antibacterial and antioxidant | | | | |
| | therapies.<br><br>Overall, droplet microfluidics offers a powerful tool for tissue engineering research by enabling precise control over material fabrication and 3D cell culture environments, leading to advancements in understanding tissue development and regeneration. | | | | |
| How can droplet microfluidics aid in studying cell signaling pathways? | Droplet microfluidics offers a powerful tool for studying cell signaling pathways by enabling the isolation and controlled manipulation of individual cells within discrete droplets.<br><br>Here's how it aids in this research:<br><br>* **Single-Cell Resolution:** Droplets can encapsulate | Droplet microfluidics offers a powerful platform for studying cell signaling pathways by enabling the precise isolation and manipulation of individual cells within controlled environments. This technology allows for the encapsulation of single cells in droplets, providing a controlled space to study cell-cell interactions and signaling events. | The response provides a clear and concise explanation of how droplet microfluidics aids in studying cell signaling pathways, highlighting its benefits such as single-cell resolution, controlled environment, and cell-cell communication studies. It also includes | 76 | 76.0 |

| | single cells, allowing researchers to study signaling events at the individual cell level, overcoming the limitations of bulk analysis which averages responses across a population of cells. This is crucial because cells within a population can exhibit significant heterogeneity in their signaling behavior.<br>* **Controlled Environment:** Each droplet acts as a microenvironment where specific signaling molecules and conditions can be precisely controlled. Researchers can introduce different stimuli or inhibitors into individual droplets to investigate how they influence signaling pathways.<br>* **Cell-Cell Communication | The ability to merge droplets containing different cell types or reagents facilitates the investigation of cell communication, such as paracrine signaling, in a way that mimics in vivo conditions. Additionally, droplet-based microfluidics supports high-throughput screening and single-cell genomic studies, offering insights into phenotypic variations and signaling dynamics within genetically identical cell populations. This capability is particularly valuable for understanding complex cellular behaviors in development, differentiation, and disease. | relevant quotes from the provided text to support the points made. However, it lacks some depth and detail compared to the ground truth answer, particularly regarding high-throughput screening and single-cell genomic studies.<br>SCORE: 76 | | |
|---|---|---|---|---|

| | Studies:** Droplets can be merged, bringing cells into close proximity without direct contact. This allows for the study of paracrine signaling, where cells communicate through secreted factors, mimicking in vivo conditions more accurately.<br><br>The provided text highlights these benefits:<br><br>* "This functionality may provide a new approach for studying cell signaling at the single cell level, offering potential for a wide range of studies involving cell–cell communication and paracrine signaling..."<br>* "...the droplets put the cells in close vicinity, offering potential for a wide range of | | | | |
|---|---|---|---|---|---|

| | Studies:** Droplets can be merged, bringing cells into close proximity without direct contact. This allows for the study of paracrine signaling, where cells communicate through secreted factors, mimicking in vivo conditions more accurately.<br><br>The provided text highlights these benefits:<br><br>* "This functionality may provide a new approach for studying cell signaling at the single cell level, offering potential for a wide range of studies involving cell–cell communication and paracrine signaling..."<br>* "...the droplets put the cells in close vicinity, offering potential for a wide range of | | | | |
|---|---|---|---|---|---|
| | cell–cell communication studies involving contact-independent secretory mechanisms such as paracrine signaling."<br><br>In summary, droplet microfluidics provides a unique platform to dissect complex cell signaling pathways with high precision and control, enabling researchers to gain deeper insights into cellular communication and its role in various biological processes. | | | | |
| What is the potential of droplet microfluidics in personalized medicine? | Droplet microfluidics holds significant potential for advancing personalized medicine.<br><br>Here's how, based on the provided | Droplet microfluidics holds great promise in personalized medicine by enabling precise, high-throughput analyses of biological samples. The | The response provides a clear and concise overview of the potential of droplet microfluidics in personalized medicine, | 56 | 56.0000000 0000001 |

| | | | | |
|---|---|---|---|---|
| | context:<br><br>* **Drug Discovery and Validation:** Droplet microfluidics allows for precise testing of drug combinations on individual cells, enabling the identification of effective treatments tailored to a patient's specific tumor profile. This is particularly valuable for pancreatic cancer, where personalized treatment strategies are crucial due to the disease's complexity.<br>* **Single-Cell Analysis:** The ability to analyze individual cells within droplets allows researchers to understand the heterogeneity of tumors and identify unique cellular characteristics that can guide personalized | technology allows for the optimization of treatments by testing drug combinations on specific cell lines, such as pancreatic cancer cells, to identify the most effective therapies. It also facilitates the early detection of cancers through highly sensitive detection of tumor markers in bodily fluids, offering a minimally invasive approach for patient monitoring. Additionally, droplet-based digital PCR (dPCR) enables the detection of rare sequences and precise quantification of nucleic acids, which is crucial for cancer research and tailoring treatments based on a patient's genetic background and tumor profile. Overall, droplet microfluidics can significantly advance personalized | highlighting its applications in drug discovery, single-cell analysis, and early cancer detection. It also summarizes these points into three key benefits (optimizing drug selection, understanding tumor heterogeneity, enabling early disease detection). However, it lacks specific details and technical depth compared to the Ground Truth Answer, which provides more comprehensive information on droplet microfluidics' capabilities and its potential impact on personalized | | |

| | | | | |
|---|---|---|---|---|
| | therapies.<br>* **Early Cancer Detection:** Droplet microfluidics enables highly sensitive detection of tumor markers in bodily fluids like blood, urine, or feces. This minimally invasive approach could facilitate early cancer detection, allowing for timely intervention and improved patient outcomes.<br><br>**In essence, droplet microfluidics empowers personalized medicine by:**<br><br>1. **Optimizing drug selection:** Identifying the most effective drug combinations for individual patients.<br>2. **Understanding tumor heterogeneity:** | medicine by providing tailored treatment strategies and improving early detection for better patient outcomes. | medicine.<br><br>SCORE: 56 | | |
| | Analyzing the unique characteristics of individual cells within a tumor.<br>3. **Enabling early disease detection:** Detecting cancer at earlier stages through sensitive analysis of bodily fluids.<br><br>The context highlights these points, emphasizing the potential of droplet microfluidics to revolutionize personalized medicine by providing precise and targeted approaches to diagnosis and treatment. | | | | |
| How are droplet microfluidic devices used for nanoparticle synthesis? | Droplet microfluidic devices offer a unique and powerful approach to nanoparticle synthesis due to their ability to precisely control reaction conditions within spatially | Droplet microfluidic devices revolutionize nanoparticle synthesis by enabling precise control over reaction conditions within isolated microreactors. These droplets | The response is clear, concise, and directly addresses the question about how droplet microfluidic devices are used for nanoparticle synthesis. It | 96 | 96.0 |

| | | | | |
|---|---|---|---|---|
| isolated microreactors (droplets).<br><br>Here's how they work:<br><br>1. **Reagent Encapsulation:** Reagents are brought together in a co-flowing stream just before droplet formation. Each droplet acts as a miniature reactor, containing well-defined quantities of materials.<br>2. **Fast Mixing and Controlled Environment:** The advection flow field within the droplets accelerates mixing, ensuring rapid and homogeneous reaction initiation. This confinement also prevents contamination from channel walls and allows for the use of toxic or volatile chemicals.<br>3. **Precise Control over | encapsulate well-defined quantities of reagents, ensuring rapid mixing and preventing contamination, even when using toxic or volatile chemicals. This approach addresses limitations like diffusion-limited mixing found in traditional systems, resulting in nanoparticles with controlled size, shape, and narrow size distribution. The method supports multi-step synthesis on millisecond timescales, allowing for the scalable production of high-quality nanoparticles with applications in fields such as drug delivery, imaging, and catalysis. | provides a detailed explanation of the process, advantages, and applications, aligning well with the ground truth answer. SCORE: 96 | | |

| | Synthesis Conditions:** Droplet-based systems enable fine-tuning of parameters like temperature, pH, and reagent concentrations within each droplet, leading to controlled nanoparticle size, shape, and crystal structure.<br><br>**Advantages of Droplet Microfluidics for Nanoparticle Synthesis:**<br><br>* **Miniaturization:** Reactions occur in tiny droplets, reducing reagent consumption and enabling high throughput synthesis.<br>* **Enhanced Mass and Heat Transfer:** The small droplet size promotes rapid mass and heat transfer, accelerating reaction kinetics.<br>* **Reduced Time and Reagent Consumption:** | | | | |
| | Precise control over reaction conditions minimizes waste and shortens synthesis times. * **Narrow Size Distribution:** Controlled mixing and residence time within droplets lead to a narrower size distribution of nanoparticles. **Applications:** Droplet microfluidics has been successfully used to synthesize a wide range of nanoparticles, including inorganic (CdS, CdSe), organic, and hybrid materials. These nanoparticles find applications in diverse fields such as biomedical imaging, drug delivery, catalysis, and electronics. | | | | |
| What role do droplets play in | Droplets serve as the fundamental | Droplet-based microfluidics provides a | The response accurately describes the | 85 | 85.0 |

| polymer particle production? | building blocks for polymer particle production in droplet-based microfluidics.

Here's a breakdown of their role:

* **Templates:** Monomers or oligomers are emulsified into droplets, breaking them down into tiny liquid threads. These droplets act as individual templates for forming polymer particles.
* **Controlled Environment:** The confined space within each droplet allows for precise control over the reaction environment, influencing factors like concentration and mixing.
* **Solidification:** Chemical and physical crosslinking processes solidify the | versatile platform for producing monodisperse polymer particles by encapsulating monomers or oligomers into droplets and solidifying them through chemical or physical crosslinking. This approach enables the creation of diverse particle types, including solid, hollow, asymmetric, and multicompartment al structures with precise control over size, shape, and morphology. Techniques such as UV-initiated polymerization and solvent evaporation facilitate the transformation of droplets into solid particles. Additionally, functional micro- or nanostructures can be incorporated within or on the surface of these particles by introducing specific ingredients into the droplets, expanding their potential | role of droplets in polymer particle production, providing a clear explanation of how they serve as templates, control the reaction environment, and undergo solidification. It also highlights the influence of droplet size and flow parameters on the final particle properties. However, it lacks some details present in the ground truth answer regarding the diversity of particle types that can be created and their potential applications. SCORE: 85 | | |

| | | | | | |
|---|---|---|---|---|---|
| | droplets, transforming them into solid polymer particles. This can be achieved through various methods like UV-initiated polymerization or solvent evaporation.<br><br>The size and shape of the final polymer particles are directly influenced by the droplet size and flow parameters used in the microfluidic system.<br><br>Essentially, droplets provide a highly controlled and scalable platform for generating monodisperse polymer particles with diverse properties and morphologies. | applications in areas like drug delivery, tissue engineering, and advanced materials. | | | |
| How can droplet microfluidics improve the reproducibility of | Droplet microfluidics significantly improves the reproducibility | Droplet microfluidics enhances the reproducibility of chemical synthesis | The response accurately captures the key benefits of droplet | 86 | 86.0 |
| chemical synthesis? | of chemical synthesis through several key mechanisms:<br><br>* **Accurate Manipulation:** Droplets offer precise control over reagent volumes and mixing conditions, eliminating variations often encountered in traditional batch reactors.<br>* **Better Reproducibility:** The defined environment within each droplet ensures consistent reaction parameters, leading to more uniform product formation and reducing batch-to-batch variability.<br>* **Reliable Automation:** Microfluidic systems can be readily automated, minimizing human error and ensuring consistent execution of synthesis | by providing precise control over reaction conditions and minimizing variability. Droplets act as isolated microreactors, ensuring accurate manipulation of parameters such as temperature and concentration, critical for consistent results. This system enables optimal mixing intensity and parallel processing in separate droplet channels, reducing batch-to-batch variability and ensuring uniform product quality. Automation further minimizes human error, making the process highly reliable. These advantages are particularly beneficial for synthesizing complex nanomaterials and exploring new reaction parameter spaces, leading to the production of high-quality, reproducible materials. | microfluidics in improving chemical synthesis reproducibility, including precise control over reagent volumes, consistent reaction parameters, and reliable automation. However, it could be more concise and directly address the question without introducing a separate statement that is not part of the main answer. SCORE: 86 | | |
| | protocols.<br><br>The text highlights that "Droplet reactors provide obvious advantages over conventional batch reactors, such as accurate manipulation, better reproducibility, and reliable automation." This suggests that the controlled nature of droplet microfluidics directly addresses the challenges associated with achieving reproducible results in traditional chemical synthesis methods. | | | | |
| What are the applications of droplet microfluidics in catalysis studies? | Droplet microfluidics offers several promising applications for catalysis studies.<br><br>Here's a breakdown based on the provided | Droplet microfluidics offers transformative applications in catalysis studies by enabling precise, efficient, and high-throughput investigations at | The response provides a detailed breakdown of the applications of droplet microfluidics in catalysis studies, including | 76 | 76.0 |

| | | |
|---|---|---|
| context:<br><br>* **High-Throughput Screening:** Droplets allow for the rapid screening of thousands of reaction conditions, gene mutations, and chemicals in a high-throughput manner with low sample/reagent consumption. This is particularly useful for catalyst discovery and optimization.<br>* **Single-Particle Analysis:** The ability to encapsulate individual catalysts or catalytic reactions within droplets enables single-particle analysis. This allows researchers to study the behavior of individual catalysts and understand their activity at a fundamental level. | the nanoscale. These systems facilitate enantioselective biocatalysis using nanoliter-scale devices that integrate microflow reactors and electrophoresis for enantiomeric differentiation. Droplets serve as isolated reaction containers, allowing for controlled catalytic transformations at the single-particle or cell level, with reaction volumes as small as a few picoliters. This confinement minimizes diffusion effects and enhances reaction efficiency. Applications span organic synthesis, biochemical analysis, and biotechnology, showcasing droplet microfluidics as a powerful tool for studying rapid reaction dynamics, optimizing catalysts, and exploring reaction | high-throughput screening, single-particle analysis, controlled reaction environment, and rapid reaction dynamics. It also mentions specific examples such as protein crystallization screening, drug screening, catalyst screening, and directed evolution screening. However, it lacks the depth and precision of the ground truth answer, which provides a more comprehensive overview of the transformative applications of droplet microfluidics in catalysis studies. |

| | | | | |
|---|---|---|---|---|
| | * **Controlled Reaction Environment:** Droplets provide a confined reaction environment that minimizes diffusion and dilution effects, allowing for precise control over reaction conditions. This is crucial for studying sensitive catalytic processes.<br><br>* **Rapid Reaction Dynamics:** Enhanced mixing and mass transfer within droplets facilitate the study of rapid reaction dynamics. Researchers can monitor catalytic reactions in real-time and gain insights into the mechanisms involved.<br><br>**Specific Examples Mentioned:**<br><br>* The text mentions | mechanisms in real time.<br><br>SCORE: 76 | | |

| | droplet microfluidics being used for protein crystallization screening, drug screening, catalyst screening, and directed evolution screening. These applications highlight the versatility of the technology for various catalytic processes.<br><br>\* The use of droplet microfluidics in studying enantioselective biocatalysis at the nanoliter scale demonstrates its potential for precise control and analysis of catalytic reactions.<br><br>\*\*Future Directions:\*\*<br><br>The text suggests that further downsizing reaction volumes to a few picoliters could enable investigations into selective | | | | |
|---|---|---|---|---|---|
| | catalytic transformations at the single particle or cell level, opening up new avenues in catalysis research.

Overall, droplet microfluidics presents a powerful toolset for advancing catalysis studies by enabling high-throughput screening, single-particle analysis, precise control over reaction conditions, and real-time monitoring of rapid reactions. | | | | |
| How do droplets enhance the study of reaction kinetics? | Droplets enhance the study of reaction kinetics in several ways:

* **Rapid Mixing:** Droplet-based microfluidics often induce chaotic flow within the droplets, leading to extremely rapid mixing of reactants. This allows for the | Droplet-based microfluidics revolutionizes reaction kinetics studies by offering rapid, precise, and high-throughput analysis with minimal sample consumption. The chaotic mixing within picoliter droplets ensures millisecond-level time resolution for fast reactions. These systems minimize reagent | The response is clear and concise, accurately addressing the question by highlighting four key benefits of using droplet-based microfluidics in studying reaction kinetics. It provides | 83 | 83.0 |

| | | | | |
|---|---|---|---|---|
| | study of fast reactions that would be difficult to observe with conventional methods.<br>* **Small Volume Consumption:** Droplets are incredibly small (nanoliter to picoliter volumes), significantly reducing the amount of sample and reagents required for an experiment. This is particularly beneficial for bioanalysis where reagents can be expensive or in limited supply.<br>* **Precise Control:** Droplet-based systems allow for precise control over the composition and concentration of reactants within each droplet. This enables the creation of concentration gradients and facilitates the study of reaction | use, making them ideal for bioanalysis where resources are costly or scarce. Precise control over droplet composition facilitates the creation of concentration gradients, enabling the measurement of kinetic parameters. High-throughput capabilities allow simultaneous monitoring of numerous reactions, significantly expanding experimental diversity. Additionally, extended tracking of individual droplets supports long-term monitoring of enzyme-inhibitor kinetics, providing deeper insights into reaction mechanisms. | specific examples and details to support each point, demonstrating a good understanding of the topic. However, it could be improved with more comprehensive coverage of the subject matter, as the ground truth answer includes additional points such as extended tracking of individual droplets for long-term monitoring of enzyme-inhibitor kinetics.<br>SCORE: 83 | | |

| | kinetics under a wide range of conditions.<br>* **High Throughput:** Droplets can be generated at high frequencies (hundreds to thousands per second), allowing for the simultaneous analysis of numerous reactions. This significantly increases the throughput of kinetic experiments.<br><br>The combination of these factors makes droplet-based microfluidics a powerful tool for studying reaction kinetics, particularly for fast reactions and bioanalytical applications. | | | | |
|---|---|---|---|---|---|
| What are the benefits of using droplets for emulsion polymerization? | The provided text highlights several benefits of using droplets in emulsion polymerization: | Droplet-based systems offer significant advantages for emulsion polymerization, | The answer is mostly relevant to the question but lacks focus on the | 38 | 38.0 |

| | | | | |
|---|---|---|---|---|
| | * **High Molar Mass Polymer Chains:** Emulsion polymerization allows for the production of polymer chains with high molar mass at relatively fast reaction rates compared to other free-radical polymerization methods like bulk, solution, or suspension.<br>* **Controlled Particle Size and Morphology:** Emulsion polymerization offers precise control over the size and morphology of the resulting polymer particles (typically submicron). This is crucial for tailoring the properties of the final product.<br>* **Versatility in Monomer Selection:** A wide range of monomers can be used in emulsion polymerization, | enhancing efficiency, precision, and scalability. They enable high production rates and precise control over polymer particle size and morphology, facilitating the synthesis of high-molar-mass polymers at faster reaction rates. Droplet parallelization allows the screening of diverse chemicals or concentrations and supports the encapsulation of millions of gene variants for quantitative emulsion PCR. These systems also empower directed enzyme or microorganism evolution through successive mutagenesis and selection rounds. Furthermore, droplet isolation prevents cross-contamination, adsorption, and solvent evaporation, making them ideal for applications like in vitro | specific benefits of using droplets for emulsion polymerization. It also includes unrelated information about directed enzyme or microorganism evolution. SCORE: 38 | | |
| | enabling the synthesis of diverse polymers with varying characteristics. * **Ease of Processing:** The latex produced in emulsion polymerization is easily removable from the reactor and can be directly used in applications like coatings or further processed.<br><br>Essentially, the droplet-based nature of emulsion polymerization allows for efficient monomer dispersion, controlled reaction conditions, and ultimately, the production of high-quality polymers with desirable properties. | compartmentalizat ion (IVC). This unique platform provides a robust, scalable approach for both analytical and biological studies. | | | |
| How can microfluidics be | Microfluidics offers several advantages for | Microfluidics enables the creation of smart | The response provides a clear and | 73 | 73.0 |

| | | | | |
|---|---|---|---|---|
| used to generate smart materials? | generating smart materials, particularly those at the nanoscale.<br><br>Here's how:<br><br>* **Precise Control over Reaction Conditions:** Microfluidic devices allow for the creation and manipulation of tiny droplets containing reactants. This enables precise control over factors like temperature, concentration gradients, and mixing times, which are crucial for directing the formation of desired material properties.<br>* **Template-Based Synthesis:** Droplets can act as templates for synthesizing particles with specific shapes and sizes. Techniques like single emulsion templating allow for the | materials through precise control of reaction conditions, rapid synthesis, and innovative templating methods. Droplet templating allows monodisperse droplets to act as structural templates, facilitating the synthesis of uniform particles, such as hydrogels for single-cell analysis. By manipulating solute and temperature gradients on short timescales, microfluidics supports the production of nanoscale materials, including semiconducting polymers, catalysts, and nanoparticles with unique optical and electronic properties. Its ability to rapidly explore complex reaction parameter spaces, coupled with advanced analytics and machine learning, accelerates material discovery | concise overview of how microfluidics can be used to generate smart materials, highlighting its advantages and specific techniques such as precise control over reaction conditions, template-based synthesis, rapid synthesis, and integration with analytics. However, it lacks some depth in explaining the underlying principles and mechanisms, and there are minor inaccuracies in the examples provided (e.g., "biomimetic materials" is not explicitly | | |

| | | | | |
|---|---|---|---|---|
| | compartmentalization of reagents within droplets, leading to controlled gelation or solidification into uniform particles.<br>* **Rapid Synthesis and Exploration:** Microfluidic platforms enable rapid synthesis on a millisecond timescale, allowing for the exploration of vast reaction parameter spaces. This accelerates the discovery of new materials with tailored properties.<br>* **Integration with Analytics and Machine Learning:** Microfluidic systems can be integrated with sensors and analytical tools to monitor reactions in real-time. Coupled with machine learning algorithms, this facilitates the optimization of synthesis parameters and | and optimization. This approach offers transformative potential in biotechnology, energy, and electronics by enabling the generation of materials with tailored properties and enhanced functionality. | mentioned in the ground truth answer). The response also does not fully capture the transformative potential of microfluidics in various fields as described in the ground truth.<br>SCORE: 73 | | |

| | the design of novel smart materials. **Examples:** The provided text highlights examples of using microfluidics for synthesizing: * **Semiconducting polymers:** Used in electronics and optoelectronics. * **Catalysts:** Essential for accelerating chemical reactions in various industries. * **Biomimetic materials:** Mimicking natural structures and functions for applications in biomedicine and beyond. * **Nanoscale materials:** Including inorganic semiconductor nanoparticles, metal halide perovskite nanocrystals, and rare earth upconversion | | | | |
|---|---|---|---|---|---|
| | nanoparticles, with potential applications in solar energy, lighting, and sensing.<br><br>By combining precise control over reaction conditions, template-based synthesis, rapid synthesis capabilities, and integration with analytical tools, microfluidics empowers the development of innovative smart materials with diverse functionalities. | | | | |
| What are the applications of droplet microfluidics in environmental sensing? | Droplet microfluidics shows great promise for environmental sensing, offering several advantages over traditional methods.<br><br>Here's what the provided text highlights about its applications in this field:<br><br>* **Miniaturized samplers and | Droplet microfluidics offers transformative applications in environmental sensing by enabling precise, miniaturized, and efficient monitoring systems. It supports on-site environmental surveillance through portable devices for rapid toxicity assessments and detection of | The response provides some relevant information about the applications of droplet microfluidics in environmental sensing, but it lacks depth and specific examples compared to the ground truth answer. It also fails | 36 | 36.0 |

| | | | | |
|---|---|---|---|---|
| | analytical devices:** Droplet microfluidics enables the creation of small, portable devices for on-site environmental monitoring. This allows for rapid toxicity assessments and real-time data collection in the field. * **Controllable microreactors:** These devices can be used to study complex interfacial processes occurring in the environment. Researchers can create miniature models of these processes on a chip, allowing for precise control and observation. * **Integration with analytical techniques:** Droplet microfluidics can be combined with other analytical methods like absorption spectroscopy | contaminants, such as mercury ions. As controllable microreactors, droplet microfluidics facilitates the study of complex interfacial processes, aiding in chemical synthesis and degradation studies. Furthermore, it advances environmental sensing by enabling in situ visualization and precise flow control, enhancing our understanding of environmental processes. These capabilities position droplet microfluidics as a valuable tool for environmental science and engineering. | to mention key aspects such as detection of contaminants like mercury ions and in situ visualization. SCORE: 36 | | |
| | and electron microscopy to provide comprehensive environmental data.<br><br>**Specific examples mentioned in the text include:**<br><br>* Using droplet microfluidics for on-site environmental surveillance and rapid toxicity assessment of chemicals.<br>* Exploring complex environmental interfacial processes using controllable microreactors.<br><br>Overall, droplet microfluidics offers a powerful toolset for advancing environmental sensing by enabling miniaturization, precise control, and integration with advanced analytical techniques. | | | | |
| How can microfluidic | Microfluidic droplets offer | Microfluidic droplets | The response is clear, | 96 | 96.0 |

| droplets improve combinatorial chemistry? | several advantages for combinatorial chemistry, significantly enhancing efficiency and reducing reagent consumption compared to traditional methods like microtiter plates.

Here's how:

* **Miniaturization:** Droplets act as tiny reactors, each containing picoliter volumes of reagents. This allows for the use of six to eight orders of magnitude less starting material per reaction, drastically reducing reagent costs and waste.
* **High Throughput:** Microfluidic systems can generate droplets at high frequencies (e.g., 2.3 kHz), enabling the rapid creation of large libraries of | revolutionize combinatorial chemistry by drastically reducing reagent consumption and enhancing efficiency. Each droplet, acting as a picoliter-scale microreactor, requires six to eight orders of magnitude less material than traditional methods, minimizing waste and costs. This miniaturization enables high-throughput synthesis of large compound libraries, supporting applications like drug discovery and material synthesis. Additionally, droplet manipulation techniques allow for precise reaction control, combinatorial fusion of reagents, and multi-stage experiments with online analysis. These advantages make microfluidic droplets a powerful tool for automating | concise, and directly addresses the question by providing detailed explanations of how microfluidic droplets improve combinatorial chemistry. It highlights various advantages such as miniaturization, high throughput, automation, compartmentalization, and online analysis, with specific examples from a paper on Ugi-type reactions. SCORE: 96 | | |
|---|---|---|---|---|---|

| | compounds.<br>* **Automation and Online Analysis:** Droplet-based microfluidics allows for automation of synthesis steps and integration with online analytical techniques, streamlining the process and facilitating multi-stage experiments.<br>* **Compartmentalization:** Each droplet acts as a separate compartment, ensuring precise control over individual reactions and minimizing cross-contamination. This is crucial for combinatorial synthesis where numerous variations are tested simultaneously.<br><br>The paper specifically highlights the use of microfluidic | complex chemical workflows and optimizing reaction conditions in a highly efficient and scalable manner. | | | |

| | droplets for a three-component Ugi-type reaction to synthesize a library of potential thrombin inhibitors. The system produced $10^6$ droplets, each containing a unique combination of reagents, demonstrating its potential for high-throughput combinatorial synthesis.<br><br>Beyond medicinal chemistry, this droplet-based approach holds promise for other applications like:<br><br>* **Optimizing Reaction Conditions:** Simultaneously testing various parameters within individual droplets allows for efficient identification of optimal reaction conditions. | | | | |
| --- | --- | --- | --- | --- | --- |
| | * **Combinatorial Synthesis of Novel Materials:** Exploring diverse chemical combinations in a controlled manner can lead to the discovery of new materials with unique properties.<br><br>Overall, microfluidic droplets offer a powerful platform for combinatorial chemistry, enabling efficient synthesis, high throughput screening, and precise control over individual reactions. | | | | |
| What are the challenges of scaling up droplet-based chemical synthesis? | Scaling up droplet-based chemical synthesis presents several key challenges:<br><br>**1. Device Fabrication and Large-Scale Production:**<br><br>* **Fluid Distribution:** | Scaling up droplet-based chemical synthesis faces several challenges. Achieving uniform fluid distribution in microfluidic devices remains difficult despite progress in parallel droplet generation, with | The response provides a clear and structured overview of the challenges associated with scaling up droplet-based chemical synthesis, addressing | 86 | 86.0 |

| | While microfluidic reactors allow for parallelized droplet generation, achieving uniform fluid distribution remains a challenge. 3D printing techniques can help fabricate complex channel structures for better control. * **Integration and Control:** Integrating industrial-scale interfaces and built-in control systems into the fabrication process is crucial for scalability.

**2. Multistep Reactions:**

* **Process Simplification:** * Simplifying fundamental functional units like droplet generation, merging, and phase separation is essential for multistep reactions (synthesis, | 3D printing offering potential solutions for complex channel fabrication. Simplifying functional units for multistep reactions, such as droplet generation, merging, and phase separation, is vital for seamless continuous flow processes. Developing low-cost materials and standardizing chip designs are necessary for commercial viability, especially for high-throughput production in diverse fluids. Furthermore, creating complex droplet emulsions and expanding nanomaterial diversity are critical to meeting industrial demands. Addressing these challenges will enable safe, sustainable, and scalable microfluidic applications. | four key areas (device fabrication, multistep reactions, cost and standardization, complexity and diversity) with specific examples and potential solutions. It demonstrates a good understanding of the subject matter and effectively conveys relevant information in a concise manner. SCORE: 86 | | |

| | functionalizatio n, purification). * **Integrated Processes:** Developing continuous flow processes that combine synthesis and post-treatments in a single step holds significant potential.

**3. Cost and Standardization:**

* **Low-Cost Fabrication:** Developing low-cost microfabrication methods and materials is crucial for commercial viability.
* **Standardized Designs:** Lack of well-established design rules for high-throughput generation in non-Newtonian fluids hinders standardization.

**4. Complexity and Diversity:**

* **Multiple Emulsions:** Current scaling | | | | |
|---|---|---|---|---|

| | up primarily focuses on single emulsions, while practical applications often require multiple or complex droplet emulsions. * **Material Diversity:** The diversity and functionalities of nanomaterials produced via droplet microfluidics are still limited compared to microscale particles.<br><br>Addressing these challenges through collaborative efforts and innovative solutions will be crucial for realizing the full potential of droplet-based chemical synthesis on an industrial scale. | | | | |
|---|---|---|---|---|---|
| How can artificial intelligence enhance droplet microfluidic research? | Artificial intelligence (AI) has the potential to significantly enhance droplet microfluidic research in | Machine learning can be used to optimize the design of microfluidic devices. Deep Learning will be | The response is well-structured and provides several specific examples of | 86 | 86.0 |

| | several ways: | integrated to sort, monitor, and process the images. Furthermore, the multimodal agents can be used to study and design the research of droplet microfluidics, a possible excuse robot to perform fabrication and experiments automatically. | how artificial intelligence can enhance droplet microfluidic research. It covers various aspects such as data analysis, system control, droplet characterization, novel applications, and automation efficiency. The answer also concludes with a general statement about the potential of AI in advancing scientific discovery and technological innovation. SCORE: 86 | | |
|---|---|---|---|---|---|
| | * **Data Analysis and Interpretation:** Droplet microfluidics generates vast amounts of data. AI algorithms, particularly deep learning and reinforcement learning, can analyze this data rapidly and effectively, identifying patterns and insights that would be difficult or impossible for humans to discern. This can lead to a better understanding of complex biological and chemical processes occurring within the droplets. | | | | |
| | * **System Control and Optimization:** AI can be integrated into droplet microfluidic systems to control and optimize various | | | | |

| | parameters in real-time. By analyzing data from sensors and feedback loops, AI algorithms can adjust flow rates, temperature, and other variables to enhance the performance and efficiency of the system. This can lead to more accurate and reproducible results.

* **Droplet Characterization :** AI can be used to develop sophisticated detection techniques for characterizing the chemical components within droplets. Machine learning models can be trained to recognize complex flow patterns and spectral signatures, enabling the identification and quantification of specific molecules | | | | |
|---|---|---|---|---|---|

| | within the droplets. This opens up possibilities for high-throughput screening of drug compounds and other applications.<br><br>* **Novel Applications:** The integration of AI with droplet microfluidics can lead to the development of novel applications in areas such as drug discovery, personalized medicine, and environmental monitoring. For example, AI-powered droplet microfluidic systems could be used to screen for potential drug candidates, analyze individual cells for disease diagnosis, or detect pollutants in water samples.<br><br>* **Automation and Efficiency:** AI can automate | | | | |
|---|---|---|---|---|---|
| | many aspects of droplet microfluidic research, reducing manual labor and minimizing human error. This can significantly increase throughput and efficiency, allowing researchers to conduct more experiments and generate data faster.<br><br>Overall, the combination of AI and droplet microfluidics holds immense promise for advancing scientific discovery and technological innovation in various fields. | | | | |
| How can large language models enhance droplet microfluidic research? | Large language models (LLMs) are emerging as powerful tools to enhance droplet microfluidic research in several ways:<br><br>**1. Data | LLMs can be used to build the Agents and Multi-Agents to synthesize knowledge of droplet microfluidics from literature such as research articles and books. Then, | The response provides a clear and concise overview of how large language models (LLMs) can enhance droplet | 56 | 56.0000000<br>0000001 |

| | Processing and Feature Extraction:** LLMs can process and extract contextual information from tabular data formats commonly generated in droplet microfluidics experiments. This allows for more efficient analysis and identification of patterns within the data.<br><br>**2. Predictive Modeling:** LLMs, when combined with deep neural networks, can be used to predict droplet diameter, generation rate, and even classify different operating regimes with high accuracy. This predictive capability can optimize experimental conditions and improve throughput.<br><br>**3. Droplet | the idea hypotheses will be proposed by the agents, and the novelty and feasibility will be assessed. The agents will provide the autonomous platform to design and fabricate droplet microfluidic devices. Furthermore, LLMs have the researchers accelerate knowledge of droplet microfluidics. | microfluidic research, covering data processing, predictive modeling, droplet characterization, and automation. However, it does not fully align with the ground truth answer, which suggests a more comprehensive integration of LLMs in designing and fabricating droplet microfluidic devices, proposing hypotheses, assessing novelty and feasibility, and accelerating knowledge acquisition. The response is mostly relevant but lacks focus on these specific | | |

| | Characterization:** LLMs can analyze complex flow patterns within droplets, enabling more sophisticated characterization of chemical components. This opens up possibilities for screening large libraries of drug compounds and understanding droplet behavior at a deeper level.

**4. Automation and Control:** LLMs can be integrated into digital microfluidic platforms to automate droplet manipulation and control processes. By recognizing droplet states through semantic segmentation, LLMs can trigger sorting mechanisms and adjust voltage states of actuating electrodes for precise droplet | | aspects.
SCORE: 56 | | |

| | | | | | |
|---|---|---|---|---|---|
| | handling.<br><br>The integration of LLMs in droplet microfluidics research holds immense potential for accelerating discoveries in various fields, including biomedical research, drug discovery, and materials science. | | | | |
| How can autonomous droplet microfluidic research? | Autonomous droplet microfluidics holds immense potential for scientific discovery, but several challenges need to be addressed.<br><br>Here's a breakdown based on the provided text:<br><br>**Key Challenges:**<br><br>* **Fast Detection Methods:** Probing small volume droplets dynamically requires advanced | Autonomous droplet microfluidics is an emerging field with significant potential for scientific discovery, aiming to create highly integrated, automated systems for bio/chemical functions like PCR, sample preparation, and detection. Key challenges include the need for fast detection methods to probe small volume droplets, precise control over droplet dynamics in complex microchannel | The response attempts to address the question but lacks clarity and coherence; mostly irrelevant details are included. SCORE: 22 | 22 | 22.0 |

| | | | | |
|---|---|---|---|---|
| | detection techniques like single-molecule detection, fluorescence lifetime imaging, and spectroscopy. These are areas of ongoing research.<br>* **Droplet Control:** Precisely controlling droplet behavior within complex microchannel networks is crucial for automation. This demands sophisticated design simulation tools that integrate fluid dynamics, chemistry, biology, physics, software engineering, and high-performance computing.<br><br>**Potential Solutions & Future Directions:**<br><br>* **"Droplet Intelligence/Computing":** The text suggests | networks, and the development of virtual design platforms for optimizing system designs. Solutions under exploration include using advanced detection techniques such as single-molecule detection and fluorescence lifetime imaging, as well as integrating microelectronic actuators for droplet manipulation. Additionally, the development of virtual design and experiment platforms could streamline the creation of optimized microfluidic systems. The potential applications are vast, including directed evolution, closed-loop synthesis, and automated drug discovery, where systems could autonomously optimize compound combinations. Collaboration across disciplines | | |

| | integrating computational capabilities directly into the microfluidic system to enable on-chip control of droplet behavior. This could lead to truly autonomous systems capable of adapting and responding to experimental conditions.<br>* **Virtual Design and Experiment Platforms (VDEP):** These platforms could revolutionize the design process by allowing researchers to simulate and optimize microfluidic systems virtually before physical fabrication.<br><br>**Example Application:**<br><br>The text provides an example of a future smart droplet-based platform for | like fluid dynamics, chemistry, and software engineering will be key to overcoming challenges and advancing this technology, paving the way for groundbreaking scientific advances. | | | |
| --- | --- | --- | --- | --- | --- |

| | | | | |
|---|---|---|---|---|
| | autonomous scientific discovery:<br><br>*  **Evolutionary Algorithms for Drug Discovery:**  Imagine a system that uses evolutionary algorithms to automatically screen and optimize combinations of small molecules for anti-inflammatory effects. The microfluidic device would handle the synthesis, testing, and selection of compounds based on feedback from detection systems.<br><br>**In essence, autonomous droplet microfluidics aims to create self-contained, intelligent laboratories capable of performing complex experiments | | | |

| | | | | | |
|---|---|---|---|---|---|
| | with minimal human intervention.** This has the potential to accelerate scientific discovery in fields like drug development, materials science, and biological research. | | | | |
| What is the potential for fully integrated droplet microfluidic platforms? | Fully integrated droplet microfluidic platforms hold immense potential across various fields.<br><br>Here's a breakdown based on the provided context:<br><br>**Enhanced Control and Precision:**<br><br>* **Payload Exchange:** Recent advancements like stable droplets and fluorinated Pickering emulsions allow for precise control over both intra-droplet and inter-droplet | Droplet microfluidics offers transformative solutions to global health challenges by enabling rapid, cost-effective, and high-throughput analysis of biological samples. By generating and manipulating tiny droplets as individual reaction chambers, this technology can revolutionize diagnostics, particularly in resource-limited settings. For example, it facilitates point-of-care testing for infectious diseases like tuberculosis, malaria, and COVID-19, providing results quickly and | The response provides a detailed and structured analysis of the potential of fully integrated droplet microfluidic platforms, covering enhanced control and precision, applications, future directions, and challenges. It demonstrates a good understanding of the subject matter and presents relevant examples. However, it does not directly | 76 | 76.0 |

| | molecular transport. This opens doors for high-throughput screening of small molecules and complex synthetic cells mimicking natural cell functions. * **Automation:** Integrating microfluidic design with automation tools could significantly improve performance metrics and streamline workflows.<br><br>**Applications:**<br><br>* **Biochemical Experiments:** Droplet microfluidics excels in complex biochemical experiments due to its advantages like decreased diffusion distance, faster mixing, laminar flow, and parallel processing capabilities. | accurately. Droplet microfluidics is also crucial for multiplexed screening, enabling simultaneous detection of multiple biomarkers, which enhances early disease detection and monitoring. Additionally, it supports the development of low-cost diagnostic devices, reducing dependence on expensive laboratory infrastructure. In drug discovery, it accelerates the screening of therapeutic compounds, improving the treatment of diseases. The technology also advances personalized medicine by analysing single cells and their responses to treatments. Overall, droplet microfluidics addresses global health disparities by delivering accessible, | address the transformative solutions to global health challenges mentioned in the ground truth answer. SCORE: 76 | | |

| | | | | |
|---|---|---|---|---|
| | * **Diagnostics:** * Droplet-based platforms are being used for developing diagnostic assays, integrating optical methods like confocal fluorescence spectroscopy for detection.<br><br>**Future Directions:**<br><br>* **Smart Microfluidic Devices:** Combining microfluidics with droplet intelligence/computing functions could lead to self-regulating devices capable of making logical decisions based on experimental results, optimizing processes like enzyme creation, drug screening, and PCR conditions.<br>* **Virtual Design and Experiment Platforms:** Integrating monitoring and | scalable, and efficient solutions for disease diagnosis, monitoring, and treatment. | | | |

| | control systems within the platform will enable rapid information acquisition for decision-making in follow-on experiments, paving the way for highly automated and integrated systems.<br><br>**Challenges:**<br><br>* **Design Complexity:** Designing microfluidic circuits is significantly more challenging than designing integrated circuits, requiring specialized tools and expertise.<br>* **Droplet Size Control:** While progress has been made, generating sub-femtoliter droplets for mimicking biological syntheses remains a challenge.<br><br>Overall, fully | | | | |
|---|---|---|---|---|---|

| | integrated droplet microfluidic platforms have the potential to revolutionize various fields by enabling precise control over molecular transport, automation of workflows, and development of sophisticated diagnostic and synthetic biology tools. | | | | |
|---|---|---|---|---|---|
| How can droplet microfluidics contribute to global health challenges? | Droplet microfluidics offers several promising avenues for addressing global health challenges.<br><br>Here's how:<br><br>* **Rapid and Sensitive Pathogen Detection:** As highlighted in the provided text, droplet microfluidics combined with techniques like LAMP (Loop-mediated isothermal amplification) allows for rapid and sensitive | Droplet microfluidics has significant potential to address global health challenges by enhancing diagnostic and therapeutic processes. It enables rapid and sensitive pathogen detection, which is critical for early disease diagnosis and containment. The technology's high-throughput capabilities allow simultaneous analysis of multiple samples, streamlining disease diagnosis and treatment. Additionally, droplet | The response provides a clear and detailed explanation of how droplet microfluidics can contribute to addressing global health challenges. It highlights the potential benefits in terms of rapid pathogen detection, high-throughput screening, point-of-care diagnostics, and personalized | 76 | 76.0 |

| | | | | |
|---|---|---|---|---|
| | detection of pathogenic bacteria like *Salmonella typhimurium*. This is crucial for early diagnosis and containment of infectious diseases, especially in resource-limited settings where access to sophisticated laboratory equipment is limited.

* **High-Throughput Screening:** The high-throughput nature of droplet microfluidics enables the screening of a large number of samples simultaneously. This is particularly valuable for epidemiological surveillance, identifying outbreaks, and drug discovery efforts targeting infectious diseases.
* **Point-of-Care Diagnostics:** | microfluidics can be integrated into point-of-care platforms, making it feasible for use in resource-limited settings. This approach can improve healthcare accessibility and efficiency. Furthermore, droplet microfluidics facilitates precise single-cell analysis, enabling advances in personalized medicine and more effective treatment strategies. While challenges such as cost, standardization, and integration with existing systems remain, continued development of this technology could significantly impact global health by improving detection, diagnostics, and treatment capabilities. | medicine. The discussion also acknowledges the challenges that need to be addressed for wider adoption. However, it would benefit from a more concise summary statement at the beginning and a clearer conclusion that ties together the main points. SCORE: 76 | | |

| | Droplet microfluidics platforms can be miniaturized and made portable, facilitating point-of-care diagnostics. This brings testing capabilities closer to patients, reducing reliance on centralized laboratories and enabling faster treatment decisions.

* **Personalized Medicine:** While not explicitly mentioned in the text, droplet microfluidics can contribute to personalized medicine by enabling precise analysis of individual patient samples. This could lead to tailored treatment strategies for infectious diseases based on specific pathogen profiles and | | | | |
| --- | --- | --- | --- | --- | --- |

| | patient characteristics. | | | | |
|---|---|---|---|---|---|
| | **Challenges and Future Directions:** | | | | |
| | Despite its potential, wider adoption of droplet microfluidics in global health faces challenges: | | | | |
| | * **Cost-Effectiveness:** While the technology is advancing, making it more affordable for widespread use in low-resource settings remains a key challenge. * **Standardizati on and Regulation:** Establishing standardized protocols and regulatory frameworks for droplet microfluidics-based diagnostics is crucial for ensuring reliability and acceptance. * **Integration | | | | |
| | with Existing Healthcare Systems:** Seamless integration of droplet microfluidics platforms into existing healthcare infrastructure is essential for maximizing their impact.<br><br>Continued research and development efforts are needed to overcome these challenges and fully realize the transformative potential of droplet microfluidics in addressing global health challenges. | | | | |
| How can open-source designs impact the future of microfluidic devices? | Open-source designs have the potential to revolutionize the field of microfluidics by making it more accessible, affordable, and collaborative.<br><br>Here's how:<br><br>* **Reduced | Open-source designs for microfluidic devices hold the potential to revolutionize the field by increasing accessibility, reducing costs, and fostering collaboration. These designs allow for the creation of custom | The response is clear, concise, and directly addresses the question about the impact of open-source designs on the future of microfluidic devices. It provides | 83 | 83.0 |

| | | | | |
|---|---|---|---|---|
| | Costs:** Open-source designs eliminate the need for expensive proprietary software and hardware, allowing researchers and developers to fabricate microfluidic devices using readily available tools like desktop 3D printers and common household equipment. This democratizes access to microfluidics technology, enabling wider adoption in research, education, and even point-of-care diagnostics. * **Increased Collaboration:** Sharing designs openly fosters a collaborative environment where researchers can build upon each other's work, leading to faster innovation and development of more | predictive models without requiring extensive expertise, enabling broader participation in device development. The use of open-source tools, such as Design Automation of Fluid Dynamics (DAFD), accelerates design and prototyping, enabling rapid iteration and optimization. Additionally, the integration of machine learning algorithms can drive the development of new microfluidic tools and optimize existing ones. This democratization of design expertise and the collaborative, community-driven approach will facilitate the creation of affordable, specialized devices, ultimately leading to breakthroughs in applications like single-cell analysis and global health diagnostics. The | specific examples and details about reduced costs, increased collaboration, rapid prototyping, and customization, which aligns with the ground truth answer. However, it lacks some depth in discussing the integration of machine learning algorithms and the potential for breakthroughs in applications like single-cell analysis and global health diagnostics. SCORE: 83 | | |

| | | | | |
|---|---|---|---|---|
| | sophisticated devices.<br>* **Rapid Prototyping:** Open-source platforms like Flui3d enable rapid prototyping, allowing researchers to quickly iterate on designs and test different configurations. This accelerates the research process and facilitates the exploration of novel applications for microfluidics.<br>* **Customization and Specialization:** Open-source designs can be easily customized to meet specific needs, enabling researchers to tailor devices for particular applications or experimental setups.<br><br>The article highlights how open-source microfluidic designs are already being | open-source model promises to transform microfluidics into a more inclusive and innovative field. | | |

| | | | | | |
|---|---|---|---|---|---|
| | used to develop low-cost diagnostic testing platforms that can be deployed in resource-limited settings. This has the potential to significantly improve healthcare access and outcomes globally.

Overall, open-source designs have the power to unlock the full potential of microfluidics by making it a more inclusive, collaborative, and accessible field. | | | | |
| What are the challenges of scaling up microfluidic devices for industrial use? | Scaling up microfluidic devices for industrial use presents several significant challenges:

* **Throughput Limitations:** Micrometric channel sizes inherent to microfluidics restrict production efficiency. | Scaling up microfluidic devices for industrial use presents several challenges, including throughput limitations, lack of standardization, and integration complexities. The small micrometric channel sizes restrict production efficiency, and increasing | The response provides a clear and concise overview of the challenges associated with scaling up microfluidic devices for industrial use, including throughput limitations, | 76 | 76.0 | |

| | While increasing channel size and parallelization can boost throughput, these strategies must be carefully weighed against fabrication costs and potential impacts on reaction kinetics. * **Lack of Standardization:** The absence of standardized design and fabrication methods across microfluidic platforms makes it difficult to compare results and hinders widespread adoption in industrial settings. * **Scaling Difficulties:** While microfluidics excels at small-scale operations, transitioning to industrial production volumes remains a significant hurdle. * **Testing and Reliability:** | parallelization or channel size may impact fabrication costs and reaction kinetics. There is also a lack of standardized design methods, making it difficult to compare results and adopt microfluidics in industrial settings. Additionally, scaling from laboratory to industrial production remains a hurdle, with issues in cost-effective microfabrication, generating complex droplet emulsions, and ensuring reliable fluid control. The integration of microfluidic devices with industrial interfaces, testing methods, and continuous processes is essential for successful commercialization. Overcoming these challenges requires innovative solutions, such as simplifying multistep reactions and | lack of standardization, integration complexities, testing and reliability issues. It also acknowledges the importance of overcoming these challenges for realizing the full potential of microfluidics in industrial settings. However, it lacks some depth and specific examples compared to the ground truth answer, which provides more detailed information on the challenges and potential solutions. SCORE: 76 | | |
|---|---|---|---|---|

| | The microfluidic industry lacks established testing strategies, methods, and reliability models. This is particularly crucial for heterogeneous systems where understanding fault modes requires extensive research.<br><br>* **Integration Complexity:** Integrating diverse functional components within a single device presents a challenge. Precise control over fluid flow rate, pressure, and temperature is also essential for successful industrial applications.<br><br>Overcoming these challenges is crucial for realizing the full potential of microfluidics in industrial settings. | developing standardized, low-cost fabrication methods. | | | |
|---|---|---|---|---|---|

| How might droplet microfluidics impact agriculture? | Droplet microfluidics has the potential to significantly impact agriculture in several ways:<br><br>* **Precision Fertilization and Pest Control:** Droplet-templated capsules can be used to deliver fertilizers and pesticides directly to plant roots, minimizing waste and environmental impact. This targeted approach ensures that crops receive the precise nutrients they need while reducing the overall amount of chemicals used.<br>* **Enhanced Crop Growth:** By precisely controlling the release of nutrients and pesticides, droplet microfluidics can optimize plant growth and yield. | Droplet microfluidics has the potential to transform agriculture by enhancing crop management and food security. It enables the precise release of fertilizers and pesticides, reducing waste and optimizing plant growth. The technology also supports improved crop monitoring through long-term studies, offering insights into plant development. By analyzing individual plant cells, droplet microfluidics can help uncover genetic and phenotypic traits, enhancing disease resistance and drought tolerance in crops. Additionally, it aids in disease detection, allowing for early intervention to prevent crop loss. Despite challenges such as scalability, cost, and integration with existing agricultural practices, ongoing | The response is well-structured and provides several specific examples of how droplet microfluidics can impact agriculture. It addresses the question thoroughly, covering various aspects such as precision fertilization, enhanced crop growth, improved disease detection, and single-cell analysis. The discussion on challenges and future directions adds depth to the answer, showing an understanding of the limitations and potential solutions. SCORE: 86 | 86 | 86.0 |

| | | | | |
|---|---|---|---|---|
| | * **Improved Disease Detection:** Microfluidic devices can be used to rapidly and accurately detect plant diseases, allowing for early intervention and preventing widespread crop losses. This is particularly important in the face of climate change and emerging pathogens.<br><br>* **Single-Cell Analysis:** Droplet microfluidics enables the analysis of individual plant cells, providing insights into their genetic makeup, physiology, and response to environmental stresses. This information can be used to develop new varieties of crops that are more resistant to disease, drought, and other | research aims to overcome these barriers, making droplet microfluidics a promising tool for precision agriculture and sustainable food production. | | | |

| | challenges. | | | | |
|---|---|---|---|---|---|
| | **Challenges and Future Directions:** | | | | |
| | While droplet microfluidics holds great promise for agriculture, there are still some challenges to overcome: | | | | |
| | * **Scalability:** Current microfluidic devices are often limited in their throughput, making it difficult to scale up production for large-scale agricultural applications.<br>* **Cost:** Microfluidic devices can be expensive to manufacture and operate, which may limit their adoption by smallholder farmers.<br>* **Integration with Existing Systems:** Integrating microfluidic technology into existing agricultural | | | | |

| | practices requires careful consideration of factors such as infrastructure, labor, and regulatory requirements.

Despite these challenges, ongoing research and development efforts are addressing these limitations and paving the way for wider adoption of droplet microfluidics in agriculture. | | | | |
|---|---|---|---|---|---|
| What are the current limitations of droplet microfluidics, and how can they be overcome? | Droplet microfluidics, while promising for high-throughput biological assays, faces several limitations:

**1. Sample Size:** Droplet microfluidics is most efficient for assays involving at least 10^5 samples. For smaller sample sizes, traditional methods like | Droplet microfluidics has immense potential for high-throughput biological assays but faces several limitations that hinder its broader adoption. One key challenge is its efficiency for assays involving small sample sizes; droplet microfluidics excels with large volumes (at least 10^5 samples), making traditional methods like | The response provides a clear and concise overview of the current limitations of droplet microfluidics and offers suggestions on how they can be addressed. It covers key challenges such as sample size, interface stability, spatial | 76 | 76.0 | |

| | | | | |
|---|---|---|---|---|
| | microtiter plates are more practical.<br><br>**2. Interface Stability:** The oil-water interface in each droplet needs to be stable and inert. Finding the right surfactant can be challenging.<br><br>**3. Spatial Barcoding:** Droplets lack the spatial barcoding capabilities of microtiter plates. This requires developing alternative barcoding systems, which can be complex.<br><br>**4. Long-Term Stability & Cross-Talk:** Imperfect partitioning between phases can lead to droplet shrinkage, coalescence, and inter-droplet cross-talk.<br><br>**Overcoming these limitations:** | microtiter plates more suitable for lower sample numbers. Another critical issue is maintaining a stable and inert oil-water interface, which is essential for droplet formation. Finding the right surfactant to keep this interface stable remains a complex chemistry challenge. Furthermore, droplet microfluidics lacks the spatial barcoding available in microtiter plates, making it difficult to organize and track multiple droplet libraries. As a result, developing alternative barcoding systems is necessary but challenging. The automation of droplet movement is another obstacle, as it requires precise control, often through techniques like electrowetting and dielectrophoresis. Lastly, droplet | barcoding, and long-term stability & cross-talk, and proposes potential solutions like improved surfactants, advanced barcoding systems, and enhanced droplet stability. The response also touches upon future directions for the field, including integrating more functional components into single devices and utilizing machine learning techniques. However, it could be further improved by providing more detailed explanations of these challenges and solutions, as well as specific examples or | | |

| | | | | |
|---|---|---|---|---|
| | * **Improved Surfactants:** Research is ongoing to develop surfactants that provide more stable and inert oil-water interfaces.<br>* **Advanced Barcoding Systems:** Researchers are exploring new barcoding systems suitable for droplet libraries.<br>* **Enhanced Droplet Stability:** Studies on oil phase and surfactant formulations aim to improve droplet isolation properties, reducing shrinkage, coalescence, and cross-talk.<br><br>**Future Directions:**<br><br>* Integrating more functional components into single devices for complex experiments.<br>* Utilizing machine | stability over time can be compromised, with issues such as shrinkage, coalescence, and cross-talk between droplets. To address these issues, research is underway to develop improved surfactants, more reliable barcoding systems, and novel materials and techniques to enhance droplet control. These advancements, coupled with better analytical tools, are expected to overcome current limitations and expand droplet microfluidics' applications in chemical and biological research. | references to support the claims made.<br><br>SCORE: 76 | | |

| | learning techniques to create "intelligent" systems that can direct and automate the experimentation process.<br><br>The field of droplet microfluidics is rapidly evolving, with ongoing research addressing these limitations and paving the way for wider applications in chemical and biological research. | | | | |
|---|---|---|---|---|---|
| How can interdisciplinary collaboration accelerate innovation in droplet microfluidics? | The provided text highlights the crucial role of interdisciplinary collaboration in accelerating innovation within droplet microfluidics.<br><br>Here's how, according to the text:<br><br>* **Bridging Knowledge Gaps:** Droplet microfluidics benefits from a | Interdisciplinary collaboration is essential for advancing droplet microfluidics by combining expertise from various fields such as chemistry, biology, material science, and engineering. By uniting these diverse areas, researchers can address complex challenges such as cost-effective fabrication, scalable | The response provides a clear and concise explanation of how interdisciplinary collaboration can accelerate innovation in droplet microfluidics. It accurately highlights the benefits of combining knowledge | 76 | 76.0 |

| | convergence of knowledge in fields like chemistry, biology, material science, and engineering. Collaboration allows experts from these diverse areas to share insights, identify synergies, and develop novel solutions that wouldn't be possible within isolated disciplines. * **Addressing Complex Challenges:** Scaling up droplet microfluidics for commercial applications requires overcoming multifaceted challenges, such as cost-effective fabrication, complex pattern control, and the synthesis of diverse nanomaterials. Interdisciplinary teams can bring together specialized expertise to tackle these intricate | production, and novel interfacial reaction systems. For example, biologists and chemists can leverage droplet microfluidics for high-throughput analysis, while material scientists and engineers work on developing more robust and affordable microfabrication methods. Furthermore, integrating technologies like sensitive detectors and machine learning algorithms can enhance the analytical capabilities of droplet-based systems. Collaborative initiatives, such as the EMBL Microfluidics Conference, foster knowledge exchange and innovation by bringing together professionals from different disciplines. This cooperation is crucial for overcoming technical barriers, | from various fields, addressing complex challenges, and fostering innovation through collaborative initiatives. However, it lacks some depth and specific examples compared to the Ground Truth Answer. SCORE: 76 | | |
|---|---|---|---|---|
| | problems effectively. * **Fostering Innovation:** Conferences and meetings like the EMBL Microfluidics Conference series and Single Cell Genomics (SCG) meetings already promote interdisciplinary interaction. However, sustained collaborative initiatives from academia and industry are essential to further drive innovation and accelerate the development of new applications in fields like drug discovery, single-cell analysis, and nanomaterial synthesis.<br><br>In essence, interdisciplinary collaboration fosters a fertile ground for innovation by breaking down silos, promoting knowledge | reducing manufacturing costs, and developing new applications in fields like drug discovery, single-cell analysis, and nanomaterial synthesis. Ultimately, interdisciplinary collaboration accelerates the evolution of droplet microfluidics, pushing the boundaries of what is achievable in this promising field. | | | |

| | exchange, and enabling the development of holistic solutions that push the boundaries of droplet microfluidics technology. | | | | |
|---|---|---|---|---|---|

| question | answer | ground_truth | feedback | score | percentage |
|---|---|---|---|---|---|
| What is microfluidics? | The precise control of fluids at the microscale. | Microfluidics is a multidisciplinary field that involves the study and manipulation of fluids at the microscale, typically within channels with dimensions ranging from tens to hundreds of micrometers. It is a multidisciplinary field that involves molecular analysis, molecular biology, and microelectronics. Microfluidics has been increasingly used in the biological sciences | The response is concise but lacks detail and context; it only partially addresses the question by providing a definition without explaining the broader scope and applications of microfluidics. SCORE: 41 | 41 | 41.0 |

| | | because precise and controlled experiments can be conducted at a lower cost and faster pace. Lab on a Chip devices use microfluidics for applications such as Point of Care testing of diseases, or Organ on a Chip studies. | | | |
|---|---|---|---|---|---|
| What is droplet microfluidics? | Droplet microfluidics is a technology that uses tiny droplets of liquid to manipulate and process small volumes of fluids. | Droplet-based microfluidics is a branch of microfluidics focusing on the creation and manipulation of discrete volumes of liquids through adoption of immiscible fluid flows. Microdroplets allow for handling miniature volumes ($\mu$L to fL) of fluids conveniently, provide better mixing, encapsulation, sorting, and sensing, and suit high throughput experiments. Compared with continuous-flow systems, droplet microfluidic systems enable isolation and manipulation of samples in separate | The response provides a basic definition of droplet microfluidics, but lacks detail and context compared to the ground truth answer. It does not mention key aspects such as immiscible fluid flows, advantages like better mixing and encapsulation, or the ability to process large numbers of samples in parallel. | 16 | 16.0 |
| | | compartments. The droplets generated with uniform size and shape provide a stable microenvironment for biochemical reactions. Processing of a large number of droplets in parallel facilitates to achieve ultrahigh throughput. Confinement of individual targets within the droplets and independent processing of each of them allow high-precision analyses. | SCORE: 16 | | |
| How to generate microdroplets? | Microfluidic devices, acoustic waves, electrospraying, and inkjet printing. | Microdroplet generation in microfluidics is achieved through two main methods: active and passive. Active methods utilize external forces like piezoelectric actuators to precisely control droplet size, formation rate, and spacing. These methods enable high uniformity in droplet generation, with deviations less than 0.3%. Passive methods rely on fluid properties and channel geometry, such as T-junctions | The response provides a list of methods to generate microdroplets (microfluidic devices, acoustic waves, electrospraying, and inkjet printing), but it lacks any explanation or context. It does not address the question's | 10 | 10.0 |
| | | or cross-flow configurations, to create droplets. These methods are influenced by factors like flow rate, viscosity, and interfacial tension, allowing for fine control over droplet size and uniformity. For example, PDMS-based microfluidic devices have been used for nanoparticle synthesis, where hydrodynamic coupling of nozzles generates droplet pairs for precise mixing. Both methods are valuable in applications like nanoparticle synthesis, offering high reproducibility and control over reaction conditions in biomedical and material science research. | request for information on how to generate microdroplets, nor does it provide any details about the methods mentioned. SCORE: 10 | | |
| What materials are typically used to fabricate droplet microfluidic devices? | Polydimethyl siloxane (PDMS), glass, and polymers like polycarbonate. | Droplet microfluidic devices are commonly fabricated from a variety of materials, each offering distinct advantages for specific applications. The most widely used material is PDMS (Polydimethylsiloxane), favored for its | The answer provides a list of materials used for fabricating droplet microfluidic devices but lacks detailed information about the | 16 | 16.0 |

| | | ease of use with soft lithography and flexibility, though it has limitations such as poor solvent resistance. Alternatives to PDMS include thermoset polyesters, thermoplastics like PMMA, polycarbonate, and polystyrene, and amorphous polymers such as cyclic olefin copolymer. Fluoropolymers like PTFE and glass are also used, offering superior solvent resistance and biofouling resistance. Silicon and capillary devices are utilized in certain applications, providing precise control for microfluidic operations. These materials are chosen based on factors like solvent resistance, biofouling, and the ability to mass-produce devices using methods such as injection molding or hot embossing, ensuring broad applicability in | advantages and limitations of each material. It also doesn't mention certain key materials like thermoset polyesters, amorphous polymers, fluoropoly mers, silicon, and capillary devices that are mentioned in the ground truth answer. The response is somewhat relevant but incomplete and unclear. SCORE: 16 | | |
|---|---|---|---|---|---|

| | | research and industrial settings. | | | |
|---|---|---|---|---|---|
| What are the applications of droplet microfluidics? | Droplet microfluidics has applications in drug discovery, diagnostics, single-cell analysis, and chemical synthesis. | Droplet microfluidics has diverse applications across biomedical research, diagnostics, and industrial fields. It enables high-throughput single-cell analysis, such as genomics, transcriptomics, and drug screening, while also facilitating nanoparticle synthesis, enzyme evolution, and chemical reactions. In healthcare, it supports personalized medicine, point-of-care diagnostics, and vaccine development. Beyond biology, it is used for environmental monitoring, food safety testing, and the development of cosmetics and pharmaceuticals. Its ability to create isolated, uniform reaction chambers in droplets makes it a powerful tool for precise, cost-effective, and scalable workflows in science and | The response is partially correct but lacks detail and scope compared to the ground truth answer; it only mentions four specific applications whereas the ground truth lists several more. SCORE: 51-55 | 51 | 51.0 |

| | | industry. | | | |
|---|---|---|---|---|---|
| What are the main benefits of using droplet microfluidics over bulk systems? | Enhanced control over reaction conditions, reduced reagent consumption, increased throughput, and improved sensitivity. | Droplet microfluidics offers significant advantages over traditional bulk systems. It enables individual compartmentalization, isolating single cells or molecules in tiny liquid compartments for precise control. This allows for single-cell analysis and early detection of rare cells or secreted molecules. The technology supports massive parallel processing, analyzing millions of droplets simultaneously, ideal for high-throughput screening and examining phenotypic or genetic variability. Additionally, iterative operations such as merging, mixing, and sorting droplets enable complex experimental protocols and continuous monitoring of populations. The system's highly controllable local environments | The answer provides some relevant information about the benefits of using droplet microfluidics over bulk systems, but it lacks detail and context. It only mentions a few general advantages without explaining how they compare to traditional methods or providing specific examples. SCORE: 41 | 41 | 41.0 |
| | | ensure consistent, reproducible results. These benefits make droplet microfluidics a versatile tool in drug discovery, medical diagnostics, single-cell analysis, and other fields, offering powerful insights and precise manipulation at the individual level. | | | |
| What are the advantages of droplet microfluidic systems compared with continuous-flow systems? | Droplet microfluidic systems offer advantages such as enhanced reaction control, reduced reagent consumption, and improved scalability compared to continuous-flow systems. | Droplet-based microfluidic systems offer significant advantages over traditional continuous-flow systems. These include the ability to overcome issues such as Taylor dispersion, slow mixing, and surface-molecule interactions. Droplet systems allow for ultra-small assay volumes, reducing reagent consumption and improving sensitivity. They also minimize cross-contamination and sample adsorption by compartmentalizing individual samples. The system's chaotic advection enhances mixing | The response provides a concise summary of the advantages of droplet microfluidic systems compared to continuous-flow systems, but lacks specific details and examples. It does not fully capture the nuances and benefits mentioned in the ground truth answer, such as overcomin | 41 | 41.0 |

| | | efficiency, enabling faster reactions. With droplet generation frequencies up to hundreds of kHz, droplet microfluidics supports high-throughput experimentation. Furthermore, the rapid and precise control of droplet payloads and sub-millisecond mixing times make it ideal for applications requiring fast, controlled analysis. These benefits make droplet-based systems particularly valuable for cell-based assays, enzyme kinetics, and high-throughput screening. | g Taylor dispersion, minimizing cross-contaminat ion, and supporting high-throughput experiment ation. SCORE: 41 | | |
|---|---|---|---|---|---|
| What are the basic principles behind droplet generation in microfluidics? | The basic principles behind droplet generation in microfluidics involve controlling the flow of immiscible liquids, typically an oil and a water-based solution containing the desired reagents. | Droplet generation in microfluidics relies on emulsion processes, where an oil phase (continuous phase) encapsulates an aqueous phase (dispersed phase) to form droplets. Surfactants are added to lower interfacial tension, stabilize the droplets, and prevent coalescence. The process requires | The response provides a clear and concise overview of the basic principles behind droplet generation in microfluidi cs. It correctly identifies the importance | 46 | 46.0 |

| | Droplets are formed by manipulating the interfacial tension between these liquids and using geometric features within microfluidic channels to create controlled breakup events. | precise control over the flow of both liquids, often achieved using syringe pumps or pressure-driven systems. Droplets are generated through surface-induced instabilities, influenced by microchannel geometries such as T-junctions or flow-focusing, which control droplet size and frequency. The design and flow control of the system determine the size uniformity and production rate of droplets. These principles enable high-throughput applications such as drug delivery, high-throughput screening, and microreactors for bio-catalysis, ensuring reproducibility and productivity in biological experiments. | of controlling interfacial tension and using geometric features within microfluidic channels to create controlled breakup events, which is partially aligned with the ground truth answer. However, it lacks specific details about surfactants, surface-induced instabilities, and the role of microchannel geometries in determining droplet size and frequency. The response also does not mention high- | | |
|---|---|---|---|---|---|

| | | | throughput application s or the importance of system design and flow control for reproducibi lity and productivit y.

SCORE: 46 | | |
|---|---|---|---|---|---|
| How is droplet size controlled in a droplet generator? | Droplet size can be controlled by adjusting parameters such as flow rate, nozzle diameter, and applied voltage. | Droplet size control in microfluidic generators can be achieved through several methods. In passive droplet generation, adjusting microchannel dimensions or controlling the flow rates of immiscible liquids allows for precise control over droplet size, shape, and monodispersity. Tree-splitters, which bisect large droplets into smaller ones, also influence droplet size and uniformity, with splitting angles affecting droplet diameter. In active droplet generation, technologies like piezoelectric actuators independently control droplet size, | The response provides some relevant informatio n about factors that can influence droplet size, but it lacks detail and context compared to the ground truth. It does not mention specific methods like microchan nel dimensions or tree-splitters, which are crucial for precise | 41 | 41.0 |
| | | formation rate, and spacing, offering high precision. Additionally, the velocity of the continuous and dispersed phases plays a significant role in controlling droplet size. These methods enable the generation of uniform droplets for various applications, such as high-throughput screening and drug delivery. | control over droplet size in passive generation. SCORE: 41 | | |
| What is the role of surface tension in droplet formation? | Surface tension causes the liquid to minimize its surface area, leading to the spherical shape of droplets. | Surface tension plays a critical role in droplet formation in microfluidic systems by influencing droplet size, shape, and formation mechanisms. It acts to minimize surface area, favoring spherical droplet shapes. At low Weber numbers (We < 1), where surface tension dominates, droplets form through dripping or squeezing. As the Weber number increases, fluid inertia competes with surface tension, leading to jetting behavior. The capillary number (Ca) also influences droplet | The response provides a clear explanation of how surface tension affects droplet formation, specifically mentioning its role in minimizing surface area and favoring spherical shapes. However, it lacks the comprehensive details provided by the ground truth | 45 | 45.0 |
| | | size, with surface tension controlling droplet formation at low Ca. Techniques like electrowetting can measure surface tension in microfluidic systems, providing insights into the interfacial properties. Additionally, surface tension is a key parameter in T-junctions and flow focusing devices, where droplet size measurements offer an indirect way to assess surface tension at the time-scale of droplet production. Understanding the interplay between surface tension and other forces is essential for precise droplet control. | answer regarding the influence of Weber numbers, capillary number, and other factors on droplet formation mechanisms. SCORE: 45 | | |
| How does a flow-focusing device differ from a T-junction in droplet formation? | A flow-focusing device uses three inlets to create a constricted channel, while a T-junction only uses two. This constriction in flow-focusing allows for more precise | Flow-focusing devices and T-junctions are two common microfluidic designs for droplet formation, differing in geometry and mechanism. A flow-focusing device uses a central dispersed phase channel and side continuous phase channels converging at a | The response correctly identifies a key difference between flow-focusing devices and T-junctions (the number of inlets), but lacks detail | 41 | 41.0 |

| | control over droplet size and shape. | narrow orifice, where hydrodynamic focusing pinches off droplets. It produces smaller, highly uniform droplets with precise control over size and high throughput. In contrast, a T-junction has perpendicular channels where the continuous phase shears the dispersed phase at the intersection, forming larger droplets with less uniformity and lower throughput. Flow-focusing is ideal for applications requiring monodisperse droplets, while T-junctions suit simpler or larger-droplet applications. | on the actual mechanism of droplet formation, size control, and application suitability. It also simplifies the comparison without providing a comprehensive explanation. SCORE: 41 | | |
|---|---|---|---|---|---|
| What role does the geometry of the microchannel play in droplet formation? | The geometry of the microchannel, such as its width, depth, and surface properties, influences droplet size, shape, and frequency. | The geometry of a microchannel plays a critical role in droplet formation, influencing the size, shape, and generation frequency of droplets. Key geometric factors include the channel width, depth, and junction design, such as T-junctions | The response provides a basic understanding of the role of microchannel geometry in droplet formation, mentioning its | 41 | 41.0 |

| | | or flow-focusing geometries. These features govern the interaction between the continuous and dispersed phases, affecting the shear forces and pressure gradients that drive droplet breakup. For instance, narrower channels or smaller orifices promote the formation of smaller droplets due to increased shear stress. Similarly, the junction type controls droplet detachment mechanics; in T-junctions, droplets form due to the balance of interfacial tension and shear force, while in flow-focusing designs, they form under pinching at the constriction. Channel geometry also determines flow regimes ranging from dripping to jetting further impacting droplet characteristics. Thus, optimizing microchannel geometry is crucial for precise control over droplet formation in applications like | influence on size, shape, and frequency. However, it lacks detail and specific examples, making it somewhat unclear and incomplete compared to the ground truth answer. SCORE: 41 | | |
|---|---|---|---|---|---|

| | | microfluidics, emulsification, and lab-on-a-chip systems. | | | |
|---|---|---|---|---|---|
| What factors affect droplet coalescence in microfluidic devices? | Surface tension, viscosity, flow rate, channel geometry, and the presence of surfactants. | Droplet coalescence in microfluidic devices is influenced by several factors. Surfactant concentration plays a key role, with higher concentrations preventing droplet fusion, while lower concentrations promote coalescence. The duration droplets remain in contact within the pairing channel, known as pairing time, also impacts coalescence, with longer times favoring fusion. The geometry of the pairing channel, specifically its cross-sectional area, controls the number of droplets that can come into contact and coalesce. Additionally, droplet size and volume affect how efficiently droplets merge. Finally, electrocoalescence, which applies electric fields to induce droplet | The response provides a concise list of factors affecting droplet coalescence but lacks the depth and detail provided by the ground truth answer, particularly regarding the specific effects of surfactant concentration and pairing time. SCORE: 41 | 41 | 41.0 |

| | | fusion, offers an alternative method for on-demand coalescence. By optimizing these factors—surfactant concentration, pairing time, channel geometry, and droplet size—reliable and selective droplet coalescence can be achieved in microfluidic systems. | | | |
|---|---|---|---|---|---|
| What design elements are critical for producing double or multiple emulsions? | Droplet size, interfacial tension, viscosity of phases, surfactant type and concentration, homogenization method, and temperature. | To generate double or multiple emulsions, several design elements are essential. Co-flow and flow-focusing configurations enable precise control over fluid flow, allowing for stable droplet formation, especially in double emulsions. Sequential emulsification steps, where inner droplets are emulsified into intermediate phases and then into outer phases, are key for multiple emulsions. Capillary microfluidic devices, such as consecutive capillary and coaxial configurations, are | The response provides some relevant information about the design elements of producing double or multiple emulsions, but it lacks detail and context. It only mentions a few general factors (droplet size, interfacial tension, viscosity of phases, surfactant type and concentrati | 16 | 16.0 |

| | | used to control droplet size, number, and monodispersity. Additionally, precise flow control and shear force management are crucial for stable and consistent droplet formation. Microchannel geometry optimization and material compatibility further enhance emulsification processes. By carefully managing these factors, researchers can produce double and multiple emulsions with controlled internal structures and compositions for various applications. | on, homogeniz ation method, and temperatur e) without explaining their specific roles in generating these types of emulsions. SCORE: 16 | | |
|---|---|---|---|---|---|
| What key factors influence the design of flow-focusing droplet microfluidics optimized using machine learning? | Droplet size, viscosity ratio, flow rates, channel geometry, and surface tension. | The design of flow-focusing droplet microfluidics, optimized using machine learning, depends on several critical factors. Channel geometry, including the shape and size, controls droplet generation by defining the break-off point where the dispersed phase meets the continuous phase. Flow rates of both | The response is partially correct but lacks detail and context; it only mentions a few key factors without explaining their significanc e or relationshi | 41 | 41.0 |

| | | phases affect droplet size and frequency, with higher continuous phase flow reducing droplet size. Viscosity influences flow behavior and shear forces, crucial for stable droplet formation. Shear forces between phases are controlled by adjusting channel geometry or flow rates, ensuring consistent droplet generation. Machine learning algorithms predict and optimize droplet diameter and generation rate by analyzing these design parameters, fluid properties, and large datasets, automating the design process for more efficient droplet production in life sciences applications. | p to machine learning optimizatio n. SCORE: 41 | | |
|---|---|---|---|---|---|
| How do channel wall properties (hydrophobicity/hyd rophilicity) influence droplet dynamics? | The hydrophobicit y or hydrophilicity of a channel wall influences droplet dynamics by affecting the interfacial tension | The channel wall properties, specifically hydrophobicity or hydrophilicity, significantly influence droplet dynamics in microfluidic systems. These properties determine the | The response provides a clear explanatio n of how channel wall properties influence droplet dynamics, | 56 | 56.00000000 000001 |

| | between the droplet and the channel surface. Hydrophobic walls tend to repel water droplets, leading to smaller droplets with higher mobility, while hydrophilic walls attract water droplets, resulting in larger droplets that are less mobile. | wettability of the channel walls, affecting the interaction between the continuous and dispersed phases. In hydrophobic channels, the dispersed phase (typically water) tends to form droplets as it minimizes contact with the channel walls. This enhances droplet stability and facilitates easier detachment during droplet formation. Hydrophobic surfaces are ideal for applications requiring water-in-oil emulsions or precise droplet control. Conversely, hydrophilic channels promote wetting of the dispersed phase, making it more likely to spread along the walls instead of forming discrete droplets. This is useful for oil-in-water systems or applications involving liquid films. Wall properties also impact droplet size, breakup, coalescence, and transport. | mentioning the effects on interfacial tension, droplet size, and mobility. However, it lacks detailed information about the impact on droplet breakup, coalescence, and transport, as well as specific examples or applications. SCORE: 56 | | |
|---|---|---|---|---|---|
| | | Hydrophobicity favors smaller droplets and reduces the likelihood of droplet adherence, while hydrophilicity can lead to higher flow resistance and altered droplet generation behavior. Tailoring wall properties is crucial for optimizing microfluidic processes. | | | |
| How can you incorporate active control mechanisms (e.g., electrical, magnetic) into droplet devices? | By integrating electrodes or magnetic coils within the device structure to manipulate droplet movement and behavior through electric fields or magnetic forces. | Active control mechanisms such as electrical, magnetic, thermal, and mechanical methods can precisely manipulate droplet formation in microfluidic devices. Electrical control methods include direct current (DC), which uses electrodes to control droplet size, and alternating current (AC), which utilizes electrowetting-on-dielectric (EWOD) and dielectrophoresis (DEP) to influence droplet behavior. Magnetic control employs ferrofluids and external magnets to non-invasively guide | The response attempts to address the question but lacks detail and specificity; it only mentions two types of control mechanisms (electrical and magnetic) without elaborating on their application or other relevant methods. SCORE: 16 | 16 | 16.0 |

| | | droplet formation. Thermal control involves heating the entire device or applying localized heating to adjust fluid parameters like viscosity and interfacial tension. Mechanical control techniques, such as pneumatic and piezoelectric components, can also regulate droplet dynamics. These active control strategies enable precise and flexible droplet generation, allowing for tailored applications in complex microfluidic systems, including droplet manipulation and multicomponent emulsion creation. | | | |
|---|---|---|---|---|---|
| How do flow rates affect droplet size and frequency? | Higher flow rates generally result in larger droplet sizes and increased droplet frequency. | Flow rates significantly influence droplet size and frequency in microfluidic systems. As the flow rate ratio ($\varphi = Qd/Qc$) increases, the shearing force from the continuous phase dominates, leading to smaller droplets due to rapid breakup of the dispersed phase. Additionally, higher flow rates in the | The response is concise and directly addresses the question, but lacks detail and nuance compared to the ground truth. It provides a clear | 41 | 41.0 |

| | | continuous phase can result in larger droplets, as the dispersed phase doesn't break up until it reaches the downstream section. The frequency of droplet generation also increases with higher flow rate ratios, demonstrating a positive correlation. However, droplet diameter growth trends vary depending on the model and input conditions. Overall, increasing the flow rate ratio enhances droplet formation frequency and reduces droplet size due to stronger shearing forces, with complex interactions influenced by factors such as interfacial tension and channel geometry. | statement about the relationship between flow rates and droplet size/freque ncy, but doesn't delve into the underlying mechanism s or complexiti es mentioned in the ground truth. SCORE: 41 | | |
| What techniques are available for droplet splitting? | Droplet microfluidics, acoustic droplet ejection, inkjet printing. | Droplet splitting techniques can be categorized into passive and active methods. Passive techniques rely on the microchannel geometry and fluid properties, such as T and Y junctions, flow focusing | The answer provides a list of techniques (droplet microfluidi cs, acoustic droplet ejection, inkjet printing) | 10 | 10.0 |

| | | configurations, or introducing ribs/obstructions in the flow path. These methods are simple but lack dynamic control over splitting ratios and are sensitive to flow conditions. Active techniques utilize external forces like electric, magnetic, acoustic, or thermal fields to precisely control droplet splitting. Examples include electric field-based splitting, magnetic fields for ferrofluids, acoustic radiation force, and thermocapillary methods. Active techniques offer more flexibility, enabling on-demand droplet splitting with controlled ratios, overcoming the limitations of passive methods. However, each technique has its own set of advantages and constraints depending on the application and droplet type. | but lacks explanation or context. It does not address the distinction between passive and active methods as mentioned in the ground truth. SCORE: 10 | | |
| What advantages does droplet formation by passive methods | Passive methods are typically less energy-intensive and | Passive droplet generation methods offer several advantages over active methods. | The response only mentions two | 16 | 16.0 |

| have over active methods? | more cost-effective. | They simplify device design by relying on flow rate manipulation and microchannel geometry, eliminating the need for complex external forces. This leads to cost-effectiveness and easier fabrication. Passive methods also provide precise control over droplet size, shape, and monodispersity, making them ideal for applications requiring uniform droplets. Additionally, they protect droplets from external interference and cross-contamination, ensuring integrity. Passive methods can produce continuous droplet streams with uniform size distributions (1%-3% dispersity) without requiring external actuation. While they are not suitable for dynamic manipulation or "droplet-on-demand" applications, passive methods are favored for their | advantages of passive droplet formation methods (less energy-intensive and more cost-effective), whereas the ground truth answer highlights several additional benefits such as simplified device design, precise control over droplet size and shape, protection from external interferenc e, and ability to produce continuous droplet streams. SCORE: 16 | | |
|---|---|---|---|---|---|

| | | simplicity, precision, and cost-efficiency in controlled environments. | | | |
|---|---|---|---|---|---|
| How do you design a device for monodisperse droplet generation? | A common method is using a microfluidic chip with a T-junction where one channel carries the liquid to be emulsified and another carries the continuous phase. Precise control of flow rates and geometry allows for monodisperse droplet formation. | To design a device for monodisperse droplet generation, two common methods are used: cross-flow and flow-focusing designs. In cross-flow designs, the continuous phase flows around the dispersed phase in a constriction channel, with droplet size controlled by adjusting channel dimensions and shear stress. This method is simple but can lead to droplet damage due to shear forces or wall interactions. Flow-focusing designs, on the other hand, use sheath flows to stabilize the interface between phases, enabling smaller droplet formation with precise control over size by adjusting flow rates and orifice dimensions. This method offers high monodispersity and is less prone to droplet damage. | The response provides a basic understanding of the concept but lacks detail and accuracy compared to the ground truth answer. It only mentions one method (microfluidic chip with T-junction) and does not discuss the limitations or alternatives, such as cross-flow and flow-focusing designs. SCORE: 41 | 41 | 41.0 |
| | | Both designs require careful optimization of flow rates, channel dimensions, and surface tension to achieve uniform droplet sizes. The choice of design depends on specific needs such as droplet size, throughput, and resistance to damage during formation. | | | |
| What materials are most commonly used for fabricating droplet microfluidic devices? | Polydimethyl siloxane (PDMS) and glass. | The most commonly used materials for fabricating droplet microfluidic devices are PDMS (polydimethylsiloxane) and silicon-based materials. PDMS is preferred due to its ease of fabrication using soft lithography, which allows for the creation of precise and versatile microfluidic devices. This material provides high transparency, efficient sealing, and the ability to mold complex microchannels, making it ideal for droplet generation and manipulation. While PDMS is dominant, alternatives such as | The answer is partially correct but lacks detail and context; it does not provide a comprehensive explanation of the materials used in fabricating droplet microfluidic devices. SCORE: 41 | 41 | 41.0 |

| | | thermoplastics (e.g., PMMA, polycarbonate), fluoropolymers (e.g., PTFE), and glass materials are also used, each offering unique benefits like better solvent resistance or mass production capabilities. Despite the variety of materials available, PDMS remains the material of choice for most droplet microfluidics applications due to its simplicity and versatility. | | | |
|---|---|---|---|---|---|
| What are the advantages and limitations of using PDMS for microfluidic devices? | Advantages: Biocompatibility, optical transparency, ease of fabrication, chemical inertness. Limitations: Gas permeability, low thermal stability, susceptibility to swelling in organic solvents. | Polydimethylsiloxane (PDMS) is widely used for microfluidic devices due to its several advantages, including low toxicity, biocompatibility, optical transparency, and high gas permeability, which make it ideal for biomedical applications. Its flexibility and cost-effectiveness also contribute to its popularity, as it allows for easy fabrication using soft lithography. PDMS enables real-time observation | The response provides a clear and concise list of advantages (biocompatibility, optical transparency, ease of fabrication, chemical inertness) and limitations (gas permeability, low thermal stability, susceptibility to swelling in | 41 | 41.0 |

| | | and is well-suited for flexible and wearable electronics. However, PDMS has limitations, such as its hydrophobic nature, which can lead to fouling in microchannels, affecting performance. Additionally, its soft elastomeric properties limit mechanical strength, and large-scale production can be challenging. Despite these drawbacks, PDMS remains a versatile and widely used material, though further research is needed to address its limitations for broader applications. | organic solvents) of PDMS for microfluidic devices. However, it lacks the depth and detail provided by the ground truth answer, particularly regarding its biomedical applications, flexibility, cost-effectiveness, hydrophobic nature, and mechanical strength. The response does not address the broader context or potential future research directions related to PDMS.

SCORE: 41 | | |
|---|---|---|---|---|---|

| How does 3D printing facilitate microfluidic device fabrication? | 3D printing enables the precise fabrication of complex microfluidic channels and structures with high resolution, allowing for customized designs and intricate geometries. | 3D printing revolutionizes microfluidic device fabrication by enabling the creation of complex and precise channel structures that are challenging to achieve with traditional methods like soft lithography. Techniques such as Stereolithography (SL), Multijet Modeling (MJM), and Fused Deposition Modeling (FDM) offer high resolution, complex geometries, and the integration of various components like valves and pumps. 3D printing also allows rapid prototyping, customization, and the use of diverse materials for biocompatibility and transparency. Advantages over traditional methods include reduced fabrication time and labor, increased accuracy, and environmental friendliness due to the additive manufacturing process. With its ability to create | The response provides a clear and concise answer that addresses the core aspects of how 3D printing facilitates microfluidic device fabrication. It highlights the precision, customization, and high resolution capabilities of 3D printing in creating complex channel structures. However, it lacks some depth and details compared to the ground truth answer, which covers additional techniques, advantages, and potential | 46 | 46.0 |
| --- | --- | --- | --- | --- | --- |

| | | hierarchical 3D configurations and integrate multiple functions, 3D printing holds great potential for advancing microfluidic device design and performance. | benefits of 3D printing in this context. SCORE: 46 | | |
|---|---|---|---|---|---|
| What alternative materials can be used for applications involving organic solvents? | Glass, ceramics, fluoropolymers, and certain metals. | Several sustainable alternatives to traditional organic solvents are gaining attention for various applications. Supercritical CO2 is a non-toxic, environmentally friendly solvent, particularly useful in extraction processes. Ionic liquids, which are liquid salts at room temperature, offer low vapor pressure and broad application in chemical reactions. Organic carbonates, like dimethyl carbonate, are biodegradable and non-toxic, ideal for fine chemical synthesis. Biosolvents, derived from renewable resources such as plants and bacteria, provide a greener option for extraction and cleaning. Other alternatives include | The answer provides a list of materials that can be used as alternatives to organic solvents but lacks the detailed explanation and context provided in the ground truth. It does not address the broader applications, properties, or environmental benefits of these alternatives. SCORE: 21 | 21 | 21.0 |

| | | fluorous solvents, non-toxic liquid polymers like polyethylene glycol (PEG), water, ethanol, methanol, and ethyl acetate. These alternatives are chosen based on their specific properties such as polarity, boiling point, and environmental impact, promoting more sustainable chemical processes and reducing reliance on hazardous petroleum-based solvents. | | | |
|---|---|---|---|---|---|
| How does device surface treatment affect droplet formation? | Device surface treatment can influence droplet formation by altering the surface energy and wettability. | Surface treatment of PDMS microfluidic devices plays a critical role in determining droplet formation and stability. PDMS is naturally hydrophobic, suitable for generating water-in-oil (W/O) droplets but not oil-in-water (O/W) droplets. To generate O/W droplets, selective hydrophilic treatment is required. The researchers employed two methods: manual PVA deposition after plasma | The response provides some relevant information about device surface treatment and its effect on droplet formation, but lacks detail and context compared to the ground truth answer. It does not mention | 16 | 16.0 |

| | | treatment, and a simultaneous approach combining PVA injection with on-chip plasma treatment. The latter method enables precise wettability patterning, allowing for both W/O and O/W droplet formation within the same device. The plasma treatment increases surface energy, making the PDMS hydrophilic, while PVA deposition stabilizes the coating, ensuring long-term device use. This technique provides controllable droplet sizes and stable double emulsion formation, with the ability to manipulate droplet characteristics by adjusting flow rates. The method also allows for reliable, reproducible results, making it a powerful tool for droplet microfluidic applications. | specific methods of surface treatment or their effects on PDMS devices, nor does it discuss the importance of surface energy and wettability in droplet formation. SCORE: 16 | | |
|---|---|---|---|---|---|
| What bonding techniques are used to seal microfluidic channels? | UV curing, thermal bonding, plasma bonding, epoxy | Several bonding techniques are employed to seal microfluidic channels, each suited to different materials. Thermal | The response provides a list of specific techniques (UV | 45 | 45.0 |

| | | | | |
|---|---|---|---|---|
| | adhesives. | fusion bonding involves compressing thermoplastic substrates at temperatures near their glass transition point, allowing them to fuse upon cooling. Chemical bonding creates covalent bonds between substrate surfaces through chemical reactions, while solvent bonding uses solvents to dissolve the materials partially, promoting adhesion through interdiffusion. For PDMS devices, oxygen plasma treatment is commonly used to make surfaces reactive, enabling irreversible bonding when in contact with another PDMS or glass piece. Additionally, corona wand bonding offers a less expensive alternative to oxygen plasma treatment, though it is less consistent. The choice of bonding technique depends on material properties, bond strength requirements, and | curing, thermal bonding, plasma bonding, epoxy adhesives) that are relevant to sealing microfluidic channels. However, it lacks detailed explanations and context as provided in the ground truth answer. SCORE: 45 | | |
| | | available equipment, all of which are crucial for the successful fabrication of microfluidic devices. | | | |
| What challenges arise when fabricating complex microfluidic geometries? | Challenges include precise material deposition, maintaining channel alignment and dimensions, integrating multiple materials and functionalities, and achieving leak-proof connections. | Fabricating complex microfluidic geometries presents several challenges, including controlling wettability in intricate channel structures, which is critical for fluid flow and droplet formation. Additionally, scaling up microfabrication processes for commercial applications requires cost-effective techniques and materials. Current droplet generation methods mainly produce single emulsions, limiting their use in applications that require multiple or complex emulsions. Moreover, droplet microfluidics faces limitations in producing diverse and functional nano-scale materials compared to other methods. Conventional | The response touches on some relevant aspects such as material deposition, channel alignment, and leak-proof connections but misses key points like wettability control, scaling up microfabrication processes, limitations in droplet generation, and challenges with 3D geometries. SCORE: 16 | 16 | 16.0 |

| | | lithography techniques, which are restricted to rectangular cross-sections, struggle with creating complex 3D geometries. Although advancements like 3D printing and stacking methods offer solutions, effective mixing strategies and versatile fabrication of 3D networks with the required features remain significant hurdles in the field. Addressing these challenges will require innovative approaches and collaboration across various disciplines. | | | |
|---|---|---|---|---|---|
| How do you fabricate devices capable of handling multiphase flows? | Fabrication techniques for multiphase flow devices typically involve microfabrication processes like photolithography, etching, and deposition. Specific methods depend on the device's design and | Fabricating devices for handling multiphase flows involves several techniques and considerations. Key approaches include advanced micromanufacturing methods such as 3D printing, which eliminates the need for bonding processes, and flow-focusing techniques, which use nozzles and sheath fluids to control droplet | The response provides a basic understanding of the topic by mentioning microfabrication processes like photolithography, etching, and deposition. However, it lacks | 41 | 41.0 |

| | materials. | formation. Controlling wettability is crucial for generating specific droplet types, while parallelization of droplet generators enhances production scalability. Challenges include ensuring even fluid distribution, integrating various components like valves and connectors, and addressing the lack of standardized design rules. Additionally, methods like consecutive capillary assemblies and external forces (e.g., pneumatic or electrical control) can be employed to create complex emulsions. Through careful design and optimization, efficient fabrication of multiphase flow devices for applications such as droplet generation can be achieved. | depth and specific details about handling multiphase flows, and does not address key considerations such as controlling wettability or parallelization of droplet generators. SCORE: 41 | | |
|---|---|---|---|---|---|
| What is the impact of channel roughness on droplet formation and behavior? | Channel roughness influences droplet formation by affecting surface | The impact of channel roughness on droplet formation and behavior is a key factor in microfluidics. | The response attempts to address the question but lacks depth and | 31 | 31.0 |

| | | | | |
|---|---|---|---|---|
| | tension, contact angle, and flow patterns. Rough surfaces can promote droplet coalescence and alter their trajectory and size distribution. | Studies indicate that surface roughness can significantly affect droplet dynamics, such as movement, spreading, and stability. For example, increased roughness can hinder droplet spreading and prevent secondary droplet formation by altering flow dynamics and the contact line. Additionally, vibration and surface structure play roles in influencing droplet impact and splash behavior. Rougher surfaces tend to reduce spreading, while smoother surfaces may enhance it. Surface wettability also interacts with roughness to further complicate droplet behavior. Overall, channel roughness introduces complexity to droplet dynamics, with its effects varying based on factors like roughness level, fluid type, and external conditions, highlighting the need for further | clarity compared to the ground truth. It mentions some relevant factors such as surface tension, contact angle, and flow patterns, but does not provide a comprehensive explanation of the impact of channel roughness on droplet formation and behavior. SCORE: 31 | | |

| | | research in this area for better understanding and optimization in microfluidic applications. | | | |
|---|---|---|---|---|---|
| How can sensors be integrated into droplet microfluidic devices during fabrication? | Sensors can be integrated into droplet microfluidic devices during fabrication through various methods, including:<br><br>* **Microfabrication techniques:** Sensors can be patterned and etched onto the device substrate alongside the microchannels using photolithography, etching, and deposition processes.<br>* **Embedding sensors within the channels:** Sensors can be incorporated directly into the | Integrating sensors into droplet microfluidic devices during fabrication is a complex process that requires advanced microfabrication techniques and careful material selection. For successful integration, precision in manufacturing is crucial, as it often involves combining different microfluidic units, such as droplet generation and detection units, on a single chip. Various techniques like photolithography, soft lithography, and direct integration methods can be used, depending on the type of sensor (e.g., optical or electrochemical). The fabrication process also needs to account for material compatibility, sensor functionality, and alignment to | The response provides a clear and concise overview of methods to integrate sensors into droplet microfluidic devices during fabrication. It lists specific techniques such as microfabrication, embedding sensors within channels, surface functionalization, and integration with external sensing systems. However, the answer lacks depth in explaining the complexities involved | 51 | 51.0 |
| | microfluidic channels by embedding them in the material used to construct the channels. * **Surface functionalization:** Sensors can be attached to the surface of the microfluidic channels or chambers through chemical bonding or physical adsorption. * **Integration with external sensing systems:** External sensors can be coupled to the droplet microfluidic device via optical fibers, electrical connections, or other interfaces. | ensure reliable performance. Additionally, sensor integration may face challenges such as maintaining stability within the microfluidic environment and minimizing interference with fluid flow. By considering these factors, sensors can be effectively integrated into droplet microfluidic devices for applications like chemical reactions and biological detection. | in this process, such as material selection, precision manufacturing, and potential challenges like maintaining stability and minimizing interference. SCORE: 51 | | |
| How are droplet microfluidic devices used for single-cell analysis? | Droplet microfluidic devices encapsulate individual | Droplet microfluidic devices are widely used for single-cell analysis due to their ability | The response is mostly relevant and clear, | 58 | 57.99999999 999999 |

| | | | | |
|---|---|---|---|---|
| | cells within tiny water droplets, allowing for parallel processing and analysis of thousands of cells simultaneously. This enables the study of gene expression, protein production, and cellular behavior at the single-cell level. | to encapsulate individual cells within tiny droplets, allowing for high-throughput and precise experimentation. These devices work by generating microdroplets in a fluid flow, each containing a single cell along with reagents needed for analysis, such as enzymes, antibodies, or primers. This allows for the isolation of cells in a controlled environment, enabling high-resolution studies of gene expression, protein profiling, and cellular responses. The compartmentalized nature of droplets also reduces cross-contamination between samples. Moreover, the small size and scalability of droplet microfluidics make it possible to perform large-scale, parallel analysis of thousands of single cells in a short amount of time. This technology has applications in genomics, proteomics, and | but lacks focus or includes unnecessary details; broadly acceptable. SCORE: 58 | | |
| | | drug screening, facilitating insights into cell heterogeneity and the identification of rare cell populations. | | | |
| What are the advantages of droplet microfluidics for high-throughput screening? | Droplet microfluidics enables massively parallel processing, reduced reagent consumption, and precise control over reaction conditions. | Droplet microfluidics offers several advantages for high-throughput screening, including reduced sample volumes, which lowers costs, and increased assay sensitivity through compartmentalization. This method allows for the screening of up to $10^8$ samples per day at kilohertz speeds, significantly improving throughput and efficiency. The small droplet volumes enhance effective concentrations and signal-to-noise ratios, critical for sensitive detection of low-concentration biological molecules such as antibodies, enzymes, and single genomes. Additionally, droplet microfluidics enables single-cell | The response provides some relevant information about droplet microfluidics but lacks detail and accuracy compared to the ground truth. It mentions advantages such as massively parallel processing, reduced reagent consumption, and precise control over reaction conditions, which are partially correct but do not fully capture the | 41 | 41.0 |

| | | and single-molecule assays, providing detailed insights into cellular behaviors and allowing for novel experimental approaches. This technology is particularly beneficial for applications like drug discovery, enzyme evolution, and diagnostics, offering a versatile and cost-effective platform for advanced biological research. | benefits of this technology for high-throughput screening. SCORE: 41 | | |
|---|---|---|---|---|---|
| How can droplets be used to encapsulate cells or biomolecules? | Droplets can encapsulate cells or biomolecules through a process called microfluidics. | Droplet microfluidics provides versatile methods for encapsulating cells and biomolecules. Limiting dilution is a traditional approach that randomly distributes single cells or biomolecules in droplets, though it has low efficiency. Inertial microfluidics leverages spiral geometries and inertial forces to guide cells into droplets, improving encapsulation efficiency by ensuring single-cell droplets. Bi- | The response attempts to address the question but lacks detail and accuracy; it only mentions a process called microfluidics without elaborating on its specifics or other methods for encapsulating cells or biomolecules. SCORE: 16 | 16 | 16.0 |

| | | disperse emulsions create an emulsion of empty and cell-containing droplets, with the latter selectively recovered using methods like pinched flow fractionation. For biomolecules, limiting dilution is commonly used, adjusting concentration and drop size to achieve single-molecule encapsulation. Additionally, droplets can encapsulate cells in lipid vesicles for biological applications or serve as microreactors in processes like DNA analysis. These techniques enable precise control over encapsulation, enhancing their applicability in cell-based assays, drug discovery, and genetic research | | | |
|---|---|---|---|---|---|
| What are the applications of droplet microfluidics in drug discovery? | Droplet microfluidics enables high-throughput screening, single-cell analysis, and personalized medicine approaches in | Droplet microfluidics plays a significant role in drug discovery by enabling high-throughput screening and precise analysis of drug interactions at the single-cell level. | The response is concise and directly addresses one aspect of droplet microfluidics' | 41 | 41.0 |

| | | | | |
|---|---|---|---|---|
| | drug discovery. | This technology allows the generation of picoliter-sized droplets, each encapsulating a single cell or a small group of cells, which can be exposed to various drug compounds. This compartmentalization facilitates parallel testing of numerous drug candidates, speeding up the identification of promising therapeutics. Droplet microfluidics is particularly useful in identifying drug efficacy, optimizing drug formulations, and studying complex cellular responses such as gene expression, protein production, or metabolic changes in response to treatment. Additionally, it enables the screening of rare cell populations or targets that would be difficult to isolate using traditional methods. The ability to rapidly test and analyse thousands | applications in drug discovery. It mentions high-throughput screening and single-cell analysis but lacks detail and context compared to the ground truth answer, which provides a more comprehensive overview of the technology's capabilities and benefits. SCORE: 41 | | |
| | | of drug interactions in a controlled environment accelerates drug discovery, reduces costs, and improves the chances of identifying novel therapeutic agents. | | | |
| How can droplet microfluidics enable precise control of chemical reactions? | Droplet microfluidics enables precise control of chemical reactions by encapsulating individual reactants within tiny droplets, allowing for independent manipulation and reaction conditions. | Droplet microfluidics offers precise control over chemical reactions by dispensing reactants into individual droplets with accurate volume control. This enables the mixing of reactants within confined spaces, enhancing reaction efficiency. Coalesced droplets act as micro-reactors, providing fast mixing and long operation times without clogging, even with sticky or precipitating reactants. The system allows for optimization of reaction parameters, such as temperature, flow rate, and pressure, ensuring the production of specific products, like mesoporous silica microspheres with superior surface area and | The response is mostly relevant but lacks focus on the question's core aspect of "precise control" and includes minor inaccuracies regarding the specifics of how droplet microfluidics enables this control. SCORE: 49 | 49 | 49.0 |
| | | narrow pore distributions. This capability to finely control reaction conditions makes droplet microfluidics a powerful tool for precise, high-throughput chemical synthesis. | | | |
| What are the key challenges of using droplet microfluidics in diagnostics? | Cost, scalability, and integration with existing diagnostic platforms. | Droplet microfluidics presents significant challenges in diagnostics, including preventing wetting and cross-contamination between aqueous droplets and immiscible organic liquids, which can compromise results. Monitoring oxygen levels within droplets is difficult due to the sensitivity of the liquid-liquid interface to mechanical perturbations and mass transfer. Developing stable detection chemistries is also complex, as the confined droplet environment can alter chemical reactions and impact accuracy. Additionally, challenges in | The response only partially addresses the question by mentioning cost, scalability, and integration with existing diagnostic platforms, but misses the key technical challenges specific to droplet microfluidics. SCORE: 21 | 21 | 21.0 |
| | | automation and accessibility limit the widespread use of droplet microfluidics, particularly for high-throughput assays. Overcoming these issues requires advancements in surfactants, microfluidic engineering, and automated techniques, which could enhance the precision and applicability of droplet systems in diagnostics. | | | |
| How are droplets used for DNA sequencing applications? | Droplets encapsulate individual DNA molecules, allowing for parallel sequencing and amplification. | Droplet microfluidics plays a key role in DNA sequencing by enabling high-throughput analysis and precise targeting of genomic regions. This includes single-cell sequencing, where individual cells are encapsulated in droplets for independent gene expression profiling. Barcode-based sequencing allows for the labeling of nucleic acids with unique identifiers, enabling the tracking of specific genomes during analysis. | The response provides a concise and accurate description of how droplets are used for DNA sequencing applications, but it lacks detail and context compared to the ground truth answer. It only mentions one aspect of droplet microfluidi | 41 | 41.0 |

| | | Multiplexed PCR amplification is achieved by performing PCR in separate droplets, ensuring uniform amplification of multiple targets. Additionally, emulsion PCR combines droplet merger with PCR to amplify specific genomic regions, facilitating efficient sequencing. These techniques offer a miniaturized, efficient platform for analyzing complex genetic information, improving accuracy and throughput in genomic studies. | cs in DNA sequencing , whereas the ground truth answer covers multiple techniques and their benefits. SCORE: 41 | | |
|---|---|---|---|---|---|
| What role do droplets play in tissue engineering research? | Droplets are used to create microenviron ments for cells, mimicking the natural conditions they experience in the body. | Droplet microfluidics plays a pivotal role in tissue engineering by enabling precise fabrication of biomaterials and 3D cell culture systems. Droplets serve as templates for creating hydrogel-based materials, which can encapsulate cells for more accurate modeling of tissue structures. This technology facilitates the production of microcarriers and | The response attempts to address the question but lacks detail and accuracy compared to the ground truth answer; it only mentions one aspect of droplets in tissue engineerin g research, without | 16 | 16.0 |

| | | complex biomaterials, such as hollow bacterial cellulose (BC) microspheres, which form injectable scaffolds for wound healing, promoting high cell proliferation and tissue regeneration in vivo. Additionally, droplet-based microfluidics allows for the creation of core-shell structures, such as alginate shells or metal-organic frameworks, which offer controlled release of therapeutic agents. These advancements provide a powerful platform for studying tissue development and regenerative therapies. | providing comprehensive information. SCORE: 16 | | |
|---|---|---|---|---|---|
| How can droplet microfluidics aid in studying cell signaling pathways? | Droplet microfluidics allows for the encapsulation of single cells or small groups of cells within individual droplets, enabling precise control over their | Droplet microfluidics offers a powerful platform for studying cell signaling pathways by enabling the precise isolation and manipulation of individual cells within controlled environments. This technology allows for the encapsulation of | The response accurately conveys the main idea of how droplet microfluidics can aid in studying cell signaling pathways, | 46 | 46.0 |

| | environment and facilitating the study of cell signaling pathways in isolation. | single cells in droplets, providing a controlled space to study cell-cell interactions and signaling events. The ability to merge droplets containing different cell types or reagents facilitates the investigation of cell communication, such as paracrine signaling, in a way that mimics in vivo conditions. Additionally, droplet-based microfluidics supports high-throughput screening and single-cell genomic studies, offering insights into phenotypic variations and signaling dynamics within genetically identical cell populations. This capability is particularly valuable for understanding complex cellular behaviors in development, differentiation, and disease. | but lacks detail and specific examples compared to the ground truth. It also doesn't mention high-throughput screening and single-cell genomic studies. SCORE: 46 | | |
| --- | --- | --- | --- | --- | --- |
| What is the potential of droplet microfluidics in personalized medicine? | Droplet microfluidics enables high-throughput, single-cell | Droplet microfluidics holds great promise in personalized medicine by | The response provides a clear and concise | 46 | 46.0 |

| | analysis and drug screening, facilitating personalized treatment strategies and disease diagnosis. | enabling precise, high-throughput analyses of biological samples. The technology allows for the optimization of treatments by testing drug combinations on specific cell lines, such as pancreatic cancer cells, to identify the most effective therapies. It also facilitates the early detection of cancers through highly sensitive detection of tumor markers in bodily fluids, offering a minimally invasive approach for patient monitoring. Additionally, droplet-based digital PCR (dPCR) enables the detection of rare sequences and precise quantification of nucleic acids, which is crucial for cancer research and tailoring treatments based on a patient's genetic background and tumor profile. Overall, droplet microfluidics can significantly advance personalized medicine by | overview of the potential of droplet microfluidics in personalized medicine, but lacks specific examples and details compared to the ground truth. It also doesn't mention digital PCR or its applications. SCORE: 46 | | |
|---|---|---|---|---|---|

| | | providing tailored treatment strategies and improving early detection for better patient outcomes. | | | |
|---|---|---|---|---|---|
| How are droplet microfluidic devices used for nanoparticle synthesis? | Droplet microfluidic devices create thousands of tiny water droplets, each acting as a reaction chamber. Nanoparticles can be synthesized within these droplets by controlling the reagents and conditions inside each droplet. | Droplet microfluidic devices revolutionize nanoparticle synthesis by enabling precise control over reaction conditions within isolated microreactors. These droplets encapsulate well-defined quantities of reagents, ensuring rapid mixing and preventing contamination, even when using toxic or volatile chemicals. This approach addresses limitations like diffusion-limited mixing found in traditional systems, resulting in nanoparticles with controlled size, shape, and narrow size distribution. The method supports multi-step synthesis on millisecond timescales, allowing for the scalable production of high-quality nanoparticles with applications in | The response provides a clear and concise explanation of how droplet microfluidic devices are used for nanoparticle synthesis. It accurately describes the process of creating tiny water droplets that act as reaction chambers and controlling reagents within them to synthesize nanoparticles. However, it lacks detail on the benefits and advantages of this method compared | 51 | 51.0 |

| | | fields such as drug delivery, imaging, and catalysis. | to traditional systems, such as preventing contamination and achieving controlled size distribution. SCORE: 51 | | |
|---|---|---|---|---|---|
| What role do droplets play in polymer particle production? | Droplets act as microreactors, encapsulating monomers and initiating polymerization. | Droplet-based microfluidics provides a versatile platform for producing monodisperse polymer particles by encapsulating monomers or oligomers into droplets and solidifying them through chemical or physical crosslinking. This approach enables the creation of diverse particle types, including solid, hollow, asymmetric, and multicompartmental structures with precise control over size, shape, and morphology. Techniques such as UV-initiated polymerization and solvent evaporation facilitate the transformation of droplets into solid | The response provides a concise and accurate description of the role of droplets in polymer particle production, but lacks detail and context compared to the ground truth. It does not mention the variety of particle types that can be produced or the specific techniques used for solidification. | 41 | 41.0 |
| | | particles. Additionally, functional micro- or nanostructures can be incorporated within or on the surface of these particles by introducing specific ingredients into the droplets, expanding their potential applications in areas like drug delivery, tissue engineering, and advanced materials. | SCORE: 41 | | |
| How can droplet microfluidics improve the reproducibility of chemical synthesis? | Droplet microfluidics allows for precise control over reaction volumes and conditions, minimizing variations between experiments. | Droplet microfluidics enhances the reproducibility of chemical synthesis by providing precise control over reaction conditions and minimizing variability. Droplets act as isolated microreactors, ensuring accurate manipulation of parameters such as temperature and concentration, critical for consistent results. This system enables optimal mixing intensity and parallel processing in separate droplet channels, reducing batch-to-batch variability and ensuring uniform product quality. | The response is concise and directly addresses the question by highlighting a key benefit of droplet microfluidics in chemical synthesis. However, it lacks some details and nuances present in the ground truth answer, such as the specific advantages of precise | 51 | 51.0 |
| | | Automation further minimizes human error, making the process highly reliable. These advantages are particularly beneficial for synthesizing complex nanomaterials and exploring new reaction parameter spaces, leading to the production of high-quality, reproducible materials. | control over temperature and concentration, optimal mixing intensity, and parallel processing. SCORE: 51 | | |
| What are the applications of droplet microfluidics in catalysis studies? | Droplet microfluidics enables high-throughput screening of catalysts, precise control over reaction conditions, and the study of heterogeneous catalysis. | Droplet microfluidics offers transformative applications in catalysis studies by enabling precise, efficient, and high-throughput investigations at the nanoscale. These systems facilitate enantioselective biocatalysis using nanoliter-scale devices that integrate microflow reactors and electrophoresis for enantiomeric differentiation. Droplets serve as isolated reaction containers, allowing for controlled catalytic transformations at the single-particle or cell level, with | The response provides some relevant information about the applications of droplet microfluidics in catalysis studies, but it is incomplete and lacks depth. It only mentions three general points without elaboration or context, whereas the ground truth | 41 | 41.0 |

| | | reaction volumes as small as a few picoliters. This confinement minimizes diffusion effects and enhances reaction efficiency. Applications span organic synthesis, biochemical analysis, and biotechnology, showcasing droplet microfluidics as a powerful tool for studying rapid reaction dynamics, optimizing catalysts, and exploring reaction mechanisms in real time. | answer provides a detailed explanation of the transformative capabilities of droplet microfluidics in catalysis research. SCORE: 41 | | |
|---|---|---|---|---|---|
| How do droplets enhance the study of reaction kinetics? | Droplets provide isolated microenvironments that allow for precise control over reactant concentrations and reaction conditions, enabling researchers to study reaction kinetics in a more detailed and accurate manner. | Droplet-based microfluidics revolutionizes reaction kinetics studies by offering rapid, precise, and high-throughput analysis with minimal sample consumption. The chaotic mixing within picoliter droplets ensures millisecond-level time resolution for fast reactions. These systems minimize reagent use, making them ideal for bioanalysis where resources are costly or scarce. Precise control over droplet | The response is mostly relevant and clear but lacks focus or includes unnecessary details; broadly acceptable. SCORE: 58 | 58 | 57.99999999 999999 |

| | | composition facilitates the creation of concentration gradients, enabling the measurement of kinetic parameters. High-throughput capabilities allow simultaneous monitoring of numerous reactions, significantly expanding experimental diversity. Additionally, extended tracking of individual droplets supports long-term monitoring of enzyme-inhibitor kinetics, providing deeper insights into reaction mechanisms. | | | |
| --- | --- | --- | --- | --- | --- |
| What are the benefits of using droplets for emulsion polymerization? | Improved control over particle size and morphology, enhanced stability of the emulsion, increased reaction rate. | Droplet-based systems offer significant advantages for emulsion polymerization, enhancing efficiency, precision, and scalability. They enable high production rates and precise control over polymer particle size and morphology, facilitating the synthesis of high-molar-mass | The response is partially correct but lacks detail and context; it only mentions three benefits of using droplets for emulsion polymeriza tion without providing any | 51 | 51.0 |

| | | polymers at faster reaction rates. Droplet parallelization allows the screening of diverse chemicals or concentrations and supports the encapsulation of millions of gene variants for quantitative emulsion PCR. These systems also empower directed enzyme or microorganism evolution through successive mutagenesis and selection rounds. Furthermore, droplet isolation prevents cross-contamination, adsorption, and solvent evaporation, making them ideal for applications like in vitro compartmentalization (IVC). This unique platform provides a robust, scalable approach for both analytical and biological studies. | additional information or explanation. SCORE: 51 | | |
|---|---|---|---|---|---|
| How can microfluidics be used to generate smart materials? | Microfluidics can be used to create patterned and controlled deposition of materials, | Microfluidics enables the creation of smart materials through precise control of reaction conditions, rapid synthesis, and | The response provides a clear and concise answer that addresses | 41 | 41.0 |

| | allowing for the fabrication of smart materials with specific functionalities. | innovative templating methods. Droplet templating allows monodisperse droplets to act as structural templates, facilitating the synthesis of uniform particles, such as hydrogels for single-cell analysis. By manipulating solute and temperature gradients on short timescales, microfluidics supports the production of nanoscale materials, including semiconducting polymers, catalysts, and nanoparticles with unique optical and electronic properties. Its ability to rapidly explore complex reaction parameter spaces, coupled with advanced analytics and machine learning, accelerates material discovery and optimization. This approach offers transformative potential in biotechnology, energy, and electronics by enabling the generation of | the question about microfluidics and smart materials, but it lacks depth and specific examples compared to the ground truth. It also doesn't mention some key aspects such as reaction conditions, rapid synthesis, templating methods, or advanced analytics and machine learning. SCORE: 41 | | |
|---|---|---|---|---|---|

| | | materials with tailored properties and enhanced functionality. | | | |
|---|---|---|---|---|---|
| What are the applications of droplet microfluidics in environmental sensing? | Droplet microfluidics enables sensitive and rapid detection of pollutants, pathogens, and biomarkers in environmental samples. Applications include water quality monitoring, air pollution analysis, and soil contamination assessment. | Droplet microfluidics offers transformative applications in environmental sensing by enabling precise, miniaturized, and efficient monitoring systems. It supports on-site environmental surveillance through portable devices for rapid toxicity assessments and detection of contaminants, such as mercury ions. As controllable microreactors, droplet microfluidics facilitates the study of complex interfacial processes, aiding in chemical synthesis and degradation studies. Furthermore, it advances environmental sensing by enabling in situ visualization and precise flow control, enhancing our understanding of environmental processes. These capabilities position droplet | The response provides some relevant information about the applications of droplet microfluidics in environmental sensing, but it lacks depth and detail compared to the ground truth answer. It does not mention specific examples or details about the technology's capabilities, such as on-site surveillance, miniaturization, and precise flow control. SCORE: 41 | 41 | 41.0 |

| | | microfluidics as a valuable tool for environmental science and engineering. | | | |
|---|---|---|---|---|---|
| How can microfluidic droplets improve combinatorial chemistry? | Microfluidic droplets enable the rapid and efficient screening of vast chemical libraries by encapsulating individual reactions in tiny, isolated compartments. This allows for parallel synthesis and testing of numerous combinations, accelerating the discovery of novel compounds. | Microfluidic droplets revolutionize combinatorial chemistry by drastically reducing reagent consumption and enhancing efficiency. Each droplet, acting as a picoliter-scale microreactor, requires six to eight orders of magnitude less material than traditional methods, minimizing waste and costs. This miniaturization enables high-throughput synthesis of large compound libraries, supporting applications like drug discovery and material synthesis. Additionally, droplet manipulation techniques allow for precise reaction control, combinatorial fusion of reagents, and multi-stage experiments with online analysis. These advantages make microfluidic | The response accurately captures the essence of how microfluidic droplets improve combinatorial chemistry, highlighting their ability to enable rapid screening of vast chemical libraries. However, it lacks some details present in the ground truth answer regarding reagent consumption reduction and precise reaction control. SCORE: 76 | 76 | 76.0 |

| | | droplets a powerful tool for automating complex chemical workflows and optimizing reaction conditions in a highly efficient and scalable manner. | | | |
|---|---|---|---|---|---|
| What are the challenges of scaling up droplet-based chemical synthesis? | Cost, throughput, reproducibility, and integration with downstream processes. | Scaling up droplet-based chemical synthesis faces several challenges. Achieving uniform fluid distribution in microfluidic devices remains difficult despite progress in parallel droplet generation, with 3D printing offering potential solutions for complex channel fabrication. Simplifying functional units for multistep reactions, such as droplet generation, merging, and phase separation, is vital for seamless continuous flow processes. Developing low-cost materials and standardizing chip designs are necessary for commercial viability, especially for high-throughput production in diverse fluids. Furthermore, creating complex droplet emulsions | The response is concise but lacks detail and context; it only mentions four general points without elaboration or connection to the broader topic of scaling up droplet-based chemical synthesis. SCORE: 16 | 16 | 16.0 |

| | | and expanding nanomaterial diversity are critical to meeting industrial demands. Addressing these challenges will enable safe, sustainable, and scalable microfluidic applications. | | | |
|---|---|---|---|---|---|
| How can artificial intelligence enhance droplet microfluidic research? | AI can automate data analysis, optimize experimental parameters, and accelerate the design of new microfluidic devices. | Machine learning can be used to optimize the design of microfluidic devices. Deep Learning will be integrated to sort, monitor, and process the images. Furthermore, the multimodal agents can be used to study and design the research of droplet microfluidics, a possible excuse robot to perform fabrication and experiments automatically. | The response partially addresses the question by mentioning automation and optimization in droplet microfluidic research, but misses the more advanced aspects mentioned in the ground truth answer such as deep learning for image processing and multimodal agents. SCORE: 51 | 51 | 51.0 |
| How can large language models | Large language | LLMs can be used to build the Agents | The response | 42 | 42.0 |
| enhance droplet microfluidic research? | models can analyze and interpret complex experimental data from droplet microfluidics, accelerate the design of new experiments and devices, and generate synthetic data for model training. | and Multi-Agents to synthesize knowledge of droplet microfluidics from literature such as research articles and books. Then, the idea hypotheses will be proposed by the agents, and the novelty and feasibility will be assessed. The agents will provide the autonomous platform to design and fabricate droplet microfluidic devices. Furthermore, LLMs have the researchers accelerate knowledge of droplet microfluidics. | partially addresses the question by mentioning some potential benefits of large language models (LLMs) in droplet microfluidic research, but it lacks depth and clarity compared to the ground truth answer. It also doesn't fully capture the comprehensive capabilities of LLMs as described in the ground truth. SCORE: 42 | | |
| How can autonomous droplet microfluidic research? | Advance our understanding of biological and chemical processes, enable high-throughput screening for | Autonomous droplet microfluidics is an emerging field with significant potential for scientific discovery, aiming to create highly | The response attempts to address the question but lacks clarity, relevance, | 16 | 16.0 |

| | drug discovery, and lead to the development of novel diagnostic tools. | integrated, automated systems for bio/chemical functions like PCR, sample preparation, and detection. Key challenges include the need for fast detection methods to probe small volume droplets, precise control over droplet dynamics in complex microchannel networks, and the development of virtual design platforms for optimizing system designs. Solutions under exploration include using advanced detection techniques such as single-molecule detection and fluorescence lifetime imaging, as well as integrating microelectronic actuators for droplet manipulation. Additionally, the development of virtual design and experiment platforms could streamline the creation of optimized microfluidic systems. The potential applications are vast, including | and specificity; it provides a general statement about the potential of autonomous droplet microfluidic research without discussing how it can be conducted. SCORE: 16 | | |

| | | directed evolution, closed-loop synthesis, and automated drug discovery, where systems could autonomously optimize compound combinations. Collaboration across disciplines like fluid dynamics, chemistry, and software engineering will be key to overcoming challenges and advancing this technology, paving the way for groundbreaking scientific advances. | | | |
|---|---|---|---|---|---|
| What is the potential for fully integrated droplet microfluidic platforms? | High potential for applications in diagnostics, drug discovery, and personalized medicine. | Droplet microfluidics offers transformative solutions to global health challenges by enabling rapid, cost-effective, and high-throughput analysis of biological samples. By generating and manipulating tiny droplets as individual reaction chambers, this technology can revolutionize diagnostics, particularly in resource-limited settings. For example, it facilitates point-of-care testing for | The response is concise but lacks depth and detail compared to the ground truth answer; it only mentions a few potential applications without elaboration. SCORE: 41 | 41 | 41.0 |

| | | infectious diseases like tuberculosis, malaria, and COVID-19, providing results quickly and accurately. Droplet microfluidics is also crucial for multiplexed screening, enabling simultaneous detection of multiple biomarkers, which enhances early disease detection and monitoring. Additionally, it supports the development of low-cost diagnostic devices, reducing dependence on expensive laboratory infrastructure. In drug discovery, it accelerates the screening of therapeutic compounds, improving the treatment of diseases. The technology also advances personalized medicine by analysing single cells and their responses to treatments. Overall, droplet microfluidics addresses global | | | |
|---|---|---|---|---|---|

| | | health disparities by delivering accessible, scalable, and efficient solutions for disease diagnosis, monitoring, and treatment. | | | |
|---|---|---|---|---|---|
| How can droplet microfluidics contribute to global health challenges? | Droplet microfluidics enables rapid, low-cost diagnostics and personalized medicine delivery in resource-limited settings. | Droplet microfluidics has significant potential to address global health challenges by enhancing diagnostic and therapeutic processes. It enables rapid and sensitive pathogen detection, which is critical for early disease diagnosis and containment. The technology's high-throughput capabilities allow simultaneous analysis of multiple samples, streamlining disease diagnosis and treatment. Additionally, droplet microfluidics can be integrated into point-of-care platforms, making it feasible for use in resource-limited settings. This approach can improve healthcare accessibility and efficiency. Furthermore, | The response is concise but lacks detail and context; it only touches on two aspects (diagnostics and personalized medicine) without discussing the broader implications or challenges. SCORE: 16 | 16 | 16.0 |

| | | droplet microfluidics facilitates precise single-cell analysis, enabling advances in personalized medicine and more effective treatment strategies. While challenges such as cost, standardization, and integration with existing systems remain, continued development of this technology could significantly impact global health by improving detection, diagnostics, and treatment capabilities. | | | |
|---|---|---|---|---|---|
| How can open-source designs impact the future of microfluidic devices? | Open-source designs can accelerate innovation, reduce costs, and increase accessibility to microfluidic devices. | Open-source designs for microfluidic devices hold the potential to revolutionize the field by increasing accessibility, reducing costs, and fostering collaboration. These designs allow for the creation of custom predictive models without requiring extensive expertise, enabling broader participation in device development. The use of open-source tools, such as | The response provides some relevant information about the impact of open-source designs on microfluidic devices, but it lacks depth and detail compared to the ground truth answer. It only | 41 | 41.0 |
| | | Design Automation of Fluid Dynamics (DAFD), accelerates design and prototyping, enabling rapid iteration and optimization. Additionally, the integration of machine learning algorithms can drive the development of new microfluidic tools and optimize existing ones. This democratization of design expertise and the collaborative, community-driven approach will facilitate the creation of affordable, specialized devices, ultimately leading to breakthroughs in applications like single-cell analysis and global health diagnostics. The open-source model promises to transform microfluidics into a more inclusive and innovative field. | mentions three benefits (accelerating innovation, reducing costs, and increasing accessibility) without providing any specific examples or context. SCORE: 41 | | |
| What are the challenges of scaling up microfluidic devices for industrial use? | Cost, throughput, integration with existing infrastructure, and material compatibility. | Scaling up microfluidic devices for industrial use presents several challenges, including throughput limitations, lack of | The response provides a concise list of challenges but lacks depth and | 16 | 16.0 |

| | | standardization, and integration complexities. The small micrometric channel sizes restrict production efficiency, and increasing parallelization or channel size may impact fabrication costs and reaction kinetics. There is also a lack of standardized design methods, making it difficult to compare results and adopt microfluidics in industrial settings. Additionally, scaling from laboratory to industrial production remains a hurdle, with issues in cost-effective microfabrication, generating complex droplet emulsions, and ensuring reliable fluid control. The integration of microfluidic devices with industrial interfaces, testing methods, and continuous processes is essential for successful commercialization. Overcoming these challenges requires innovative | context, failing to address the complexities and nuances involved in scaling up microfluidic devices for industrial use. SCORE: 16 | | |
|---|---|---|---|---|---|

| | | solutions, such as simplifying multistep reactions and developing standardized, low-cost fabrication methods. | | | |
|---|---|---|---|---|---|
| How might droplet microfluidics impact agriculture? | Droplet microfluidics can enable precise and efficient delivery of fertilizers, pesticides, and other agricultural inputs, leading to increased crop yields and reduced environmental impact. | Droplet microfluidics has the potential to transform agriculture by enhancing crop management and food security. It enables the precise release of fertilizers and pesticides, reducing waste and optimizing plant growth. The technology also supports improved crop monitoring through long-term studies, offering insights into plant development. By analyzing individual plant cells, droplet microfluidics can help uncover genetic and phenotypic traits, enhancing disease resistance and drought tolerance in crops. Additionally, it aids in disease detection, allowing for early intervention to prevent crop loss. Despite challenges such as scalability, cost, and integration | The response is concise and addresses the core aspect of how droplet microfluidics can impact agriculture. However, it lacks depth and details compared to the ground truth answer, and some relevant points such as disease detection and genetic trait analysis are not mentioned. SCORE: 51 | 51 | 51.0 |

| | | with existing agricultural practices, ongoing research aims to overcome these barriers, making droplet microfluidics a promising tool for precision agriculture and sustainable food production. | | | |
| --- | --- | --- | --- | --- | --- |
| What are the current limitations of droplet microfluidics, and how can they be overcome? | Droplet microfluidics currently faces limitations in throughput, scalability, complex reactions, and material compatibility. These can be overcome through advancements in droplet generation techniques, high-throughput platforms, integration with other technologies like optofluidic devices, and the development of new materials. | Droplet microfluidics has immense potential for high-throughput biological assays but faces several limitations that hinder its broader adoption. One key challenge is its efficiency for assays involving small sample sizes; droplet microfluidics excels with large volumes (at least 10^5 samples), making traditional methods like microtiter plates more suitable for lower sample numbers. Another critical issue is maintaining a stable and inert oil-water interface, which is essential for droplet formation. Finding the right surfactant to keep this interface stable remains a complex | The response attempts to address the question but lacks specificity and detail regarding the limitations of droplet microfluidics. It mentions some general challenges, such as throughput, scalability, complex reactions, and material compatibility, but does not delve into the specific issues mentioned in the | 16 | 16.0 |

| | | chemistry challenge. Furthermore, droplet microfluidics lacks the spatial barcoding available in microtiter plates, making it difficult to organize and track multiple droplet libraries. As a result, developing alternative barcoding systems is necessary but challenging. The automation of droplet movement is another obstacle, as it requires precise control, often through techniques like electrowetting and dielectrophoresis. Lastly, droplet stability over time can be compromised, with issues such as shrinkage, coalescence, and cross-talk between droplets. To address these issues, research is underway to develop improved surfactants, more reliable barcoding systems, and novel materials and techniques to enhance droplet control. These | ground truth answer (efficiency for small sample sizes, stable oil-water interface, spatial barcoding, automation of droplet movement, and droplet stability). The response also fails to provide a comprehensive overview of how these limitations can be overcome. SCORE: 16 | | |
| --- | --- | --- | --- | --- | --- |

| | | advancements, coupled with better analytical tools, are expected to overcome current limitations and expand droplet microfluidics' applications in chemical and biological research. | | | |
|---|---|---|---|---|---|
| How can interdisciplinary collaboration accelerate innovation in droplet microfluidics? | Interdisciplinary collaboration can accelerate innovation in droplet microfluidics by bringing together expertise from diverse fields such as engineering, biology, chemistry, and computer science. This allows for the development of novel devices, materials, and applications that would be difficult to achieve through a single discipline approach. | Interdisciplinary collaboration is essential for advancing droplet microfluidics by combining expertise from various fields such as chemistry, biology, material science, and engineering. By uniting these diverse areas, researchers can address complex challenges such as cost-effective fabrication, scalable production, and novel interfacial reaction systems. For example, biologists and chemists can leverage droplet microfluidics for high-throughput analysis, while material scientists and engineers work on developing more robust and affordable microfabrication methods. | The response provides a basic understanding of the importance of interdisciplinary collaboration in accelerating innovation in droplet microfluidics. However, it lacks depth and specific examples compared to the ground truth answer. It also does not mention key challenges such as cost- | 41 | 41.0 |

| | | Furthermore, integrating technologies like sensitive detectors and machine learning algorithms can enhance the analytical capabilities of droplet-based systems. Collaborative initiatives, such as the EMBL Microfluidics Conference, foster knowledge exchange and innovation by bringing together professionals from different disciplines. This cooperation is crucial for overcoming technical barriers, reducing manufacturing costs, and developing new applications in fields like drug discovery, single-cell analysis, and nanomaterial synthesis. Ultimately, interdisciplinary collaboration accelerates the evolution of droplet microfluidics, pushing the boundaries of what is achievable in this promising field. | effective fabrication, scalable production, or novel interfacial reaction systems that are crucial for advancing the field. SCORE: 41 | | |
|---|---|---|---|---|---|

## S7: QA by MISTRAL-based Scientific Mentor

| question | answer | ground_truth | feedback | score | percentage |
|----------|--------|--------------|----------|-------|------------|
| What is microfluidics? | Microfluidics is a versatile technological field that enables studies of dynamic conditions, particularly in the areas of | Microfluidics is a multidisciplinary field that involves the study and manipulation of fluids at the microscale, typically within channels with | The response accurately describes the concept of microfluidics, its goals, and various | 85 | 85.0 |

| | | | | |
|---|---|---|---|---|
| | biotechnology and chemistry. It involves manipulating small volumes (less than a millimeter) of fluids within microscopic channels or droplets. The main goal of microfluidics is to bring traditional chemical and biological systems onto a small device that can analyze samples quickly, affordably, and effectively. This technology has gained popularity due to its potential applications in various scientific fields such as engineering, physics, agriculture, material synthesis, miniaturized biotechnology and nanotechnology, biochemistry, chemistry, material science and metallurgy, chemical engineering, water treatment, | dimensions ranging from tens to hundreds of micrometers. It is a multidisciplinary field that involves molecular analysis, molecular biology, and microelectronics. Microfluidics has been increasingly used in the biological sciences because precise and controlled experiments can be conducted at a lower cost and faster pace. Lab on a Chip devices use microfluidics for applications such as Point of Care testing of diseases, or Organ on a Chip studies. | applications in different scientific fields. It also mentions an emerging technology within microfluidics, droplet-based microfluidics, and its potential uses. However, it does not explicitly mention the multidisciplinary nature of microfluidics or its connection to molecular analysis, biology, and microelectronics as stated in the ground truth answer. SCORE: 85 | | |

| | life sciences, among others. One emerging technology within microfluidics is droplet-based microfluidics, which involves manipulating discrete fluid micropackets of controlled volume and composition. This technology has a wide range of applications in micrototal analysis systems (μTAS). | | | |
|---|---|---|---|---|
| What is droplet microfluidics? | Droplet microfluidics is a technology that allows the controlled manipulation of small volumes of liquid in microchannels. It consists of networks of channels with diameters typically measuring 10–100μm, which enable efficient and rapid partitioning of single molecules (or single cells or beads) from complex samples. The use of droplet-based | Droplet-based microfluidics is a branch of microfluidics focusing on the creation and manipulation of discrete volumes of liquids through adoption of immiscible fluid flows. Microdroplets allow for handling miniature volumes (μL to fL) of fluids conveniently, provide better mixing, encapsulation, sorting, and sensing, and suit high throughput experiments. Com | The response accurately describes the concept of droplet microfluidics, its principles, and applications, aligning well with the ground truth. It provides a clear and concise explanation of the technology, including its advantages over | 96 | 96.0 |
| | microfluidics extends the capabilities of these procedures in terms of efficiency, throughput, and sensitivity. Highly monodispersed droplets can be produced at high frequency and are stabilized by the use of surfactants. These droplets act as independent microreactors, which can then be manipulated in microchannels through the implementation of different modules for merging, splitting, incubating, or sorting. This technology has potential applications in biomedical research and biotechnology, including advanced diagnostics and therapeutics. | pared with continuous-flow systems, droplet microfluidic systems enable isolation and manipulation of samples in separate compartments. The droplets generated with uniform size and shape provide a stable microenvironment for biochemical reactions. Processing of a large number of droplets in parallel facilitates to achieve ultrahigh throughput. Confinement of individual targets within the droplets and independent processing of each of them allow high-precision analyses. | continuous-flow systems. SCORE: 96 | | |
| How to generate microdroplets? | Microdroplets can be generated using two main methods based on microfluidic devices: active | Microdroplet generation in microfluidics is achieved through two main methods: active | The response provides a clear explanation of | 56 | 56.0000000 0000001 |

| | | | | |
|---|---|---|---|---|
| | and passive. Active methods typically use a device that employs an external force, such as a piezoelectric actuator, to generate droplets. Passive methods rely on the inherent properties of the fluids and channel geometry to create droplets.<br><br>In the context provided, microdroplets were generated using two PDMS-based microfluidic devices for bulk syntheses. The first device was used for gold nanostar synthesis, while the second device was designed for magnetic iron oxide nanomaterials. The gold nanostars were grown inside the droplets, allowing for local concentrations and good control over reagent mixing during particle | and passive. Active methods utilize external forces like piezoelectric actuators to precisely control droplet size, formation rate, and spacing. These methods enable high uniformity in droplet generation, with deviations less than 0.3%. Passive methods rely on fluid properties and channel geometry, such as T-junctions or cross-flow configurations, to create droplets. These methods are influenced by factors like flow rate, viscosity, and interfacial tension, allowing for fine control over droplet size and uniformity. For example, PDMS-based microfluidic devices have been used for nanoparticle synthesis, where hydrodynamic coupling of nozzles generates droplet pairs for precise mixing. Both methods are | microdroplet generation using active and passive methods based on microfluidic devices. It also includes specific examples from studies involving PDMS-based microfluidic devices for nanoparticle synthesis, demonstrating a good understanding of the topic. However, it lacks a comprehensive overview of the general principles behind microdroplet generation and does not explicitly state that there are two main methods (active and passive) as | | |

| | formation. The monodispersity of the product demonstrated the potential for continuous synthesis of high-quality anisotropic NPs with improved reproducibility.

For magnetic iron oxide nanomaterials, a PDMS-based microfluidic chip was designed that enabled droplet pairs to be produced based on hydrodynamic coupling of two separated nozzles. One of the droplets contained an ammonium hydroxide solution, while the other contained $Fe^{2+}/Fe^{3+}$ mixture. The two droplets merged when passing through two energized electrodes, and a precipitate of iron oxide nanomaterials appeared.

In another study, a simple and | valuable in applications like nanoparticle synthesis, offering high reproducibility and control over reaction conditions in biomedical and material science research. | mentioned in the ground truth answer. The response also focuses more on specific examples rather than providing a broad explanation of the methods. SCORE: 56 | | |
|---|---|---|---|---|---|

| | robust droplet generator based on a piezoelectric actuator was described, which is integrated into a microfluidic device. This generator can independently control the droplet size, rate of formation, and distance between droplets, with high uniformity. The cross-flow and T-junction configurations were tested, showing no significant differences, but the inlet to main channel ratio was found to be important. As this ratio increases, droplets tend to be generated in bursts instead of individually. | | | | |
|---|---|---|---|---|---|
| What materials are typically used to fabricate droplet microfluidic devices? | The materials typically used to fabricate droplet microfluidic devices include PDMS (Polydimethylsilo xane), thermoset polyesters, poly(methylmeth acrylate) (PMMA), polycarbonate, polystyrene, | Droplet microfluidic devices are commonly fabricated from a variety of materials, each offering distinct advantages for specific applications. The most widely used material is PDMS (Polydimethylsilo | The response is mostly accurate and provides a comprehens ive list of materials used for fabricating droplet microfluidi c devices. It | 84 | 84.0 |

| | cyclic olefin copolymer, fluoropolymers such as polytetrafluoroethylene, and glasses. These materials are chosen for their specific advantages such as outstanding solvent resistance, negligible biofouling, and the ability to mass produce devices using techniques such as injection molding or hot embossing. PDMS is the most commonly used material in droplet microfluidics due to its stability and precision when fabricated using soft lithography. Other methods for fabricating microfluidic chips include 3D printing and capillary devices. | xane), favored for its ease of use with soft lithography and flexibility, though it has limitations such as poor solvent resistance. Alternatives to PDMS include thermoset polyesters, thermoplastics like PMMA, polycarbonate, and polystyrene, and amorphous polymers such as cyclic olefin copolymer. Fluoropolymers like PTFE and glass are also used, offering superior solvent resistance and biofouling resistance. Silicon and capillary devices are utilized in certain applications, providing precise control for microfluidic operations. These materials are chosen based on factors like solvent resistance, biofouling, and the ability to mass-produce devices using methods such as injection molding | correctly identifies the advantages of these materials and mentions specific techniques for mass production. However, it slightly overemphasizes the benefits of PDMS and does not fully acknowledge its limitations as mentioned in the ground truth answer. SCORE: 84 | | |
|---|---|---|---|---|---|

| | | or hot embossing, ensuring broad applicability in research and industrial settings. | | | |
|---|---|---|---|---|---|
| What are the applications of droplet microfluidics? | The applications of droplet microfluidics are extensive and can be categorized into five areas: single-cell analysis, medical diagnostics, drug discovery, food and feed industry, and environmental engineering. In the context of single-cell analysis, droplet microfluidics enables handling and analysis of individual cells, which is crucial for revealing cell heterogeneity and interrogating life activities and pathologic development. This technology also plays a significant role in biological screening, segment labelling, cell packaging, virus detection, specific proteins or tissue remodeling, and more, particularly | Droplet microfluidics has diverse applications across biomedical research, diagnostics, and industrial fields. It enables high-throughput single-cell analysis, such as genomics, transcriptomics, and drug screening, while also facilitating nanoparticle synthesis, enzyme evolution, and chemical reactions. In healthcare, it supports personalized medicine, point-of-care diagnostics, and vaccine development. Beyond biology, it is used for environmental monitoring, food safety testing, and the development of cosmetics and pharmaceuticals. Its ability to create isolated, uniform reaction chambers in droplets makes | The response provides a clear and detailed overview of the applications of droplet microfluidics, covering various fields such as single-cell analysis, medical diagnostics, drug discovery, food and feed industry, and environmental engineering. It accurately describes the capabilities of the technology in handling individual cells, biological screening, and high-throughput | 83 | 83.0 |
| | in the field of biomedical analysis. Examples of applications include absolute quantification of samples with extremely low concentrations, distinguishing differences in enzyme activity, genes, and protein expression at the single-cell level, and high-throughput screening of gene mutations, chemicals, and reactions. Additionally, droplet microfluidics is suitable for in vivo sampling, measurement of dynamic release of organs, and collection of separated components in an electrophoresis system due to its ability to preserve temporal concentration information with high resolution and fidelity. | it a powerful tool for precise, cost-effective, and scalable workflows in science and industry. | screening of gene mutations, chemicals, and reactions. The response also highlights its potential for preserving temporal concentration information with high resolution and fidelity. However, it could be improved by providing more specific examples or real-world applications to further illustrate the breadth of droplet microfluidics' uses.<br><br>SCORE: 83 | | |
| What are the main benefits of using droplet | The main benefits of using droplet microfluidics over bulk systems | Droplet microfluidics offers significant advantages over | The response is clear and concise, | 85 | 85.0 |

| microfluidics over bulk systems? | include:

1. Single-cell manipulation and analysis: Droplet microfluidics allows for the isolation of single cells from a bulk, each into its own tiny liquid compartment. This feature opens up possibilities such as early detection of cells and secreted molecules, and the ability to isolate and culture rare individuals.

2. Handling large numbers of individual droplets: Droplet microfluidics enables the analysis of massively large numbers (even millions) of individual droplets. This allows for the examination of phenotypic and genetic variabilities at the level of cells or small populations, as well as high-throughput | traditional bulk systems. It enables individual compartmentalization, isolating single cells or molecules in tiny liquid compartments for precise control. This allows for single-cell analysis and early detection of rare cells or secreted molecules. The technology supports massive parallel processing, analyzing millions of droplets simultaneously, ideal for high-throughput screening and examining phenotypic or genetic variability. Additionally, iterative operations such as merging, mixing, and sorting droplets enable complex experimental protocols and continuous monitoring of populations. The system's highly controllable local environments ensure consistent, reproducible | providing detailed explanations of the main benefits of using droplet microfluidics over bulk systems. It covers various aspects such as single-cell manipulation, handling large numbers of individual droplets, executing iterative operations, application in different fields, and advantages in drug discovery and validation. The answer aligns well with the ground truth, although it could be slightly condensed for better conciseness. SCORE: 85 | | |
|---|---|---|---|---|
| | screening.<br><br>3. Executing iterative operations on droplets: Droplet microfluidics allows for controlled formation of droplets, merging them and mixing with additional reagents, splitting, sorting, incubation even over hundreds of generations and extended periods of time. This enables conducting multiple measurements on the same droplets or tracking the evolution of a population in controllably changing chemical environments.<br><br>4. Application in various fields: Droplet microfluidics has extensive application coverage, including single-cell analysis, medical diagnostics, drug discovery, food and feed industry, | results. These benefits make droplet microfluidics a versatile tool in drug discovery, medical diagnostics, single-cell analysis, and other fields, offering powerful insights and precise manipulation at the individual level. | | | |
| | and environmental engineering.  5. Advantages in drug discovery and validation: In the context of drug discovery and validation, droplet microfluidics offers distinct advantages such as high single-cell resolution, rapid experimental reaction rates, sensitivity, the capability to identify subpopulations and unravel cellular heterogeneity, and compatibility with AI and robotic systems. | | | | |
| What are the advantages of droplet microfluidic systems compared with continuous-flow systems? | Droplet microfluidic systems offer several advantages over continuous-flow (or single-phase) systems. Here are some key points:  1. Low sample volumes: Droplet microfluidics allows for the use of very small amounts of samples, which | Droplet-based microfluidic systems offer significant advantages over traditional continuous-flow systems. These include the ability to overcome issues such as Taylor dispersion, slow mixing, and surface-molecule interactions. Droplet systems allow for ultra- | The response is comprehensive and accurately lists the advantages of droplet microfluidic systems compared to continuous-flow systems. It covers various | 95 | 95.0 |

| | can be beneficial in terms of cost and resource efficiency.<br><br>2. Facile integration of functionalities: Droplet microfluidic systems can easily incorporate different functions, such as heat and mass transport control, which can enhance experimental flexibility.<br><br>3. Exquisite control of heat and mass transport: This feature allows for precise manipulation of the system's properties, leading to improved experiment outcomes.<br><br>4. Overcoming problems related to Taylor dispersion, surface-molecule interactions, and slow mixing: These issues are common in continuous-flow systems but are | small assay volumes, reducing reagent consumption and improving sensitivity. They also minimize cross-contamination and sample adsorption by compartmentalizing individual samples. The system's chaotic advection enhances mixing efficiency, enabling faster reactions. With droplet generation frequencies up to hundreds of kHz, droplet microfluidics supports high-throughput experimentation. Furthermore, the rapid and precise control of droplet payloads and sub-millisecond mixing times make it ideal for applications requiring fast, controlled analysis. These benefits make droplet-based systems particularly valuable for cell-based assays, enzyme kinetics, | aspects such as low sample volumes, facile integration of functionalities, exquisite control of heat and mass transport, overcoming problems related to Taylor dispersion, surface-molecule interactions, and slow mixing, ability to define and form ultra-small assay volumes, negligible cross-contamination and sample adsorption, enhanced mixing due to chaotic advection, high droplet generation frequencies, control over the droplet payload, and access | | |

| | minimized or eliminated in droplet microfluidics.

5. Ability to define and form ultra-small assay volumes: This feature allows for the performance of experiments on very small scales, which can be useful in various scientific fields.

6. Negligible cross-contamination and sample adsorption: Droplet microfluidic systems have minimal issues with contamination from surrounding environments or sample adsorption, making them more reliable for certain experiments.

7. Enhanced mixing due to chaotic advection: The unique mixing properties of droplet microfluidics can lead to faster and | and high-throughput screening. | to sub-millisecond mixing times. The response is well-structured and provides detailed explanations for each point. It also aligns with the ground truth answer in terms of content and key points discussed. SCORE: 95 | | |
|---|---|---|---|---|---|
| | more efficient experimentation.<br><br>8. High droplet generation frequencies: Droplet-based microfluidic systems can generate droplets at very high frequencies, enabling ultra-high-throughput experimentation.<br><br>9. Control over the droplet payload: The payload (contents) of each droplet can be controlled rapidly and efficiently in a droplet-based system.<br><br>10. Access to sub-millisecond mixing times: This rapid mixing capability is another advantage of droplet microfluidics, making it suitable for experiments requiring fast mixing times. | | | | |
| What are the basic principles behind droplet generation in microfluidics? | The basic principles behind droplet generation in microfluidics involve the | Droplet generation in microfluidics relies on emulsion processes, where an oil phase (continuous | The response accurately describes the basic principles | 76 | 76.0 |

| | production of droplets with a controlled size at high frequency. This is achieved by controlling the flow of at least two liquids, either by volume using syringe pumps or by pressure using hydrostatic reservoirs. The droplets are typically generated by a surface-induced instability using different device geometries depending on the required droplet frequency, monodispersity, or variability in droplet size.

Surfactants are added to the continuous phase to lower the interfacial energy and facilitate the formation of new interfaces and stabilize the formed emulsion droplets from coalescence by adsorbing at the liquid/liquid interface. The choice of surfactant, construction | phase) encapsulates an aqueous phase (dispersed phase) to form droplets. Surfactants are added to lower interfacial tension, stabilize the droplets, and prevent coalescence. The process requires precise control over the flow of both liquids, often achieved using syringe pumps or pressure-driven systems. Droplets are generated through surface-induced instabilities, influenced by microchannel geometries such as T-junctions or flow-focusing, which control droplet size and frequency. The design and flow control of the system determine the size uniformity and production rate of droplets. These principles enable high-throughput applications such as drug delivery, high-throughput screening, and microreactors for | behind droplet generation in microfluidics, including the role of surfactants, emulsion processes, and control over liquid flow. It also mentions specific geometries such as T-junctions and flow-focusing, which is consistent with the ground truth answer. However, it could be more concise and focused on the core principles without providing additional information about PDMS-based devices. SCORE: 76 | | |
|---|---|---|---|---|
| | materials, and geometries that generate different routes of droplet generation impact the feasibility of the process of droplet formation in microfluidic platforms.<br><br>The continuous fluid is typically organic oil, and the dispersed fluid is generally an aqueous liquid. The dispersed phase partitions into uniform microdroplets, which can be handled, collected, incubated, split, sorted, and combined. T-junction and flow-focusing geometries are the two primary production methods of droplets in microfluidic applications among passive techniques.<br><br>The fabrication of polydimethylsiloxane (PDMS)-based devices via the conventional | bio-catalysis, ensuring reproducibility and productivity in biological experiments. | | | |

| | soft lithography method is a well-developed and straightforward process. | | | | |
|---|---|---|---|---|---|
| How is droplet size controlled in a droplet generator? | Droplet size can be controlled in a droplet generator through several methods. One method, as described in the provided context, is by using splitters to bisect large droplets into smaller ones, thereby reducing their size and scaling up experimental throughput logarithmically. The design of these splitters, such as tree-splitters, can influence the uniformity of the droplet array. Another method, not explicitly mentioned in the provided context but hinted at, is by adjusting microchannel dimensions and geometries or controlling the flow rates of two immiscible liquids. This is a characteristic of passive droplet generation | Droplet size control in microfluidic generators can be achieved through several methods. In passive droplet generation, adjusting microchannel dimensions or controlling the flow rates of immiscible liquids allows for precise control over droplet size, shape, and monodispersity. Tree-splitters, which bisect large droplets into smaller ones, also influence droplet size and uniformity, with splitting angles affecting droplet diameter. In active droplet generation, technologies like piezoelectric actuators independently control droplet size, formation rate, and spacing, offering high precision. Additionally, the velocity of the | The response provides some relevant information about droplet size control in a droplet generator, mentioning several methods such as using splitters, adjusting microchannel dimensions, controlling flow rates of immiscible liquids, and piezoelectric actuators. However, the answer contains inaccuracies regarding the scaling up experimental throughput logarithmically with splitter | 38 | 38.0 |

| | methods, which are widely used to generate continuous droplet streams of uniform size distributions. Additionally, there exists a method based on a piezoelectric actuator, as described in another document not provided here, which allows for independent control over the droplet size, rate of formation, and distance between droplets. | continuous and dispersed phases plays a significant role in controlling droplet size. These methods enable the generation of uniform droplets for various applications, such as high-throughput screening and drug delivery. | usage, and it lacks clarity on how these methods are applied in practice. The response also fails to provide a comprehensive overview of droplet size control as described in the ground truth answer. SCORE: 38 | | |
|---|---|---|---|---|---|
| What is the role of surface tension in droplet formation? | In droplet formation, surface tension plays a crucial role as it opposes the inertial forces imposed by the flow of the dispersed phase. The competition between these two forces is quantified using the Weber number (We). When the Weber number is approximately equal to one, the transition between dripping and jetting occurs. In | Surface tension plays a critical role in droplet formation in microfluidic systems by influencing droplet size, shape, and formation mechanisms. It acts to minimize surface area, favoring spherical droplet shapes. At low Weber numbers (We < 1), where surface tension dominates, droplets form through dripping or squeezing. As the Weber number | The response accurately conveys the role of surface tension in droplet formation, mentioning its opposition to inertial forces, the Weber number, and its influence on droplet size and shape. It also correctly | 76 | 76.0 |

| | microfluidic applications, surface tension measurements are feasible, and its effect on surfactant can be measured. The surface tension of the interface is a parameter controlling droplet size in a T-junction or in a flow focusing junction, and a measurement of droplet size provides an indirect way to access the surface tension at the time-scale of droplet production. | increases, fluid inertia competes with surface tension, leading to jetting behavior. The capillary number (Ca) also influences droplet size, with surface tension controlling droplet formation at low Ca. Techniques like electrowetting can measure surface tension in microfluidic systems, providing insights into the interfacial properties. Additionally, surface tension is a key parameter in T-junctions and flow focusing devices, where droplet size measurements offer an indirect way to assess surface tension at the time-scale of droplet production. Understanding the interplay between surface tension and other forces is essential for precise droplet control. | mentions microfluidic applications and techniques for measuring surface tension. However, it lacks a clear explanation of how surface tension minimizes surface area and favors spherical shapes, which is a crucial aspect of its role in droplet formation. The response also does not explicitly discuss the capillary number's influence on droplet size or provide a comprehensive overview of the interplay between | | |
|---|---|---|---|---|---|

| | | | surface tension and other forces in microfluidic systems. SCORE: 76 | | |
|---|---|---|---|---|---|
| How does a flow-focusing device differ from a T-junction in droplet formation? | A flow-focusing device differs from a T-junction in droplet formation primarily due to the geometry and mechanism of droplet generation. In a T-junction, the dispersed phase perpendicularly intersects the main channel containing the continuous phase, forming an interface at the junction. As fluid flow continues, the shear forces generated by the continuous phase cause the head of the dispersed phase to elongate into the main channel until it breaks and forms a droplet. The size of the droplets can be adjusted by altering the fluid flow rates, the channel widths, or changing the relative viscosity | Flow-focusing devices and T-junctions are two common microfluidic designs for droplet formation, differing in geometry and mechanism. A flow-focusing device uses a central dispersed phase channel and side continuous phase channels converging at a narrow orifice, where hydrodynamic focusing pinches off droplets. It produces smaller, highly uniform droplets with precise control over size and high throughput. In contrast, a T-junction has perpendicular channels where the continuous phase shears the dispersed phase at the intersection, forming larger droplets with less uniformity and | The response provides a clear explanation of the differences between flow-focusing devices and T-junctions in terms of geometry and mechanism of droplet formation. It accurately describes the process of droplet generation in both devices, including the role of shear forces and Rayleigh-Plateau instability. However, it could be improved by providing more | 76 | 76.0 |

| | | | | |
|---|---|---|---|---|
| | between the two phases.<br><br>On the other hand, in a flow-focusing device, the dispersed and continuous phases are forced through a narrow region in the microfluidic device. The streams of co-flowing liquids undergo large elongational flow as they pass the small orifice, generating smaller droplets due to the Rayleigh–Plateau instability. This method allows for more flexibility in generated droplet sizes because the effective geometry can be adjusted by changing the flow rates of dispersed and continuous phases. By focusing and thinning the stream of dispersed phase through an orifice, very small droplets can also be generated at high frequencies. | lower throughput. Flow-focusing is ideal for applications requiring monodisperse droplets, while T-junctions suit simpler or larger-droplet applications. | concise language and a clearer comparison of the two methods, as well as highlighting the specific advantages of flow-focusing devices for certain applications. SCORE: 76 | | |

| What role does the geometry of the microchannel play in droplet formation? | The geometry of the microchannel plays a crucial role in droplet formation in microfluidics. There are primarily two methods for generating droplets, T-junction and flow-focusing, which both depend on the channel geometry to control droplet generation.

In the T-junction configuration, the dispersed phase (containing the material of interest) intersects the main channel containing the continuous phase. The shear forces generated by the continuous phase and the subsequent pressure gradient cause the head of the dispersed phase to elongate into the main channel until it thins and breaks, forming a droplet. The size of the droplets can be changed by altering the | The geometry of a microchannel plays a critical role in droplet formation, influencing the size, shape, and generation frequency of droplets. Key geometric factors include the channel width, depth, and junction design, such as T-junctions or flow-focusing geometries. These features govern the interaction between the continuous and dispersed phases, affecting the shear forces and pressure gradients that drive droplet breakup. For instance, narrower channels or smaller orifices promote the formation of smaller droplets due to increased shear stress. Similarly, the junction type controls droplet detachment mechanics; in T-junctions, droplets form due to the balance of interfacial tension | The response accurately addresses the question by providing a detailed explanation of how the geometry of the microchannel influences droplet formation. It correctly identifies key geometric factors such as channel width, depth, junction design, and flow rates that affect droplet size, shape, and generation frequency. The answer also provides specific examples of how different geometries impact droplet formation, including the effects | 86 | 86.0 | |
|---|---|---|---|---|---|---|

| | fluid flow rates, the channel widths, or by changing the relative viscosity between the two phases.

In the flow-focusing configuration, the dispersed and continuous phases are forced through a narrow region in the microfluidic device. The size of droplets formed can be controlled according to the water fraction values (wf), where wf is the ratio of the relative volumetric flow rates of the water phase and the immiscible oil phase.

The size of droplets formed in T-junctions is affected by several factors, including the structure of the channel, the two-phase flow rates, and some physical properties such as the | and shear force, while in flow-focusing designs, they form under pinching at the constriction. Channel geometry also determines flow regimes ranging from dripping to jetting further impacting droplet characteristics. Thus, optimizing microchannel geometry is crucial for precise control over droplet formation in applications like microfluidics, emulsification, and lab-on-a-chip systems. | of narrower channels and different junction types. However, it could be improved by providing a more concise summary statement at the end to directly address the question's focus on the role of geometry in droplet formation.

SCORE: 86 | | |

| | | | | | |
|---|---|---|---|---|---|
| | hydrophobicity of the channel surface. Modifying the PDMS channels using 1H,1H,2H,2H-perfluorooctyl-trichlorosilane in toluene solution can create hydrophobic PDMS channels, which may influence droplet formation as well.<br><br>In summary, the geometry of the microchannel determines the method of droplet formation and influences factors such as droplet size, flow rates, channel widths, and physical properties like hydrophobicity. | | | | |
| What factors affect droplet coalescence in microfluidic devices? | Droplet coalescence in microfluidic devices can be affected by several factors. The most significant factors are the pairing time (s), the concentration of the surfactant, and the geometry of the pairing channel. | Droplet coalescence in microfluidic devices is influenced by several factors. Surfactant concentration plays a key role, with higher concentrations preventing droplet fusion, while lower concentrations | The response provides a clear and detailed explanation of the factors affecting droplet coalescence in microfluidic devices. It accurately | 76 | 76.0 |

| | | | | |
|---|---|---|---|---|
| | 1. Pairing Time (s): The pairing time is the duration for which two droplets stay in physical contact within the pairing channel before entering the coalescence channel. If this time is less than 4 ms, droplet pairs will not coalesce. However, if it's more than 8 ms, reliable and selective coalescence can occur.<br><br>2. Surfactant Concentration: The concentration of the surfactant used in the continuous phase affects the interfacial stabilization and thus the droplet coalescence. For example, using 1.5 mM EA-surfactant ensures that droplet pairs always coalesce, while using 7.5 mM EA-surfactant prevents droplets from fusing. | promote coalescence. The duration droplets remain in contact within the pairing channel, known as pairing time, also impacts coalescence, with longer times favoring fusion. The geometry of the pairing channel, specifically its cross-sectional area, controls the number of droplets that can come into contact and coalesce. Additionally, droplet size and volume affect how efficiently droplets merge. Finally, electrocoalescence, which applies electric fields to induce droplet fusion, offers an alternative method for on-demand coalescence. By optimizing these factors—surfactant concentration, pairing time, channel geometry, and droplet size—reliable and selective droplet coalescence can be achieved in | identifies pairing time, surfactant concentration, and geometry of the pairing channel as key factors, and provides specific examples to support these claims. However, it does not mention droplet size and volume, which are also mentioned in the ground truth answer as important factors. Additionally, the response does not discuss electrocoalescence as an alternative method for on-demand coalescence. SCORE: 76 | | |

| | 3. Geometry of the Pairing Channel: The geometry of the pairing channel plays a crucial role in achieving selective multiple coalescence events. The cross-sectional area of the pairing channel should be close to the cross-sectional area of a single spherical droplet, so that only one re-injected emulsion droplet can contact each droplet generated on-chip. This ensures that only paired droplets fuse in the coalescence channel.

In addition, the design and fabrication of the device should be simple and robust compared to other electric coalescence methods, enabling on-demand droplet coalescence without precise synchronization in spacing and time between the droplets. This | microfluidic systems. | | | |
|---|---|---|---|---|---|

| | | | | |
|---|---|---|---|---|
| | method also does not require precise electrode alignment and a metal deposition for the electrode. | | | | |
| What design elements are critical for producing double or multiple emulsions? | To produce double or multiple emulsions, the following design elements are critical:<br><br>1. Co-flow and flow-focusing configurations: These configurations can be combined to generate double-emulsion droplets. In this setup, two cylindrical capillaries are aligned end-to-end within a square capillary. The innermost phase flows through the inner cylindrical capillary, while the intermediate phase flows through the interstitial space between the inner and outer capillaries in the same direction. The outermost phase flows through the interstitial space | To generate double or multiple emulsions, several design elements are essential. Co-flow and flow-focusing configurations enable precise control over fluid flow, allowing for stable droplet formation, especially in double emulsions. Sequential emulsification steps, where inner droplets are emulsified into intermediate phases and then into outer phases, are key for multiple emulsions. Capillary microfluidic devices, such as consecutive capillary and coaxial configurations, are used to control droplet size, number, and monodispersity. Additionally, precise flow control and shear | The response provides a detailed description of specific design elements used to produce double or multiple emulsions, including co-flow and flow-focusing configurations, elementary micro-fluidic devices, coaxial capillary microfluidic devices, step emulsification channels, and accurate control of internal structure and composition. However, the answer | 58 | 57.9999999 9999999 |

| | | | | |
|---|---|---|---|---|
| | from the opposite end in the opposite direction and hydrodynamically focuses the two co-flowing liquid phases into double-emulsion droplets as they pass through the orifice.

2. Elementary micro-fluidic devices: These devices are skillfully assembled in series to precisely control the monodispersity, number, and size of the inner droplets of multiple emulsions. For example, a consecutive capillary micro-fluidic device can be used to generate multiple emulsions containing control-numbered daughter and granddaughter droplets.

3. Coaxial capillary microfluidic devices: These devices are used for the generation | force management are crucial for stable and consistent droplet formation. Microchannel geometry optimization and material compatibility further enhance emulsification processes. By carefully managing these factors, researchers can produce double and multiple emulsions with controlled internal structures and compositions for various applications. | lacks a clear summary statement that directly addresses the question and does not explicitly mention the importance of precise flow control and shear force management, which are critical factors in producing stable double or multiple emulsions. The response also includes some minor inaccuracies and lacks depth in comparison to the ground truth answer.

SCORE: 58 | | |

| | of different emulsions, such as water-in-oil-in-water (W/O/W) emulsions.<br><br>4. Step emulsification channel: This channel can be parallelized with drop markers to produce highly monodisperse multiple emulsions.<br><br>5. Accurate control of internal structure and composition: This is critical for the versatility of multiple emulsions, allowing accurate manipulation of transport kinetics and the loading and release levels of encapsulated substances. | | | | |
|---|---|---|---|---|---|
| What key factors influence the design of flow-focusing droplet microfluidics optimized using machine learning? | The key factors that influence the design of flow-focusing droplet microfluidics optimized using machine learning are primarily related to the fluid dynamics and geometry of the microfluidic device. | The design of flow-focusing droplet microfluidics, optimized using machine learning, depends on several critical factors. Channel geometry, including the shape and size, controls droplet generation by | The response is clear and concise, providing a detailed explanation of the key factors influencing the design of flow-focusing droplet | 83 | 83.0 |

| | | | | |
|---|---|---|---|---|
| | 1. Channel Geometry: The shape and size of the channel play a critical role in controlling droplet generation. In flow-focusing designs, the dispersed phase is injected through the channel, and the continuous phase forms an outer shell around it. The orifice where the two phases meet determines the break-off point for droplets.<br><br>2. Flow Rate: Both the flow rate of the dispersed phase and the continuous phase significantly impact the size and frequency of droplet generation. Increasing the flow rate of the continuous phase can decrease the size of the droplets, while increasing the oil flow rate also increases the frequency of droplet generation. | defining the break-off point where the dispersed phase meets the continuous phase. Flow rates of both phases affect droplet size and frequency, with higher continuous phase flow reducing droplet size. Viscosity influences flow behavior and shear forces, crucial for stable droplet formation. Shear forces between phases are controlled by adjusting channel geometry or flow rates, ensuring consistent droplet generation. Machine learning algorithms predict and optimize droplet diameter and generation rate by analyzing these design parameters, fluid properties, and large datasets, automating the design process for more efficient droplet production in life sciences applications. | microfluidics optimized using machine learning. It accurately addresses all relevant points mentioned in the ground truth answer, including channel geometry, flow rates, viscosity, shear rate, and design method. However, it could be improved by adding more depth to some points and providing a clearer connection between these factors and their impact on droplet generation. SCORE: 83 | | |

| | 3. Viscosity: The viscosity of both phases is another important factor that governs droplet generation in flow-focusing designs.<br><br>4. Shear Rate: In the flow-focusing configuration, the dispersed phase is subjected to shear by the continuous phase, which enables more controlled and stable droplet generation. The design employs symmetric shearing by the continuous phase on the dispersed phase, and an extension of this concept is shear-focusing, which aims to create a singular point of highest shear for consistent droplet formation.<br><br>5. Design Method: Flow-focusing channels can be fabricated using various methods such as soft lithography or the insertion of capillary sheathes | | | | |
|---|---|---|---|---|---|
| | into microdevices. The choice of design method can influence the performance and characteristics of the droplet generator.<br><br>Machine learning algorithms are used to predict the droplet diameter and rate based on these factors, enabling design automation and reducing the need for microfluidic expertise and design iterations in life sciences applications. | | | | |
| How do channel wall properties (hydrophobicity/hydrophilicity) influence droplet dynamics? | Channel wall properties, specifically hydrophobicity and hydrophilicity, play a crucial role in influencing droplet dynamics within microfluidic devices.<br><br>In the case of hydrophobic channel walls, they repel water and allow oil to lift the aqueous phase from the walls, surround it, | The channel wall properties, specifically hydrophobicity or hydrophilicity, significantly influence droplet dynamics in microfluidic systems. These properties determine the wettability of the channel walls, affecting the interaction between the continuous and dispersed phases. In hydrophobic channels, the | The response accurately addresses the question by providing a detailed explanation of how channel wall properties influence droplet dynamics, including specific examples of hydrophobic and | 83 | 83.0 |

| | | | | |
|---|---|---|---|---|
| | and encapsulate it into drops, thus forming water-in-oil (W/O) emulsions. This is essential for ultrahigh-throughput biological applications that require W/O emulsions.<br><br>On the other hand, hydrophilic channel walls are necessary to generate oil-in-water (O/W) emulsions. In O/W emulsions, the oil phase adheres to the hydrophilic channel walls, while the aqueous phase forms droplets.<br><br>The wetting properties of the channel walls can be controlled through various methods such as plasma bonding, chemical treatments, and surface coatings with materials like PEG or PVA. These treatments can temporarily convert a naturally hydrophobic | dispersed phase (typically water) tends to form droplets as it minimizes contact with the channel walls. This enhances droplet stability and facilitates easier detachment during droplet formation. Hydrophobic surfaces are ideal for applications requiring water-in-oil emulsions or precise droplet control. Conversely, hydrophilic channels promote wetting of the dispersed phase, making it more likely to spread along the walls instead of forming discrete droplets. This is useful for oil-in-water systems or applications involving liquid films. Wall properties also impact droplet size, breakup, coalescence, and transport. Hydrophobicity favors smaller droplets and reduces the likelihood of droplet adherence, | hydrophilic channels and their effects on emulsion formation. It also discusses the importance of maintaining desired wetting properties for stable droplet formation and manipulation in microfluidics. However, some minor inaccuracies and omissions are present, such as the lack of discussion on the impact of wall properties on droplet size, breakup, coalescence, and transport, which is mentioned in the ground | | |

| | surface into a hydrophilic one or vice versa. The addition of surfactants can also alter the surface wettability, making it more hydrophobic or oleophobic as demonstrated by Xu et al.

It's important to note that maintaining the desired wetting properties is crucial for stable droplet formation and manipulation in microfluidics. The elasticity of PDMS, while beneficial for handling, can be unfavorable if precise channel dimensions are required for stable droplet formation or manipulation. Additionally, low aspect ratio structures may deform or collapse at high flow rates due to the high pressure levels necessary to drive the flow.

In summary, the hydrophobicity/h | while hydrophilicity can lead to higher flow resistance and altered droplet generation behavior. Tailoring wall properties is crucial for optimizing microfluidic processes. | truth answer. SCORE: 83 | | |
|---|---|---|---|---|
| | ydrophilicity of channel walls significantly influences droplet dynamics in microfluidic devices by determining whether water-in-oil or oil-in-water emulsions are formed. The choice between these two depends on the specific application and can be controlled through various surface treatments and coatings. | | | |
| How can you incorporate active control mechanisms (e.g., electrical, magnetic) into droplet devices? | Active control mechanisms such as electrical and magnetic can be incorporated into droplet devices to manipulate the generation, size, and behavior of droplets. For instance, in the case of electrical control, a direct current (DC) or alternating current (AC) can be used to electrically actuate droplet generation. This can be achieved by incorporating electrodes into the flow-focusing | Active control mechanisms such as electrical, magnetic, thermal, and mechanical methods can precisely manipulate droplet formation in microfluidic devices. Electrical control methods include direct current (DC), which uses electrodes to control droplet size, and alternating current (AC), which utilizes electrowetting-on-dielectric (EWOD) and | The response provides a clear explanation of how electrical and magnetic control mechanisms can be incorporated into droplet devices, mentioning specific methods such as electrowetting-on-dielectric (EWOD) | 51 | 51.0 |

| | | | | |
|---|---|---|---|---|
| | device, as shown in Figure 8A, where the waterflow serves as a conductor and the oil stream acts as an insulator. The size of the droplets can be controlled precisely by tuning the electric field strength, which decreases at higher voltage.

   Another electrical method is the use of electrowetting-on-dielectric (EWOD) effect, where an electrical field can be applied to reduce the contact angle between the conductive liquid flow and the channel. This controllable wettability switch allows for precise control over droplet formation.

   In magnetic control, permanent or electromagnets can be utilized to implement non-invasive | dielectrophoresis (DEP) to influence droplet behavior. Magnetic control employs ferrofluids and external magnets to non-invasively guide droplet formation. Thermal control involves heating the entire device or applying localized heating to adjust fluid parameters like viscosity and interfacial tension. Mechanical control techniques, such as pneumatic and piezoelectric components, can also regulate droplet dynamics. These active control strategies enable precise and flexible droplet generation, allowing for tailored applications in complex microfluidic systems, including droplet manipulation and multicomponent emulsion creation. | and the use of ferrofluids. However, it lacks comprehensive coverage of other active control mechanisms like thermal and mechanical methods, which are mentioned in the ground truth answer. The response also does not fully address the broader context of droplet manipulation and multicomponent emulsion creation. SCORE: 51 | | |

| | magnetic control over droplet formation. Ferrofluids with magnetic characteristics are often used as the continuous phase, and factors such as the magnet's direction and location, the density and gradient of the magnetic flux, and the magnetization curve affect droplet formation in the presence of a magnetic field.

It is important to note that these active control mechanisms require external power and actuation to modify the droplet behavior, making them different from passive control methods. Contemporary chemical applications often necessitate an exceptional parallel microfluidic network to create on-demand double or | | | | |
|---|---|---|---|---|---|

| | multicomponent emulsions. | | | | |
|---|---|---|---|---|---|
| How do flow rates affect droplet size and frequency? | The flow rate of the aqueous liquid significantly affects both the droplet size and frequency. According to the provided context, as the flow rate of the aqueous liquid increases, the droplet size or pulse width decreases (Fig. 10a). This inverse relationship is due to the fact that a higher flow rate leads to a higher drag force on the droplet, which in turn results in a higher formation frequency.<br><br>Additionally, Fig. 11(a) shows that the droplet formation frequency is proportional to the fourth power of the mean velocity of the carrier liquid (f∝U^4). This relationship suggests that increasing the flow rate will result in a higher droplet formation | Flow rates significantly influence droplet size and frequency in microfluidic systems. As the flow rate ratio (φ = Qd/Qc) increases, the shearing force from the continuous phase dominates, leading to smaller droplets due to rapid breakup of the dispersed phase. Additionally, higher flow rates in the continuous phase can result in larger droplets, as the dispersed phase doesn't break up until it reaches the downstream section. The frequency of droplet generation also increases with higher flow rate ratios, demonstrating a positive correlation. However, droplet diameter growth trends vary depending on the model and input conditions. Overall, increasing the | The response accurately addresses the question about how flow rates affect droplet size and frequency. It provides specific examples from figures (10a and 11(a)) to support its claims, and it correctly identifies the relationship between flow rate and droplet formation frequency as well as the effect of interfacial tension on droplet size. The response also acknowledges the complexity of the relationships involved and the | 85 | 85.0 |

| | | | | |
|---|---|---|---|---|
| | frequency.<br><br>On the other hand, Fig. 10(b) indicates that increasing the volume ratio between detergent and water decreases the interfacial tension to the carrier liquid, allowing droplets to form at smaller diameters. This means that a higher volume ratio of detergent to water also leads to smaller droplet sizes.<br><br>It's important to note that these relationships are based on the balance between the drag force and the interfacial tension force, with a higher drag force or flow rate leading to a higher formation frequency, and a lower interfacial tension (due to a higher surfactant concentration) also leading to a higher formation frequency.<br><br>The context | flow rate ratio enhances droplet formation frequency and reduces droplet size due to stronger shearing forces, with complex interactions influenced by factors such as interfacial tension and channel geometry. | potential for different mechanisms of droplet formation depending on the flow rates. However, it does not fully capture the nuances of the ground truth answer regarding the relationship between flow rate ratio ($\varphi = Q_d/Q_c$) and droplet size/frequency.<br><br>SCORE: 85 | | |
| | does not provide information about the relationship between flow rates of the continuous phase and droplet size/frequency, but it does mention that increasing the flow rate of the continuous phase can lead to different mechanisms of droplet formation such as the dripping mechanism or jetting flow, which may affect the droplet size and frequency. | | | | |
| What techniques are available for droplet splitting? | There are two main categories of droplet splitting techniques: passive and active. Passive techniques include using common T and Y junctions, flow focusing junction configurations, or placing external ribs/obstructions in the mother droplets' flow route. These methods primarily depend on the microchannel | Droplet splitting techniques can be categorized into passive and active methods. Passive techniques rely on the microchannel geometry and fluid properties, such as T and Y junctions, flow focusing configurations, or introducing ribs/obstructions in the flow path. These methods are simple but lack dynamic control over splitting ratios and are sensitive to flow | The response is mostly relevant and clear, but lacks focus or includes unnecessary details; broadly acceptable. SCORE: 58 | 58 | 57.9999999 9999999 |

| | geometry and working fluids. Active techniques use an electric, magnetic, acoustic, or thermal field to precisely separate the droplets. Examples of active droplet splitting techniques include using a pneumatic actuator, electric and magnetic droplet splitting systems, thermocapillary droplet splitting methods, and acoustic methods for label-free droplet splitting based on acoustic radiation force. Each technique has its advantages and limitations in terms of operational flexibility, sensitivity to fluid properties and flow conditions, selective or on-demand operation, control over the droplet splitting ratio, and applicability to different types of droplets. | conditions. Active techniques utilize external forces like electric, magnetic, acoustic, or thermal fields to precisely control droplet splitting. Examples include electric field-based splitting, magnetic fields for ferrofluids, acoustic radiation force, and thermocapillary methods. Active techniques offer more flexibility, enabling on-demand droplet splitting with controlled ratios, overcoming the limitations of passive methods. However, each technique has its own set of advantages and constraints depending on the application and droplet type. | | | |
|---|---|---|---|---|---|

| What advantages does droplet formation by passive methods have over active methods? | Passive droplet generation methods offer several advantages over active methods. One of the main benefits is that they can increase deformation and instability of flow field interface for preparing monodisperse droplets simply by adjusting microchannel dimensions and geometries or controlling the flow rates of two immiscible liquids without additional driving forces. This simplicity in setup and operation makes passive methods more cost-effective and easier to use compared to active methods, which typically require complex control equipment.

Another advantage is that passive methods are less likely to interfere with or cross-contaminate the | Passive droplet generation methods offer several advantages over active methods. They simplify device design by relying on flow rate manipulation and microchannel geometry, eliminating the need for complex external forces. This leads to cost-effectiveness and easier fabrication. Passive methods also provide precise control over droplet size, shape, and monodispersity, making them ideal for applications requiring uniform droplets. Additionally, they protect droplets from external interference and cross-contamination, ensuring integrity. Passive methods can produce continuous droplet streams with uniform size distributions (1%-3% dispersity) without requiring external actuation. While they are not suitable for | The response accurately conveys the advantages of passive droplet generation methods over active methods, including simplified device design, cost-effectiveness, precise control over droplet characteristics, protection from external interference, and ability to produce uniform droplet streams. However, it includes a minor inaccuracy regarding the limitations of passive methods, stating they are not suitable for dynamic manipulation or | 87 | 87.0 |

| | | | | |
|---|---|---|---|---|
| | synthesized droplets due to their ability to provide precise control over parameters such as size, shape, monodispersity, and components inside the droplets. This makes passive methods ideal for generating continuous droplet streams of uniform size distributions (1%–3% dispersity).<br><br>Lastly, passive methods are widely used to generate droplets in a simple, quick, and efficient manner. However, it's important to note that while passive methods have limitations in independently manipulating droplets in a dynamic and bespoke manner, active methods can create user-defined droplets in a "droplet-on-demand" fashion. The choice of droplet generation | dynamic manipulation or "droplet-on-demand" applications, passive methods are favored for their simplicity, precision, and cost-efficiency in controlled environments. | "droplet-on-demand" applications, which is partially correct but could be more precisely stated. The response also provides a clear and concise explanation with minimal verbosity. SCORE: 87 | | |

| | method should be based on the specific experimental requirements. | | | | |
|---|---|---|---|---|---|
| How do you design a device for monodisperse droplet generation? | To design a device for monodisperse droplet generation, two common methods are typically used: cross-flow designs and flow-focusing designs.

1. Cross-flow designs: In this design, the continuous phase flows into two side channels and surrounds the dispersed phase in a constriction channel. The dispersed phase is squeezed between the two streams of the continuous phase, and spherical droplets are generated in the dripping or jetting regimes. The size of the droplets can be adjusted by changing the width and length of the constriction channel, which affects the shear stress imposed on | To design a device for monodisperse droplet generation, two common methods are used: cross-flow and flow-focusing designs. In cross-flow designs, the continuous phase flows around the dispersed phase in a constriction channel, with droplet size controlled by adjusting channel dimensions and shear stress. This method is simple but can lead to droplet damage due to shear forces or wall interactions. Flow-focusing designs, on the other hand, use sheath flows to stabilize the interface between phases, enabling smaller droplet formation with precise control over size by adjusting flow rates and orifice dimensions. This method offers high monodispersity | The response accurately describes the two common methods for monodisperse droplet generation (cross-flow and flow-focusing designs), their principles, advantages, and limitations. It also provides specific examples of high-performance devices and considerations for choosing between these methods. The answer is clear, concise, and well-structured, with a good balance of technical | 94 | 94.0 |

| | the dispersed phase. Cross-flow designs offer advantages such as high monodispersity and ease of combination with other microfluidic units. However, the generated droplets may be damaged due to shear stress or adhering to the channel walls.

2. Flow-focusing designs: This design is commonly used for generating monodisperse droplets in liquids. It leverages the instability between shear forces and surface tension of sheath flows and a central flow. The emulsion droplet size can be controlled by adjusting the flow velocity ratios of the two liquids and the orifice width. Notably, the droplet sizes can be much smaller than the orifice using this method. | and is less prone to droplet damage. Both designs require careful optimization of flow rates, channel dimensions, and surface tension to achieve uniform droplet sizes. The choice of design depends on specific needs such as droplet size, throughput, and resistance to damage during formation. | detail and general information. SCORE: 94 | | |
|---|---|---|---|---|---|

| | Additionally, monodispersed bubbles in liquid with sizes ranging from 10 to 1000 micrometers can also be produced by this flow-focusing method.<br><br>In recent years, high-performance flow-focusing devices for spontaneous generation of monodispersed droplets have been reported. For example, a circular orifice integrated inside a silicon-based microchannel can generate droplets at high rates while protecting them from damage caused by wetting at the microchannel wall.<br><br>In summary, both cross-flow and flow-focusing designs are effective for monodisperse droplet generation. The choice between these two methods depends on specific | | | | |
| --- | --- | --- | --- | --- | --- |
| | requirements such as desired droplet size, throughput rate, and resistance to damage during droplet formation. | | | | |
| What materials are most commonly used for fabricating droplet microfluidic devices? | The most commonly used materials for fabricating droplet microfluidic devices are PDMS (Polydimethylsilo xane) and glass. These materials are often used due to their high precision, efficient sealing, and good transparency. PDMS is particularly popular because it can be easily molded using soft lithography, allowing for the repeated fabrication of multiple chips with sophisticated networks for precise droplet manipulations. However, it's important to note that other materials such as thermoplastics (like poly(methyl | The most commonly used materials for fabricating droplet microfluidic devices are PDMS (polydimethylsilo xane) and silicon-based materials. PDMS is preferred due to its ease of fabrication using soft lithography, which allows for the creation of precise and versatile microfluidic devices. This material provides high transparency, efficient sealing, and the ability to mold complex microchannels, making it ideal for droplet generation and manipulation. While PDMS is dominant, alternatives such as thermoplastics (e.g., PMMA, polycarbonate), fluoropolymers (e.g., PTFE), and glass materials are also used, each | The response accurately addresses the question by listing common materials used in fabricating droplet microfluidi c devices. It provides a detailed explanation of each material's properties, including their advantages and disadvantag es. However, it slightly deviates from the ground truth answer by mentioning glass as one of the most commonly used materials, | 85 | 85.0 |

| | | | | |
|---|---|---|---|---|
| | methacrylate), polycarbonate, and polystyrene), amorphous polymers (like cyclic olefin copolymer), fluoropolymers (like polytetrafluoroethylene), and glasses can also be used as substrate materials for droplet-based microfluidic systems. Each material has its own advantages, such as outstanding solvent resistance or the ability to mass produce devices using techniques like injection molding or hot embossing. | offering unique benefits like better solvent resistance or mass production capabilities. Despite the variety of materials available, PDMS remains the material of choice for most droplet microfluidics applications due to its simplicity and versatility. | whereas the correct answer is silicon-based materials. The response also includes additional information about other materials that can be used for droplet microfluidic devices.  SCORE: 85 | | |
| What are the advantages and limitations of using PDMS for microfluidic devices? | Advantages of using PDMS for microfluidic devices include its low cost, adaptability through chemical surface modifications and capability to reversibly and irreversibly bond to other materials. Other advantages are its thermal and chemical | Polydimethylsiloxane (PDMS) is widely used for microfluidic devices due to its several advantages, including low toxicity, biocompatibility, optical transparency, and high gas permeability, which make it ideal for biomedical | The response provides a clear and concise list of advantages and limitations of using PDMS for microfluidic devices. It covers most of the key points mentioned | 76 | 76.0 |

| | | | | |
|---|---|---|---|---|
| | stability, flexibility, reversible adhesion to other surfaces, nontoxicity, high gas permeability with known diffusion coefficients for oxygen and carbon dioxide, and optical transparency which permits real-time monitoring by microscopy techniques.

Limitations of using PDMS for microfluidic devices include its elastomeric nature, which is not well suitable for certain techniques. Another limitation is the fouling in the microchannels that arises from the hydrophobic character of the PDMS polymer. The further development of flexible electronics is dependent on further research to improve the flexible and smart microfluidic | applications. Its flexibility and cost-effectiveness also contribute to its popularity, as it allows for easy fabrication using soft lithography. PDMS enables real-time observation and is well-suited for flexible and wearable electronics. However, PDMS has limitations, such as its hydrophobic nature, which can lead to fouling in microchannels, affecting performance. Additionally, its soft elastomeric properties limit mechanical strength, and large-scale production can be challenging. Despite these drawbacks, PDMS remains a versatile and widely used material, though further research is needed to address its limitations for broader applications. | in the ground truth answer, including low cost, adaptability, thermal and chemical stability, flexibility, and optical transparency as advantages, and elastomeric nature, fouling, and limited mechanical strength as limitations. However, it lacks some details and nuances present in the ground truth answer, such as PDMS's biocompatibility and its suitability for biomedical applications. The response also does not fully address the | | |

| | | | | | |
|---|---|---|---|---|---|
| | devices due to these limitations in large-scale production and cost-effective manufacture. | | need for further research to improve large-scale production and cost-effective manufacture.<br><br>SCORE: 76 | | |
| How does 3D printing facilitate microfluidic device fabrication? | 3D printing facilitates microfluidic device fabrication by allowing for the direct generation of 3D structures in a single step. This process, also known as additive manufacturing, builds successive layers of materials automatically under computer control. The geometric information is stored in a CAD file and processed by a computer-controlled printer system.<br><br>This method offers several advantages over traditional fabrication techniques:<br>1. It satisfies | 3D printing revolutionizes microfluidic device fabrication by enabling the creation of complex and precise channel structures that are challenging to achieve with traditional methods like soft lithography. Techniques such as Stereolithography (SL), Multijet Modeling (MJM), and Fused Deposition Modeling (FDM) offer high resolution, complex geometries, and the integration of various components like valves and pumps. 3D printing also allows rapid prototyping, customization, and | The response provides a clear explanation of how 3D printing facilitates microfluidic device fabrication, mentioning the advantages of additive manufacturing, specific techniques like Stereolithography (SL), and benefits such as rapid prototyping and customization. However, it lacks some depth in discussing the impact on | 76 | 76.0 |

| | | | | |
|---|---|---|---|---|
| | arbitrary structural design via a "mail-order" service, meaning that complex structures can be easily designed and produced. 2. It enables more accurate fabrication with high resolution and less time and labor costs. 3. As an additive manufacturing method without removal of redundant parts, it is environmentally friendly. 4. The fabrication route can be commercialized with integration of industrial-grade user interfaces and embedded control systems. One of the best established 3D printing techniques for microfluidic chip fabrication is stereolithography (SL). In this process, photocurable resin liquids are polymerized layer by layer through laser | the use of diverse materials for biocompatibility and transparency. Advantages over traditional methods include reduced fabrication time and labor, increased accuracy, and environmental friendliness due to the additive manufacturing process. With its ability to create hierarchical 3D configurations and integrate multiple functions, 3D printing holds great potential for advancing microfluidic device design and performance. | traditional methods and the potential for advancing microfluidic device design and performance. SCORE: 76 | | |

| | | | | |
|---|---|---|---|---|
| | irradiation. This method has been shown to be convenient and efficient in terms of time and cost for microfluidic device fabrication. | | | |
| What alternative materials can be used for applications involving organic solvents? | The alternative materials that can be used for applications involving organic solvents are primarily inorganic materials such as glass and silicon. Glass materials are recognized as the most advantageous material for microfluidic device fabrication because of their inherent transparency, compatibility with most solvents (including organic solutions), and superior resistance to acetones, benzene, and alcohols compared to other materials like PDMS and PMMA. In addition, a much smaller channel | Several sustainable alternatives to traditional organic solvents are gaining attention for various applications. Supercritical CO2 is a non-toxic, environmentally friendly solvent, particularly useful in extraction processes. Ionic liquids, which are liquid salts at room temperature, offer low vapor pressure and broad application in chemical reactions. Organic carbonates, like dimethyl carbonate, are biodegradable and non-toxic, ideal for fine chemical synthesis. Biosolvents, derived from renewable resources such as plants and bacteria, provide a greener option for | The response attempts to address the question but provides a narrow view of alternative materials by only mentioning glass and silicon, which is not in line with the ground truth that highlights various other sustainable alternatives. SCORE: 16 | 16 | 16.0 |
| | can be achieved by using glass materials compared to other devices. However, it's important to note that these inorganic materials are generally expensive and difficult to achieve commercial usage. | extraction and cleaning. Other alternatives include fluorous solvents, non-toxic liquid polymers like polyethylene glycol (PEG), water, ethanol, methanol, and ethyl acetate. These alternatives are chosen based on their specific properties such as polarity, boiling point, and environmental impact, promoting more sustainable chemical processes and reducing reliance on hazardous petroleum-based solvents. | | | |
| How does device surface treatment affect droplet formation? | The device surface treatment plays a crucial role in determining the type of droplets that can be generated within the microfluidic system. The original PDMS material is hydrophobic, which makes it suitable for water-in-oil (W/O) droplet formation but not oil-in-water | Surface treatment of PDMS microfluidic devices plays a critical role in determining droplet formation and stability. PDMS is naturally hydrophobic, suitable for generating water-in-oil (W/O) droplets but not oil-in-water (O/W) droplets. To generate O/W droplets, selective hydrophilic | The response is mostly relevant and clear, but lacks focus or includes unnecessary details; broadly acceptable. SCORE: 58 | 58 | 57.9999999 9999999 |

| | (O/W) droplets. To generate O/W droplets, a selective treatment is required to amend the PDMS hydrophobicity.

   The authors of this study introduced an approach for selectively treating the PDMS channels to make them hydrophilic, which allows for the generation of O/W droplets. This was achieved by performing the whole selective patterning of PDMS channels in one step using simultaneous PVA injection and on-chip plasma treatment. The plasma treatment increases the surface energy and generates alcoholic hydroxyl, silanol, and carboxylic acid radicals on the PDMS surface, making it hydrophilic. An irreversible PVA coating is then | treatment is required. The researchers employed two methods: manual PVA deposition after plasma treatment, and a simultaneous approach combining PVA injection with on-chip plasma treatment. The latter method enables precise wettability patterning, allowing for both W/O and O/W droplet formation within the same device. The plasma treatment increases surface energy, making the PDMS hydrophilic, while PVA deposition stabilizes the coating, ensuring long-term device use. This technique provides controllable droplet sizes and stable double emulsion formation, with the ability to manipulate droplet characteristics by adjusting flow rates. The method also allows for | | | |

| | adsorbed to the PDMS surface as a result of covalent bonding of PVA molecules to plasma-generated radical species.<br><br>However, if all channels are filled with this treatment, the whole device becomes hydrophilic and W/O droplets are not generated. The authors found that their proposed protocol could reliably generate O/W droplets, while the method reported by Trantidou et al. often resulted in the whole microfluidic network turning hydrophilic due to uncontrollable flow issues.<br><br>This on-chip surface modification technique enables controlled wettability patterning of complex microfluidic networks and allows for the | reliable, reproducible results, making it a powerful tool for droplet microfluidic applications. | | | |
|---|---|---|---|---|---|

| | controllable formation of double emulsion microdroplets, which can be manipulated by tuning the ratio of disperse and continuous phase flow rates. The stability of the PVA coating on PDMS surfaces also allows the devices to be stored for several weeks and applied for double emulsion generation for several hours. | | | | |
|---|---|---|---|---|---|
| What bonding techniques are used to seal microfluidic channels? | The bonding techniques used to seal microfluidic channels, as described in the provided context, include oxygen plasma treatment and thermal fusion bonding. Oxygen plasma treatment is a common method for bonding PDMS channels for ultrahigh-throughput biological applications. In this approach, a fully cured PDMS replicate is treated with oxygen plasma | Several bonding techniques are employed to seal microfluidic channels, each suited to different materials. Thermal fusion bonding involves compressing thermoplastic substrates at temperatures near their glass transition point, allowing them to fuse upon cooling. Chemical bonding creates covalent bonds between substrate surfaces through chemical reactions, while solvent bonding uses solvents to | The response provides a clear description of two specific bonding techniques (oxygen plasma treatment and thermal fusion bonding) used to seal microfluidic channels, with examples provided. However, it lacks comprehensive | 46 | 46.0 | |

| | | | | | |
|---|---|---|---|---|---|
| | immediately before bonding it to another piece of material. Thermal fusion bonding involves compressing and bonding thermoplastic substrates at a temperature around the material's glass transition temperature. For instance, a PMMA microfluidic chip can be made by bonding at 80 °C with a pressure of 2.7 MPa for 30 minutes, followed by at 110 °C with a pressure of 0.02 MPa for 1 hour. | dissolve the materials partially, promoting adhesion through interdiffusion. For PDMS devices, oxygen plasma treatment is commonly used to make surfaces reactive, enabling irreversible bonding when in contact with another PDMS or glass piece. Additionally, corona wand bonding offers a less expensive alternative to oxygen plasma treatment, though it is less consistent. The choice of bonding technique depends on material properties, bond strength requirements, and available equipment, all of which are crucial for the successful fabrication of microfluidic devices. | information on other methods such as chemical bonding, solvent bonding, and corona wand bonding, which are mentioned in the ground truth answer. The response also does not discuss the factors that influence the choice of bonding technique, such as material properties and bond strength requirements. SCORE: 46 | | |
| What challenges arise when fabricating complex microfluidic geometries? | The main challenge in fabricating complex microfluidic geometries is that conventional lithography | Fabricating complex microfluidic geometries presents several challenges, including controlling | The response provides a clear and concise overview of the challenges | 56 | 56.0000000 0000001 |

| | | | | |
|---|---|---|---|---|
| | methods are primarily limited to rectangular cross-sections. This limitation makes it difficult to create channels with complex 3D geometries. Another challenge is the development of fast, versatile methods for fabricating 3D microfluidic networks with necessary geometries and feature sizes.<br><br>Recent progress includes methods for layer-by-layer stacking of 2D microfluidic platforms and semi-automated weaving of microfluidic tubing. However, fabrication of 3D microstructures with arbitrary geometries remains a significant challenge in the field of micromechanics and microfabrication.<br><br>In addition to these technical challenges, | wettability in intricate channel structures, which is critical for fluid flow and droplet formation. Additionally, scaling up microfabrication processes for commercial applications requires cost-effective techniques and materials. Current droplet generation methods mainly produce single emulsions, limiting their use in applications that require multiple or complex emulsions. Moreover, droplet microfluidics faces limitations in producing diverse and functional nano-scale materials compared to other methods. Conventional lithography techniques, which are restricted to rectangular cross-sections, struggle with creating complex 3D geometries. Although advancements like | associated with fabricating complex microfluidic geometries, including limitations of conventional lithography methods and difficulties in designing effective mixing strategies. However, it does not fully address the ground truth answer's points about controlling wettability, scaling up microfabrication processes for commercial applications, and limitations in producing diverse and functional nano-scale | | |

| | designing effective mixing strategies for complex 3D structures can also be difficult due to the unique phenomena that may occur in such structures. | 3D printing and stacking methods offer solutions, effective mixing strategies and versatile fabrication of 3D networks with the required features remain significant hurdles in the field. Addressing these challenges will require innovative approaches and collaboration across various disciplines. | materials. The response also lacks specific examples or details to support its claims. SCORE: 56 | | |
|---|---|---|---|---|---|
| How do you fabricate devices capable of handling multiphase flows? | Devices capable of handling multiphase flows can be fabricated using several methods. One approach is the use of an annular capillary array as the injection tube, where inner and middle fluids flow coaxially through the inner layer and sheath layer of the tube, respectively, into a surrounding outer fluid. When the three phases meet at the tip of the injection tube, double emulsions are prepared. By injecting multiple inner phases into the capillary | Fabricating devices for handling multiphase flows involves several techniques and considerations. Key approaches include advanced micromanufacturing methods such as 3D printing, which eliminates the need for bonding processes, and flow-focusing techniques, which use nozzles and sheath fluids to control droplet formation. Controlling wettability is crucial for generating specific droplet | The response provides a clear and detailed explanation of methods for fabricating devices capable of handling multiphase flows, including the use of annular capillary arrays, parallel channels, and advanced micromanufacturing techniques. It also | 76 | 76.0 |

| | array, multicomponent double emulsions with different inner cores can be obtained from the same device.

Another method involves designing parallel channels for different inner or middle phase fluids to fabricate double-emulsion droplets with multiple cores or shells. This design can be coupled with the previous one, allowing for the fabrication of more complex emulsions.

It's worth noting that in some cases, double-emulsion generation may need to be mediated by external forces such as pneumatic control, electrical force, and mechanical vibration.

Advanced micromanufacturing techniques can also be used to | types, while parallelization of droplet generators enhances production scalability. Challenges include ensuring even fluid distribution, integrating various components like valves and connectors, and addressing the lack of standardized design rules. Additionally, methods like consecutive capillary assemblies and external forces (e.g., pneumatic or electrical control) can be employed to create complex emulsions. Through careful design and optimization, efficient fabrication of multiphase flow devices for applications such as droplet generation can be achieved. | mentions the importance of controlling wettability and parallelization of droplet generators for scalability. However, it does not address some key aspects mentioned in the ground truth answer, such as 3D printing, flow-focusing techniques, and the challenges of ensuring even fluid distribution and integrating various components. SCORE: 76 | | |

| | improve the fabrication process, making it more efficient and precise. For instance, photolithographic processes can be employed to fabricate hundreds or even thousands of droplet generators, enabling massive parallelization for scaling up production. | | | | |
|---|---|---|---|---|---|
| What is the impact of channel roughness on droplet formation and behavior? | The impact of channel roughness on droplet formation and behavior can be significant. Roughness in the capillary channel can prevent secondary droplet formation and decrease the number of fingers formed around the primary droplet during normal droplet impact. This is because increasing surfaces roughness alters the flow dynamics, affecting the contact line and droplet stability. Additionally, | The impact of channel roughness on droplet formation and behavior is a key factor in microfluidics. Studies indicate that surface roughness can significantly affect droplet dynamics, such as movement, spreading, and stability. For example, increased roughness can hinder droplet spreading and prevent secondary droplet formation by altering flow dynamics and the contact line. Additionally, vibration and | The response provides a clear explanation of how channel roughness affects droplet formation and behavior, mentioning its impact on flow dynamics, contact line, and stability. It also discusses the influence of surface structures and gas properties | 71 | 71.0 |
| | surface structures and the surrounding gas (such as air) can influence the droplet impact dynamics. However, these factors are often neglected in studies due to the large density and viscosity ratios between the fluid and the surrounding gas. In this specific study, a direct numerical simulation is performed in a three-dimensional channel to explore the dynamic behavior of droplets moving across a rough capillary channel. | surface structure play roles in influencing droplet impact and splash behavior. Rougher surfaces tend to reduce spreading, while smoother surfaces may enhance it. Surface wettability also interacts with roughness to further complicate droplet behavior. Overall, channel roughness introduces complexity to droplet dynamics, with its effects varying based on factors like roughness level, fluid type, and external conditions, highlighting the need for further research in this area for better understanding and optimization in microfluidic applications. | on droplet impact dynamics. However, it lacks detailed discussion on the specific effects of roughness levels, fluid types, and external conditions, which are crucial aspects mentioned in the ground truth answer. The response could benefit from a more comprehensive exploration of these factors to provide a complete understanding of the topic. SCORE: 71 | | |
| How can sensors be integrated into droplet microfluidic devices during fabrication? | Sensors can be integrated into droplet microfluidic devices during fabrication through several | Integrating sensors into droplet microfluidic devices during fabrication is a complex process | The response provides a clear and concise overview of methods to | 51 | 51.0 |

| | methods. One common approach is microfabrication and soft lithography, where the sensor components are fabricated using photolithography or other microfabrication techniques and then integrated with the microfluidic channels. Another method is direct integration, where the sensor components are directly patterned onto the substrate along with the microfluidic channels. In some cases, sensors can also be assembled after the microfluidic device has been fabricated by using methods such as drop-casting or inkjet printing to deposit the sensor material onto the device. The choice of method depends on factors such as the type of sensor, the desired sensitivity and | that requires advanced microfabrication techniques and careful material selection. For successful integration, precision in manufacturing is crucial, as it often involves combining different microfluidic units, such as droplet generation and detection units, on a single chip. Various techniques like photolithography, soft lithography, and direct integration methods can be used, depending on the type of sensor (e.g., optical or electrochemical). The fabrication process also needs to account for material compatibility, sensor functionality, and alignment to ensure reliable performance. Additionally, sensor integration may face challenges such as maintaining | integrate sensors into droplet microfluidic devices during fabrication. It correctly identifies various techniques such as microfabrication, soft lithography, direct integration, drop-casting, and inkjet printing. However, it lacks the depth and nuance provided in the ground truth answer regarding the complexity of the process, precision in manufacturing, material selection, and potential challenges like stability and interference with fluid | | |

| | selectivity, and the compatibility of the sensor material with the microfluidic device. | stability within the microfluidic environment and minimizing interference with fluid flow. By considering these factors, sensors can be effectively integrated into droplet microfluidic devices for applications like chemical reactions and biological detection. | flow. The response also does not explicitly mention the importance of considering factors such as sensor type, sensitivity, selectivity, and compatibility with the microfluidic device.<br><br>SCORE: 51 | | |
|---|---|---|---|---|---|
| How are droplet microfluidic devices used for single-cell analysis? | Droplet microfluidic devices are used for single-cell analysis by encapsulating individual cells in oil-based droplets. This allows researchers to manipulate the droplets and perform assays that provide genetic, epigenetic, and functional information about each cell. The technique enables the examination of interactions between encapsulated | Droplet microfluidic devices are widely used for single-cell analysis due to their ability to encapsulate individual cells within tiny droplets, allowing for high-throughput and precise experimentation. These devices work by generating microdroplets in a fluid flow, each containing a single cell along with reagents needed for analysis, such as enzymes, antibodies, or | The response is clear and concise, accurately describing the use of droplet microfluidic devices for single-cell analysis. It provides specific examples and applications, such as high-throughput fluorescence imaging and single-cell RNA | 83 | 83.0 |

| | | |
|---|---|---|
| cells, cytotoxicity, kinetics, and other important factors. Additionally, it facilitates downstream omics analysis, uncovering crucial insights into the relationship between cell function and phenotype. For instance, high-throughput fluorescence imaging can be conducted within these oil droplets to evaluate cytotoxicity effectively. This method offers distinct advantages over conventional methods of drug discovery and validation, such as single-cell resolution, high-throughput screening, and compatibility with AI and robotic systems. Examples of successful applications include the development of platforms for single cell RNA | primers. This allows for the isolation of cells in a controlled environment, enabling high-resolution studies of gene expression, protein profiling, and cellular responses. The compartmentalized nature of droplets also reduces cross-contamination between samples. Moreover, the small size and scalability of droplet microfluidics make it possible to perform large-scale, parallel analysis of thousands of single cells in a short amount of time. This technology has applications in genomics, proteomics, and drug screening, facilitating insights into cell heterogeneity and the identification of rare cell populations. | sequencing platforms like inDrop and Drop-seq. However, it could be improved by providing more technical details about the encapsulation process and the benefits of droplet microfluidics over conventional methods. The response also slightly deviates from the ground truth answer by not mentioning the controlled environment and compartmentalized nature of the droplets as much. SCORE: 83 |

| | | | | | |
|---|---|---|---|---|---|
| | sequencing like inDrop and Drop-seq. | | | | |
| What are the advantages of droplet microfluidics for high-throughput screening? | Droplet microfluidics offers several significant advantages for high-throughput screening applications. These include:<br><br>1. Reduced sample volumes: The use of droplets allows for a significant reduction in the volume of samples, leading to concomitant reductions in cost.<br><br>2. Increased assay sensitivity: Compartmentalization in droplets increases assay sensitivity by increasing the effective concentration of rare species and decreasing the time required to reach detection thresholds.<br><br>3. High-speed manipulation and measurement: Manipulation and measurement at kilohertz speeds | Droplet microfluidics offers several advantages for high-throughput screening, including reduced sample volumes, which lowers costs, and increased assay sensitivity through compartmentalization. This method allows for the screening of up to $10^8$ samples per day at kilohertz speeds, significantly improving throughput and efficiency. The small droplet volumes enhance effective concentrations and signal-to-noise ratios, critical for sensitive detection of low-concentration biological molecules such as antibodies, enzymes, and single genomes. Additionally, droplet microfluidics enables single-cell and single-molecule assays, | The response provides a clear and comprehensive list of advantages of droplet microfluidics for high-throughput screening, including reduced sample volumes, increased assay sensitivity, high-speed manipulation and measurement, ultra-high throughput, single-cell studies, high sensitivity, and linking phenotype to genotype. However, the answer could be improved by condensing some points and providing a | 58 | 57.9999999 9999999 |

| | enable up to 10^8 samples to be screened in one day, making it possible to conduct high-throughput assays of complex chemical, biochemical, or pharmaceutical analytical processes in a rapid, automated, and reproducible manner.

4. Ultra-high throughput: The ultra-high throughput attainable by droplet microfluidics is particularly significant for cost when screening large libraries of precious small compounds. Small volumes lead to an enormous increase in effective concentrations and signal-to-noise ratios, which is critical for performing single-cell assays.

5. Single-cell studies: Droplet microfluidics not | providing detailed insights into cellular behaviors and allowing for novel experimental approaches. This technology is particularly beneficial for applications like drug discovery, enzyme evolution, and diagnostics, offering a versatile and cost-effective platform for advanced biological research. | more cohesive narrative that aligns with the ground truth answer. The response is mostly clear and relevant but lacks focus in some areas. SCORE: 58 | | |
|---|---|---|---|---|---|

| | | | | |
|---|---|---|---|---|
| | only magnifies the scale of existing high-throughput screens but also enables new types of experiments, such as single-cell population studies. These studies can provide valuable insights into bacterial persistence, virus–host interactions, and cell growth in general.<br><br>6. High sensitivity: Droplet microfluidics is useful for detecting and analyzing small amounts of biological molecules, such as antibodies, enzymes, and singlegenomes.<br><br>7. Linking phenotype to genotype: Compartmentaliz ation in droplets can be used as a method for linking phenotype to genotype, in place of genetic | | | |

| | | | | | |
|---|---|---|---|---|---|
| | techniques such as surface display. | | | | |
| How can droplets be used to encapsulate cells or biomolecules? | Droplets can be used to encapsulate cells or biomolecules in several ways. The most common method is through limiting dilution, where the concentration and drop size are tuned to precisely set the fraction of drops containing single molecules or cells. This approach is widely used due to its simplicity but has a limitation that it requires high flow rates or close packing of the cells.<br><br>Another method involves exploiting inertial microfluidics, where a spiral geometry couples Dean forces with inertial lift to order the cells prior to encapsulation. The periodicity of the regularly spaced cells can be matched to that of the drop | Droplet microfluidics provides versatile methods for encapsulating cells and biomolecules. Limiting dilution is a traditional approach that randomly distributes single cells or biomolecules in droplets, though it has low efficiency. Inertial microfluidics leverages spiral geometries and inertial forces to guide cells into droplets, improving encapsulation efficiency by ensuring single-cell droplets. Bi-disperse emulsions create an emulsion of empty and cell-containing droplets, with the latter selectively recovered using methods like pinched flow fractionation. For biomolecules, limiting dilution is commonly used, adjusting | The response provides a clear and detailed explanation of various methods for encapsulating cells or biomolecules using droplets, including limiting dilution, inertial microfluidics, bi-disperse emulsions, lipid vesicles, and slugs as microreactors. It accurately describes the advantages and limitations of each method, aligning with the ground truth answer. The response is well-structured, concise, | 85 | 85.0 |

| | formation, filling most drops with single cells. This approach may potentially be more valuable for biological applications because it does not require high flow rates or close packing of the cells but is yet to be widely adopted due to the complexity of integrating the cell triggering and sorting devices into a microfluidics system.

Additionally, bi-disperse emulsions consisting of small, empty drops and large drops containing single cells can be created. The cell-containing drops can then be selectively recovered from the emulsion using passive methods like pinched flow fractionation. This encapsulation approach does not require toxic solvents and is a | concentration and drop size to achieve single-molecule encapsulation. Additionally, droplets can encapsulate cells in lipid vesicles for biological applications or serve as microreactors in processes like DNA analysis. These techniques enable precise control over encapsulation, enhancing their applicability in cell-based assays, drug discovery, and genetic research | and easy to follow, making it a solid example of a clear and relevant answer. SCORE: 85 | | |
| --- | --- | --- | --- | --- | --- |

| | single-step process.

Furthermore, lipid vesicles encapsulating biological macromolecules or cells can be generated by injecting an EtOH/H2O mixture into a water–lipid emulsion containing oleic acid and phospholipids as the carrier fluid. Encapsulation of biological molecules or cells is achieved by using an aqueous solution containing the encapsulation targets to generate the initial droplets, which then rearrange to form vesicles. Cells encapsulated in these vesicles remained viable for up to 2 hours.

Lastly, slugs can be used as microreactors for DNA analysis and handling nanoliter volumes of solutions in slugs. The sample | | | | |
|---|---|---|---|---|---|
| | and reagents are combined and heated in the reaction chamber, and when the reaction is finished, the slug is moved by pressure to an electrophoresis channel for on-chip separation and detection. Polymerase chain reaction of DNA has also been performed in continuous flows of slugs and plugs. | | | | |
| What are the applications of droplet microfluidics in drug discovery? | Droplet microfluidics has several applications in drug discovery. One application is drug dissociation assays, which are crucial for evaluating the drug release rate and extent in microparticle- and nanoparticle-based drug delivery systems. Another significant application is the use of single-cell microfluidic devices for drug discovery studies. These devices allow for the | Droplet microfluidics plays a significant role in drug discovery by enabling high-throughput screening and precise analysis of drug interactions at the single-cell level. This technology allows the generation of picoliter-sized droplets, each encapsulating a single cell or a small group of cells, which can be exposed to various drug compounds. This compartmentalization facilitates parallel testing of | The response provides a clear and concise overview of droplet microfluidics applications in drug discovery, including drug dissociation assays and single-cell analysis. However, it lacks some depth and specific details compared to the ground | 56 | 56.0000000 0000001 |

| | investigation of cells individually by encapsulating them in oil-based droplets, manipulating the individual droplets, and performing assays that provide genetic, epigenetic, and functional information about individual cells. This technique enables the examination of interactions between encapsulated cells, cytotoxicity, kinetics, and other important factors. Additionally, it facilitates the isolation of individual cells for downstream omics analysis, uncovering crucial insights into the relationship between cell function and phenotype. For instance, high-throughput fluorescence imaging can be conducted within these oil droplets to evaluate | numerous drug candidates, speeding up the identification of promising therapeutics. Droplet microfluidics is particularly useful in identifying drug efficacy, optimizing drug formulations, and studying complex cellular responses such as gene expression, protein production, or metabolic changes in response to treatment. Additionally, it enables the screening of rare cell populations or targets that would be difficult to isolate using traditional methods. The ability to rapidly test and analyse thousands of drug interactions in a controlled environment accelerates drug discovery, reduces costs, and improves the chances of identifying novel therapeutic agents. | truth answer, which highlights the technology's ability to accelerate drug discovery, reduce costs, and improve therapeutic agent identification. The response also does not explicitly mention the generation of picoliter-sized droplets or the parallel testing of numerous drug candidates, which are key aspects of droplet microfluidics in drug discovery. SCORE: 56 | | |
|---|---|---|---|---|

| | cytotoxicity effectively. | | | | |
|---|---|---|---|---|---|
| How can droplet microfluidics enable precise control of chemical reactions? | Droplet microfluidics enables precise control of chemical reactions by dispensing different reactants into individual droplets with precise volume control. The chemical reaction is activated by coalescing pairs of droplets inside the microfluidic device, which acts as a reaction container. This approach allows for long operation times even for reactants that would otherwise clog microfluidic devices. The presented microfluidic scheme overcomes the limitations of single-phase microfluidics due to its inherent capability of precise process and volume control. Additionally, suitable pre-processing can be added to the microfluidic | Droplet microfluidics offers precise control over chemical reactions by dispensing reactants into individual droplets with accurate volume control. This enables the mixing of reactants within confined spaces, enhancing reaction efficiency. Coalesced droplets act as micro-reactors, providing fast mixing and long operation times without clogging, even with sticky or precipitating reactants. The system allows for optimization of reaction parameters, such as temperature, flow rate, and pressure, ensuring the production of specific products, like mesoporous silica microspheres with superior surface area and narrow pore distributions. This capability to finely control | The response is clear and concise, accurately addressing the question about how droplet microfluidics enables precise control of chemical reactions. It provides specific examples and details about the benefits of this technology, such as precise volume control, long operation times, and optimization of reaction parameters. However, it could be improved by being even more concise and directly aligning with the ground | 85 | 85.0 |

| | device, such as giving droplets some time at elevated temperature and reduced flow velocity to develop the gel network and start the evaporation of the solvent. This method allows for the performance of complex chemical reactions, as demonstrated by the production of silica particles using combined acid-catalyzed hydrolysis and base-catalyzed condensation for rapid gelation. All microfluidic processing steps can be optimized to meet the particular requirements of the desired product, resulting in superior properties such as a high surface area and narrow pore radius distribution. | reaction conditions makes droplet microfluidics a powerful tool for precise, high-throughput chemical synthesis. | truth answer in terms of phrasing and emphasis on key points. SCORE: 85 | | |
|---|---|---|---|---|---|
| What are the key challenges of using droplet microfluidics in diagnostics? | The key challenges of using droplet microfluidics in diagnostics, as outlined in the | Droplet microfluidics presents significant challenges in diagnostics, | The response provides a detailed and accurate list of | 96 | 96.0 |

| | provided context, include:

1. Prevention of wetting and cross-contamination: Since the droplets are aqueous and the continuous liquids are immiscible organic liquids, finding the right surfactant that will maintain a stable and inert interface is often challenging. Additionally, developing a suitable barcoding system for droplet libraries can be difficult.

2. Monitoring and controlling the level of oxygen in the droplets: The chemistry inside the droplets may require specific conditions or generate chemistries that can alter properties such as surface tension or mass transfer rates across the interface, which can impact the mechanics of | including preventing wetting and cross-contamination between aqueous droplets and immiscible organic liquids, which can compromise results. Monitoring oxygen levels within droplets is difficult due to the sensitivity of the liquid-liquid interface to mechanical perturbations and mass transfer. Developing stable detection chemistries is also complex, as the confined droplet environment can alter chemical reactions and impact accuracy. Additionally, challenges in automation and accessibility limit the widespread use of droplet microfluidics, particularly for high-throughput assays. Overcoming these issues requires advancements in surfactants, microfluidic engineering, and | challenges associated with using droplet microfluidics in diagnostics. It covers various aspects such as preventing wetting and cross-contamination, monitoring oxygen levels, developing stable detection chemistries, automating techniques for manipulating droplets, and addressing limitations compared to bulk assays. The answer is well-structured, clear, and concise, providing a comprehensive overview of the challenges faced by | | |

| | handling the droplets or the physico-chemistry of the assays.<br><br>3. Providing stable and reliable detection chemistries: The high surface area-to-volume ratios of droplets necessitate the development of suitable detection systems for these small volumes.<br><br>4. Widening the portfolio of automated techniques for manipulation of droplets in situ in microfluidic chips: This involves making pivotal techniques such as generation, detection, or sorting of droplets more easily available to users without a background in microfluidics.<br><br>5. Addressing limitations compared to bulk assays: While droplet microfluidics offers enormous | automated techniques, which could enhance the precision and applicability of droplet systems in diagnostics. | this technology. SCORE: 96 | | |
|---|---|---|---|---|---|

| | opportunities for high-throughput biological assays, it also has some constraints that limit its applications. For example, finding the right surfactant that will maintain a stable and inert interface is often a challenging chemistry problem. Similarly, while a library of a billion droplets can easily be handled in a single microtube, droplets cannot take advantage of the spatial barcoding that is available in the two-dimensional arrays of microtiter plates.<br><br>These challenges pose good opportunities for advancements in each area, which could broaden the use of droplet systems in microbiology and diagnostics. | | | | |
|---|---|---|---|---|---|
| How are droplets used for DNA sequencing applications? | Droplet microfluidics is a technique used in high-throughput DNA sequencing. | Droplet microfluidics plays a key role in DNA sequencing by enabling high- | The response accurately describes the | 96 | 96.0 |

| | In this method, individual droplets containing single biological targets (such as cells, bacteria, or viruses) are encapsulated and processed independently for precise and high-throughput analyses. The genotypes of the encapsulated single targets can be analyzed using barcodes, which link with target DNA and RNA molecules. These barcodes consist of unique sequences of oligonucleotides that can be read out through sequencing. A library of barcodes is separately encapsulated into the droplets with single targets, thus labeling the nucleic acids from the same targets with unique primers. The processed samples are then pooled and sequenced to profile gene expression at a | throughput analysis and precise targeting of genomic regions. This includes single-cell sequencing, where individual cells are encapsulated in droplets for independent gene expression profiling. Barcode-based sequencing allows for the labeling of nucleic acids with unique identifiers, enabling the tracking of specific genomes during analysis. Multiplexed PCR amplification is achieved by performing PCR in separate droplets, ensuring uniform amplification of multiple targets. Additionally, emulsion PCR combines droplet merger with PCR to amplify specific genomic regions, facilitating efficient sequencing. These techniques offer a miniaturized, efficient platform for analyzing complex genetic | application of droplets in DNA sequencing, including high-throughput analysis, single-cell sequencing, barcode-based labeling, and multiplexed PCR amplification. It provides a clear explanation of the process and its benefits. SCORE: 96 | | |
|---|---|---|---|---|---|
| | single-target level. This technology allows for high-throughput single-cell sequencing, enabling screening of cell heterogeneity in chromatin states, somatic mutations, and transcriptional states. | information, improving accuracy and throughput in genomic studies. | | | |
| What role do droplets play in tissue engineering research? | Droplets play a significant role in tissue engineering research, particularly through the use of microfluidics. In this context, droplets are used to create small-scale structures such as core-shell microparticles or hollow bacterial cellulose (BC) microspheres. These structures can be used as building blocks for creating injectable scaffolds, which are highly porous and demonstrate proper in vitro cell culture with improved depth distribution and high cell proliferation rates | Droplet microfluidics plays a pivotal role in tissue engineering by enabling precise fabrication of biomaterials and 3D cell culture systems. Droplets serve as templates for creating hydrogel-based materials, which can encapsulate cells for more accurate modeling of tissue structures. This technology facilitates the production of microcarriers and complex biomaterials, such as hollow bacterial cellulose (BC) microspheres, which form injectable scaffolds for | The response is clear, concise, and accurately addresses the question about the role of droplets in tissue engineering research. It provides specific examples and details about the use of microfluidics, core-shell microparticles, hollow bacterial cellulose microspheres, and metal-organic | 86 | 86.0 |

| | | | | |
|---|---|---|---|---|
| | compared to bulk BC scaffolds.<br><br>The use of droplets allows for precise control over the size, composition, and structure of these building blocks, enabling the creation of functional elements that can be assembled into complex structures. For example, the injectable scaffold generated using hollow BC microspheres demonstrated tissue regeneration in vivo, leading to a faster wound-healing process in a rat skin model.<br><br>Additionally, other studies have used droplets to encapsulate metal-organic frameworks (MOFs) for controlled release of ions such as calcium, copper, and zinc. These MOFs provide antibacterial and antioxidant | wound healing, promoting high cell proliferation and tissue regeneration in vivo. Additionally, droplet-based microfluidics allows for the creation of core-shell structures, such as alginate shells or metal-organic frameworks, which offer controlled release of therapeutic agents. These advancements provide a powerful platform for studying tissue development and regenerative therapies. | frameworks, which align with the ground truth answer. The response demonstrates a good understanding of the topic and effectively conveys relevant information. SCORE: 86 | | |
| | properties, making them useful in tissue engineering applications.<br><br>    In summary, droplets are essential tools in tissue engineering research, enabling the creation of complex structures with precise control over their composition and properties, which can be used for tissue regeneration and other applications. | | | | |
| How can droplet microfluidics aid in studying cell signaling pathways? |    Droplet microfluidics can aid in studying cell signaling pathways by providing well-defined environments for individual cells. The droplet content can be systematically varied during droplet generation and modified subsequently by fusion with another droplet, allowing for the manipulation of | Droplet microfluidics offers a powerful platform for studying cell signaling pathways by enabling the precise isolation and manipulation of individual cells within controlled environments. This technology allows for the encapsulation of single cells in droplets, providing a controlled space to study cell-cell | The response provides a clear and detailed explanation of how droplet microfluidi cs can aid in studying cell signaling pathways, including the use of controlled environmen ts, single-cell encapsulati | 82 | 82.0 |

| | | | | |
|---|---|---|---|---|
| | specific combinations of cells. This microfluidic assay seems to be much faster than conventional assay techniques.

In single cell-based microdroplet assays, the expression of proteins and activity of enzymes can be analyzed using fluorescence techniques. As long as the volume of the droplet is kept constant during the experiment, the fluorescence intensity allows for a quantitative analysis. On-chip lysis of cells in droplets is an alternative way to access the cell contents, but also opens a route to introduce material into cells. Electroporation of cells in microfluidic droplets can be achieved when cell-containing droplets flow through a pair of | interactions and signaling events. The ability to merge droplets containing different cell types or reagents facilitates the investigation of cell communication, such as paracrine signaling, in a way that mimics in vivo conditions. Additionally, droplet-based microfluidics supports high-throughput screening and single-cell genomic studies, offering insights into phenotypic variations and signaling dynamics within genetically identical cell populations. This capability is particularly valuable for understanding complex cellular behaviors in development, differentiation, and disease. | on, and high-throughput screening. However, it could be improved by providing more concise language and a clearer connection to the broader context of cell signaling research. The response also slightly deviates from the ground truth answer in some minor points, but overall provides accurate information on the topic. SCORE: 82 | | |

| | microelectrodes with a constant voltage established in between.<br><br>    Studies on stationary droplets containing single cells show the potential of droplet-based microfluidics for studies into stochastic phenotypic variations within populations of genetically identical cells. This could be particularly beneficial for fundamental studies in the field of cell–cell communication and paracrine signaling, as mentioned in the provided context.<br><br>    Furthermore, droplet-based microfluidics can also be employed to study single-cell genomics at high throughput, which is attracting increasing attention due to its significance in development, | | | | |
|---|---|---|---|---|---|
| | differentiation, signaling, and disease. For example, genomic heterogeneity is of major importance in cancer therapy, since it can cause the selection of resistant cells. Droplet-based microfluidic technology that can allow for compartmentaliza tion not only of single cells but also singlegenomes is thus an ideal technique to study single-cell genomics at high throughput. | | | | |
| What is the potential of droplet microfluidics in personalized medicine? | Droplet microfluidics has significant potential in personalized medicine. The technology allows for the optimization of treatments for specific cell lines, such as pancreatic cancer cells, by conducting a combination of drugs that induce apoptosis. This is particularly useful in | Droplet microfluidics holds great promise in personalized medicine by enabling precise, high-throughput analyses of biological samples. The technology allows for the optimization of treatments by testing drug combinations on specific cell lines, such as pancreatic cancer cells, to | The response accurately addresses the question about the potential of droplet microfluidi cs in personalize d medicine, providing specific examples and details that align with the ground truth. It | 94 | 94.0 |

| | validating particular drugs and drug combinations to determine their efficacy in inducing cytotoxicity toward primary human tumors. | identify the most effective therapies. It also facilitates the early detection of cancers through highly sensitive detection of tumor markers in bodily fluids, offering a minimally invasive approach for patient monitoring. Additionally, droplet-based digital PCR (dPCR) enables the detection of rare sequences and precise quantification of nucleic acids, which is crucial for cancer research and tailoring treatments based on a patient's genetic background and tumor profile. Overall, droplet microfluidics can significantly advance personalized medicine by providing tailored treatment strategies and improving early detection for better patient outcomes. | demonstrates a clear understanding of the technology's capabilities and its applications in drug discovery, cancer research, and patient monitoring. The language is technical but coherent, and the structure is logical and easy to follow. SCORE: 94 | | |
| | Furthermore, the high-throughput capabilities and accuracy of single-cell microfluidics and droplet-based microenvironments have been demonstrated in quantitatively determining foodborne pathogens like Salmonella. This sensitivity and rapid detection speak to the potential of microfluidics in personalized medicine, where rare sequences within biological samples can be detected with high precision. | | | | |
| | In addition, the miniaturization of microfluidic tools has enabled the development of robust digital | | | | |

| | Polymerase Chain Reaction (dPCR) systems, which are ideal for cancer research and clinical follow-up. These systems allow for the precise detection of target nucleic acids within samples, even in minimally invasive manners such as detecting tumor markers in body effluents like blood, urine, or feces.

In conclusion, droplet microfluidics provides a sophisticated platform for modeling tumors and the TME (tumor microenvironment), with promising future applications in drug discovery and personalized medicine. The technology's potential is enormous, and it is anticipated that it could become an essential tool for personalized or precision | | | | |
|---|---|---|---|---|---|

| | | | | |
|---|---|---|---|---|
| | medicine, with the objective of detecting cancers in early stages and treating patients with regimens that best fit their genetic background as well as the genome of their tumor. | | | | |
| How are droplet microfluidic devices used for nanoparticle synthesis? | Droplet microfluidic devices are utilized for nanoparticle synthesis due to their unique advantages such as miniaturization, enhanced mass and heat transfer, and reduced time and reagent consumption. The reaction is encapsulated in confined droplets, allowing the use of toxic or volatile chemicals without contaminating the channel walls. The advection flow field within the droplets accelerates mixing, resulting in a well-defined starting point and an even residence | Droplet microfluidic devices revolutionize nanoparticle synthesis by enabling precise control over reaction conditions within isolated microreactors. These droplets encapsulate well-defined quantities of reagents, ensuring rapid mixing and preventing contamination, even when using toxic or volatile chemicals. This approach addresses limitations like diffusion-limited mixing found in traditional systems, resulting in nanoparticles with controlled size, shape, and | The response is mostly relevant and clear, but lacks focus or includes unnecessary details; broadly acceptable. SCORE: 58 | 58 | 57.9999999 9999999 |

| | | | | |
|---|---|---|---|---|
| | time, which contribute to a narrower size distribution of the final nanoparticles.<br><br>    Local control over the synthetic environments can be exerted on separate droplet reactors, enabling homogeneous synthesis and quantity production. The reaction parameters scale up linearly, allowing for homogeneous synthesis and large-scale production. Typically, nanoparticles can be synthesized in a homogeneous manner, where seeds emerge in situ and grow in the droplet reactors, or they can be generated in a seed-mediated manner, where seeds are preformed and then added into the droplet reactors for growth.<br><br>    The flexibility of droplet | narrow size distribution. The method supports multi-step synthesis on millisecond timescales, allowing for the scalable production of high-quality nanoparticles with applications in fields such as drug delivery, imaging, and catalysis. | | | |

| | microfluidic techniques has enabled the synthesis of a variety of nanoparticles, including inorganic, organic, and hybrid or complex nanomaterials, with excellent control over the size, size distribution, shape, and crystal structure. | | | | |
|---|---|---|---|---|---|
| What role do droplets play in polymer particle production? | Droplets play a crucial role in the synthesis and fabrication of monodisperse polymer particles using droplet-based microfluidic systems. The emulsification of monomers or oligomers into droplets breaks up their liquid threads, which can then be solidified by chemical and physical crosslinking processes. This process allows for the production of a wide variety of materials, including gels, polymers, and | Droplet-based microfluidics provides a versatile platform for producing monodisperse polymer particles by encapsulating monomers or oligomers into droplets and solidifying them through chemical or physical crosslinking. This approach enables the creation of diverse particle types, including solid, hollow, asymmetric, and multicompartmental structures with precise control over size, shape, and morphology. Techniques such as UV-initiated | The response accurately addresses the question about the role of droplets in polymer particle production, providing a clear explanation of the process and its benefits. It correctly identifies the importance of uniformity in droplet size for implementing functionaliti | 76 | 76.0 |

| | | | | |
|---|---|---|---|---|
| | polymer-doped particles with controlled size, shape, and composition. The uniformity of these droplets is essential for implementing the functionalities of the resulting polymer particles in various fields such as biochemical analysis, tissue engineering, and controlled release applications. | polymerization and solvent evaporation facilitate the transformation of droplets into solid particles. Additionally, functional micro- or nanostructures can be incorporated within or on the surface of these particles by introducing specific ingredients into the droplets, expanding their potential applications in areas like drug delivery, tissue engineering, and advanced materials. | es in various fields. However, it lacks some details mentioned in the ground truth answer, such as specific techniques like UV-initiated polymerization and solvent evaporation, and the variety of particle types that can be created. SCORE: 76 | | |
| How can droplet microfluidics improve the reproducibility of chemical synthesis? | Droplet microfluidics improves the reproducibility of chemical synthesis by offering several advantages over conventional batch reactors. Firstly, droplet reactors allow for accurate manipulation and better reproducibility due to their precise control over solution | Droplet microfluidics enhances the reproducibility of chemical synthesis by providing precise control over reaction conditions and minimizing variability. Droplets act as isolated microreactors, ensuring accurate manipulation of parameters such as temperature and concentration, | The response is clear, concise, and directly addresses the question by highlighting the specific advantages of droplet microfluidics in improving chemical synthesis reproducibility. It | 78 | 78.0 |

| | volume, mixing intensity, and reaction conditions. This precision ensures consistent results, reducing variability between batches. | critical for consistent results. This system enables optimal mixing intensity and parallel processing in separate droplet channels, reducing batch-to-batch variability and ensuring uniform product quality. Automation further minimizes human error, making the process highly reliable. These advantages are particularly beneficial for synthesizing complex nanomaterials and exploring new reaction parameter spaces, leading to the production of high-quality, reproducible materials. | provides a detailed explanation of how precision, encapsulation, accelerated mixing, and local control contribute to consistent results. The answer aligns well with the ground truth, although it could be slightly more concise and focused on the core aspects. SCORE: 78 | | |
|---|---|---|---|---|---|
| | Secondly, the encapsulation of reactions within individual droplets prevents contamination and blocking of the channel walls, which can be a significant issue in batch reactors. This feature also allows for the use of toxic or volatile chemicals without fear of contaminating the reactor. | | | | |
| | Thirdly, the advection flow field within the droplets accelerates mixing, providing a well-defined starting point and even residence time. This contributes to a narrower size distribution of the final nanoparticles, further improving | | | | |
| | reproducibility.<br><br>    Lastly, local control over the synthetic environment can be exerted on separate droplet reactors, enabling homogeneous synthesis and quantity production. The reaction parameters scale up linearly, ensuring consistent results across different batches.<br><br>    In summary, the precision, encapsulation, accelerated mixing, and local control offered by droplet microfluidics significantly improve the reproducibility of chemical synthesis compared to conventional batch reactors. | | | | |
| What are the applications of droplet microfluidics in catalysis studies? | Droplet microfluidics has several applications in catalysis studies. Some of these include catalyzed reactions, photochemical | Droplet microfluidics offers transformative applications in catalysis studies by enabling precise, efficient, and high- | The response provides some relevant information about the applications of droplet | 46 | 46.0 |

| | reactions, crystal advancement, and gaseous synthesis. Additionally, droplet-based systems are becoming a prime choice for measuring kinetics of various enzymes (proteins that assist in speeding up chemical reactions) and immunoassay (a procedure that detects or extracts molecules, proteins of interest). The latest advancements focus on creating biochemical networks which can perform numerous chemical or biological enzymatic functions/reactions with nonequilibrium conditions. These reaction networks find applications in protein crystallization, performing self-regulation and amplification, reaction interaction control, and | throughput investigations at the nanoscale. These systems facilitate enantioselective biocatalysis using nanoliter-scale devices that integrate microflow reactors and electrophoresis for enantiomeric differentiation. Droplets serve as isolated reaction containers, allowing for controlled catalytic transformations at the single-particle or cell level, with reaction volumes as small as a few picoliters. This confinement minimizes diffusion effects and enhances reaction efficiency. Applications span organic synthesis, biochemical analysis, and biotechnology, showcasing droplet microfluidics as a powerful tool for studying rapid reaction dynamics, optimizing catalysts, and | microfluidics in catalysis studies. However, it lacks precision and clarity compared to the ground truth answer. The provided answer includes some correct concepts such as catalyzed reactions, enzyme kinetics, and biochemical networks, but they are not accurately described or connected to the broader context of droplet microfluidics in catalysis. SCORE: 46 | | |
|---|---|---|---|---|

| | autocatalytic reaction detection. | exploring reaction mechanisms in real time. | | | |
|---|---|---|---|---|---|
| How do droplets enhance the study of reaction kinetics? | Droplet-based microfluidic systems significantly enhance the study of reaction kinetics due to several reasons. First, they offer high time resolution, allowing for the investigation of rapid reactions on a timescale of milliseconds or even microseconds. This is achieved through techniques such as flow focusing, which can reduce diffusive mixing times to less than 10 microseconds, and chaotic flow induction within droplets, as demonstrated by Song et al.

Second, these systems enable the probing of a large number of discrete reactions while consuming minimal sample volumes. This is particularly advantageous in bioanalysis where | Droplet-based microfluidics revolutionizes reaction kinetics studies by offering rapid, precise, and high-throughput analysis with minimal sample consumption. The chaotic mixing within picoliter droplets ensures millisecond-level time resolution for fast reactions. These systems minimize reagent use, making them ideal for bioanalysis where resources are costly or scarce. Precise control over droplet composition facilitates the creation of concentration gradients, enabling the measurement of kinetic parameters. High-throughput capabilities allow simultaneous monitoring of numerous reactions, significantly expanding experimental | The response is clear and concise, accurately addressing the question with relevant details from a scientific perspective. It provides a comprehensive explanation of how droplet-based microfluidic systems enhance the study of reaction kinetics, including high time resolution, low sample consumption, precise control over droplet payload, and creation of concentration gradients. However, it could benefit | 85 | 85.0 |

| | reagents are often expensive or in limited supply.

Third, the compartmentalization of reactions within nanoliter to picoliter droplets provides a means for precisely controlling the droplet payload and creating concentration gradients. This expansion in experimental diversity facilitates the study of enzyme kinetics by allowing for the continuous addition of an inhibitor to control the rate of the specific enzyme-catalyzed process.

Lastly, when coupled with fluorescence microscopy, droplet-based microfluidic systems provide a powerful platform for enzyme kinetics studies due to their rapid and chaotic mixing of reagents within | diversity. Additionally, extended tracking of individual droplets supports long-term monitoring of enzyme-inhibitor kinetics, providing deeper insights into reaction mechanisms. | from minor adjustments to align more closely with the ground truth answer in terms of phrasing and emphasis on key points. SCORE: 85 | | |
|---|---|---|---|---|---|

| | picoliter droplets. This feature enables the facile creation of concentration gradients and thus a massive expansion in experimental diversity.

In summary, droplet-based microfluidic systems offer several unique advantages for the study of reaction kinetics, including high time resolution, low sample consumption, precise control over the droplet payload, and the ability to create concentration gradients for a wide range of experiments. | | | | |
|---|---|---|---|---|---|
| What are the benefits of using droplets for emulsion polymerization? | The benefits of using droplets for emulsion polymerization include:

1. Small sample volume: Droplet-based microfluidics allows for the use of small sample volumes on the order of picoliters and nanoliters, | Droplet-based systems offer significant advantages for emulsion polymerization, enhancing efficiency, precision, and scalability. They enable high production rates and precise control over polymer particle | The response is mostly relevant to the question but lacks focus on the specific benefits of using droplets for emulsion polymerization. It includes | 36 | 36.0 |

| | | |
|---|---|---|
| | which can be advantageous in various applications such as chemical analysis and protein crystalization.

2. Formation of emulsions: Dispersed droplets of one liquid in a second liquid can form an emulsion, which has numerous applications in industries like food and cosmetics. Emulsions are important for packaging small amounts of fluid and other active ingredients.

3. Unique properties: Emulsion polymerization allows for the production of materials with unique properties due to the formation of polymer particles in a continuous aqueous phase. This method is particularly useful because it offers tolerance | size and morphology, facilitating the synthesis of high-molar-mass polymers at faster reaction rates. Droplet parallelization allows the screening of diverse chemicals or concentrations and supports the encapsulation of millions of gene variants for quantitative emulsion PCR. These systems also empower directed enzyme or microorganism evolution through successive mutagenesis and selection rounds. Furthermore, droplet isolation prevents cross-contamination, adsorption, and solvent evaporation, making them ideal for applications like in vitro compartmentalization (IVC). This unique platform provides a robust, scalable approach for both analytical and biological studies. | some accurate information about emulsion polymerization in general, but does not directly address the advantages of using droplet-based systems specifically. SCORE: 36 | | |

| | to a wide range of monomers, the ability to control submicron particle size and morphology, and the possibility to produce polymers that are easily processable.

4. High molar mass at relatively high reaction rates: Emulsion polymerization enables the production of polymer chains with high molar mass at relatively high reaction rates in comparison to other methods of free-radical polymerization such as bulk, solution or suspension.

5. Stability and optically clear solutions: Micro-emulsions, which are systems with monomer droplets in water, are thermodynamically stable and optically a one-phase solution. This stability makes emulsion polymerization | | | | |
|---|---|---|---|---|---|
| | more practical and efficient for large-scale applications. | | | | |
| How can microfluidics be used to generate smart materials? | Microfluidics can be utilized to generate smart materials by leveraging the unique advantages it offers in material synthesis. The primary benefits include precision, control, and speed. In the microscale, phenomena such as droplet templating allow for the production of materials with properties that could not be achieved otherwise.

For instance, single emulsion templating can combine and compartmentalize distinct reagents within monodisperse droplets of a desired size. This method is particularly useful in creating hydrogel particles, which are valuable in single-cell analysis but | Microfluidics enables the creation of smart materials through precise control of reaction conditions, rapid synthesis, and innovative templating methods. Droplet templating allows monodisperse droplets to act as structural templates, facilitating the synthesis of uniform particles, such as hydrogels for single-cell analysis. By manipulating solute and temperature gradients on short timescales, microfluidics supports the production of nanoscale materials, including semiconducting polymers, catalysts, and nanoparticles with unique optical and electronic properties. Its ability to rapidly explore complex | The response demonstrates a good understanding of microfluidics' role in generating smart materials, covering key aspects such as precision control, rapid synthesis, templating methods, and integration of analytics and machine learning. However, it lacks some depth and specificity compared to the ground truth answer, particularly regarding the types of materials that can be synthesized and their | 76 | 76.0 |

| | | | | |
|---|---|---|---|---|
| | require a controllable chemical composition and size. By using microfluidic networks like co-flow, picoinjection, and drop-let fusion, reagents can be combined in defined sequences, enabling the synthesis of complex materials not possible with bulk mixing.

Furthermore, microfluidics allows for the exploration of complex reaction parameter spaces on timescales many orders of magnitude shorter than conventional methods. This rapid experimentation is crucial for generating smart materials, as it enables researchers to quickly optimize and refine their synthesis processes.

Lastly, the | reaction parameter spaces, coupled with advanced analytics and machine learning, accelerates material discovery and optimization. This approach offers transformative potential in biotechnology, energy, and electronics by enabling the generation of materials with tailored properties and enhanced functionality. | applications. SCORE: 76 | | |

| | | | | | |
|---|---|---|---|---|---|
| | ability to integrate sensitive analytics, reaction control architectures, and efficient machine learning algorithms in microfluidic systems further enhances their potential for creating smart materials. These advanced tools can help analyze and optimize the properties of the materials being synthesized, ultimately leading to the development of more sophisticated and intelligent materials. | | | | |
| What are the applications of droplet microfluidics in environmental sensing? | The applications of droplet microfluidics in environmental sensing include, but are not limited to, chemical detection systems for bioengineering and environmental monitoring. Researchers have combined quantum sensing with droplet | Droplet microfluidics offers transformative applications in environmental sensing by enabling precise, miniaturized, and efficient monitoring systems. It supports on-site environmental surveillance through portable devices for rapid toxicity | The response is mostly relevant to the question and provides some accurate information about the applications of droplet microfluidics in environmental sensing, but it | 52 | 52.0 |

| | | | | |
|---|---|---|---|---|
| | microfluidics to develop a highly sensitive chemical detection system. This platform uses quantum sensors made of nanodiamonds to detect chemicals with exceptional sensitivity.

In addition, the use of microfluidic devices in environmental science provides unique platforms to access microscopic, in situ visualized, dynamic outcomes by building reaction micromodels on-chip and integrating complementary analytical techniques like absorption spectroscopy, electron microscopy, etc. This technology is particularly beneficial for miniaturized samplers or analytical devices for on-site environmental surveillance or rapid toxicity | assessments and detection of contaminants, such as mercury ions. As controllable microreactors, droplet microfluidics facilitates the study of complex interfacial processes, aiding in chemical synthesis and degradation studies. Furthermore, it advances environmental sensing by enabling in situ visualization and precise flow control, enhancing our understanding of environmental processes. These capabilities position droplet microfluidics as a valuable tool for environmental science and engineering. | contains minor inaccuracies and lacks focus on certain aspects. It also includes some additional details that are not mentioned in the ground truth answer. SCORE: 52 | | |

| | | | | | |
|---|---|---|---|---|---|
| | assessment of particular chemicals.

Moreover, controllable microreactors for exploring complex environmental interfacial processes are one of the most advanced yet underexplored frontiers of microfluidic technology for environmental science. This area is expected to benefit significantly from microfluidic technology both currently and in the near future. | | | | |
| How can microfluidic droplets improve combinatorial chemistry? | Microfluidic droplets offer several advantages over conventional microtiter plate techniques for combinatorial chemistry. First, the use of picolitre droplets as microreactors allows for individual experiments using six to eight orders of magnitude less starting material | Microfluidic droplets revolutionize combinatorial chemistry by drastically reducing reagent consumption and enhancing efficiency. Each droplet, acting as a picoliter-scale microreactor, requires six to eight orders of magnitude less material than traditional methods, | The response is clear, concise, and accurately addresses the question about how microfluidic droplets improve combinatorial chemistry. It provides specific examples of advantages | 78 | 78.0 |

| | than microtiter plate approaches, significantly reducing reagent consumption and improving efficiency through automation and online analysis.

Second, the miniaturization afforded by droplet-based microfluidics increases throughput, enabling the screening of large numbers of compounds at high speed. This is particularly beneficial for early stage drug discovery where success rates are inherently low.

Third, a large repertoire of downstream microfluidic droplet manipulation and detection techniques allows for complex multi-stage experiments with multi-dimensional analysis. This means that the platform could be | minimizing waste and costs. This miniaturization enables high-throughput synthesis of large compound libraries, supporting applications like drug discovery and material synthesis. Additionally, droplet manipulation techniques allow for precise reaction control, combinatorial fusion of reagents, and multi-stage experiments with online analysis. These advantages make microfluidic droplets a powerful tool for automating complex chemical workflows and optimizing reaction conditions in a highly efficient and scalable manner. | such as reduced reagent consumption, increased throughput, and complex multi-stage experiments. However, it could be slightly improved by adding more depth or details to some points, but overall, it is a solid response that aligns with the ground truth answer. SCORE: 78 | | |

| | | | | | |
|---|---|---|---|---|---|
| | used for applications outside of medicinal chemistry, such as combinatorial synthesis aimed at producing novel materials.<br><br>Lastly, this approach could provide an efficient way to screen different reaction conditions, allowing for many parameters to be manipulated simultaneously to determine the ideal combination of reagents. In summary, microfluidic droplets allow large areas of chemical parameter space to be sampled with minimal reagent consumption. | | | | |
| What are the challenges of scaling up droplet-based chemical synthesis? | The challenges of scaling up droplet-based chemical synthesis can be summarized as follows:<br><br>1. Device fabrication and large-scale production: The | Scaling up droplet-based chemical synthesis faces several challenges. Achieving uniform fluid distribution in microfluidic devices remains difficult despite progress in | The response provides a detailed and structured answer that addresses the question about the challenges of scaling up droplet- | 83 | 83.0 |

| | | | | |
|---|---|---|---|---|
| | microfluidic reactor has achieved parallelization of droplet generation, but the issue of fluid distribution remains a challenge. A 3D-printing technique can be developed to fabricate small and complex channel structures for better control over fluid distribution. Additionally, the 3D-printing technique can be further developed to integrate industrial-scale interfaces and built-in control systems.<br><br>2. Multistep reactions: Simplification of fundamental functional units, including droplet generation, droplet merging, and phase separation, is necessary. An integrated process combining synthesis and post-treatments in | parallel droplet generation, with 3D printing offering potential solutions for complex channel fabrication. Simplifying functional units for multistep reactions, such as droplet generation, merging, and phase separation, is vital for seamless continuous flow processes. Developing low-cost materials and standardizing chip designs are necessary for commercial viability, especially for high-throughput production in diverse fluids. Furthermore, creating complex droplet emulsions and expanding nanomaterial diversity are critical to meeting industrial demands. Addressing these challenges will enable safe, sustainable, and scalable microfluidic applications. | based chemical synthesis. It covers various aspects such as device fabrication, multistep reactions, tailored microfluidic chip designs, trade-off between throughput and microchannels, translation to industrial applications, and environmental concerns. The answer is clear, concise, and relevant, with some minor inaccuracies in the specific details provided. SCORE: 83 | | |

| | a continuous flow process deserves particular attention. In-situ purification can be further developed for practical applications.

3. Tailored microfluidic chip designs and materials: There is a need for tailored microfluidic chip designs and materials for different target applications, which currently lack well-established standardization design rules, particularly for high-throughput generation in non-Newtonian fluids with high viscosity.

4. Trade-off between throughput and the number of microchannels: There is a limitation for industrialization and commercialization of micro-reaction systems due to the trade- | | | | |
|---|---|---|---|---|---|

| | | | | |
|---|---|---|---|---|
| | off between throughput and the number of microchannels and its stable operation condition.<br><br>5. Translating laboratory results into industrial applications: It is imperative to identify and address the existing challenges when translating laboratory results into industrial applications and devise solutions for these key issues to facilitate the successful scaling-up of microfluidics systems.<br><br>6. Environmental concerns: Chemical manufacturing must tend to be safe, green, and sustainable under the circumstances of carbon peaking and carbon neutrality. Safety and environmental problems during the chemical production process will | | | |

| | | | | | |
|---|---|---|---|---|---|
| | always be a concern of society. | | | | |
| How can artificial intelligence enhance droplet microfluidic research? | Artificial Intelligence (AI) can significantly enhance droplet microfluidic research in several ways. First, AI algorithms can analyze the vast amounts of data generated by these systems more efficiently than traditional analytical methods. This is particularly useful because the advanced capabilities of AI are well-suited to the rapid data generation capabilities of droplet-based technologies.

Second, deep learning and reinforcement learning algorithms have already been successfully used in conjunction with droplet-based microfluidic systems. These AI approaches can analyze characterization | Machine learning can be used to optimize the design of microfluidic devices. Deep Learning will be integrated to sort, monitor, and process the images. Furthermore, the multimodal agents can be used to study and design the research of droplet microfluidics, a possible excuse robot to perform fabrication and experiments automatically. | The response provides a clear and concise overview of how artificial intelligence can enhance droplet microfluidic research. It highlights several key points, including efficient data analysis, the use of deep learning and reinforcement learning algorithms, handling large-scale data volumes, maximizing information extraction from each droplet, and the importance of additive manufacturing techniques | 85 | 85.0 |

| | data quickly and provide instructions back into the microfluidic system to control the nature or composition of new droplets. This feedback mechanism can help achieve high degrees of accuracy and sensitivity in the analysis of pathogens, as mentioned in the context provided.<br><br>    Third, as large-scale droplet-based platforms develop, data volume and generation rates will exceed the interpretive capabilities of conventional analytics. Integrating deep learning algorithms that can rapidly analyze characterization data will be crucial to handle this data overload.<br><br>    Lastly, maximizing the amount of information that | | in creating advanced microfluidic components. However, it does not directly address the specific point about machine learning being used to optimize device design or the integration of multimodal agents for research purposes mentioned in the Ground Truth Answer. SCORE: 85 | | |
|---|---|---|---|---|---|

| | | | | |
|---|---|---|---|---|
| | can be extracted from each droplet remains a challenge. While various novel detection methods are being developed, the routine application of high-content methods (such as those based on vibrational spectroscopies and electrochemical techniques) will likely be critical in engendering important new applications in the chemical and biological sciences.<br><br>    In addition to these benefits, the use of additive manufacturing techniques will become increasingly important in creating advanced microfluidic components for 'out-of-lab' settings. This is because these techniques can help create pumpless fluid manipulation schemes, which | | | |
| | are essential for the development of more sophisticated droplet microfluidic systems. | | | | |
| How can large language models enhance droplet microfluidic research? | Large language models can significantly enhance droplet microfluidic research in several ways. First, they can help process and analyze large datasets generated by droplet microfluidics, as demonstrated in the study "MicroFluidic-LLMs." This framework effectively captures contextual information from tabular data formats and leverages pre-trained large language models for analysis, reducing errors and enhancing accuracy.

Second, large language models can be used to develop more sophisticated droplet | LLMs can be used to build the Agents and Multi-Agents to synthesize knowledge of droplet microfluidics from literature such as research articles and books. Then, the idea hypotheses will be proposed by the agents, and the novelty and feasibility will be assessed. The agents will provide the autonomous platform to design and fabricate droplet microfluidic devices. Furthermore, LLMs have the researchers accelerate knowledge of droplet microfluidics. | The response provides a clear and concise overview of how large language models can enhance droplet microfluidic research, covering data analysis, droplet characterization, scale-up strategies, and detection techniques. However, it does not fully align with the ground truth answer, which suggests using LLMs to build agents that synthesize knowledge | 46 | 46.0 |

| | | | | |
|---|---|---|---|---|
| | characterization techniques. For instance, they can recognize complex flow patterns, which is technically challenging for current detection assays. This would enable more accurate identification of pathogens and facilitate the application of droplet microfluidics in a broader range of medical tests.<br><br>   Third, large language models can aid in the development of advanced scale-up strategies to simultaneously screen multiple droplets. Currently, droplets are normally detected and sorted one by one, limiting throughput. By using machine learning algorithms, it may be possible to develop strategies that can screen and sort multiple droplets at once, thereby increasing | | from literature and design devices autonomously. The response also lacks specific examples or references to support its claims. SCORE: 46 | |

| | efficiency and throughput.<br><br>    Lastly, large language models can contribute to the integration of more developed detection techniques for characterizing and probing chemical components within droplets. This would allow for the screening of large libraries of drug compounds, which is currently prevented due to technical challenges in current detection assays.<br><br>In summary, large language models can enhance droplet microfluidic research by improving data analysis, developing sophisticated droplet characterization techniques, aiding in the development of advanced scale-up strategies, and integrating more developed | | | | |
| --- | --- | --- | --- | --- | --- |

| | | | | |
|---|---|---|---|---|
| | detection techniques for chemical component screening. | | | | |
| How can autonomous droplet microfluidic research? | Autonomous droplet microfluidic research aims to develop intelligent, automated, and integrated systems that can perform scientific discovery tasks autonomously. This is particularly relevant for high-throughput, single-cell, and single-molecule analyses.<br><br>To achieve this goal, several challenges need to be addressed:<br><br>1. Fast detection methods are required to probe small volume droplets dynamically. Advances in single-molecule detection, fluorescence lifetime imaging, spectroscopy, and other related technologies could be utilized for this purpose. | Autonomous droplet microfluidics is an emerging field with significant potential for scientific discovery, aiming to create highly integrated, automated systems for bio/chemical functions like PCR, sample preparation, and detection. Key challenges include the need for fast detection methods to probe small volume droplets, precise control over droplet dynamics in complex microchannel networks, and the development of virtual design platforms for optimizing system designs. Solutions under exploration include using advanced detection techniques such as single-molecule detection and fluorescence | The response is mostly relevant and clear, but lacks focus or includes unnecessary details; broadly acceptable. SCORE: 58 | 58 | 57.9999999 9999999 |

| | 2. Control over droplets' dynamic behavior is essential for system automation. Emerging droplet intelligence/computing functions may provide a way to achieve on-chip control of droplet behavior.

3. Designing such intelligent automated and integrated systems requires a powerful virtual design and experiment platform that can replace the present empirical design methodologies.

To illustrate the potential of microfluidic microdroplet technology for future autonomous scientific discovery, an example was given for automated drug discovery, where the system would use integrated microelectronic components as | lifetime imaging, as well as integrating microelectronic actuators for droplet manipulation. Additionally, the development of virtual design and experiment platforms could streamline the creation of optimized microfluidic systems. The potential applications are vast, including directed evolution, closed-loop synthesis, and automated drug discovery, where systems could autonomously optimize compound combinations. Collaboration across disciplines like fluid dynamics, chemistry, and software engineering will be key to overcoming challenges and advancing this technology, paving the way for groundbreaking scientific advances. | | | |
|---|---|---|---|---|

| | actuators to provide motive power and perform selection and reproduction strategies based on fitness calculations.<br><br>In conclusion, continuous-flow droplet-based microfluidic PCR has made significant progress at individual component levels but highly integrated systems are still rare. The development of smart microfluidic platforms for autonomous scientific discovery will have a major impact on artificial intelligence and biological computations, particularly in applications such as directed evolution and closed loop synthesis that require feedback-based redirection of experiments. | | | | |
| What is the potential for fully | Fully integrated droplet | Droplet microfluidics | The response | 56 | 56.0000000 0000001 |

| integrated droplet microfluidic platforms? | microfluidic platforms have significant potential in various fields, particularly in high-throughput small-molecule screening, directed evolution of proteins, PCR, and the formation of artificial cells. These platforms offer unique advantages such as decreased diffusion distance, faster mixing, laminar flow, and the ability to produce large numbers of micro-reactors for parallel processing while maintaining each reactor independent and isolated.

The integration of optical methods into these platforms allows for a wide array of biochemical diagnostic assays. With appropriate micro-electronic/optical sensors and actuators, these smart, self- | offers transformative solutions to global health challenges by enabling rapid, cost-effective, and high-throughput analysis of biological samples. By generating and manipulating tiny droplets as individual reaction chambers, this technology can revolutionize diagnostics, particularly in resource-limited settings. For example, it facilitates point-of-care testing for infectious diseases like tuberculosis, malaria, and COVID-19, providing results quickly and accurately. Droplet microfluidics is also crucial for multiplexed screening, enabling simultaneous detection of multiple biomarkers, which enhances early disease detection and monitoring. Additionally, it supports the | provides a clear and concise overview of the potential of fully integrated droplet microfluidic platforms, highlighting their advantages in various fields such as high-throughput small-molecule screening, directed evolution of proteins, and biochemical diagnostic assays. However, it lacks the transformative impact and global health implications emphasized in the Ground Truth Answer, focusing more on technical aspects rather than | | |
|---|---|---|---|---|---|

| | regulating devices can make logical decisions based on experimental results to generate subsequent experiments. They can be best used for optimizing and creating enzymes, drug screening, complex protocol sequences, and optimization of PCR conditions.

However, challenges remain in monitoring and controlling biological and chemical processes within droplets to rapidly acquire information for decision-making in follow-on experiments. To address this, the development of virtual design and experiment platforms is crucial for creating highly automated and integrated systems. | development of low-cost diagnostic devices, reducing dependence on expensive laboratory infrastructure. In drug discovery, it accelerates the screening of therapeutic compounds, improving the treatment of diseases. The technology also advances personalized medicine by analysing single cells and their responses to treatments. Overall, droplet microfluidics addresses global health disparities by delivering accessible, scalable, and efficient solutions for disease diagnosis, monitoring, and treatment. | real-world applications and societal benefits. The response also acknowledges challenges but does not fully address them or provide a comprehensive solution like the Ground Truth Answer.

SCORE: 56 | | |
|---|---|---|---|---|---|
| How can droplet microfluidics contribute to | Droplet microfluidics can contribute to | Droplet microfluidics has significant | The response provides a | 76 | 76.0 |

| global health challenges? | global health challenges in several ways. One significant application is the detection of pathogenic bacteria, such as Salmonella typhimurium, which causes a high number of human infections and deaths annually worldwide. The LAMP-assisted droplet-based microfluidic technique developed for this purpose is simple in operation, sensitive, specific, rapid, and high-throughput, making it suitable for many applications in biological assays. This technique could potentially address the challenges posed by current gold standard methods like PCR, which require bulky and expensive instruments for temperature cycling, particularly in resource-limited countries. | potential to address global health challenges by enhancing diagnostic and therapeutic processes. It enables rapid and sensitive pathogen detection, which is critical for early disease diagnosis and containment. The technology's high-throughput capabilities allow simultaneous analysis of multiple samples, streamlining disease diagnosis and treatment. Additionally, droplet microfluidics can be integrated into point-of-care platforms, making it feasible for use in resource-limited settings. This approach can improve healthcare accessibility and efficiency. Furthermore, droplet microfluidics facilitates precise single-cell analysis, enabling advances in personalized medicine and more effective | clear and relevant discussion on the potential applications of droplet microfluidics in addressing global health challenges. It highlights specific examples such as pathogen detection, high-throughput analysis, and single-cell analysis, which align with the ground truth answer. However, it could be improved by providing more comprehensive information on the broader impact of droplet microfluidics on healthcare | | |

| | | treatment strategies. While challenges such as cost, standardization, and integration with existing systems remain, continued development of this technology could significantly impact global health by improving detection, diagnostics, and treatment capabilities. | accessibility and efficiency, as well as acknowledging potential challenges and limitations. SCORE: 76 | | |
|---|---|---|---|---|---|
| | Moreover, droplet microfluidics allows the introduction of single particles in droplets, yielding important results that could not be reached without microfluidic contributions. This includes a very low limit of detection in immunoassay applications or the efficient barcoding of single cells for subsequent gene studies. The high-throughput analysis of individual cells using solid and hydrogel particles in droplet microfluidics is expected to forge new paths in the fields of biology and medical sciences, potentially contributing to future developments in global health. | | | | |
| How can open-source designs impact the future of microfluidic devices? | Open-source designs for microfluidic devices can significantly impact their future | Open-source designs for microfluidic devices hold the potential to revolutionize the field by increasing | The response provides a clear and concise overview of the | 76 | 76.0 |

| | | | | |
|---|---|---|---|---|
| | development in several ways. Firstly, the open-source nature of these designs allows for easy integration and customization by a diverse community of users, leading to more comprehensive design automation tools that cater to various applications. This collaborative approach fosters innovation and rapid advancement in the field.<br><br>Secondly, open-source designs can facilitate the development of machine learning algorithms, which can predict the performance of these devices and enable non-expert users or automated systems to design and fabricate devices for specific applications. This democratization of microfluidic device design could potentially | accessibility, reducing costs, and fostering collaboration. These designs allow for the creation of custom predictive models without requiring extensive expertise, enabling broader participation in device development. The use of open-source tools, such as Design Automation of Fluid Dynamics (DAFD), accelerates design and prototyping, enabling rapid iteration and optimization. Additionally, the integration of machine learning algorithms can drive the development of new microfluidic tools and optimize existing ones. This democratization of design expertise and the collaborative, community-driven approach will facilitate the creation of affordable, specialized devices, ultimately | potential impact of open-source designs on microfluidic devices, covering key points such as increased collaboration, democratization of access, and innovation. However, it slightly deviates from the ground truth by not explicitly mentioning specific tools like Design Automation of Fluid Dynamics (DAFD) or providing detailed examples of breakthroughs in applications like single-cell analysis and global health diagnostics. The tone is also somewhat | | |

| | reduce costs and increase accessibility, making these tools more widely available for research and commercial purposes.

Lastly, open-source designs can provide a platform for data collection and assimilation, which can be used to train machine learning algorithms and create entirely new microfluidic tools. This approach could lead to the development of innovative devices that address specific needs in various fields, such as single-cell analysis or chemical and biological research.

In summary, open-source designs for microfluidic devices have the potential to transform the design process, increase | leading to breakthroughs in applications like single-cell analysis and global health diagnostics. The open-source model promises to transform microfluidics into a more inclusive and innovative field. | more optimistic than the ground truth answer. SCORE: 76 | | |
|---|---|---|---|---|

| | | | | | |
|---|---|---|---|---|---|
| | collaboration, democratize access, and foster innovation in the field. The future of droplet-based microfluidics looks exciting indeed. | | | | |
| What are the challenges of scaling up microfluidic devices for industrial use? | The challenges of scaling up microfluidic devices for industrial use include:<br><br>1. Lack of standardized design principles and strategies for integrated microfluidic devices catering to both Newtonian and non-Newtonian droplet mass production, particularly for high-throughput generation in non-Newtonian fluids with high viscosity.<br><br>2. Achieving patterned wettability control of complex channels with high feasibility and accuracy as the scale expands.<br><br>3. The cost | Scaling up microfluidic devices for industrial use presents several challenges, including throughput limitations, lack of standardization, and integration complexities. The small micrometric channel sizes restrict production efficiency, and increasing parallelization or channel size may impact fabrication costs and reaction kinetics. There is also a lack of standardized design methods, making it difficult to compare results and adopt microfluidics in industrial settings. Additionally, scaling from laboratory to industrial production remains a hurdle, with issues in | The answer provides a detailed list of specific challenges related to scaling up microfluidic devices for industrial use, including issues with design principles, cost considerations, droplet generation, and integration complexities. However, the response could be improved by providing more concise language and directly addressing the question's | 52 | 52.0 |

| | | | | |
|---|---|---|---|---|
| | consideration when pushing a technology to the commercial market, which requires the development of low-cost microfabrication methods and materials.<br><br>4. Current scaling up of droplet generation is restricted to single emulsions; however, for practical applications, multiple or complex droplet emulsions need to be generated, and progressively more novel interfacial reaction systems are expected to be incorporated.<br><br>5. Compared with the microscale particles, the diversity and functionalities of the droplet microfluidics-derived nano-scale materials are still lacking.<br><br>6. The current scale-up is still limited to laboratory | cost-effective microfabrication, generating complex droplet emulsions, and ensuring reliable fluid control. The integration of microfluidic devices with industrial interfaces, testing methods, and continuous processes is essential for successful commercialization. Overcoming these challenges requires innovative solutions, such as simplifying multistep reactions and developing standardized, low-cost fabrication methods. | core points, such as throughput limitations, standardizat ion, and integration complexitie s. The answer also includes some minor inaccuracie s and lacks a clear summary or conclusion to tie together the various challenges discussed. SCORE: 52 | | |

| | investigation which is a long way to the commercial-scale application.

7. The simplification of multistep reactions, such as synthesis, functionalization, and purification, in a consecutive flow process. Efforts should be devoted to the simplification of fundamental functional units, including droplet generation, droplet merging, and phase separation. An integrated process combining synthesis and post-treatments in a continuous flow process deserves particular attention. Notably, in-situ purification can be further developed. | | | | |
|---|---|---|---|---|---|
| How might droplet microfluidics impact agriculture? | Droplet microfluidics could significantly benefit agriculture by enabling the controlled release | Droplet microfluidics has the potential to transform agriculture by enhancing crop management and food security. It | The response demonstrates a good understanding of the potential applications | 73 | 73.0 |

| | | | | |
|---|---|---|---|---|
| | of fertilizers and pesticides, thereby enhancing crop growth. This is achieved through droplet-templated capsules that regulate the release of these substances. Additionally, it offers a compact and convenient approach for detecting plant diseases, which are major factors impacting crop production and food security. The technology allows for high-sensitivity detection by dividing highly diluted samples into large numbers of micron-scaled droplets, although the detection is currently limited to fluorescence. However, as research in this field continues to advance, it promises to provide more versatile methods for uniquely coding each droplet and decoding the codes, enabling a | enables the precise release of fertilizers and pesticides, reducing waste and optimizing plant growth. The technology also supports improved crop monitoring through long-term studies, offering insights into plant development. By analyzing individual plant cells, droplet microfluidics can help uncover genetic and phenotypic traits, enhancing disease resistance and drought tolerance in crops. Additionally, it aids in disease detection, allowing for early intervention to prevent crop loss. Despite challenges such as scalability, cost, and integration with existing agricultural practices, ongoing research aims to overcome these barriers, making droplet microfluidics a promising tool for precision agriculture and | of droplet microfluidics in agriculture. It highlights the benefits of controlled fertilizer and pesticide release, disease detection, and analysis of individual plant cells to uncover genetic traits. However, it lacks some details and specific examples compared to the ground truth answer. The tone is clear and concise, but there are minor inaccuracies regarding the current limitations of droplet microfluidics in detecting diseases (limited to | | |

| | precise linking of genotypic and phenotypic characteristics of single cells. This could reveal mechanisms of cell heterogeneity and dysfunction, ultimately accelerating the development of innovative agricultural solutions. | sustainable food production. | fluorescenc e) and the potential for versatile methods of coding and decoding droplets.

SCORE: 73 | | |
|---|---|---|---|---|---|
| What are the current limitations of droplet microfluidics, and how can they be overcome? | The current limitations of droplet microfluidics include:

1. Efficiency for low sample numbers: Droplet microfluidics is more efficient for assays involving at least 10^5 samples, while tools such as 1536-well microtiter plates are generally sufficient for fewer samples.

2. Stable and inert oil–water interface: The oil–water interface that forms each droplet must be both stable and inert. Finding the right surfactant | Droplet microfluidics has immense potential for high-throughput biological assays but faces several limitations that hinder its broader adoption. One key challenge is its efficiency for assays involving small sample sizes; droplet microfluidics excels with large volumes (at least 10^5 samples), making traditional methods like microtiter plates more suitable for lower sample numbers. Another critical issue is maintaining a stable and inert oil-water interface, which is essential for | The response provides a clear and detailed list of the current limitations of droplet microfluidi cs, including efficiency for low sample numbers, stable oil-water interface, lack of spatial barcoding, automated control of droplets, and long-term droplet stability. It also offers suggestions | 76 | 76.0 |

| | that will maintain such an interface is often a challenging chemistry problem.<br><br>3. Lack of spatial barcoding: Droplets cannot take advantage of the spatial barcoding that is available in the two-dimensional arrays of microtiter plates. Thus, droplet libraries require a suitable barcoding system, which can be difficult to develop.<br><br>4. Automated control of droplets: The automated control of the movement of many droplets is challenging and current progress is being made in methods for the active control of droplets by means of electrowetting on dielectric (EWOD), dielectrophoretic, and electric forces. | droplet formation. Finding the right surfactant to keep this interface stable remains a complex chemistry challenge. Furthermore, droplet microfluidics lacks the spatial barcoding available in microtiter plates, making it difficult to organize and track multiple droplet libraries. As a result, developing alternative barcoding systems is necessary but challenging. The automation of droplet movement is another obstacle, as it requires precise control, often through techniques like electrowetting and dielectrophoresis. Lastly, droplet stability over time can be compromised, with issues such as shrinkage, coalescence, and cross-talk between droplets. To address these issues, research is | on how to overcome these limitations through research in new materials, surface patterning, analytical tools, integrated platforms, understanding droplet behavior, and new technologies based on droplets. The response is well-structured and provides relevant information. However, it could be improved by providing more depth and clarity in some areas, such as the specific challenges of developing a stable oil-water interface | | |

| | 5. Long term droplet stability: Imperfect partitioning between the continuous and discrete phases can lead to issues such as shrinkage, coalescence, and inter-droplet cross-talk. Studies on oil phase and surfactant formulations with improved isolation properties are being carried out to tackle this problem.<br><br>To overcome these limitations, research is being conducted in several areas:<br><br>1. Development of new materials for fabricating devices with desired properties.<br>2. Techniques for surface patterning.<br>3. Analytical tools with higher sensitivity and accuracy.<br>4. Integrated microfluidic platforms. | underway to develop improved surfactants, more reliable barcoding systems, and novel materials and techniques to enhance droplet control. These advancements, coupled with better analytical tools, are expected to overcome current limitations and expand droplet microfluidics' applications in chemical and biological research. | and the potential applications of droplet microfluidics beyond biological assays.<br><br>SCORE: 76 | | |
|---|---|---|---|---|---|
| | 5. Studies to understand the behavior of droplets better. 6. New technologies based on droplets will also benefit many areas of science, such as the synthesis of functional reaction networks and the investigation of the influence of mixing on nucleation of protein crystals. | | | | |
| How can interdisciplinary collaboration accelerate innovation in droplet microfluidics? | Interdisciplinary collaboration can significantly accelerate innovation in droplet microfluidics by combining knowledge from different disciplines. This is because the challenges and opportunities in this field span across various areas such as chemistry, biology, materials science, and engineering. For instance, chemists and biologists can collaborate to rethink and | Interdisciplinary collaboration is essential for advancing droplet microfluidics by combining expertise from various fields such as chemistry, biology, material science, and engineering. By uniting these diverse areas, researchers can address complex challenges such as cost-effective fabrication, scalable production, and novel interfacial reaction systems. For example, biologists and chemists can leverage droplet | The response demonstrates a clear understanding of the importance of interdisciplinary collaboration in accelerating innovation in droplet microfluidics. It provides specific examples and details from various fields such as chemistry, biology, | 83 | 83.0 |

| | reimagine experimental workflows, leveraging the unique capabilities of droplet microfluidics for high-throughput analysis and precise control over chemical and biological reactions. On the other hand, material scientists and engineers can contribute by developing low-cost microfabrication methods and materials, improving the scalability and robustness of droplet microfluidic devices.

Moreover, interdisciplinary collaboration can help in addressing specific challenges such as achieving high-feasibility and accuracy in patterned wettability control, reducing costs for commercial-scale manufacturing, | microfluidics for high-throughput analysis, while material scientists and engineers work on developing more robust and affordable microfabrication methods. Furthermore, integrating technologies like sensitive detectors and machine learning algorithms can enhance the analytical capabilities of droplet-based systems. Collaborative initiatives, such as the EMBL Microfluidics Conference, foster knowledge exchange and innovation by bringing together professionals from different disciplines. This cooperation is crucial for overcoming technical barriers, reducing manufacturing costs, and developing new applications in fields like drug discovery, single- | materials science, and engineering, showcasing how collaboration can address complex challenges and lead to exciting accomplishments. However, some minor inaccuracies and differences in wording compared to the ground truth answer are noted. SCORE: 83 | | |
| --- | --- | --- | --- | --- |

| | developing novel methods for generating multiple or complex droplet emulsions, and synthesizing nanomaterials with unprecedented features.

In conclusion, the EMBL Microfluidics Conference series and Single Cell Genomics (SCG) meetings are examples of initiatives that aim to foster interdisciplinary collaboration in droplet microfluidics. Such initiatives, along with continued efforts from academia and industry, are expected to drive exciting accomplishments in this field over the next few years. | cell analysis, and nanomaterial synthesis. Ultimately, interdisciplinary collaboration accelerates the evolution of droplet microfluidics, pushing the boundaries of what is achievable in this promising field. | | | |
|---|---|---|---|---|---|

| question | answer | ground_truth | feedback | score | percentage |
|----------|--------|--------------|----------|-------|------------|
| What is microfluidics? | Microfluidics is a technology that deals with the manipulation of fluids, typically on a scale of microliters ($10^{-6}$ liters) to picoliters ($10^{-12}$ liters), within a lab-on-a-chip or similar device. It's used in various fields such as biology, chemistry, and | Microfluidics is a multidisciplinary field that involves the study and manipulation of fluids at the microscale, typically within channels with dimensions ranging from tens to hundreds of micrometers. It is a multidisciplinary field that involves | The response is clear and concise, accurately describing the definition and scope of microfluidics, its applications, and relevance to various fields. It also provides specific examples of its | 86 | 86.0 |

| | engineering for applications like single cell analysis, drug discovery, and diagnostics. | molecular analysis, molecul ar biology, and microelectronics. Microfluidics has been increasingly used in the biological sciences because precise and controlled experiments can be conducted at a lower cost and faster pace. Lab on a Chip devices use microfluidics for applications such as Point of Care testing of diseases, or Organ on a Chip studies. | use in lab-on-a-chip devices. SCORE: 86 | | |
|---|---|---|---|---|---|
| What is droplet microfluidics? | Droplet microfluidics is a laboratory technique that allows for the generation, manipulation, and analysis of thousands to millions of tiny liquid droplets in a continuous flow. It's often used in biology and chemistry for applications such as single-cell analysis, drug screening, and DNA sequencing due to its high | Droplet-based microfluidics is a branch of microfluidics focusing on the creation and manipulation of discrete volumes of liquids through adoption of immiscible fluid flows. Microdroplets allow for handling miniature volumes ($\mu$L to fL) of fluids conveniently, provide better mixing, encapsulation, | The response is mostly relevant, clear, and concise, but lacks some depth and includes minor inaccuracies regarding the specific details of droplet microfluidics. It also doesn't fully capture the nuances of the technique as described in the ground truth answer. SCORE: 52 | 52 | 52.0 |

| | throughput, precision, and miniaturization capabilities. | sorting, and sensing, and suit high throughput experiments. Co mpared with continuous-flow systems, droplet microfluidic systems enable isolation and manipulation of samples in separate compartments. The droplets generated with uniform size and shape provide a stable microenvironmen t for biochemical reactions. Processing of a large number of droplets in parallel facilitates to achieve ultrahigh throughput. Confinement of individual targets within the droplets and independent processing of each of them allow high-precision analyses. | | | |
| How to generate microdroplets? | Microdroplets can be generated using various methods, but one common technique is the | Microdroplet generation in microfluidics is achieved through two main methods: active | The response attempts to address the question but lacks clarity and precision | 16 | 16.0 |

| | droplet generation by digital microfluidics (DMF). In this method, a hydrophobic substrate with electrodes is used. A voltage pulse applied to the electrodes causes a small area of the substrate to become hydrophilic, allowing a droplet to form at that location. The process can be automated and allows for precise control over the volume and positioning of each droplet. Other methods include acoustic droplet ejection devices (ADEPT) and microvalves. | and passive. Active methods utilize external forces like piezoelectric actuators to precisely control droplet size, formation rate, and spacing. These methods enable high uniformity in droplet generation, with deviations less than 0.3%. Passive methods rely on fluid properties and channel geometry, such as T-junctions or cross-flow configurations, to create droplets. These methods are influenced by factors like flow rate, viscosity, and interfacial tension, allowing for fine control over droplet size and uniformity. For example, PDMS-based microfluidic devices have been used for nanoparticle synthesis, where hydrodynamic coupling of nozzles generates droplet pairs for | compared to the ground truth answer. It mentions a specific method (digital microfluidics) but does not provide a comprehensive overview of microdroplet generation methods as stated in the ground truth. SCORE: 16 | | |
| --- | --- | --- | --- | --- | --- |

| | | precise mixing. Both methods are valuable in applications like nanoparticle synthesis, offering high reproducibility and control over reaction conditions in biomedical and material science research. | | | |
|---|---|---|---|---|---|
| What materials are typically used to fabricate droplet microfluidic devices? | The typical materials used for fabricating droplet microfluidic devices include PDMS (Polydimethylsiloxane), glass, and metals such as stainless steel or gold. These materials are chosen for their compatibility with biological samples, chemical resistance, and ease of fabrication. | Droplet microfluidic devices are commonly fabricated from a variety of materials, each offering distinct advantages for specific applications. The most widely used material is PDMS (Polydimethylsiloxane), favored for its ease of use with soft lithography and flexibility, though it has limitations such as poor solvent resistance. Alternatives to PDMS include thermoset polyesters, thermoplastics like PMMA, polycarbonate, and polystyrene, and amorphous | The answer provides a clear and concise list of materials used for fabricating droplet microfluidic devices, including PDMS, glass, and metals. However, it lacks the depth and detail provided by the ground truth answer, which lists specific alternatives to PDMS and their advantages. SCORE: 41 | 41 | 41.0 |

| | | polymers such as cyclic olefin copolymer. Fluoropolymers like PTFE and glass are also used, offering superior solvent resistance and biofouling resistance. Silicon and capillary devices are utilized in certain applications, providing precise control for microfluidic operations. These materials are chosen based on factors like solvent resistance, biofouling, and the ability to mass-produce devices using methods such as injection molding or hot embossing, ensuring broad applicability in research and industrial settings. | | | |
|---|---|---|---|---|---|
| What are the applications of droplet microfluidics? | Droplet microfluidics has a wide range of applications, particularly in biology and chemistry. Some key areas include single- | Droplet microfluidics has diverse applications across biomedical research, diagnostics, and industrial fields. It enables high- | The response provides a good overview of the applications of droplet microfluidics, but lacks some depth and | 46 | 46.0 |

| | cell analysis, drug discovery and screening, diagnostics, and genetic engineering. It allows for precise control over the environment within each droplet, enabling researchers to study cells or reactions under controlled conditions. Additionally, it can be used for high-throughput experiments, making it a valuable tool in scientific research. | throughput single-cell analysis, such as genomics, transcriptomics, and drug screening, while also facilitating nanoparticle synthesis, enzyme evolution, and chemical reactions. In healthcare, it supports personalized medicine, point-of-care diagnostics, and vaccine development. Beyond biology, it is used for environmental monitoring, food safety testing, and the development of cosmetics and pharmaceuticals. Its ability to create isolated, uniform reaction chambers in droplets makes it a powerful tool for precise, cost-effective, and scalable workflows in science and industry. | specificity compared to the ground truth answer. It mentions key areas such as single-cell analysis, drug discovery, diagnostics, and genetic engineering, which are all relevant, but does not cover the full range of applications mentioned in the ground truth answer, including nanoparticle synthesis, enzyme evolution, personalized medicine, point-of-care diagnostics, vaccine development, environmental monitoring, food safety testing, and cosmetics/pharmaceuticals development. SCORE: 46 | | |
|---|---|---|---|---|---|
| What are the main benefits of | 1. Increased control and | Droplet microfluidics | The answer provides 7 | 46 | 46.0 |
| using droplet microfluidics over bulk systems? | precision: Droplet microfluidics allows for precise manipulation of small volumes (picoliters to nanoliters) of fluids, which is difficult in bulk systems.<br><br>2. Reduced reagent consumption: Due to the small volume of each droplet, reagents are used more efficiently, leading to cost savings and less waste.<br><br>3. Improved safety: Handling hazardous or toxic substances in bulk can be risky, but with droplet microfluidics, these substances are contained within individual droplets, reducing the risk of exposure.<br><br>4. Enhanced mixing and reaction kinetics: The high surface-to-volume ratio in | offers significant advantages over traditional bulk systems. It enables individual compartmentalization, isolating single cells or molecules in tiny liquid compartments for precise control. This allows for single-cell analysis and early detection of rare cells or secreted molecules. The technology supports massive parallel processing, analyzing millions of droplets simultaneously, ideal for high-throughput screening and examining phenotypic or genetic variability. Additionally, iterative operations such as merging, mixing, and sorting droplets enable complex experimental protocols and continuous monitoring of populations. The | specific points that highlight the advantages of using droplet microfluidics over bulk systems. However, upon comparison with the ground truth answer, it is clear that while some points are accurate (e.g., increased control and precision, reduced reagent consumption), others lack depth or relevance to the broader context of droplet microfluidics (e.g., scalability, simplified sample preparation). The response does not fully capture the essence of the technology's capabilities in terms of single-cell analysis, high-throughput screening, and | | |

| | droplets promotes rapid and efficient mixing, leading to improved reaction kinetics compared to bulk systems.<br><br>5. Scalability: Droplet microfluidics can be easily scaled up or down by adjusting the number of droplets generated, making it suitable for both laboratory-scale experiments and large-scale industrial applications.<br><br>6. Simplified sample preparation: The small volumes required in droplet microfluidics make it easier to prepare and handle complex samples compared to bulk systems.<br><br>7. Increased throughput: Droplet microfluidics can process | system's highly controllable local environments ensure consistent, reproducible results. These benefits make droplet microfluidics a versatile tool in drug discovery, medical diagnostics, single-cell analysis, and other fields, offering powerful insights and precise manipulation at the individual level. | precise manipulation at the individual level. SCORE: 46 | | |
|---|---|---|---|---|---|
| | multiple reactions simultaneously, leading to increased throughput and faster results compared to traditional bulk methods. | | | | |
| What are the advantages of droplet microfluidic systems compared with continuous-flow systems? | 1. Scalability: Droplet microfluidics allows for the creation of thousands to millions of individual droplets, each containing a small volume of fluid. This enables parallel processing and miniaturization of reactions, which can be difficult or impossible in continuous-flow systems.

2. Reduced reagent consumption: Due to the small volumes involved, droplet microfluidics can significantly reduce the amount of reagents required for a given reaction compared to | Droplet-based microfluidic systems offer significant advantages over traditional continuous-flow systems. These include the ability to overcome issues such as Taylor dispersion, slow mixing, and surface-molecule interactions. Droplet systems allow for ultra-small assay volumes, reducing reagent consumption and improving sensitivity. They also minimize cross-contamination and sample adsorption by compartmentalizing individual samples. The system's chaotic advection enhances mixing efficiency, enabling faster | The response provides a clear and detailed explanation of the advantages of droplet microfluidic systems compared to continuous-flow systems, covering various aspects such as scalability, reagent consumption, reaction control, safety, downstream processing, and mixing efficiency. However, it does not fully capture the nuances and specific benefits mentioned in the ground truth answer, such as overcoming Taylor | 76 | 76.0 |

| | continuous-flow systems. This can lead to cost savings and improved experimental efficiency.

3. Improved control over reactions: The encapsulation of reactants within individual droplets allows for precise control over reaction conditions, such as temperature, pH, and concentration. This is difficult to achieve in continuous-flow systems where these parameters are often uniform throughout the system.

4. Increased safety: Droplet microfluidics can help contain potentially hazardous or toxic reagents within individual droplets, reducing the risk of exposure during experiments. | reactions. With droplet generation frequencies up to hundreds of kHz, droplet microfluidics supports high-throughput experimentation. Furthermore, the rapid and precise control of droplet payloads and sub-millisecond mixing times make it ideal for applications requiring fast, controlled analysis. These benefits make droplet-based systems particularly valuable for cell-based assays, enzyme kinetics, and high-throughput screening. | dispersion, minimizing cross-contamination, and supporting high-throughput experimentation at kHz frequencies. The response is mostly accurate but lacks some depth and detail compared to the comprehensive description provided by the ground truth answer.

SCORE: 76 | | |
|---|---|---|---|---|

| | | | | | |
|---|---|---|---|---|---|
| | 5. Simplified downstream processing: The encapsulation of products within individual droplets can facilitate their separation and purification, as well as simplify downstream analysis techniques such as PCR amplification or protein quantification.<br><br>6. Enhanced mixing efficiency: In droplet microfluidics, the high surface-to-volume ratio and shear forces generated during droplet formation can lead to improved mixing compared to continuous-flow systems, which may be beneficial for certain reactions that require thorough mixing of reactants. | | | | |
| What are the basic principles behind droplet generation in microfluidics? | The basic principles behind droplet generation in microfluidics are | Droplet generation in microfluidics relies on emulsion | The response accurately describes the three main methods of | 35 | 35.0 |

| | primarily based on three phenomena: the T-junction, the flow focusing, and the coaxial flow.

1. T-Junction: In this method, two immiscible fluids meet at a T-shaped junction. The pressure difference between the two channels causes droplets to be generated at the junction.

2. Flow Focusing: This method uses a channel with an inner and outer flow. The inner flow is the dispersed phase (droplet-forming fluid), while the outer flow is the continuous phase (carrier fluid). The geometry of the channel focuses the inner flow into droplets.

3. Coaxial Flow: In this method, two fluids are pumped through concentric | processes, where an oil phase (continuous phase) encapsulates an aqueous phase (dispersed phase) to form droplets. Surfactants are added to lower interfacial tension, stabilize the droplets, and prevent coalescence. The process requires precise control over the flow of both liquids, often achieved using syringe pumps or pressure-driven systems. Droplets are generated through surface-induced instabilities, influenced by microchannel geometries such as T-junctions or flow-focusing, which control droplet size and frequency. The design and flow control of the system determine the size uniformity and production rate of droplets. These principles enable high-throughput applications such | droplet generation in microfluidics (T-junction, flow focusing, and coaxial flow), but it lacks a comprehensive explanation of the underlying principles and mechanisms. It also does not fully address the role of surfactants, surface-induced instabilities, or the importance of precise control over liquid flow. The response seems to focus more on listing methods rather than explaining the fundamental principles behind droplet generation in microfluidics.

SCORE: 35 | | |

| | channels. The inner fluid forms the droplets, and the outer fluid acts as a sheath around it, which helps in stabilizing the droplet and preventing coalescence.

Each of these methods can be adjusted to control the size, number, and composition of the droplets generated. | as drug delivery, high-throughput screening, and microreactors for bio-catalysis, ensuring reproducibility and productivity in biological experiments. | | | |
|---|---|---|---|---|---|
| How is droplet size controlled in a droplet generator? | Droplet size in a droplet generator can be controlled by adjusting several parameters such as the flow rate of the fluid, pressure, voltage (in the case of electrospray), and the geometry of the nozzle or capillary tube. The smaller the nozzle diameter, the smaller the droplets produced. Increasing the pressure or voltage can also lead to smaller droplet sizes due to an increase in | Droplet size control in microfluidic generators can be achieved through several methods. In passive droplet generation, adjusting microchannel dimensions or controlling the flow rates of immiscible liquids allows for precise control over droplet size, shape, and monodispersity. Tree-splitters, which bisect large droplets into smaller ones, also influence droplet size and uniformity, with | The response provides some relevant information about factors that can influence droplet size in a droplet generator, but it lacks specificity and accuracy regarding the methods used to control droplet size. It also does not address the specific context of microfluidic generators as mentioned in the ground | 36 | 36.0 |

| | kinetic energy. Additionally, the type of fluid used and its surface tension play a role in determining the final droplet size. | splitting angles affecting droplet diameter. In active droplet generation, technologies like piezoelectric actuators independently control droplet size, formation rate, and spacing, offering high precision. Additionally, the velocity of the continuous and dispersed phases plays a significant role in controlling droplet size. These methods enable the generation of uniform droplets for various applications, such as high-throughput screening and drug delivery. | truth answer. SCORE: 36 | | |
|---|---|---|---|---|---|
| What is the role of surface tension in droplet formation? | Surface tension plays a crucial role in droplet formation. It is the force that acts at the interface between two phases, such as liquid and gas, tending to minimize the surface area of the liquid phase. | Surface tension plays a critical role in droplet formation in microfluidic systems by influencing droplet size, shape, and formation mechanisms. It acts to minimize surface area, favoring | The response provides a clear explanation of the role of surface tension in droplet formation, mentioning its influence on droplet size, shape, and stability. However, it | 56 | 56.000000 00000001 |

| | | |
|---|---|---|
| | In the case of droplet formation, this means that when a liquid is sprayed or drips, it tends to form spherical droplets due to the balance between the forces of gravity pulling the droplet downward and the surface tension pulling it inward to form a sphere. This property allows for consistent droplet size and stability, which is important in many practical applications like inkjet printing, fog formation, and cloud physics. | spherical droplet shapes. At low Weber numbers (We < 1), where surface tension dominates, droplets form through dripping or squeezing. As the Weber number increases, fluid inertia competes with surface tension, leading to jetting behavior. The capillary number (Ca) also influences droplet size, with surface tension controlling droplet formation at low Ca. Techniques like electrowetting can measure surface tension in microfluidic systems, providing insights into the interfacial properties. Additionally, surface tension is a key parameter in T-junctions and flow focusing devices, where droplet size measurements offer an indirect way to assess surface tension at the time-scale of | lacks specific details about microfluidic systems, Weber numbers, capillary numbers, and other technical aspects mentioned in the ground truth answer. SCORE: 56 |

| | | | | | |
|---|---|---|---|---|---|
| | | droplet production. Understanding the interplay between surface tension and other forces is essential for precise droplet control. | | | |
| How does a flow-focusing device differ from a T-junction in droplet formation? | A flow-focusing device and a T-junction differ in the way they influence droplet formation. In a T-junction, two fluid streams meet at a right angle, causing turbulence that leads to the spontaneous formation of droplets due to surface tension. On the other hand, a flow-focusing device uses a converging and diverging channel geometry to create a stable, laminar flow. Here, the Taylor cone instability is induced by the pressure difference between the two main channels, resulting in the formation of monodisperse droplets with | Flow-focusing devices and T-junctions are two common microfluidic designs for droplet formation, differing in geometry and mechanism. A flow-focusing device uses a central dispersed phase channel and side continuous phase channels converging at a narrow orifice, where hydrodynamic focusing pinches off droplets. It produces smaller, highly uniform droplets with precise control over size and high throughput. In contrast, a T-junction has perpendicular channels where the continuous phase shears the dispersed phase | The response accurately describes the key differences between flow-focusing devices and T-junctions in terms of their geometry and mechanism of droplet formation. It correctly explains how flow-focusing devices produce smaller, highly uniform droplets with precise control over size and high throughput, whereas T-junctions form larger droplets with less uniformity and lower throughput. The response also highlights the suitability of each device for specific | 95 | 95.0 |
| | controlled size and regularity. This makes flow-focusing devices more suitable for applications requiring precise control over droplet size, such as microfluidics and lab-on-a-chip technologies. | at the intersection, forming larger droplets with less uniformity and lower throughput. Flow-focusing is ideal for applications requiring monodisperse droplets, while T-junctions suit simpler or larger-droplet applications. | applications. SCORE: 95 | | |
| What role does the geometry of the microchannel play in droplet formation? | The geometry of a microchannel plays a crucial role in droplet formation. Factors such as channel width, height, and length, as well as the surface texture, can significantly influence the size, number, and stability of the droplets generated through hydrodynamic focusing or other methods. For instance, narrow channels tend to produce smaller droplets due to increased shear stress, while wider channels may result in | The geometry of a microchannel plays a critical role in droplet formation, influencing the size, shape, and generation frequency of droplets. Key geometric factors include the channel width, depth, and junction design, such as T-junctions or flow-focusing geometries. These features govern the interaction between the continuous and dispersed phases, affecting the shear forces and pressure gradients that | The response provides a clear explanation of the role of microchannel geometry in droplet formation, mentioning specific factors such as channel width, height, surface texture, and their effects on droplet size, number, and stability. However, it lacks detailed discussion on junction design (e.g., T-junctions or flow-focusing geometries) and the impact of these | 56 | 56.000000 00000001 |

| | larger droplets. Additionally, the presence of textured surfaces can alter the interfacial tension between the fluids and promote more uniform droplet formation. | drive droplet breakup. For instance, narrower channels or smaller orifices promote the formation of smaller droplets due to increased shear stress. Similarly, the junction type controls droplet detachment mechanics; in T-junctions, droplets form due to the balance of interfacial tension and shear force, while in flow-focusing designs, they form under pinching at the constriction. Channel geometry also determines flow regimes ranging from dripping to jetting further impacting droplet characteristics. Thus, optimizing microchannel geometry is crucial for precise control over droplet formation in applications like microfluidics, emulsification, and lab-on-a-chip | features on droplet breakup mechanics. The response also does not explicitly mention the importance of optimizing microchannel geometry for precise control over droplet formation in applications like microfluidics, emulsification, and lab-on-a-chip systems. SCORE: 56 | | |

| | | systems. | | | |
|---|---|---|---|---|---|
| What factors affect droplet coalescence in microfluidic devices? | Droplet coalescence in microfluidic devices can be influenced by several factors, including:<br><br>1. Flow rates: The rate at which the fluids are flowing through the device can affect droplet-droplet interactions and thus coalescence.<br>2. Surface tension: This is a key factor as it determines the interfacial energy between the droplets and the surrounding fluid.<br>3. Interfacial properties: The properties of the interface between the oil and water phases, such as wettability and surface roughness, can impact coalescence.<br>4. Droplet size: Smaller droplets are more susceptible to coalescence due | Droplet coalescence in microfluidic devices is influenced by several factors. Surfactant concentration plays a key role, with higher concentrations preventing droplet fusion, while lower concentrations promote coalescence. The duration droplets remain in contact within the pairing channel, known as pairing time, also impacts coalescence, with longer times favoring fusion. The geometry of the pairing channel, specifically its cross-sectional area, controls the number of droplets that can come into contact and coalesce. Additionally, droplet size and volume affect how efficiently droplets merge. Finally, electrocoalescence, which applies | The response provides a clear and concise list of factors that affect droplet coalescence in microfluidic devices. However, it lacks the depth and detail provided by the ground truth answer, particularly regarding the specific interactions between these factors and their relative importance. SCORE: 51 | 51 | 51.0 |
| | to their higher surface-to-volume ratio.<br>5. Shear rate: The rate at which the fluid is being sheared can affect the stability of the droplets and thus influence coalescence.<br>6. Presence of surfactants or additives: These substances can alter interfacial properties, affecting the stability of the droplets and their tendency to coalesce. | electric fields to induce droplet fusion, offers an alternative method for on-demand coalescence. By optimizing these factors—surfactant concentration, pairing time, channel geometry, and droplet size—reliable and selective droplet coalescence can be achieved in microfluidic systems. | | | |
| What design elements are critical for producing double or multiple emulsions? | 1. Proper choice of oil and water phases: The oil phase should be immiscible with the continuous aqueous phase, and both phases should be stable on their own.<br><br>2. Emulsifying agent or surfactant: A suitable emulsifying agent is essential to stabilize the interface between the oil and water phases. The choice of | To generate double or multiple emulsions, several design elements are essential. Co-flow and flow-focusing configurations enable precise control over fluid flow, allowing for stable droplet formation, especially in double emulsions. Sequential emulsification steps, where inner droplets are emulsified into | The response provides a clear and detailed explanation of the design elements necessary for producing double or multiple emulsions, including proper choice of oil and water phases, emulsifying agent, high-energy input, control over droplet size and distribution, | 45 | 45.0 |

| | | | | |
|---|---|---|---|---|
| | surfactant depends on factors such as the type of emulsion (oil-in-water or water-in-oil), the desired size distribution, and the stability requirements.<br><br>3. High-energy input: Double or multiple emulsions are typically produced by high-energy methods such as homogenization, ultrasonication, or microfluidizatio n to create small droplets of one phase within another.<br><br>4. Control over droplet size and distribution: The size and distribution of the inner and outer droplets play a crucial role in the final properties of the emulsion. This can be achieved through adjusting the energy input, surfactant concentration, | intermediate phases and then into outer phases, are key for multiple emulsions. Capillary microfluidic devices, such as consecutive capillary and coaxial configurations, are used to control droplet size, number, and monodispersity. Additionally, precise flow control and shear force management are crucial for stable and consistent droplet formation. Microchannel geometry optimization and material compatibility further enhance emulsification processes. By carefully managing these factors, researchers can produce double and multiple emulsions with controlled internal structures and compositions for various applications. | and stability. However, it does not specifically mention co-flow and flow-focusing configurations, sequential emulsification steps, capillary microfluidic devices, precise flow control, shear force management, microchannel geometry optimization, or material compatibility, which are mentioned in the ground truth answer as key factors for double and multiple emulsions. SCORE: 45 | | |

| | and flow rates of the phases.<br><br>5. Stability: To maintain the structure of double or multiple emulsions over time, it is essential to ensure their stability against coalescence, creaming, and phase separation. This can be achieved by optimizing the composition, using appropriate surfactants, and adjusting processing conditions. | | | | |
|---|---|---|---|---|---|
| What key factors influence the design of flow-focusing droplet microfluidics optimized using machine learning? | The design of flow-focusing droplet microfluidics optimized using machine learning is influenced by several key factors. These include:<br><br>1. Fluid properties: Viscosity, surface tension, and density of the fluids used in the system play a significant role in determining | The design of flow-focusing droplet microfluidics, optimized using machine learning, depends on several critical factors. Channel geometry, including the shape and size, controls droplet generation by defining the break-off point where the dispersed phase meets the continuous phase. | The response provides a clear and concise list of factors influencing flow-focusing droplet microfluidics optimized using machine learning. However, it lacks depth and specific details about how these factors interact with each other and the | 46 | 46.0 |

| | the behavior of the droplets.

2. Geometry and dimensions: The design of the microchannels, including their length, width, and aspect ratio, can greatly affect the generation and stability of droplets.

3. Flow rates: The flow rates of the continuous and dispersed phases must be carefully controlled to ensure proper droplet formation and stability.

4. Machine learning algorithms: The choice of machine learning techniques, such as neural networks or genetic algorithms, can impact the optimization process and the final design of the microfluidic device.

5. Data | Flow rates of both phases affect droplet size and frequency, with higher continuous phase flow reducing droplet size. Viscosity influences flow behavior and shear forces, crucial for stable droplet formation. Shear forces between phases are controlled by adjusting channel geometry or flow rates, ensuring consistent droplet generation. Machine learning algorithms predict and optimize droplet diameter and generation rate by analyzing these design parameters, fluid properties, and large datasets, automating the design process for more efficient droplet production in life sciences applications. | machine learning algorithms used for optimization. The answer also does not fully align with the ground truth in terms of providing detailed explanations of the relationships between design parameters and fluid properties. SCORE: 46 | | |

| | | | | |
|---|---|---|---|---|
| | availability: The quality and quantity of data used for training the machine learning models are crucial for achieving accurate predictions and optimal designs.<br><br>6. Validation and testing: It is essential to validate the designs using experimental verification to ensure their practicality and reliability. | | | |
| How do channel wall properties (hydrophobicity/hydrophilicity) influence droplet dynamics? | Channel wall hydrophobicity/hydrophilicity significantly influences droplet dynamics. In hydrophobic channels, the water-repelling surface causes droplets to minimize their contact area with the channel walls, leading to a tendency for spherical droplets and reduced coalescence. This can result in stable, discrete droplets that | The channel wall properties, specifically hydrophobicity or hydrophilicity, significantly influence droplet dynamics in microfluidic systems. These properties determine the wettability of the channel walls, affecting the interaction between the continuous and dispersed phases. In hydrophobic channels, the dispersed phase (typically water) tends to form | The response accurately addresses the question by providing a clear explanation of how channel wall properties influence droplet dynamics, including the effects of hydrophobicity and hydrophilicity on droplet behavior. It also provides relevant examples and applications for each type | 85 | 85.0 |

| | move along the channel more easily due to lower friction.

In contrast, hydrophilic channels encourage wetting, causing droplets to spread out over the surface. This can lead to larger, less stable droplets and increased coalescence, making it harder for individual droplets to move through the channel. The properties of the channel walls thus play a crucial role in controlling droplet behavior within microfluidic systems. | droplets as it minimizes contact with the channel walls. This enhances droplet stability and facilitates easier detachment during droplet formation. Hydrophobic surfaces are ideal for applications requiring water-in-oil emulsions or precise droplet control. Conversely, hydrophilic channels promote wetting of the dispersed phase, making it more likely to spread along the walls instead of forming discrete droplets. This is useful for oil-in-water systems or applications involving liquid films. Wall properties also impact droplet size, breakup, coalescence, and transport. Hydrophobicity favors smaller droplets and reduces the likelihood of droplet adherence, while hydrophilicity | of surface property. However, it does not explicitly mention the impact on droplet size, breakup, coalescence, and transport as stated in the ground truth answer. SCORE: 85 | | |
|---|---|---|---|---|

| | | | | | |
|---|---|---|---|---|---|
| | | can lead to higher flow resistance and altered droplet generation behavior. Tailoring wall properties is crucial for optimizing microfluidic processes. | | | |
| How can you incorporate active control mechanisms (e.g., electrical, magnetic) into droplet devices? | Active control mechanisms can be incorporated into droplet devices by utilizing actuators and sensors to manipulate the behavior of the droplets. For instance, electrical or magnetic actuation can be employed to generate forces on the droplets, which in turn allows for precise positioning, merging, splitting, or deformation of the droplets. Sensors can provide feedback about the state of the droplet system and help optimize the control strategy. | Active control mechanisms such as electrical, magnetic, thermal, and mechanical methods can precisely manipulate droplet formation in microfluidic devices. Electrical control methods include direct current (DC), which uses electrodes to control droplet size, and alternating current (AC), which utilizes electrowetting-on-dielectric (EWOD) and dielectrophoresis (DEP) to influence droplet behavior. Magnetic control employs ferrofluids and external magnets | The response provides a general idea of incorporating active control mechanisms into droplet devices but lacks specific details about the types of electrical and magnetic methods that can be used. It also doesn't mention thermal, mechanical, or other control methods as mentioned in the ground truth answer. SCORE: 41 | 41 | 41.0 |
| | Integrating these components requires careful design considerations to ensure compatibility with the specific properties of the droplet devices, such as their size, material, and surface tension. | to non-invasively guide droplet formation. Thermal control involves heating the entire device or applying localized heating to adjust fluid parameters like viscosity and interfacial tension. Mechanical control techniques, such as pneumatic and piezoelectric components, can also regulate droplet dynamics. These active control strategies enable precise and flexible droplet generation, allowing for tailored applications in complex microfluidic systems, including droplet manipulation and multicomponent emulsion creation. | | | |
| How do flow rates affect droplet size and frequency? | Flow rate significantly impacts both the size and frequency of droplets in a fluid system. A higher flow rate | Flow rates significantly influence droplet size and frequency in microfluidic systems. As the flow rate ratio ($\varphi$ | The response accurately describes how flow rates affect droplet size and frequency, aligning well | 85 | 85.0 |

| | generally results in larger droplet sizes due to increased kinetic energy, which causes the fluid to break up into larger droplets. Conversely, lower flow rates produce smaller droplets because there is less kinetic energy available for breaking up the fluid.

Regarding frequency, as the flow rate increases, the time between droplet formation also tends to decrease, leading to an increase in droplet frequency. This is because a higher flow rate means more fluid is being pushed through the system at any given moment, resulting in more frequent droplet formation events. | = Qd/Qc) increases, the shearing force from the continuous phase dominates, leading to smaller droplets due to rapid breakup of the dispersed phase. Additionally, higher flow rates in the continuous phase can result in larger droplets, as the dispersed phase doesn't break up until it reaches the downstream section. The frequency of droplet generation also increases with higher flow rate ratios, demonstrating a positive correlation. However, droplet diameter growth trends vary depending on the model and input conditions. Overall, increasing the flow rate ratio enhances droplet formation frequency and reduces droplet size due to stronger shearing | with the ground truth. It provides a clear explanation of the relationship between flow rate and droplet characteristics, including the impact on size and frequency. However, it does not delve into the complexities mentioned in the ground truth answer regarding the influence of factors such as interfacial tension and channel geometry. SCORE: 85 | | |

| | | forces, with complex interactions influenced by factors such as interfacial tension and channel geometry. | | | |
|---|---|---|---|---|---|
| What techniques are available for droplet splitting? | 1. Manual Droplet Splitting: This method involves manually dividing a larger droplet into smaller ones using a pipette. It is time-consuming and requires careful handling to avoid contamination.<br><br>2. Automated Droplet Splitting: This technique uses specialized equipment like the Fluidigm C1 or BioMark HD systems, which can automatically split droplets based on the volume of each droplet's content. This method is faster and more precise than manual splitting.<br><br>3. Acoustic Droplet | Droplet splitting techniques can be categorized into passive and active methods. Passive techniques rely on the microchannel geometry and fluid properties, such as T and Y junctions, flow focusing configurations, or introducing ribs/obstructions in the flow path. These methods are simple but lack dynamic control over splitting ratios and are sensitive to flow conditions. Active techniques utilize external forces like electric, magnetic, acoustic, or thermal fields to precisely control droplet splitting. Examples include electric field-based splitting, | The response provides a detailed list of various techniques for droplet splitting, including manual, automated, acoustic, electrical, thermal, magnetic, and optical methods. It accurately describes each technique's characteristics, advantages, and potential applications. However, the answer could be improved by providing a clear categorization of these techniques into passive and active methods as mentioned in the ground truth, and highlighting their respective | 85 | 85.0 |

| | Splitting: This technique uses sound waves to split droplets. The system, such as the Acouloop, generates high-frequency sound waves that cause the droplet to divide into two or more smaller droplets. This method is non-contact and can be used for a wide range of volumes and viscosities.<br><br>4. Electrical Droplet Splitting: This technique uses an electric field to split droplets. The system, such as the Digital Microfluidics EDC, generates an electric field that causes the droplet to divide into two or more smaller droplets. This method is also non-contact and can be used for a wide range of volumes and viscosities.<br><br>5. Thermal Droplet Splitting: This technique uses | magnetic fields for ferrofluids, acoustic radiation force, and thermocapillary methods. Active techniques offer more flexibility, enabling on-demand droplet splitting with controlled ratios, overcoming the limitations of passive methods. However, each technique has its own set of advantages and constraints depending on the application and droplet type. | strengths and limitations.<br>SCORE: 85 | | |

| | heat to split droplets. The system, such as the TTP Labtech Mosquito, generates heat that causes the droplet to evaporate, leaving behind two or more smaller droplets. This method is contact-based and can be used for a wide range of volumes and viscosities.

6. Magnetic Droplet Splitting: This technique uses magnetic fields to split droplets. The system, such as the Magnetic Particle Technology (MPT) system, generates a magnetic field that causes the droplet to divide into two or more smaller droplets. This method is non-contact and can be used for a wide range of volumes and viscosities.

7. Optical Droplet Splitting: This | | | | |
|---|---|---|---|---|---|
| | technique uses light to split droplets. The system, such as the OptiDrops system, generates light that causes the droplet to divide into two or more smaller droplets. This method is non-contact and can be used for a wide range of volumes and viscosities. | | | | |
| What advantages does droplet formation by passive methods have over active methods? | Passive droplet formation methods have several advantages over active methods. Firstly, they are simpler and less expensive as they do not require external energy sources or complex equipment. Secondly, passive methods often produce smaller, more uniform droplets which can lead to better mixing and reaction control in various applications such as microfluidics and chemical synthesis. Lastly, passive methods | Passive droplet generation methods offer several advantages over active methods. They simplify device design by relying on flow rate manipulation and microchannel geometry, eliminating the need for complex external forces. This leads to cost-effectiveness and easier fabrication. Passive methods also provide precise control over droplet size, shape, and monodispersity, making them ideal for applications requiring uniform | The response accurately identifies the advantages of passive droplet formation methods over active ones, including simplicity, lower cost, smaller and more uniform droplets, and safety. However, it lacks some details provided in the ground truth answer, such as precise control over droplet size and shape, protection from external interference, and the ability | 76 | 76.0 |
| | tend to be safer and more environmentally friendly because they avoid the potential hazards associated with high-energy sources or chemicals used in active methods. | droplets. Additionally, they protect droplets from external interference and cross-contamination, ensuring integrity. Passive methods can produce continuous droplet streams with uniform size distributions (1%-3% dispersity) without requiring external actuation. While they are not suitable for dynamic manipulation or "droplet-on-demand" applications, passive methods are favored for their simplicity, precision, and cost-efficiency in controlled environments. | to produce continuous streams with uniform size distributions. SCORE: 76 | | |
| How do you design a device for monodisperse droplet generation? | Designing a device for monodisperse droplet generation involves several steps. Here's a simplified process:<br><br>1. Choose the appropriate | To design a device for monodisperse droplet generation, two common methods are used: cross-flow and flow-focusing designs. In cross-flow designs, the continuous phase | The response provides a detailed and structured approach to designing a device for monodisperse droplet generation, including the selection of | 76 | 76.0 |

| | | | | |
|---|---|---|---|---|
| | method for droplet generation, such as T-Tojette, Taylor cone, or flow focusing. Each method has its advantages and disadvantages in terms of droplet size distribution, throughput, and compatibility with different fluids.<br><br>2. Design the geometry of the device based on the chosen method. For example, for a T-Tojette generator, you would need to design a tee junction with specific angles and dimensions. For Taylor cone generation, you would need to create a capillary tube that forms a cone when fluid is expelled from it.<br><br>3. Select materials for the device that are compatible with the fluids you plan to use. The materials should | flows around the dispersed phase in a constriction channel, with droplet size controlled by adjusting channel dimensions and shear stress. This method is simple but can lead to droplet damage due to shear forces or wall interactions. Flow-focusing designs, on the other hand, use sheath flows to stabilize the interface between phases, enabling smaller droplet formation with precise control over size by adjusting flow rates and orifice dimensions. This method offers high monodispersity and is less prone to droplet damage. Both designs require careful optimization of flow rates, channel dimensions, and surface tension to achieve uniform droplet sizes. The choice of design depends on | methods, materials, control systems, testing, validation, and iteration. It covers various aspects relevant to the question, such as geometry design, material compatibility, flow rate regulation, and performance optimization. However, it does not directly address the two common methods mentioned in the ground truth answer (cross-flow and flow-focusing designs) or their specific characteristics, which are crucial for a comprehensive understanding of monodisperse droplet generation. SCORE: 76 | | |

| | be non-reactive, biocompatible if necessary, and able to withstand the pressure and flow conditions in the device.

4. Implement a control system to regulate the flow rates of the fluids and the pressure within the device. This could involve pumps, valves, and pressure sensors.

5. Test the device under various conditions to optimize droplet size, monodispersity, and throughput. This may require adjusting the geometry, flow rates, or pressure settings.

6. Validate the performance of the device by comparing the generated droplets with theoretical predictions or standards. This could involve microscopic imaging, particle | specific needs such as droplet size, throughput, and resistance to damage during formation. | | | |
|---|---|---|---|---|---|
| | sizing, and other analytical techniques.<br><br>7. Iterate on the design based on the results of the testing and validation steps to improve the device's performance. | | | | |
| What materials are most commonly used for fabricating droplet microfluidic devices? | The most commonly used materials for fabricating droplet microfluidic devices are PDMS (Polydimethylsiloxane), glass, and metals such as stainless steel or gold. These materials are chosen due to their biocompatibility, chemical resistance, and ease of fabrication. | The most commonly used materials for fabricating droplet microfluidic devices are PDMS (polydimethylsiloxane) and silicon-based materials. PDMS is preferred due to its ease of fabrication using soft lithography, which allows for the creation of precise and versatile microfluidic devices. This material provides high transparency, efficient sealing, and the ability to mold complex microchannels, making it ideal for droplet generation and manipulation. While PDMS is | The response is mostly accurate in listing common materials used for fabricating droplet microfluidic devices, but it lacks specificity regarding the preferred material (PDMS) and does not mention silicon-based materials as a key component. It also includes metals which are not typically used for this purpose. SCORE: 41 | 41 | 41.0 |

| | | dominant, alternatives such as thermoplastics (e.g., PMMA, polycarbonate), fluoropolymers (e.g., PTFE), and glass materials are also used, each offering unique benefits like better solvent resistance or mass production capabilities. Despite the variety of materials available, PDMS remains the material of choice for most droplet microfluidics applications due to its simplicity and versatility. | | | |
|---|---|---|---|---|---|
| What are the advantages and limitations of using PDMS for microfluidic devices? | Advantages of using Polydimethylsiloxane (PDMS) for microfluidic devices include its low cost, ease of fabrication through soft lithography, biocompatibility, and optical transparency. PDMS is also flexible, which allows for the creation of complex geometries and channels. | Polydimethylsiloxane (PDMS) is widely used for microfluidic devices due to its several advantages, including low toxicity, biocompatibility, optical transparency, and high gas permeability, which make it ideal for biomedical applications. Its flexibility and cost-effectiveness | The response accurately lists the advantages of using PDMS for microfluidic devices but misses some key points mentioned in the ground truth answer, such as low toxicity and high gas permeability. It also correctly identifies the limitations of PDMS, | 51 | 51.0 |
| | Limitations of using PDMS for microfluidic devices are primarily related to its mechanical properties. PDMS has a relatively low Young's modulus, making it prone to deformation under pressure or flow conditions. Additionally, PDMS is not chemically resistant to many solvents and reagents commonly used in laboratory settings, which can lead to device failure over time. Lastly, the bonding process of PDMS devices can be challenging and may result in leaks if not performed correctly. | also contribute to its popularity, as it allows for easy fabrication using soft lithography. PDMS enables real-time observation and is well-suited for flexible and wearable electronics. However, PDMS has limitations, such as its hydrophobic nature, which can lead to fouling in microchannels, affecting performance. Additionally, its soft elastomeric properties limit mechanical strength, and large-scale production can be challenging. Despite these drawbacks, PDMS remains a versatile and widely used material, though further research is needed to address its limitations for broader applications. | including its mechanical properties, chemical resistance, and bonding process, but does not mention its hydrophobic nature or soft elastomeric properties. The response is clear and concise but lacks comprehensive coverage of the topic. SCORE: 51 | | |
| How does 3D printing facilitate microfluidic device fabrication? | 3D printing facilitates microfluidic device fabrication by enabling the | 3D printing revolutionizes microfluidic device fabrication by enabling the creation of | The response accurately conveys the benefits of 3D printing in microfluidic | 56 | 56.000000 00000001 |

| | creation of complex, three-dimensional structures with high precision and reproducibility. This is particularly useful in microfluidics as these devices often require intricate channels and features for fluid manipulation at small scales. The use of 3D printing allows for rapid prototyping, customization, and integration of functional components into the device design, making it a versatile tool for microfluidic research and development. | complex and precise channel structures that are challenging to achieve with traditional methods like soft lithography. Techniques such as Stereolithography (SL), Multijet Modeling (MJM), and Fused Deposition Modeling (FDM) offer high resolution, complex geometries, and the integration of various components like valves and pumps. 3D printing also allows rapid prototyping, customization, and the use of diverse materials for biocompatibility and transparency. Advantages over traditional methods include reduced fabrication time and labor, increased accuracy, and environmental friendliness due to the additive manufacturing | device fabrication, including precision, customization, and rapid prototyping. However, it lacks specific details about the techniques used (SL, MJM, FDM) and the advantages over traditional methods such as reduced fabrication time and labor, increased accuracy, and environmental friendliness. SCORE: 56 | | |
|---|---|---|---|---|---|

| | | process. With its ability to create hierarchical 3D configurations and integrate multiple functions, 3D printing holds great potential for advancing microfluidic device design and performance. | | | |
|---|---|---|---|---|---|
| What alternative materials can be used for applications involving organic solvents? | Some alternative materials that can be used for applications involving organic solvents are:<br><br>1. Ionic liquids: These are salts with melting points below 100 degrees Celsius. They have low volatility, high thermal stability, and excellent solvent properties.<br><br>2. Supercritical fluids: These are substances that are above their critical temperature and pressure. Carbon dioxide is a common supercritical fluid used as an alternative to | Several sustainable alternatives to traditional organic solvents are gaining attention for various applications. Supercritical CO2 is a non-toxic, environmentally friendly solvent, particularly useful in extraction processes. Ionic liquids, which are liquid salts at room temperature, offer low vapor pressure and broad application in chemical reactions. Organic carbonates, like dimethyl carbonate, are biodegradable and non-toxic, | The answer provides a clear and concise list of alternative materials that can be used for applications involving organic solvents. It covers various types such as ionic liquids, supercritical fluids, deep eutectic solvents, green solvents, and water, with some details on their properties and advantages. However, the response could benefit from more comprehensive information and a broader scope of alternatives, as | 56 | 56.000000 00000001 |

| | organic solvents due to its low toxicity and cost-effectiveness.<br><br>3. Deep eutectic solvents (DES): These are a class of ionic liquids that are formed by the combination of a hydrogen bond donor (HBD) and a hydrogen bond acceptor (HBA). They have lower costs and easier synthesis compared to traditional ionic liquids.<br><br>4. Green solvents: These are solvents derived from renewable resources, such as vegetable oils or sugars. Examples include ethyl lactate and butyl levulinate.<br><br>5. Water: In some cases, water can be used as a green alternative to organic solvents, especially in reactions that are | ideal for fine chemical synthesis. Biosolvents, derived from renewable resources such as plants and bacteria, provide a greener option for extraction and cleaning. Other alternatives include fluorous solvents, non-toxic liquid polymers like polyethylene glycol (PEG), water, ethanol, methanol, and ethyl acetate. These alternatives are chosen based on their specific properties such as polarity, boiling point, and environmental impact, promoting more sustainable chemical processes and reducing reliance on hazardous petroleum-based solvents. | well as a clearer connection to the specific applications they serve. The answer also lacks direct references or citations to support its claims.<br><br>SCORE: 56 | | |

| | hydrophilic or where catalysts are water-soluble. | | | | |
|---|---|---|---|---|---|
| How does device surface treatment affect droplet formation? | Device surface treatment significantly influences droplet formation. Surface properties such as wettability, roughness, and chemical composition can all affect how liquids spread and form droplets on a surface. For example, hydrophilic surfaces (those that attract water) tend to cause liquids to spread out more evenly, while hydrophobic surfaces (those that repel water) cause liquids to form smaller, more spherical droplets due to the water's tendency to minimize contact with the surface. Additionally, rougher surfaces can trap air pockets and create larger, irregular | Surface treatment of PDMS microfluidic devices plays a critical role in determining droplet formation and stability. PDMS is naturally hydrophobic, suitable for generating water-in-oil (W/O) droplets but not oil-in-water (O/W) droplets. To generate O/W droplets, selective hydrophilic treatment is required. The researchers employed two methods: manual PVA deposition after plasma treatment, and a simultaneous approach combining PVA injection with on-chip plasma treatment. The latter method enables precise wettability patterning, allowing for both W/O and O/W droplet formation within the same | The response provides a clear and concise explanation of how device surface treatment affects droplet formation, discussing various factors such as wettability, roughness, and chemical composition. It also mentions the specific effects on hydrophilic and hydrophobic surfaces, as well as the impact of surface roughness. However, it does not specifically address the context of PDMS microfluidic devices or the methods used to achieve selective hydrophilic treatment, which is a key | 55 | 55.000000 00000001 |
| | droplets. The specific effects depend on the type of treatment and material of the device surface. | device. The plasma treatment increases surface energy, making the PDMS hydrophilic, while PVA deposition stabilizes the coating, ensuring long-term device use. This technique provides controllable droplet sizes and stable double emulsion formation, with the ability to manipulate droplet characteristics by adjusting flow rates. The method also allows for reliable, reproducible results, making it a powerful tool for droplet microfluidic applications. | aspect of the ground truth answer. SCORE: 55 | | |
| What bonding techniques are used to seal microfluidic channels? | The bonding techniques used to seal microfluidic channels include:<br><br>1. Thermal Bonding: This is the most common method, where two parts are | Several bonding techniques are employed to seal microfluidic channels, each suited to different materials. Thermal fusion bonding involves compressing thermoplastic substrates at temperatures near | The response accurately lists various bonding techniques used to seal microfluidic channels (Thermal Bonding, UV-Curing or Photochemical Bonding, | 46 | 46.0 |

| | heated above their glass transition temperature (Tg) and then pressed together.<br><br>2. UV-Curing or Photochemical Bonding: A photosensitive adhesive is applied to the channel walls, and the device is exposed to ultraviolet light, causing the adhesive to cure and bond the channels together.<br><br>3. Solvent Bonding: This method involves the use of a solvent that dissolves a layer on one side of the channel, allowing the two sides to bond when they are brought into contact.<br><br>4. Plasma Bonding: A plasma is used to activate the surface of the channels, promoting chemical bonds between them | their glass transition point, allowing them to fuse upon cooling. Chemical bonding creates covalent bonds between substrate surfaces through chemical reactions, while solvent bonding uses solvents to dissolve the materials partially, promoting adhesion through interdiffusion. For PDMS devices, oxygen plasma treatment is commonly used to make surfaces reactive, enabling irreversible bonding when in contact with another PDMS or glass piece. Additionally, corona wand bonding offers a less expensive alternative to oxygen plasma treatment, though it is less consistent. The choice of bonding technique depends on material properties, bond | Solvent Bonding, Plasma Bonding, and Click Chemistry) and provides a brief description of each. However, the answer lacks depth in explaining the advantages and disadvantages of these methods, as well as their specific applications and limitations. The response also does not provide a clear connection to the ground truth, which discusses the suitability of different techniques for various materials and the importance of material properties in choosing the appropriate bonding method. SCORE: 46 | | |
|---|---|---|---|---|
| | upon contact.<br><br>5. Click Chemistry: This method uses reversible or irreversible chemical reactions to bond the channel walls together.<br><br>Each technique has its advantages and disadvantages in terms of cost, speed, precision, and compatibility with different materials. | strength requirements, and available equipment, all of which are crucial for the successful fabrication of microfluidic devices. | | |
| What challenges arise when fabricating complex microfluidic geometries? | 1. Precision and Repeatability: Fabricating complex microfluidic geometries requires high precision and repeatability, as even small deviations can significantly impact the performance of the device.<br><br>2. Material Selection: Choosing appropriate materials that are biocompatible, optically transparent, and | Fabricating complex microfluidic geometries presents several challenges, including controlling wettability in intricate channel structures, which is critical for fluid flow and droplet formation. Additionally, scaling up microfabrication processes for commercial applications requires cost-effective techniques and | The response provides a clear and concise list of challenges associated with fabricating complex microfluidic geometries, including precision and repeatability, material selection, integration of multiple components, scalability, cost-effectiveness, and functional verification. However, the | 56 | 56.000000 00000001 |

| | mechanically robust is crucial for applications in life sciences.

3. Integration of Multiple Components: Incorporating multiple functional components such as valves, pumps, sensors, and actuators into a single microfluidic device can be challenging due to the need for precise alignment and integration.

4. Scalability: Scaling up the fabrication process to produce large numbers of devices while maintaining consistency in geometry and performance is another challenge.

5. Cost-Effectiveness: Achieving cost-effective mass production of microfluidic devices remains a significant | materials. Current droplet generation methods mainly produce single emulsions, limiting their use in applications that require multiple or complex emulsions. Moreover, droplet microfluidics faces limitations in producing diverse and functional nano-scale materials compared to other methods. Conventional lithography techniques, which are restricted to rectangular cross-sections, struggle with creating complex 3D geometries. Although advancements like 3D printing and stacking methods offer solutions, effective mixing strategies and versatile fabrication of 3D networks with the required features remain significant hurdles in the field. Addressing | answer lacks depth and context compared to the ground truth answer, which provides more nuanced information on specific challenges such as controlling wettability in intricate channel structures, limitations in droplet generation methods, and difficulties with creating complex 3D geometries using conventional lithography techniques.

SCORE: 56 | | |
|---|---|---|---|---|

| | | | | | |
|---|---|---|---|---|---|
| | hurdle, as many current fabrication methods are labor-intensive and expensive.<br><br>6. Functional Verification: Ensuring the proper functioning of the device after fabrication is essential, which may require specialized equipment and techniques for testing and characterization. | these challenges will require innovative approaches and collaboration across various disciplines. | | | |
| How do you fabricate devices capable of handling multiphase flows? | Fabricating devices for handling multiphase flows involves several steps. First, the design of the device should consider the specific properties and behavior of each phase (liquid, gas, or solid) in the flow. Second, materials suitable for handling these phases must be selected. Third, manufacturing techniques such as 3D printing, casting, | Fabricating devices for handling multiphase flows involves several techniques and considerations. Key approaches include advanced micromanufacturing methods such as 3D printing, which eliminates the need for bonding processes, and flow-focusing techniques, which use nozzles and sheath fluids to control droplet formation. Controlling wettability is crucial for | The response provides a clear and concise overview of the steps involved in fabricating devices capable of handling multiphase flows. It correctly identifies key considerations, materials, and manufacturing techniques, including 3D printing, casting, machining, and assembly. However, it | 46 | 46.0 |

| | machining, or assembly are used to create the device. Lastly, testing and optimization are crucial to ensure the device functions effectively under various multiphase flow conditions. | generating specific droplet types, while parallelization of droplet generators enhances production scalability. Challenges include ensuring even fluid distribution, integrating various components like valves and connectors, and addressing the lack of standardized design rules. Additionally, methods like consecutive capillary assemblies and external forces (e.g., pneumatic or electrical control) can be employed to create complex emulsions. Through careful design and optimization, efficient fabrication of multiphase flow devices for applications such as droplet generation can be achieved. | lacks specific details on advanced micromanufact uring methods such as flow-focusing techniques, controlling wettability, parallelization of droplet generators, and the challenges associated with multiphase flow device fabrication. The response also does not address the importance of standardized design rules or the use of consecutive capillary assemblies and external forces in creating complex emulsions. SCORE: 46 | | |
|---|---|---|---|---|---|

| What is the impact of channel roughness on droplet formation and behavior? | Channel roughness significantly affects droplet formation and behavior. Rough surfaces can induce instabilities in the fluid flow, leading to the breakup of the fluid into droplets. The size, number, and distribution of these droplets can be influenced by the magnitude and spatial pattern of the surface roughness. Additionally, the interaction between the droplets and the rough surface can alter the trajectory and velocity of the droplets, affecting their deformation, coalescence, and evaporation. | The impact of channel roughness on droplet formation and behavior is a key factor in microfluidics. Studies indicate that surface roughness can significantly affect droplet dynamics, such as movement, spreading, and stability. For example, increased roughness can hinder droplet spreading and prevent secondary droplet formation by altering flow dynamics and the contact line. Additionally, vibration and surface structure play roles in influencing droplet impact and splash behavior. Rougher surfaces tend to reduce spreading, while smoother surfaces may enhance it. Surface wettability also interacts with roughness to further | The response provides a clear explanation of how channel roughness affects droplet formation and behavior, including its impact on fluid flow instability, droplet size, distribution, deformation, coalescence, and evaporation. However, it lacks specific details about the interaction between surface wettability and roughness, as well as the effects of vibration and surface structure on droplet dynamics. The response also does not mention the varying effects of roughness level, fluid type, and external conditions on droplet behavior. SCORE: 56 | 56 | 56.000000 00000001 |

| | | complicate droplet behavior. Overall, channel roughness introduces complexity to droplet dynamics, with its effects varying based on factors like roughness level, fluid type, and external conditions, highlighting the need for further research in this area for better understanding and optimization in microfluidic applications. | | | |
|---|---|---|---|---|---|
| How can sensors be integrated into droplet microfluidic devices during fabrication? | Sensors can be integrated into droplet microfluidic devices during fabrication through several methods. One common approach is to incorporate the sensors directly into the device's design, such as using photolithography or 3D printing techniques to create channels and reservoirs for the sensor components. Another method is to use micro- | Integrating sensors into droplet microfluidic devices during fabrication is a complex process that requires advanced microfabrication techniques and careful material selection. For successful integration, precision in manufacturing is crucial, as it often involves combining different microfluidic units, such as droplet | The response provides a clear overview of methods to integrate sensors into droplet microfluidic devices during fabrication, including direct integration, micro-assembly, and modular components. However, it lacks the depth and nuance provided in the ground truth answer regarding the | 56 | 56.000000 00000001 |

| | assembly techniques to place prefabricated sensors onto the device after initial fabrication. Additionally, some devices can be designed with modular components that allow for easy integration of sensors post-fabrication. The specific approach used depends on factors such as the type of sensor, the desired functionality, and the fabrication method employed for the microfluidic device. | generation and detection units, on a single chip. Various techniques like photolithography, soft lithography, and direct integration methods can be used, depending on the type of sensor (e.g., optical or electrochemical). The fabrication process also needs to account for material compatibility, sensor functionality, and alignment to ensure reliable performance. Additionally, sensor integration may face challenges such as maintaining stability within the microfluidic environment and minimizing interference with fluid flow. By considering these factors, sensors can be effectively integrated into droplet microfluidic devices for applications like chemical reactions and | complexity of the process, material selection, and potential challenges. SCORE: 56 | | |

| | | biological detection. | | | |
|---|---|---|---|---|---|
| How are droplet microfluidic devices used for single-cell analysis? | Droplet microfluidic devices are utilized for single-cell analysis by encapsulating individual cells in tiny droplets. This process, known as droplet-based microfluidics, allows for the isolation and manipulation of single cells within a large population. Each cell is contained in its own droplet, which can then be transported, sorted, mixed, or analyzed independently.

These devices are particularly useful for applications such as gene expression analysis, protein quantification, and cell-to-cell variation studies. They enable high-throughput single-cell analysis with minimal sample manipulation, | Droplet microfluidic devices are widely used for single-cell analysis due to their ability to encapsulate individual cells within tiny droplets, allowing for high-throughput and precise experimentation. These devices work by generating microdroplets in a fluid flow, each containing a single cell along with reagents needed for analysis, such as enzymes, antibodies, or primers. This allows for the isolation of cells in a controlled environment, enabling high-resolution studies of gene expression, protein profiling, and cellular responses. The compartmentalized nature of droplets also reduces cross-contamination | The response accurately describes the use of droplet microfluidic devices for single-cell analysis, including encapsulation, isolation, and manipulation of individual cells. It also highlights their applications in gene expression analysis, protein quantification, and cell-to-cell variation studies. However, some minor details are missing compared to the ground truth answer, such as the specific reagents used within the droplets and the technology's application in genomics, proteomics, and drug screening. The response is | 58 | 57.999999 99999999 |

| | | | | | |
|---|---|---|---|---|---|
| | reducing the risk of contamination and ensuring accurate results. Additionally, droplet microfluidics can be combined with other technologies like flow cytometry or next-generation sequencing to provide comprehensive insights into cellular behavior at a single-cell level. | between samples. Moreover, the small size and scalability of droplet microfluidics make it possible to perform large-scale, parallel analysis of thousands of single cells in a short amount of time. This technology has applications in genomics, proteomics, and drug screening, facilitating insights into cell heterogeneity and the identification of rare cell populations. | clear, concise, and relevant but lacks a bit of depth and detail. SCORE: 58 | | |
| What are the advantages of droplet microfluidics for high-throughput screening? | 1. Precision and Reproducibility: Droplet microfluidics allows for the precise control of reaction volumes, typically in the picoliter range, which is crucial for high-throughput screening applications. The consistency of these reactions leads to reproducible results. | Droplet microfluidics offers several advantages for high-throughput screening, including reduced sample volumes, which lowers costs, and increased assay sensitivity through compartmentalization. This method allows for the screening of up to $10^8$ samples per day at kilohertz speeds, | The response accurately addresses the question by listing specific advantages of droplet microfluidics for high-throughput screening. It covers key points such as precision and reproducibility, scalability, cost-effectiveness, safety and containment, | 76 | 76.0 |

| | | | | |
|---|---|---|---|---|
| | 2. Scalability: The ability to generate thousands to millions of droplets per second makes droplet microfluidics highly scalable for large-scale screening experiments.<br><br>3. Cost-Effective: By reducing the amount of reagents required, droplet microfluidics can significantly lower costs associated with high-throughput screening compared to traditional methods.<br><br>4. Safety and Containment: Droplets are physically isolated from each other, minimizing the risk of cross-contamination between reactions. This is particularly beneficial when working with hazardous or | significantly improving throughput and efficiency. The small droplet volumes enhance effective concentrations and signal-to-noise ratios, critical for sensitive detection of low-concentration biological molecules such as antibodies, enzymes, and single genomes. Additionally, droplet microfluidics enables single-cell and single-molecule assays, providing detailed insights into cellular behaviors and allowing for novel experimental approaches. This technology is particularly beneficial for applications like drug discovery, enzyme evolution, and diagnostics, offering a versatile and cost-effective platform for advanced | and flexibility. However, it does not fully capture the depth and breadth of benefits mentioned in the ground truth answer, particularly regarding assay sensitivity, effective concentrations, signal-to-noise ratios, single-cell and single-molecule assays, and its applications in drug discovery, enzyme evolution, and diagnostics.<br><br>SCORE: 76 | | |

| | precious materials. | biological research. | | | |
|---|---|---|---|---|---|
| | 5. Flexibility: Droplet microfluidics can be easily adapted to a wide range of applications, including cell biology, chemistry, and biochemistry, making it a versatile tool for high-throughput screening. | | | | |
| How can droplets be used to encapsulate cells or biomolecules? | Droplets, specifically microdroplets, can be used for encapsulation of cells or biomolecules through a process called droplet digital PCR (ddPCR) or microfluidics. In these methods, tiny water droplets containing a single cell or molecule are generated and isolated in oil droplets using hydrodynamic focusing. This allows for precise manipulation and analysis of individual cells | Droplet microfluidics provides versatile methods for encapsulating cells and biomolecules. Limiting dilution is a traditional approach that randomly distributes single cells or biomolecules in droplets, though it has low efficiency. Inertial microfluidics leverages spiral geometries and inertial forces to guide cells into droplets, improving encapsulation efficiency by ensuring single- | The response provides a clear explanation of how droplets can be used to encapsulate cells or biomolecules, mentioning specific methods such as droplet digital PCR (ddPCR) and microfluidics. However, it lacks the comprehensive detail and variety of techniques provided in the Ground Truth Answer, particularly regarding different | 46 | 46.0 |

| | or biomolecules at a very small scale. | cell droplets. Bi-disperse emulsions create an emulsion of empty and cell-containing droplets, with the latter selectively recovered using methods like pinched flow fractionation. For biomolecules, limiting dilution is commonly used, adjusting concentration and drop size to achieve single-molecule encapsulation. Additionally, droplets can encapsulate cells in lipid vesicles for biological applications or serve as microreactors in processes like DNA analysis. These techniques enable precise control over encapsulation, enhancing their applicability in cell-based assays, drug discovery, and genetic research | approaches for cell and biomolecule encapsulation. The response also does not mention some key concepts like bi-disperse emulsions or pinched flow fractionation. SCORE: 46 | | |
|---|---|---|---|---|---|
| What are the applications of droplet microfluidics in drug discovery? | Droplet microfluidics has several applications in drug discovery, | Droplet microfluidics plays a significant role in drug discovery by | The response is clear and concise, accurately addressing the | 76 | 76.0 |

| including high-throughput screening of potential drugs, optimization of drug formulations, and studying the behavior of cells exposed to drugs. This technology allows for precise control over the volume and composition of each droplet, enabling the creation of thousands of miniature reactors containing a single cell or a few cells. This makes it possible to test many different compounds on individual cells quickly and efficiently, which can significantly reduce the time and cost associated with drug discovery. Additionally, droplet microfluidics can be used to study the effects of drugs on cells in a more physiologically | enabling high-throughput screening and precise analysis of drug interactions at the single-cell level. This technology allows the generation of picoliter-sized droplets, each encapsulating a single cell or a small group of cells, which can be exposed to various drug compounds. This compartmentalization facilitates parallel testing of numerous drug candidates, speeding up the identification of promising therapeutics. Droplet microfluidics is particularly useful in identifying drug efficacy, optimizing drug formulations, and studying complex cellular responses such as gene expression, protein production, or metabolic changes in response to treatment. | question about the applications of droplet microfluidics in drug discovery. It provides relevant details on high-throughput screening, optimization of drug formulations, and studying cell behavior. However, it lacks some depth and specific examples compared to the ground truth answer, which provides more comprehensive information on the technology's capabilities and its advantages over traditional methods. SCORE: 76 | | |
|---|---|---|---|---|

| | relevant environment than traditional cell culture methods. | Additionally, it enables the screening of rare cell populations or targets that would be difficult to isolate using traditional methods. The ability to rapidly test and analyse thousands of drug interactions in a controlled environment accelerates drug discovery, reduces costs, and improves the chances of identifying novel therapeutic agents. | | | |
|---|---|---|---|---|---|
| How can droplet microfluidics enable precise control of chemical reactions? | Droplet microfluidics allows for precise control of chemical reactions by confining and manipulating small volumes (typically picoliters) of reactants within individual droplets. This confinement reduces the reaction volume, thereby minimizing the amount of reagents required and reducing the | Droplet microfluidics offers precise control over chemical reactions by dispensing reactants into individual droplets with accurate volume control. This enables the mixing of reactants within confined spaces, enhancing reaction efficiency. Coalesced droplets act as micro-reactors, | The response accurately addresses the question by providing detailed information on how droplet microfluidics enables precise control of chemical reactions. It highlights the benefits of confinement, rapid mixing, and parallel processing in microfluidic channels. However, it | 85 | 85.0 |

|  |  |  |  |  |
|---|---|---|---|---|
|  | likelihood of unwanted side reactions or contamination. Additionally, the continuous flow of reactants through the microfluidic channels enables rapid mixing and reaction kinetics, which can be further controlled by adjusting the flow rates of the reactants. Furthermore, droplet-based microfluidics allows for parallel processing of multiple reactions within a single device, increasing throughput and reducing experimental time. | providing fast mixing and long operation times without clogging, even with sticky or precipitating reactants. The system allows for optimization of reaction parameters, such as temperature, flow rate, and pressure, ensuring the production of specific products, like mesoporous silica microspheres with superior surface area and narrow pore distributions. This capability to finely control reaction conditions makes droplet microfluidics a powerful tool for precise, high-throughput chemical synthesis. | could be improved with more emphasis on the specific advantages of droplet-based systems over traditional methods and a clearer connection to the ground truth answer's points about optimization of reaction parameters and production of specific products. SCORE: 85 |  |  |
| What are the key challenges of using droplet microfluidics in diagnostics? | 1. Scalability: Droplet microfluidics require precise control over small volumes, which can be challenging to scale up for high-throughput applications in | Droplet microfluidics presents significant challenges in diagnostics, including preventing wetting and cross-contamination | The response provides a clear and concise list of challenges associated with using droplet microfluidics in diagnostics. However, it does not fully | 41 | 41.0 |

| | | |
|---|---|---|
| diagnostics.<br><br>2. Integration with readout systems: The small size of droplets and the need for sensitive detection methods pose challenges when integrating droplet microfluidics with existing diagnostic platforms.<br><br>3. Cost: The fabrication of microfluidic devices can be expensive, which may limit their widespread adoption in resource-limited settings.<br><br>4. Complexity: The design and operation of droplet microfluidics systems are complex, requiring specialized expertise for development and maintenance.<br><br>5. Contamination: The small size | between aqueous droplets and immiscible organic liquids, which can compromise results. Monitoring oxygen levels within droplets is difficult due to the sensitivity of the liquid-liquid interface to mechanical perturbations and mass transfer. Developing stable detection chemistries is also complex, as the confined droplet environment can alter chemical reactions and impact accuracy. Additionally, challenges in automation and accessibility limit the widespread use of droplet microfluidics, particularly for high-throughput assays. Overcoming these issues requires advancements in surfactants, microfluidic engineering, and automated techniques, which | capture the nuances and complexities mentioned in the ground truth answer, such as preventing wetting and cross-contamination, monitoring oxygen levels within droplets, developing stable detection chemistries, and overcoming issues in automation and accessibility. The response seems to focus more on general challenges rather than the specific technical difficulties highlighted in the ground truth answer. SCORE: 41 | | |
| | and high sensitivity of the system make it susceptible to contamination, which can affect the accuracy of results.<br><br>6. Regulatory approval: As with any new technology, regulatory approval for use in diagnostics can be a lengthy and costly process.<br><br>7. Standardization: There is currently no standardized protocol or platform for droplet microfluidics, making it difficult to compare results between different labs and systems. | could enhance the precision and applicability of droplet systems in diagnostics. | | |
| How are droplets used for DNA sequencing applications? | Droplets are used in DNA sequencing applications through a method called Digital PCR (dPCR) or Droplet Digital PCR (ddPCR). This technology | Droplet microfluidics plays a key role in DNA sequencing by enabling high-throughput analysis and precise targeting of genomic regions. This | The response accurately describes the use of droplets in DNA sequencing applications through Digital PCR (dPCR) or Droplet Digital PCR | 56 | 56.000000 00000001 |

| | | | | |
|---|---|---|---|---|
| | allows for the simultaneous analysis of thousands of individual reactions, each containing a single molecule of DNA.<br><br>In this process, a sample is partitioned into tens of thousands of nanoliter-sized droplets using oil-in-water emulsion or microfluidic technology. Each droplet contains either zero or one target molecule (e.g., DNA). The droplets are then subjected to PCR amplification, and the presence or absence of the target molecule is determined by measuring the fluorescence intensity in each droplet.<br><br>The advantage of using droplets for DNA sequencing applications is that it provides a more accurate | includes single-cell sequencing, where individual cells are encapsulated in droplets for independent gene expression profiling. Barcode-based sequencing allows for the labeling of nucleic acids with unique identifiers, enabling the tracking of specific genomes during analysis. Multiplexed PCR amplification is achieved by performing PCR in separate droplets, ensuring uniform amplification of multiple targets. Additionally, emulsion PCR combines droplet merger with PCR to amplify specific genomic regions, facilitating efficient sequencing. These techniques offer a miniaturized, efficient platform for analyzing complex genetic information, | (ddPCR), highlighting its advantages over traditional bulk PCR methods. However, it lacks detailed information on single-cell sequencing, barcode-based sequencing, multiplexed PCR amplification, and emulsion PCR, which are key aspects of droplet microfluidics in DNA sequencing as per the ground truth answer. SCORE: 56 | | |
| | quantification of the target molecules compared to traditional bulk PCR methods. Additionally, it reduces the risk of cross-contamination between samples due to the physical separation of reactions into individual droplets. | improving accuracy and throughput in genomic studies. | | | |
| What role do droplets play in tissue engineering research? | Droplets are used extensively in tissue engineering research for the production of biomimetic micro- and nanostructures. These structures can mimic the extracellular matrix found in native tissues, providing a suitable environment for cell growth and differentiation. The controlled fabrication of these structures using droplet-based microfluidics allows for precise control over the composition, | Droplet microfluidics plays a pivotal role in tissue engineering by enabling precise fabrication of biomaterials and 3D cell culture systems. Droplets serve as templates for creating hydrogel-based materials, which can encapsulate cells for more accurate modeling of tissue structures. This technology facilitates the production of microcarriers and complex biomaterials, such as hollow bacterial cellulose (BC) | The response accurately conveys the role of droplets in tissue engineering research, highlighting their use in creating biomimetic micro- and nanostructures, controlled fabrication using droplet-based microfluidics, and precise control over scaffold composition, size, and spatial arrangement. However, it lacks specific details about the production | 76 | 76.0 |

| | size, and spatial arrangement of the scaffolds, which is crucial for tissue engineering applications. | microspheres, which form injectable scaffolds for wound healing, promoting high cell proliferation and tissue regeneration in vivo. Additionally, droplet-based microfluidics allows for the creation of core-shell structures, such as alginate shells or metal-organic frameworks, which offer controlled release of therapeutic agents. These advancements provide a powerful platform for studying tissue development and regenerative therapies. | of hydrogel-based materials, microcarriers, core-shell structures, and their applications in tissue development and regenerative therapies mentioned in the ground truth answer. SCORE: 76 | | |
|---|---|---|---|---|---|
| How can droplet microfluidics aid in studying cell signaling pathways? | Droplet microfluidics can aid in studying cell signaling pathways by providing a controlled environment for single-cell analysis. This technology allows researchers to | Droplet microfluidics offers a powerful platform for studying cell signaling pathways by enabling the precise isolation and manipulation of individual cells within controlled environments. This technology | The response demonstrates a good understanding of the topic and provides relevant information about how droplet microfluidics can aid in studying cell signaling | 52 | 52.0 |

| | encapsulate individual cells within micron-sized droplets, which can be manipulated and analyzed independently. By isolating cells from each other, it is possible to study the effects of specific signaling molecules on cell behavior without interference from neighboring cells. Additionally, the ability to precisely control the composition of each droplet allows for the investigation of various signaling pathways under different conditions. This level of control and precision can lead to a deeper understanding of cellular processes and potential therapeutic targets. | allows for the encapsulation of single cells in droplets, providing a controlled space to study cell-cell interactions and signaling events. The ability to merge droplets containing different cell types or reagents facilitates the investigation of cell communication, such as paracrine signaling, in a way that mimics in vivo conditions. Additionally, droplet-based microfluidics supports high-throughput screening and single-cell genomic studies, offering insights into phenotypic variations and signaling dynamics within genetically identical cell populations. This capability is particularly valuable for understanding complex cellular behaviors in development, | pathways. However, it lacks some depth and specific details compared to the ground truth answer. The language is clear, but there are minor inaccuracies and omissions in comparison to the expected level of detail. SCORE: 52 | | |
|---|---|---|---|---|---|

| | | differentiation, and disease. | | | |
|---|---|---|---|---|---|
| What is the potential of droplet microfluidics in personalized medicine? | Droplet microfluidics has a significant potential in personalized medicine due to its ability to perform high-throughput, label-free, and cost-effective analysis of individual cells or biomolecules. This technology allows for the creation of thousands of tiny droplets containing a single cell or biomolecule, which can be manipulated, analyzed, and sorted based on their specific characteristics.<br><br>In personalized medicine, this technology could be used to develop diagnostic tests tailored to an individual's genetic makeup, as well as for the development of targeted therapies based on the unique molecular profile | Droplet microfluidics holds great promise in personalized medicine by enabling precise, high-throughput analyses of biological samples. The technology allows for the optimization of treatments by testing drug combinations on specific cell lines, such as pancreatic cancer cells, to identify the most effective therapies. It also facilitates the early detection of cancers through highly sensitive detection of tumor markers in bodily fluids, offering a minimally invasive approach for patient monitoring. Additionally, droplet-based digital PCR (dPCR) enables the detection of rare sequences and precise quantification of | The response accurately captures the potential of droplet microfluidics in personalized medicine, highlighting its ability to perform high-throughput analysis, develop targeted therapies, and enable precise diagnosis. However, it lacks specific examples, such as pancreatic cancer cells or tumor markers, which are mentioned in the ground truth answer. SCORE: 85 | 85 | 85.0 |
| | of a patient's cells or tissues. Additionally, droplet microfluidics can help in drug screening and toxicity testing, which are crucial steps in the development of personalized treatments.<br><br>Overall, droplet microfluidics has the potential to revolutionize the field of personalized medicine by enabling more precise, efficient, and cost-effective diagnosis and treatment of diseases. | nucleic acids, which is crucial for cancer research and tailoring treatments based on a patient's genetic background and tumor profile. Overall, droplet microfluidics can significantly advance personalized medicine by providing tailored treatment strategies and improving early detection for better patient outcomes. | | | |
| How are droplet microfluidic devices used for nanoparticle synthesis? | Droplet microfluidic devices are utilized in nanoparticle synthesis due to their ability to create monodisperse droplets, which ensures uniformity in the size and composition of the particles produced. The process typically involves mixing | Droplet microfluidic devices revolutionize nanoparticle synthesis by enabling precise control over reaction conditions within isolated microreactors. These droplets encapsulate well-defined quantities of reagents, ensuring rapid mixing and | The response accurately describes the use of droplet microfluidic devices in nanoparticle synthesis, highlighting their ability to create uniform particles. However, it lacks some details present in the ground truth answer, such as the | 76 | 76.0 |

| | reactants inside these droplets, followed by the addition of energy (either thermal or electrical) to initiate the chemical reaction that leads to nanoparticle formation. This method allows for precise control over the particle size, shape, and composition, making it a valuable tool in nanotechnology research and applications. | preventing contamination, even when using toxic or volatile chemicals. This approach addresses limitations like diffusion-limited mixing found in traditional systems, resulting in nanoparticles with controlled size, shape, and narrow size distribution. The method supports multi-step synthesis on millisecond timescales, allowing for the scalable production of high-quality nanoparticles with applications in fields such as drug delivery, imaging, and catalysis. | specific advantages over traditional systems and the potential applications. SCORE: 76 | | |
| --- | --- | --- | --- | --- | --- |
| What role do droplets play in polymer particle production? | Droplets play a crucial role in the production of polymer particles. In this process, also known as emulsion polymerization, monomers are dispersed in water using surfactants to form stable | Droplet-based microfluidics provides a versatile platform for producing monodisperse polymer particles by encapsulating monomers or oligomers into droplets and solidifying them through chemical or physical | The response provides a clear explanation of the role of droplets in polymer particle production, but it lacks detail and nuance compared to the ground truth answer. It | 46 | 46.0 |

| | droplets. The monomers inside these droplets then undergo polymerization, resulting in the formation of polymer particles or latex particles. These particles can be further processed into various products such as paints, adhesives, and coatings. | crosslinking. This approach enables the creation of diverse particle types, including solid, hollow, asymmetric, and multicompartmental structures with precise control over size, shape, and morphology. Techniques such as UV-initiated polymerization and solvent evaporation facilitate the transformation of droplets into solid particles. Additionally, functional micro- or nanostructures can be incorporated within or on the surface of these particles by introducing specific ingredients into the droplets, expanding their potential applications in areas like drug delivery, tissue engineering, and advanced materials. | also does not mention some key aspects such as the diversity of particle types that can be produced or the incorporation of functional micro- or nanostructures within or on the surface of these particles. SCORE: 46 | | |
|---|---|---|---|---|---|
| How can droplet microfluidics improve the reproducibility of | Droplet microfluidics can significantly enhance the reproducibility | Droplet microfluidics enhances the reproducibility of chemical | The response is clear and concise, accurately addressing the | 85 | 85.0 |

| chemical synthesis? | of chemical synthesis by providing precise control over reaction conditions, such as temperature, pressure, and mixing. This is achieved through the continuous flow of reactants in small, monodisperse droplets that are generated within a microchannel network. The small volume of each droplet ensures uniformity of reaction conditions, minimizing variations that can occur in traditional batch reactions. Additionally, the ability to perform multiple parallel reactions within a single device further increases reproducibility and throughput. | synthesis by providing precise control over reaction conditions and minimizing variability. Droplets act as isolated microreactors, ensuring accurate manipulation of parameters such as temperature and concentration, critical for consistent results. This system enables optimal mixing intensity and parallel processing in separate droplet channels, reducing batch-to-batch variability and ensuring uniform product quality. Automation further minimizes human error, making the process highly reliable. These advantages are particularly beneficial for synthesizing complex nanomaterials and exploring new reaction parameter spaces, leading to the | question about how droplet microfluidics improves the reproducibility of chemical synthesis. It provides specific examples of the benefits of droplet microfluidics, such as precise control over reaction conditions and uniformity of reaction conditions. However, it could be improved by providing more detail on the automation aspect and its impact on human error, which is highlighted in the ground truth answer. SCORE: 85 | | |
|---|---|---|---|---|---|
| | | production of high-quality, reproducible materials. | | | |
| What are the applications of droplet microfluidics in catalysis studies? | Droplet microfluidics has several applications in catalysis studies, including:<br><br>1. High-throughput screening of catalysts: The small reaction volumes and precise control over the reaction conditions allow for rapid testing of a large number of catalyst candidates.<br><br>2. Studying heterogeneous catalysis: Droplet microfluidics can be used to encapsulate catalyst particles within droplets, allowing for the study of their behavior in a controlled environment.<br><br>3. Investigating homogeneous catalysis: By creating monodisperse | Droplet microfluidics offers transformative applications in catalysis studies by enabling precise, efficient, and high-throughput investigations at the nanoscale. These systems facilitate enantioselective biocatalysis using nanoliter-scale devices that integrate microflow reactors and electrophoresis for enantiomeric differentiation. Droplets serve as isolated reaction containers, allowing for controlled catalytic transformations at the single-particle or cell level, with reaction volumes as small as a few picoliters. This confinement minimizes diffusion effects and enhances reaction | The response provides a clear and concise list of specific applications of droplet microfluidics in catalysis studies. However, it lacks the depth and detail provided by the ground truth answer, particularly in terms of the transformative impact on catalysis research and the integration with other technologies such as electrophoresis. SCORE: 51 | 51 | 51.0 |
| | droplets containing reactants and catalysts, it is possible to study the kinetics and mechanisms of homogeneous catalytic reactions.

4. Catalyst optimization: The ability to easily vary reaction conditions such as temperature, pressure, and composition allows for the optimization of catalyst performance.

5. Studying catalyst deactivation and regeneration: Droplet microfluidics can be used to simulate the aging and deactivation of catalysts under various conditions, as well as investigate methods for their regeneration or replacement. | efficiency. Applications span organic synthesis, biochemical analysis, and biotechnology, showcasing droplet microfluidics as a powerful tool for studying rapid reaction dynamics, optimizing catalysts, and exploring reaction mechanisms in real time. | | | |
| How do droplets enhance the study | Droplets, particularly microdroplets, | Droplet-based microfluidics revolutionizes | The response provides a clear and | 76 | 76.0 |

| of reaction kinetics? | are useful in the study of reaction kinetics because they allow for rapid mixing and high surface-to-volume ratio. This means that reactions can be initiated and monitored quickly, and the small size of the droplets ensures a homogeneous mixture, which is crucial for accurate kinetic measurements. Additionally, the ability to generate and manipulate droplets on demand allows for precise control over reaction conditions such as temperature, pressure, and composition. This makes droplet-based systems an effective tool for studying a wide range of chemical reactions, from simple ones to complex biochemical processes. | reaction kinetics studies by offering rapid, precise, and high-throughput analysis with minimal sample consumption. The chaotic mixing within picoliter droplets ensures millisecond-level time resolution for fast reactions. These systems minimize reagent use, making them ideal for bioanalysis where resources are costly or scarce. Precise control over droplet composition facilitates the creation of concentration gradients, enabling the measurement of kinetic parameters. High-throughput capabilities allow simultaneous monitoring of numerous reactions, significantly expanding experimental diversity. Additionally, extended tracking of individual droplets supports | concise explanation of how droplets enhance the study of reaction kinetics, highlighting their benefits in terms of rapid mixing, high surface-to-volume ratio, and precise control over reaction conditions. However, it lacks some details present in the ground truth answer, such as the specific mention of millisecond-level time resolution, concentration gradients, and long-term monitoring of enzyme-inhibitor kinetics. The response also uses slightly more general language compared to the ground truth answer.

SCORE: 76 | | |

| | | long-term monitoring of enzyme-inhibitor kinetics, providing deeper insights into reaction mechanisms. | | | |
|---|---|---|---|---|---|
| What are the benefits of using droplets for emulsion polymerization? | Droplets in emulsion polymerization offer several advantages. They serve as the sites of polymerization, allowing for a more controlled and uniform particle size distribution. This is beneficial because it can lead to improved physical properties of the final product, such as better mechanical strength and rheological behavior. Additionally, droplets can be stabilized using surfactants, which can also act as initiators for the polymerization process, further enhancing control over the reaction. Lastly, the use of droplets allows | Droplet-based systems offer significant advantages for emulsion polymerization, enhancing efficiency, precision, and scalability. They enable high production rates and precise control over polymer particle size and morphology, facilitating the synthesis of high-molar-mass polymers at faster reaction rates. Droplet parallelization allows the screening of diverse chemicals or concentrations and supports the encapsulation of millions of gene variants for quantitative emulsion PCR. These systems also empower directed enzyme or microorganism | The response accurately conveys the benefits of using droplets for emulsion polymerization, including controlled particle size distribution, improved physical properties, and ease of processing. However, it lacks depth and specificity compared to the ground truth answer, which highlights additional advantages such as high production rates, precise control over polymer particle size and morphology, and applications in directed enzyme or | 56 | 56.000000 00000001 |

| | for the production of latex particles, which are water-dispersible and can be easily processed into a variety of products. | evolution through successive mutagenesis and selection rounds. Furthermore, droplet isolation prevents cross-contamination, adsorption, and solvent evaporation, making them ideal for applications like in vitro compartmentaliza tion (IVC). This unique platform provides a robust, scalable approach for both analytical and biological studies. | microorganism evolution. SCORE: 56 | | |
|---|---|---|---|---|---|
| How can microfluidics be used to generate smart materials? | Microfluidics, a technology that manipulates fluids at the sub-millimeter scale, can be utilized to generate smart materials by creating controlled environments for chemical reactions. The small volumes and precise control over flow rates, temperature, and pressure in microfluidic systems enable the synthesis of | Microfluidics enables the creation of smart materials through precise control of reaction conditions, rapid synthesis, and innovative templating methods. Droplet templating allows monodisperse droplets to act as structural templates, facilitating the synthesis of uniform particles, such as hydrogels for single-cell analysis. By | The response provides a clear explanation of how microfluidics can be used to generate smart materials, mentioning specific examples such as photonic crystals and shape-memory polymers. However, it lacks some details about the precise control of reaction | 61 | 61.0 |

| | | | | |
|---|---|---|---|---|
| | complex structures and properties within these materials.<br><br>For instance, microfluidics can facilitate the production of photonic crystals, which are smart materials with unique optical properties due to their periodic structure. Additionally, it can be used for the creation of shape-memory polymers, liquid crystals, and other responsive materials that change their physical state in response to external stimuli such as temperature, light, or electric fields.<br><br>By leveraging the advantages of microfluidics, researchers can design and fabricate smart materials with tailored properties for various applications, including energy | manipulating solute and temperature gradients on short timescales, microfluidics supports the production of nanoscale materials, including semiconducting polymers, catalysts, and nanoparticles with unique optical and electronic properties. Its ability to rapidly explore complex reaction parameter spaces, coupled with advanced analytics and machine learning, accelerates material discovery and optimization. This approach offers transformative potential in biotechnology, energy, and electronics by enabling the generation of materials with tailored properties and enhanced functionality. | conditions, rapid synthesis, and innovative templating methods that are mentioned in the ground truth answer.<br>SCORE: 61 | | |

| | harvesting, sensing, drug delivery, and biomedical devices. | | | | |
|---|---|---|---|---|---|
| What are the applications of droplet microfluidics in environmental sensing? | Droplet microfluidics has several applications in environmental sensing, particularly for the detection and analysis of pollutants, pathogens, and biomarkers. Here are some examples:<br><br>1. Water Quality Monitoring: Droplet microfluidics can be used to rapidly analyze water samples for contaminants such as heavy metals, pesticides, or bacteria. This is particularly useful in real-time monitoring of water resources.<br><br>2. Air Pollution Analysis: The technology can also be applied to detect and quantify air pollutants like volatile organic | Droplet microfluidics offers transformative applications in environmental sensing by enabling precise, miniaturized, and efficient monitoring systems. It supports on-site environmental surveillance through portable devices for rapid toxicity assessments and detection of contaminants, such as mercury ions. As controllable microreactors, droplet microfluidics facilitates the study of complex interfacial processes, aiding in chemical synthesis and degradation studies. Furthermore, it advances environmental sensing by enabling in situ visualization and | The response provides some relevant information about the applications of droplet microfluidics in environmental sensing, but it lacks depth and clarity compared to the ground truth answer. It also includes examples that are not explicitly mentioned in the ground truth answer. SCORE: 46 | 46 | 46.0 |

| | compounds (VOCs) or particulate matter. | precise flow control, enhancing our understanding of environmental processes. These capabilities position droplet microfluidics as a valuable tool for environmental science and engineering. | | | |
|---|---|---|---|---|---|
| | 3. Biological Monitoring: Droplet microfluidics can help in the detection and identification of pathogens, such as bacteria or viruses, in environmental samples. This is crucial for monitoring potential health risks associated with water or air pollution. | | | | |
| | 4. Bio-monitoring of Soil and Sediment: The technology can be used to analyze soil and sediment samples for contaminants, biomarkers, or microbial communities, providing insights into the environmental impact of industrial activities or natural processes. | | | | |

| How can microfluidic droplets improve combinatorial chemistry? | Microfluidic droplets can significantly enhance combinatorial chemistry by allowing for the precise and independent encapsulation, mixing, and reaction of multiple reagents in a small volume. This miniaturization enables high-throughput screening of chemical reactions, reducing waste and increasing efficiency. Additionally, microfluidics allows for better control over reaction conditions such as temperature, pressure, and mixing rates, leading to more consistent results and faster discovery of new compounds. | Microfluidic droplets revolutionize combinatorial chemistry by drastically reducing reagent consumption and enhancing efficiency. Each droplet, acting as a picoliter-scale microreactor, requires six to eight orders of magnitude less material than traditional methods, minimizing waste and costs. This miniaturization enables high-throughput synthesis of large compound libraries, supporting applications like drug discovery and material synthesis. Additionally, droplet manipulation techniques allow for precise reaction control, combinatorial fusion of reagents, and multi-stage experiments with online analysis. These advantages make | The response accurately conveys the benefits of microfluidic droplets in combinatorial chemistry, including miniaturization, high-throughput screening, and precise control over reaction conditions. However, it lacks some details mentioned in the Ground Truth Answer, such as the specific reduction in reagent consumption (six to eight orders of magnitude) and the ability for multi-stage experiments with online analysis. SCORE: 76 | 76 | 76.0 |

| | | microfluidic droplets a powerful tool for automating complex chemical workflows and optimizing reaction conditions in a highly efficient and scalable manner. | | | |
|---|---|---|---|---|---|
| What are the challenges of scaling up droplet-based chemical synthesis? | Scaling up droplet-based chemical synthesis presents several challenges, including:<br><br>1. Control over reaction conditions: Maintaining consistent reaction conditions across a large number of droplets can be difficult due to variations in temperature, pressure, and mixing.<br><br>2. Mass transfer limitations: As the scale increases, ensuring efficient mass transfer between the droplet's interior and exterior becomes | Scaling up droplet-based chemical synthesis faces several challenges. Achieving uniform fluid distribution in microfluidic devices remains difficult despite progress in parallel droplet generation, with 3D printing offering potential solutions for complex channel fabrication. Simplifying functional units for multistep reactions, such as droplet generation, merging, and phase separation, is vital for seamless continuous flow processes. Developing low- | The response provides a clear and concise list of specific challenges associated with scaling up droplet-based chemical synthesis, including control over reaction conditions, mass transfer limitations, equipment scaling, product isolation, cost-effectiveness, and safety concerns. However, the answer lacks depth and detail in addressing these challenges, and some points seem to | 41 | 41.0 |

| | more challenging, potentially leading to incomplete reactions or product contamination.

3. Scaling up equipment: Developing large-scale equipment that can handle thousands of droplets simultaneously while maintaining precise control over each droplet is a significant engineering challenge.

4. Product isolation and purification: Separating the desired product from the reaction mixture and other byproducts can be difficult, especially when dealing with small quantities in each droplet.

5. Cost-effectiveness: Scaling up the process to be cost-effective | cost materials and standardizing chip designs are necessary for commercial viability, especially for high-throughput production in diverse fluids. Furthermore, creating complex droplet emulsions and expanding nanomaterial diversity are critical to meeting industrial demands. Addressing these challenges will enable safe, sustainable, and scalable microfluidic applications. | be more relevant than others. The response also does not fully align with the ground truth answer, which provides a more comprehensive overview of the challenges and potential solutions. SCORE: 41 | | |
|---|---|---|---|---|
| | requires optimizing the use of resources such as energy, solvents, and reagents while maintaining high yields and product quality.<br><br>6. Safety and environmental concerns: Handling large numbers of potentially hazardous chemicals and managing waste products can pose safety risks and contribute to environmental pollution if not properly managed. | | | |
| How can artificial intelligence enhance droplet microfluidic research? | Artificial Intelligence (AI) can significantly enhance droplet microfluidic research in several ways. First, AI algorithms can be used for precise control and automation of the fluidic system, ensuring consistent and reproducible results. Second, machine learning techniques can analyze large | Machine learning can be used to optimize the design of microfluidic devices. Deep Learning will be integrated to sort, monitor, and process the images. Furthermore, the multimodal agents can be used to study and design the research of droplet microfluidics, a possible excuse | The response provides some relevant information about how AI can enhance droplet microfluidic research, but it lacks depth and accuracy compared to the Ground Truth Answer. It mentions three potential applications of AI in this field, but they are not as | 41 | 41.0 |

| | datasets generated from these experiments to identify patterns, trends, and correlations that might not be apparent through manual analysis. Lastly, AI can aid in the optimization of experimental conditions, such as flow rates, pressures, and temperatures, by predicting the optimal settings for specific applications based on historical data. | robot to perform fabrication and experiments automatically. | comprehensive or specific as those mentioned in the Ground Truth Answer. SCORE: 41 | | |
|---|---|---|---|---|---|
| How can large language models enhance droplet microfluidic research? | Large language models, such as myself, can assist in enhancing droplet microfluidic research by providing insights and explanations about complex concepts, helping to interpret experimental results, suggesting new methodologies or techniques, and even generating | LLMs can be used to build the Agents and Multi-Agents to synthesize knowledge of droplet microfluidics from literature such as research articles and books. Then, the idea hypotheses will be proposed by the agents, and the novelty and feasibility will be assessed. The agents will provide the autonomous platform to | The response provides a clear and concise overview of how large language models can assist in enhancing droplet microfluidic research, including providing insights, interpreting results, suggesting methodologies, generating hypotheses, | 55 | 55.000000 00000001 |

| | hypotheses for further investigation. They can also help in writing research papers, summarizing findings, and communicating results effectively to a broader audience. Furthermore, they can be used as educational tools to teach students and researchers about the principles of droplet microfluidics. | design and fabricate droplet microfluidic devices. Furthermore, LLMs have the researchers accelerate knowledge of droplet microfluidics. | and educational purposes. However, it does not fully capture the potential of LLMs to build autonomous platforms for designing and fabricating devices or accelerating researchers' knowledge as mentioned in the ground truth answer. SCORE: 55 | | |
| How can autonomous droplet microfluidic research? | Autonomous droplet microfluidic research can be conducted by utilizing automated systems that control the movement, manipulation, and analysis of droplets within a microfluidic channel. This can involve using sensors, machine learning algorithms, and feedback mechanisms to perform tasks such as mixing, | Autonomous droplet microfluidics is an emerging field with significant potential for scientific discovery, aiming to create highly integrated, automated systems for bio/chemical functions like PCR, sample preparation, and detection. Key challenges include the need for fast detection methods to probe small volume droplets, precise | The response attempts to address the question but lacks depth and clarity in its explanation of autonomous droplet microfluidic research. It touches on some relevant points such as automation and machine learning algorithms, but fails to provide a comprehensive overview of the field and | 31 | 31.0 |

| | sorting, and analyzing biological samples with minimal human intervention. The goal is to improve efficiency, accuracy, and throughput in microfluidic research while reducing the potential for errors and contamination. | control over droplet dynamics in complex microchannel networks, and the development of virtual design platforms for optimizing system designs. Solutions under exploration include using advanced detection techniques such as single-molecule detection and fluorescence lifetime imaging, as well as integrating microelectronic actuators for droplet manipulation. Additionally, the development of virtual design and experiment platforms could streamline the creation of optimized microfluidic systems. The potential applications are vast, including directed evolution, closed-loop synthesis, and automated drug discovery, where systems | its challenges. SCORE: 31 | | |
|---|---|---|---|---|---|
| | | could autonomously optimize compound combinations. Collaboration across disciplines like fluid dynamics, chemistry, and software engineering will be key to overcoming challenges and advancing this technology, paving the way for groundbreaking scientific advances. | | | |
| What is the potential for fully integrated droplet microfluidic platforms? | Fully integrated droplet microfluidic platforms have significant potential in various fields, particularly in biology and chemistry. They offer advantages such as high throughput, precise control over reaction conditions, reduced reagent consumption, and improved safety compared to traditional laboratory methods. These platforms can be | Droplet microfluidics offers transformative solutions to global health challenges by enabling rapid, cost-effective, and high-throughput analysis of biological samples. By generating and manipulating tiny droplets as individual reaction chambers, this technology can revolutionize diagnostics, particularly in | The response provides a general overview of the potential of fully integrated droplet microfluidic platforms but lacks the depth and specificity of the ground truth answer. It mentions various applications such as single-cell analysis, drug discovery, and diagnostics, but does not highlight the transformative | 46 | 46.0 |

| | | | | |
|---|---|---|---|---|
| | used for single-cell analysis, drug discovery, diagnostics, and more. However, challenges remain in terms of scalability, cost, and ease of use, which are areas of ongoing research and development. | resource-limited settings. For example, it facilitates point-of-care testing for infectious diseases like tuberculosis, malaria, and COVID-19, providing results quickly and accurately. Droplet microfluidics is also crucial for multiplexed screening, enabling simultaneous detection of multiple biomarkers, which enhances early disease detection and monitoring. Additionally, it supports the development of low-cost diagnostic devices, reducing dependence on expensive laboratory infrastructure. In drug discovery, it accelerates the screening of therapeutic compounds, improving the treatment of diseases. The technology also | impact on global health challenges or the specific benefits for resource-limited settings. The response also fails to emphasize the technology's potential for point-of-care testing, multiplexed screening, and personalized medicine. SCORE: 46 | | |
| | | advances personalized medicine by analysing single cells and their responses to treatments. Overall, droplet microfluidics addresses global health disparities by delivering accessible, scalable, and efficient solutions for disease diagnosis, monitoring, and treatment. | | | |
| How can droplet microfluidics contribute to global health challenges? | Droplet microfluidics can significantly contribute to global health challenges in several ways. Firstly, it allows for the miniaturization and standardization of biological assays, making them faster, cheaper, and more accessible. This is particularly important for resource-limited settings where diagnostic tools are often lacking.<br><br>Secondly, | Droplet microfluidics has significant potential to address global health challenges by enhancing diagnostic and therapeutic processes. It enables rapid and sensitive pathogen detection, which is critical for early disease diagnosis and containment. The technology's high-throughput capabilities allow simultaneous analysis of multiple samples, streamlining disease diagnosis and treatment. | The response provides a clear and concise overview of the potential contributions of droplet microfluidics to global health challenges. It highlights three key areas where the technology can make an impact: miniaturization of biological assays, single-cell analysis, and large-scale production of vaccines and therapeutic proteins. | 56 | 56.000000 00000001 |

| | | | | |
|---|---|---|---|---|
| | droplet microfluidics can be used to create single-cell suspensions, enabling researchers to study individual cells in a controlled environment. This can lead to a better understanding of diseases at the cellular level and potentially the development of more targeted treatments.

Lastly, droplet microfluidics can facilitate the production of vaccines and therapeutic proteins on a large scale with high precision and purity. This is crucial for global health as it can help in the rapid response to outbreaks and ensure the availability of life-saving drugs. | Additionally, droplet microfluidics can be integrated into point-of-care platforms, making it feasible for use in resource-limited settings. This approach can improve healthcare accessibility and efficiency. Furthermore, droplet microfluidics facilitates precise single-cell analysis, enabling advances in personalized medicine and more effective treatment strategies. While challenges such as cost, standardization, and integration with existing systems remain, continued development of this technology could significantly impact global health by improving detection, diagnostics, and treatment capabilities. | However, it lacks some depth and nuance compared to the ground truth answer, which provides a more comprehensive discussion of the benefits and potential applications of droplet microfluidics in global health.

SCORE: 56 | | |

| | | | | | |
|---|---|---|---|---|---|
| How can open-source designs impact the future of microfluidic devices? | Open-source designs for microfluidic devices can significantly influence the future in several ways. First, they promote collaboration and knowledge sharing among researchers, allowing for faster innovation and problem-solving. Second, they reduce the cost of development by eliminating the need to reinvent the wheel, as scientists can build upon existing designs. Third, open-source designs can lead to increased accessibility, making it easier for smaller labs, educational institutions, and even DIY enthusiasts to develop their own microfluidic devices. Lastly, open-source designs can foster transparency and reproducibility in research, as | Open-source designs for microfluidic devices hold the potential to revolutionize the field by increasing accessibility, reducing costs, and fostering collaboration. These designs allow for the creation of custom predictive models without requiring extensive expertise, enabling broader participation in device development. The use of open-source tools, such as Design Automation of Fluid Dynamics (DAFD), accelerates design and prototyping, enabling rapid iteration and optimization. Additionally, the integration of machine learning algorithms can drive the development of new microfluidic tools and optimize existing ones. This | The response provides a clear and concise overview of the potential impact of open-source designs on microfluidic devices, highlighting collaboration, cost reduction, accessibility, and transparency. However, it lacks depth and specific examples compared to the ground truth answer, which provides more detailed information about the tools and technologies involved (e.g., Design Automation of Fluid Dynamics) and their applications (e.g., single-cell analysis and global health diagnostics). SCORE: 56 | 56 | 56.000000 00000001 |

| | others can easily replicate experiments using the same design. | democratization of design expertise and the collaborative, community-driven approach will facilitate the creation of affordable, specialized devices, ultimately leading to breakthroughs in applications like single-cell analysis and global health diagnostics. The open-source model promises to transform microfluidics into a more inclusive and innovative field. | | | |
|---|---|---|---|---|---|
| What are the challenges of scaling up microfluidic devices for industrial use? | Scaling up microfluidic devices for industrial use presents several challenges, including:<br><br>1. Cost: The fabrication of microfluidic devices can be expensive due to the need for specialized equipment and materials. As the scale increases, the cost per unit may become | Scaling up microfluidic devices for industrial use presents several challenges, including throughput limitations, lack of standardization, and integration complexities. The small micrometric channel sizes restrict production efficiency, and increasing | The response provides a comprehensive list of challenges associated with scaling up microfluidic devices for industrial use, covering various aspects such as cost, reproducibility, integration, scalability, regulatory compliance, durability, maintenance | 76 | 76.0 |

| | | | | |
|---|---|---|---|---|
| | prohibitive without significant economies of scale.<br><br>2. Reproducibility: Ensuring consistent performance across a large number of devices can be difficult due to variations in fabrication processes and material properties. This can lead to inconsistencies in results, which is unacceptable for industrial applications.<br><br>3. Integration: Microfluidic devices need to be integrated with other systems, such as sensors, pumps, and data analysis software. This integration can be complex and time-consuming, especially when scaling up to an industrial level.<br><br>4. Scalability: The current methods for | parallelization or channel size may impact fabrication costs and reaction kinetics. There is also a lack of standardized design methods, making it difficult to compare results and adopt microfluidics in industrial settings. Additionally, scaling from laboratory to industrial production remains a hurdle, with issues in cost-effective microfabrication, generating complex droplet emulsions, and ensuring reliable fluid control. The integration of microfluidic devices with industrial interfaces, testing methods, and continuous processes is essential for successful commercializatio n. Overcoming these challenges requires innovative solutions, such as | and repair, data handling, energy consumption, and intellectual property. However, the answer could be improved by providing more specific examples or details to support each point, and also by aligning with the ground truth answer in terms of tone and focus on key challenges such as throughput limitations, lack of standardization , and integration complexities. SCORE: 76 | | |

| | | | | |
|---|---|---|---|---|
| | fabricating microfluidic devices may not be scalable enough to meet the demands of industrial production. New manufacturing techniques may need to be developed to address this issue.<br><br>5. Regulatory Compliance: For medical or pharmaceutical applications, microfluidic devices must comply with various regulations. Ensuring compliance can add significant time and cost to the development process.<br><br>6. Durability: Industrial use requires devices that are robust and durable. Many current microfluidic devices are not designed for long-term use, which can limit their applicability in an industrial | simplifying multistep reactions and developing standardized, low-cost fabrication methods. | | |

| | setting. | | | | |
|---|---|---|---|---|---|
| | 7. Maintenance and Repair: As the number of devices increases, so does the need for maintenance and repair. This can be a significant challenge, especially if the devices are complex or require specialized knowledge to maintain. | | | | |
| | 8. Data Handling and Analysis: Large-scale microfluidic systems will generate vast amounts of data. Developing efficient methods for handling and analyzing this data is crucial for industrial applications. | | | | |
| | 9. Energy Consumption: Microfluidic devices often consume less energy than traditional laboratory equipment. However, when | | | | |

| | scaling up to an industrial level, energy consumption can become a significant concern, especially in the context of sustainability.

10. Intellectual Property and Patent Issues: As microfluidics is a relatively new field, there are many patents and intellectual property issues that can complicate the development and commercializati on of large-scale systems. | | | |
|---|---|---|---|---|---|
| How might droplet microfluidics impact agriculture? | Droplet microfluidics could potentially revolutionize agriculture by enabling precise and efficient manipulation of plant cells. This technology allows for the creation of uniform, small-scale droplets containing a single cell or a few cells, which can be genetically modified, | Droplet microfluidics has the potential to transform agriculture by enhancing crop management and food security. It enables the precise release of fertilizers and pesticides, reducing waste and optimizing plant growth. The technology also supports improved crop monitoring through long- | The response demonstrates a good understanding of the potential impact of droplet microfluidics on agriculture. It highlights several key benefits, including improved breeding programs, disease resistance, and efficient use of agrochemicals. | 78 | 78.0 |

| | screened, or analyzed in a high-throughput manner.<br><br>In agriculture, this could lead to faster and more accurate breeding programs, as well as the development of crops that are more resistant to diseases, pests, and environmental stressors. Additionally, droplet microfluidics might facilitate the production of genetically modified organisms (GMOs) in a safer and more controlled environment, reducing the risk of unintended genetic contamination.<br><br>Moreover, this technology could also be used for the rapid screening of plant growth regulators, fertilizers, and other agrochemicals, | term studies, offering insights into plant development. By analyzing individual plant cells, droplet microfluidics can help uncover genetic and phenotypic traits, enhancing disease resistance and drought tolerance in crops. Additionally, it aids in disease detection, allowing for early intervention to prevent crop loss. Despite challenges such as scalability, cost, and integration with existing agricultural practices, ongoing research aims to overcome these barriers, making droplet microfluidics a promising tool for precision agriculture and sustainable food production. | However, some minor inaccuracies and differences in emphasis compared to the ground truth answer are noted. SCORE: 78 | | |
|---|---|---|---|---|---|
| | helping to optimize their use and minimize negative environmental impacts. Overall, droplet microfluidics has the potential to significantly improve agricultural productivity, sustainability, and resilience. | | | | |
| What are the current limitations of droplet microfluidics, and how can they be overcome? | Droplet microfluidics has several limitations, including:<br><br>1. Complexity in design and fabrication: The creation of microfluidic devices requires specialized equipment and expertise, making it difficult for non-experts to design and build their own devices.<br><br>2. Scalability issues: It can be challenging to scale up droplet microfluidics from laboratory-scale experiments to industrial | Droplet microfluidics has immense potential for high-throughput biological assays but faces several limitations that hinder its broader adoption. One key challenge is its efficiency for assays involving small sample sizes; droplet microfluidics excels with large volumes (at least 10^5 samples), making traditional methods like microtiter plates more suitable for lower sample numbers. Another critical issue is maintaining a stable and inert oil-water | The response provides a list of limitations of droplet microfluidics but does not fully address the question by mentioning how these limitations can be overcome. It only mentions that researchers are focusing on developing simpler fabrication methods, digital microfluidics, automation, and integration with other technologies as potential solutions, without providing | 35 | 35.0 |

| | | | | |
|---|---|---|---|---|
| | production due to the need for precise control over fluid flow rates and droplet sizes.<br><br>3. Limited throughput: The number of droplets that can be generated per unit time is relatively low compared to other microfluidic techniques, which may limit their applicability in high-throughput screening applications.<br><br>4. Contamination risks: Droplet microfluidics involves the manipulation of small volumes of fluid, increasing the risk of contamination from external sources or cross-contamination between droplets.<br><br>To overcome these limitations, researchers are focusing on developing | interface, which is essential for droplet formation. Finding the right surfactant to keep this interface stable remains a complex chemistry challenge. Furthermore, droplet microfluidics lacks the spatial barcoding available in microtiter plates, making it difficult to organize and track multiple droplet libraries. As a result, developing alternative barcoding systems is necessary but challenging. The automation of droplet movement is another obstacle, as it requires precise control, often through techniques like electrowetting and dielectrophoresis. Lastly, droplet stability over time can be compromised, with issues such | specific details or examples. The response also lacks a clear connection to the ground truth answer, which provides more comprehensive information on the current limitations of droplet microfluidics and potential solutions.<br><br>SCORE: 35 | | |

| | simpler and more accessible fabrication methods for microfluidic devices. Additionally, advancements in digital microfluidics aim to address scalability issues by using electronic control over droplet generation and manipulation. Improvements in automation and integration with other technologies can help increase throughput and reduce contamination risks. | as shrinkage, coalescence, and cross-talk between droplets. To address these issues, research is underway to develop improved surfactants, more reliable barcoding systems, and novel materials and techniques to enhance droplet control. These advancements, coupled with better analytical tools, are expected to overcome current limitations and expand droplet microfluidics' applications in chemical and biological research. | | | |
|---|---|---|---|---|---|
| How can interdisciplinary collaboration accelerate innovation in droplet microfluidics? | Interdisciplinary collaboration in droplet microfluidics can accelerate innovation by combining expertise from various fields such as engineering, biology, chemistry, and physics. This synergy allows for the | Interdisciplinary collaboration is essential for advancing droplet microfluidics by combining expertise from various fields such as chemistry, biology, material science, and engineering. By uniting these diverse areas, researchers can | The response demonstrates a good understanding of the topic by highlighting the benefits of interdisciplinary collaboration in accelerating innovation in droplet microfluidics. It correctly identifies the combination of | 56 | 56.000000 00000001 |

| | development of novel techniques, improved device designs, and innovative applications that leverage the unique properties of droplets in microfluidic systems. Additionally, collaboration fosters the exchange of ideas, best practices, and resources, which can lead to faster problem-solving and more rapid prototyping and validation of new technologies. | address complex challenges such as cost-effective fabrication, scalable production, and novel interfacial reaction systems. For example, biologists and chemists can leverage droplet microfluidics for high-throughput analysis, while material scientists and engineers work on developing more robust and affordable microfabrication methods. Furthermore, integrating technologies like sensitive detectors and machine learning algorithms can enhance the analytical capabilities of droplet-based systems. Collaborative initiatives, such as the EMBL Microfluidics Conference, foster knowledge exchange and innovation by bringing together professionals from different | expertise from various fields as a key factor, and mentions specific examples such as novel techniques, improved device designs, and innovative applications. However, it lacks some depth and detail compared to the ground truth answer, which provides more specific examples and emphasizes the importance of addressing complex challenges like cost-effective fabrication and scalable production. The response also does not mention the EMBL Microfluidics Conference or the integration of technologies like sensitive detectors and machine learning algorithms.

SCORE: 56 | | |

| | | disciplines. This cooperation is crucial for overcoming technical barriers, reducing manufacturing costs, and developing new applications in fields like drug discovery, single-cell analysis, and nanomaterial synthesis. Ultimately, interdisciplinary collaboration accelerates the evolution of droplet microfluidics, pushing the boundaries of what is achievable in this promising field. | | | |
|---|---|---|---|---|---|

**S9: Automation Designer**

**history:**
```python
```python
import pandas as pd
from sklearn.model_selection import train_test_split
from sklearn.ensemble import GradientBoostingRegressor
from sklearn.metrics import mean_squared_error
from xgboost import XGBRegressor
from sklearn.model_selection import GridSearchCV
from sklearn.preprocessing import StandardScaler

# Load the data
df = pd.read_csv('data.csv')

# Define features and target
X = df.drop(['Observed generation rate (Hz)'], axis=1)
y = df['Observed generation rate (Hz)']
```



```
# Split the data into training and testing sets
X_train, X_test, y_train, y_test = train_test_split(X, y, test_size=0.2, random_state=42)

# Scale the features
scaler = StandardScaler()
X_train_scaled = scaler.fit_transform(X_train)
X_test_scaled = scaler.transform(X_test)

# Define hyperparameters for GridSearchCV
param_grid = {
    'max_depth': [3, 5, 7],
    'learning_rate': [0.1, 0.05, 0.01],
    'n_estimators': [50, 100, 200]
}

# Initialize the XGBoost model and GridSearchCV
xgb_model = XGBRegressor()
grid_search = GridSearchCV(estimator=xgb_model, param_grid=param_grid, cv=5, n_jobs=-1)

# Perform hyperparameter tuning
grid_search.fit(X_train_scaled, y_train)
best_params = grid_search.best_params_
print("Best parameters:", best_params)

# Train the model with the best parameters
xgb_model.set_params(**best_params)
xgb_model.fit(X_train_scaled, y_train)

# Make predictions on the test set
y_pred = xgb_model.predict(X_test_scaled)

# Evaluate the model
mse = mean_squared_error(y_test, y_pred)
print("Mean squared error:", mse)
```
 REVIEWER:
* The `XGBRegressor` instance should be initialized with a dictionary of hyperparameters instead of an empty dictionary. It should look like this: `xgb_model = XGBRegressor(max_depth=3, learning_rate=0.1, n_estimators=50)`.
* The `GridSearchCV` instance should also be initialized with the same dictionary of hyperparameters for the estimator.
* The `param_grid` dictionary contains redundant values (e.g., 'max_depth' has a value of 7 which is not present in the list). It should be updated to only include the specified values.
* There's no error handling when loading the data from the CSV file. Consider adding a try-except block to handle potential errors.



* The `StandardScaler` instance should be used with caution, as it can introduce bias into the model if not used correctly. Consider using `MinMaxScaler` or `RobustScaler` instead.
* The code does not check for overfitting. Consider adding techniques like cross-validation or early stopping to prevent overfitting.
* The variable names are not following PEP8 guidelines (e.g., 'Observed generation rate (Hz)' should be 'observed_generation_rate_hz').
* There's no docstring in the code, which is a good practice for documenting functions and classes.
 CODER:

```python
import pandas as pd
from sklearn.model_selection import train_test_split
from sklearn.ensemble import GradientBoostingRegressor
from sklearn.metrics import mean_squared_error
from xgboost import XGBRegressor
from sklearn.model_selection import GridSearchCV
from sklearn.preprocessing import MinMaxScaler
from sklearn.model_selection import cross_val_score

def load_data(file_path):
    """
    Load the data from a CSV file.

    Args:
        file_path (str): The path to the CSV file.

    Returns:
        pd.DataFrame: The loaded data.
    """
    try:
        return pd.read_csv(file_path)
    except Exception as e:
        print(f"Error loading data: {e}")
        return None

def prepare_data(df):
    """
    Prepare the data by splitting it into features and target, and scaling the features.

    Args:
        df (pd.DataFrame): The loaded data.

    Returns:
        tuple: A tuple containing the scaled features and target.
    """
    X = df.drop(['observed_generation_rate_hz'], axis=1)
```



```python
    y = df['observed_generation_rate_hz']

    # Split the data into training and testing sets
    X_train, X_test, y_train, y_test = train_test_split(X, y, test_size=0.2, random_state=42)

    # Scale the features using MinMaxScaler
    scaler = MinMaxScaler()
    X_train_scaled = scaler.fit_transform(X_train)
    X_test_scaled = scaler.transform(X_test)

    return X_train_scaled, X_test_scaled, y_train, y_test

def train_model(X_train_scaled, y_train):
    """
    Train an XGBoost model with the best hyperparameters.

    Args:
        X_train_scaled (np.ndarray): The scaled training features.
        y_train (pd.Series): The target values.

    Returns:
        XGBRegressor: The trained model.
    """
    # Define hyperparameters for GridSearchCV
    param_grid = {
        'max_depth': [3, 5],
        'learning_rate': [0.1, 0.05],
        'n_estimators': [50, 100]
    }

    # Initialize the XGBoost model and GridSearchCV with the specified hyperparameters
    xgb_model = XGBRegressor(max_depth=3, learning_rate=0.1, n_estimators=50)
    grid_search = GridSearchCV(estimator=xgb_model, param_grid=param_grid, cv=5, n_jobs=-1)

    # Perform hyperparameter tuning with cross-validation
    grid_search.fit(X_train_scaled, y_train)
    best_params = grid_search.best_params_
    print("Best parameters:", best_params)

    # Train the model with the best parameters and evaluate it using cross-validation
    xgb_model.set_params(**best_params)
    scores = cross_val_score(xgb_model, X_train_scaled, y_train, cv=5, scoring='neg_mean_squared_error')
    mse = -scores.mean()
    print("Mean squared error:", mse)
```



```
    return xgb_model

def main():
    # Load the data
    df = load_data('data.csv')

    if df is not None:
        # Prepare the data
        X_train_scaled, X_test_scaled, y_train, y_test = prepare_data(df)

        # Train the model
        xgb_model = train_model(X_train_scaled, y_train)

        # Make predictions on the test set
        y_pred = xgb_model.predict(X_test_scaled)

        # Evaluate the model
        mse = mean_squared_error(y_test, y_pred)
        print("Mean squared error:", mse)

if __name__ == "__main__":
    main()
```
 REVIEWER:
* The function `load_data` does not handle the case where the CSV file is empty.
* The function `prepare_data` does not check if the data contains any missing values before splitting it into features and target.
* In the function `train_model`, the XGBoost model is initialized with a fixed set of hyperparameters, but then GridSearchCV is used to tune these parameters. This could lead to inconsistent results.
* The variable names in the functions are not consistent (e.g., `X_train_scaled` vs `xgb_model`).
* The function `train_model` does not handle any exceptions that might occur during model training or hyperparameter tuning.
* The function `main` does not check if the loaded data is empty before calling `prepare_data`.
* The variable names in the `main` function are not consistent (e.g., `df` vs `X_train_scaled`).
* The code uses both `print` statements and f-strings, which can make it harder to read. It's better to stick with one style throughout.
* The code does not handle any potential errors that might occur during model prediction or evaluation.
* The variable `mse` is used in multiple places without being redefined. It would be better to use a more descriptive name for the mean squared error value.
* The function `train_model` returns an XGBoost model, but it's not clear what this model is supposed to do. A more descriptive return type or comment would be helpful.
* The code does not include any docstrings for the functions in the `main` function.



* The variable `best_params` is printed to the console without being used anywhere else in the code. It might be better to store it in a variable or use it for something.
* The code uses `cross_val_score` with 5 folds, but it's not clear why this specific number was chosen. A comment explaining the reasoning behind this choice would be helpful.
* The function `train_model` trains an XGBoost model and then evaluates it using cross-validation, but it does not return any metrics or predictions. It might be better to include these in the return value of the function.
 CODER:
```python
import pandas as pd
from sklearn.model_selection import train_test_split
from sklearn.ensemble import GradientBoostingRegressor
from sklearn.metrics import mean_squared_error
from xgboost import XGBRegressor
from sklearn.model_selection import GridSearchCV
from sklearn.preprocessing import MinMaxScaler
from sklearn.model_selection import cross_val_score

def load_data(file_path):
    """
    Load the data from a CSV file.

    Args:
        file_path (str): The path to the CSV file.

    Returns:
        pd.DataFrame: The loaded data.
    """
    try:
        df = pd.read_csv(file_path)
        if df.empty:
            print("Error loading data: File is empty.")
            return None
        return df
    except Exception as e:
        print(f"Error loading data: {e}")
        return None

def prepare_data(df):
    """
    Prepare the data by splitting it into features and target, checking for missing values,
    and scaling the features.

    Args:
        df (pd.DataFrame): The loaded data.
```



```python
    Returns:
        tuple: A tuple containing the scaled features and target.
    """
    if df is None or df.empty:
        print("Error preparing data: Data is empty.")
        return None

    X = df.drop(['observed_generation_rate_hz'], axis=1)
    y = df['observed_generation_rate_hz']

    # Check for missing values
    if X.isnull().values.any() or y.isnull().values.any():
        print("Error preparing data: Data contains missing values.")
        return None

    # Split the data into training and testing sets
    X_train, X_test, y_train, y_test = train_test_split(X, y, test_size=0.2, random_state=42)

    # Scale the features using MinMaxScaler
    scaler = MinMaxScaler()
    X_train_scaled = scaler.fit_transform(X_train)
    X_test_scaled = scaler.transform(X_test)

    return X_train_scaled, X_test_scaled, y_train, y_test

def train_model(X_train_scaled, y_train):
    """
    Train an XGBoost model with the best hyperparameters.

    Args:
        X_train_scaled (np.ndarray): The scaled training features.
        y_train (pd.Series): The target values.

    Returns:
        tuple: A tuple containing the trained model and its performance metrics.
    """
    try:
        # Define hyperparameters for GridSearchCV
        param_grid = {
            'max_depth': [3, 5],
            'learning_rate': [0.1, 0.05],
            'n_estimators': [50, 100]
        }

        # Initialize the XGBoost model and GridSearchCV with the specified hyperparameters
        xgb_model = XGBRegressor()
```



```python
        grid_search = GridSearchCV(estimator=xgb_model, param_grid=param_grid, cv=5,
n_jobs=-1)

        # Perform hyperparameter tuning with cross-validation
        grid_search.fit(X_train_scaled, y_train)
        best_params = grid_search.best_params_
        print("Best parameters:", best_params)

        # Train the model with the best parameters and evaluate it using cross-validation
        xgb_model.set_params(**best_params)
        scores = cross_val_score(xgb_model, X_train_scaled, y_train, cv=5,
scoring='neg_mean_squared_error')
        mse = -scores.mean()
        print("Mean squared error:", mse)

        return xgb_model, best_params, mse
    except Exception as e:
        print(f"Error training model: {e}")
        return None

def main():
    """
    The main function that loads the data, prepares it, trains a model, and evaluates its
performance.
    """
    # Load the data
    df = load_data('data.csv')

    if df is not None:
        # Prepare the data
        X_train_scaled, X_test_scaled, y_train, y_test = prepare_data(df)

        if X_train_scaled is not None:
            # Train the model
            xgb_model, best_params, mse = train_model(X_train_scaled, y_train)

            if xgb_model is not None:
                # Make predictions on the test set
                y_pred = xgb_model.predict(X_test_scaled)

                # Evaluate the model
                final_mse = mean_squared_error(y_test, y_pred)
                print("Final Mean squared error:", final_mse)

if __name__ == "__main__":
    main()
```



```

 REVIEWER:
* The function `load_data` does not handle the case where the file path is incorrect or the file does not exist. It should raise a more specific exception, such as `FileNotFoundError`.
* In the `prepare_data` function, the line `X = df.drop(['observed_generation_rate_hz'], axis=1)` assumes that the column 'observed_generation_rate_hz' exists in the dataframe. If it doesn't, this will throw an error. It should check if the column exists before trying to drop it.
* The `prepare_data` function does not handle the case where the data is not numeric and cannot be scaled by MinMaxScaler. It should add a check for non-numeric columns before scaling.
* In the `train_model` function, the line `param_grid = { ... }` defines hyperparameters that are not necessarily optimal for XGBoost models. These values may need to be adjusted based on the specific problem being solved.
* The `train_model` function does not handle the case where the GridSearchCV process fails due to an exception. It should catch and log any exceptions that occur during grid search.
* In the `main` function, there is no error handling for cases where the data loading or preparation steps fail. It should add checks to ensure that these steps complete successfully before proceeding with model training.
* The variable names in some places are not following PEP8 naming conventions (e.g., `X_train_scaled`, `y_train`, etc.). They should be renamed to follow the convention of using lowercase letters and underscores for separation.
* There is no docstring for the `main` function. It should have a docstring that describes its purpose and behavior.
* The code does not handle any potential memory issues when loading large datasets into memory. It may need to use more efficient data structures or processing techniques to avoid running out of memory.
* The code does not include any checks for overfitting or underfitting in the model. It should consider adding metrics like cross-validation score, R-squared value, etc., to evaluate the performance of the model.
* There is no logging mechanism implemented in the code. It would be beneficial to add a logging system to track important events and errors during execution.
* The `train_model` function does not return any information about the best parameters found by GridSearchCV. It should consider returning this information along with the trained model.
* The variable names are not descriptive enough, e.g., `xgb_model`, `best_params`, etc. They should be renamed to something more meaningful.
* There is no check for the case where the data is empty or has only one row after splitting it into training and testing sets in the `prepare_data` function. It should add a check for this scenario and handle it accordingly.
 CODER:
```python
import pandas as pd
from sklearn.model_selection import train_test_split
from sklearn.ensemble import GradientBoostingRegressor
from sklearn.metrics import mean_squared_error
from xgboost import XGBRegressor
from sklearn.model_selection import GridSearchCV
from sklearn.preprocessing import MinMaxScaler
```

```python
from sklearn.model_selection import cross_val_score
import logging

# Set up logging
logging.basicConfig(level=logging.INFO)
logger = logging.getLogger(__name__)

def load_data(file_path):
    """
    Load the data from a CSV file.

    Args:
        file_path (str): The path to the CSV file.

    Returns:
        pd.DataFrame: The loaded data.
    """
    try:
        df = pd.read_csv(file_path)
        if df.empty:
            logger.error("Error loading data: File is empty.")
            raise FileNotFoundError
        return df
    except Exception as e:
        logger.error(f"Error loading data: {e}")
        raise

def prepare_data(df):
    """
    Prepare the data by splitting it into features and target, checking for missing values,
    and scaling the features.

    Args:
        df (pd.DataFrame): The loaded data.

    Returns:
        tuple: A tuple containing the scaled features and target.
    """
    if df is None or df.empty:
        logger.error("Error preparing data: Data is empty.")
        return None

    X = df.drop(['observed_generation_rate_hz'], axis=1, errors='ignore')
    y = df['observed_generation_rate_hz']

    # Check for missing values
```

```python
    if X.isnull().values.any() or y.isnull().values.any():
        logger.error("Error preparing data: Data contains missing values.")
        return None

    # Split the data into training and testing sets
    X_train, X_test, y_train, y_test = train_test_split(X, y, test_size=0.2, random_state=42)

    if len(y_train) < 10 or len(y_test) < 10:
        logger.error("Error preparing data: Data is too small.")
        return None

    # Scale the features using MinMaxScaler
    scaler = MinMaxScaler()
    X_train_scaled = scaler.fit_transform(X_train)
    X_test_scaled = scaler.transform(X_test)

    return X_train_scaled, X_test_scaled, y_train, y_test

def train_model(X_train_scaled, y_train):
    """
    Train an XGBoost model with the best hyperparameters.

    Args:
        X_train_scaled (np.ndarray): The scaled training features.
        y_train (pd.Series): The target values.

    Returns:
        tuple: A tuple containing the trained model and its performance metrics.
    """
    try:
        # Define hyperparameters for GridSearchCV
        param_grid = {
            'max_depth': [3, 5],
            'learning_rate': [0.1, 0.05],
            'n_estimators': [50, 100]
        }

        # Initialize the XGBoost model and GridSearchCV with the specified hyperparameters
        xgb_model = XGBRegressor()
        grid_search = GridSearchCV(estimator=xgb_model, param_grid=param_grid, cv=5,
n_jobs=-1)

        # Perform hyperparameter tuning with cross-validation
        grid_search.fit(X_train_scaled, y_train)
        best_params = grid_search.best_params_
        logger.info("Best parameters: %s", best_params)
```

```python
        # Train the model with the best parameters and evaluate it using cross-validation
        xgb_model.set_params(**best_params)
        scores = cross_val_score(xgb_model, X_train_scaled, y_train, cv=5,
scoring='neg_mean_squared_error')
        mse = -scores.mean()
        logger.info("Mean squared error: %f", mse)

        return xgb_model, best_params, mse
    except Exception as e:
        logger.error(f"Error training model: {e}")
        return None

def main():
    """
    The main function that loads the data, prepares it, trains a model, and evaluates its
performance.
    """
    # Load the data
    try:
        df = load_data('data.csv')
    except FileNotFoundError:
        logger.error("Error loading data: File not found.")
        return

    if df is not None:
        # Prepare the data
        X_train_scaled, X_test_scaled, y_train, y_test = prepare_data(df)

        if X_train_scaled is not None:
            # Train the model
            xgb_model, best_params, mse = train_model(X_train_scaled, y_train)

            if xgb_model is not None:
                # Make predictions on the test set
                y_pred = xgb_model.predict(X_test_scaled)

                # Evaluate the model
                final_mse = mean_squared_error(y_test, y_pred)
                logger.info("Final Mean squared error: %f", final_mse)

if __name__ == "__main__":
    main()
```

REVIEWER:

* The function `load_data` does not handle the case where the file path is empty or None. It should raise a ValueError in this case.
* In the `prepare_data` function, the line `X = df.drop(['observed_generation_rate_hz'], axis=1, errors='ignore')` can be simplified to `X = df.drop('observed_generation_rate_hz', axis=1)`.
* The `prepare_data` function does not handle the case where the data is too small (less than 10 samples). It should raise a ValueError in this case.
* In the `train_model` function, the line `param_grid = { ... }` can be simplified to `param_grid = {'max_depth': [3, 5], 'learning_rate': [0.1, 0.05], 'n_estimators': [50, 100]}` (remove unnecessary quotes).
* The `train_model` function does not handle the case where the model cannot be trained or evaluated. It should raise a ValueError in this case.
* In the `main` function, the line `try: df = load_data('data.csv')` can be simplified to `df = load_data('data.csv')`.
* The `main` function does not handle the case where the data cannot be loaded or prepared. It should raise a ValueError in this case.
* There are no type hints for the function parameters and return types.
* The variable names are not descriptive (e.g., `X_train_scaled`, `y_train`, etc.). They can be renamed to something more meaningful.
* The logging level is set to INFO, but there are no log messages with a higher level (WARNING, ERROR, CRITICAL). It would be better to use the standard Python logging levels.
* There are no docstrings for the functions. It would be helpful to add docstrings to explain what each function does and what it returns.
* The `train_model` function uses GridSearchCV with 5-fold cross-validation, but the number of folds is not a power of 2 (which is recommended for unbiased estimates). It might be better to use a different number of folds or a different cross-validation strategy.
* The `train_model` function trains an XGBoost model and then evaluates it using cross-validation. However, it does not compare the performance of the trained model with other models (e.g., Linear Regression, Random Forest). It would be helpful to add this comparison to get a better understanding of the model's performance.
* The `main` function calls the `train_model` function and then evaluates its performance using mean squared error. However, it does not save the best model or its parameters. It might be helpful to save the best model and its parameters for future use.
* There are no tests for the functions. It would be helpful to add unit tests to ensure that each function works correctly.
 CODER:
```python
import pandas as pd
from sklearn.model_selection import train_test_split
from sklearn.ensemble import GradientBoostingRegressor
from sklearn.metrics import mean_squared_error
from xgboost import XGBRegressor
from sklearn.model_selection import GridSearchCV
from sklearn.preprocessing import MinMaxScaler
from sklearn.model_selection import cross_val_score
import logging
```

```python
# Set up logging
logging.basicConfig(level=logging.INFO)
logger = logging.getLogger(__name__)

def load_data(file_path: str) -> pd.DataFrame:
    """
    Load the data from a CSV file.

    Args:
        file_path (str): The path to the CSV file.

    Returns:
        pd.DataFrame: The loaded data.

    Raises:
        ValueError: If the file path is empty or None.
    """
    if not file_path:
        raise ValueError("File path cannot be empty or None.")
    try:
        df = pd.read_csv(file_path)
        if df.empty:
            logger.error("Error loading data: File is empty.")
            raise FileNotFoundError
        return df
    except Exception as e:
        logger.error(f"Error loading data: {e}")
        raise

def prepare_data(df: pd.DataFrame) -> tuple:
    """
    Prepare the data by splitting it into features and target, checking for missing values,
    and scaling the features.

    Args:
        df (pd.DataFrame): The loaded data.

    Returns:
        tuple: A tuple containing the scaled features and target.

    Raises:
        ValueError: If the data is empty or too small.
    """
    if not isinstance(df, pd.DataFrame) or df.empty:
        logger.error("Error preparing data: Data is empty.")
        return None
```

```python
    X = df.drop('observed_generation_rate_hz', axis=1)
    y = df['observed_generation_rate_hz']

    if len(y) < 10:
        logger.error("Error preparing data: Data is too small.")
        return None

    # Split the data into training and testing sets
    X_train, X_test, y_train, y_test = train_test_split(X, y, test_size=0.2, random_state=42)

    # Scale the features using MinMaxScaler
    scaler = MinMaxScaler()
    X_train_scaled = scaler.fit_transform(X_train)
    X_test_scaled = scaler.transform(X_test)

    return X_train_scaled, X_test_scaled, y_train, y_test

def train_model(X_train_scaled: np.ndarray, y_train: pd.Series) -> tuple:
    """
    Train an XGBoost model with the best hyperparameters.

    Args:
        X_train_scaled (np.ndarray): The scaled training features.
        y_train (pd.Series): The target values.

    Returns:
        tuple: A tuple containing the trained model and its performance metrics.

    Raises:
        ValueError: If the data is empty or too small.
    """
    try:
        # Define hyperparameters for GridSearchCV
        param_grid = {
            'max_depth': [3, 5],
            'learning_rate': [0.1, 0.05],
            'n_estimators': [50, 100]
        }

        # Initialize the XGBoost model and GridSearchCV with the specified hyperparameters
        xgb_model = XGBRegressor()
        grid_search = GridSearchCV(estimator=xgb_model, param_grid=param_grid, cv=5, n_jobs=-1)

        # Perform hyperparameter tuning with cross-validation
```

```python
        grid_search.fit(X_train_scaled, y_train)
        best_params = grid_search.best_params_
        logger.info("Best parameters: %s", best_params)

        # Train the model with the best parameters and evaluate it using cross-validation
        xgb_model.set_params(**best_params)
        scores = cross_val_score(xgb_model, X_train_scaled, y_train, cv=5,
scoring='neg_mean_squared_error')
        mse = -scores.mean()
        logger.info("Mean squared error: %f", mse)

        return xgb_model, best_params, mse
    except Exception as e:
        logger.error(f"Error training model: {e}")
        return None

def main():
    """
    The main function that loads the data, prepares it, trains a model, and evaluates its
performance.
    """
    # Load the data
    try:
        df = load_data('data.csv')
    except FileNotFoundError:
        logger.error("Error loading data: File not found.")
        return

    if df is not None:
        # Prepare the data
        X_train_scaled, X_test_scaled, y_train, y_test = prepare_data(df)

        if X_train_scaled is not None:
            # Train the model
            xgb_model, best_params, mse = train_model(X_train_scaled, y_train)

            if xgb_model is not None:
                # Make predictions on the test set
                y_pred = xgb_model.predict(X_test_scaled)

                # Evaluate the model
                final_mse = mean_squared_error(y_test, y_pred)
                logger.info("Final Mean squared error: %f", final_mse)

if __name__ == "__main__":
    main()
```

```

REVIEWER:
* The function `load_data` does not handle the case where the file path is a directory instead of a file.
* The function `prepare_data` does not check if the data has any missing values before scaling it. It only checks for empty data.
* In the `train_model` function, the XGBoost model is initialized with default parameters and then GridSearchCV is used to find the best hyperparameters. However, this can be inefficient because GridSearchCV tries all possible combinations of hyperparameters which can take a long time for large search spaces. A more efficient approach would be to use RandomizedSearchCV or Bayesian optimization.
* The `train_model` function does not handle the case where the data is too small (less than 10 samples) as specified in the docstring.
* In the `main` function, there are no checks for potential exceptions that could occur during the execution of the code. For example, what if the file 'data.csv' cannot be found or loaded?
* The variable names `X_train_scaled`, `X_test_scaled`, `y_train`, and `y_test` in the `prepare_data` function do not follow PEP8 naming conventions (they should be lowercase with words separated by underscores).
* The variable name `best_params` in the `train_model` function does not follow PEP8 naming conventions.
* The variable name `mse` in the `train_model` function does not follow PEP8 naming conventions.
* The variable name `final_mse` in the `main` function does not follow PEP8 naming conventions.
* There are no type hints for the variables `X_train_scaled`, `y_train`, and `xgb_model` in the `train_model` function.
* The `train_model` function returns a tuple containing the trained model, its best parameters, and its mean squared error. However, it would be more consistent with Python's naming conventions to return a dictionary instead of a tuple.
* The `main` function does not handle potential exceptions that could occur during the execution of the code. For example, what if the data cannot be prepared or the model cannot be trained?
* There are no comments in the code to explain why certain decisions were made or how the code works.
* The logging level is set to INFO but there are no log messages with a level higher than INFO (e.g., WARNING, ERROR).
* The `train_model` function uses GridSearchCV with 5-fold cross-validation. However, this can be inefficient for large datasets because it requires fitting and predicting the model on each fold multiple times. A more efficient approach would be to use KFold or StratifiedKFold instead.
* The `prepare_data` function scales the features using MinMaxScaler but does not scale the target variable. This is incorrect because scaling should be done on both features and targets if they have different units or ranges.
* The `train_model` function uses XGBRegressor with default parameters but does not specify the objective function (e.g., 'reg:squarederror'). This can lead to unexpected results because the model may not be optimized for the correct loss function.
CODER:
```python
```



```python
import pandas as pd
from sklearn.model_selection import train_test_split
from sklearn.ensemble import GradientBoostingRegressor
from sklearn.metrics import mean_squared_error
from xgboost import XGBRegressor
from sklearn.model_selection import RandomizedSearchCV, KFold
from sklearn.preprocessing import MinMaxScaler
from scipy.stats import randint as sp_randint
import numpy as np
import logging

# Set up logging
logging.basicConfig(level=logging.INFO)
logger = logging.getLogger(__name__)

def load_data(file_path: str) -> pd.DataFrame:
    """
    Load the data from a CSV file.

    Args:
        file_path (str): The path to the CSV file.

    Returns:
        pd.DataFrame: The loaded data.

    Raises:
        ValueError: If the file path is empty or None, or if the file does not exist.
    """
    if not file_path:
        raise ValueError("File path cannot be empty or None.")
    try:
        df = pd.read_csv(file_path)
        if df.empty:
            logger.error("Error loading data: File is empty.")
            raise FileNotFoundError
        return df
    except Exception as e:
        logger.error(f"Error loading data: {e}")
        raise

def prepare_data(df: pd.DataFrame) -> tuple:
    """
    Prepare the data by splitting it into training and testing sets, and scaling the features.

    Args:
        df (pd.DataFrame): The input data.
```



```python
    Returns:
        tuple: A tuple containing the scaled training features, scaled test features, target values for
training, and target values for testing.

    Raises:
        ValueError: If the data is empty or too small.
    """
    if not isinstance(df, pd.DataFrame) or df.empty:
        logger.error("Error preparing data: Data is empty.")
        return None

    X = df.drop('observed_generation_rate_hz', axis=1)
    y = df['observed_generation_rate_hz']

    if len(y) < 10:
        logger.error("Error preparing data: Data is too small.")
        return None

    # Split the data into training and testing sets
    X_train, X_test, y_train, y_test = train_test_split(X, y, test_size=0.2, random_state=42)

    # Scale the features using MinMaxScaler
    scaler = MinMaxScaler()
    X_train_scaled = scaler.fit_transform(X_train)
    X_test_scaled = scaler.transform(X_test)

    return X_train_scaled, X_test_scaled, y_train, y_test

def train_model(X_train_scaled: np.ndarray, y_train: pd.Series) -> tuple:
    """
    Train an XGBoost model with the best hyperparameters.

    Args:
        X_train_scaled (np.ndarray): The scaled training features.
        y_train (pd.Series): The target values for training.

    Returns:
        tuple: A tuple containing the trained model and its performance metrics.

    Raises:
        ValueError: If the data is empty or too small.
    """
    try:
        # Define hyperparameters for RandomizedSearchCV
        param_grid = {
```



```python
        'max_depth': sp_randint(3, 10),
        'learning_rate': [0.1, 0.05],
        'n_estimators': [50, 100]
    }

    # Initialize the XGBoost model and RandomizedSearchCV with the specified
hyperparameters
    xgb_model = XGBRegressor()
    random_search = RandomizedSearchCV(estimator=xgb_model,
param_distributions=param_grid, cv=KFold(n_splits=5), n_iter=10, n_jobs=-1)

    # Perform hyperparameter tuning with cross-validation
    random_search.fit(X_train_scaled, y_train)
    best_params = random_search.best_params_
    logger.info("Best parameters: %s", best_params)

    # Train the model with the best parameters and evaluate it using cross-validation
    xgb_model.set_params(**best_params)
    scores = KFold(n_splits=5).split(X_train_scaled, y_train)
    mse = -np.mean([mean_squared_error(y_train[i], xgb_model.predict(X_train_scaled[i])) for
i in range(len(scores))])
    logger.info("Mean squared error: %f", mse)

    return xgb_model, best_params, mse
except Exception as e:
    logger.error(f"Error training model: {e}")
    return None

def main():
    """
    The main function that loads the data, prepares it, trains a model, and evaluates its
performance.
    """
    # Load the data
    try:
        df = load_data('data.csv')
    except FileNotFoundError:
        logger.error("Error loading data: File not found.")
        return

    if df is not None:
        # Prepare the data
        X_train_scaled, X_test_scaled, y_train, y_test = prepare_data(df)

        if X_train_scaled is not None:
            # Train the model
```



```
    xgb_model, best_params, mse = train_model(X_train_scaled, y_train)

    if xgb_model is not None:
        # Make predictions on the test set
        y_pred = xgb_model.predict(X_test_scaled)

        # Evaluate the model
        final_mse = mean_squared_error(y_test, y_pred)
        logger.info("Final Mean squared error: %f", final_mse)

if __name__ == "__main__":
    main()
```

REVIEWER:
* The function `load_data` does not handle the case where the file path is a directory instead of a file.
* In the `prepare_data` function, the check for empty data should be done before splitting the data into training and testing sets. This can be achieved by checking if the length of the target values (`y`) is greater than 0 before proceeding with the split.
* The `train_model` function does not handle the case where the RandomizedSearchCV fails to find a good set of hyperparameters, which could result in an empty dictionary being returned for `best_params`.
* In the `main` function, there are multiple places where exceptions are caught and logged, but then the program continues running. It would be better to re-raise these exceptions or handle them more explicitly.
* The variable names `X_train_scaled`, `X_test_scaled`, etc., could be more descriptive (e.g., `training_features_scaled`, `test_features_scaled`).
* The function `train_model` does not return the trained model, only its performance metrics. It would be better to include the model itself in the returned tuple.
* There are no type hints for the variables `param_grid` and `best_params`.
* The variable name `mse` is not very descriptive; consider renaming it to something like `cross_validation_mse`.
* In the `main` function, there is a try-except block around the call to `load_data`, but then the program continues running even if an exception is raised. It would be better to re-raise this exception or handle it more explicitly.
* The variable name `xgb_model` could be more descriptive (e.g., `best_xgboost_model`).
* There are no docstrings for the functions `train_model` and `main`.
* The function `prepare_data` does not check if the data is too small before scaling the features. It would be better to add a check for this case.
* The variable name `y_pred` could be more descriptive (e.g., `test_set_predictions`).

AutoCAD-AutoLISP

history:



```
(defun c:drawlines (/)
 (setq p1 '(0 . 0) p2 '(200 . 0) p3 '(200 . 200) p4 '(250 . 0) p5 '(250 . 200) p6 '(250 . 0) p7 '(450 .
0)
      p8 '(0 . -50) p9 '(200 . -50) p10 '(200 . -250) p11 '(250 . -50) p12 '(250 . -250) p13 '(250 . -
50) p14 '(450 . -50))

 (setq *p1* p1 *p2* p2 *p3* p3 *p4* p4 *p5* p5 *p6* p6 *p7* p7
      *p8* p8 *p9* p9 *p10* p10 *p11* p11 *p12* p12 *p13* p13 *p14* p14)

 (setq i 1)
 (while (< i 15)
  (if (eq i 1)
    (progn
     (setq p (*p1*)
          e (*p2*)
          l (list p e))
     (docladr l))
    (if (or (eq i 2) (eq i 3) (eq i 4) (eq i 5)
          (eq i 6) (eq i 7) (eq i 8) (eq i 9)
          (eq i 10) (eq i 11) (eq i 12) (eq i 13) (eq i 14))
     (progn
      (setq p (*p1*)
           e (*p2*)
           l (list p e))
      (docladr l)))
  (setq i (+ i 1)))

 (princ "\n")

 (defun docladr (l /)
  (setq p (car l) e (cadr l))
  (line p e))

 (defun *p1* (/)
  '(0 . 0))

 (defun *p2* (/)
  '(200 . 0))

 (defun *p3* (/)
  '(200 . 200))

 (defun *p4* (/)
  '(250 . 0))

 (defun *p5* (/)
```


```
  '(250 . 200))

(defun *p6* (/)
  '(250 . 0))

(defun *p7* (/)
  '(450 . 0))

(defun *p8* (/)
  '(0 . -50))

(defun *p9* (/)
  '(200 . -50))

(defun *p10* (/)
  '(200 . -250))

(defun *p11* (/)
  '(250 . -50))

(defun *p12* (/)
  '(250 . -250))

(defun *p13* (/)
  '(250 . -50))

(defun *p14* (/)
  '(450 . -50))
```

REVIEWER:
• The function `c:drawlines` has a global variable `i` which is not properly reset or cleared between calls, potentially causing incorrect results if the function is called multiple times.
• The `docladr` function uses `car` and `cadr` to access the first two elements of the list `l`, but it does not check if the list has at least two elements before doing so. This could lead to an error if the input list has less than two elements.
• The `*p1*` through `*p14*` functions are all defined as global variables, which is generally considered bad practice in AutoLisp programming. It would be better to encapsulate these values within a local scope or use a more structured approach to manage them.
• There are multiple identical function definitions for `*p6*`, `*p13*`, and `*p14*`. This could indicate a copy-paste error, and the repeated code should be refactored into a single definition.
• The `docladr` function uses the `line` function to draw a line between two points. However, it does not check if the points are valid (i.e., have both x and y coordinates) before attempting to draw the line. This could lead to an error if invalid points are passed to the function.
• The `c:drawlines` function uses a while loop with a hardcoded value of 15 as the condition, but it does not check if this value is correct or if it's intended to be dynamic. It would be better to use a more explicit and meaningful loop condition.



• There are no comments or documentation in the code to explain what the functions do, how they work, or any assumptions made about the input data. This makes it difficult for others (or even yourself) to understand the code's purpose and behavior.

CODER:

```
(defun c:drawlines ()
  (let ((i 1)
        (points '((0 . 0) (200 . 0) (200 . 200) (250 . 0) (250 . 200)
                  (250 . 0) (450 . 0) (0 . -50) (200 . -50) (200 . -250)
                  (250 . -50) (250 . -250) (250 . -50) (450 . -50))))

    (setq i 1)
    (while (< i (length points))
      (if (eq i 1)
          (progn
            (let ((p (nthcdr 0 points))
                  (e (nthcdr 1 points)))
              (docladr p e))
          )
        (if (member i '(2 3 4 5 6 7 8 9 10 11 12 13 14))
            (progn
              (let ((p (nthcdr 0 points))
                    (e (nthcdr 1 points)))
                (docladr p e))))
      )
      (setq i (+ i 1)))

    (princ "\n"))

(defun docladr (l)
  (if (< (length l) 2)
      (error "List must have at least two elements")
    (let ((p (car l))
          (e (cadr l)))
      (if (and (listp p) (listp e) (eq (length p) 2) (eq (length e) 2))
          (line p e)
        (error "Points must have x and y coordinates")))))

(defun *point-definition* (x y)
  (list x y))

(defun point1 () (*point-definition* 0 0))
(defun point2 () (*point-definition* 200 0))
(defun point3 () (*point-definition* 200 200))
(defun point4 () (*point-definition* 250 0))
(defun point5 () (*point-definition* 250 200))
(defun point6 () (*point-definition* 250 0))
```



```
(defun point7 () (*point-definition* 450 0))
(defun point8 () (*point-definition* 0 -50))
(defun point9 () (*point-definition* 200 -50))
(defun point10 () (*point-definition* 200 -250))
(defun point11 () (*point-definition* 250 -50))
(defun point12 () (*point-definition* 250 -250))
(defun point13 () (*point-definition* 250 -50))
(defun point14 () (*point-definition* 450 -50))
```

 REVIEWER:

• The function `c:drawlines` does not handle the case where `i` is equal to `(length points)` in its while loop condition.

• The `docladr` function does not check if the input list has an odd number of elements, which could lead to an error when trying to access the second element with `(nthcdr 1 points)`.

• The `point-definition*` function is not used consistently throughout the code. Some functions (e.g., `point2`) use it directly, while others (e.g., `point3`) define their own point coordinates.

• There are multiple definitions of `point6`, which could lead to confusion and unexpected behavior.

• The `line` function is not defined in this code snippet, but it's used in the `docladr` function. It should be either defined or imported from another file.

• The `princ "\n"` statement at the end of `c:drawlines` does not seem to serve any purpose and could be removed.

• There are no error checks for cases where the input list `points` is empty or contains non-list elements.

 CODER:

```
(defun c:drawlines ()
  (let ((i 1)
        (points '((0 . 0) (200 . 0) (200 . 200) (250 . 0) (250 . 200)
                  (250 . 0) (450 . 0) (0 . -50) (200 . -50) (200 . -250)
                  (250 . -50) (250 . -250) (250 . -50) (450 . -50))))

    (setq i 1)
    (while (< i (length points))
      (if (eq i (length points))
          (progn
            (let ((p (nthcdr (- i 1) points))
                  (e (nthcdr i points)))
              (docladr p e))
          )
        (if (member i '(2 3 4 5 6 7 8 9 10 11 12 13 14))
            (progn
              (let ((p (nthcdr (- i 1) points))
                    (e (nthcdr i points)))
                (docladr p e))))
      )
      (setq i (+ i 1)))
```



```
(defun docladr (points)
  (if (< (length points) 2)
      (error "List must have at least two elements")
    (let ((p (car points))
          (e (cadr points)))
      (if (and (listp p) (listp e) (eq (length p) 2) (eq (length e) 2))
          (line p e)
        (error "Points must have x and y coordinates")))))

(defun point-definition* (x y)
  (list x y))

(defun line (p1 p2)
  (format t "~A ~A~%" p1 p2))

(defun point1 () (*point-definition* 0 0))
(defun point2 () (*point-definition* 200 0))
(defun point3 () (*point-definition* 200 200))
(defun point4 () (*point-definition* 250 0))
(defun point5 () (*point-definition* 250 200))
(defun point6 () (*point-definition* 250 0)) ; removed duplicate definition
(defun point7 () (*point-definition* 450 0))
(defun point8 () (*point-definition* 0 -50))
(defun point9 () (*point-definition* 200 -50))
(defun point10 () (*point-definition* 200 -250))
(defun point11 () (*point-definition* 250 -50))
(defun point12 () (*point-definition* 250 -250))
(defun point13 () (*point-definition* 250 -50)) ; removed duplicate definition
(defun point14 () (*point-definition* 450 -50))

(defun line (p1 p2)
  (format t "~A ~A~%" p1 p2))
```
REVIEWER:
• The `line` function is defined twice, which will cause a compilation error. The second definition should be removed.
• The `docladr` function does not handle the case where the input list has more than two elements and the current index `i` is not one of the specified values (2-14). It should either skip these points or raise an error.
• The `point6` and `point13` functions are identical to `point4` and `point11`, respectively. They can be removed to avoid code duplication.
• There is no check for duplicate points in the input list, which could lead to incorrect drawing of lines.
• The `c:drawlines` function does not handle any errors that might occur during execution. It would be good to add some error handling code to make it more robust.
• The variable `i` is set to 1 at the beginning of the `c:drawlines` function, but then immediately set to 1 again in the first iteration of the while loop. This is unnecessary and can be removed.



• The `docladr` function does not check if the points are actually lists with two elements before trying to access their car and cadr. It should add a check for this case.
• There is no documentation or comments in the code to explain what it does, how it works, or any assumptions it makes about the input data. This would make it harder for others to understand and maintain the code.

CODER:

```lisp
(defun c:drawlines ()
  "Draws lines between consecutive points in the input list.
   Points are assumed to be lists of two elements, representing x and y coordinates.

   The function iterates over the input list, drawing a line for each pair of adjacent points.
   If a point is not followed by another point (i.e., it's the last point in the list),
   the function draws a line from that point to the origin (0, 0).

   The function raises an error if the input list has less than two elements or if any point
   does not have exactly two coordinates.

   It also skips drawing lines for points with indices 2-14, as per the problem statement."

  (let ((points '((0 . 0) (200 . 0) (200 . 200) (250 . 0) (250 . 200)
                  (250 . 0) (450 . 0) (0 . -50) (200 . -50) (200 . -250)
                  (250 . -50) (250 . -250) (250 . -50) (450 . -50))))
    (setq i 1)
    (while (< i (length points))
      (if (eq i (length points))
          (progn
            (let ((p (nthcdr (- i 1) points))
                  (e (point-definition* 0 0))) ; draw line to origin for last point
              (docladr p e))
          )
        (cond ((member i '(2 3 4 5 6 7 8 9 10 11 12 13 14))
               nil) ; skip drawing lines for points with indices 2-14
              (t
               (let ((p (nthcdr (- i 1) points))
                     (e (nthcdr i points)))
                 (docladr p e)))))
      (setq i (+ i 1))))

(defun docladr (points)
  "Draws a line between two points.

   The function raises an error if the input list has less than two elements or if any point
   does not have exactly two coordinates."

  (if (< (length points) 2)
```



```
     (error "List must have at least two elements")
  (let ((p (car points))
        (e (cadr points)))
    (cond ((and (listp p) (listp e) (eq (length p) 2) (eq (length e) 2))
           (line p e)
           (error "Points must have x and y coordinates")))))))

(defun point-definition* (x y)
  "Creates a list representing a point with x and y coordinates."

  (list x y))

(defun line (p1 p2)
  "Prints the points as a line.

   The function takes two points, represented as lists of two elements each,
   and prints them on the same line, separated by a space."

  (format t "~A ~A~%" p1 p2))

(defun point1 () (*point-definition* 0 0))
(defun point2 () (*point-definition* 200 0))
(defun point3 () (*point-definition* 200 200))
(defun point4 () (*point-definition* 250 0))
(defun point5 () (*point-definition* 250 200))
(defun point7 () (*point-definition* 450 0))
(defun point8 () (*point-definition* 0 -50))
(defun point9 () (*point-definition* 200 -50))
(defun point10 () (*point-definition* 200 -250))
(defun point11 () (*point-definition* 250 -50))
(defun point12 () (*point-definition* 250 -250))
(defun point14 () (*point-definition* 450 -50))
```
REVIEWER:
• The function `c:drawlines` does not check if the input list has exactly two elements, it checks if it's less than two. This might lead to incorrect behavior when drawing lines.
• The function `docladr` does not handle the case where a point is missing its y-coordinate but has an x-coordinate.
• The function `point-definition*` does not check if the input values are numbers.
• The function `line` does not handle the case where either of the points is not a list with two elements.
• There's no error handling in the `while` loop of `c:drawlines` for when `i` exceeds the length of the `points` list.
• The function `point1`, `point2`, etc. are not necessary and can be replaced with direct calls to `point-definition*`.



• The function `point-definition*` is not used consistently throughout the code, instead of calling it directly, some functions like `point1` are created which call it indirectly.
• There's no documentation for the `line` function in the context of AutoLisp.
 CODER:
```lisp
(defun c:drawlines ()
  "Draws lines between consecutive points in the input list.
   Points are assumed to be lists of two elements, representing x and y coordinates.

   The function iterates over the input list, drawing a line for each pair of adjacent points.
   If a point is not followed by another point (i.e., it's the last point in the list),
   the function draws a line from that point to the origin (0, 0).

   The function raises an error if the input list has less than two elements or if any point
   does not have exactly two coordinates.

   It also skips drawing lines for points with indices 2-14, as per the problem statement."

  (let ((points '((0 . 0) (200 . 0) (200 . 200) (250 . 0) (250 . 200)
                  (250 . 0) (450 . 0) (0 . -50) (200 . -50) (200 . -250)
                  (250 . -50) (250 . -250) (250 . -50) (450 . -50))))
    (if (< (length points) 2)
        (error "Input list must have at least two elements")
      (let ((i 1))
        (while (< i (length points))
          (cond ((member i '(2 3 4 5 6 7 8 9 10 11 12 13 14))
                 nil) ; skip drawing lines for points with indices 2-14
                (t
                 (let ((p (nthcdr (- i 1) points))
                       (e (if (< (+ i 1) (length points))
                              (nthcdr (+ i 1) points)
                            (point-definition* 0 0)))) ; draw line to origin for last point
                   (docladr p e)))
          (setq i (+ i 1)))))))

(defun docladr (points)
  "Draws a line between two points.

   The function raises an error if the input list has less than two elements or if any point
   does not have exactly two coordinates."

  (if (< (length points) 2)
      (error "List must have at least two elements")
    (let ((p (car points))
          (e (cadr points)))
      (cond ((and (listp p) (listp e) (eq (length p) 2) (eq (length e) 2))
```



```lisp
        (line (point-definition* (car p) (cadr p))
              (point-definition* (car e) (cadr e)))
        (t
         (error "Points must have x and y coordinates")))))

(defun point-definition* (x &optional y)
  "Creates a list representing a point with x and y coordinates.

  If only one argument is provided, it's assumed to be the x-coordinate,
  and the y-coordinate defaults to 0."

  (if y
      (list x y)
    (error "Point must have either two or one coordinate")))

(defun line (p1 p2)
  "Prints the points as a line.

  The function takes two points, represented as lists of two elements each,
  and prints them on the same line, separated by a space."

  (format t "~A ~A~%" p1 p2))
```
 REVIEWER:
* The `point-definition*` function does not handle the case where only one argument is provided
when called from `docladr`.
* In the `drawlines` function, the line `(cond ((member i '(2 3 4 5 6 7 8 9 10 11 12 13 14)) nil)`
should be `(cond ((or (member i '(2 3 4 5 6 7 8 9 10 11 12 13 14)) t) nil` to correctly skip
drawing lines for points with indices 2-14.
* The `point-definition*` function does not handle the case where two arguments are provided
but one of them is not a number.
* In the `drawlines` function, the line `(if (< (+ i 1) (length points))` should be `(if (> (+ i 1)
(length points))` to correctly draw a line from the last point to the origin when it's the last point
in the list.
* The `docladr` function does not handle the case where the input list has more than two
elements but one of them is not a list.
* There are no checks for potential errors in the `line` function.
 CODER:
```lisp
(defun c:drawlines ()
  "Draws lines between consecutive points in the input list.
  Points are assumed to be lists of two elements, representing x and y coordinates.

  The function iterates over the input list, drawing a line for each pair of adjacent points.
  If a point is not followed by another point (i.e., it's the last point in the list),
  the function draws a line from that point to the origin (0, 0).
```



The function raises an error if the input list has less than two elements or if any point does not have exactly two coordinates.

It also skips drawing lines for points with indices 2-14, as per the problem statement."

```
(let ((points '((0 . 0) (200 . 0) (200 . 200) (250 . 0) (250 . 200)
                (250 . 0) (450 . 0) (0 . -50) (200 . -50) (200 . -250)
                (250 . -50) (250 . -250) (250 . -50) (450 . -50))))
  (if (< (length points) 2)
      (error "Input list must have at least two elements")
    (let ((i 1))
      (while (< i (length points))
        (cond ((or (member i '(2 3 4 5 6 7 8 9 10 11 12 13 14)) t) nil)
              ; skip drawing lines for points with indices 2-14
              (t
               (let ((p (nthcdr (- i 1) points))
                     (e (if (> (+ i 1) (length points))
                            (point-definition* 0 0)
                          (point-definition* (car (nthcdr (+ i 1) points))))))
                 (docladr p e)))
              (setq i (+ i 1)))))))

(defun docladr (points)
  "Draws a line between two points.

The function raises an error if the input list has less than three elements or if any point
does not have exactly two coordinates."

  (if (< (length points) 3)
      (error "List must have at least three elements")
    (let ((p1 (car points))
          (e (cadr points)))
      (cond ((and (listp p1) (listp e) (eq (length p1) 2) (eq (length e) 2))
             (line (point-definition* (car p1) (cadr p1))
                   (point-definition* (car e) (cadr e))))
            (t
             (error "Points must have x and y coordinates")))
      (let ((p2 (nthcdr 2 points)))
        (cond ((and (listp p2) (eq (length p2) 2))
               (line (point-definition* (car p2) (cadr p2)) nil)
               t)
              (t
               (error "Points must have x and y coordinates")))))))

(defun point-definition* (x &optional y)
```

```
"Creates a list representing a point with x and y coordinates.

  If only one argument is provided, it's assumed to be the x-coordinate,
  and the y-coordinate defaults to 0."

 (if (and y (numberp x) (numberp y))
    (list x y)
   (if y
      (error "Point must have either two or one coordinate")
     (list x))))

(defun line (p1 p2)
  "Prints the points as a line.

  The function takes two points, represented as lists of two elements each,
  and prints them on the same line, separated by a space."

 (if (and (listp p1) (listp p2) (eq (length p1) 2) (eq (length p2) 2))
    (format t "~A ~A~%" p1 p2)
   (error "Invalid input: points must be lists of two elements")))
```
 REVIEWER:
* The function `c:drawlines` does not check if the input list has exactly two or more than two
elements, it only checks for less than two. It should also check for more than two.
* In the `docladr` function, the condition `(or (member i '(2 3 4 5 6 7 8 9 10 11 12 13 14)) t)` is
not necessary and can be removed as it's already handled by the while loop condition.
* The `point-definition*` function does not handle the case where only one argument is provided,
but the y-coordinate is not a number. It should raise an error in this case.
* In the `line` function, the error message "Invalid input: points must be lists of two elements"
can be improved to provide more context.
* The `c:drawlines` function does not handle the case where the input list contains non-list
elements or elements with less than two coordinates. It should raise an error in these cases.
* The `docladr` function does not check if the points are equal, it will draw a line from each
point to itself. It should skip this case.
* There is no docstring for the `c:drawlines` function that describes its purpose and behavior.